\documentclass{aastex63}
\usepackage{multirow}
\usepackage{epsfig}
\usepackage{enumitem}

\begin{document}

\title{Constrain magnetar parameters by taking into account the evolutionary effects of radius and moment of inertia with \emph{Swift}/XRT data}

\correspondingauthor{He Gao}
\email{gaohe@bnu.edu.cn}

\correspondingauthor{Li-Ping Xin}
\email{xlp@nao.cas.cn}

\author{Lin Lan}
\affiliation{CAS Key Laboratory of Space Astronomy and Technology, National Astronomical Observatories, Chinese Academy of Sciences, Beijing 100012, People's Republic of China}

\author{He Gao}
\affiliation{Institute for Frontier in Astronomy and Astrophysics, Beijing Normal University, Beijing 102206, People’s Republic of China; gaohe@bnu.edu.cn}
\affiliation{Department of Astronomy, Beijing Normal University, Beijing 100875, People's Republic of China}

\author{Shunke Ai}
\affiliation{Niels Bohr International Academy and DARK, Niels Bohr Institute, University of Copenhagen, Blegdamsvej 17, 2100, Copenhagen, Denmark}

\author{Wen-Jin Xie}
\affiliation{CAS Key Laboratory of Space Astronomy and Technology, National Astronomical Observatories, Chinese Academy of Sciences, Beijing 100012, People's Republic of China}

\author{Yong Yuan}
\affiliation{Center for Gravitational Wave Experiment, National Microgravity Laboratory, Institute of Mechanics, Chinese Academy of Sciences, Beĳing, China}

\author{Long Li}
\affiliation{Department of Physics, School of Physics and Materials Science, Nanchang University, Nanchang
330031, China}

\author{Li-Ping Xin}
\affiliation{CAS Key Laboratory of Space Astronomy and Technology, National Astronomical Observatories, Chinese Academy of Sciences, Beijing 100012, People's Republic of China}

\author{Jian-Yan Wei}
\affiliation{CAS Key Laboratory of Space Astronomy and Technology, National Astronomical Observatories, Chinese Academy of Sciences, Beijing 100012, People's Republic of China}

\begin{abstract}

A newly born millisecond magnetar has been proposed as one possible central engine of some gamma-ray bursts (GRBs) with X-ray plateau emission. In this work, we systematically analyzed the \emph{Swift}/X-Ray Telescope data of long GRBs with plateau emission that were detected before 2023 December, and estimated the physical parameters of the magnetar by considering the $R/I$ evolutionary effects, and investigated possible relationships among these parameters and their relation to the GRB jet and magnetar wind radiation. We found that neglecting the evolutionary effects of $R/I$ can lead to systematic overestimation or underestimation of magnetar physical parameters such as magnetic field strength ($B_p$), spin period ($P_0$), and ellipticity ($\epsilon$) from 20\% to 50\%. We also found that the correlations among different magnetar parameters can be approximately expressed as $\epsilon\propto P_0^{1.57\pm0.22}$, $\epsilon\propto B_p^{0.97\pm0.13}$, and $B_p\propto P_0^{1.30\pm0.16}$ for our selected equation of states (EoSs), with the 1$\sigma$ deviation included. The correlations between the GRB jet emission and the magnetar wind emission can be approximately described as $E_{\rm wind}\propto E_{\rm jet,iso}^{0.83\pm0.07}(E_{\rm jet}^{0.76\pm0.06})$, $P_0\propto E_{\rm jet,iso}^{-0.29\pm0.03}(E_{\rm jet}^{-0.26\pm0.02})$, $B_p\propto E_{\rm jet,iso}^{-0.58\pm0.06}(E_{\rm jet}^{-0.55\pm0.05})$, and $\epsilon\propto E_{\rm jet,iso}^{-0.55\pm0.07}(E_{\rm jet}^{-0.52\pm0.06})$ for our selected EoSs. The universal correlations suggest that a nascent magnetar with the faster $P_0$, lower $B_p$, as well as lower $\epsilon$ are more inclined to power a more energetic GRB jet, and the ellipticity deformation and initial spin period of newborn magnetar are likely to originate from the magnetically induced distortion mechanism and correspond to the equilibrium spin period as a result of interaction between the magnetar and its accretion disk, respectively. Finally, we found that the gravitational-wave (GW) signals from the remnants of those GW-dominated GRBs with redshift measurements cannot reach the sensitivity threshold of the current aLIGO detector, and only two cases (GRBs 150323A and 170607A) can reach the sensitivity threshold of the prospective ET detector. Future GW observations could not only offer the first smoking gun that a protomagnetar can serve as the central engine of GRBs but also play a crucial role in precisely constraining the neutron star EoS. 
\end{abstract}
\keywords{gamma-ray burst: general - magnetars}

\section{Introdution}
After close to 50 years of extensive research, the gamma-ray bursts (GRBs) have made great progress in both observations and theories \citep{Zhang2018}. However, there still exist some critical questions, such as about the GRB's central engine. Two competing types of GRB central engine candidate have been proposed based on the observations of prompt emission and early afterglow: (1) a hyperaccreting stellar-mass black hole (BH) via neutrino--antineutrino annihilation \citep{Ruffert1997,Popham1999,Chen2007,Lei2009,Lei2013,Liu2017} or Blandford--Znajek mechanism to launch the relativistic jet \citep{Blandford1977,Lee2000,Li2000}; (2) a rapidly spinning, strongly magnetized neutron star (NS, also called millisecond magnetar) via losing its rotational energy to power relativistic jet \citep{Usov1992,Thompson1994,Dai1998a,Dai1998b,Zhang2001,Metzger2008,Metzger2011,Bucciantini2012}.

Observationally, thanks to the \emph{Swift}/X-Ray Telescope (XRT) that has made abundant observations of GRBs and collected a large sample of X-ray afterglow data \citep{Gehrels2004}, a good fraction of the X-ray light curves of (long and short) GRBs were discovered to show a long-lasting plateau emission feature or a period of softer extended emission \citep[EE;][]{Norris2006,Norris2010}, which seems to be consistent with the prediction of a millisecond magnetar engine \citep{Zhang2001,Metzger2011,Metzger2014,Gompertz2013,Gompertz2014,Rowlinson2013,Rowlinson2014,Lv2014,Lv2015}. The electromagnetic (EM) dipole emission of newborn magnetar generates Poynting flux that can efficiently undergo magnetic energy dissipation processes with high efficiency \citep{Zhang2011} and inject additional energy into the GRB to induce X-ray plateaus and EE features. In particular, some GRBs exhibit an internal plateau, a nearly flat X-ray emission extending to hundreds of seconds, followed by an abrupt decay $t^{-(>3)}$ \citep{Troja2007,Rowlinson2010,Rowlinson2013,Lv2015}. Such a feature is very difficult to explain if it is powered by a BH engine, but seems to be consistent with the prediction of a millisecond magnetar. The steep decay following the X-ray plateau is usually interpreted as the supramassive NS collapsing into a BH after it spins down due to EM dipole radiation and gravitational-wave (GW) emission \citep{Usov1992,Thompson1994,Dai1998a,Dai1998b,Zhang2001,Dai2006,Gao2006,Troja2007,Metzger2008,Fan2013a,Fan2013b,Zhang2013,Zhang2014a,Ravi2014,Lv2015,Lv2017,Lv2020,Gao2016,Gao2017,Chen2017,Lan2020}.

The energy reservoir of a newly born millisecond magnetar is the total rotational energy, which could be estimated as
\begin{eqnarray}
E_{\rm rot} = \frac{1}{2} I \Omega^{2}
\label{Erot}
\end{eqnarray}
where $I$ is the moment of inertia, and $\Omega=2\pi/P$ is the angular frequency of the nascent NS. In principle, during the spin-down process, the nascent magnetar loses its rotational energy via both EM dipole radiation and GW emission, so that the spin-down law can be written as \citep{Shapiro1983,Zhang2001}
\begin{eqnarray}
\dot{E}&=-L_{\rm dip}-L_{\rm GW}&=-\frac{\eta_{\rm X} B_p^2R^6\Omega^4}{6c^3}-\frac{32GI^2\epsilon^2\Omega^6}{5c^5}.
\label{dotE}
\end{eqnarray}
where $B_p$ is the dipolar field strength at the magnetic poles on the NS surface. $R$ and $\epsilon=2(I_{\rm xx}-I_{\rm yy})/(I_{\rm xx}+I_{\rm yy})$ are the NS radius and ellipticity in terms of the principal moment of inertia, respectively. $\eta_{\rm X}$ is the radiation efficiency of converting the dipole spin-down luminosity of magnetar wind energy into X-ray radiation. Depending on the NS properties, such as the values of $B_p$ and $\epsilon$, the spin-down process will be dominated by one or the other loss term.

In principle, we can learn about the physical properties of nascent magnetars through EM observations of GRBs X-ray afterglows. Traditionally, one can estimate the dipolar magnetic field strength $B_p$, initial spin period $P_0$, ellipticity $\epsilon$, and braking index $n$ of the magnetar from the observed X-ray plateau luminosity and its ending time \citep{Troja2007,Rowlinson2010,Rowlinson2013,Fan2013a,Fan2013b,Gompertz2013,Gompertz2014,Lv2014,Lv2015,Dallosso2015,Gao2016,Gao2017,Lasky2016,Lasky2017,Lv2019,Zou2019,Zou2021a,Sarin2020a,Sarin2020b,Xie2022a,Xie2022b}. On the other hand, one can also infer the dominant energy loss mechanism (EM or GW radiation) during the spin-down of the nascent magnetar by analyzing the afterglow slope after X-ray plateau emission, which EM luminosity evolves as $L_{\rm dip}\propto t^{-\alpha}$, and $\alpha$ is -1 and -2 in the GW and EM radiation dominated scenarios, respectively. The transition from the GW emission dominated phase to EM emission dominated phase is shown as a smooth break, and the decay slope of EM luminosity changes from -1 to -2 \citep{Zhang2001,Lasky2016,Lv2018,Lv2020}.

In previous studies, the $R$ and $I$ of newborn magnetars throughout the spin-down process are usually assumed to be two fiducial values, such as $R=10^6~\rm cm$ and $I=1.5\times10^{45}~\rm g~cm^2$. However, the radius and moment of inertia should be related to the rotational speed for a rapidly rotating protomagnetar \citep{Lattimer2012,Gao2020}. This is due to the fact that the gravitational mass of a rapidly rotating NS is supported by both the neutron degeneracy pressure and the rotating centrifugal force. As the magnetar spins down, the gravitational mass supported by the rotating centrifugal force will become smaller, which requires a larger degeneracy pressure to support more gravitational mass. Thus, the nascent magnetar will be squeezed resulting in a smaller radius, and the rotational moment of inertia associated with the radius will also change. Distinctly, the specific values of $R$ and $I$ should also depend on the NS equation of states (EoSs) and the NS mass.

Due to the scale relations, $L_{\rm dip}\propto R^6$ for EM magnetic dipole radiation and $L_{\rm GW}\propto I^2$ for GW radiation. One interesting question is whether the evolutionary effects of $R$ and $I$ of a newborn magnetar have a significant impact on the observational EM luminosity. \cite{Lan2021} studied in detail how $R$ and $I$ evolve as the magnetar spins down, and how these evolution effects alter the magnetic dipole radiation behavior. They found that the $L_{\rm dip}$ could be variant within 1 to 2 orders of magnitude in different EoSs and baryonic masses, and the temporal behavior of $L_{\rm dip}$ became more complicated due to the evolution of $R$ and $I$. Ultimately, they suggested that, when using the X-ray plateau data of GRBs to diagnose the physical nature of GRB magnetars, EoS and NS mass information should be invoked as simultaneously constrained parameters. The precise constraints on these parameters by utilizing evolved $R$ and $I$ will be crucial to understand the physical nature of GRB magnetars and to confirm the connections between their physical parameters.

In this paper, by considering the evolutionary effects of $R$ and $I$ during the magnetar spin-down process, we performed a systematic study of long GRBs with X-ray plateau emission, whose central engine may be a millisecond magnetar, and derived the physical parameters of the newborn magnetar by using the X-ray plateau data and found some tight correlations between these parameters. Particularly, we focus on analyzing the possible relationships among these parameters and their relation to the GRB jet and magnetar wind radiation, which is crucial to help us to understand the mechanisms that induce the NS nonasymmetries and the energy partition between jet prompt emission and magnetar wind afterglow. Furthermore, we also discussed the magnetar dipole radiation detectability with \emph{EP} and \emph{SVOM} based on our LGRB sample, and evaluated the GW strain detection range for these GW-dominated GRBs in our sample. Throughout the paper, a concordance cosmology (flat $\Lambda$CDM) with parameters $H_0 = 67.4$ km s$^{-1}$ Mpc $^{-1}$, $\Omega_M=0.315$, and $\Omega_{\Lambda}=0.685$ has been adopted according to the {\it Planck} results \citep{Planck2020}.

\section{sample selection and data reduction}
With the successful launch of the \emph{Swift} satellite, the early time afterglow emissions of GRBs have been revealed, and a rich trove of X-ray observation data have been collected. \cite{Zhang2006} presented a canonical X-ray lightcurve based on the observational data from the \emph{Swift}/XRT, which includes a distinct steeply decaying component preceding the conventional afterglow component in many GRBs, a shallow decay component before the more normal decay component observed in a good fraction of GRBs, an X-ray flare component, and a jet break component in some GRBs. Theoretically, the steep decay component is consistent with the tail emission of jet prompt emission, and the shallow decay component is likely caused by a long-lived wind energy injection of a magnetar central engine. Traditionally, the magnetar wind energy injection signature typically exhibits a shallow decay segment (or plateau) followed by a steeper decay segment (a normal decay for canonical light curves or an abrupt decay for internal plateaus) in the X-ray afterglow when it is spinning down by losing rotational energy via EM and GW radiations. In view of such a characteristic, we performed a systematic search of long GRBs with plateau emission using all of the available \emph{Swift} GRB catalog, of which entire data include more than 1500 GRBs observed between 2005 January and 2023 November. The XRT data of which were downloaded from the UK Swift Science Data Center \footnote{http://www.swift.ac.uk/burst\_analyser/} \citep{Evans2010}. Our attention is on those LGRBs that exhibit such a transition in the XRT light curves. In order to accurately pick LGRBs with magnetar-fed energy from the available \emph{Swift} GRB catalog and precisely quantify the jet emission duration and magnetar wind emission duration, we performed an empirical fit to the XRT light curves with a smooth triple power-law function,
\begin{equation}
\label{TBPL}
F=(F_1^{-\omega_2}+F_2^{-\omega_2})^{-1/\omega_2},
\end{equation}
where $F_1$ and $F_2$ can be expressed as
\begin{equation}
F_1=F_{0}\left[\left(\frac{t}{t_{\rm
b,1}}\right)^{\omega_1\alpha_1}+\left(\frac{t}{t_{\rm
b,1}}\right)^{\omega_1\alpha_2}\right]^{-1/\omega_1},
\end{equation}
\begin{equation}\label{PL}
F_2=F_1(t_{b,2})\left(\frac{t}{t_{b,2}}\right)^{-\alpha_3},
\end{equation}
where $t_{b,1}$ and $t_{b,2}$ are the two break times, 
which are used to quantitatively describe the jet emission time and the magnetar emission time, respectively. The $\alpha_1$, $\alpha_2$, and $\alpha_3$ are the decay indices before and after $t_{b,1}$, and after $t_{b,2}$, respectively. $\omega_1$ and $\omega_2$ describe the sharpness of the break at $t_{b,1}$ and $t_{b,2}$, and are taken as $\omega_1=\omega_2=3$ in our fits \citep{Liang2007}. Three criteria are adopted for our sample selection:  
\begin{itemize}
    \item The shallow segment slope should be less than 0.3, which could ensure the generalization of the magnetar model assumption;
    \item The slope of the steeper decay segment following the plateau segment should be in the range of -1 to -2, which is the typical decay slope when GW and EM radiations dominated magnetar spin-down model, respectively \citep{Zhang2001,Lasky2016,Lv2018};
    \item The GRBs with bright X-ray flares \footnote{Bright X-ray flares are defined as $F_p/F_u>5$, where$F_p$ and $F_u$ are the flux at the peak time and the corresponding underlying flux, respectively \citep{Hu2014,Zhang2014b}} emerging during the shallow or steeper decay phase are excluded from our sample. These flares are proposed to come from later activity of the central engine \citep{Zhang2007,Peng2014}.
\end{itemize}

\begin{figure*}
\centering
\includegraphics  [angle=0,scale=0.7]   {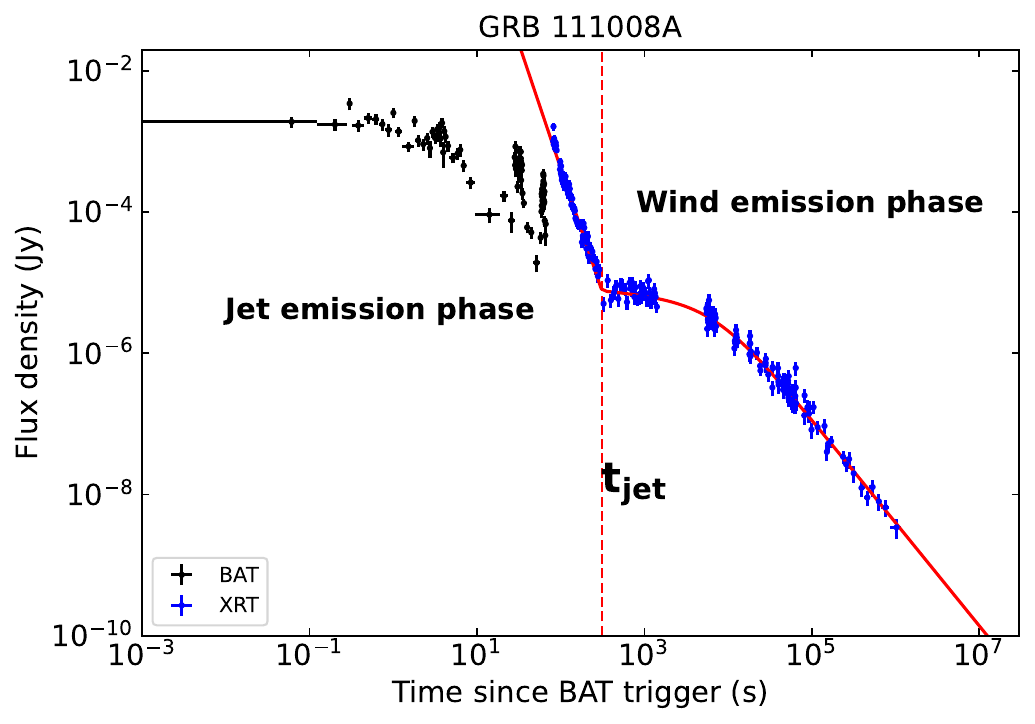}
\caption{Joint BAT+XRT light curve of GRB 111008A. The black dots are BAT data extrapolated to the XRT band (0.3–10 keV), and the blue dots are XRT data. The solid red line is the fit with a smooth triple power-law function, and the vertical dashed line is the separation of the jet emission phase and magnetar wind emission phase.}
\label{fig:111008A_LC}
\end{figure*}

\begin{figure}
\includegraphics[angle=0,scale=0.45]{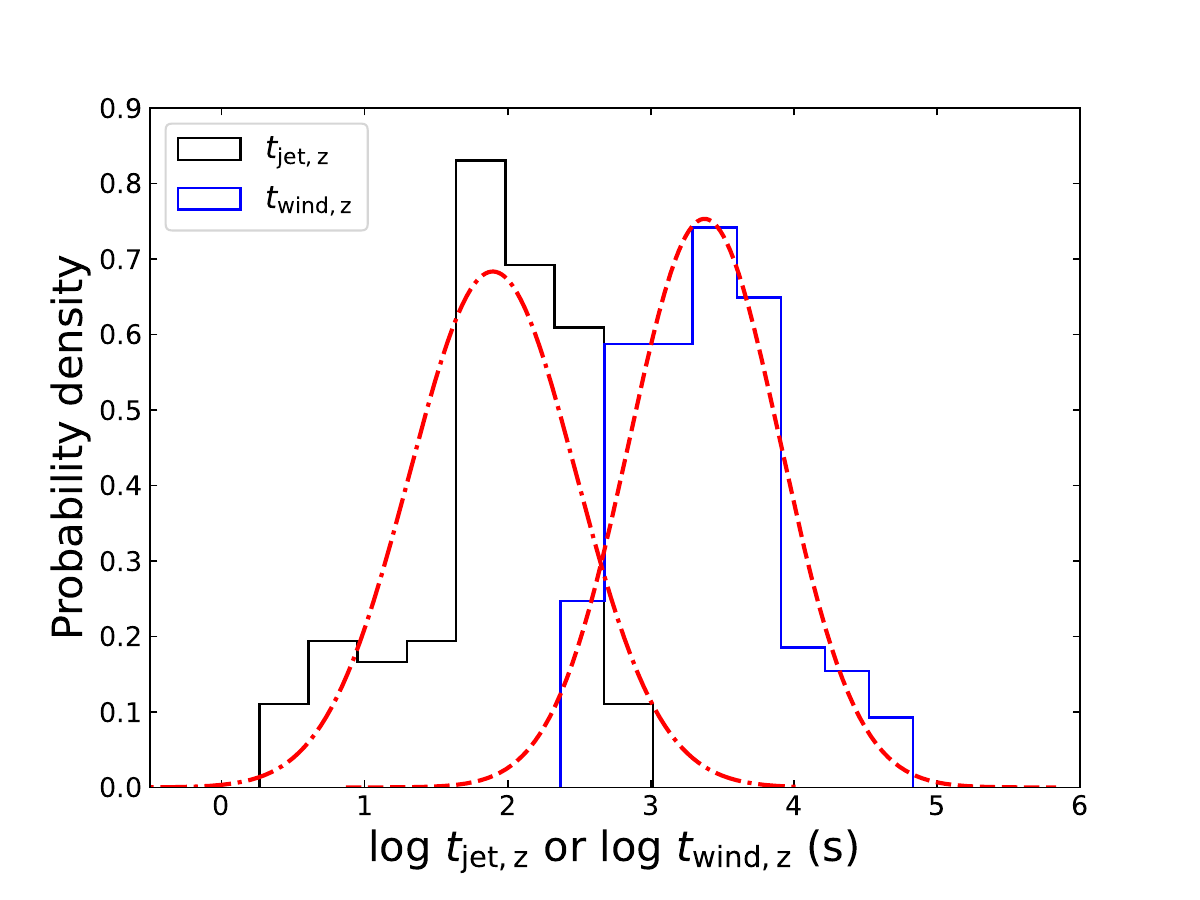}
\includegraphics[angle=0,scale=0.45]{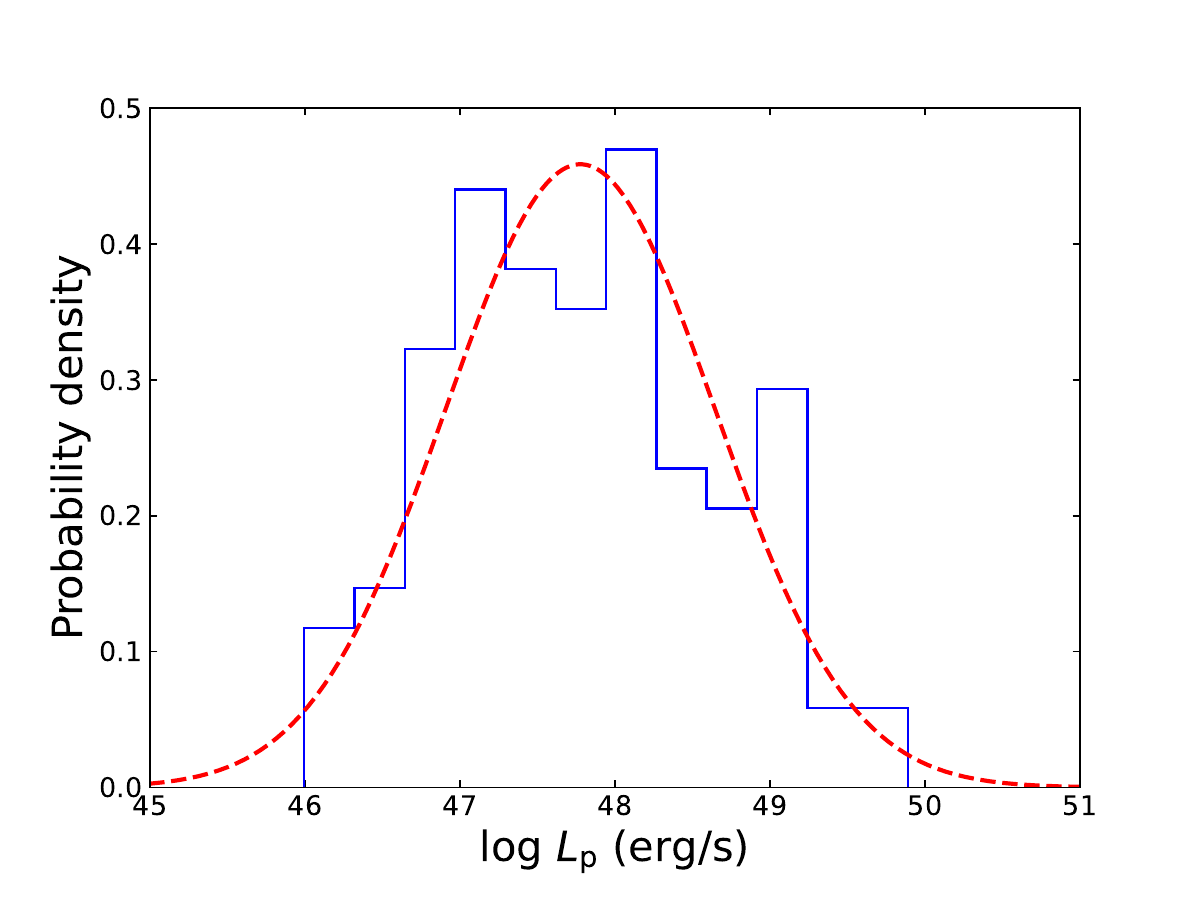}
\caption{The distributions of jet emission duration ($t_{\rm jet,z}$) and the magnetar wind emission duration ($t_{\rm wind,z}$) in the rest frame (left panel), as well as the distribution of the plateau luminosity $L_{\rm p}$ (right panel). The red dashed lines are the best Gaussian fits. We adopt $z = 1$ for the LGRBs without redshift measurement in our sample.}
\label{fig:Tb&Lp}
\end{figure}

There are 105 long GRBs that satisfy our criteria up to 2023 December, 43 of which have a measured redshift. We extensively searched for the redshift information (both spectroscopic and photometric) and prompt emission duration $T_{90}$ for each GRB in our sample from published papers or the Gamma-ray Coordinates Network Circulars\footnote{\url{https://gcn.gsfc.nasa.gov/gcn3_archive.html}} if no published paper is available. It is notable that the redshift measurement $z$ is vital to derive the intrinsic parameters (energy, luminosity, etc.), and we adopted $z=1$ for the GRBs without a redshift measurement in our sample. From an observational perspective, these GRBs can be classified into two distinct emission phases: One is dominated by prompt gamma-ray emission and early X-ray steep decay emission, and these fragments are supposed to originate from jet emission via internal shock or magnetic energy dissipation \citep{Meszaros1993,Zhang2006,Zhang2011}. Another is the long-term plateau emission phase, which may be dominated by the magnetar wind emission \citep{Zhang2001}. Based on the best-fit results of the smoothed triple power-law function, we can use $t_{\rm jet}$ and $t_{\rm wind}$ to quantitatively describe the duration of jet emission and magnetar wind emission, respectively. In Figure \ref{fig:111008A_LC}, we use GRB 111008A as an example to show the joint BAT+XRT light curve, which is divided into jet emission and magnetar wind emission segments by defining $t_{\rm jet}$ as the boundary. By dividing the jet emission and magnetar wind emission for each GRB in our sample, we performed k-correction and derived isotropic jet energies, magnetar wind energies, as well as X-ray plateau luminosities \citep{Bloom2001}, and parameters are calculated by $E_{\rm jet,iso}=4\pi D_{L}^2(k_1 \cdot S_{\gamma}+k_2 \cdot S_{\rm jet,X})/(1+z)$, $E_{\rm wind}=k_2 \cdot 4\pi D_{L}^2S_{\rm wind,X}/(1+z)$, $L_{\rm p}=k_2 \cdot 4\pi D_{L}^2F_{\rm wind}$, where $S_{\gamma}$ and $S_{\rm jet,X}$ are the prompt gamma-ray fluence and X-ray fluence before plateau emission, $S_{\rm wind,X}$ is the observed X-ray fluence of the magnetar wind (fluence of the plateau segment), $F_{\rm wind}$ is observed X-ray plateau flux, $D_L$ is the GRB luminosity distance, and the $k_1$ and $k_2$ are the k-correction factor that correct the BAT band (15-150 keV) and XRT band (0.3-10 keV) flux to a wide band in the burst rest frame (1-$10^4$ keV) and (0.1-$10^3$ keV), respectively, i.e.
\begin{equation}
k_1=\frac{\int^{10^4/{1+z}}_{1/{1+z}}EN_1(E)dE}{\int^{150}_{15}EN_1(E)dE}.
\end{equation}
\begin{equation}
k_2=\frac{\int^{10^3/{1+z}}_{0.1/{1+z}}EN_2(E)dE}{\int^{10}_{0.3}EN_2(E)dE}.
\end{equation}
Here, $N_1(E)$ and $N_2(E)$ are the observed time-dependent $\gamma$-ray photon spectrum and X-ray photon spectrum, respectively, which could be best fitted by a power-law model (spectral parameters could be obtained from the \emph{Swift} archive). The observational properties and derived results for the GRBs in our sample are summarized in Table \ref{table-1}. In Figure \ref{fig:Tb&Lp}, we showed the distributions of $t_{\rm jet}$ and $t_{\rm wind}$ in the burst frame ($t_{\rm jet,z}$ and $t_{\rm wind,z}$), as well as the $L_{\rm p}$ distribution. The log-normal distributions can be described as $\log~t_{\rm jet}/{\rm s}=1.90\pm0.58$, $\log~t_{\rm wind}/{\rm s}=3.38\pm0.53$, $\log~L_{\rm p}/{\rm erg s^{-1}}=47.78\pm0.87$, respectively. In Figure \ref{fig:Twind,jet-Lp}, we showed the scatter diagrams $L_{\rm p}-t_{\rm wind,z}$ and $L_{\rm p}-t_{\rm jet,z}$. We found that there is strong anticorrelation between $L_{\rm p}$ and $t_{\rm wind,z}$, and most of the GRBs in our sample fall into the $3\sigma$ deviation region of the best-fitting power-law model $\log L_{\rm p}=(53.21\pm0.46)-(1.59\pm0.14)\log t_{\rm wind,z}$. The anticorrelation result is consistent with the statistical results in \cite{Dainotti2008,Dainotti2013} and \cite{Lv2014}. No statistical correlation between $L_{\rm p}$ and $t_{\rm jet,z}$ can be claimed.

\begin{figure}
\includegraphics[angle=0,scale=0.45]{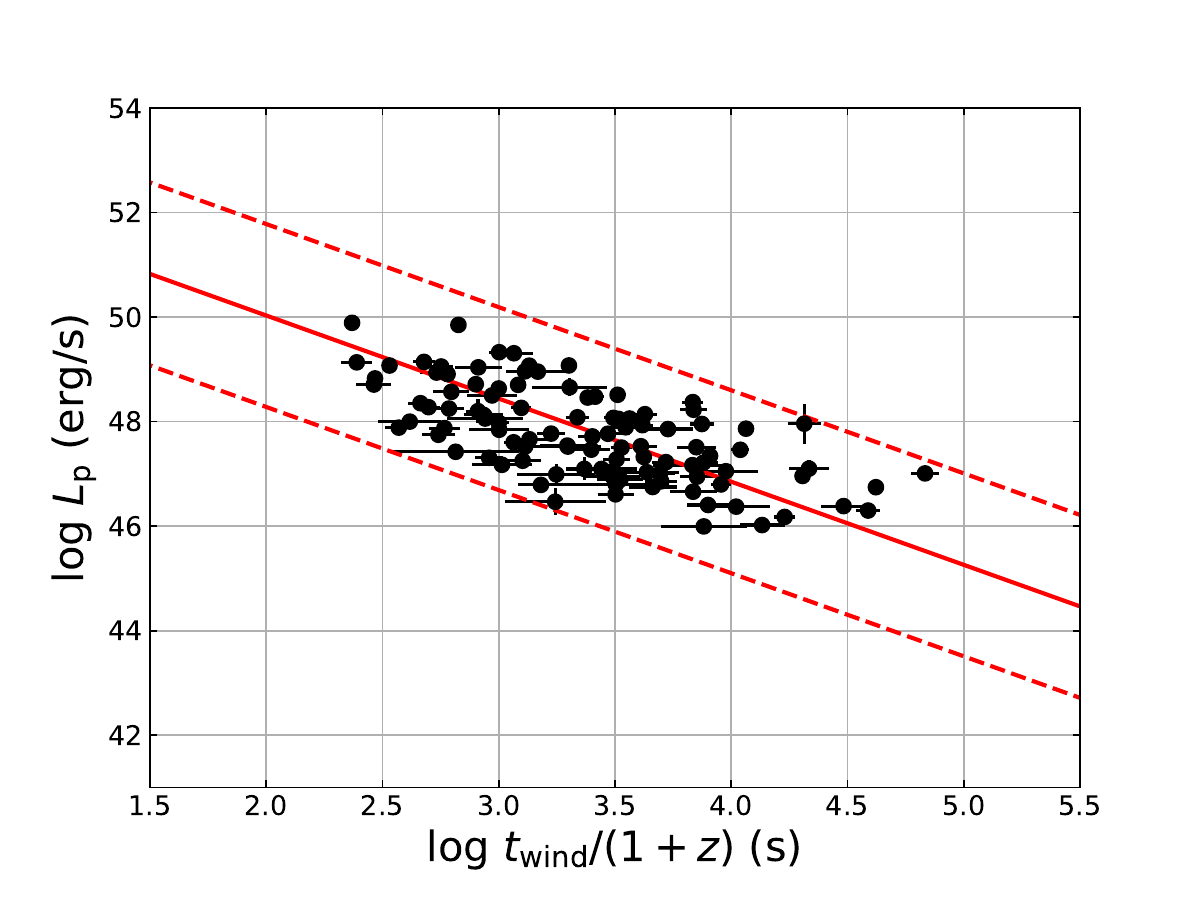}
\includegraphics[angle=0,scale=0.45]{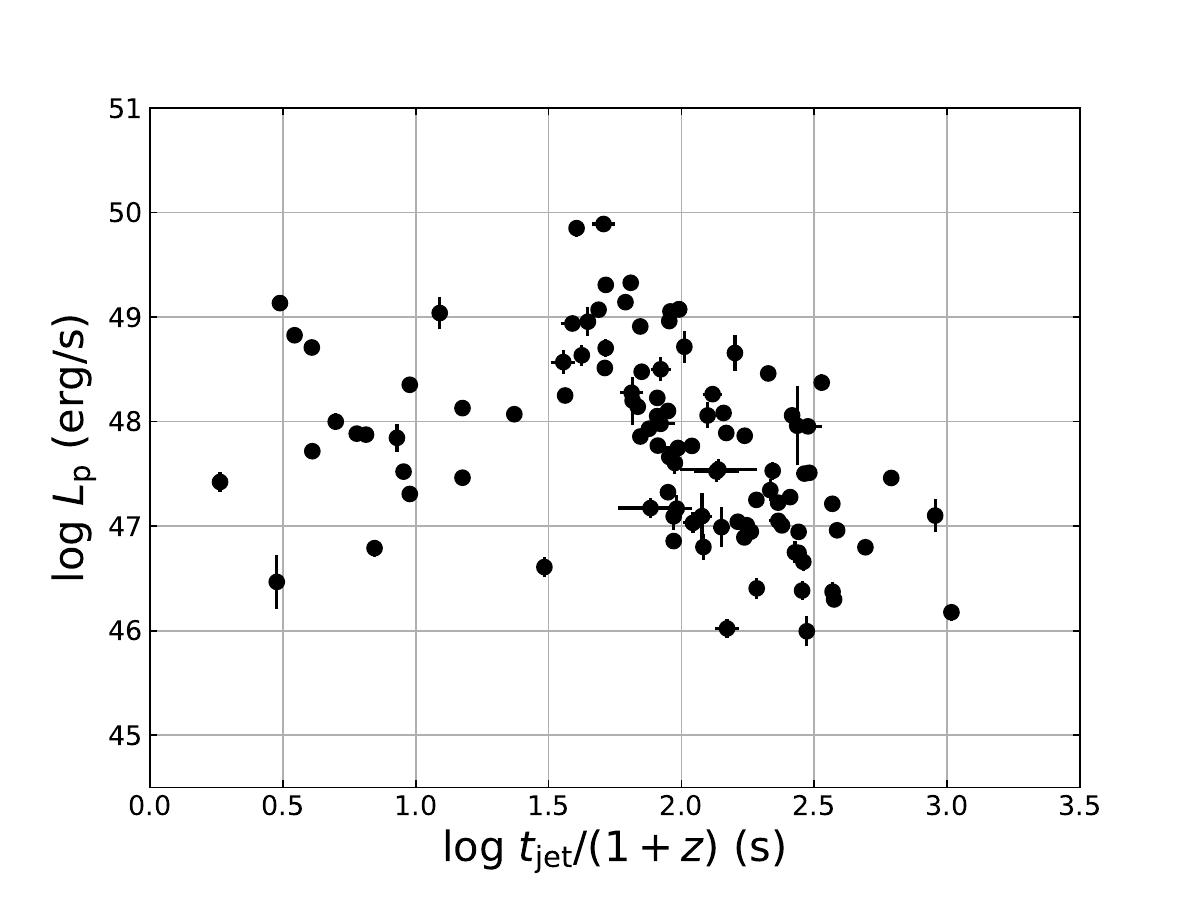}
\caption{$L_{\rm p}$ as a function of $t_b/(1+z)$ (left panel) or $t_{\rm jet}/(1+z)$ (right panel) for the LGRBs in our sample. The red solid and dashed lines are the least-squares linear fit and its 95\% confidence level, respectively.}
\label{fig:Twind,jet-Lp}
\end{figure}

\section{The spin-down characteristics of newborn magnetars with $R$ and $I$ evolution}
As the GRB's central engine, the magnetars are supposed to be rapidly rotating (spin period in the millisecond range, especially for short GRBs where the magnetars are formed from a merger), in which case its radius and moment of inertia should be related to the rotational speed \citep{Lattimer2012,Gao2020}. \cite{Lan2021} studied in detail how $R$ and $I$ evolve as the magnetar's rotational speed by using the numerical methods to treat the equilibrium equations for a stationary, axially symmetric, rigid rotating NS, within a fully general relativistic framework. In this case, the spacetime metric can be written as
\begin{eqnarray}
ds^2 &=& -e^{2\nu}\, dt^2 + r^2 \sin^2\theta B^2 e^{-2\nu}
(d\phi - \omega\, dt)^2
\nonumber \\ & & \mbox{}
+ e^{2\alpha} (dr^2 + r^2\, d\theta^2),
\end{eqnarray}
where the potentials $\nu$, $B$, $\omega$, and $\alpha$ depend only on $r$ and $\theta$, and have the following asymptotic decay \citep{Butterworth1976}:
\begin{eqnarray}
\nu &=& -\frac{M}{r} + \frac{B_0M}{3r^3} + \nu_2 P_2(\cos\theta)+\mathcal{O}\left(\frac{1}{r^{4}}\right), \nonumber \\
B &=& 1 + \frac{B_0}{r^2} +\mathcal{O}\left(\frac{1}{r^{4}}\right) , \nonumber \\
\omega &=&\frac{2I\Omega}{r^3}+\mathcal{O}\left(\frac{1}{r^{4}}\right) ,
\end{eqnarray}
where $M$ is the NS mass, and $\Omega$ is the angular frequency. $B_0$ and $\nu_2$ are real constants. When describing the interior of the NS as a perfect fluid, its energy-momentum tensor becomes
\begin{eqnarray}
T^{\mu\nu} = (\rho + p)u^{\mu}u^{\nu} + p g^{\mu\nu}, \end{eqnarray} where $\rho$ presents the energy density, $p$ denotes the pressure, and $u^{\mu}$ is the $4$-velocity.

\cite{Lan2021} investigated in detail the evolutionary behavior of the radius and moment of inertia of the newborn magnetar with rotational period from four samples of EoSs [SLy \citep{Douchin2001}, ENG \citep{Engvik1996}, AP3 \citep{Akmal1997}, WFF2 \citep{Wiringa1988}] within a range of nonrotating NS maximum mass ($2.05M_{\odot}<M_{\rm TOV}<2.39M_{\odot}$) by using public code \texttt{RNS} \citep{Stergioulas1995} to solve above field equations, and found that the $R$ and $I$ would undergo an obvious evolution as the magnetar spin-down. Due to the scale relations $L_{\rm dip}\propto R^6$ for magnetic dipole radiation and $L_{\rm GW}\propto I^2$ for GW radiation, the evolutionary effects of $R$ and $I$ would have a significant impact on the observational EM luminosity. With different EoSs and baryonic masses considered, the $L_{\rm dip}$ could be variant within 1 to 2 orders of magnitude compared to the case where $R$ and $I$ are constants. In response to this, refitting the X-ray plateau data of GRBs to correct the physical parameters of the nascent magnetar is crucial to precisely diagnose the nature of NS after considering the evolutionary effects of $R$ and $I$ in different EoSs and baryonic masses. 

When we consider the $R$ and $I$ evolutionary effects of the newborn NS, according to Equation (\ref{Erot}) and (\ref{dotE}), one can derive that a nascent magnetar loses its rotational energy through two ways: magnetic dipole torques ($\dot{E}_{\rm dip}$) and GW radiation ($\dot{E}_{\rm GW}$),
\begin{eqnarray}
\dot{E}_{\rm rot}= \frac{1}{2} \dot{I} \Omega^{2} + I\Omega \dot{\Omega} &=& \dot{E}_{\rm dip} + \dot{E}_{\rm GW} \nonumber
\\ &=& -{\beta}I\Omega^{4}-{\gamma}I\Omega^{6},
\label{E_rot dot}
\end{eqnarray}
where $\dot{I}$ and $\dot{\Omega}$ are the time derivative of the moment of inertia and the angular frequency, respectively. $\beta= \eta_{\rm X} B^2_{\rm p}R^{6}/6c^{3}I$, $\gamma=32GI\epsilon^{2}/5c^{5}$. The X-ray radiation efficiency $\eta_{\rm X}$ is quite uncertain and could evolve over time. \cite{Xiao2019} claimed that this efficiency is largely dependent on the bulk saturation Lorentz factors ($\Gamma_{\rm sat}$) of magnetar wind, and the efficiency dramatically increases with the decrease of the $\Gamma_{\rm sat}$ value. For a typical value $\Gamma_{\rm sat}\sim 300$ in GRB, the efficiency is possibly closer to 0.01. \cite{Zhong2024} suggested that the radiative efficiency evolves with time in the magnetar spin-down process, and estimated the time-averaged radiation efficiency as $\sim0.01$ for GRB 230307A. \cite{Rowlinson2014} suggested that the efficiency of the conversion of rotational energy from the magnetar wind into the observed X-ray plateau luminosity is $\leq0.2$. \cite{Gao2016} used Monte Carlo simulations and concluded that the efficiency parameter is constrained in the range of approximately 0.5. It is worth noting that, in the framework of magnetar spin-down, a low radiation efficiency, such as $\eta_{\rm X}<0.1$, would lead to the magnetar rotational speed that breaks through the Keplerian break-up limit, which may challenge the current most NS EoSs model. Therefore, we took two radiation efficiency values to evaluate our results, i.e. $\eta_{\rm X}$ = 0.1, 0.5. 

\begin{figure*}
\centering
\includegraphics  [angle=0,scale=0.6]   {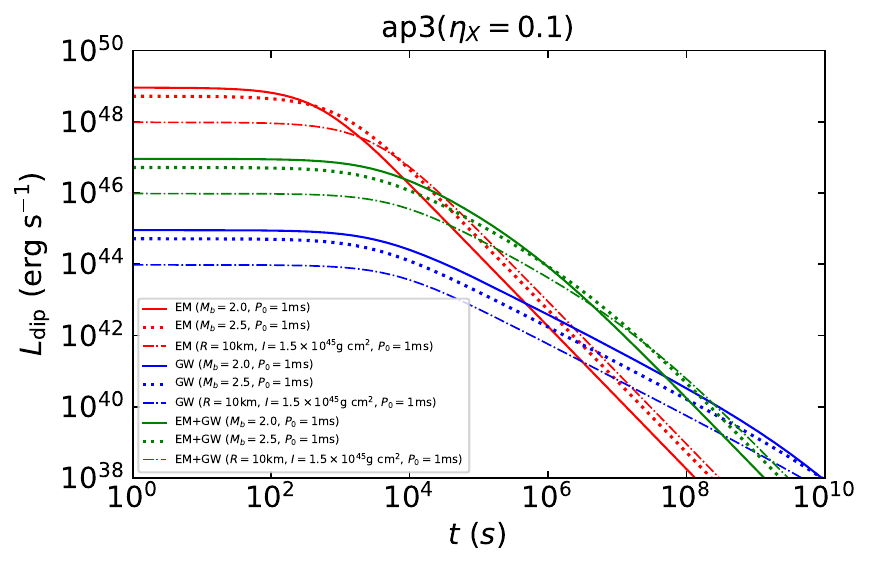}
\includegraphics  [angle=0,scale=0.6]   {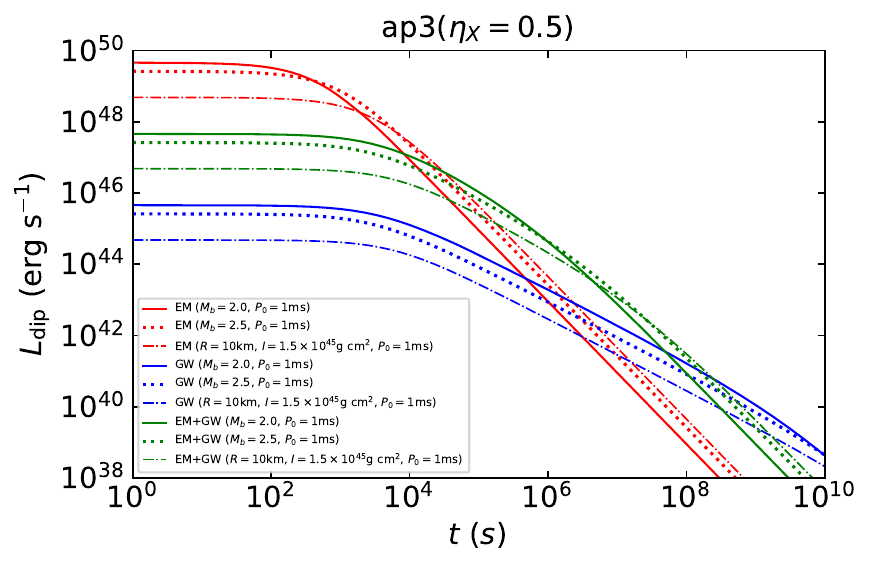}
\includegraphics  [angle=0,scale=0.6]   {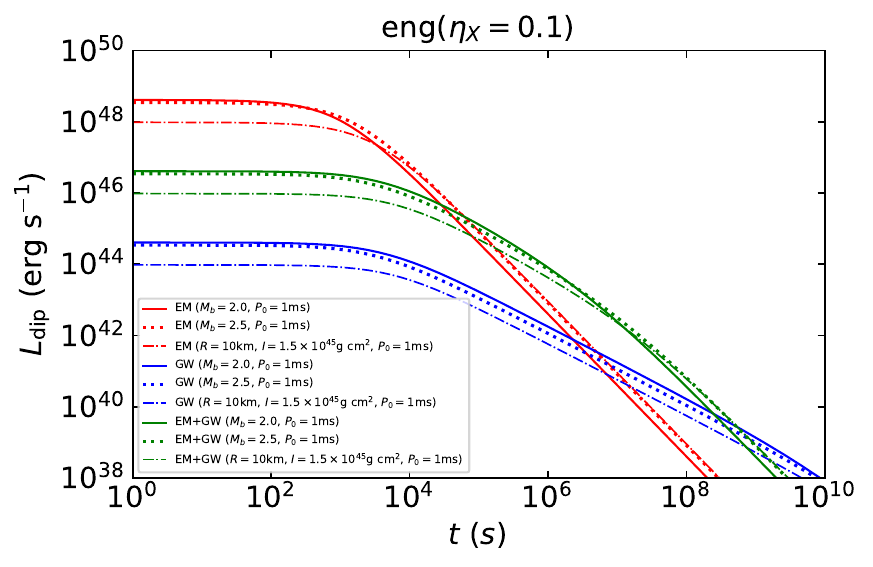}
\includegraphics  [angle=0,scale=0.6]   {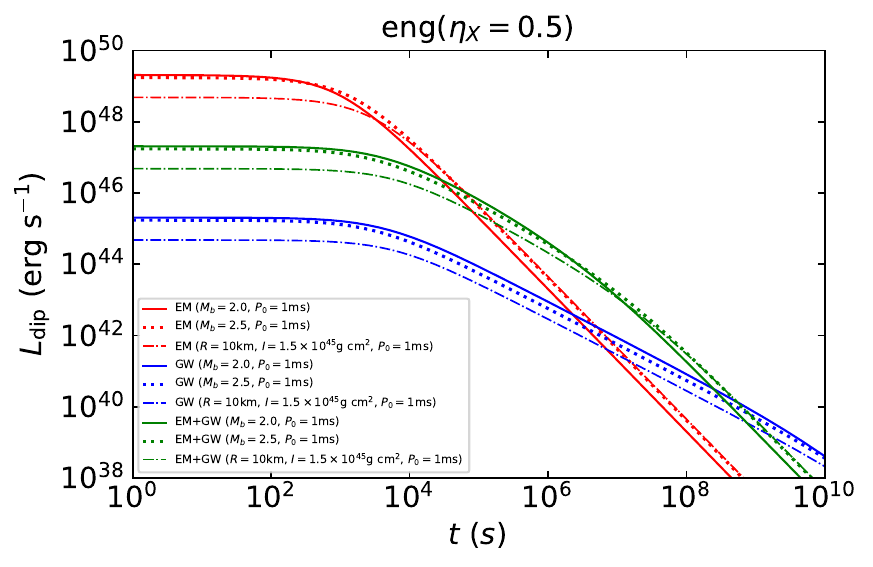}
\includegraphics  [angle=0,scale=0.6]   {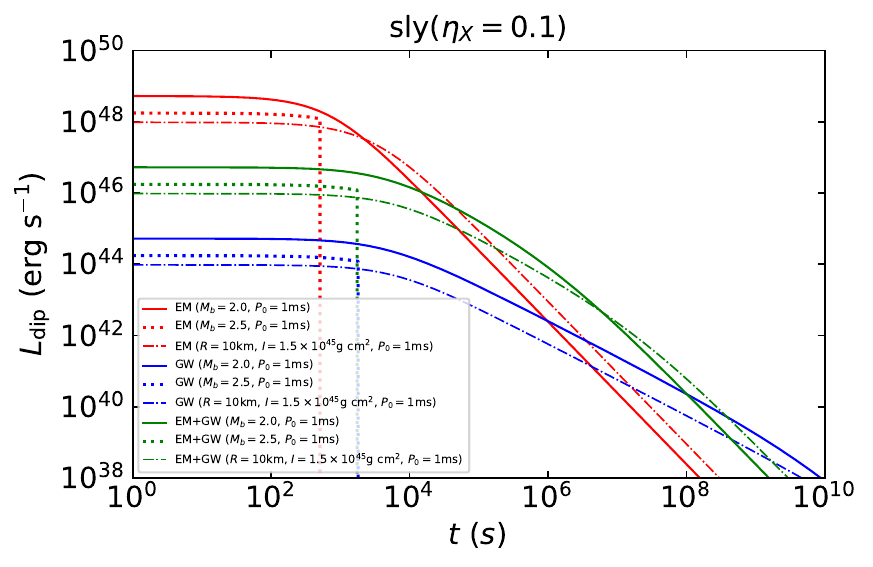}
\includegraphics  [angle=0,scale=0.6]   {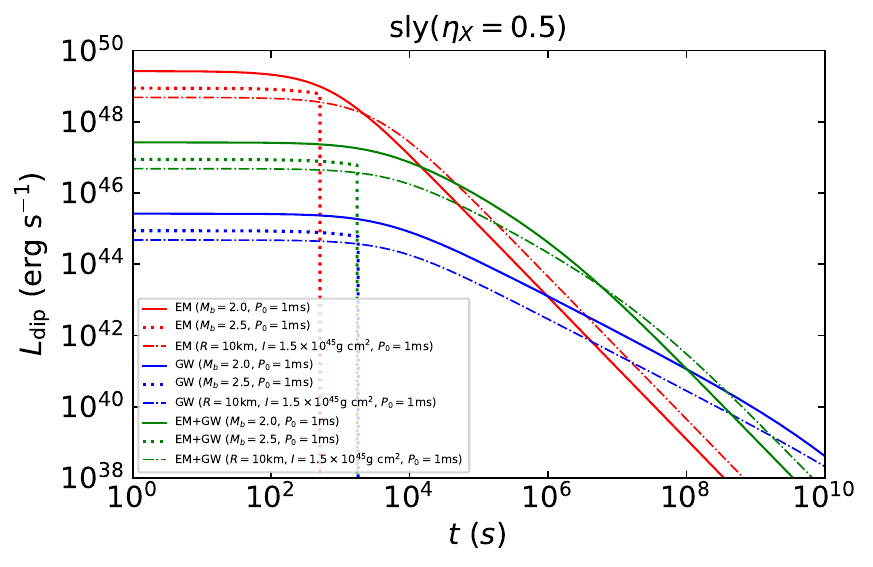}
\includegraphics  [angle=0,scale=0.6]   {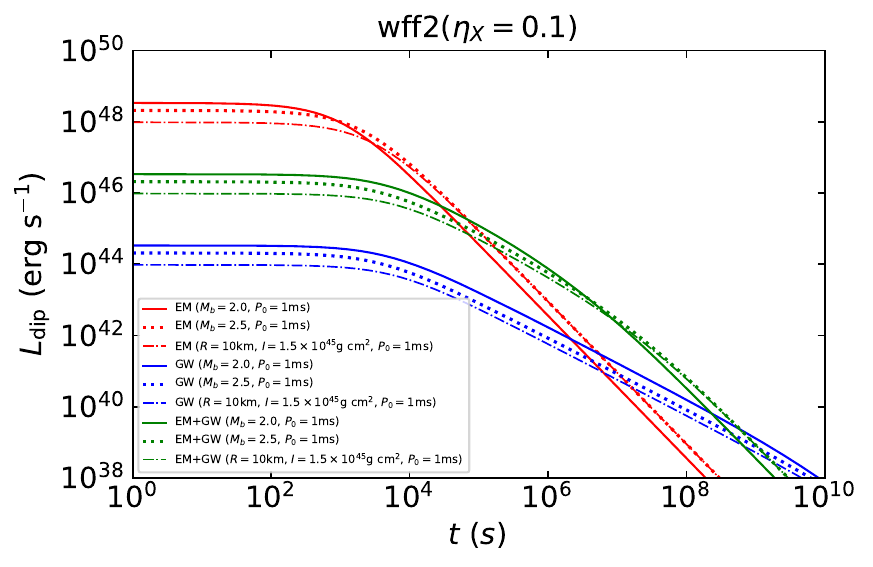}
\includegraphics  [angle=0,scale=0.6]   {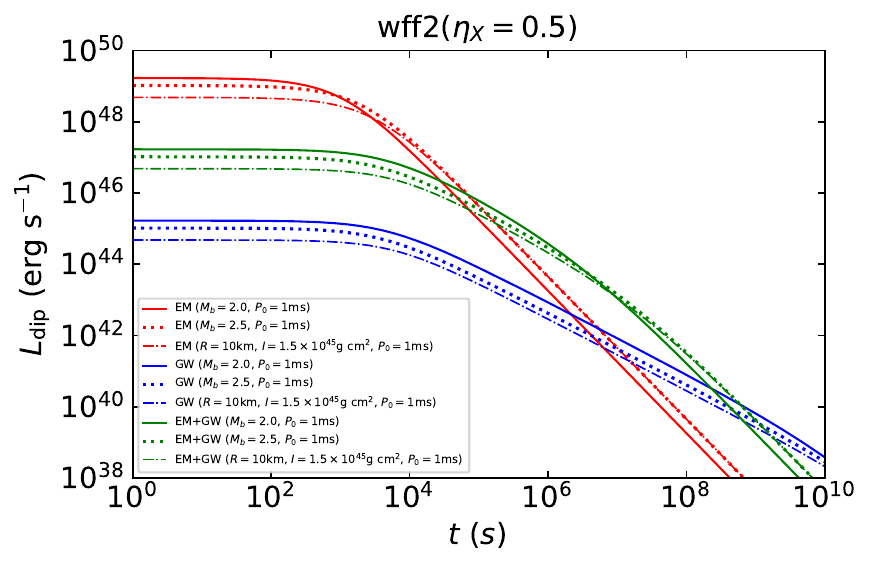}
\caption{Numerical results for dipole radiation luminosity
by considering the $R$ and $I$ evolution effects in choice of different NS EoSs, baryonic masses, radiation efficiencies, as well as the dominant energy loss term.}
\label{fig:model_LC}
\end{figure*}

With the evolutionary behaviors of $R$ and $I$ for various rotational speeds in different EoSs and baryonic masses, for an NS with given parameters (e.g., given the initial values of $B_p$, $\epsilon$, and $P_0$), we can numerically solve its deceleration history and calculate the evolutionary history of EM dipole radiation luminosity based on Equations (\ref{dotE}) and (\ref{E_rot dot}). In Figure \ref{fig:model_LC}, we showed that the evolution results of EM dipole luminosity for different EOSs, radiation efficiency, as well as the dominant energy loss term (EM radiation, GW radiation, or EM+GW radiation, modulated by configuring different initial parameters $B_p$ and $\epsilon$) after taking into account the evolutionary effects of the $R$ and $I$ during the nascent magnetar spin-down process\footnote{Here, we test two values for $M_b$ (e.g. $2.0M_{\odot}$ and $2.5M_{\odot}$), with which the NS can be supported by most of the adopted EoSs.}. In EM radiation-dominated case, we set the initial values $B_{p}=10^{15}$ G, $P_0 =1~{\rm ms}$ and $\epsilon=10^{-7}$. In GW radiation-dominated case, we set the initial values $B_{p}=10^{13}$ G, $P_0 =1~{\rm ms}$ and $\epsilon=10^{-3}$. In EM and GW radiation equivalence case, we set the initial values $B_{p}=10^{14}$ G, $P_0 =1~{\rm ms}$ and $\epsilon=10^{-3}$. Compared to the NS spin-down model with constant $R$ and $I$, the temporal behavior of $L_{\rm dip}$ would become more complicated after taking into account the $R$ and $I$ evolutionary effects and would present some new segments. For more details on the temporal behavior of $L_{\rm dip}$ after considering $R$ and $I$ evolutionary effects, please refer to our latest paper \cite{Lan2021}. More importantly, no matter how the EoS, radiation efficiency, and dominant energy loss term are changed, the $R$ and $I$ evolutionary effects always cause $L_{\rm dip}$ to vary within 1 or 2 orders of magnitude compared to the constant $R$ and $I$ scenario.

In order to analytically obtain the evolution history of the magnetic dipole radiation luminosity of the newborn magnetar under $R$ and $I$ evolutionary effects, we have to derive the evolutionary relation for the $R$ and $I$ with the rotational velocity $\Omega$. \cite{Lan2021} studied in detail how $R$ and $I$ evolve as the magnetar spins down, and found that an analytical form for $R$ and $I$ with respect to the rotational velocity $\Omega$ under certain approximations
\begin{eqnarray}
R \simeq \left\{ \begin{array}{ll} R_0(\frac{\Omega}{\Omega_{\rm k}})^{m}, & \Omega_1<\Omega<\Omega_{\rm k};\\
R_0(\frac{\Omega_1}{\Omega_{\rm k}})^{m}, &
\Omega\leq\Omega_1. \\
\end{array} \right.
\label{R_formula}
\end{eqnarray}

\begin{eqnarray}
I \simeq \left\{ \begin{array}{ll} I_0(\frac{\Omega}{\Omega_{\rm k}})^{k}, & \Omega_1<\Omega<\Omega_{\rm k};\\
I_0(\frac{\Omega_1}{\Omega_{\rm k}})^{k}, &
\Omega\leq\Omega_1, \\
\end{array} \right.
\label{I_formula}
\end{eqnarray}
where $R_{0}$, $I_{0}$, $\Omega_{\rm 1}$ and $\Omega_{\rm k}$ are the initial radius, initial moment of inertia, critical angular velocity\footnote{The critical angular velocity $\Omega_{\rm 1}$ is defined as the angular velocity below which $R$ and $I$ are no longer evolving with time.}, and Keplerian angular velocity of the newly born NS. $m$ and $k$ are the power-law index for $R$ and $I$ evolution with respect to the rotational velocity of the NS, which depend on the NS EoS and baryonic mass, with the assumption that the magnetic dipole moment $\mu\equiv B_pR^3$ is conserved during the spin-down process of the nascent magnetar.

If the spin-down process of the newborn magnetar is dominated via a single loss term alone ($\dot{E}_{\rm dip}$ or $\dot{E}_{\rm GW}$), from Equation (\ref{E_rot dot}), one has
\begin{eqnarray}
\dot{E}_{\rm rot}= \frac{1}{2} \dot{I} \Omega^{2} + I\Omega \dot{\Omega} \simeq \left\{ \begin{array}{ll} 
-{\beta}I\Omega^{4}, & \dot{E}_{\rm GW}\ll\dot{E}_{\rm dip};\\
-{\gamma}I\Omega^{6}, & \dot{E}_{\rm GW}\gg\dot{E}_{\rm dip}. \\
\end{array} \right.
\label{EM_GW dominated}
\end{eqnarray}

As the magnetar loses most of its rotational energy via EM dipole radiation, we can derive an analytical solution of $\Omega$ evolution in Equation (\ref{EM_GW dominated}), and the EM dipole luminosity and spin-down timescale in the case of EM dipole emission domination can be expressed as follows
\begin{eqnarray}
L_{\rm dip} &=& \frac{B_{p}^2R_0^6\Omega_0^{4}}{6c^3}\left[1+\frac{t}{\tau_{\rm sd,em}}\right]^\frac{4}{k-2} \nonumber \\
&=& L_{\rm sd,em}\left[1+\frac{t}{\tau_{\rm sd,em}}\right]^\frac{4}{k-2},
\label{Luminosity_EM}
\end{eqnarray}

\begin{eqnarray}
\tau_{\rm sd,em}=\frac{3(2+k)I_0c^3}{(2-k)B_{p,0}^2R_0^6\Omega_{\rm 0}^{2}}(\frac{\Omega_{\rm 0}}{\Omega_{\rm k}})^{k}.
\label{Tsdem}
\end{eqnarray}

As the magnetar loses most of its rotational energy via GW radiation, we can derive an analytical solution of $\Omega$ evolution in Equation (\ref{EM_GW dominated}), and the EM dipole luminosity and spin-down timescale in the case of GW emission domination can be expressed as follows
\begin{eqnarray}
L_{\rm dip} &=& \frac{B_{p}^2R_0^6\Omega_0^{4}}{6c^3}\left[1+\frac{t}{\tau_{\rm sd,gw}}\right]^{-\frac{4}{k+4}} \nonumber \\
&=& L_{\rm sd,em}\left[1+\frac{t}{\tau_{\rm sd,gw}}\right]^{-\frac{4}{k+4}}.
\label{Luminosity_GW}
\end{eqnarray}

\begin{eqnarray}
\tau_{\rm sd,gw}=\frac{5(k+2)c^5}{64(k+4)GI_0\epsilon^2\Omega_0^{4}}(\frac{\Omega_{\rm 0}}{\Omega_{\rm k}})^{-k}.
\label{Omega evolution}
\end{eqnarray}

Apparently, the $L_{\rm dip}$ light curves would exhibit some new segments after taking into account the $R$ and $I$ evolutionary effects, i.e., $L_{\rm dip}\propto t^{0}$ followed by $L_{\rm dip}\propto t^{-\alpha}$, and then either the GW-dominated $L_{\rm dip}\propto t^{-1}$ or the EM-dominated $L_{\rm dip}\propto t^{-2}$. The slope scale of new fragment $L_{\rm dip}\propto t^{-\alpha}$ depends on the evolutionary intensity of $R$ and $I$ at different EoSs and baryonic masses. In particular, in EM and GW radiation equivalence case, from Equation (\ref{E_rot dot}), we can see that the GW radiation is more efficient than EM dipole radiation at the early phase due to the larger rotational angular velocity $\Omega$ at this stage. Therefore, GW radiation would dominate magnetar spin-down at the early phase, and EM dipole radiation would dominate at the late phase in the GW+EM radiation case. The transition from the GW-dominated phase to the EM-dominated phase is shown as a smooth break, and the decay slope of EM dipole luminosity changes from -1 to -2. These $L_{\rm dip}$ approximate analytical results are consistent with our numerically solved results in Figure \ref{fig:model_LC}. Observationally, the spin-down EM dipole luminosity evolution behaviors have been observed in a part of long and short GRBs with X-ray plateau emission, which not only support the magnetar central engine but also are used to infer the parameters of magnetars \citep{Troja2007,Rowlinson2010,Rowlinson2013,Gompertz2013,Gompertz2014,Lv2014,Lv2015,Lv2018,Zou2019,Zou2021a,Xie2022a,Xie2022b}. In previous estimates of magnetar physical parameters, the $R$ and $I$ of magnetar in the spin-down stage are usually considered as so-called fiducial constant values. Nevertheless, from Figure \ref{fig:model_LC} we can see that the evolutionary effects of $R$ and $I$ will cause $L_{\rm dip}$ to vary within 1 or 2 orders of magnitude compared to the constant $R$ and $I$. The results suggested that, when using the X-ray plateau data of GRBs to diagnose the properties of the nascent NSs, NS's EoS, baryonic mass, and X-ray radiative efficiency information should be invoked as simultaneously constrained magnetar physical parameters. In view of this motivation, we systematically collected LGRBs with X-ray plateau emission from the \emph{Swift}/XRT catalog, and refitted the LGRBs plateau via the magnetar spin-down model after factoring in the $R$ and $I$ evolutionary effects in order to correct the physical parameters of the nascent magnetar. The precise constraints on the newborn magnetar physical parameters will not only help us to reveal the nature of the LGRBs central engine but also have important significance for us to study the process of massive star collapse, and even to identify the NS's EoS and the mechanisms that induce the NS mass nonasymmetries.

\section{Constraining the Properties of Magnetar in the Frame of $R$ and $I$ Evolution} 
Assuming that the remnant of after GRB explosion is a millisecond magnetar \citep{Bucciantini2009}, the EM dipole emission from the newborn magnetar could generate a Poynting-flux-dominated outflow, which could undergo the magnetic energy dissipation processes with high efficiency \citep{Zhang2011} and to power the X-ray plateau feature observed in GRB afterglows \citep{Troja2007,Rowlinson2010,Rowlinson2013,Gompertz2013,Gompertz2014,Lv2014,Lv2015}. Our objective is to correct magnetar physical parameters that can be inferred from the X-ray plateaus by considering the evolutionary effects of $R$ and $I$. This is crucial to understand the physical properties of the nascent magnetar. In this section, we utilized the LGRBs data with X-ray plateau emission to derive relevant physical parameters within the framework of $R$ and $I$ evolution, and performed some statistical analyses based on our curated GRB sample. It should be noted that the redshift measurement $z$ is vital to derive the intrinsic luminosity parameter, and we adopted $z=1$ for the GRBs without redshift a measurement in our sample.    

We employed the Markov Chain Monte Carlo (MCMC) method based on the emcee Python package to derive the best-fitting posterior magnetar physical parameters and their uncertainties \citep{Foreman-Mackey2013} for different NS EoSs and baryonic masses by taking two different radiation efficiency $\eta_{\rm X}$ values \footnote{We note that the SLy EoS is not sufficiently stiff, $M_{\rm TOV}=2.05~M_{\odot}$. If $M_{b}=2.5~M_{\odot}$ is adopted, the nascent magnetar will rapidly collapse during spin-down process. Therefore, we only considered the case of $M_{b}=2.0~M_{\odot}$ for the SLy EoS.}. The key parameters for defining the properties of a newly formed magnetar are the initial spin period $P_0$, the dipole magnetic field strength $B_p$, and the ellipticity $\epsilon$. The prior distributions of these parameters are set to a uniform distribution over a broad range, specifically $P_0\in [0.3-40~\rm ms]$, $B_p\in [10^{13}-10^{16}~\rm G]$, and $\epsilon \in [10^{-6}-10^{-2}]$. Based on the value of $\chi^2$/degrees of freedom (dof), we determined the best-fitting model parameters for each GRB. The MCMC best-fitting results and corner images are shown in the Appendix. In Appendix Figures \ref{fig:EM_mcmc}-\ref{fig:EM+GW_mcmc}, we showed the best-fitting results and corner plots for EM-dominated case, GW-dominated case, and EM+GW co-dominated case in four samples of EoSs with $M_{b}=2.0~M_{\odot},~2.5~M_{\odot}$ and $\eta_{\rm X}=0.1,~0.5$, respectively. In Appendix Figure \ref{fig:allGRBs_LC}, we selected the AP3 EoS with $M_{b}=2.0~M_{\odot}$ and $\eta_{\rm X}=0.5$ as an example to show the best-fitting results for X-ray light curves in our entire GRB sample. The MCMC best-fitting parameters for the magnetar spin-down model with $R$ and $I$ evolutionary effects in different EoSs, baryonic masses, and radiation efficiency values are reported in Table \ref{table-2}-\ref{table-8}. Furthermore, in order to compare the deviation for the derived magnetar physical parameters at the constant $R$ and $I$ with those we constrained by considering the $R$ and $I$ evolution correction, we also employed MCMC method to derive the best-fitting posterior magnetar physical parameters and their uncertainties by using the constant $R$ and $I$ values ($R=10^6~\rm cm$, $I=1.5\times10^{45}~\rm g~cm^2$) in the spin-down process. The MCMC best-fitting parameters for magnetar spin-down model with constant $R$ and $I$ are reported in Table \ref{table-9}.

\subsection{Statistical Properties and Correlations of Magnetar Physical Parameters}
\begin{figure*}
\centering
\includegraphics [angle=0,scale=0.4]  {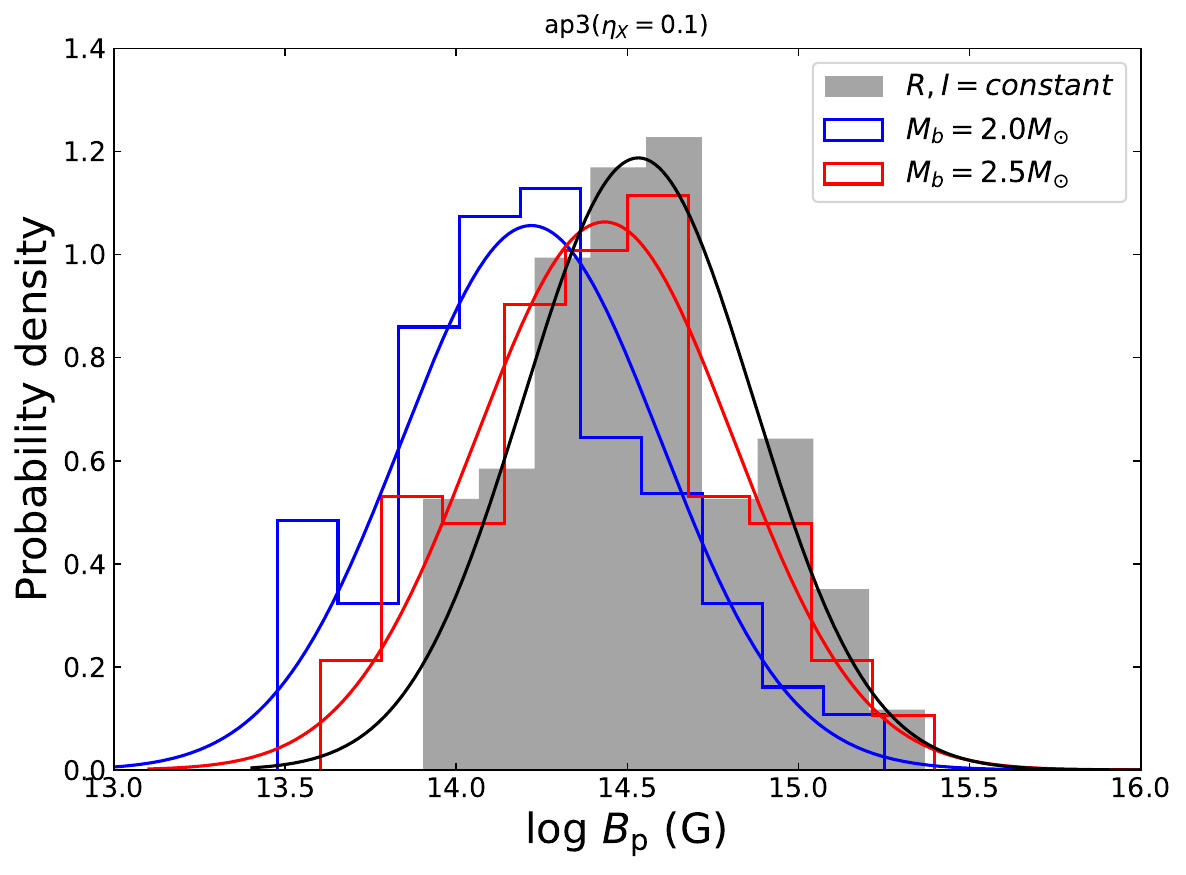}
\includegraphics [angle=0,scale=0.4]  {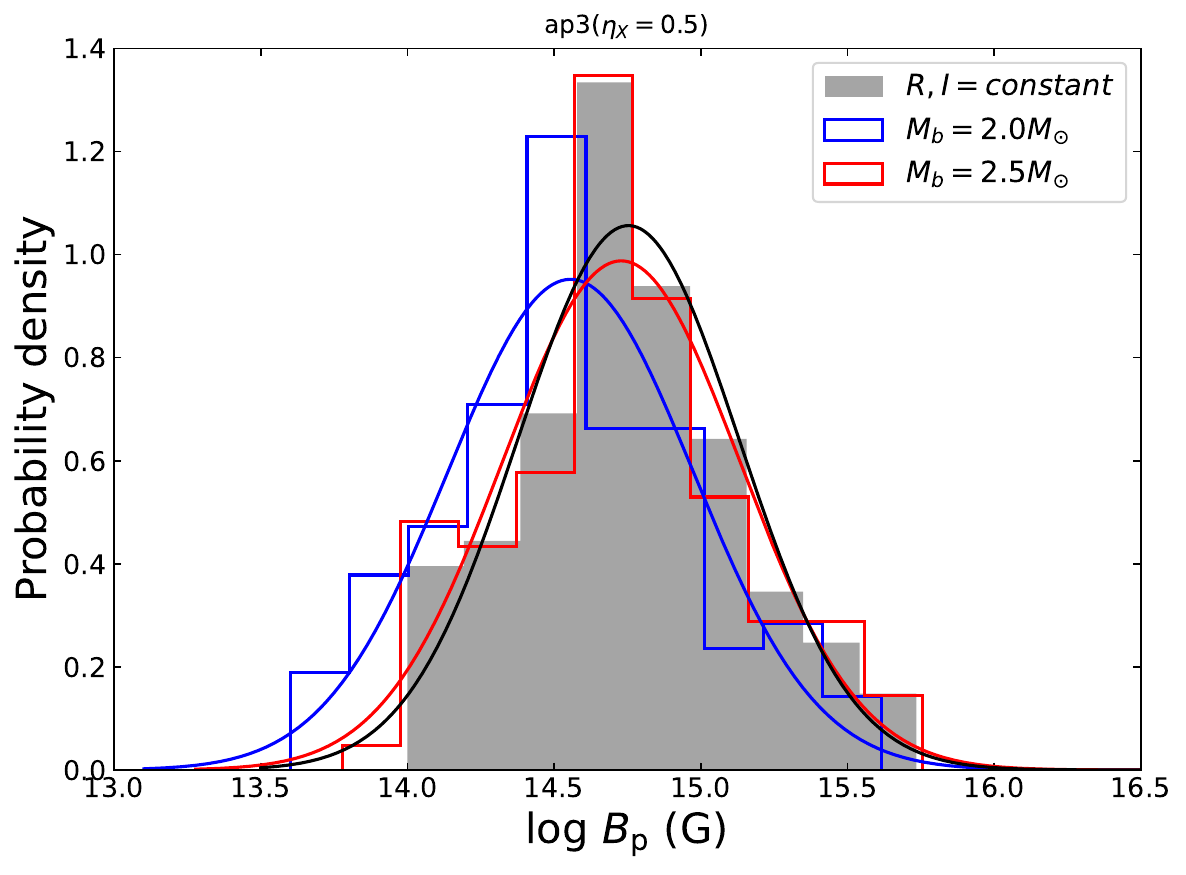}
\includegraphics [angle=0,scale=0.4]  {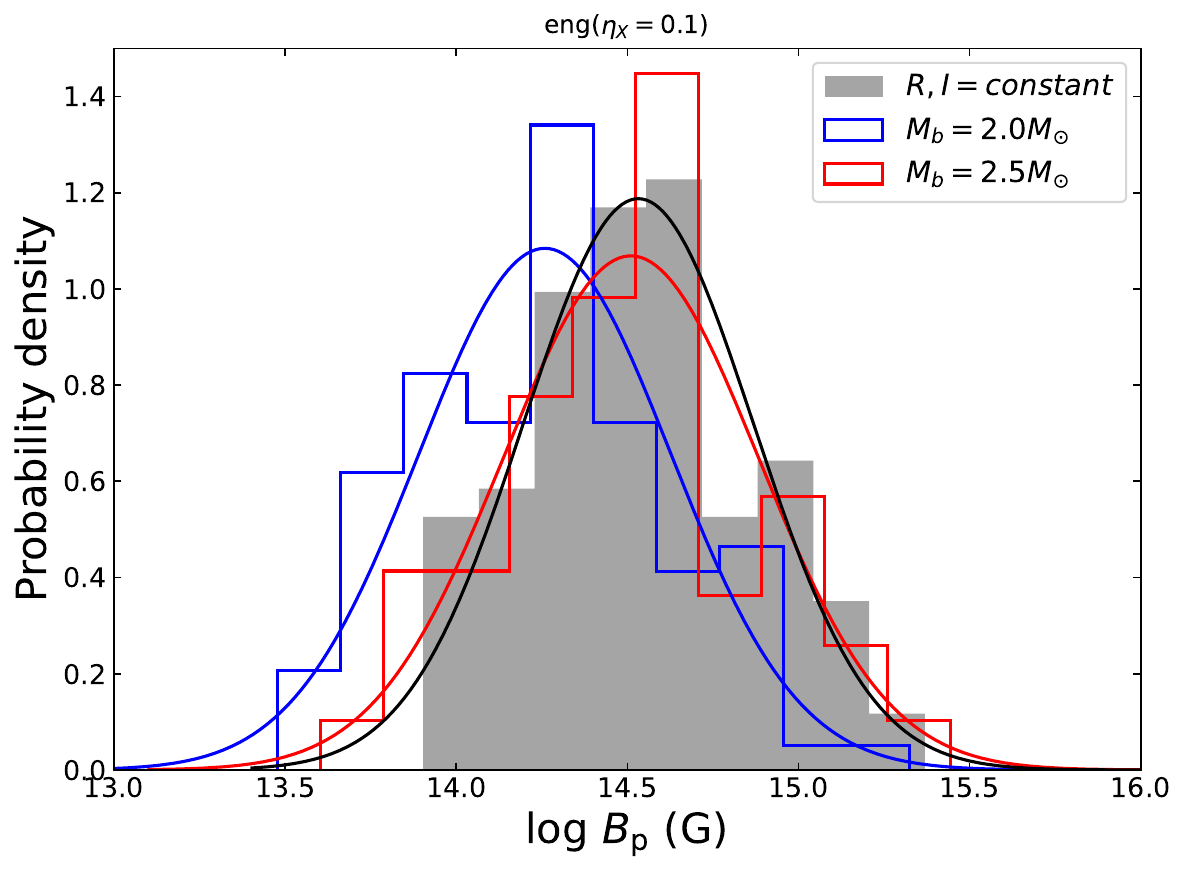}
\includegraphics [angle=0,scale=0.4]  {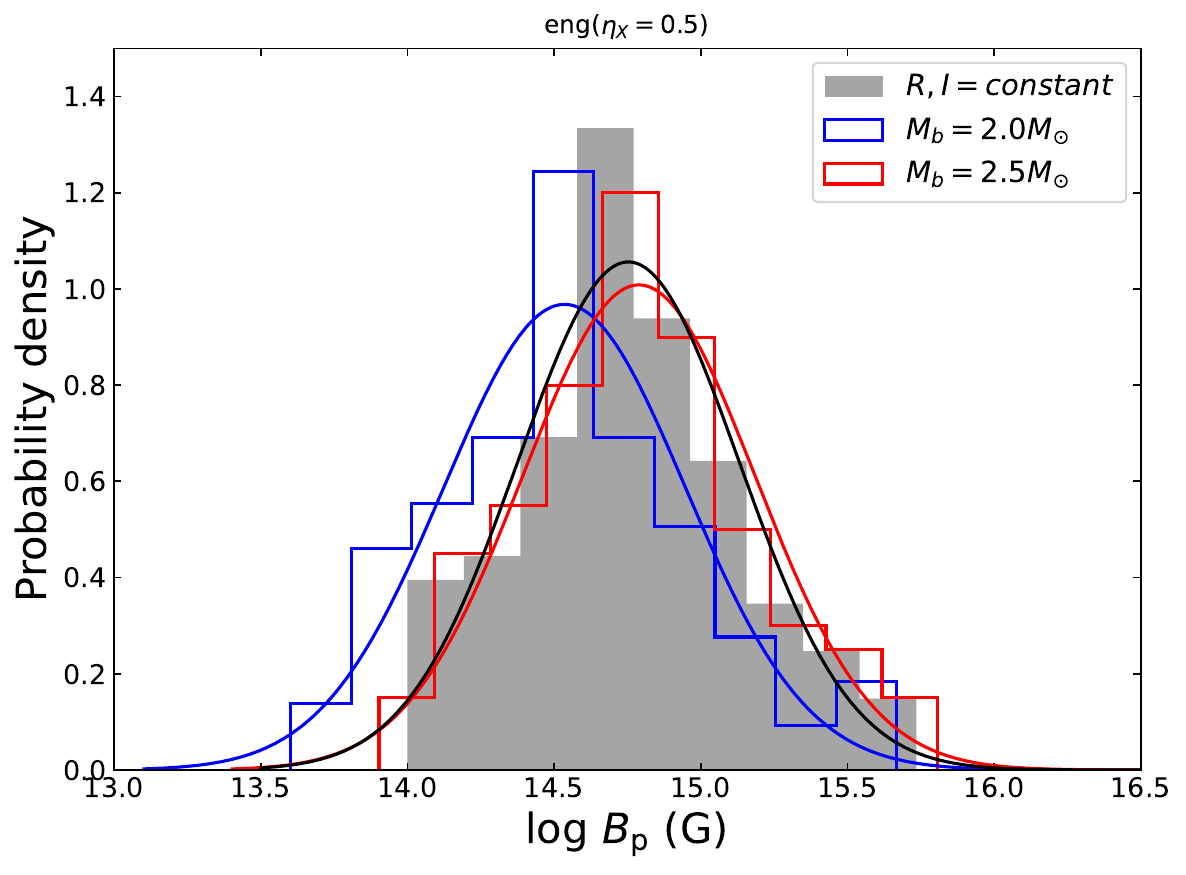}
\includegraphics [angle=0,scale=0.4]  {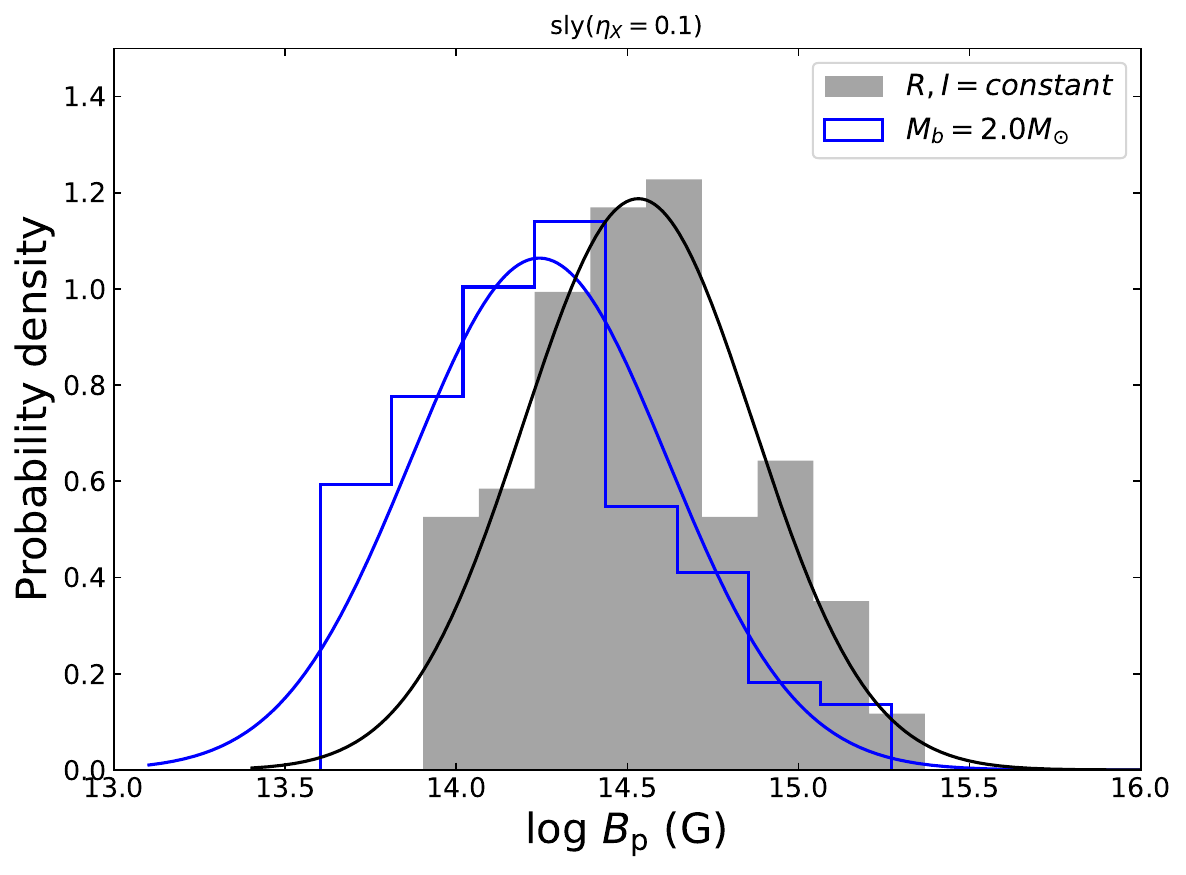}
\includegraphics [angle=0,scale=0.4]  {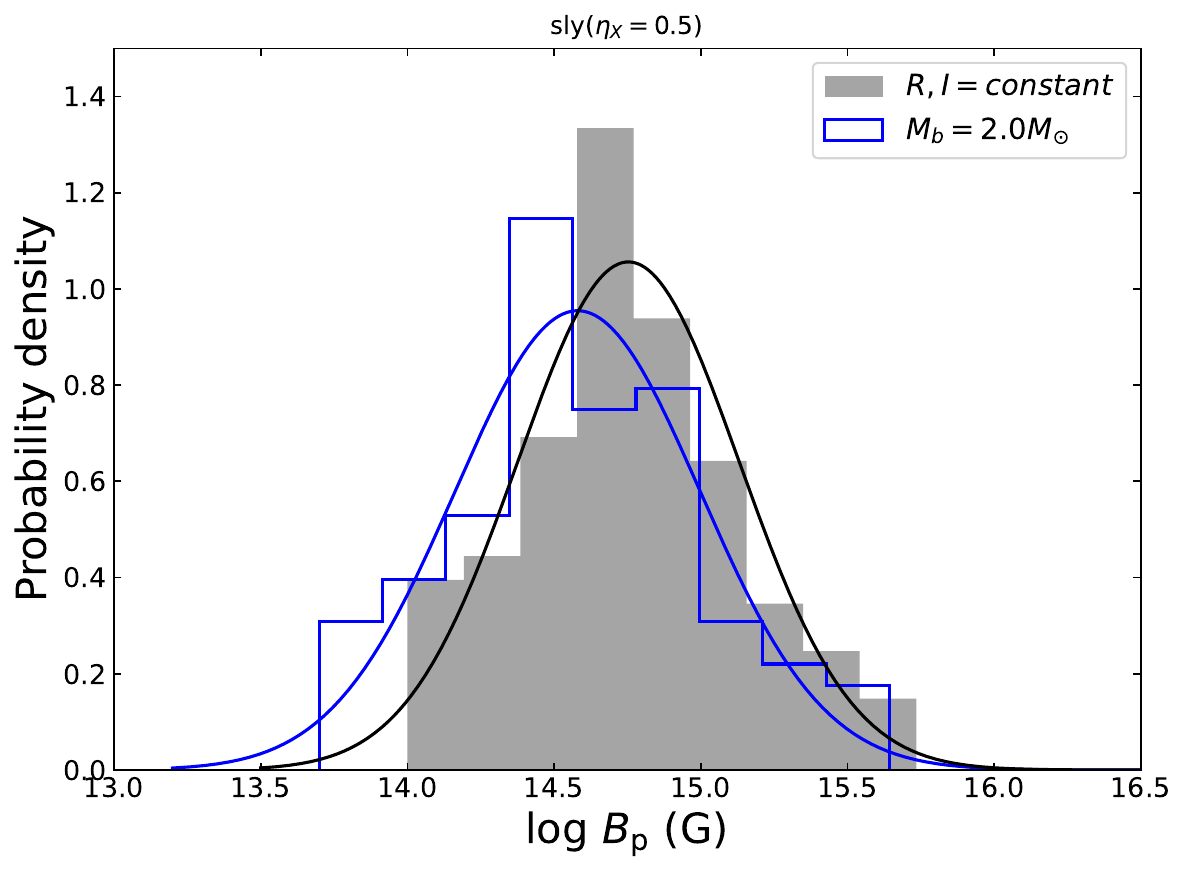}
\includegraphics [angle=0,scale=0.4]  {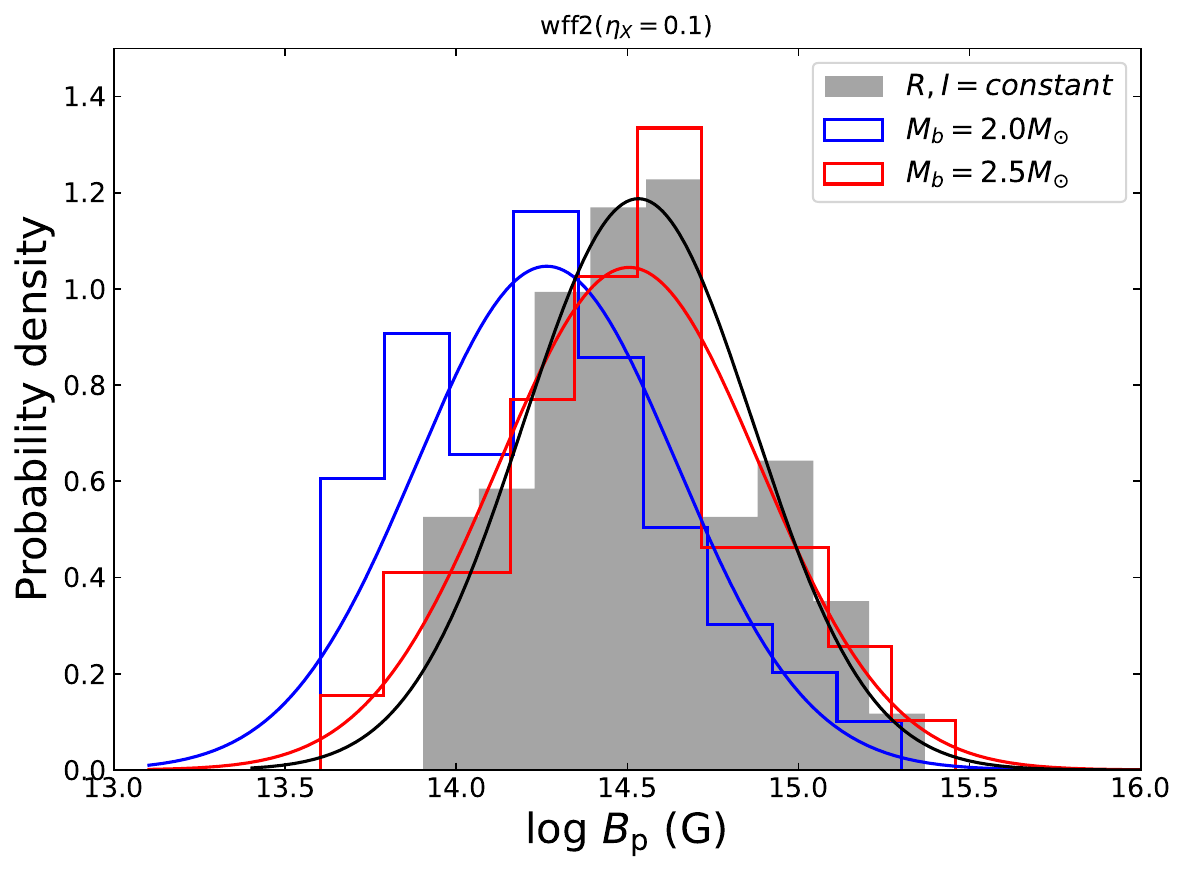}
\includegraphics [angle=0,scale=0.4]  {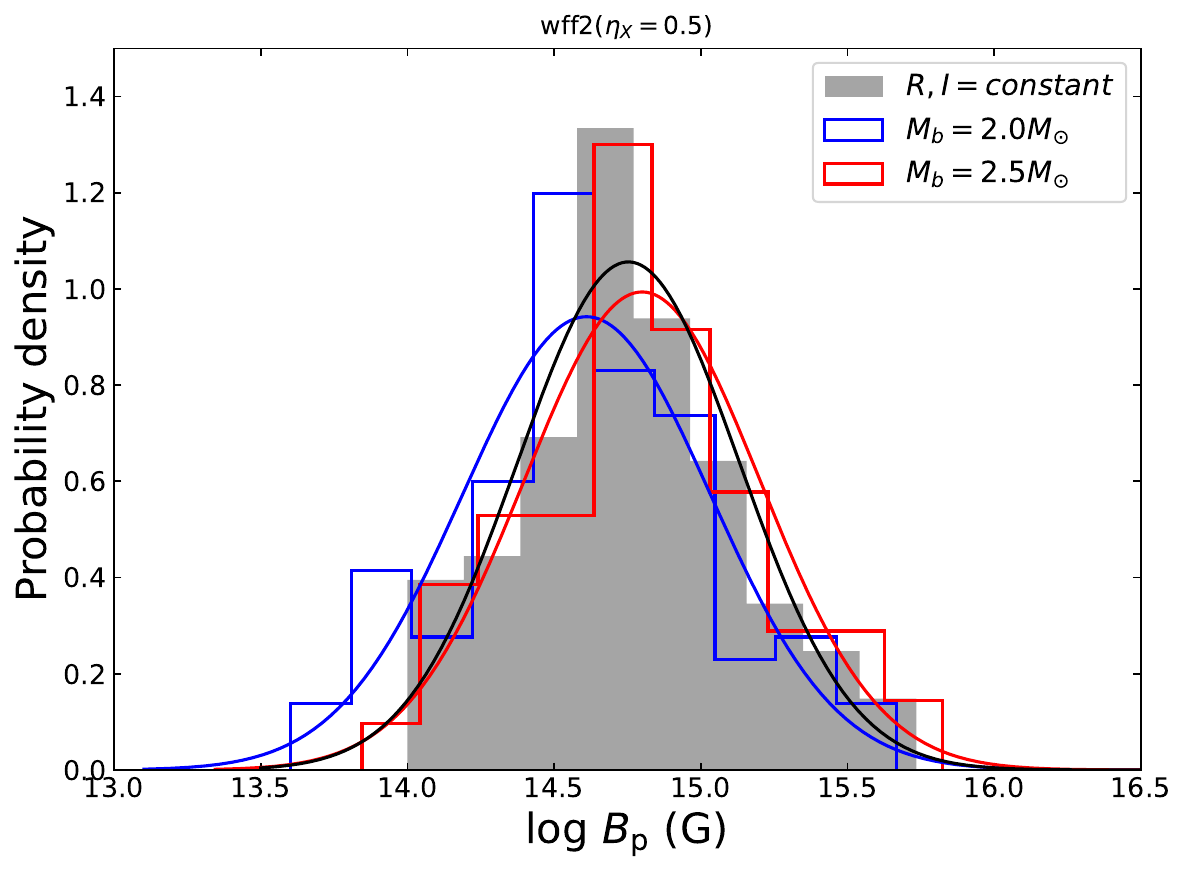}
\caption{Comparisons of the derived magnetar physical parameter $B_p$ histograms between the $R$ and $I$ evolution correction and the constant $R$ and $I$ in four samples of EoSs with $M_{b}=2.0~M_{\odot},~2.5~M_{\odot}$ and $\eta_{\rm X}=0.1,~0.5$, respectively.}
\label{fig:Bp-distribution}
\end{figure*}

\begin{figure*}
\centering
\includegraphics [angle=0,scale=0.4] {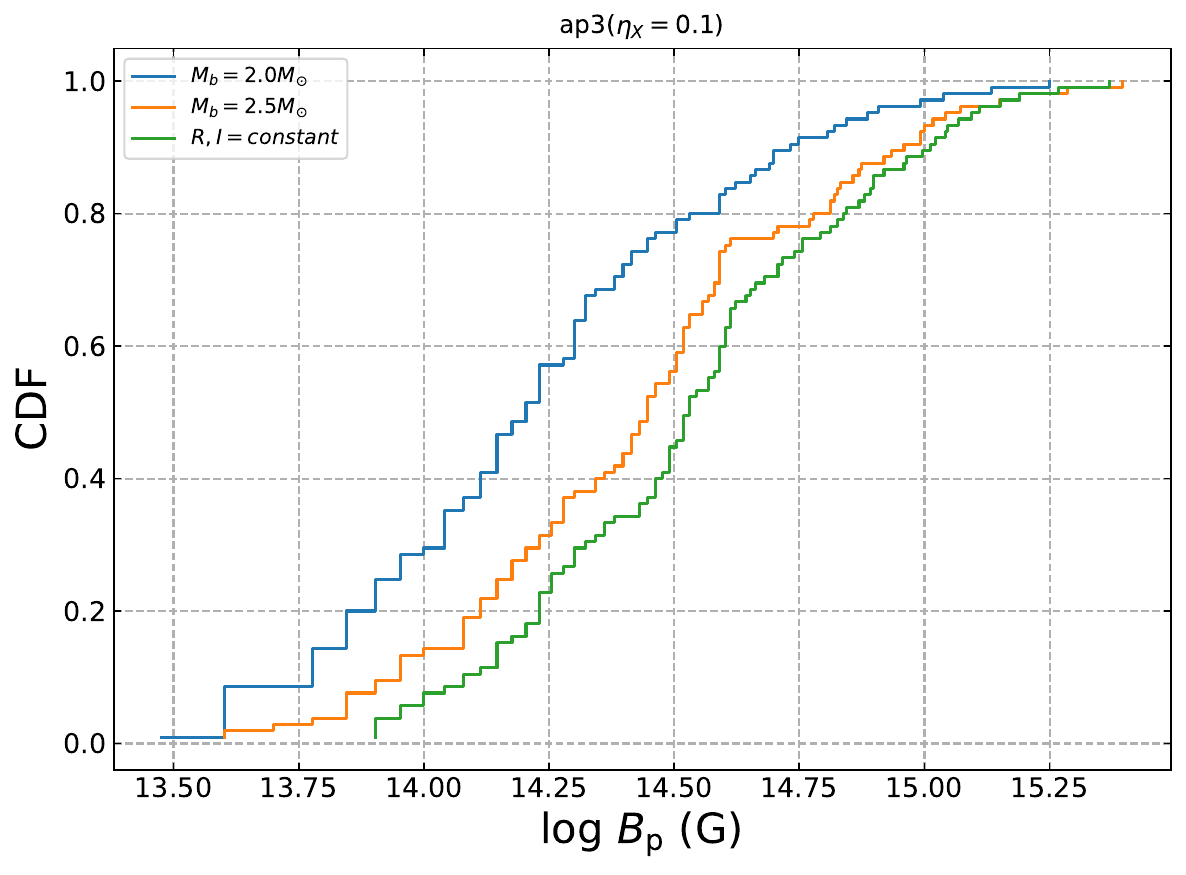}
\includegraphics [angle=0,scale=0.4] {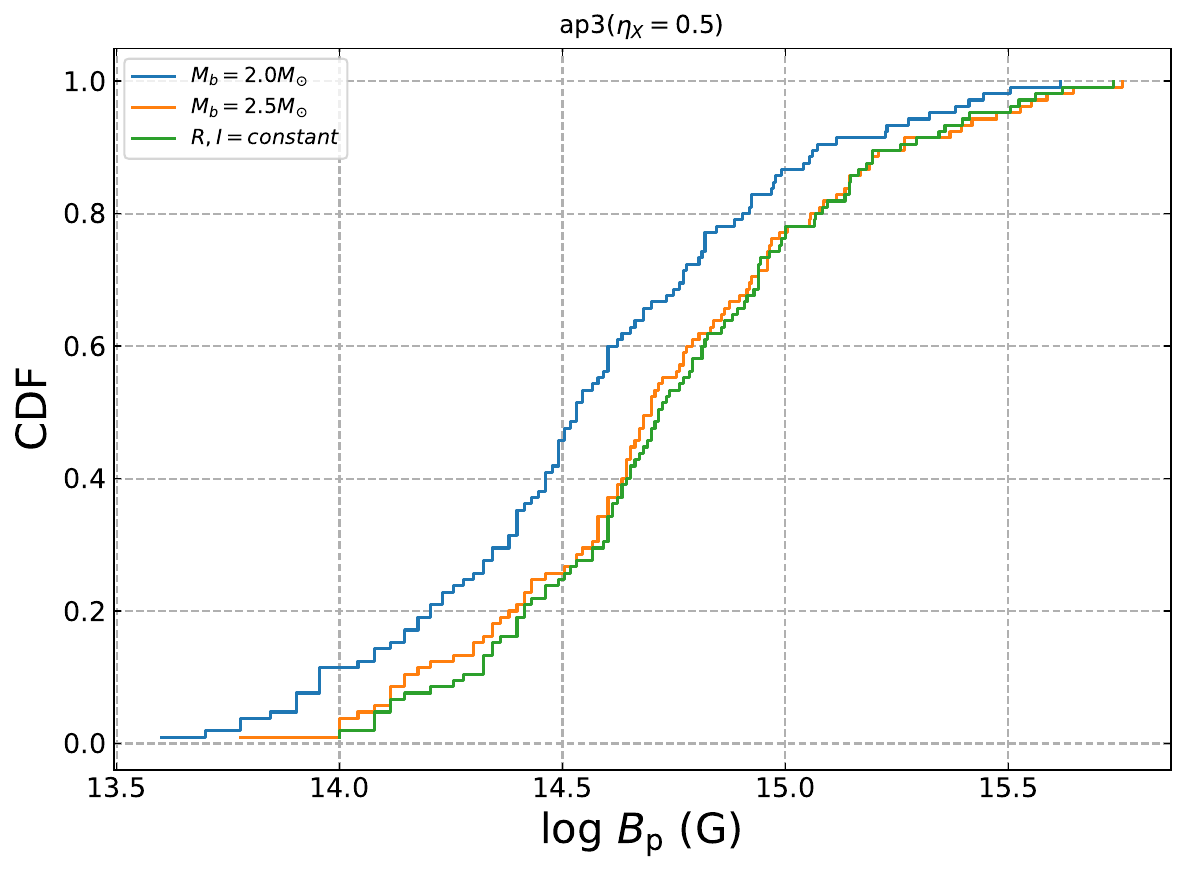}
\includegraphics [angle=0,scale=0.4] {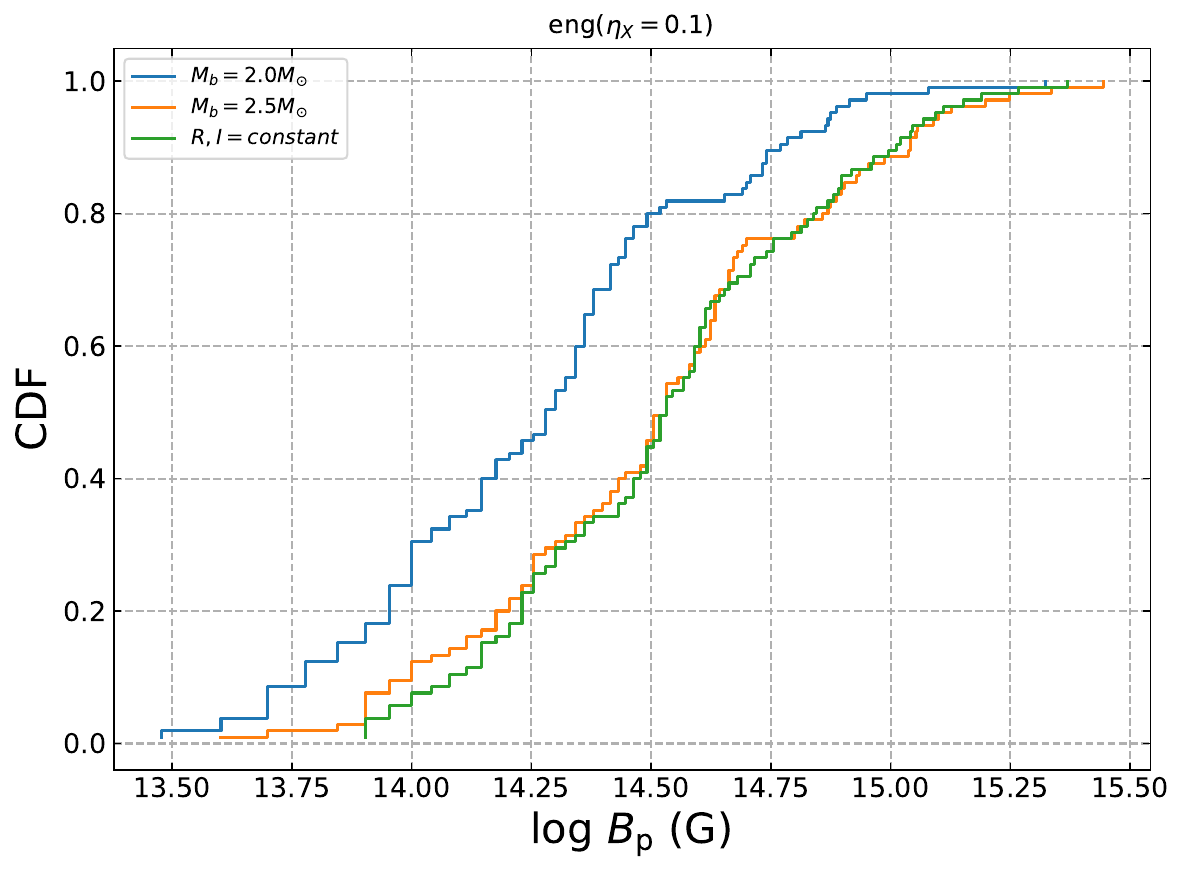}
\includegraphics [angle=0,scale=0.4] {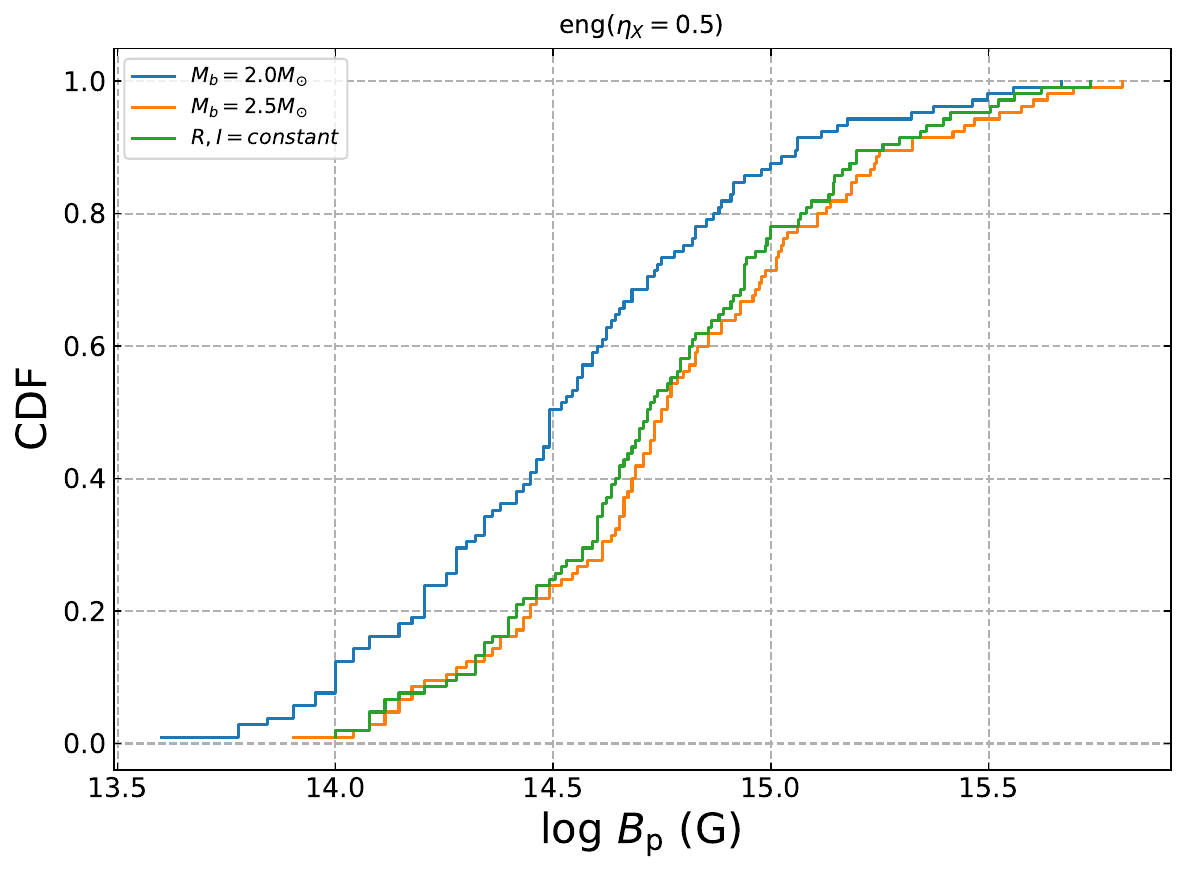}
\includegraphics [angle=0,scale=0.4] {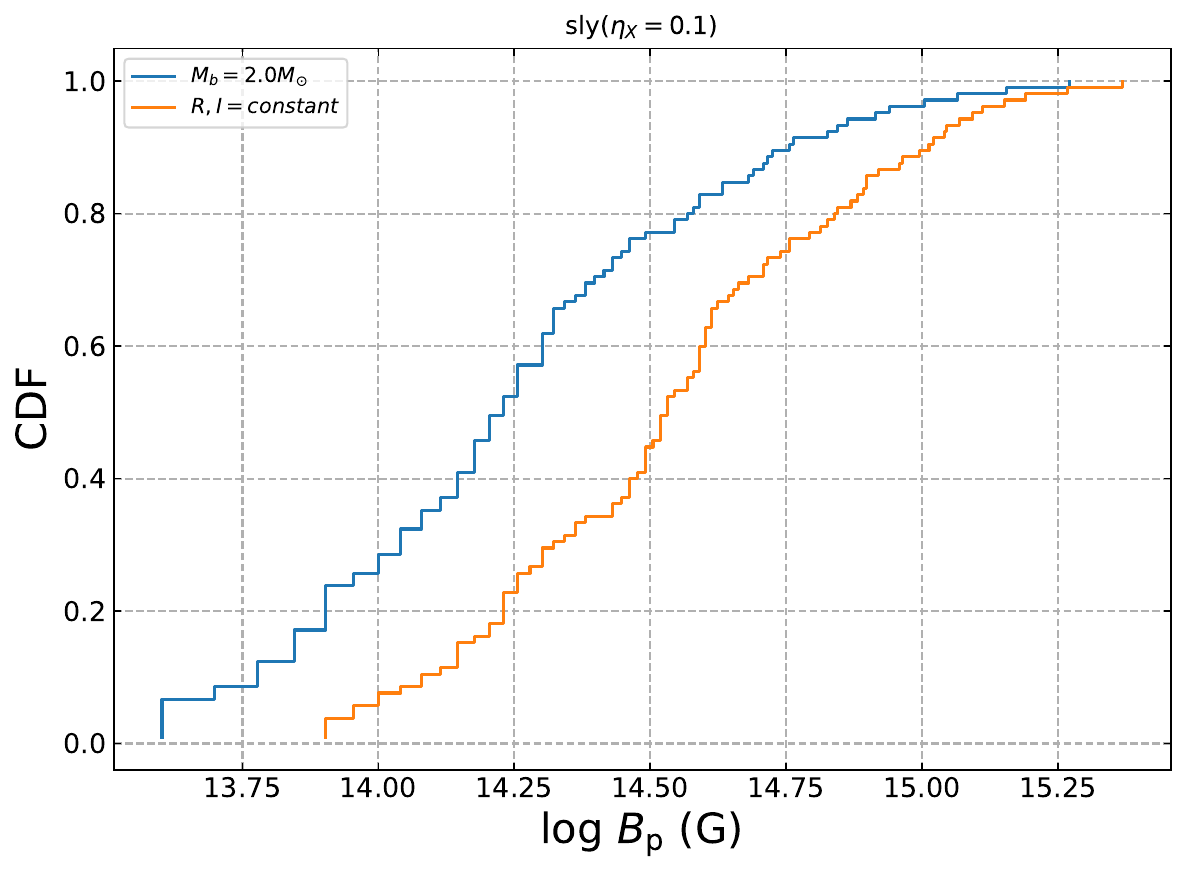}
\includegraphics [angle=0,scale=0.4] {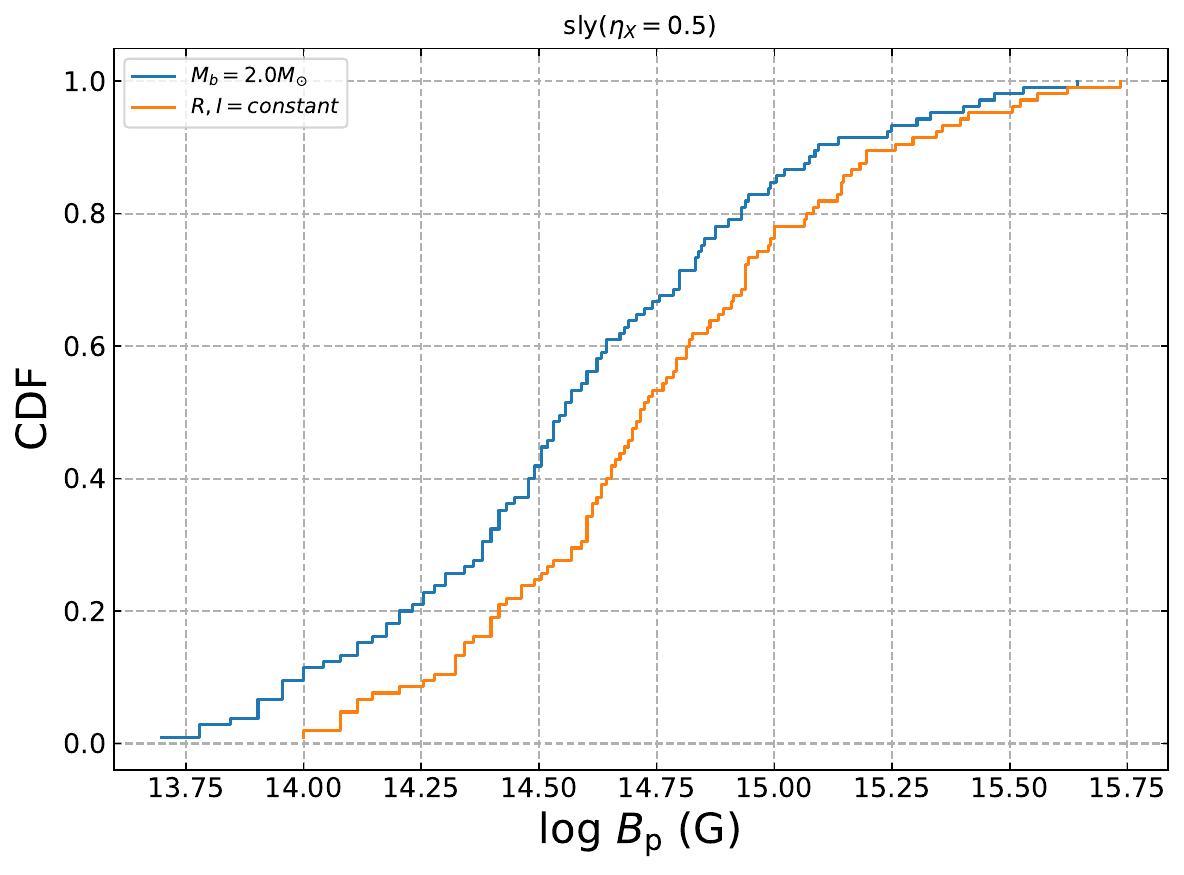}
\includegraphics [angle=0,scale=0.4]{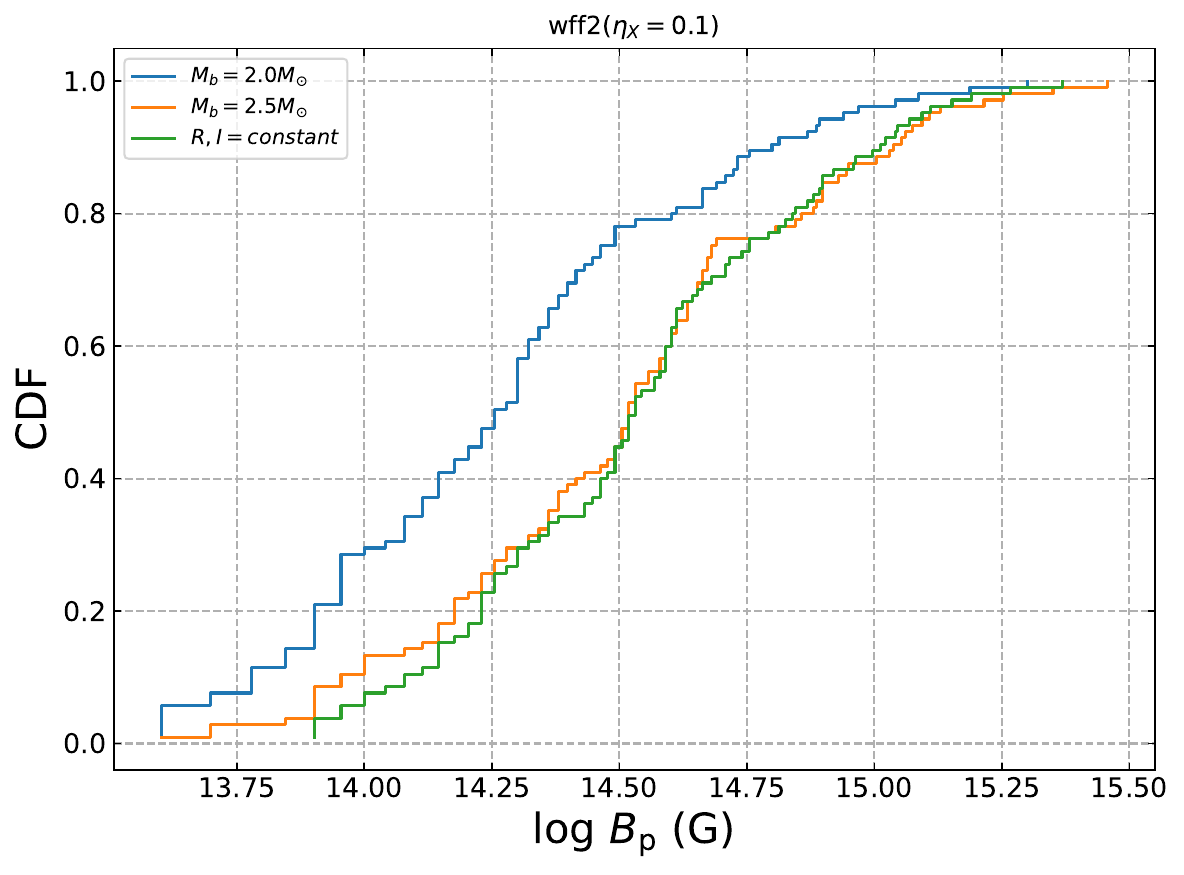}
\includegraphics [angle=0,scale=0.4]{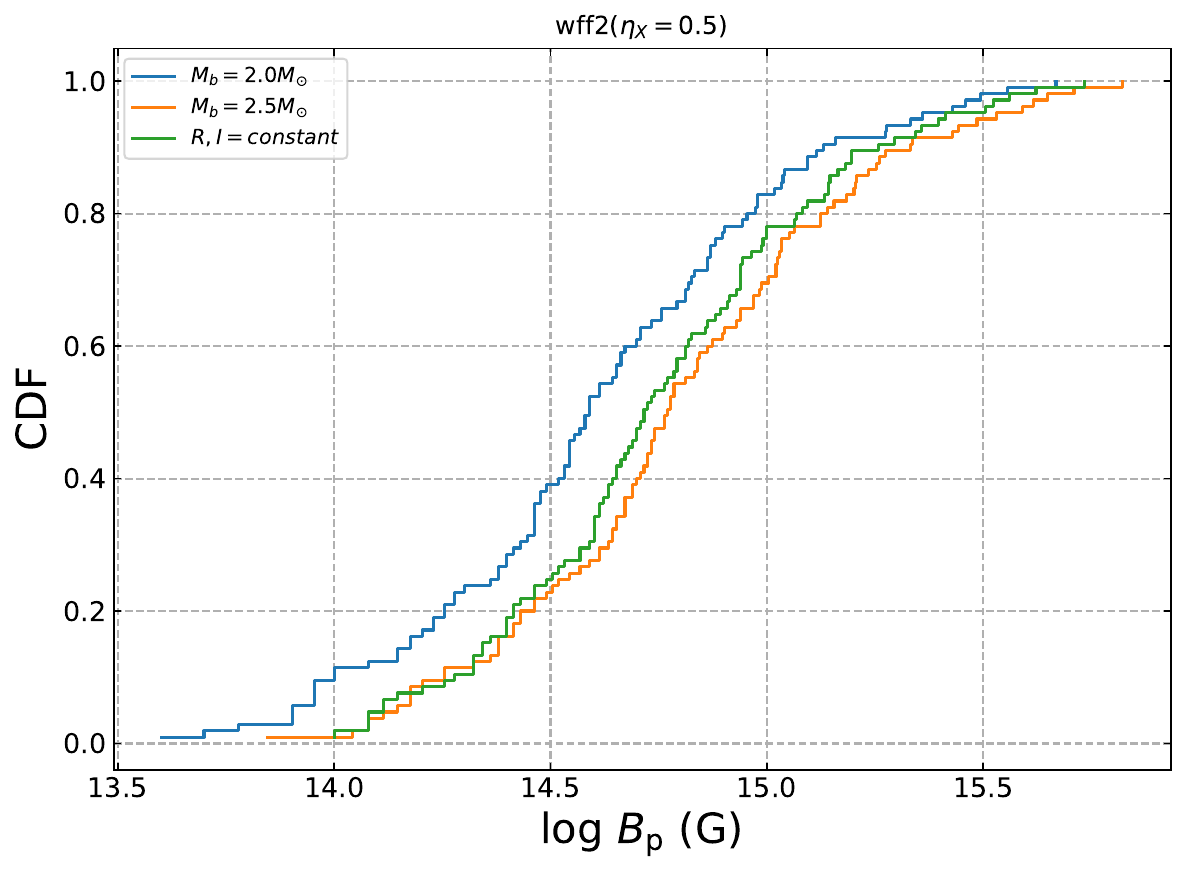}
\caption{The cumulative distributions for the derived magnetar physical parameter $B_p$ between the $R$ and $I$ evolution correction and the constant $R$ and $I$ in four samples of EoSs with $M_{b}=2.0~M_{\odot},~2.5~M_{\odot}$ and $\eta_{\rm X}=0.1,~0.5$, respectively.}
\label{fig:Bp-cdf distribution}
\end{figure*}

\begin{figure*}
\centering
\includegraphics [angle=0,scale=0.4]  {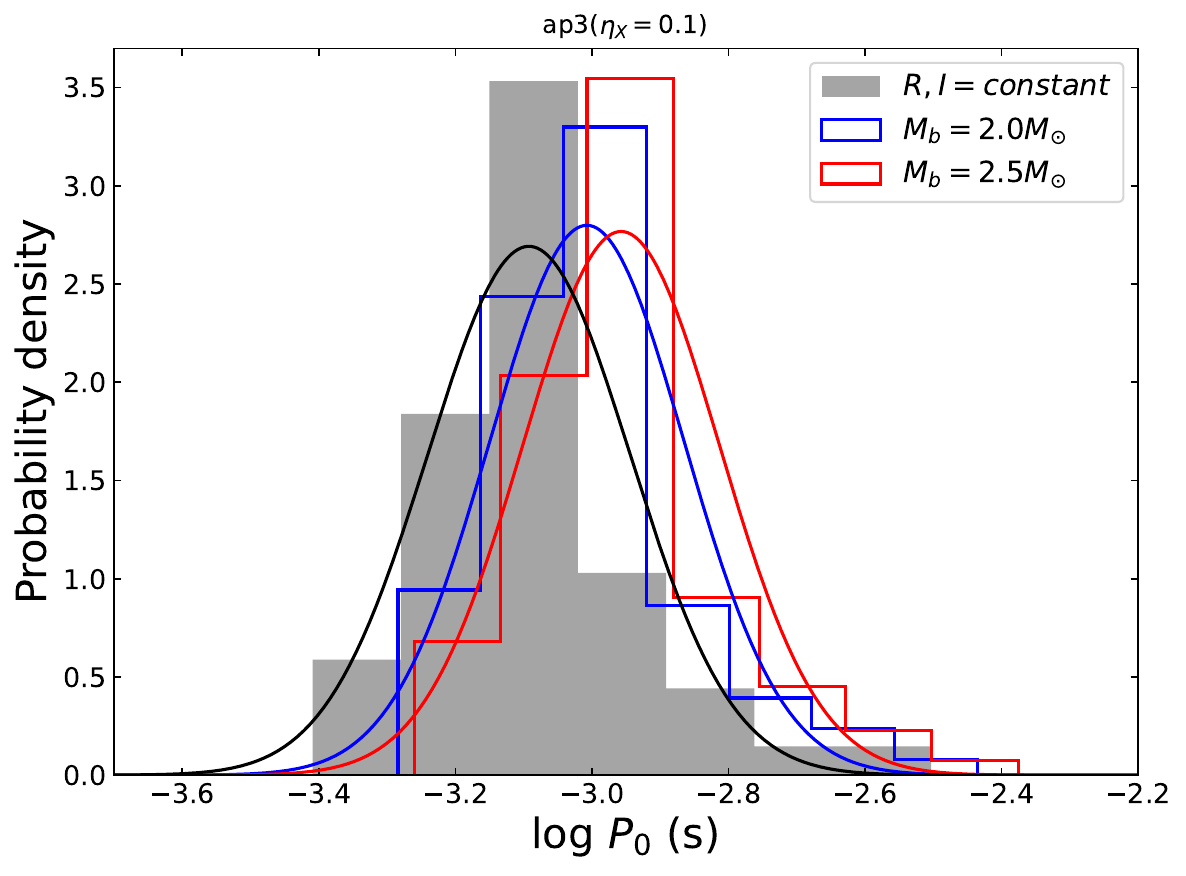}
\includegraphics [angle=0,scale=0.4]  {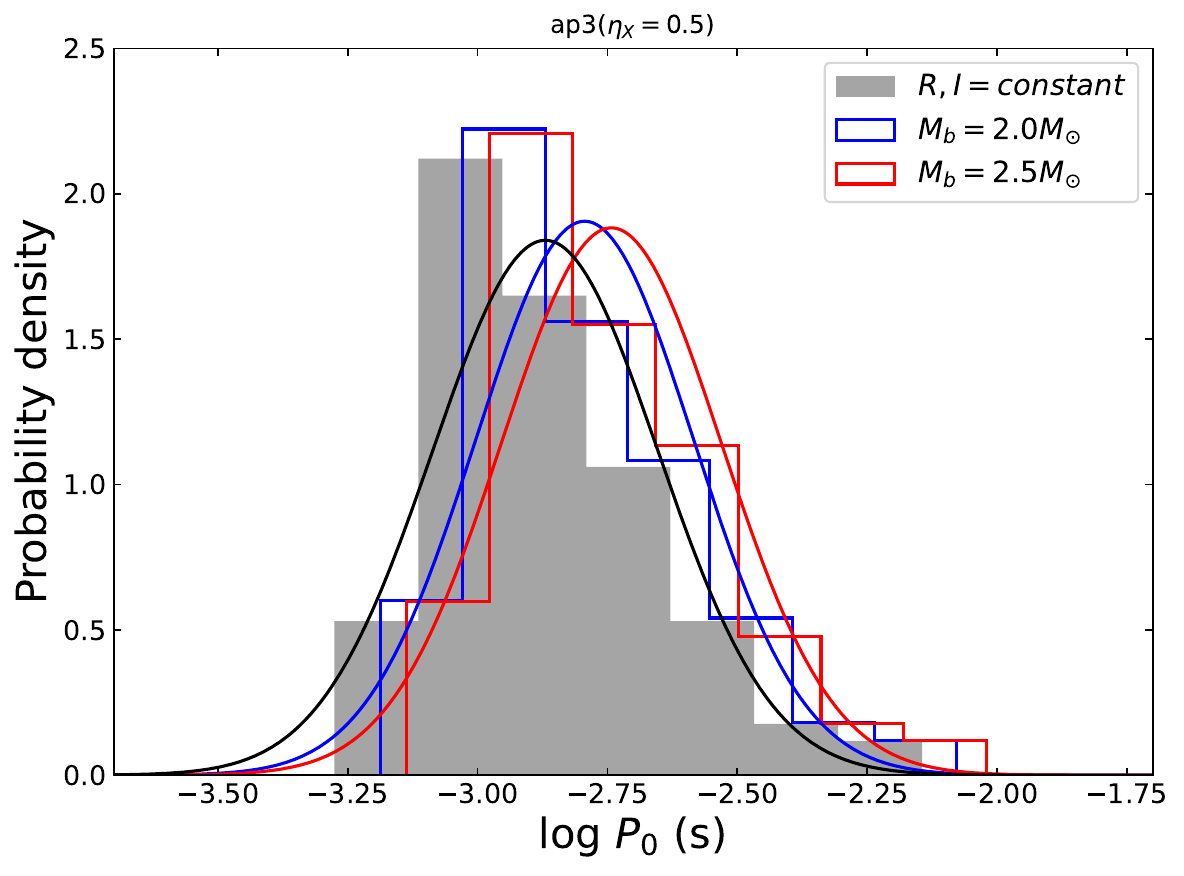}
\includegraphics [angle=0,scale=0.4]  {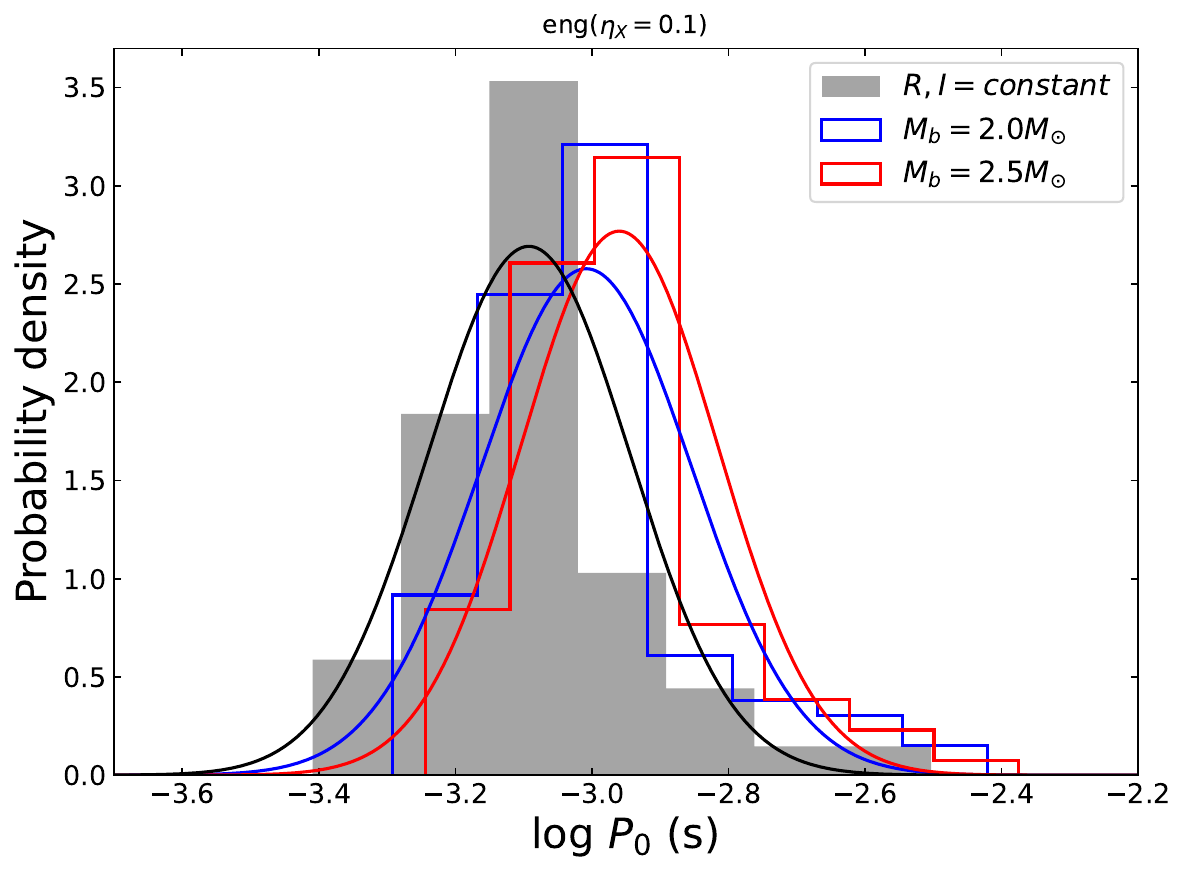}
\includegraphics [angle=0,scale=0.4]  {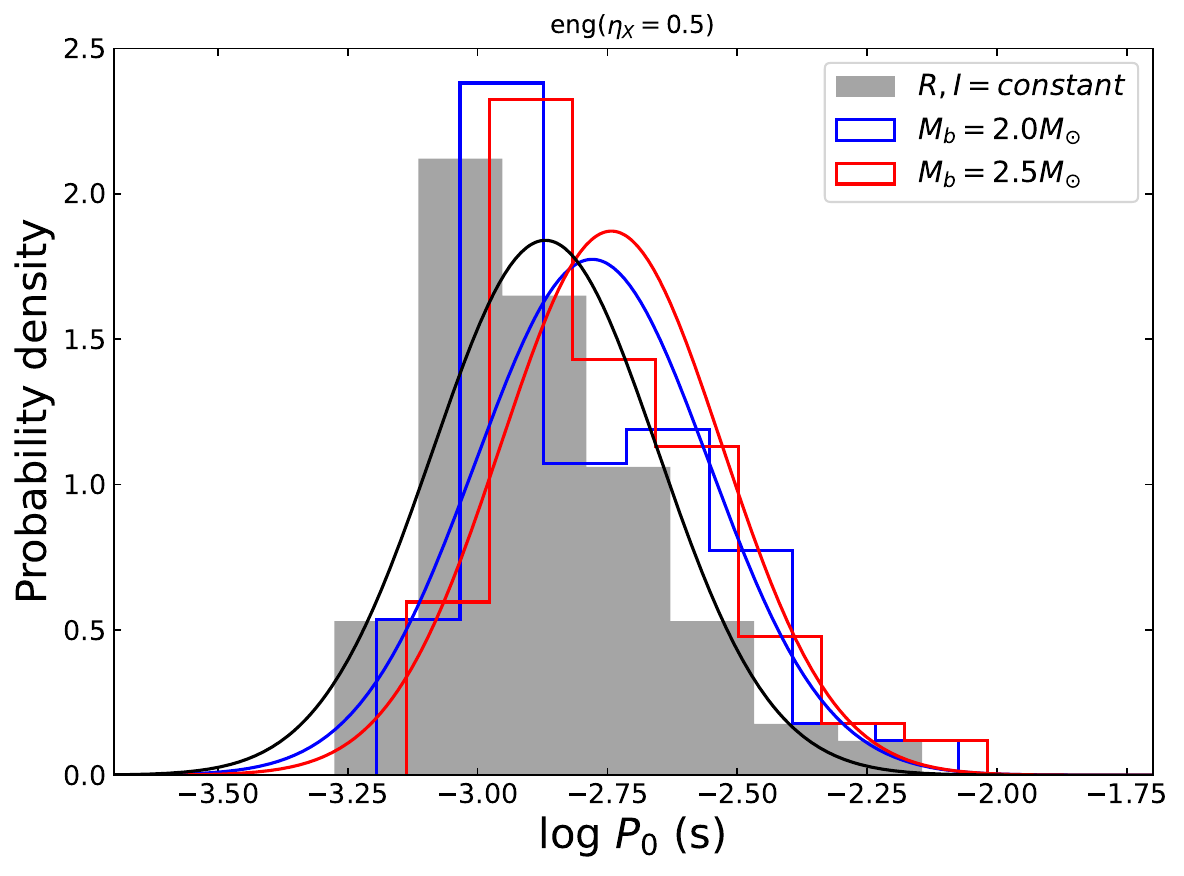}
\includegraphics [angle=0,scale=0.4]  {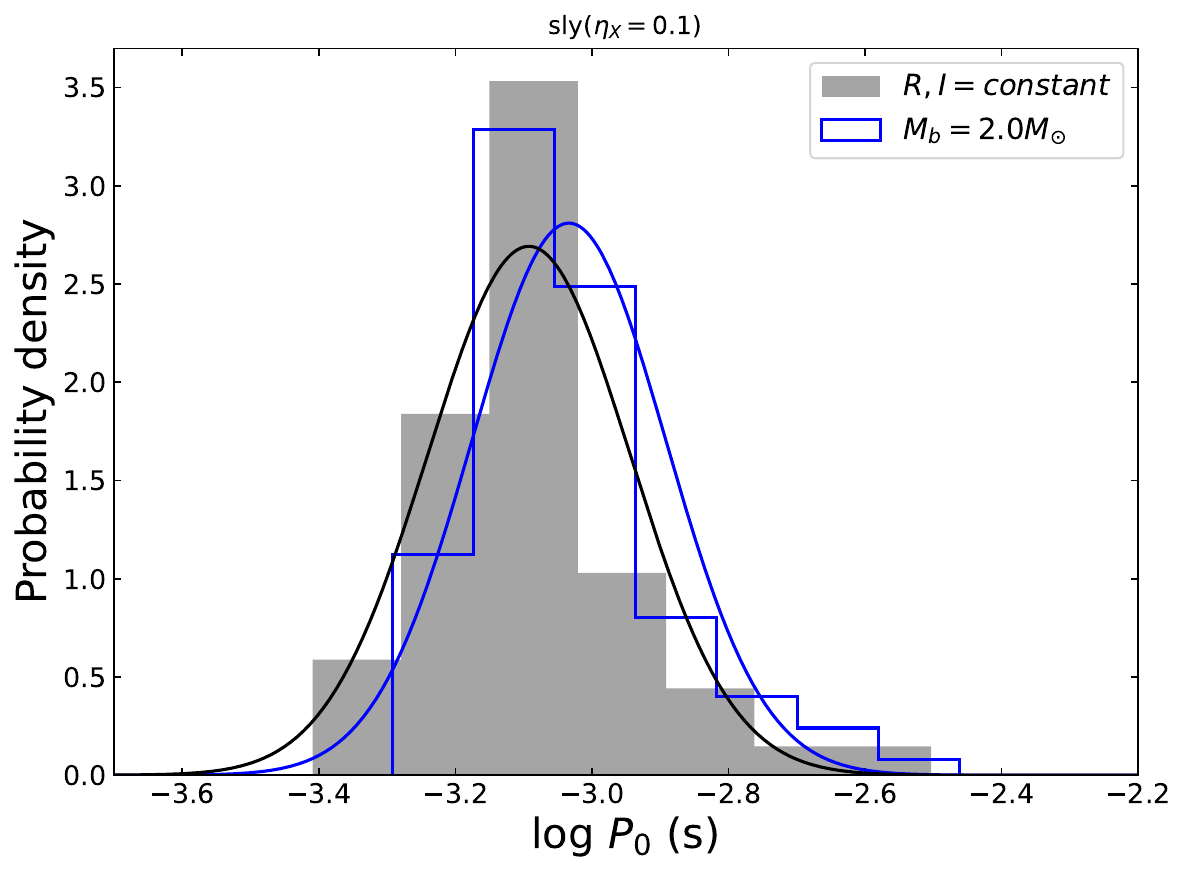}
\includegraphics [angle=0,scale=0.4]  {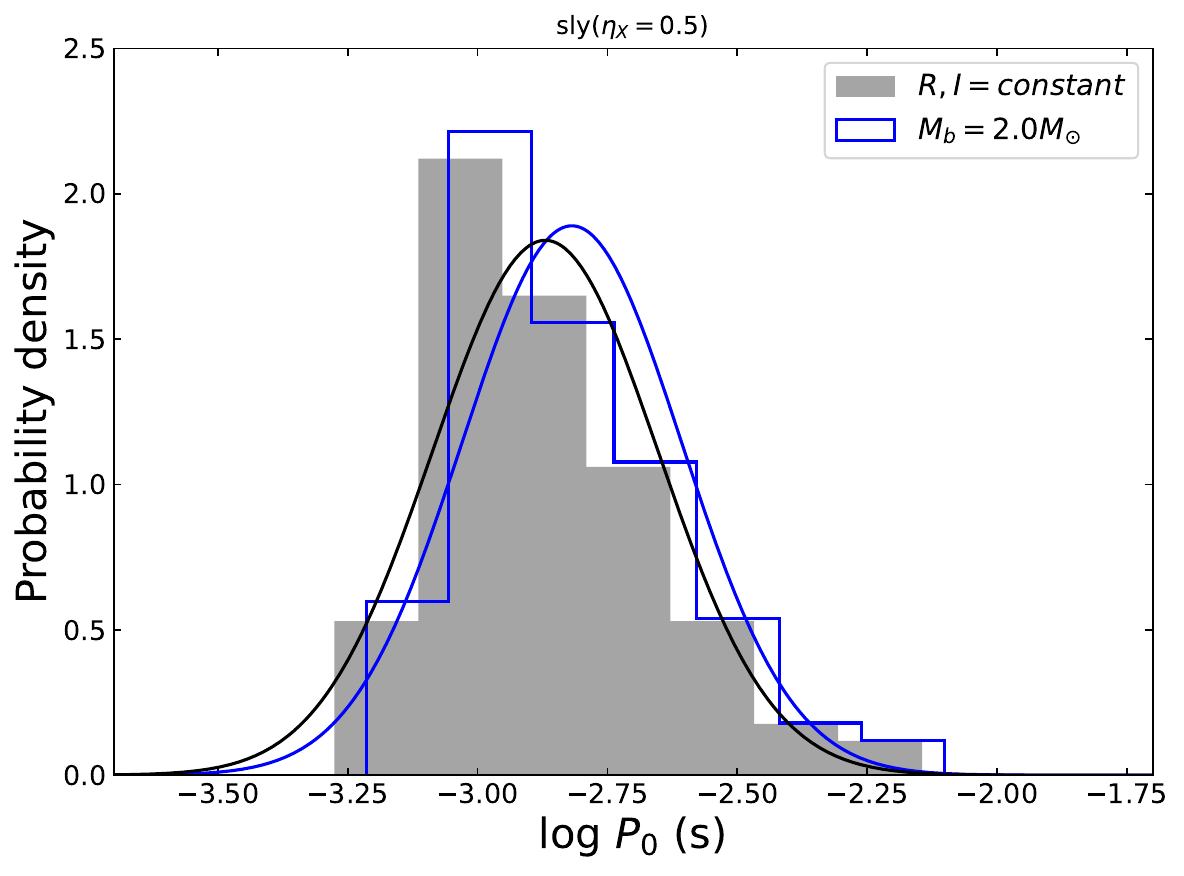}
\includegraphics [angle=0,scale=0.4]  {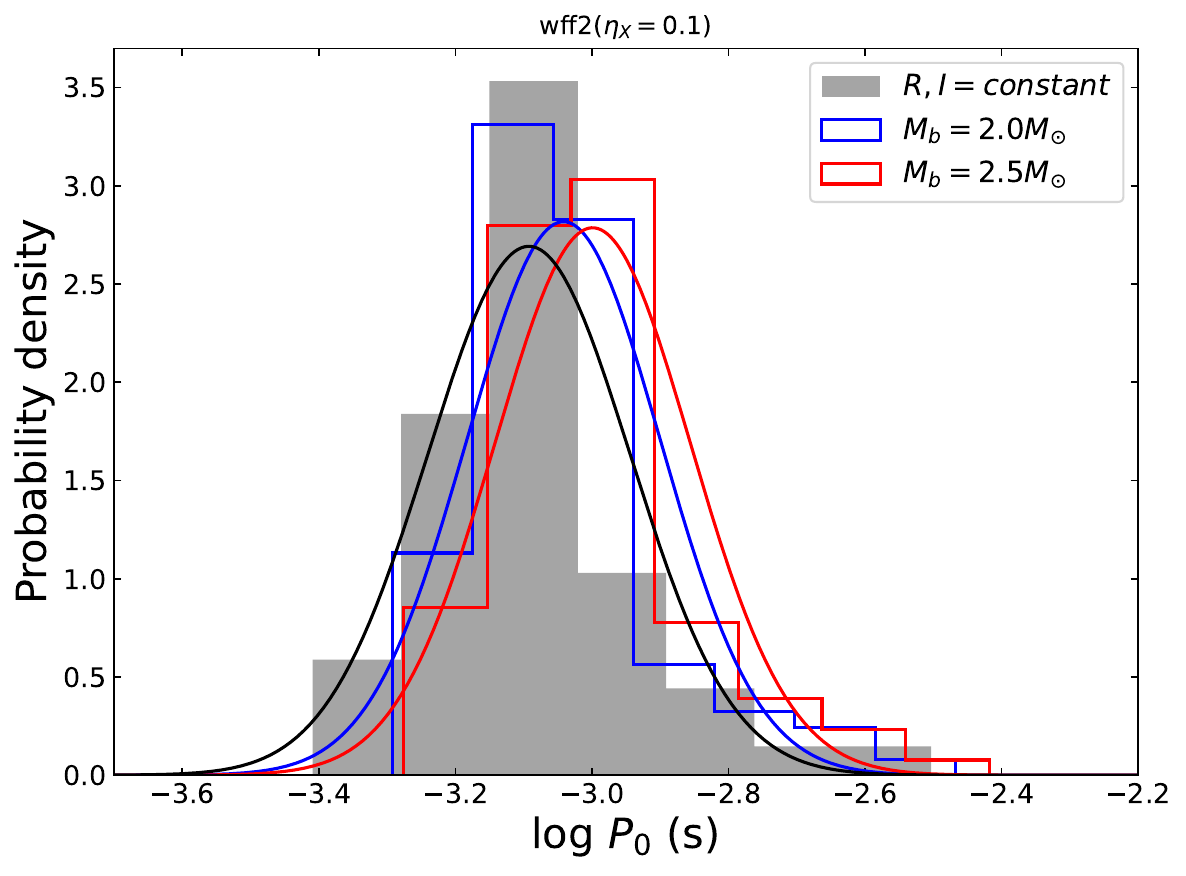}
\includegraphics [angle=0,scale=0.4]  {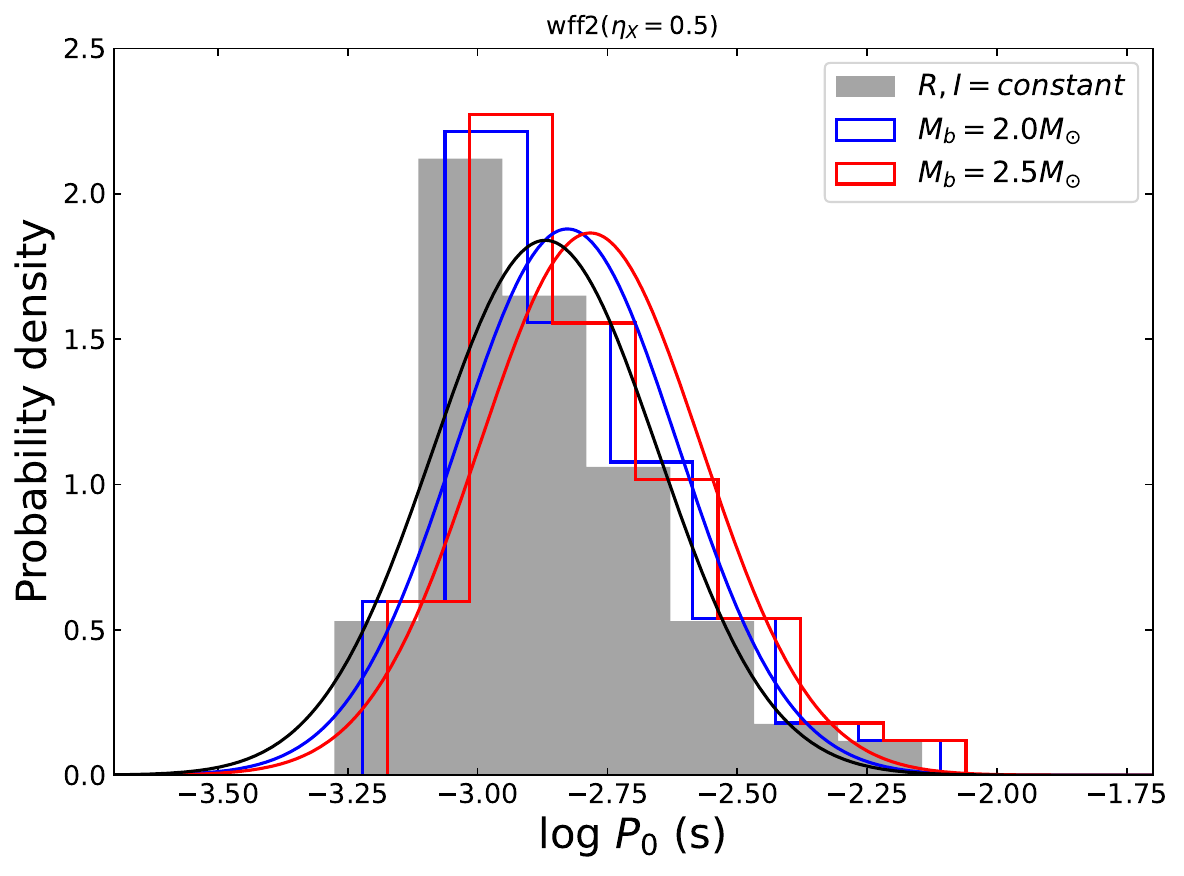}
\caption{Comparisons of the derived magnetar physical parameter $P_0$ histograms between the $R$ and $I$ evolution correction and the constant $R$ and $I$ in four samples of EoSs with $M_{b}=2.0~M_{\odot},~2.5~M_{\odot}$ and $\eta_{\rm X}=0.1,~0.5$, respectively.}
\label{fig:P0-distribution}
\end{figure*}

\begin{figure*}
\centering
\includegraphics [angle=0,scale=0.4] {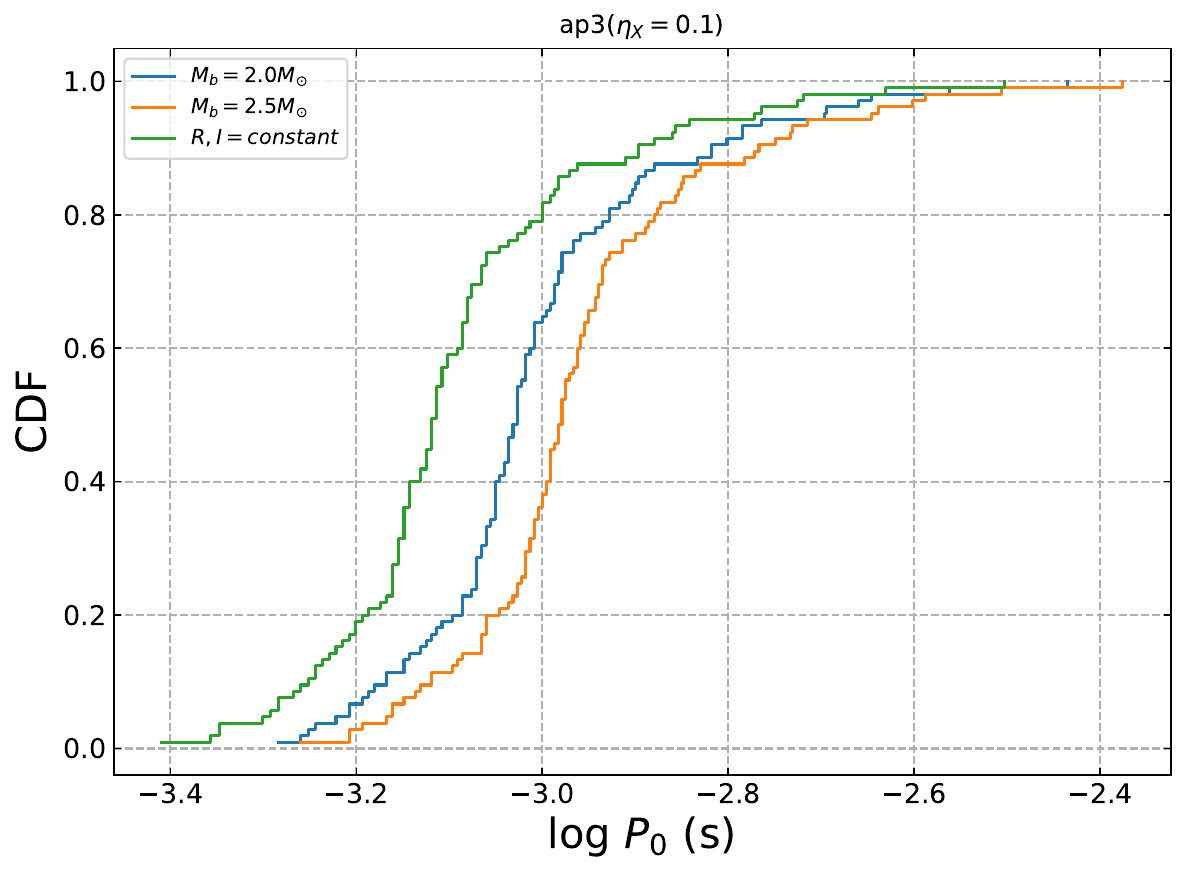}
\includegraphics [angle=0,scale=0.4] {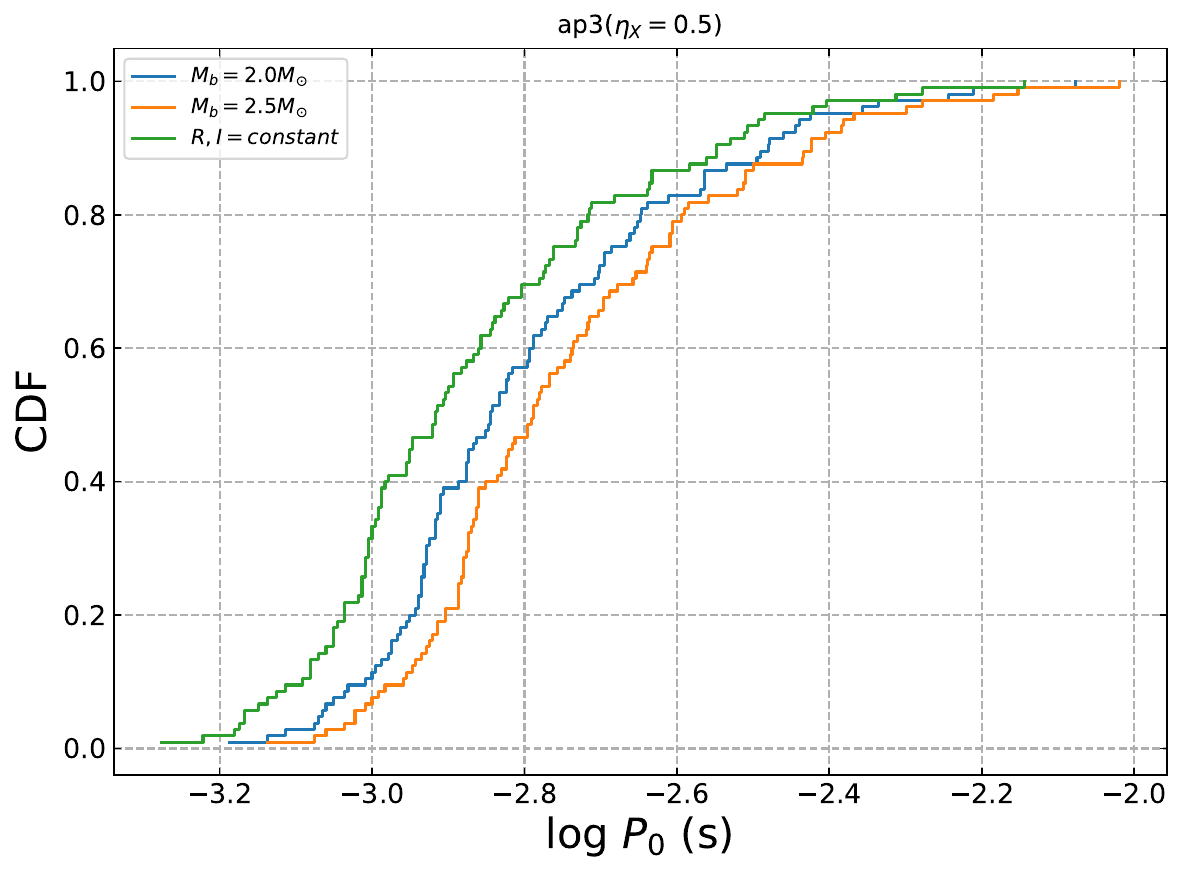}
\includegraphics [angle=0,scale=0.4] {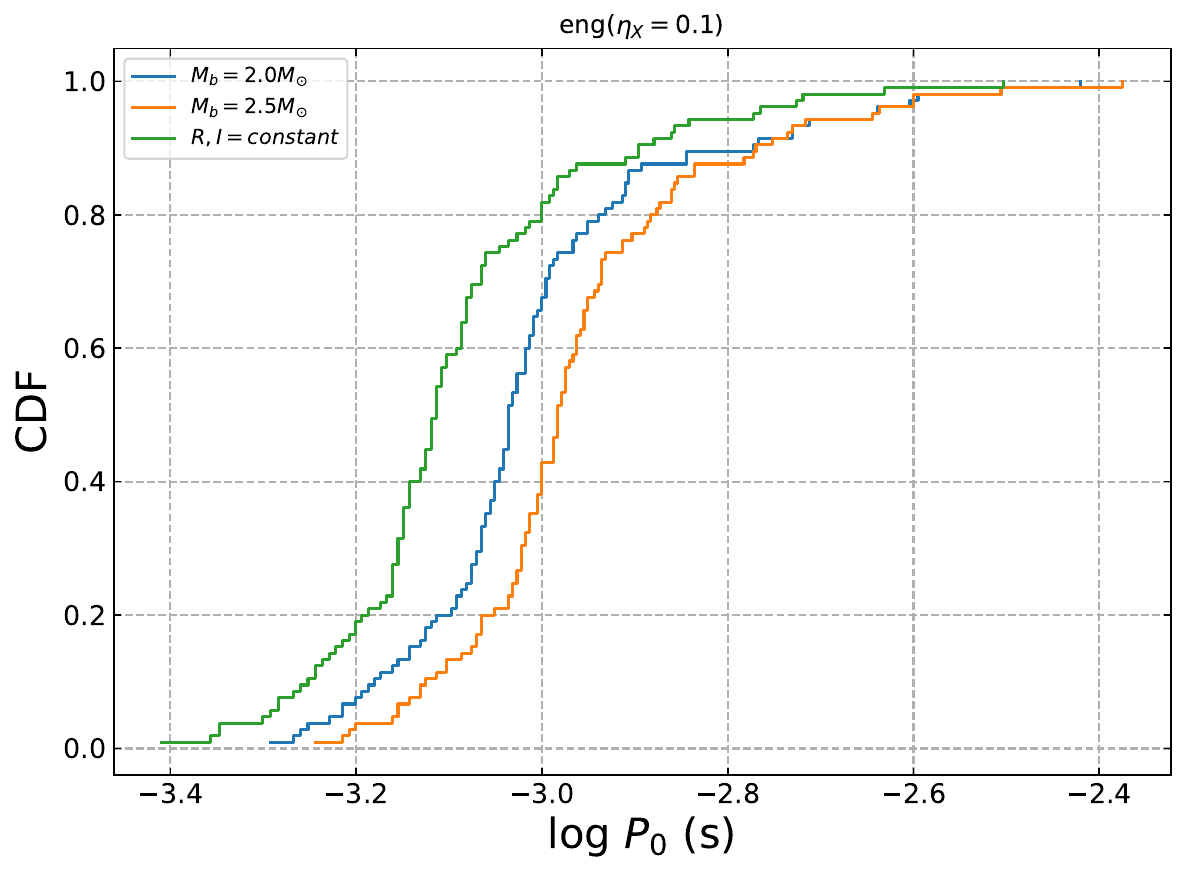}
\includegraphics [angle=0,scale=0.4] {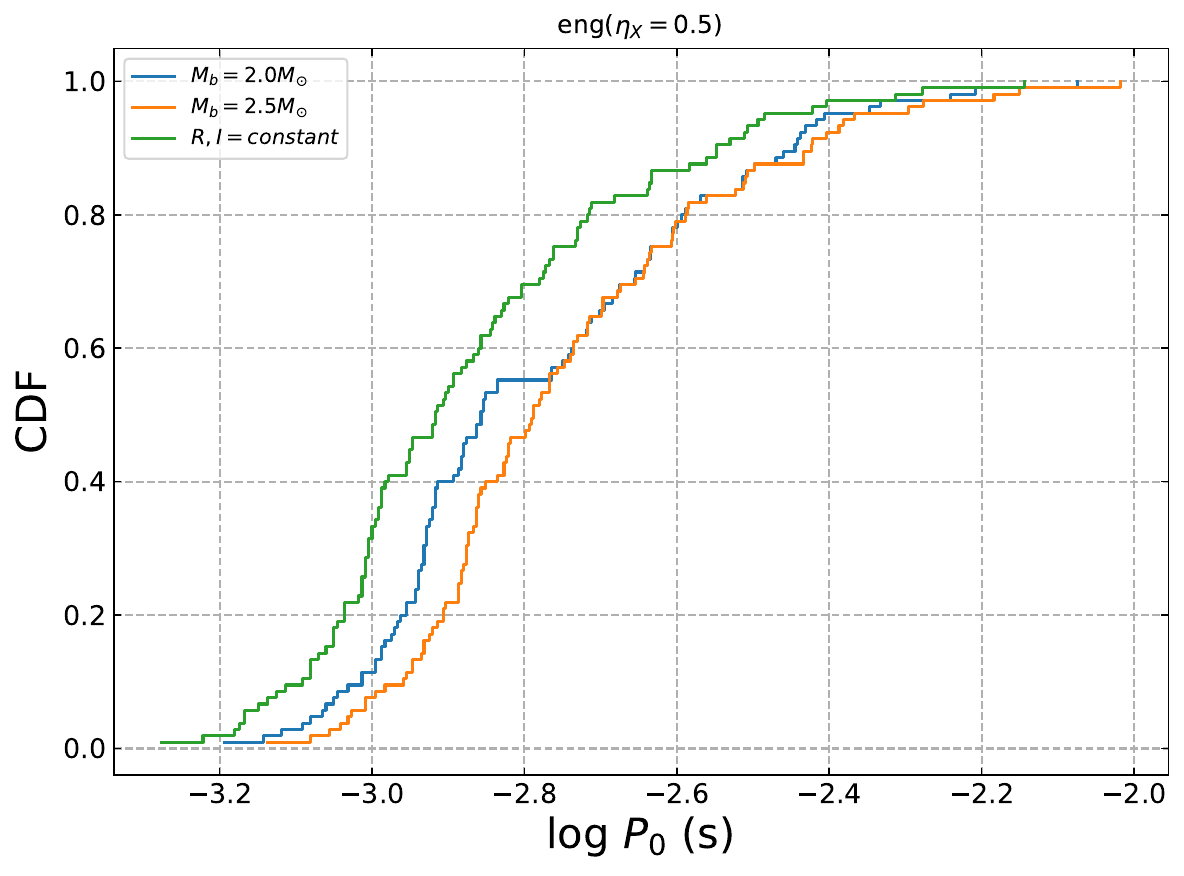}
\includegraphics [angle=0,scale=0.4] {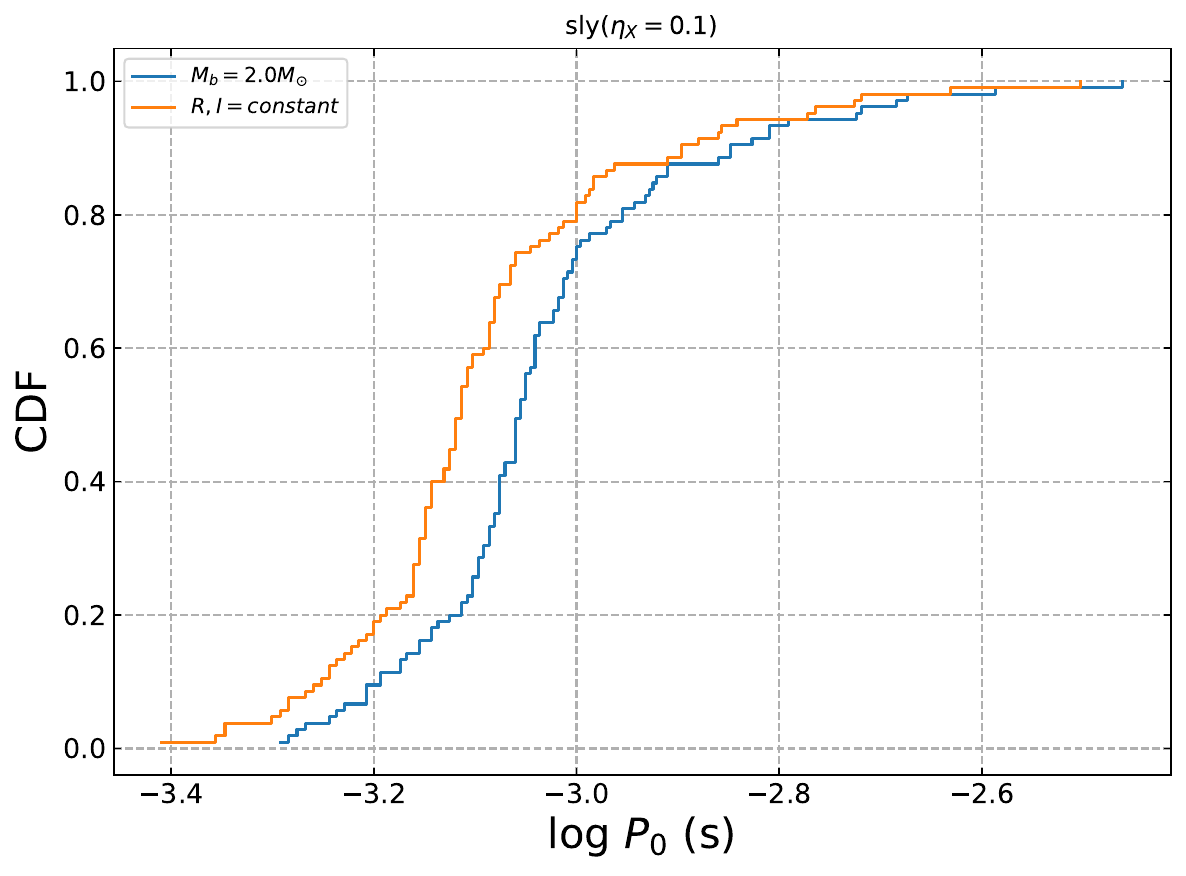}
\includegraphics [angle=0,scale=0.4] {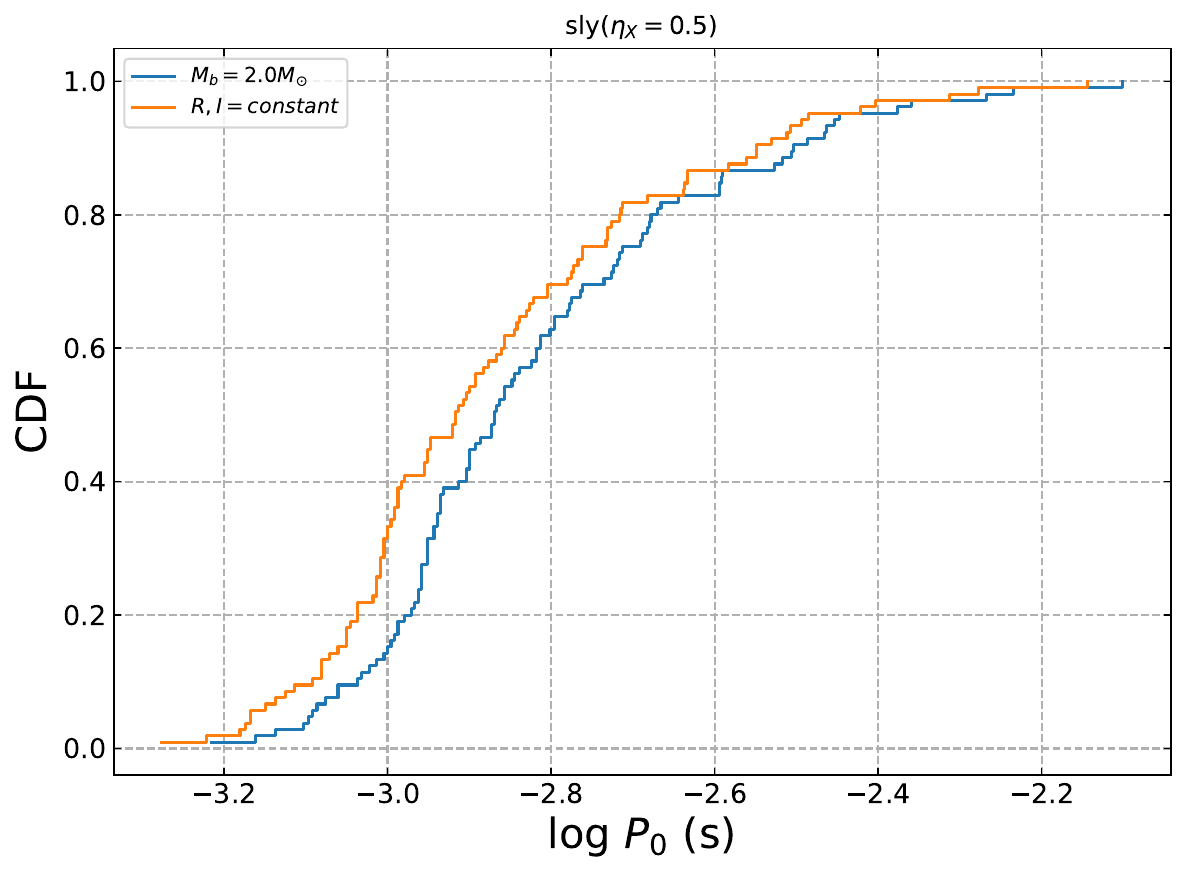}
\includegraphics [angle=0,scale=0.4]{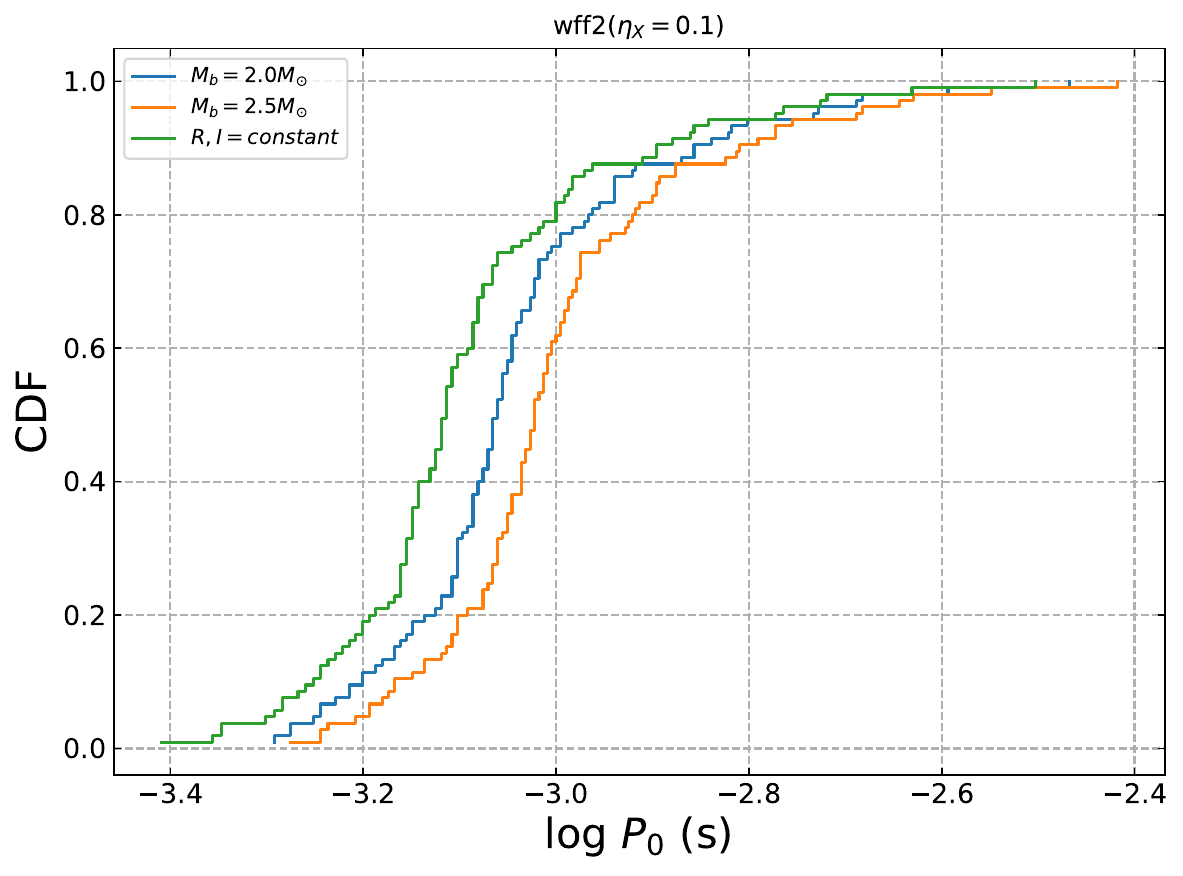}
\includegraphics [angle=0,scale=0.4]{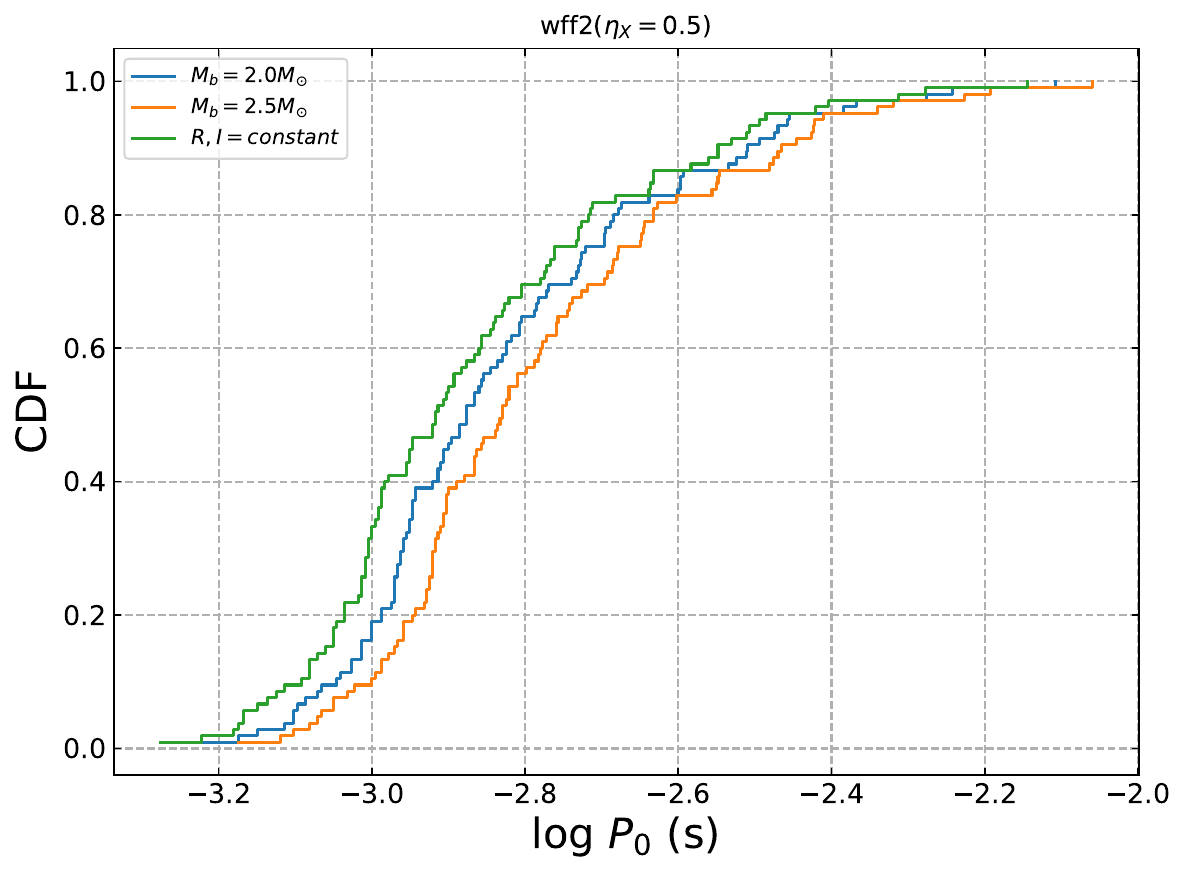}
\caption{The cumulative distributions for the derived magnetar physical parameter $P_0$ between the $R$ and $I$ evolution correction and the constant $R$ and $I$ in four samples of EoSs with $M_{b}=2.0~M_{\odot},~2.5~M_{\odot}$ and $\eta_{\rm X}=0.1,~0.5$, respectively.}
\label{fig:P0-cdf distribution}
\end{figure*}

\begin{figure*}
\centering
\includegraphics [angle=0,scale=0.4]  {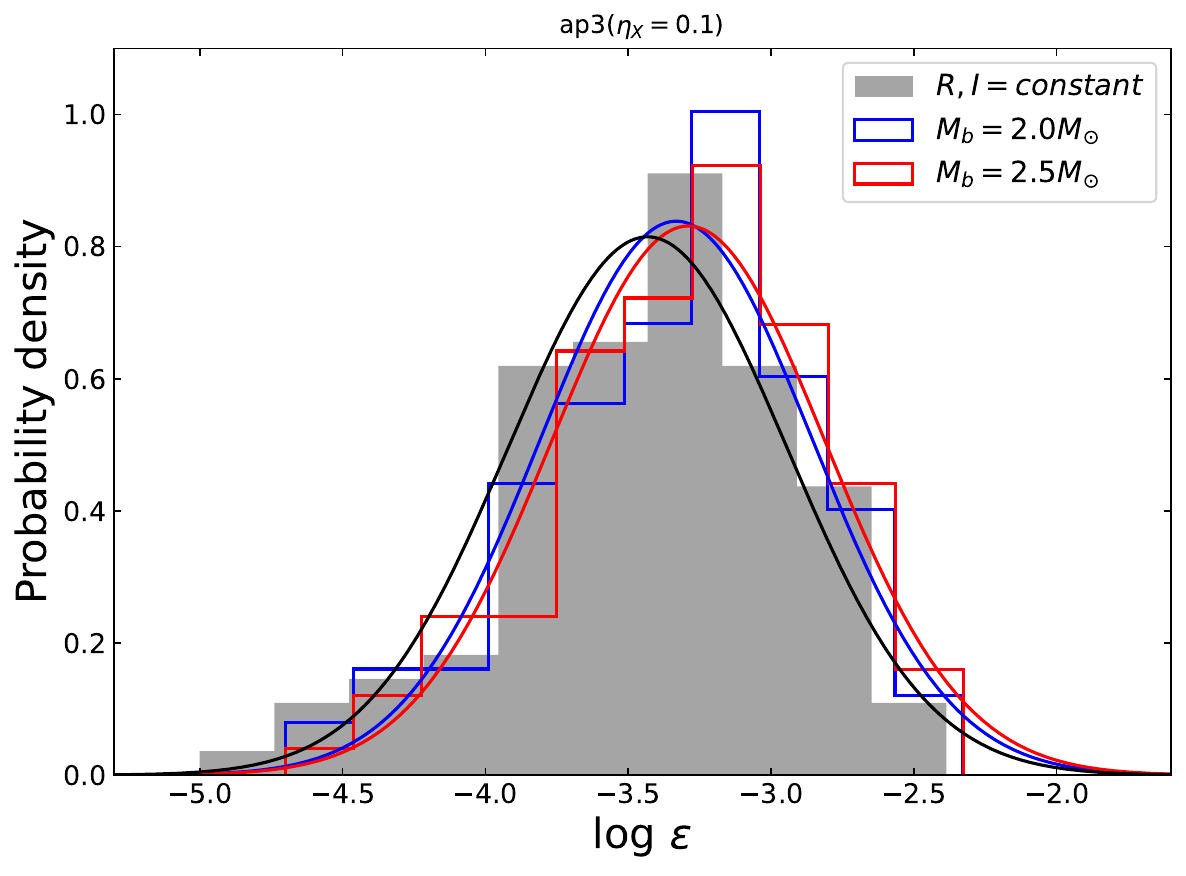}
\includegraphics [angle=0,scale=0.4]  {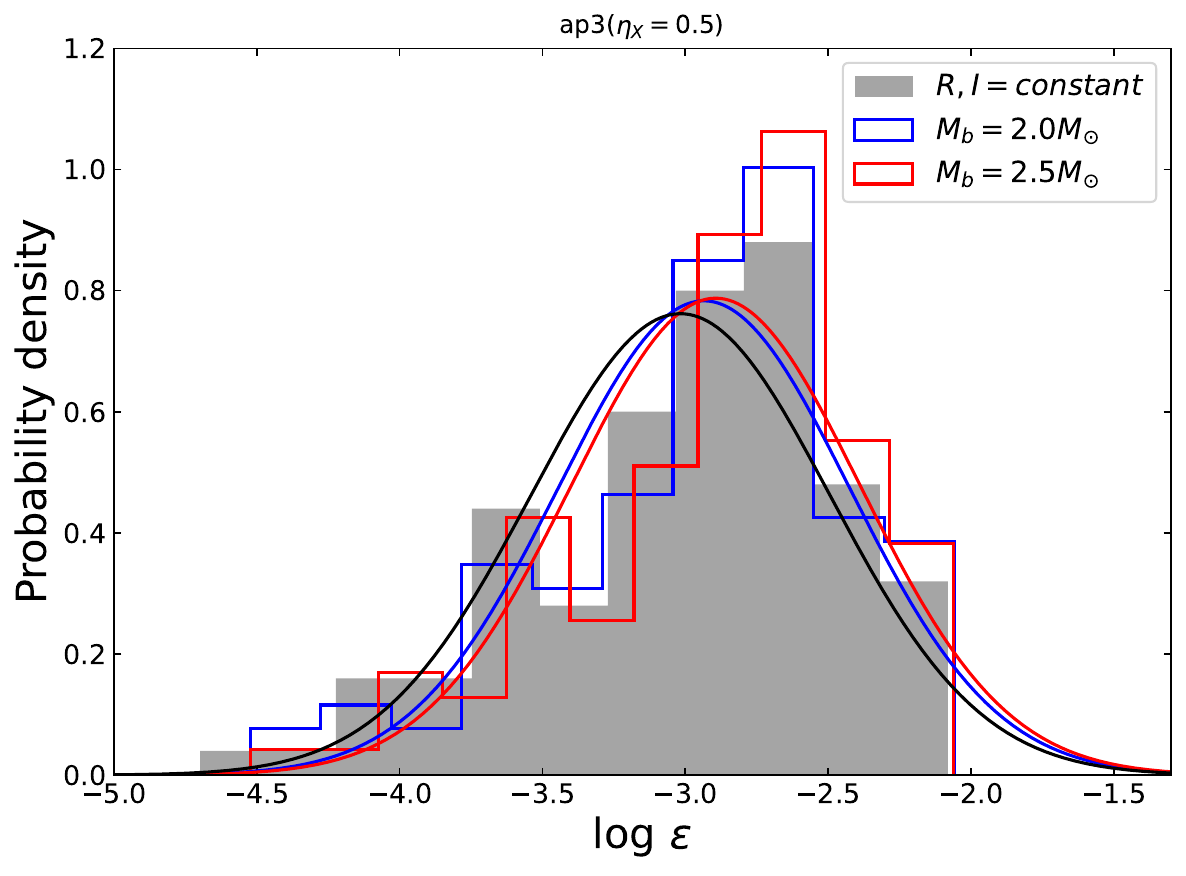}
\includegraphics [angle=0,scale=0.4]  {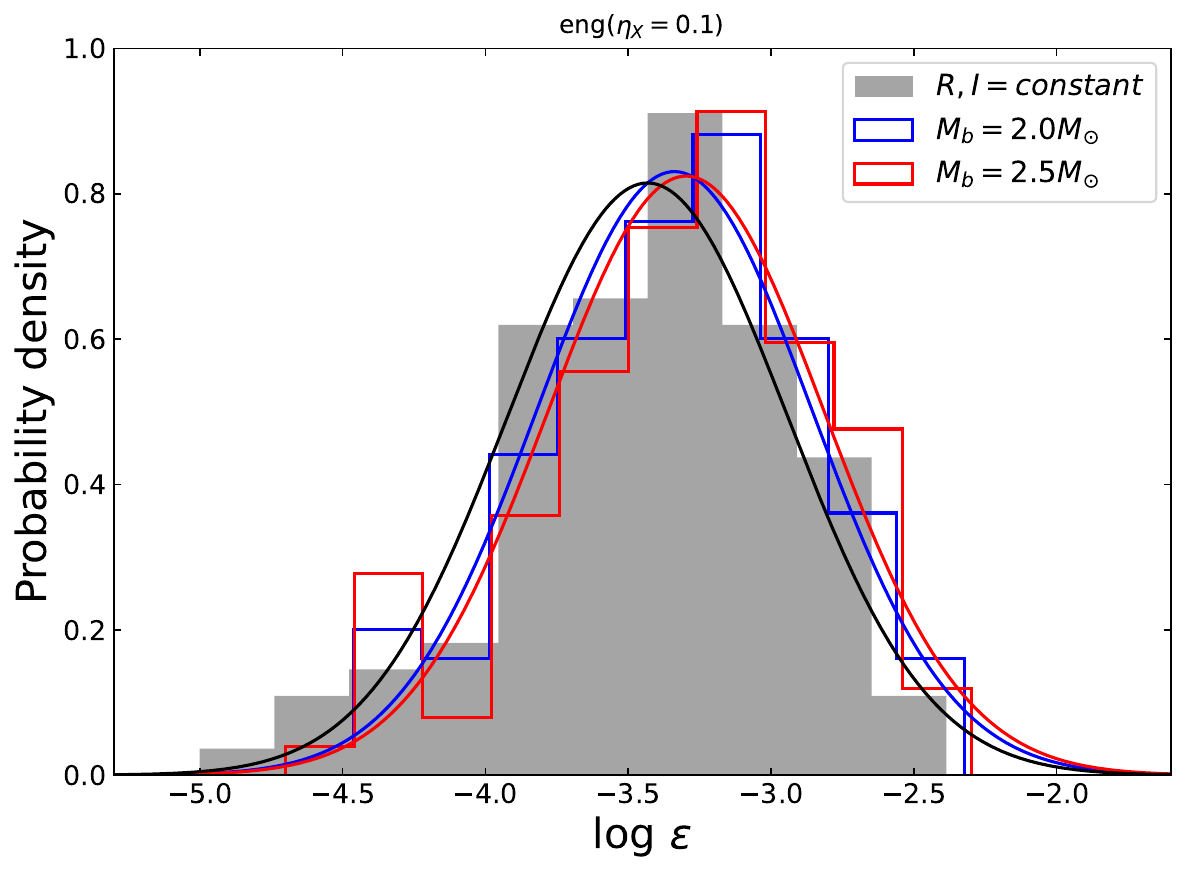}
\includegraphics [angle=0,scale=0.4]  {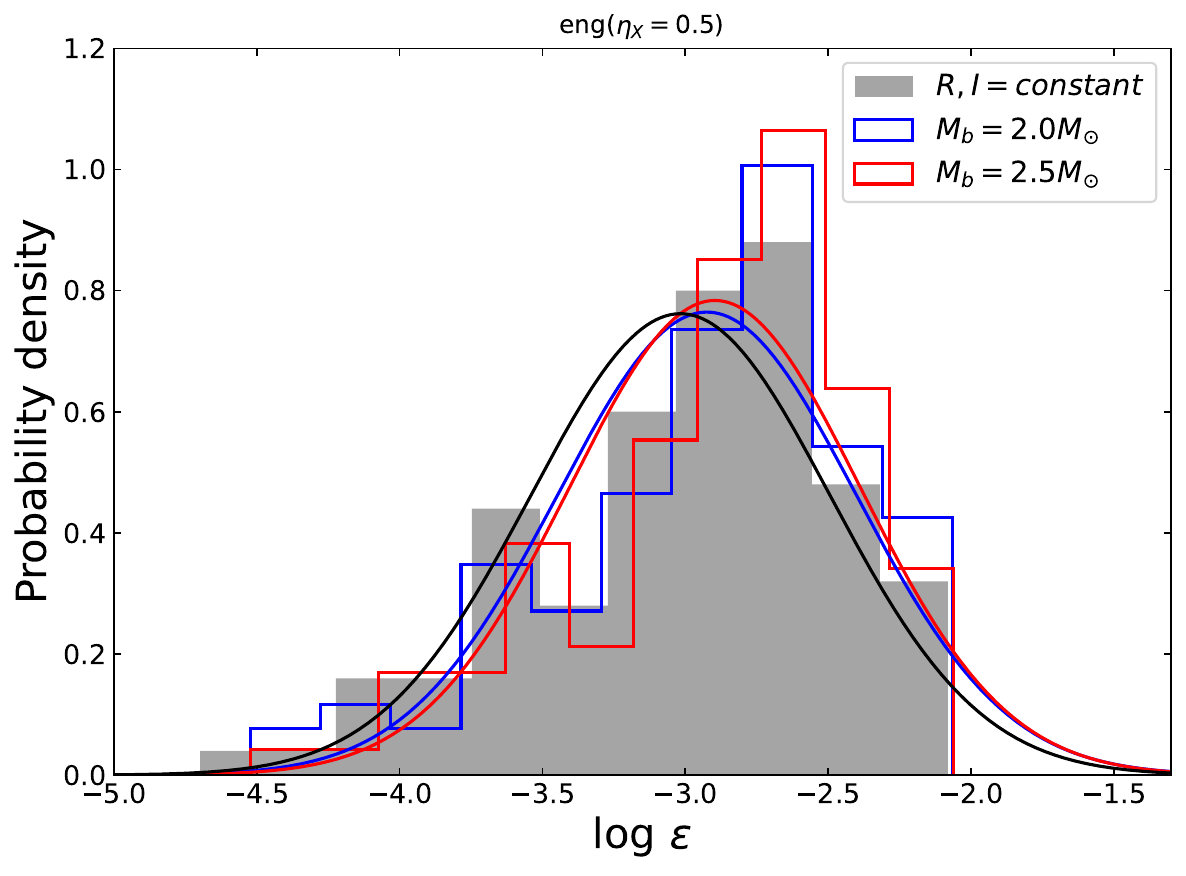}
\includegraphics [angle=0,scale=0.4]  {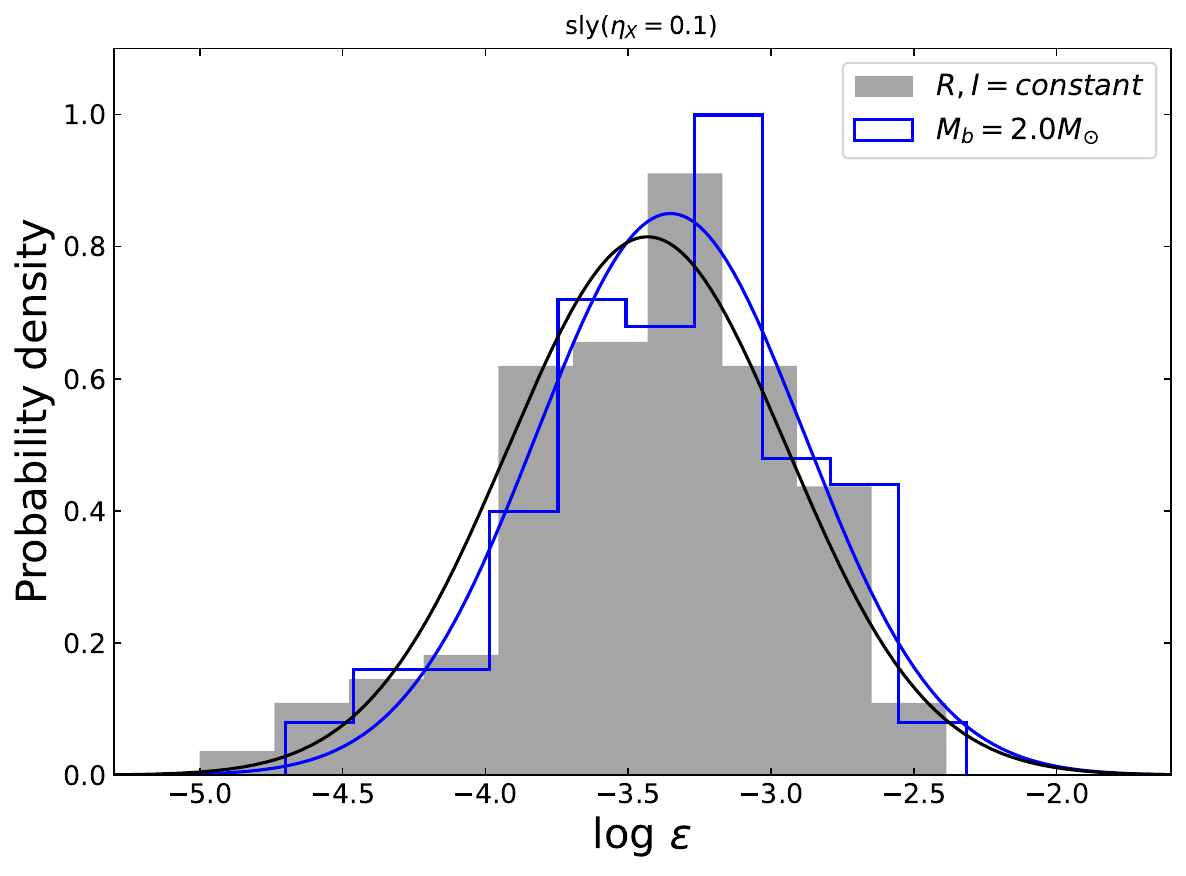}
\includegraphics [angle=0,scale=0.4]  {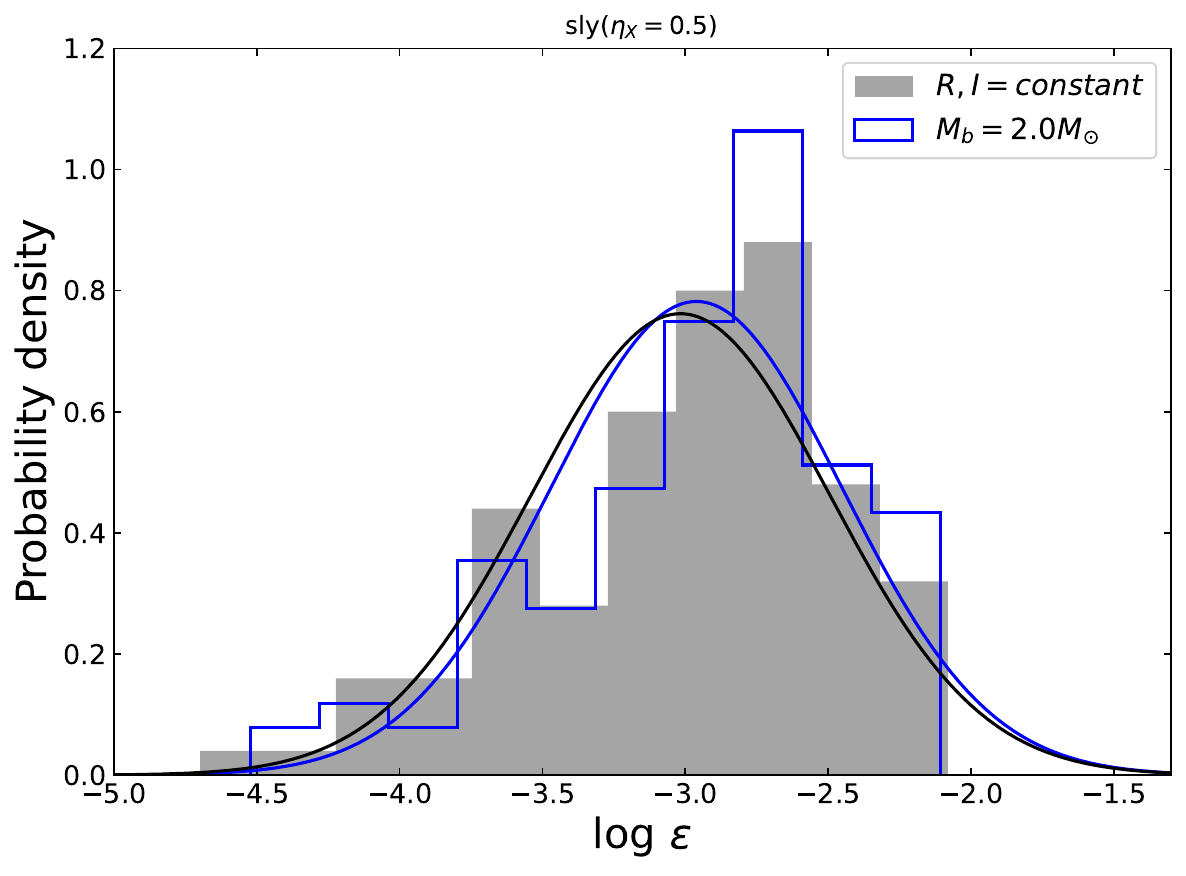}
\includegraphics [angle=0,scale=0.4]  {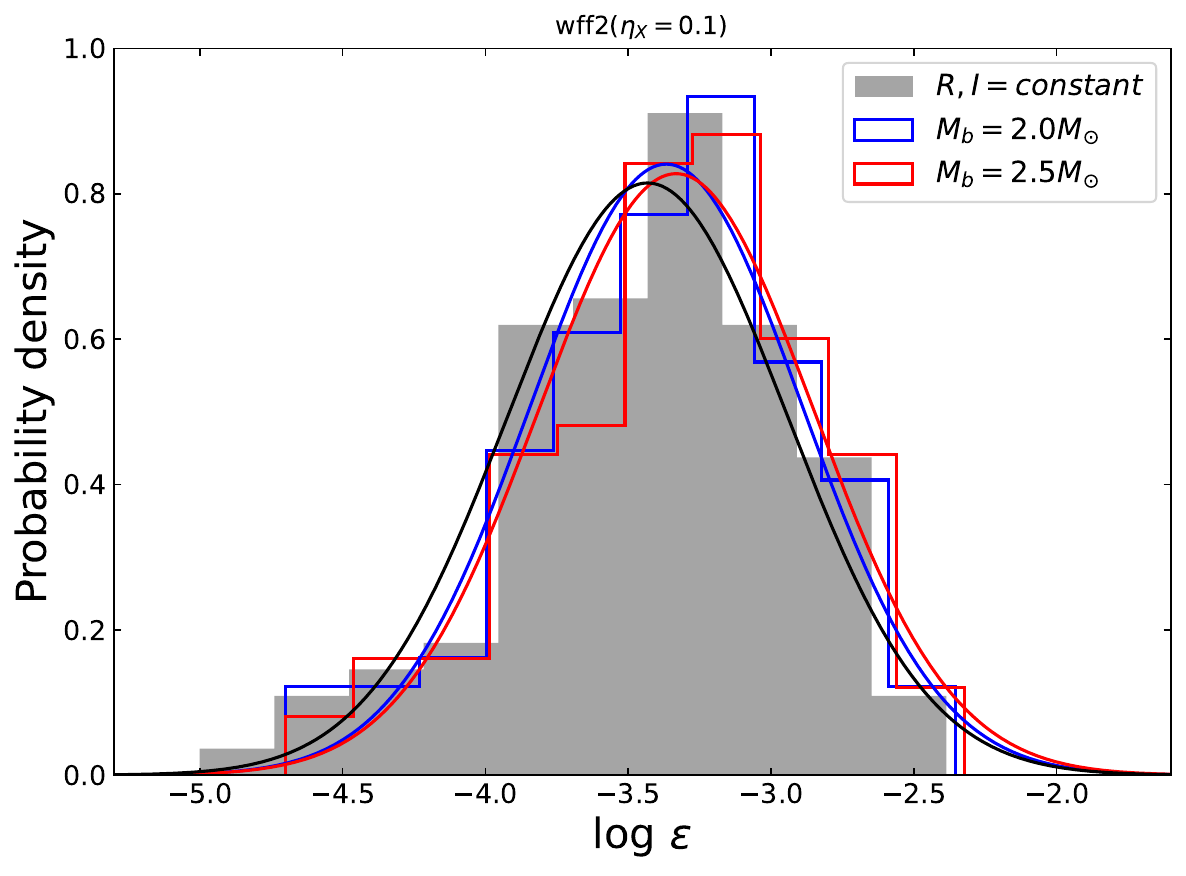}
\includegraphics [angle=0,scale=0.4]  {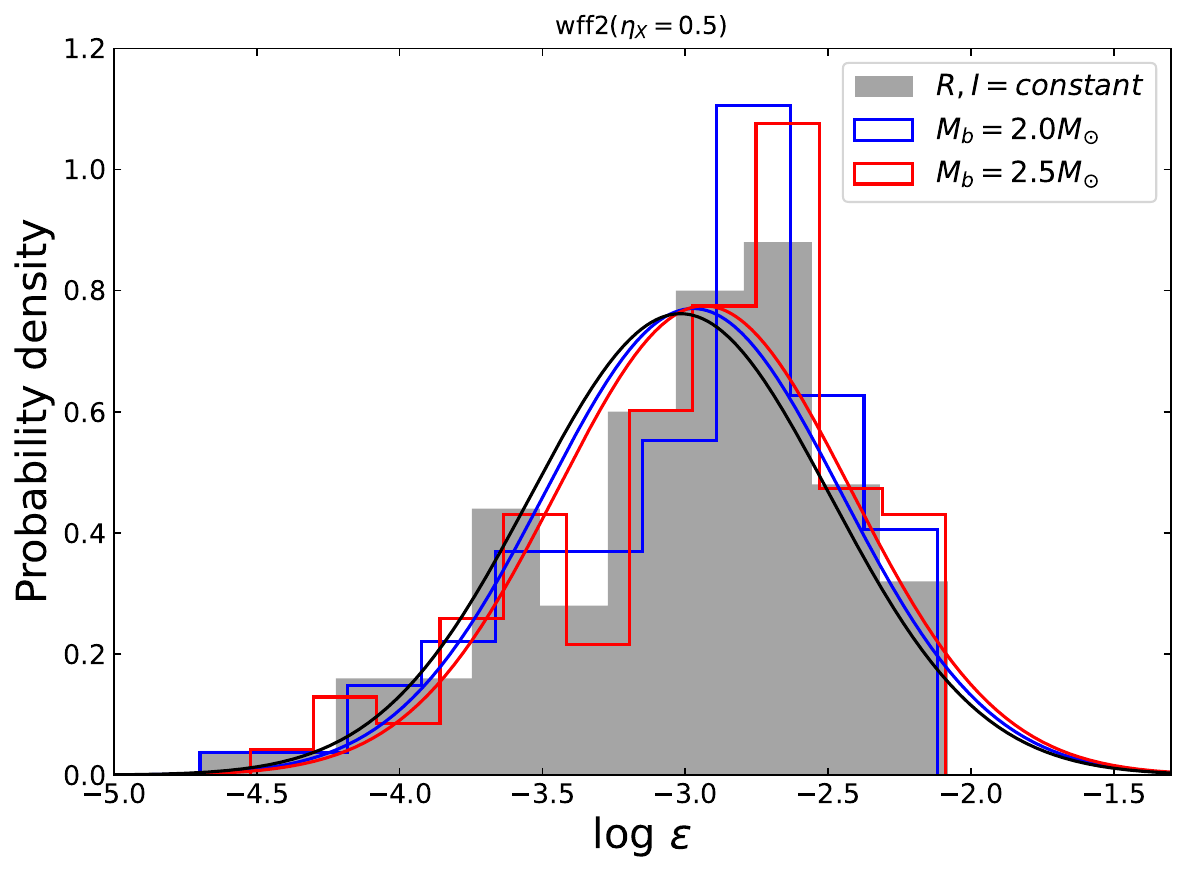}
\caption{Comparisons of the derived magnetar physical parameter $\epsilon$ histograms between the $R$ and $I$ evolution correction and the constant $R$ and $I$ in four samples of EoSs with $M_{b}=2.0~M_{\odot},~2.5~M_{\odot}$ and $\eta_{\rm X}=0.1,~0.5$, respectively.}
\label{fig:epsilon-distribution}
\end{figure*}

\begin{figure*}
\centering
\includegraphics [angle=0,scale=0.4] 
{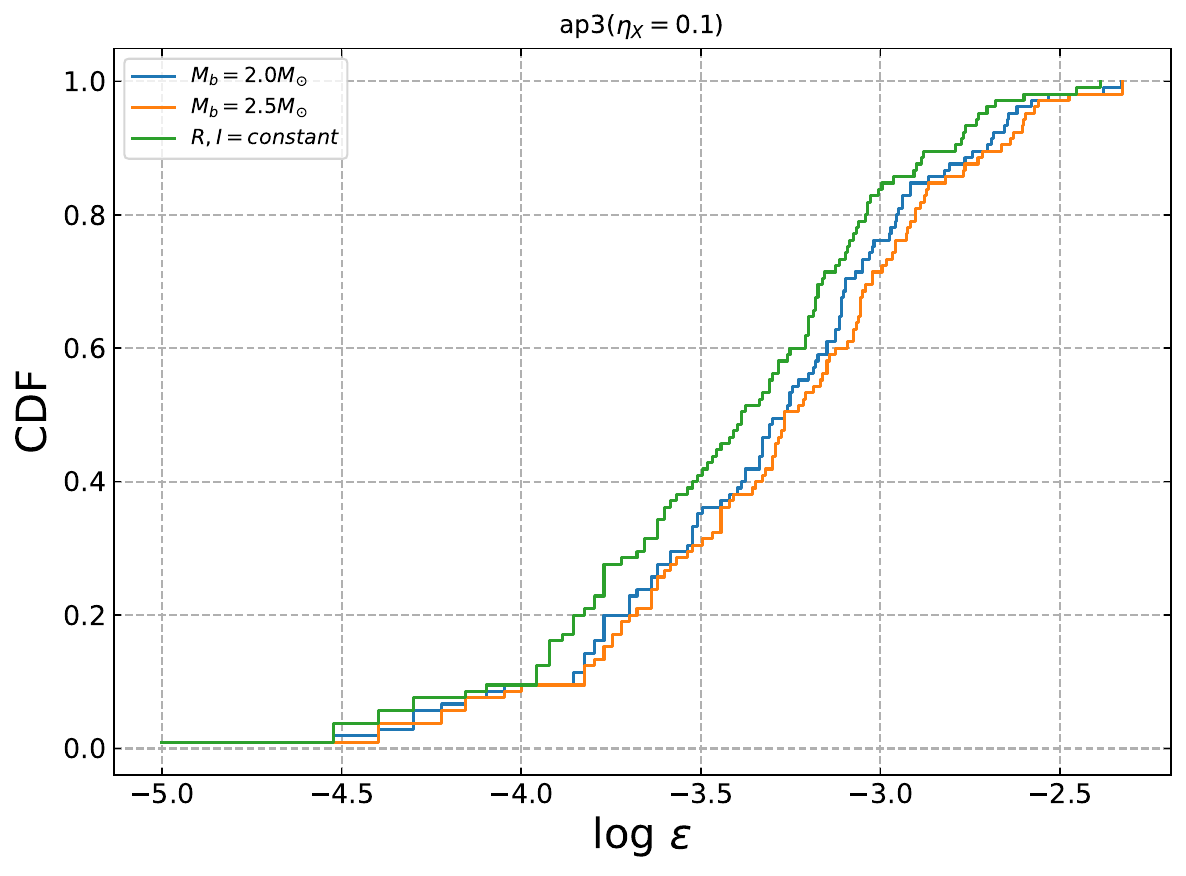}
\includegraphics [angle=0,scale=0.4] 
{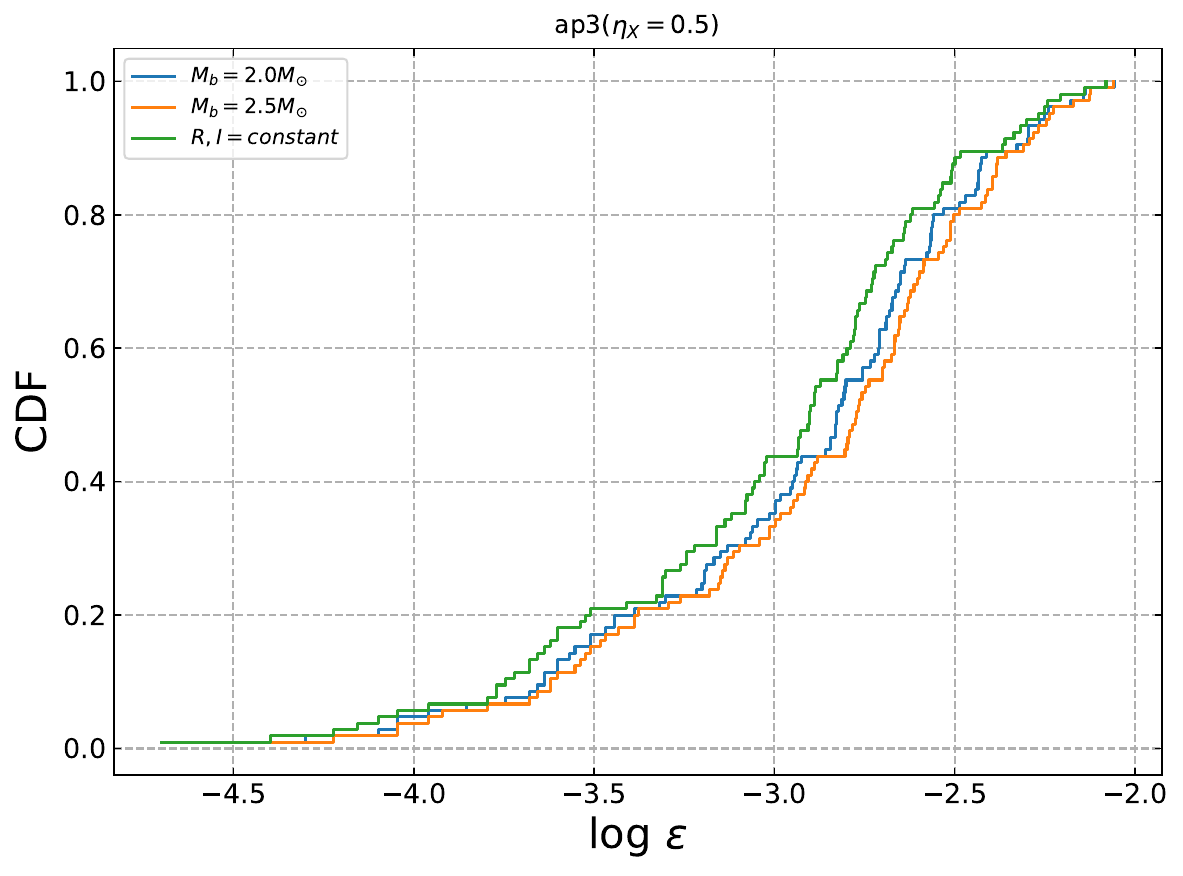}
\includegraphics [angle=0,scale=0.4] 
{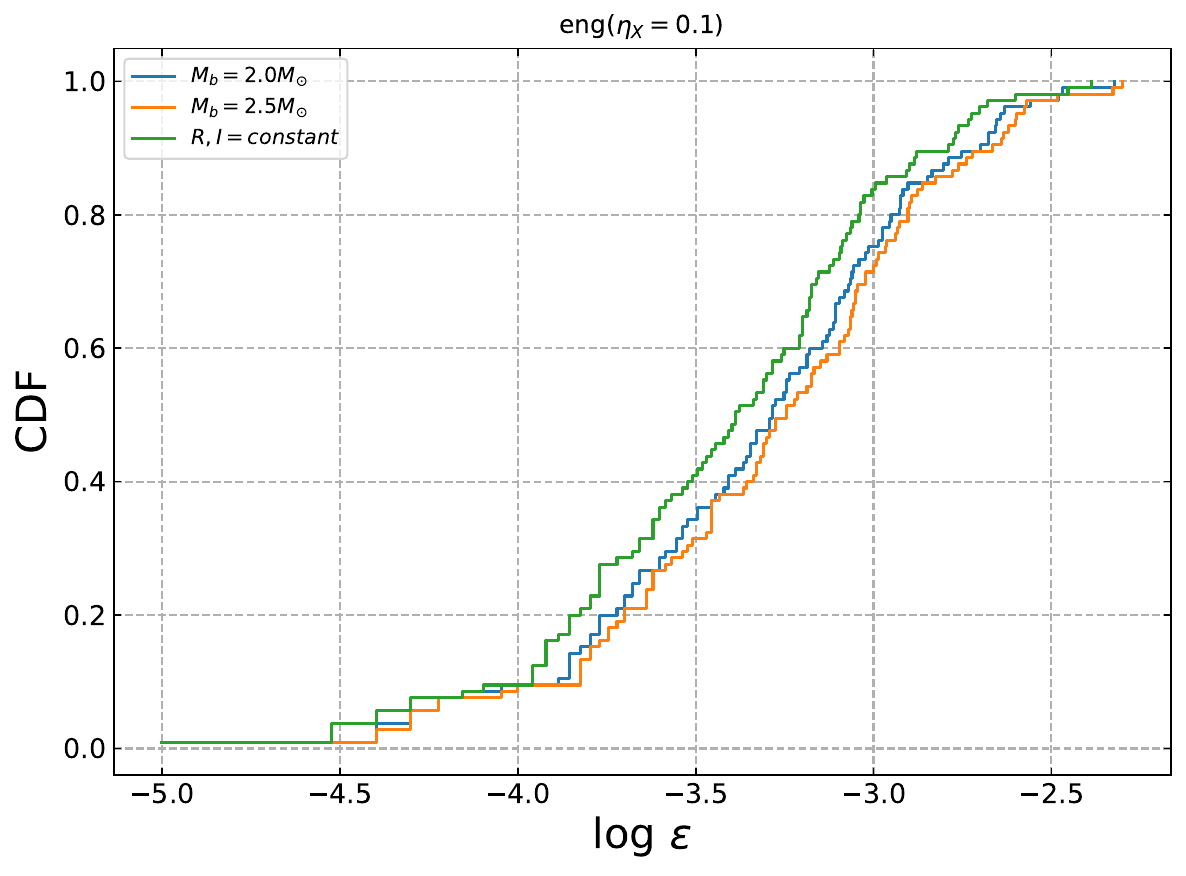}
\includegraphics [angle=0,scale=0.4] 
{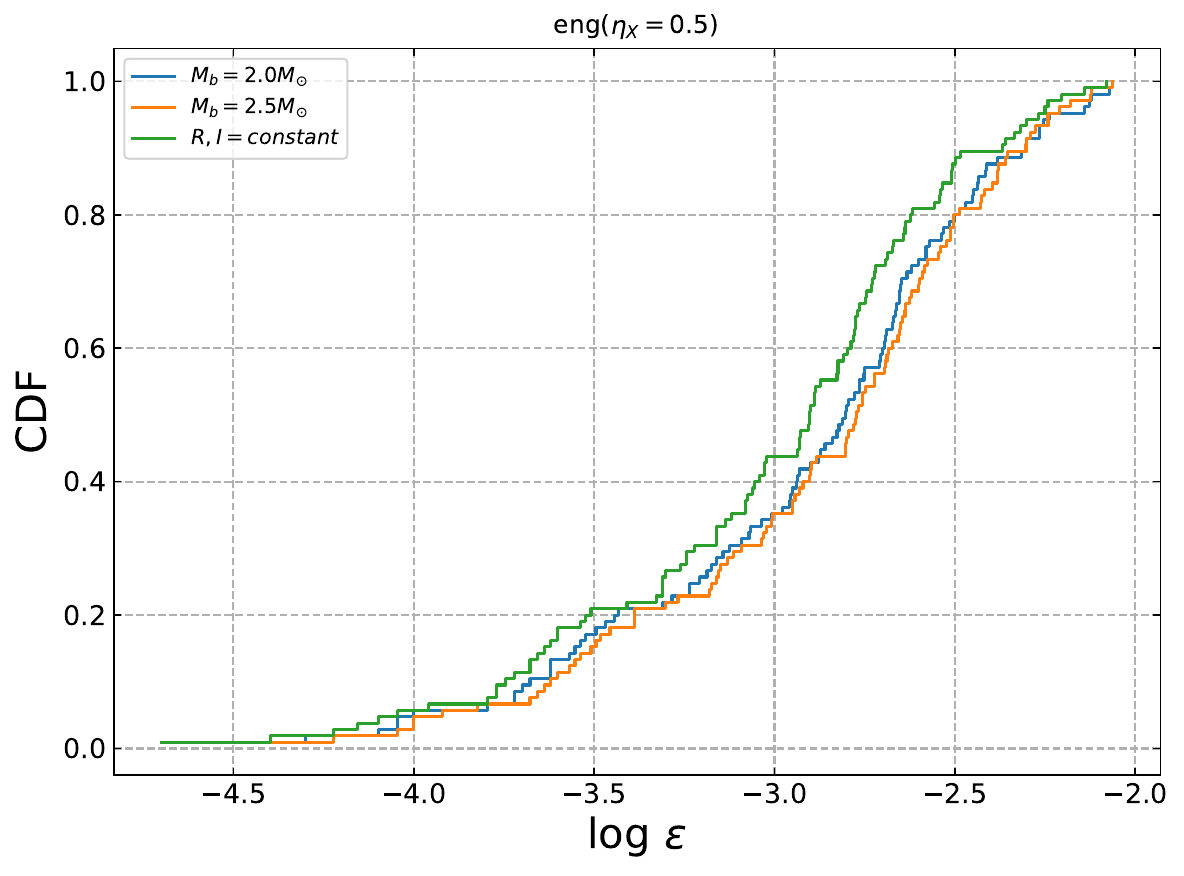}
\includegraphics [angle=0,scale=0.4] 
{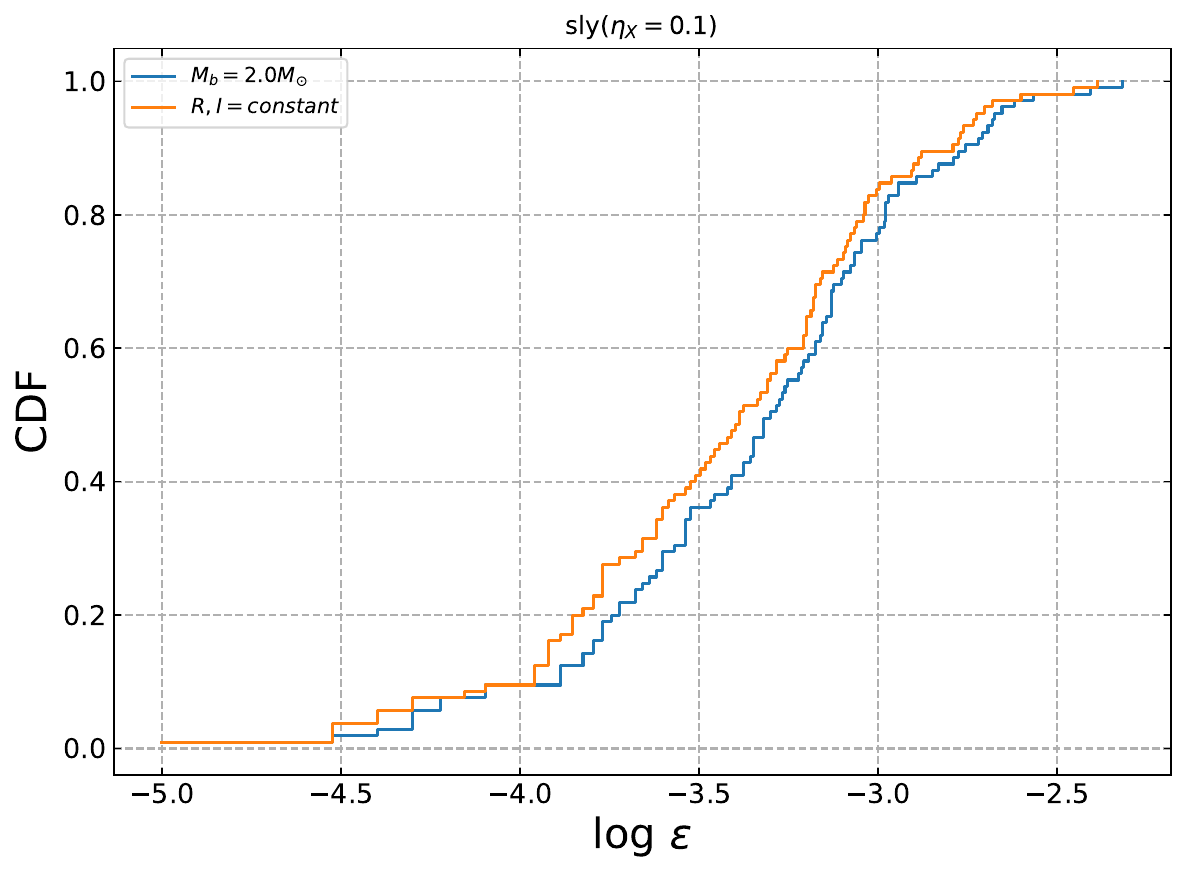}
\includegraphics [angle=0,scale=0.4] 
{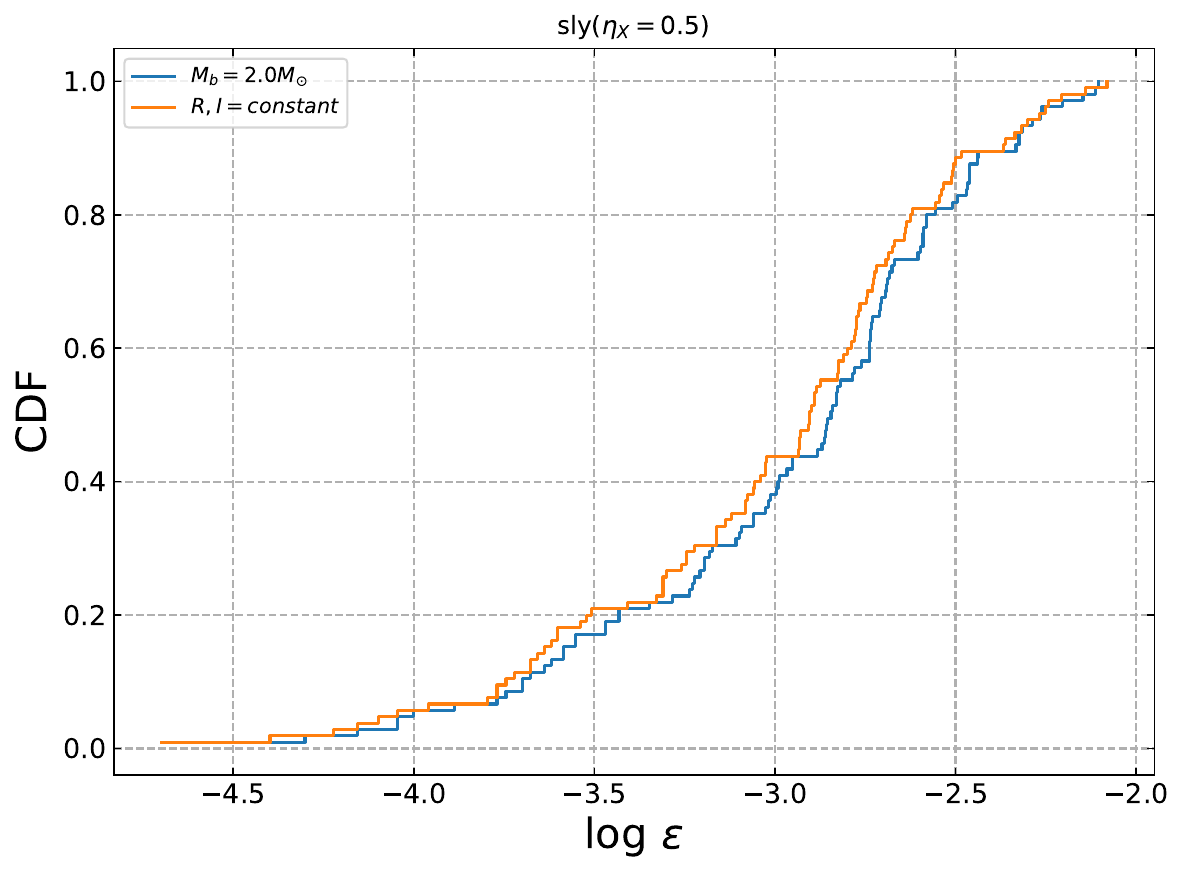}
\includegraphics [angle=0,scale=0.4]
{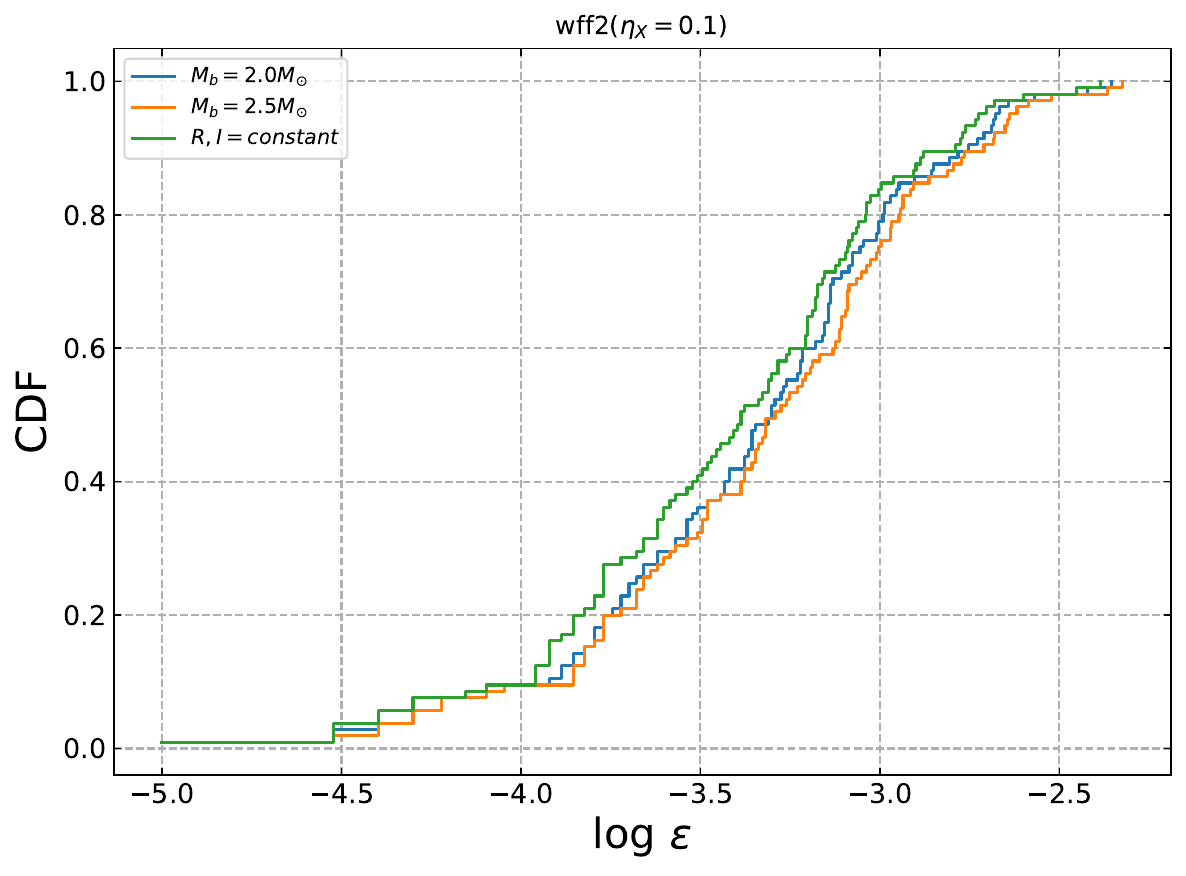}
\includegraphics [angle=0,scale=0.4]
{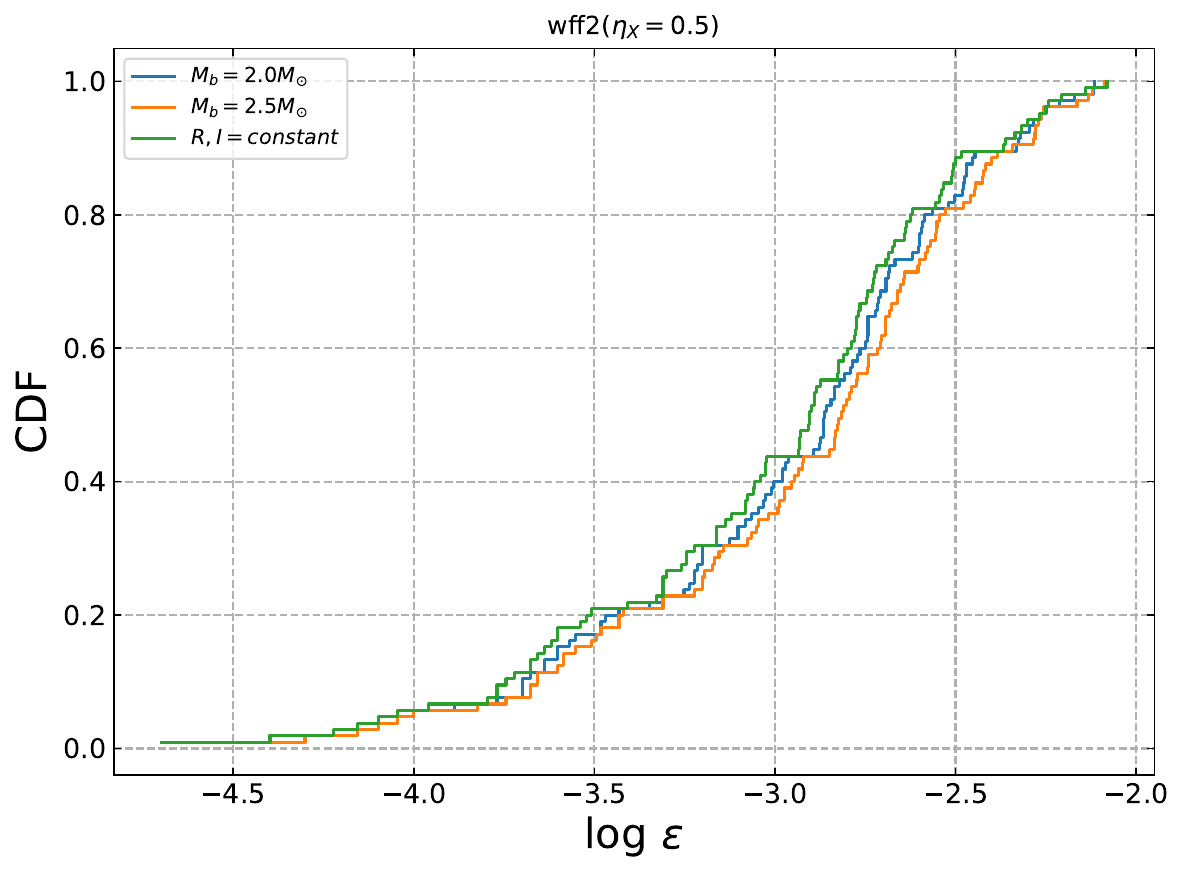}
\caption{The cumulative distributions for the derived magnetar physical parameter $\epsilon$ between the $R$ and $I$ evolution correction and the constant $R$ and $I$ in four samples of EoSs with $M_{b}=2.0~M_{\odot},~2.5~M_{\odot}$ and $\eta_{\rm X}=0.1,~0.5$, respectively.}
\label{fig:epsilon-cdf distribution}
\end{figure*}

With the above MCMC method, one can obtain the best-fitting magnetar initial parameters and perform their statistics. 
The probability density distributions and cumulative distributions of $B_p$, $P_0$, and $\epsilon$ for $\eta_{\rm X} = 0.1$ and $\eta_{\rm X} = 0.5$ in different NS EoSs, baryonic masses are shown in Figure \ref{fig:Bp-distribution}-\ref{fig:epsilon-cdf distribution}. To compare whether there is a significant difference in the magnetar parameters derived under the two scenarios of constant $R/I$ versus evolving $R/I$, we used the Kolmogorov–Smirnov (K-S) algorithm to test the deviation of the two data sets. The K-S test probability $p$-value for these magnetar parameters derived under the two scenarios of constant $R/I$ versus evolving $R/I$ is reported in Table \ref{table-10}. In general, for $M_{b}=2.0~M_{\odot},~\eta_{\rm X}=0.1$ case, one can find that the $p$-value of the K-S test for derived $B_p$, $P_0$ and $\epsilon$ between the constant $R/I$ and evolved $R/I$ are $p_{\rm KS}<10^{-7}$, $p_{\rm KS}<10^{-5}$, and $p_{\rm KS}<10^{-2}$, respectively. For $M_{b}=2.5~M_{\odot},~\eta_{\rm X}=0.1$ case, one can find that the $p$-value of the K-S test for derived $B_p$, $P_0$ and $\epsilon$ between the constant $R/I$ and evolved $R/I$ are $p_{\rm KS}<10^{-1}$, $p_{\rm KS}<10^{-10}$, and $p_{\rm KS}<10^{-2}$, respectively. For $M_{b}=2.0~M_{\odot},~\eta_{\rm X}=0.5$ case, one can find that the $p$-value of the K-S test for derived $B_p$, $P_0$ and $\epsilon$ between the constant $R/I$ and evolved $R/I$ are $p_{\rm KS}<10^{-3}$, $p_{\rm KS}<10^{-3}$, and $p_{\rm KS}<10^{-2}$, respectively. For $M_{b}=2.5~M_{\odot},~\eta_{\rm X}=0.5$ case, one can find that the $p$-value of the K-S test for derived $B_p$, $P_0$ and $\epsilon$ between the constant $R/I$ and evolved $R/I$ are $p_{\rm KS}<10^{-1}$, $p_{\rm KS}<10^{-4}$, and $p_{\rm KS}<10^{-2}$, respectively. The K-S test results indicate that the null hypothesis that the constant $R/I$ and evolving $R/I$ scenarios come from the same population can be rejected. Combining the results of Figure \ref{fig:Bp-distribution}-\ref{fig:epsilon-cdf distribution} and the K-S test, it is found that the derived $B_p$ and $P_0$ at the constant $R$ and $I$ scenarios tend to be overestimated and underestimated compared to the evolved $R$ and $I$ scenarios, respectively, and the derived $\epsilon$ are essentially equivalent in the two scenarios. In other words, the newborn magnetar properties obtained by using constant $R$ and $I$ in previous studies will present results with a larger surface magnetic field and a faster rotational velocity, which may impair our understanding of the physical nature of GRB magnetar and even demolish the connections between the physical parameters of GRB magnetar. The precise constraints on these parameters by utilizing evolved $R$ and $I$ will be crucial to understand the physical nature of GRB magnetars and to confirm the connections between their physical parameters.

In Table \ref{table-11} and \ref{table-12}, we listed the range of magnetar physical parameters and the center values and standard deviations of Gaussian fit for various scenarios, respectively. Compared to the constant $R$ and $I$ cases, we found that the real magnetar physical parameters after taking into account the NS's EoS, baryonic mass, and the radiative efficiency are tending to be significantly biased, and the derived magnetar physical parameters $B_p$ and $P_0$ values at the constant $R$ and $I$ scenarios tend to be overestimated and underestimated compared to the evolved $R$ and $I$ scenarios, respectively, and the derived $\epsilon$ values are essentially equivalent in the two scenarios. From a quantitative perspective, we can try to quantify the deviations for the derived magnetar physical parameters between constant $R/I$ and evolved $R/I$ scenarios by using the center values and standard deviations of Gaussian fit. Under certain approximations, we quantified the deviations of magnetar physical parameters between constant $R/I$ and evolved $R/I$ scenarios by considering the averaged center values and standard deviations in our selected four EoSs. For $M_{b}=2.0~M_{\odot},~\eta_{\rm X}=0.1$ case, one can find that the derived $B_p$, $P_0$ and $\epsilon$ deviations between the constant $R/I$ and evolved $R/I$ are $\bigtriangleup(\log B_p)\sim0.3~\rm G$, $\bigtriangleup P_0\sim0.14~\rm ms$, and $\bigtriangleup(\log \epsilon)\sim0.08$, which are the difference of a factor of 2.0, 0.8, and 1.2 compared to the constant $R/I$ case, respectively. For $M_{b}=2.5~M_{\odot},~\eta_{\rm X}=0.1$ scenario, we can find that the derived $B_p$, $P_0$ and $\epsilon$ deviations between the constant $R/I$ and evolved $R/I$ are $\bigtriangleup(\log B_p)\sim0.05~\rm G$, $\bigtriangleup P_0\sim0.26~\rm ms$, and $\bigtriangleup(\log \epsilon)\sim0.12$, which are the difference of a factor of 1.1, 0.7, and 1.3 compared to the constant $R/I$ case, respectively. For $M_{b}=2.0~M_{\odot},~\eta_{\rm X}=0.5$ scenario, one can find that the derived $B_p$, $P_0$ and $\epsilon$ deviations between the constant $R/I$ and evolved $R/I$ are $\bigtriangleup(\log B_p)\sim0.18~\rm G$, $\bigtriangleup P_0\sim0.20~\rm ms$, and $\bigtriangleup(\log \epsilon)\sim0.07$, which are the difference of a factor of 1.5, 0.9, and 1.2 compared to the constant $R/I$ case, respectively. For $M_{b}=2.5~M_{\odot},~\eta_{\rm X}=0.5$ scenario, one can find that the derived $B_p$, $P_0$ and $\epsilon$ deviations between the constant $R/I$ and evolved $R/I$ are $\bigtriangleup(\log B_p)\sim0.02~\rm G$, $\bigtriangleup P_0\sim0.43~\rm ms$, and $\bigtriangleup(\log \epsilon)\sim0.11$, which are the difference of a factor of 1.0, 0.7, and 1.3 compared to the constant $R/I$ case, respectively. Finally, based on the various scenarios in evolved $R/I$ frame, we also averaged the various results at evolved $R/I$ scenarios and compared them with constant $R/I$ scenarios, and found that the derived $B_p$, $P_0$ and $\epsilon$ in the constant $R/I$ cases are overestimated $\bigtriangleup(\log B_p)\sim0.17~\rm G$, underestimated $\bigtriangleup P_0\sim0.26~\rm ms$, and essentially equivalent $\bigtriangleup(\log \epsilon)\sim0.09$, which are the difference of a factor of 1.5, 0.7, and 1.2 compared to the constant $R/I$ case, respectively. Moreover, for a given EoS, different X-ray radiation efficiencies and NS baryonic masses could evidently alter the derived magnetar physical parameters. According to our results, with higher radiation efficiency and baryonic mass, the initial dipole magnetic field strength $B_p$, spin period $P_0$, as well as ellipticity deformation $\epsilon$ will become larger. For the given baryonic mass and radiation efficiency, different EoSs could also overall change the derived magnetar physical parameters, but the constraints on magnetar physical parameter magnitude seem to be independent of the EoS stiffness.

\begin{figure*}
\centering
\includegraphics [angle=0,scale=0.29]  {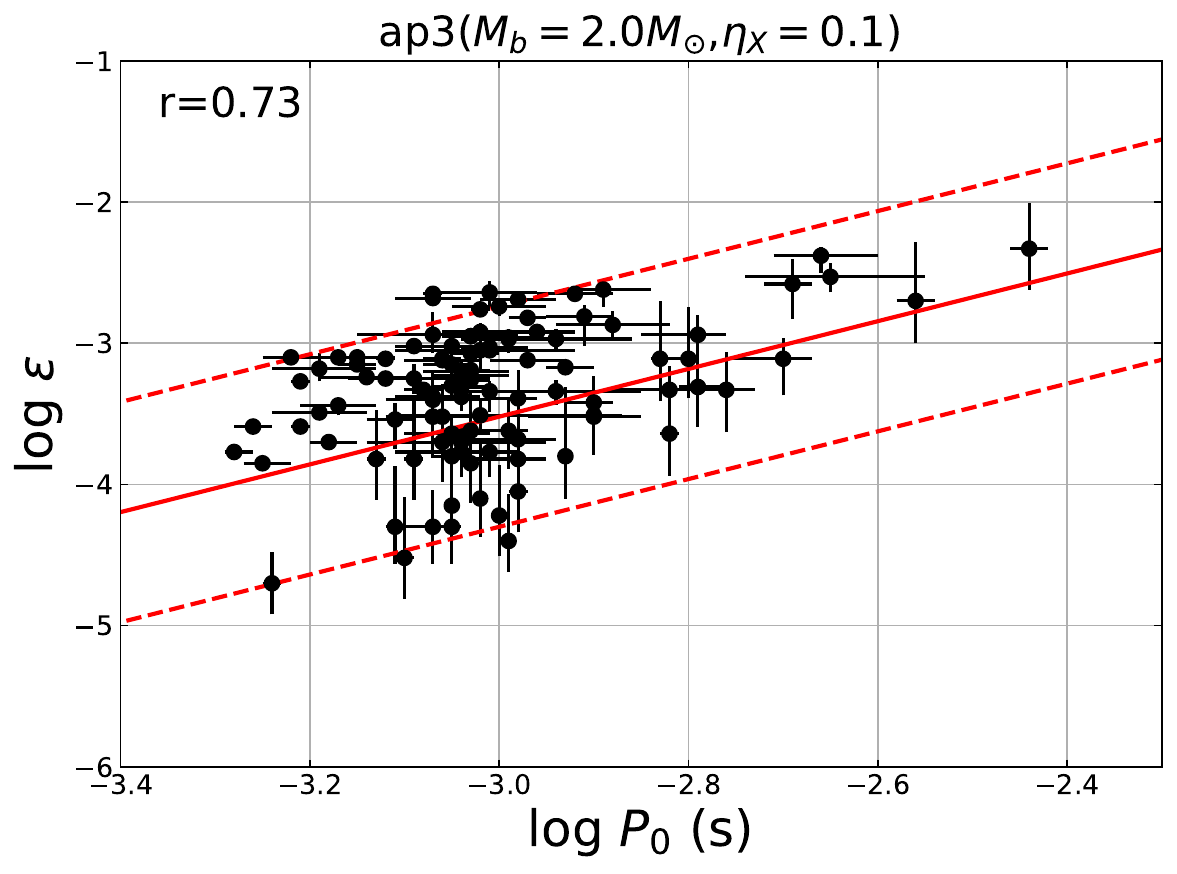}
\includegraphics [angle=0,scale=0.29]  {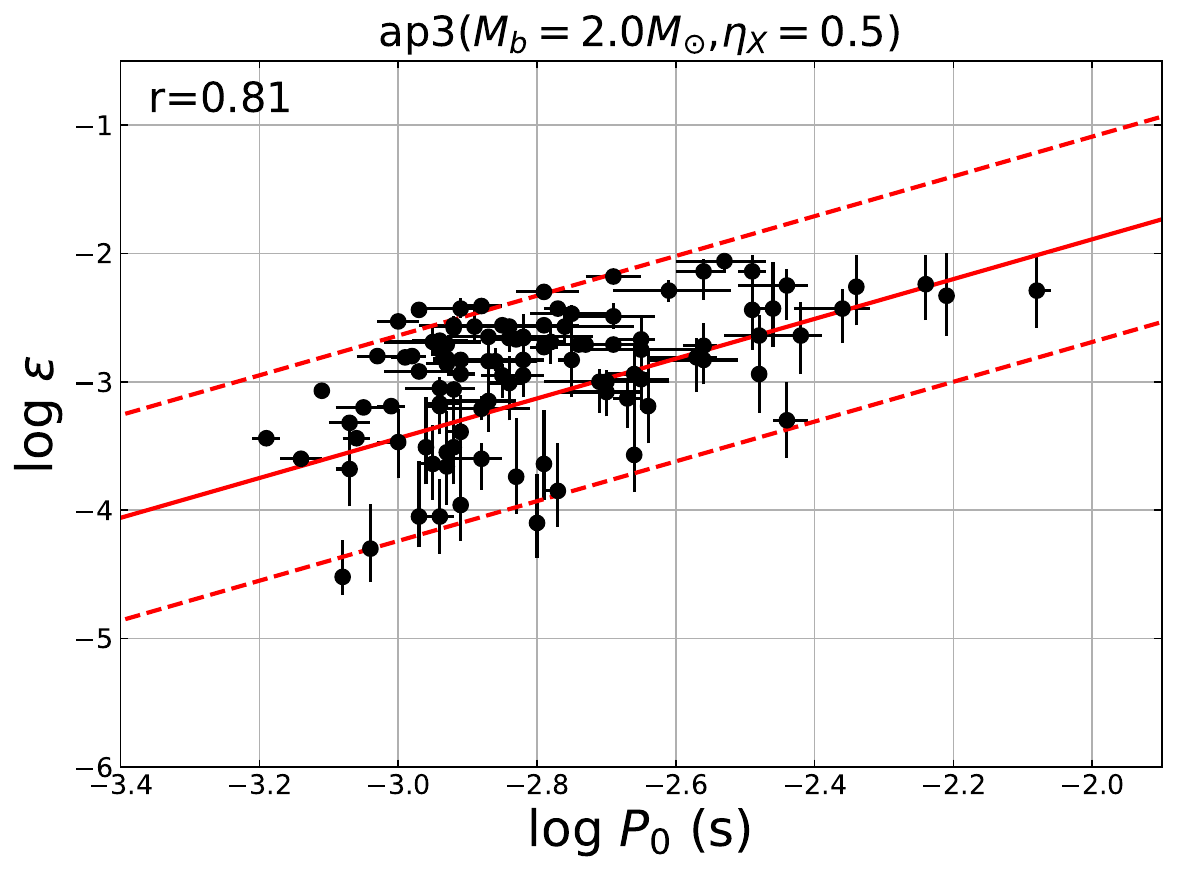}
\includegraphics [angle=0,scale=0.29]  {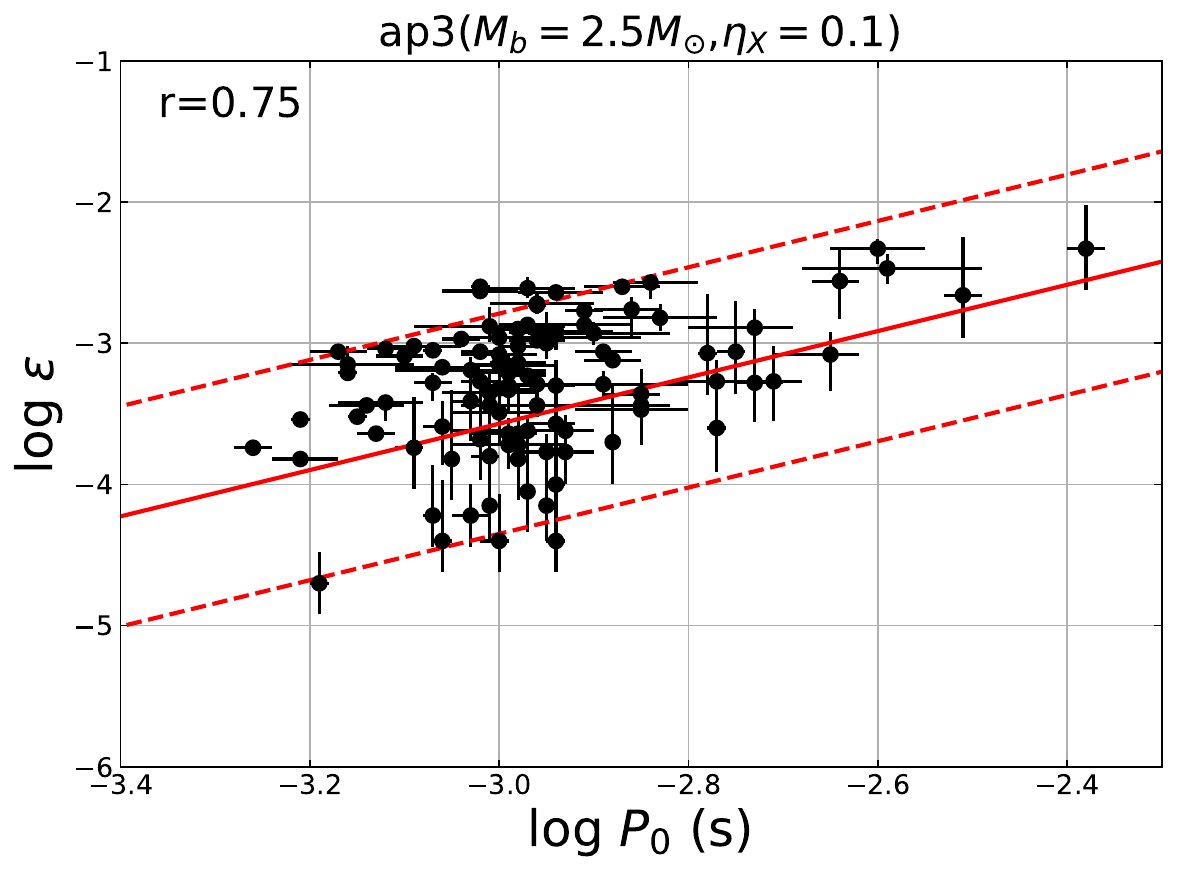}\\
\includegraphics [angle=0,scale=0.29]  {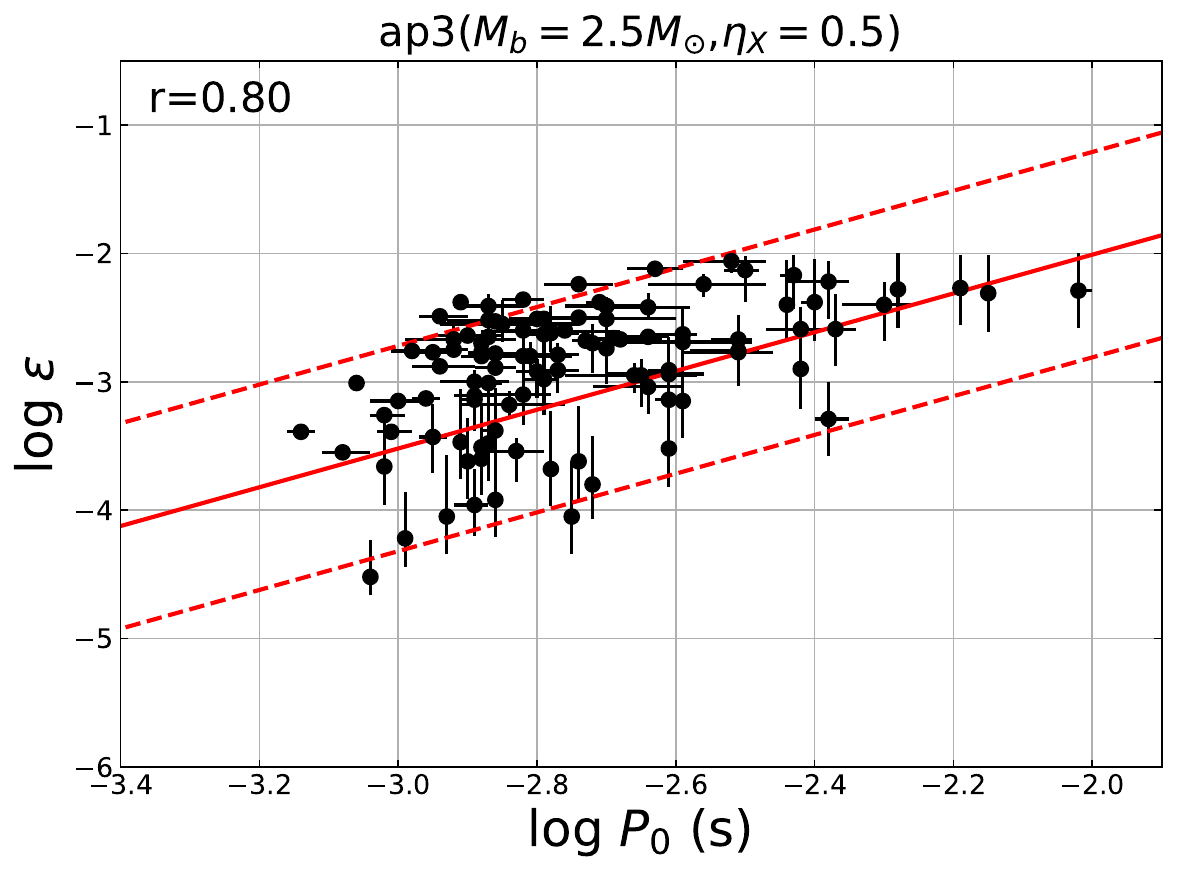}
\includegraphics [angle=0,scale=0.29]  {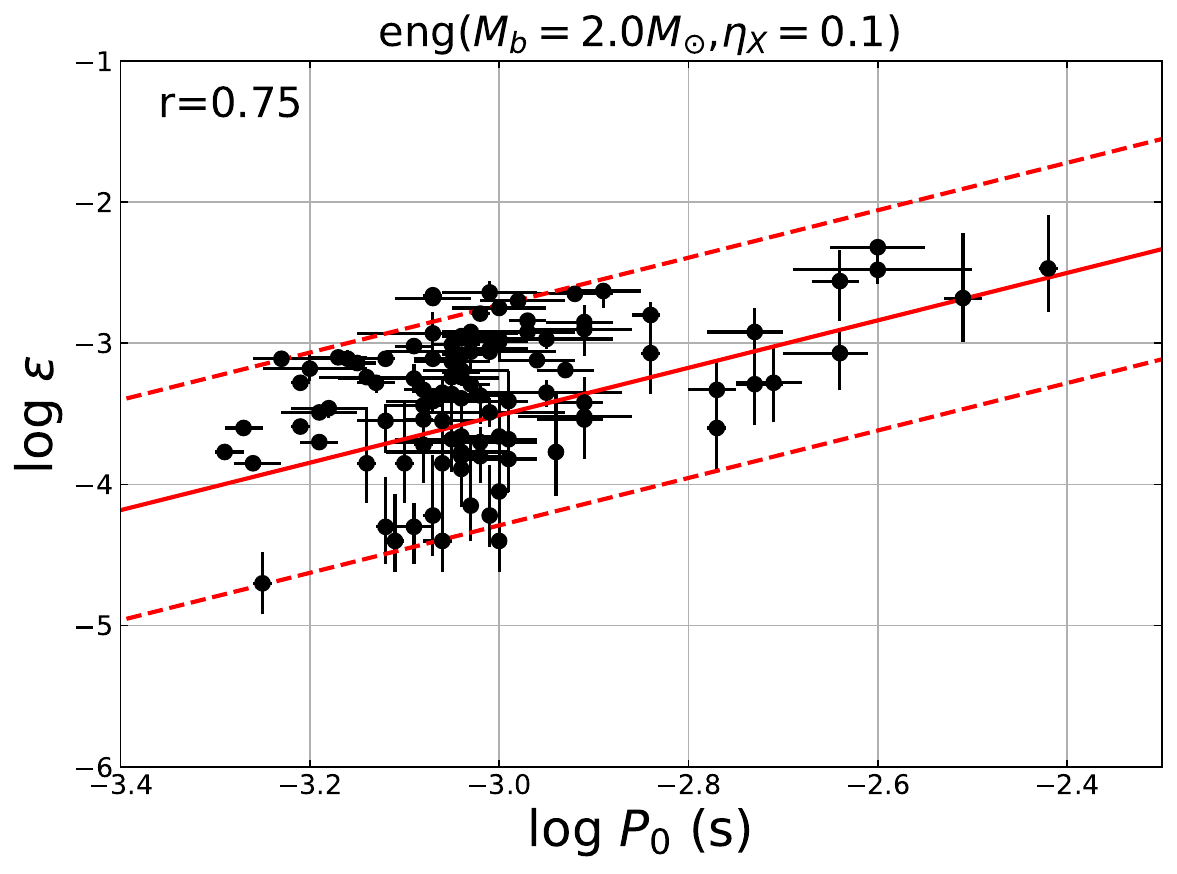}
\includegraphics [angle=0,scale=0.29]  {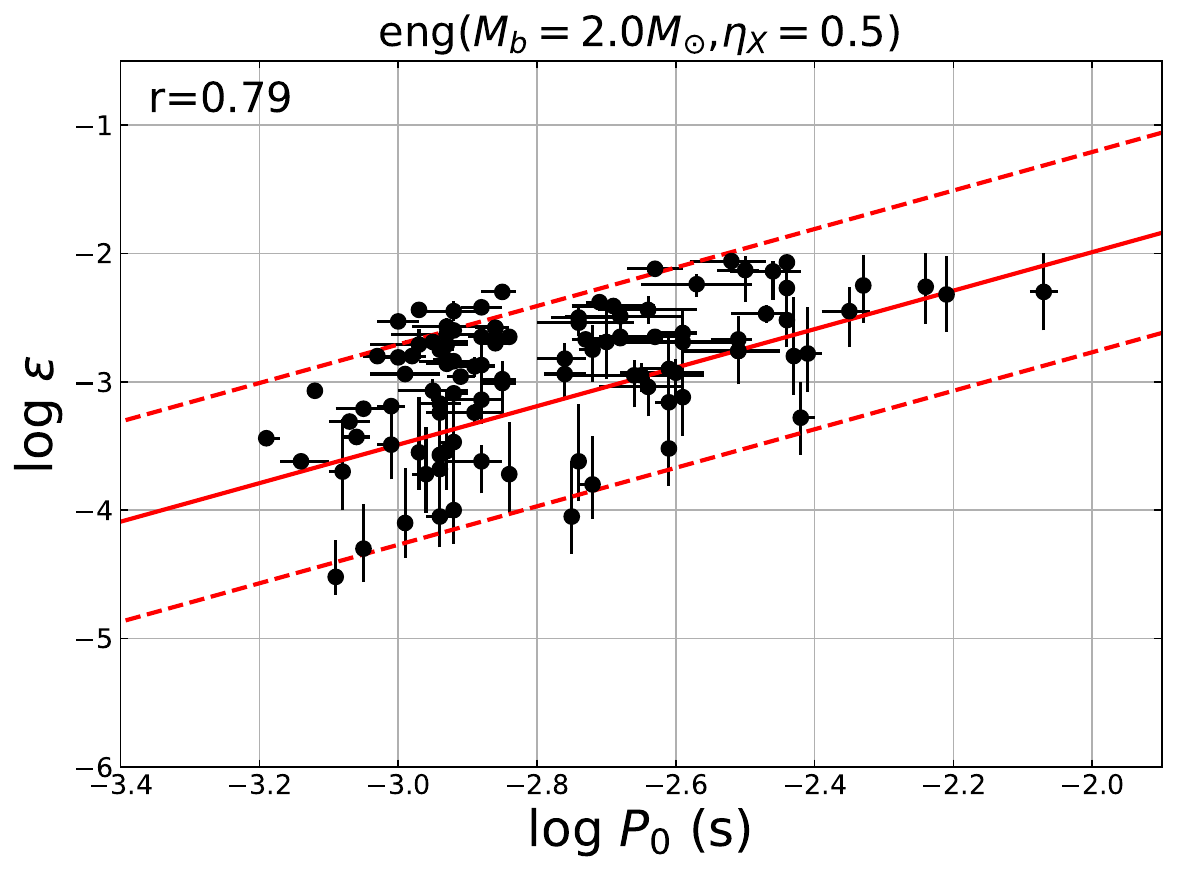}\\
\includegraphics [angle=0,scale=0.29]  {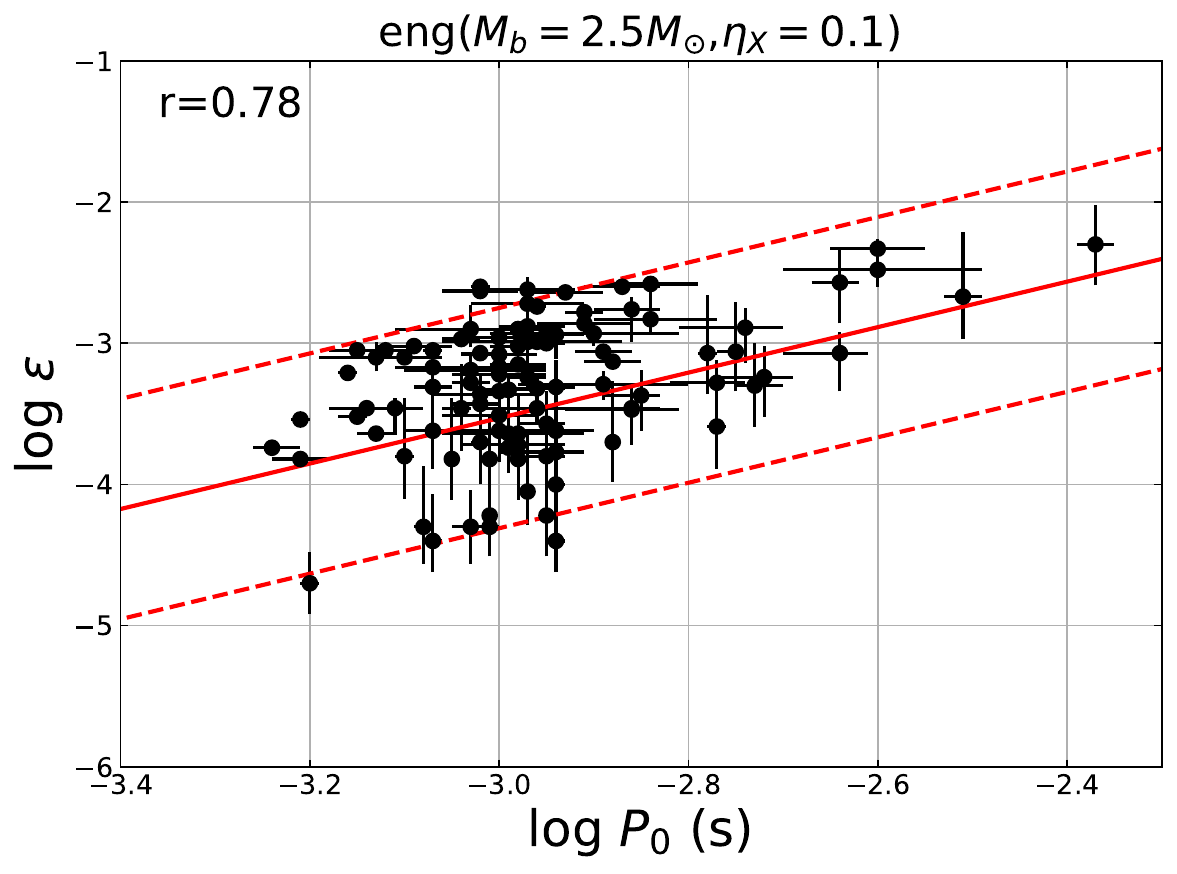}
\includegraphics [angle=0,scale=0.29]  {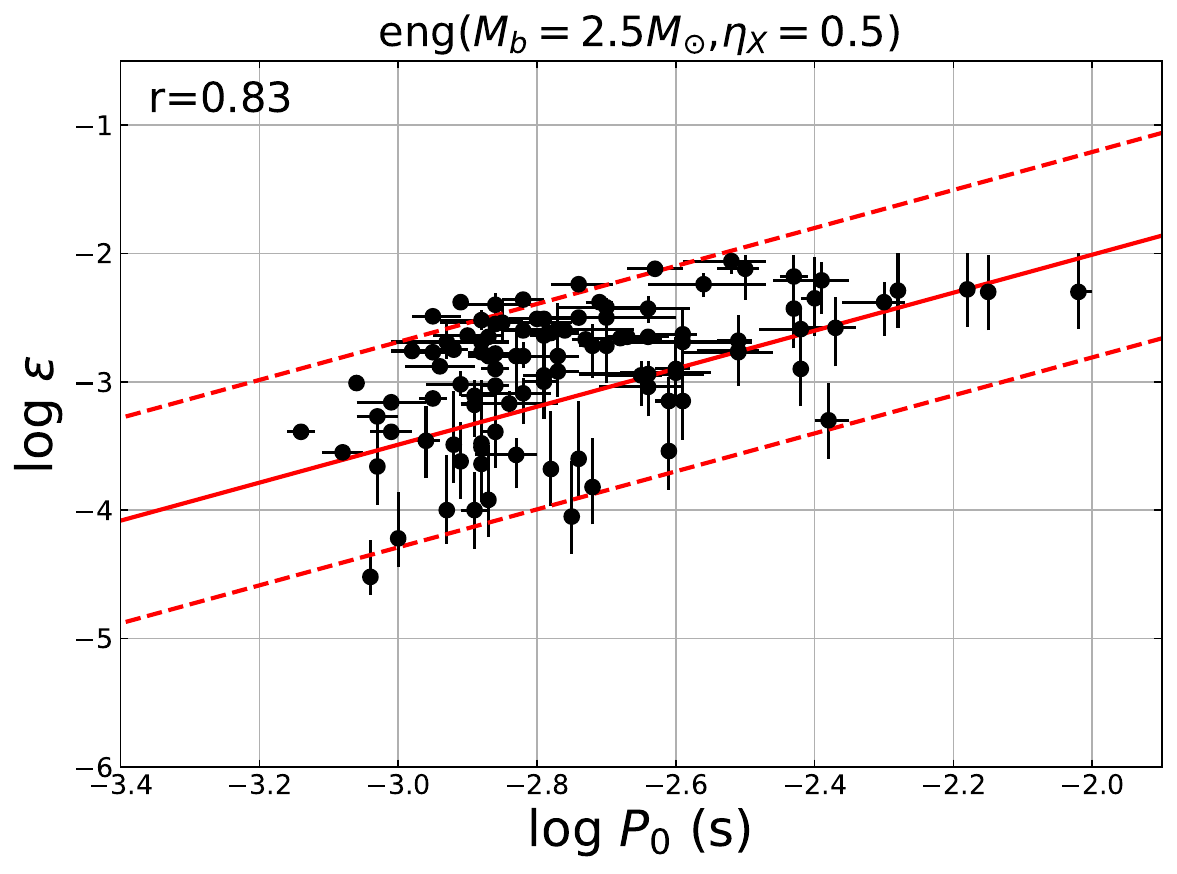}
\includegraphics [angle=0,scale=0.29]  {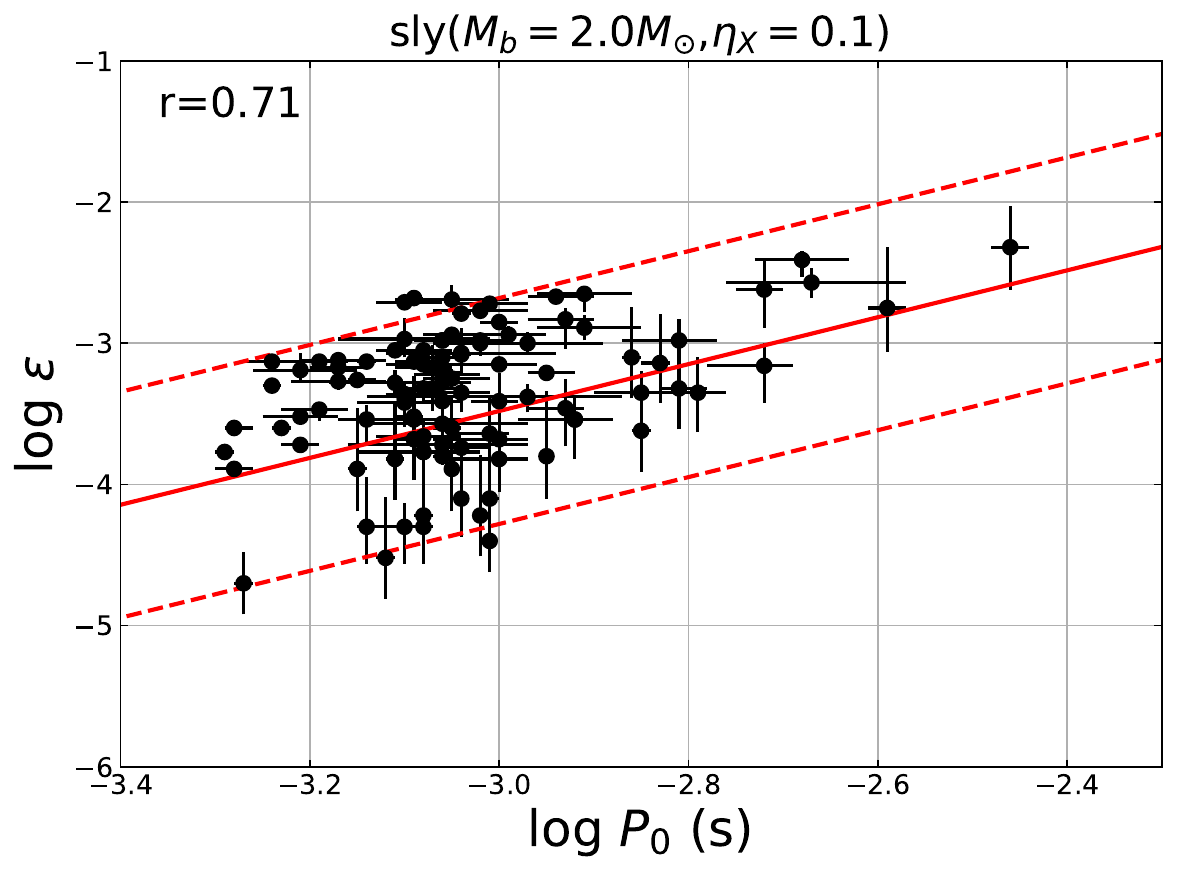}\\
\includegraphics [angle=0,scale=0.29]  {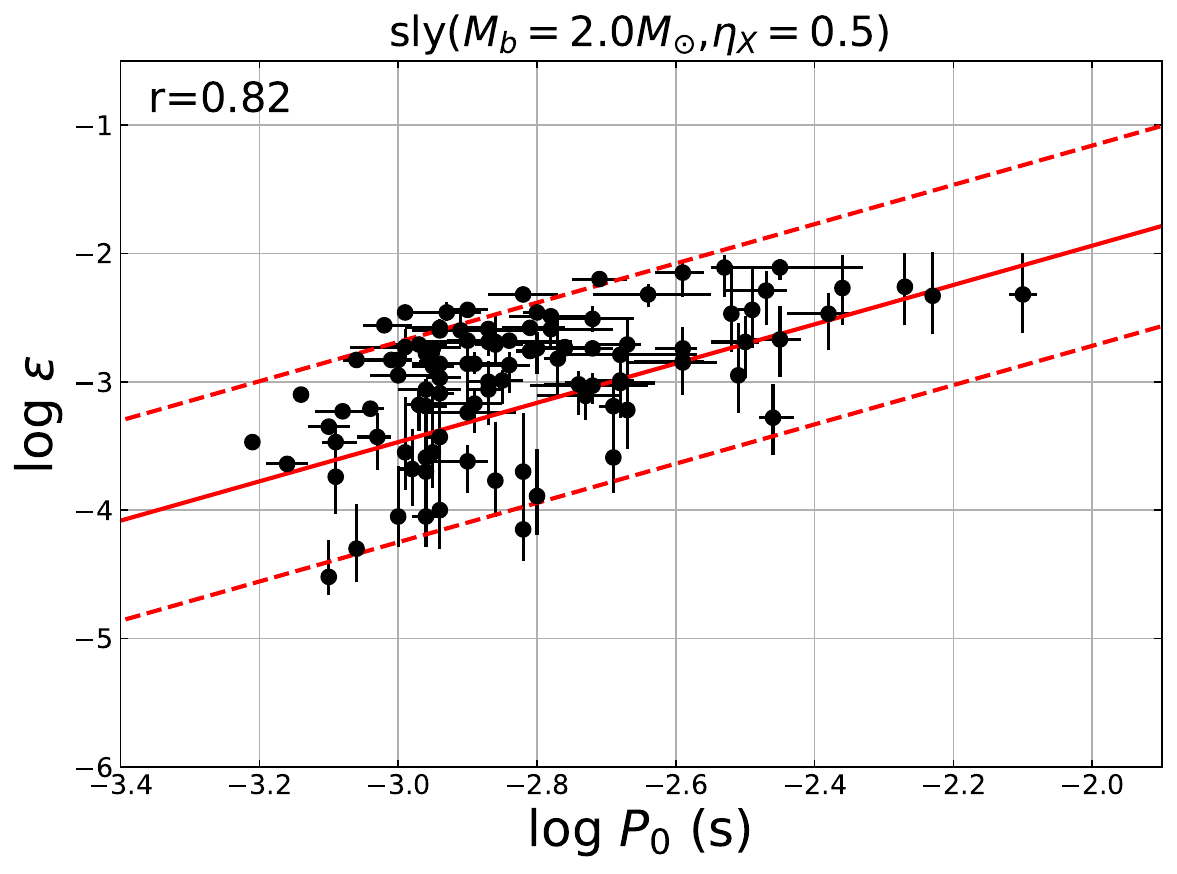}
\includegraphics [angle=0,scale=0.29]  {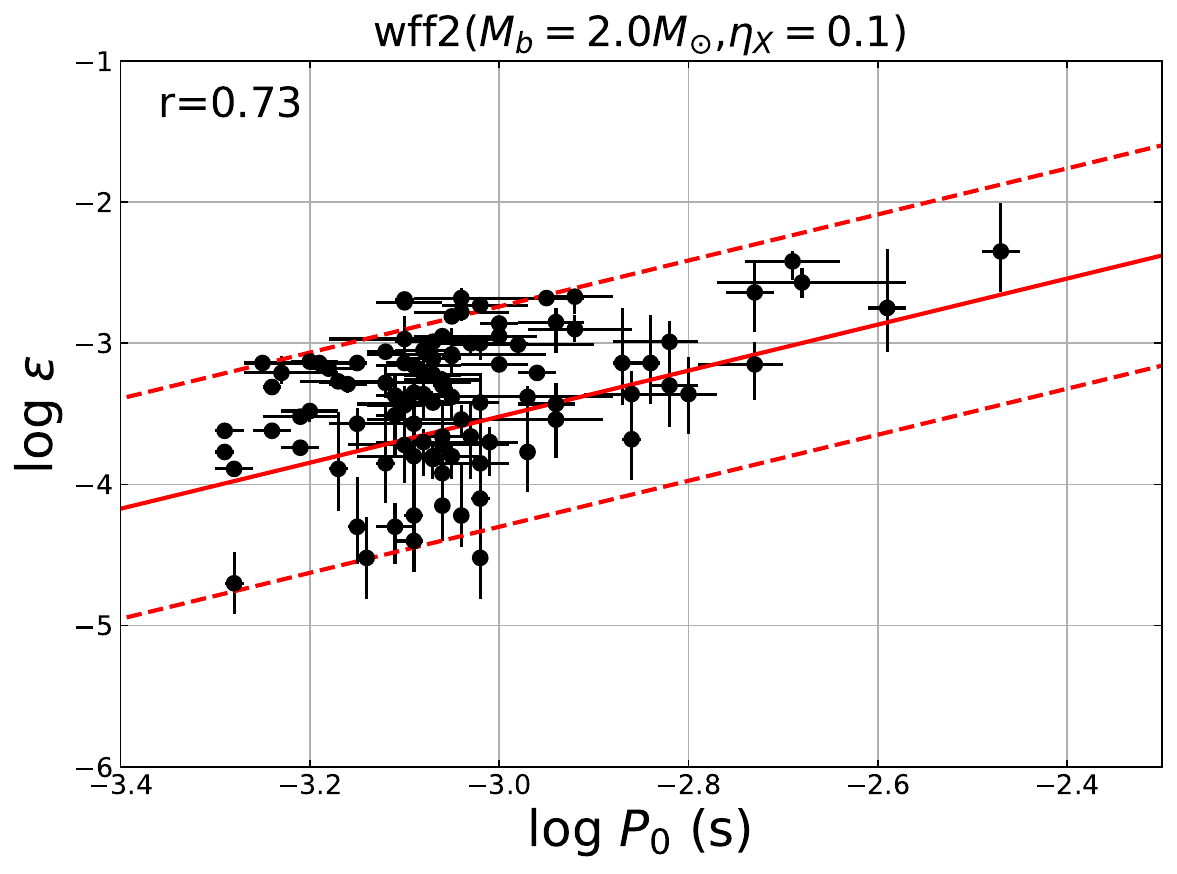}
\includegraphics [angle=0,scale=0.29]  {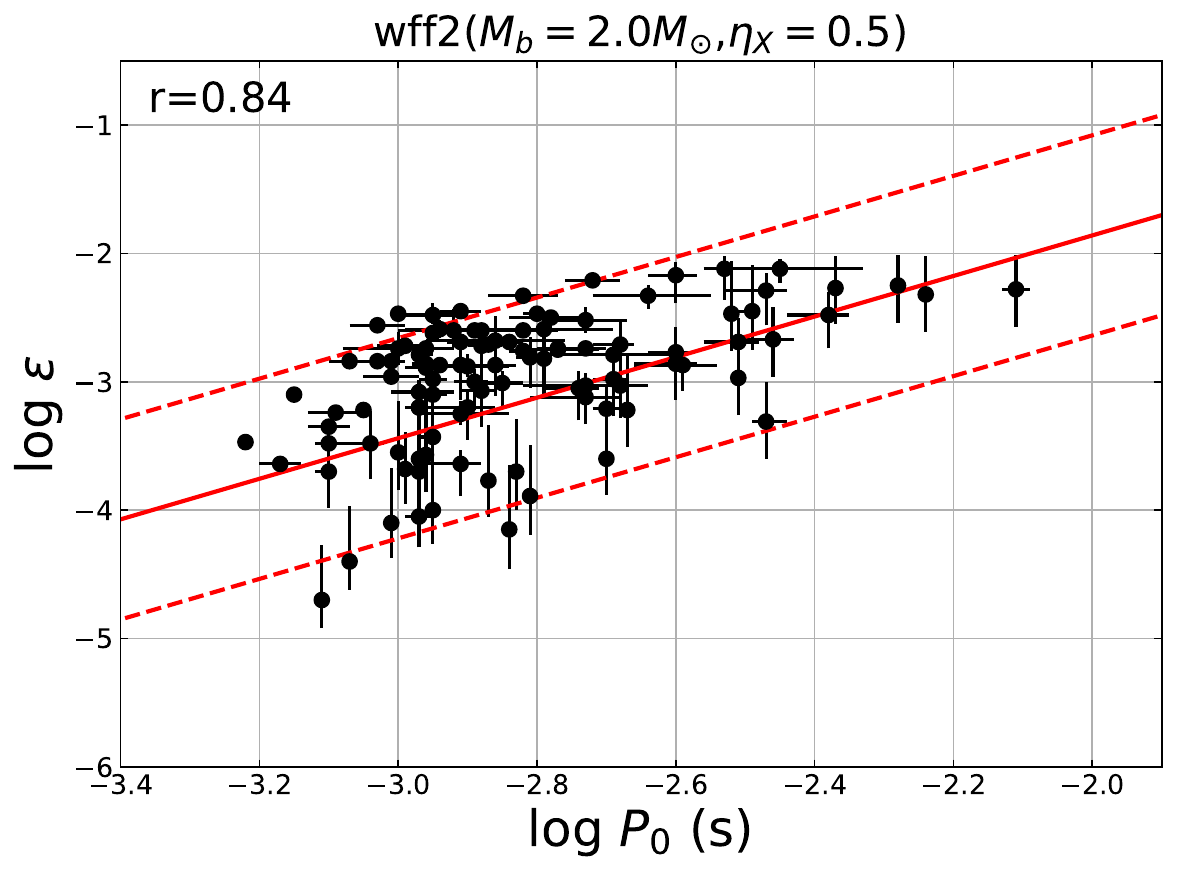}\\
\includegraphics [angle=0,scale=0.29]  {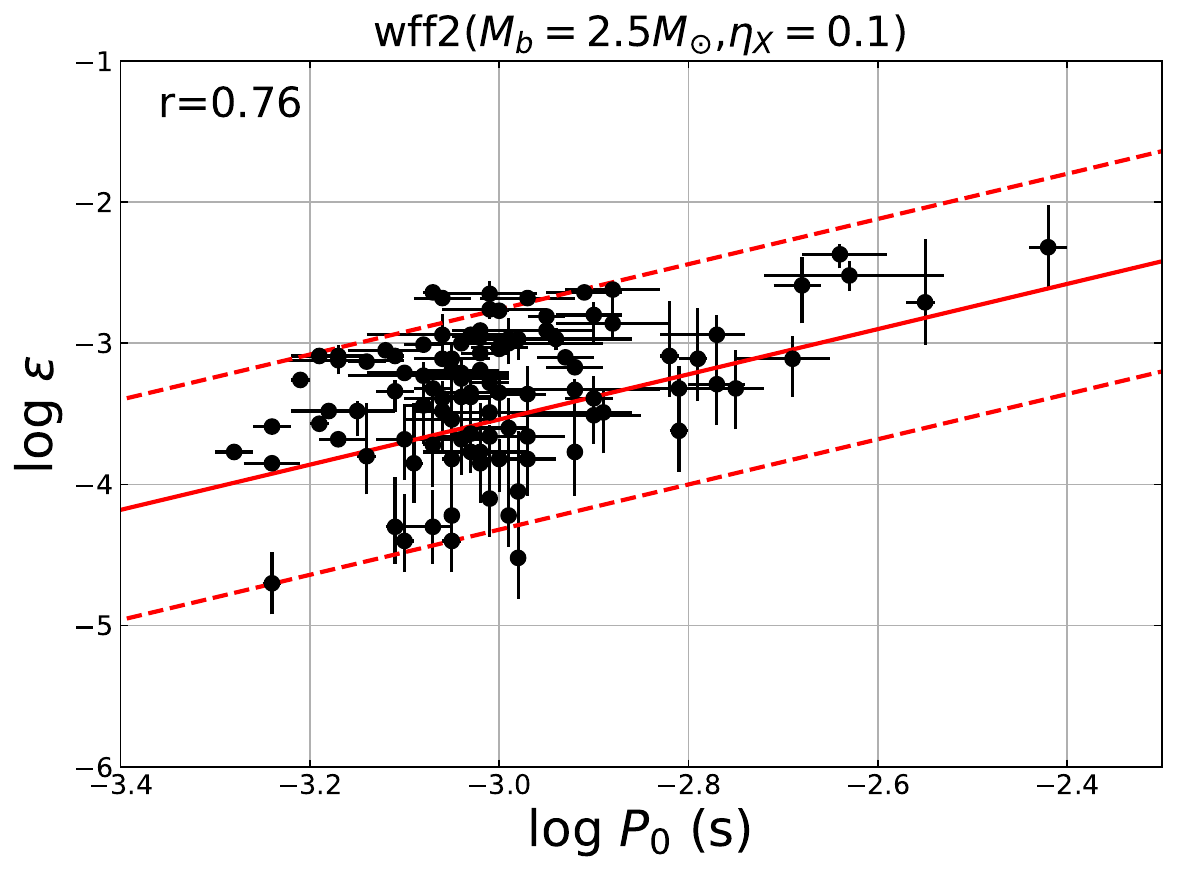}
\includegraphics [angle=0,scale=0.29]  {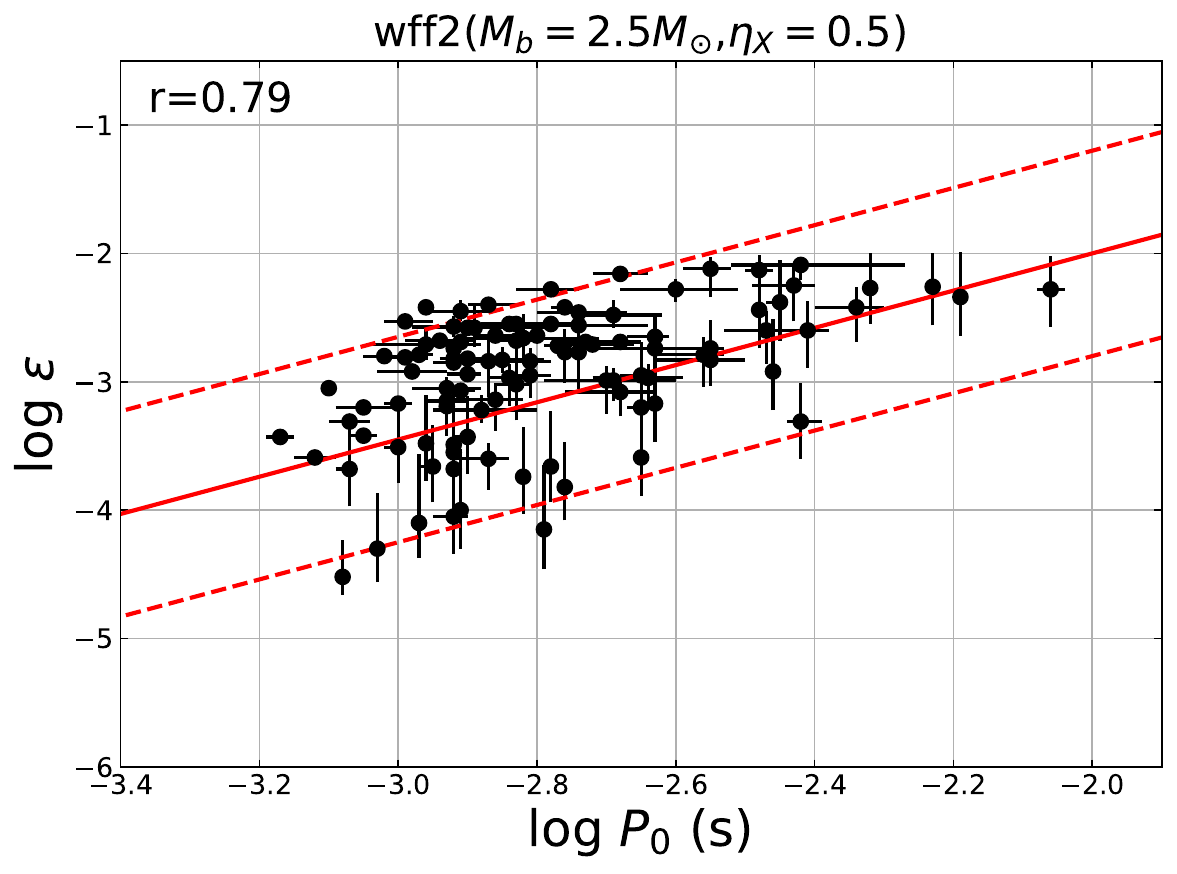}\\
\caption{The correlations between the $\epsilon$ and $P_0$ in four samples of EoSs with $M_{b}=2.0~M_{\odot},~2.5~M_{\odot}$ and $\eta_{\rm X}=0.1,~0.5$. The red solid and red dashed lines are the best-fitting results and the 95\% confidence level, respectively.}
\label{fig:P0-epsilon}
\end{figure*}

Further, we performed the least-squares regression algorithm to hunt for possible correlations among the physical parameters of the newborn magnetar. We first investigated how the newly born magnetar ellipticity deformation $\epsilon$ is related to other parameters (e.g., $\epsilon-P_0$, $\epsilon-B_p$). In Figure \ref{fig:P0-epsilon}, we showed a series of $\epsilon-P_0$ scatter diagrams for different EoSs and baryonic masses with two $\eta_{\rm x}$ values based on our LGRB sample. Interestingly, we found that there are strong correlations between $\epsilon$ and $P_0$ for our selected EoSs. Most of the GRBs in our sample fall into the $3\sigma$ deviation region of the best-fitting power-law model for all EoSs scenarios. Our best-fitting correlations for various cases are reported in Table \ref{table-13}. The strong correlations of $\epsilon-P_0$ appeared to be universal for all of our selected EoSs, which may imply that the newly born magnetar with a smaller spin period corresponds to the larger ellipticity deformation. In all of our selected EoSs, the best-fitting results can be approximately presented $\epsilon\propto P_0^{1.57\pm0.22}$, with the 1$\sigma$ deviation included. If such a $\epsilon-P_0$ relation is true, one can get $L_{\rm GW}\propto P_0^{-3}$, which suggests that GW radiation from the newborn magnetar is only effective when the magnetar is spinning fast enough. 

\begin{figure*}
\centering
\includegraphics [angle=0,scale=0.29]  {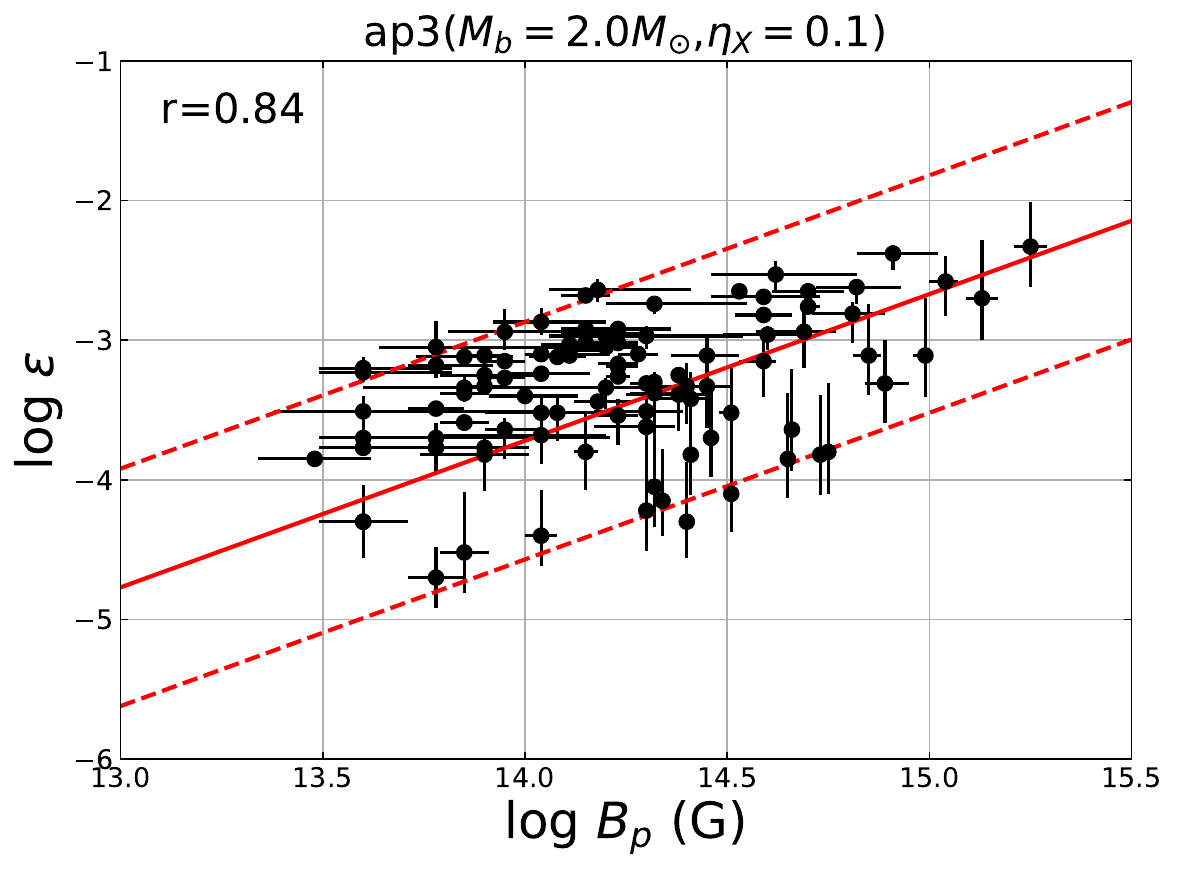}
\includegraphics [angle=0,scale=0.29]  {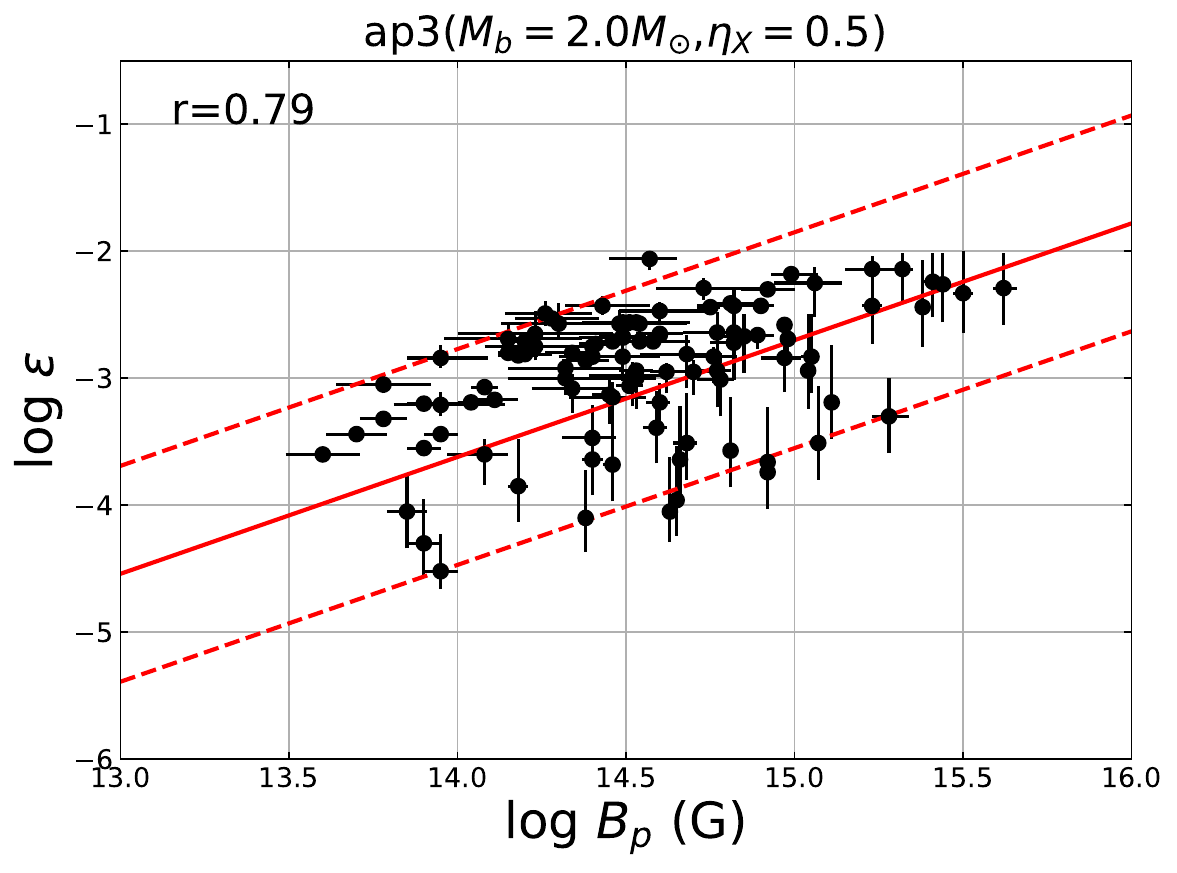}
\includegraphics [angle=0,scale=0.29]  {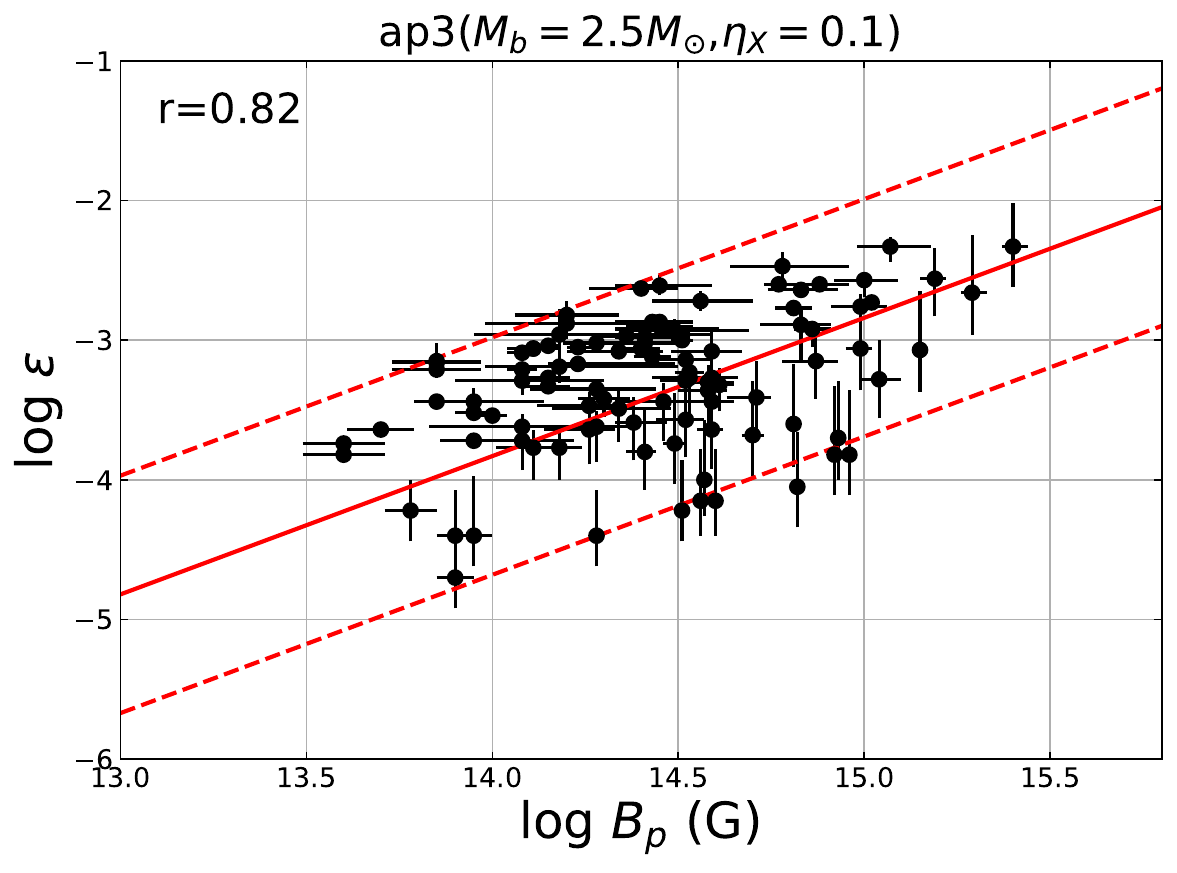}\\
\includegraphics [angle=0,scale=0.29]  {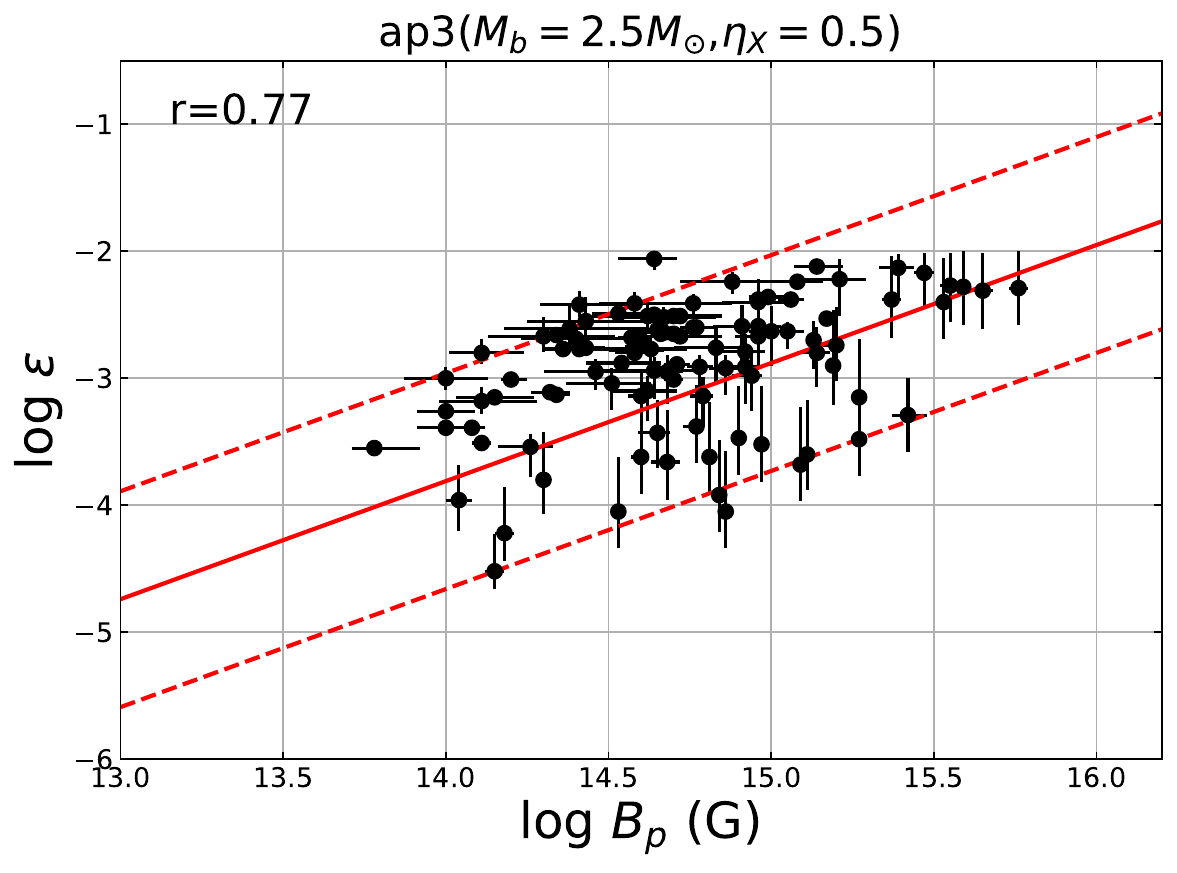}
\includegraphics [angle=0,scale=0.29]  {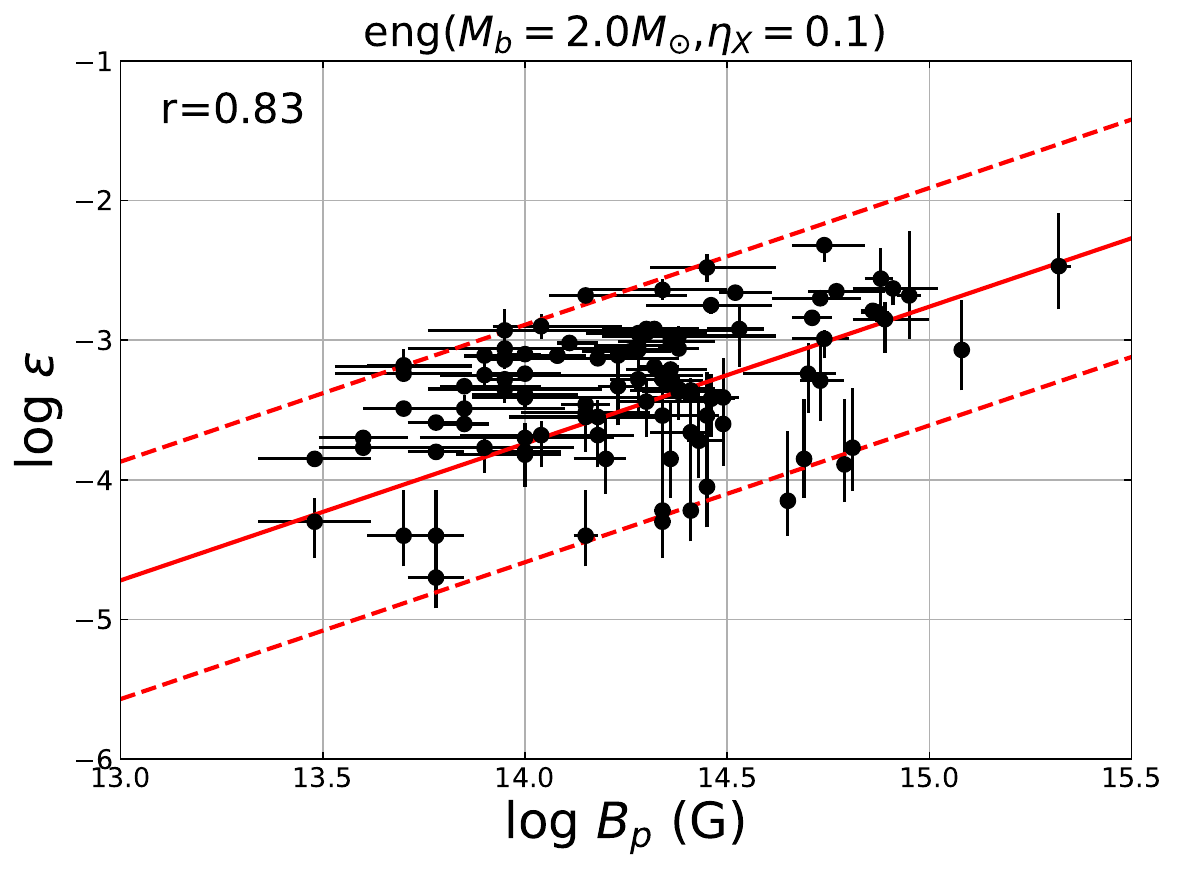}
\includegraphics [angle=0,scale=0.29]  {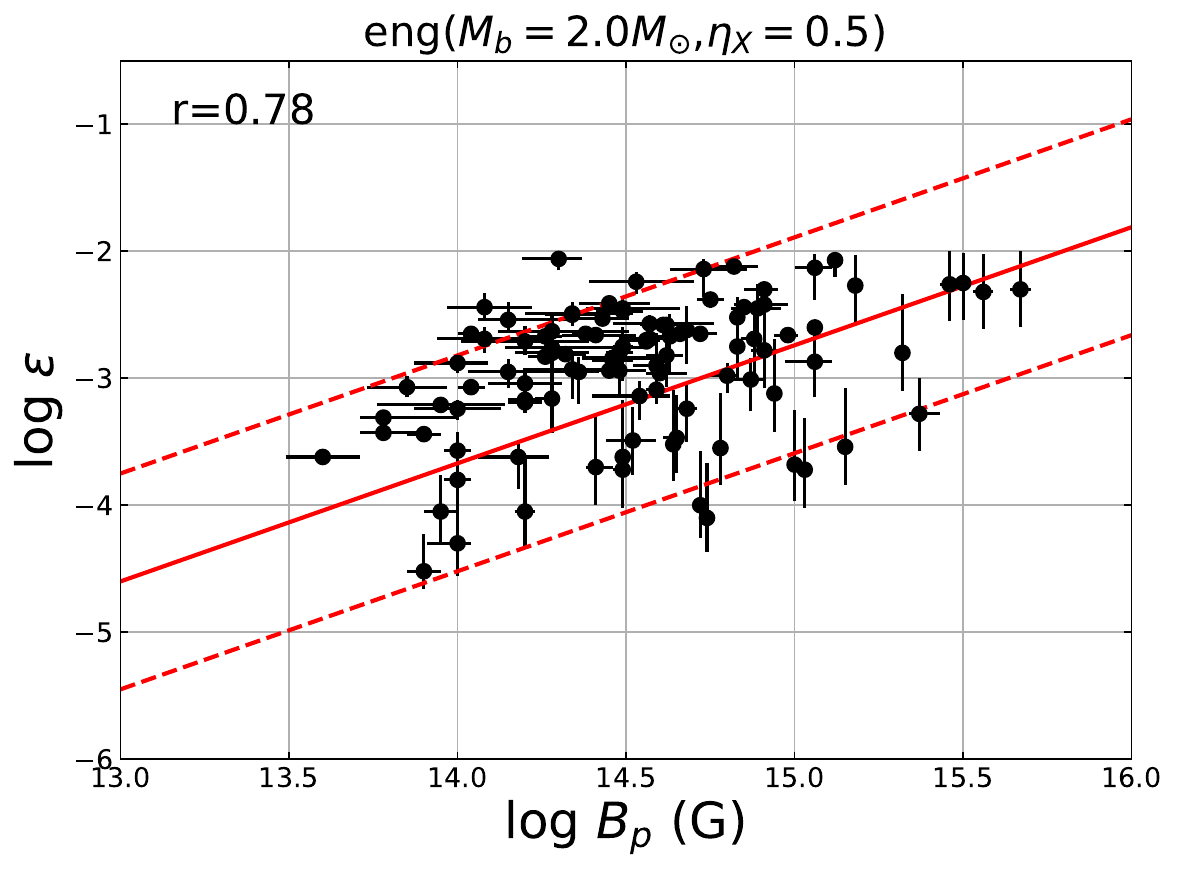}\\
\includegraphics [angle=0,scale=0.29]  {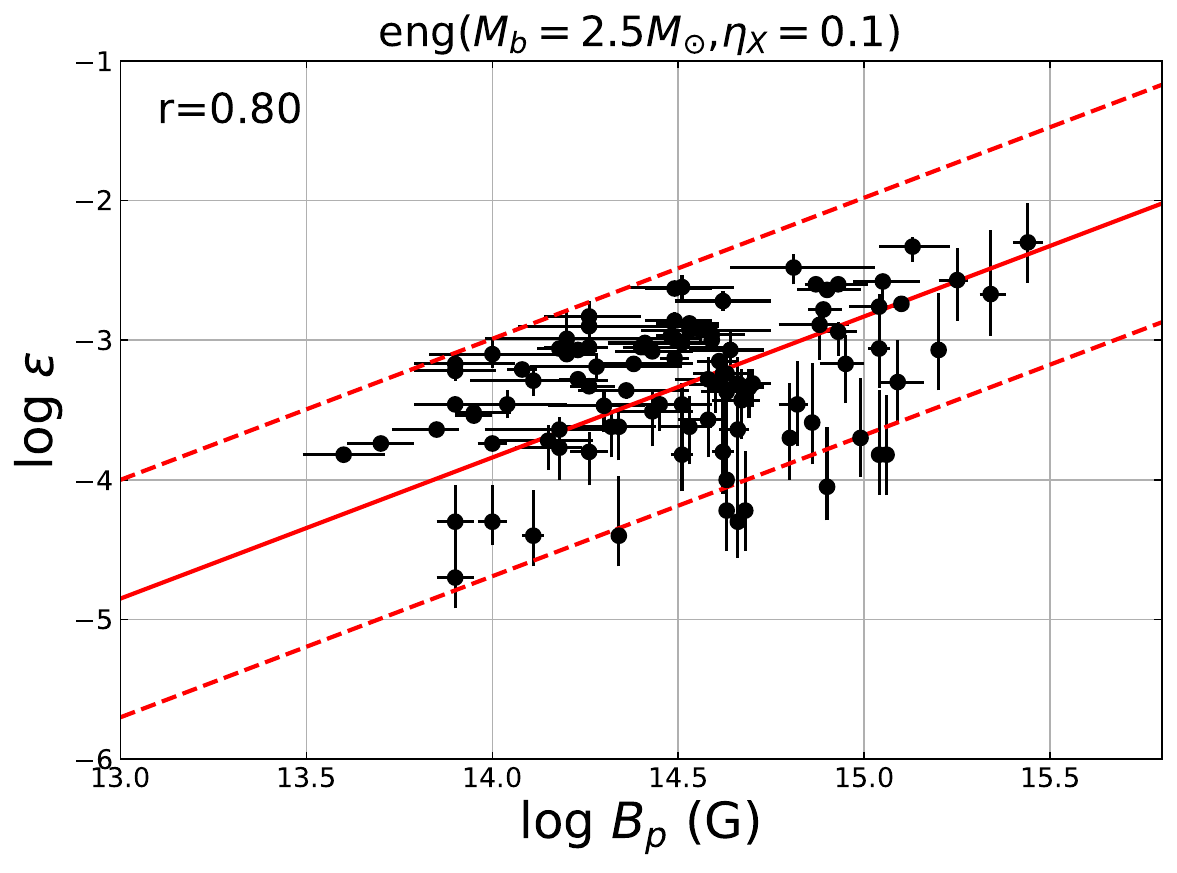}
\includegraphics [angle=0,scale=0.29]  {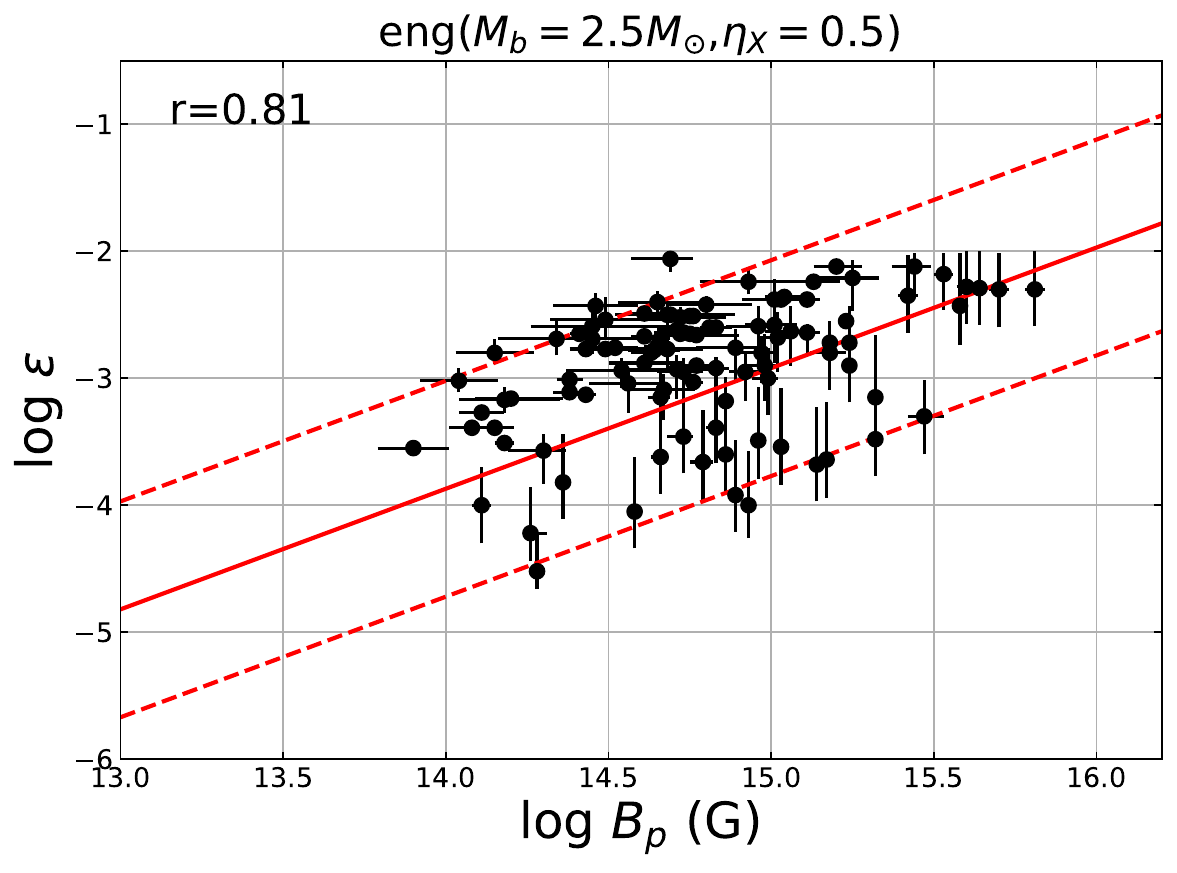}
\includegraphics [angle=0,scale=0.29]  {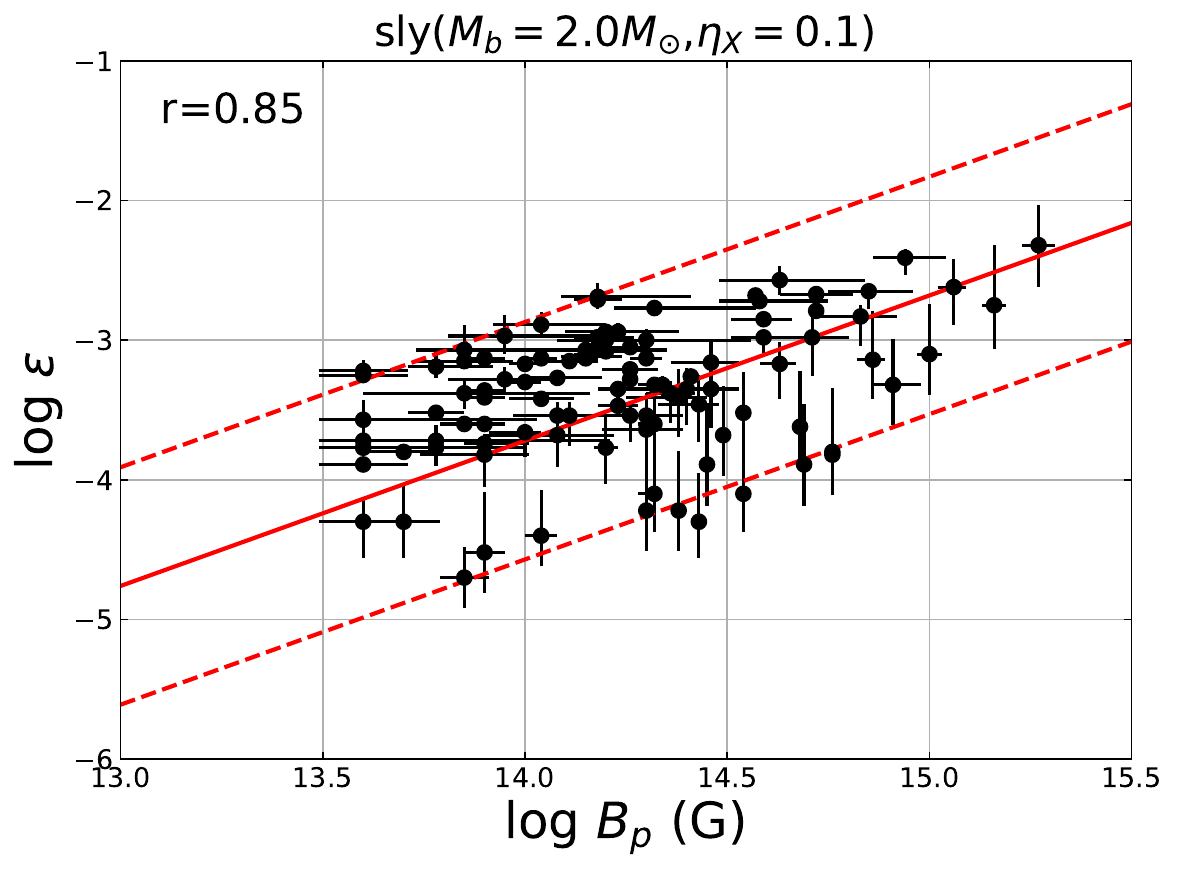}\\
\includegraphics [angle=0,scale=0.29]  {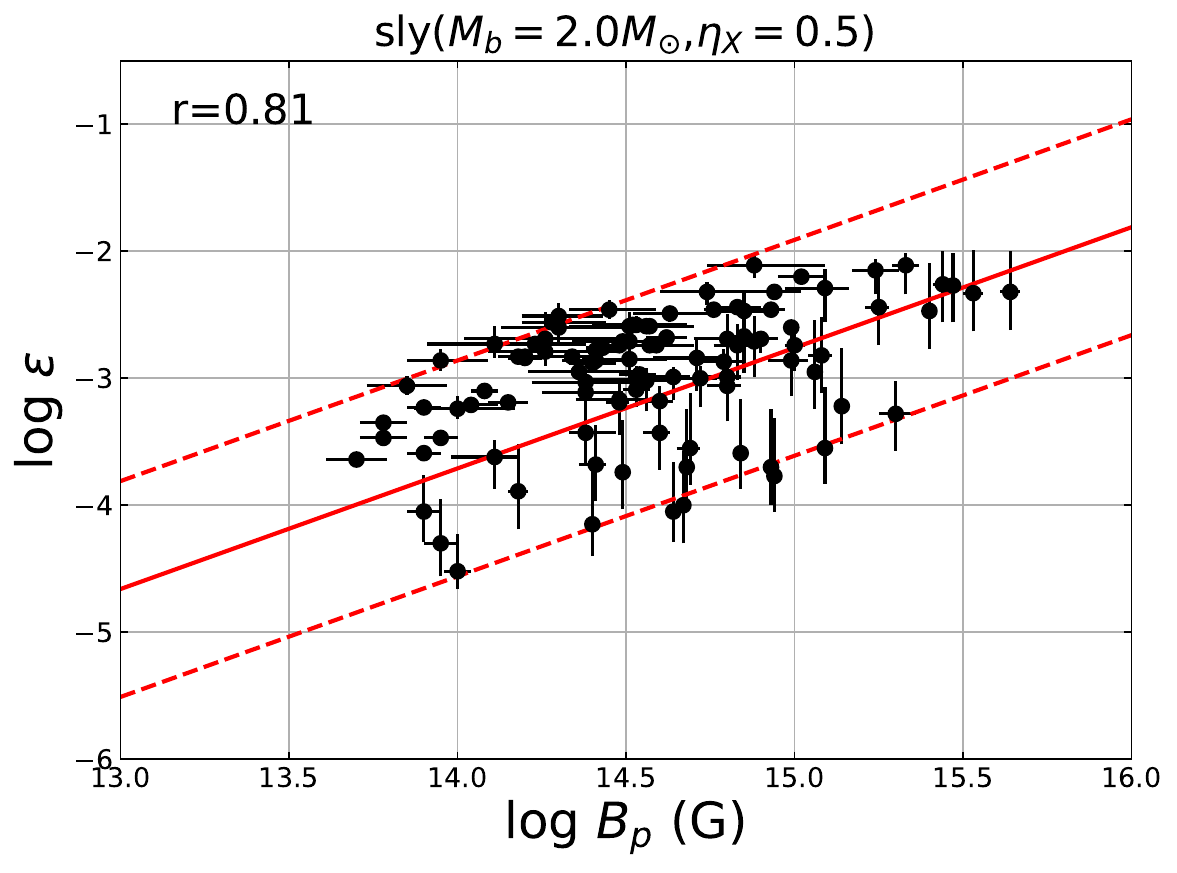}
\includegraphics [angle=0,scale=0.29]  {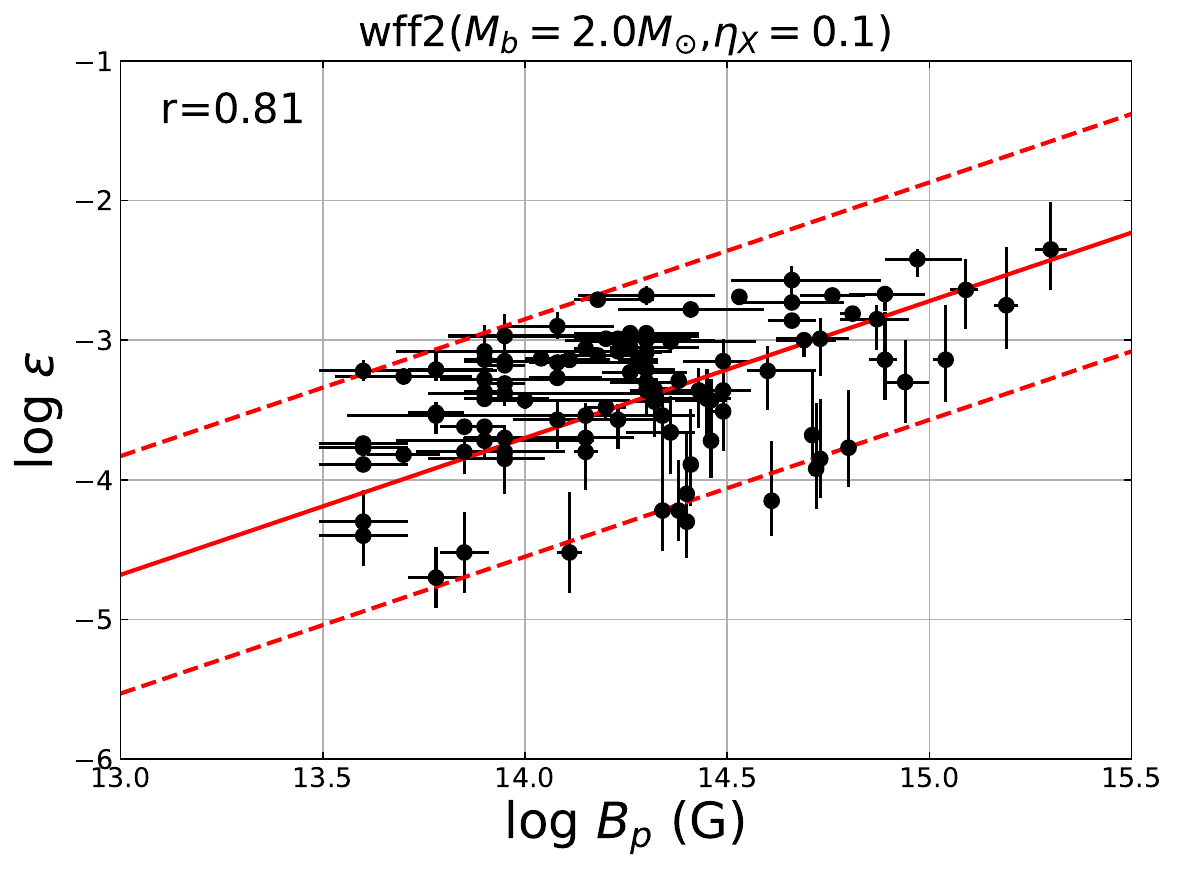}
\includegraphics [angle=0,scale=0.29]  {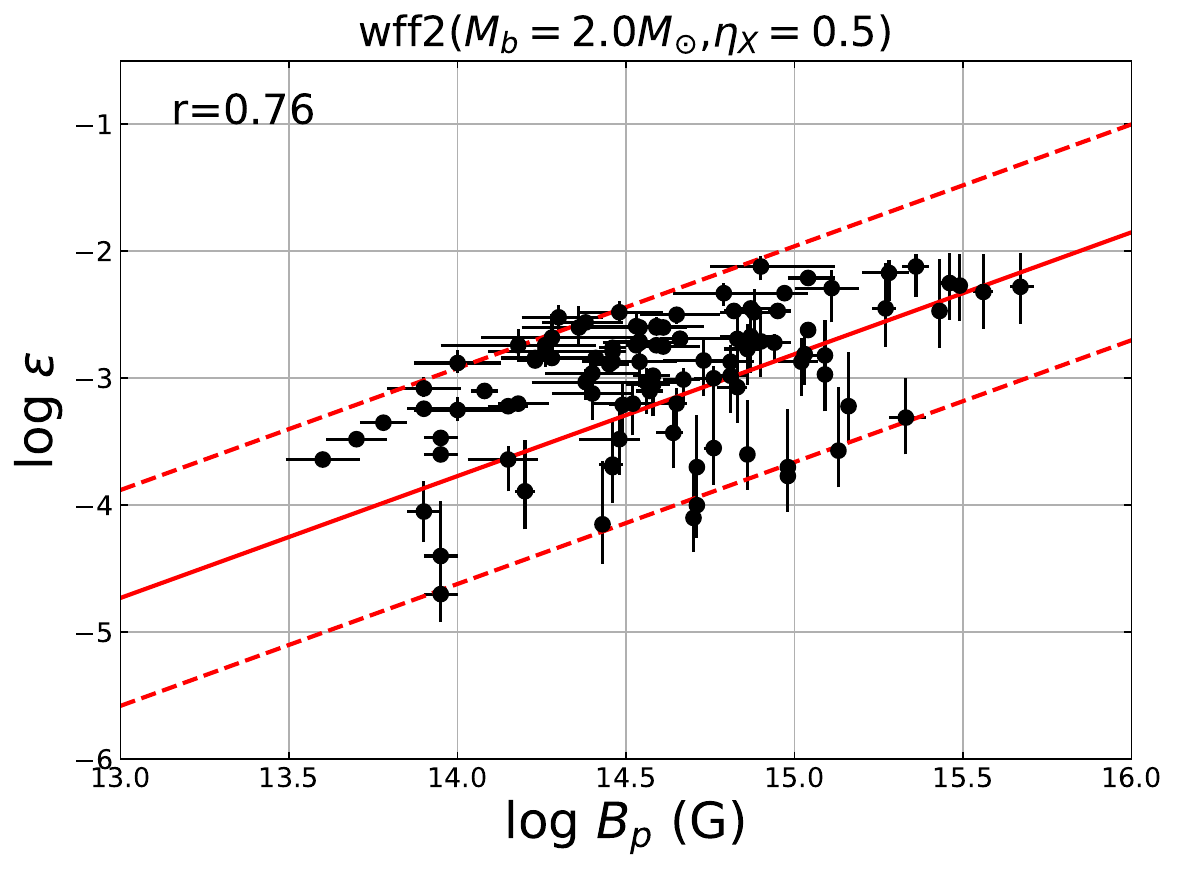}\\
\includegraphics [angle=0,scale=0.29]  {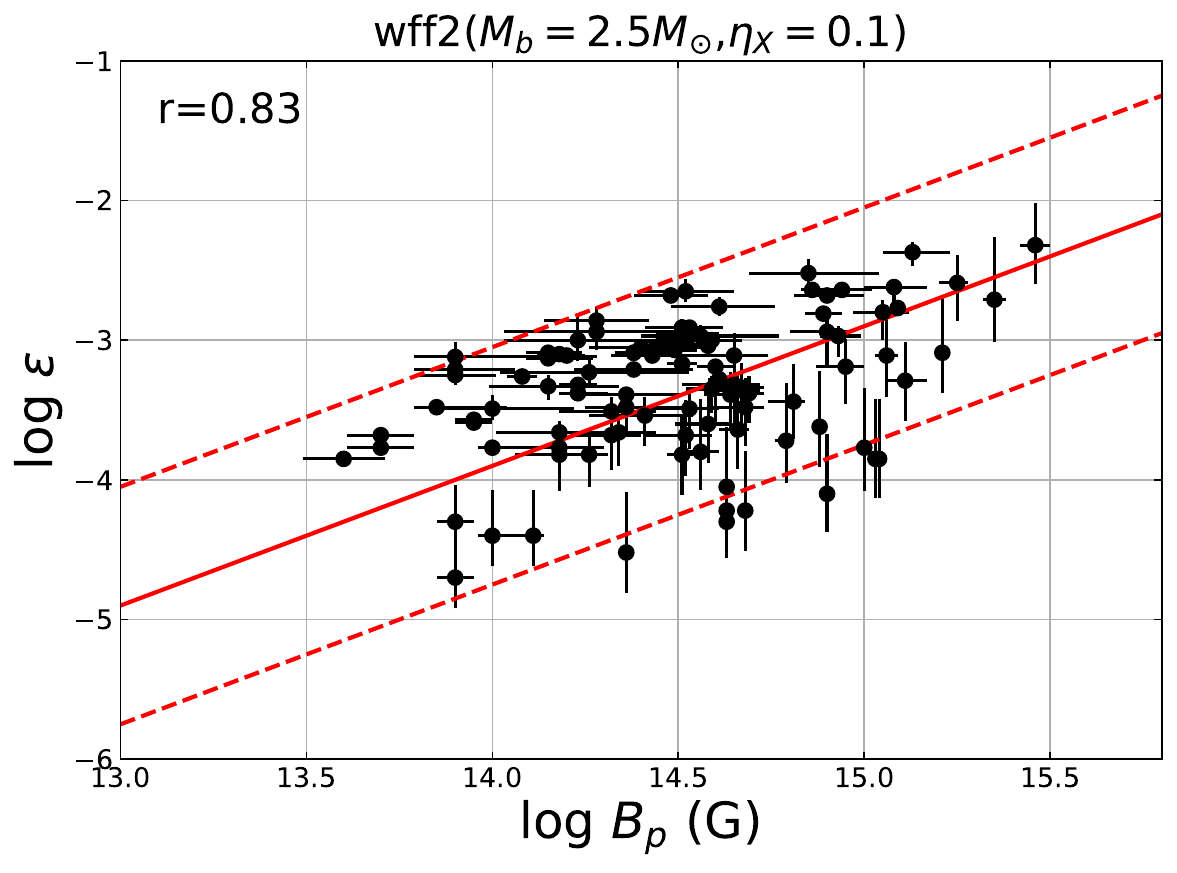}
\includegraphics [angle=0,scale=0.29]  {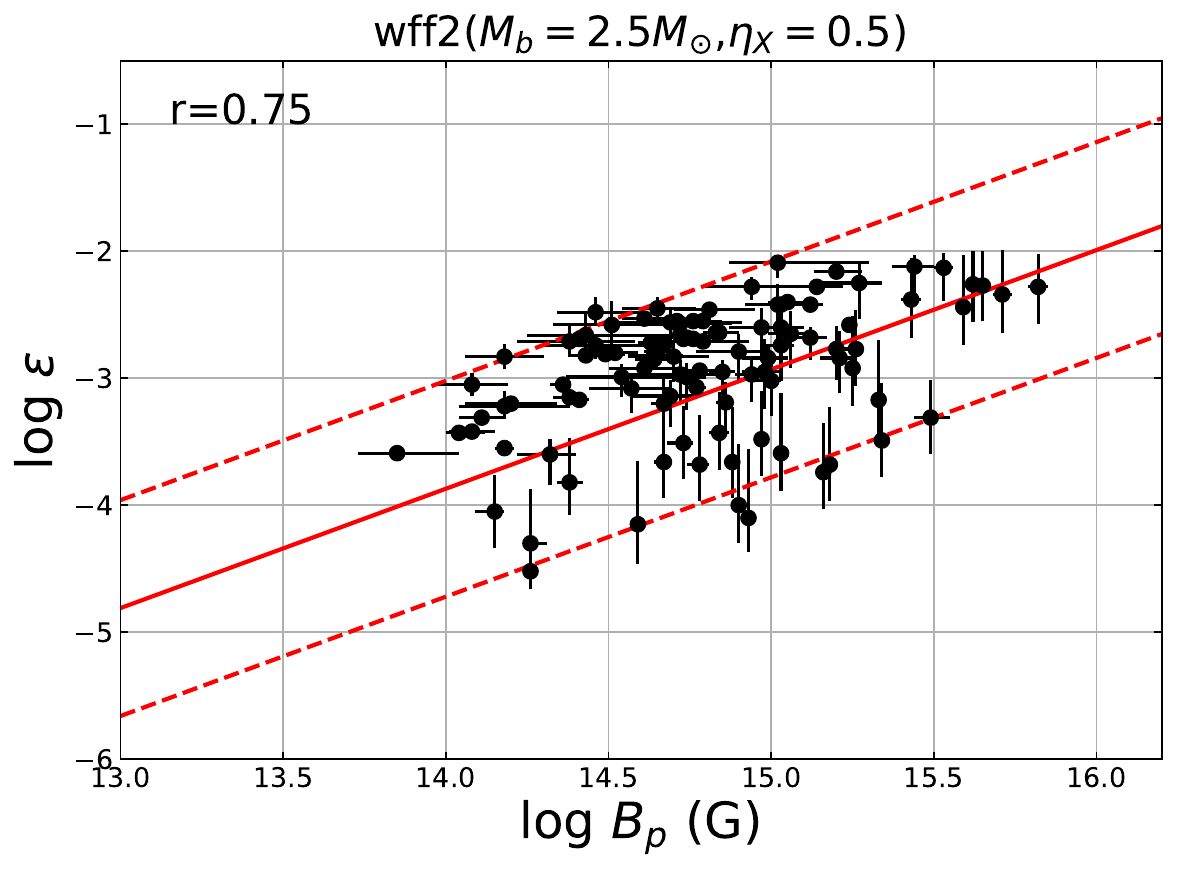}
\caption{The correlations between the $\epsilon$ and $B_p$ in four samples of EoSs with $M_{b}=2.0~M_{\odot},~2.5~M_{\odot}$ and $\eta_{\rm X}=0.1,~0.5$. The red solid and red dashed lines are the best-fitting results and the 95\% confidence level, respectively.}
\label{fig:Bp-epsilon}
\end{figure*}

In Figure \ref{fig:Bp-epsilon}, we showed a series of $\epsilon-B_p$ scatter diagrams for different EoSs and baryonic masses with two $\eta_{\rm x}$ values based on our LGRB sample. More interestingly, we found that there are also strong correlations between $\epsilon$ and $B_p$ for our selected EoSs. Most of the GRBs in our sample fall into the $3\sigma$ deviation region of the best-fitting power-law model for all EoSs scenarios. Our best-fitting correlations for various cases are reported in Table \ref{table-13}. The tight correlations of $\epsilon-B_p$ appeared to also be universal for all of our selected EoSs, which may indicate that the newly born magnetar with a stronger dipole magnetic field corresponds to the larger ellipticity, which may be an important clue for understanding the ellipticity-arising mechanisms in the GRB's magnetar central engine. The ellipticity-arising mechanisms of magnetar depend on the deformation mechanisms. To date, a number of mechanisms have been proposed to lead to NS being deformed with asymmetry and produce the NS ellipticity, including magnetically induced ellipticity, r-mode instability-induced ellipticity, starquake-induced ellipticity, and accretion mountain-induced ellipticity \citep{Ostriker1969,Bonazzola1996,Owen1998,Andersson1998,Andersson2003,Ushomirsky2000,Cutler2002,Ioka2004,Melatos2005,Haskell2006,Haskell2008,Dallosso2009,Mastrano2011,Mastrano2012,Lander2012,Lander2014,Lasky2015,de-Araujo2016,Gao2017,Zhong2019,Lin2019,Lin2020,Abbott2020b,Giliberti2022,Dallosso2022,Huang2022}. Among various mechanisms, magnetically induced deformation likely plays a dominant role in maintaining a relatively large $\epsilon$ for a newborn millisecond magnetar. The millisecond magnetar is a subset of NSs with an extremely strong magnetic ﬁeld that can exceed $10^{15}~\rm G$ at birth, and the anisotropic magnetic stress is too large for the magnetar to maintain a long-term spherical structure, which will likely produce a significant ellipticity. According to previous analytical and numerical studies, the magnetically induced ellipticity is likely to be roughly proportional to the interior magnetic field strength $B_{\rm int}$, which is determined by the NS EoSs and the twisted-torus configurations of the interior magnetic field (including the inclination angle and the magnetic energy ratio of the mixed toroidal/poloidal components) \citep{Lander2012,Lander2014,Mastrano2012,de-Araujo2016,Abbott2020b}. In general, the magnetically induced ellipticity of an NS is well approximated by the formula 
\begin{equation}
\epsilon\approx 10^{-8}\left(\frac{B_{\rm int}}{10^{12}\,{\rm G}}\right),
\label{eq:epsilon}
\end{equation}

However, in reality, very little is known about the strength and configuration of the NS interior magnetic field, which can be rapidly magnified by strong differential rotation combined with turbulent convection in the nascent NS drives of an $\alpha-\Omega$ dynamo mechanism \citep{Duncan1992,Giacomazzo2013,Zrake2013}. We expect to calculate the ellipticity from the exterior magnetic field strength $B_{\rm dip}$, which can be inferred from the EM observations of GRB X-ray afterglow. In principle, we can simply approximate that the exterior magnetic field strength of the magnetar is proportional to the interior magnetic field strength, and one can connect the interior mean magnetic field $\bar{B}_{\rm int}$ and the observationally inferred surface dipole magnetic field $B_{\rm dip}$ by defining \citep{Haskell2008,Mastrano2011,Gao2017}
\begin{eqnarray}
B_{\rm dip}=\eta\bar{B}_{\rm int},
\end{eqnarray}
where $\eta$ is a dimensionless value with $0\leq\eta\leq1$, $\eta=0$ and $\eta=1$ representing the magnetar with a purely toroidal and poloidal field component, respectively. In this context, one can get $\epsilon \propto B_p$. Coincidentally, we found that within the systematic error range, the correlations of $\epsilon-B_p$ in our GRB sample can be well described as $\epsilon\propto B_p^{0.97\pm0.13}$ for our selected EoSs, with the 1$\sigma$ deviation included. If such a relation is true, we can conclude that the ellipticity mechanism of GRB magnetar originates from magnetically induced deformation, particularly via the contribution of the strong interior magnetic field. Furthermore, we have found that, within the systematic error range, there is also a strong correlation between $\epsilon$ and $P_0$, which can be approximately described as $\epsilon\propto P_0^{1.57\pm0.22}$ for all selected EoSs. The correlation is likely to also imply that the asymmetry origin of the GRB magnetar is related to the magnetically induced distortion. This is because, in the mountain-induced deformation scenario, the NS crust is solid and elastic, and the shape of crust depends on many factors, such as the original formation history and accretion history of the NS, starquake, and the NS EoSs \citep{Ushomirsky2000,Haskell2006,Lasky2015}. In the mountain-induced deformation mechanism, the NS rotation is unable to erase the mass asymmetry and has a constant $\epsilon$ for different $P_0$. However, in the magnetically induced distortion scenario, due to the competition between the centrifugal deformation and the magnetic field deformation, the faster NS rotation velocity will likely trowel or cancel out some of the mass asymmetry, making the ellipticity smaller \citep{Cutler2000,Dallosso2009,Zanazzi2015,Huang2022}, i.e., the asymmetry deformation produced via the magnetic field distortion mechanism should satisfy the strong correlation between $\epsilon$ and $P_0$. All these statistical results seem to indicate to us that the ellipticity of GRB newborn magnetar originates from the magnetically induced distortion mechanism, and that is consistent with the theoretical derivation in \cite{Haskell2008}.

\begin{figure*}
\centering
\includegraphics [angle=0,scale=0.29]  {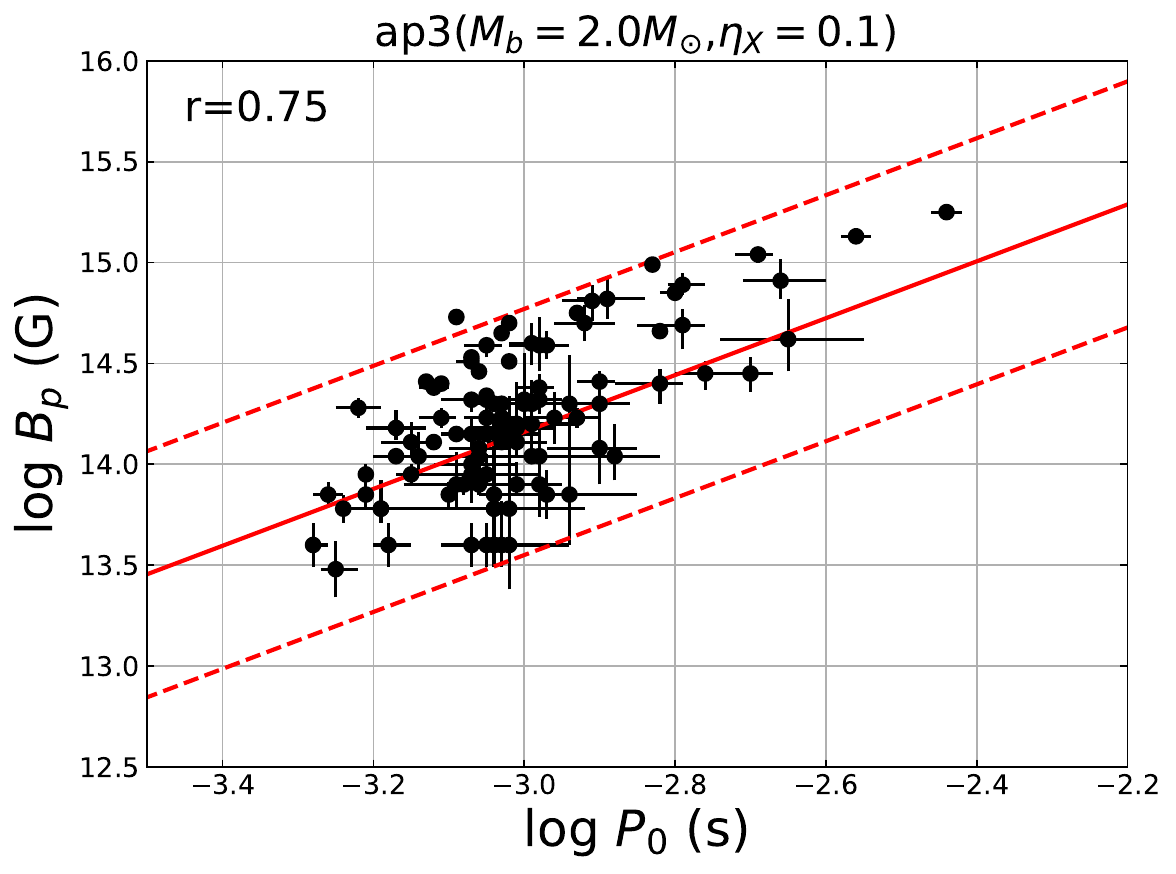}
\includegraphics [angle=0,scale=0.29]  {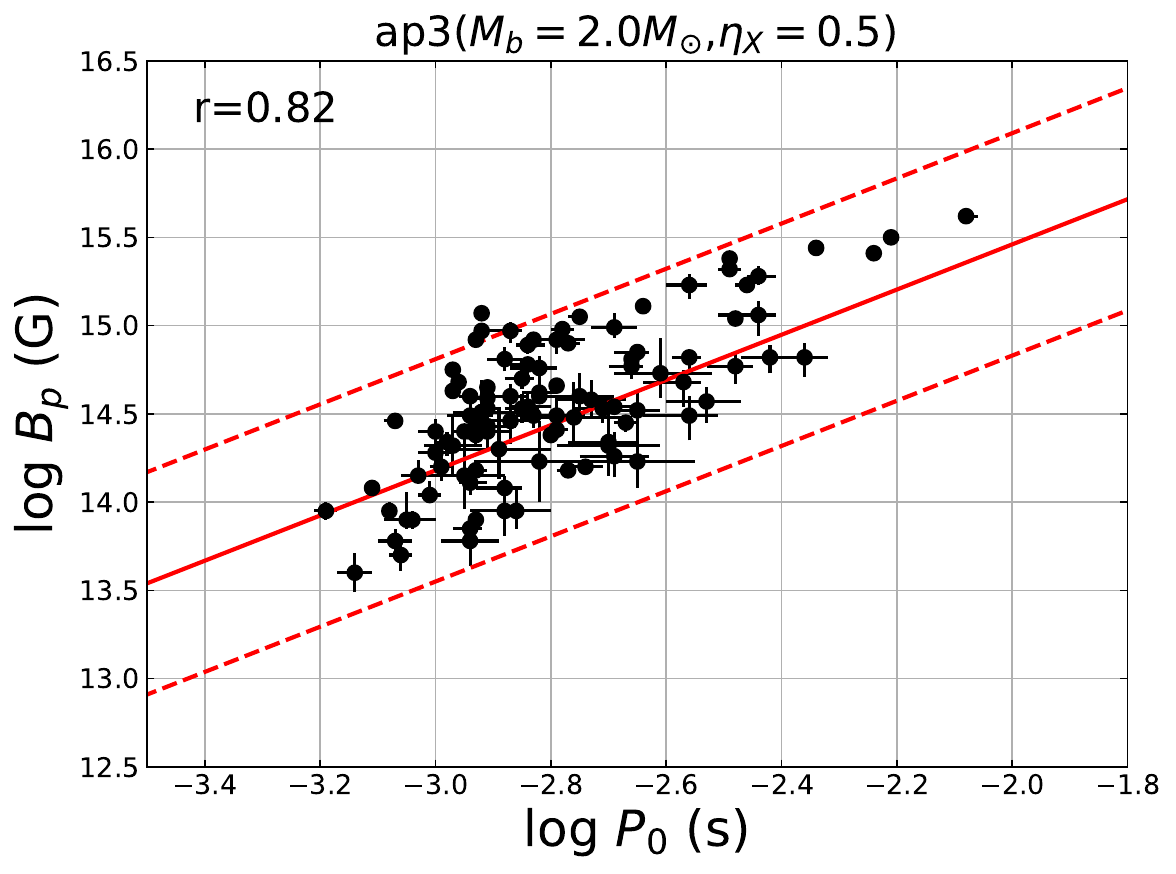}
\includegraphics [angle=0,scale=0.29]  {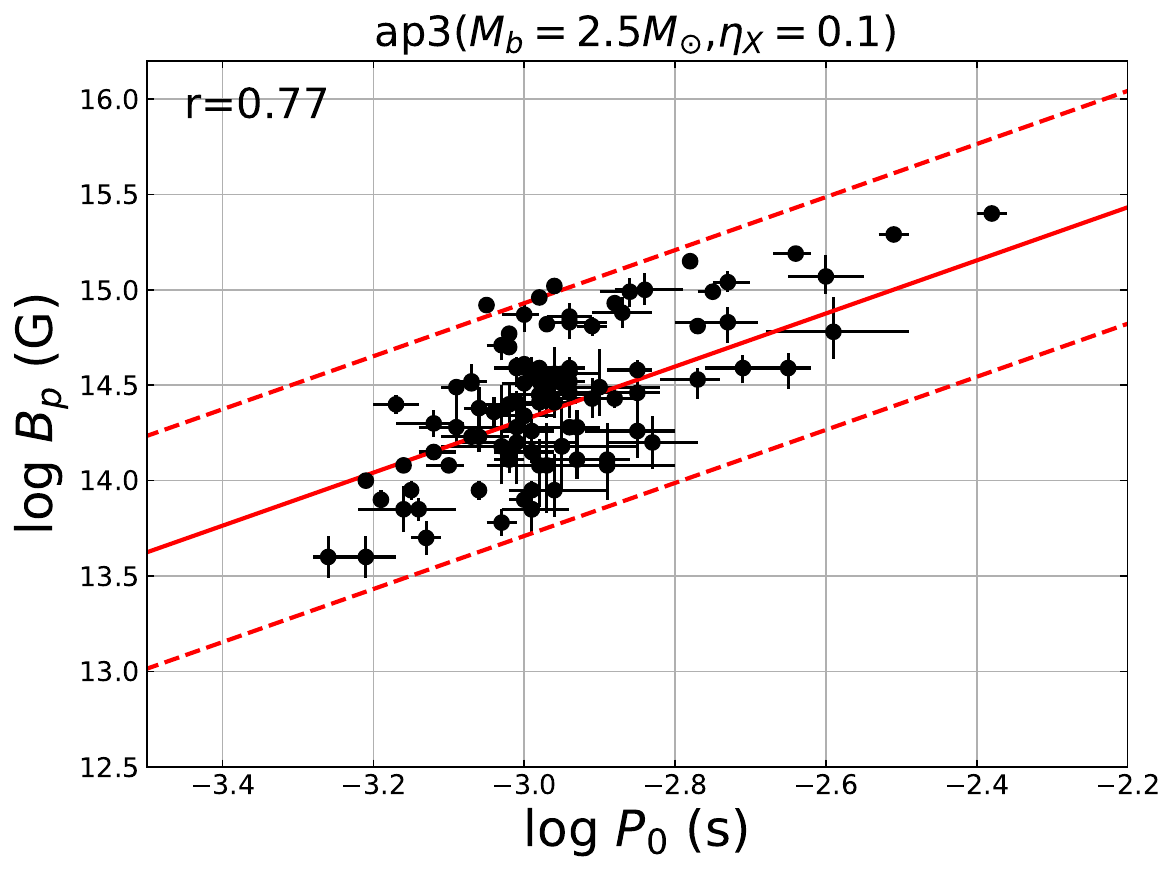}\\
\includegraphics [angle=0,scale=0.29]  {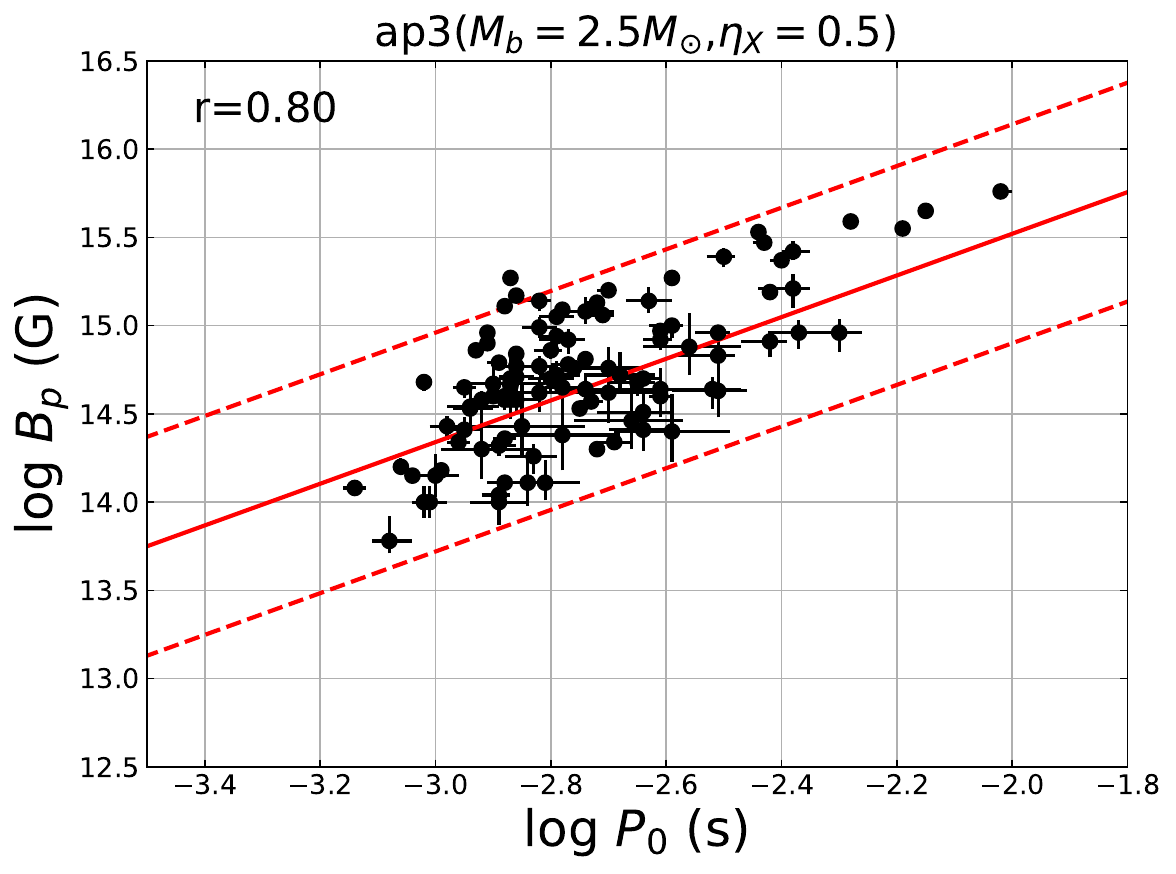}
\includegraphics [angle=0,scale=0.29]  {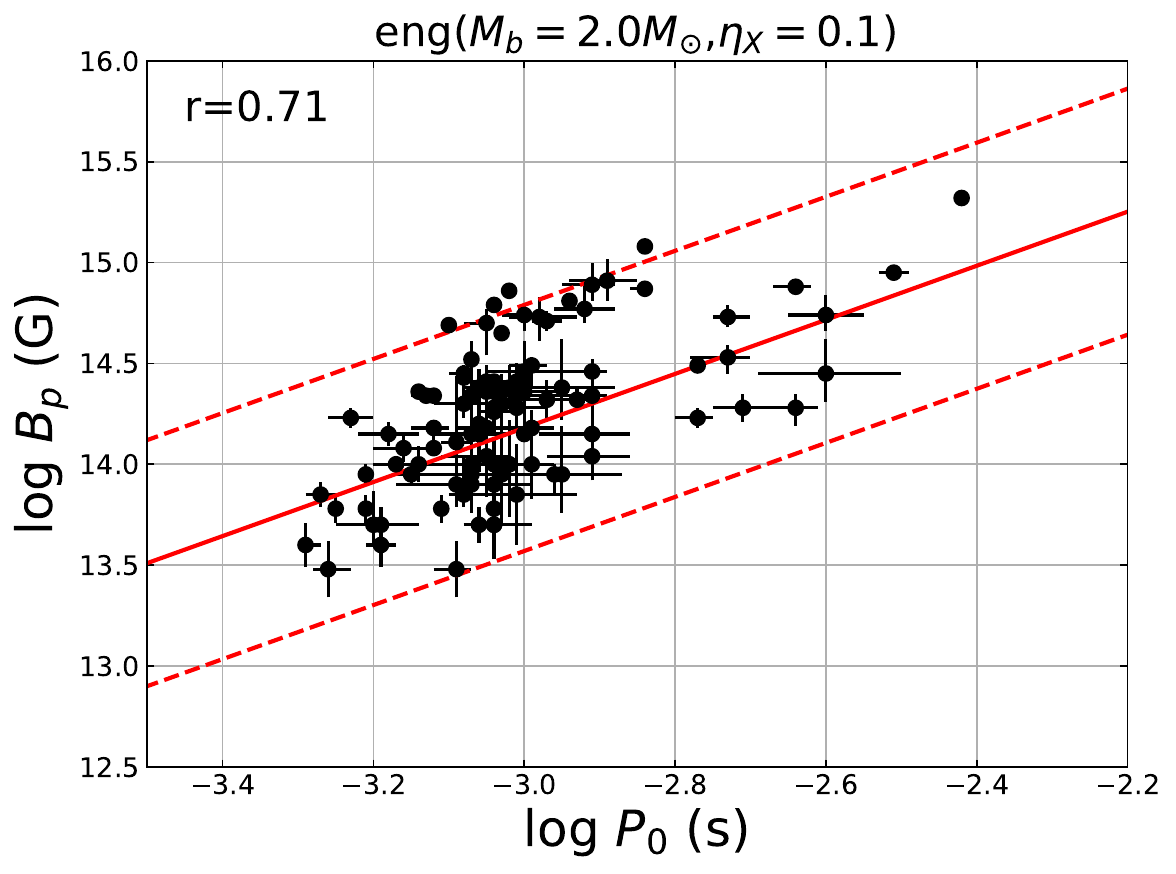}
\includegraphics [angle=0,scale=0.29]  {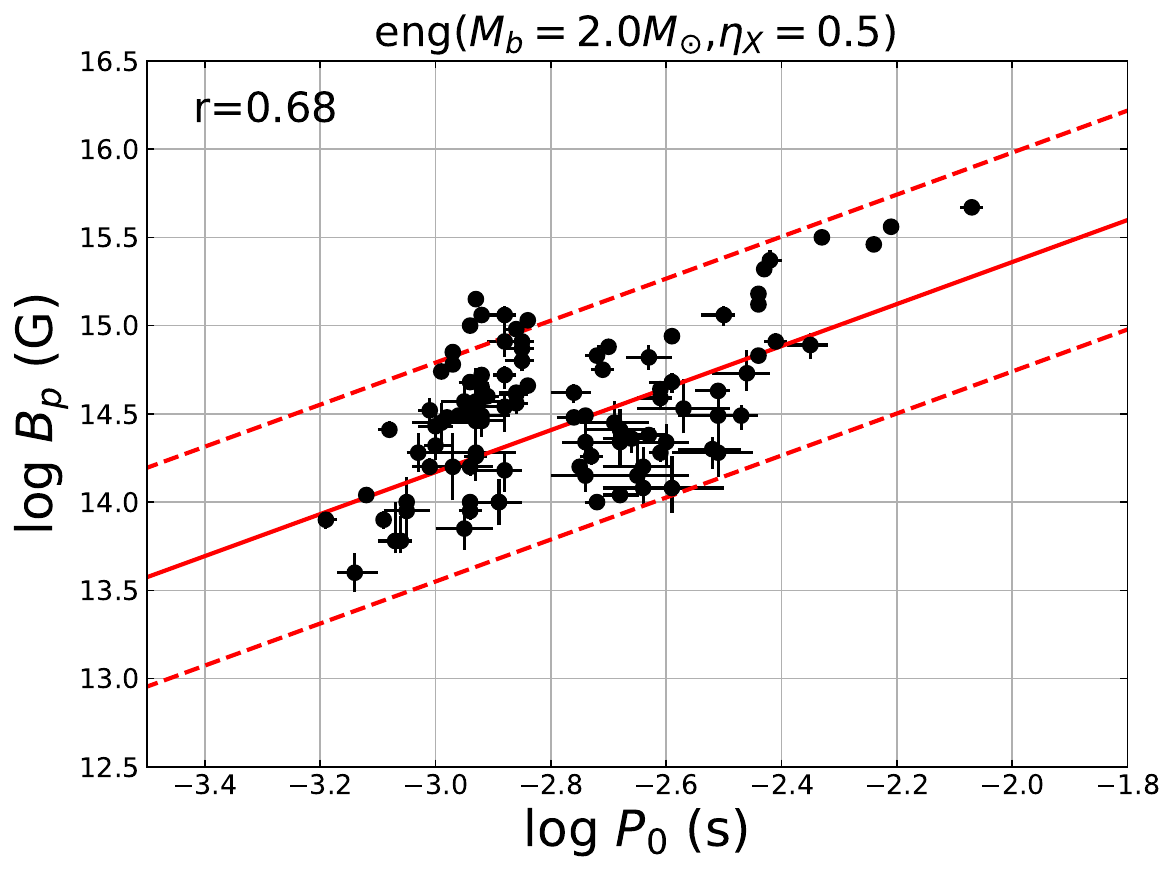}\\
\includegraphics [angle=0,scale=0.29]  {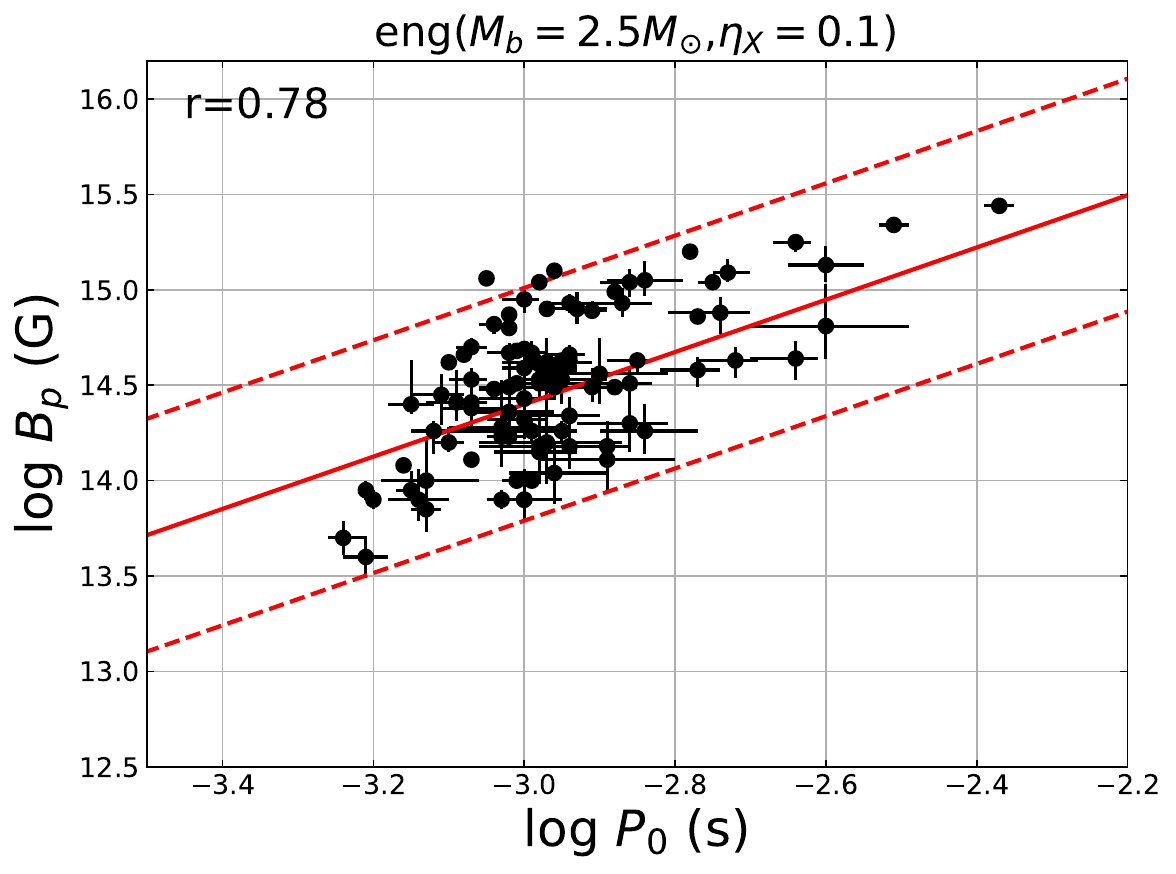}
\includegraphics [angle=0,scale=0.29]  {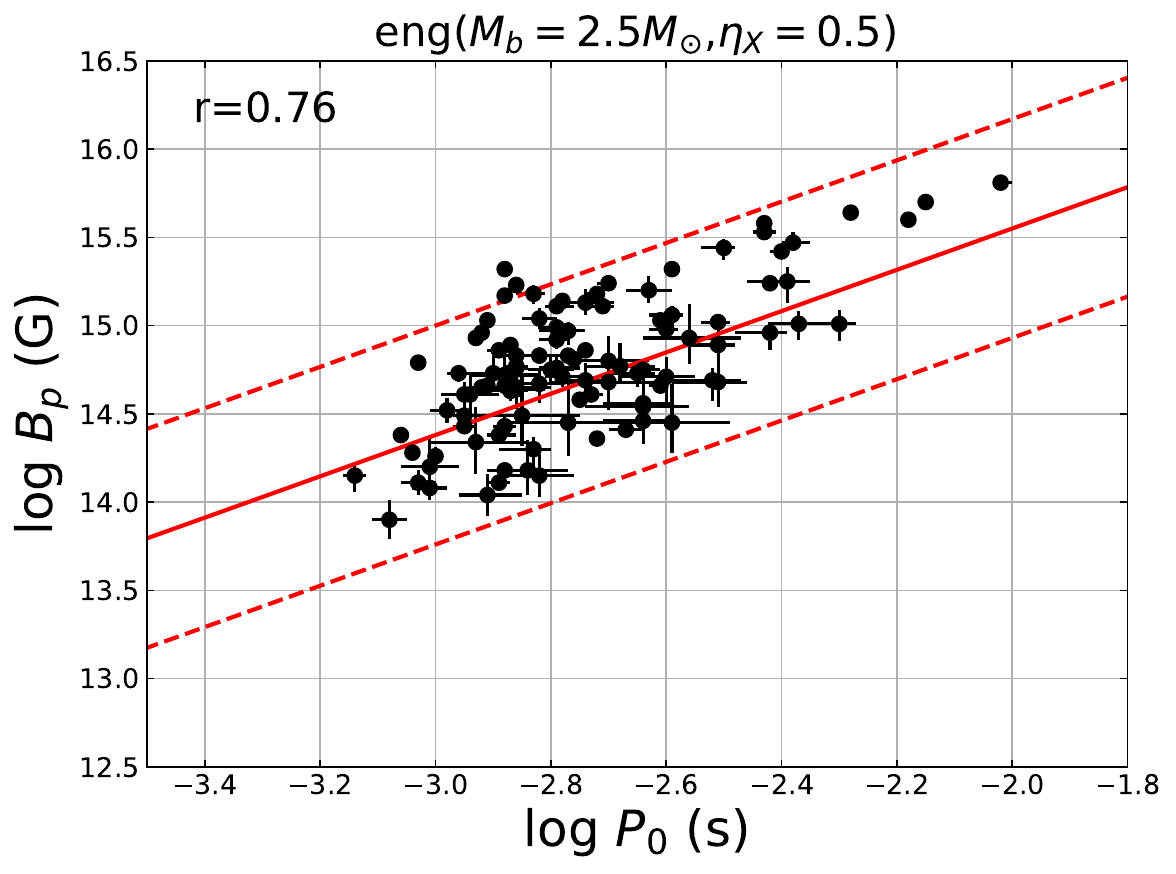}
\includegraphics [angle=0,scale=0.29]  {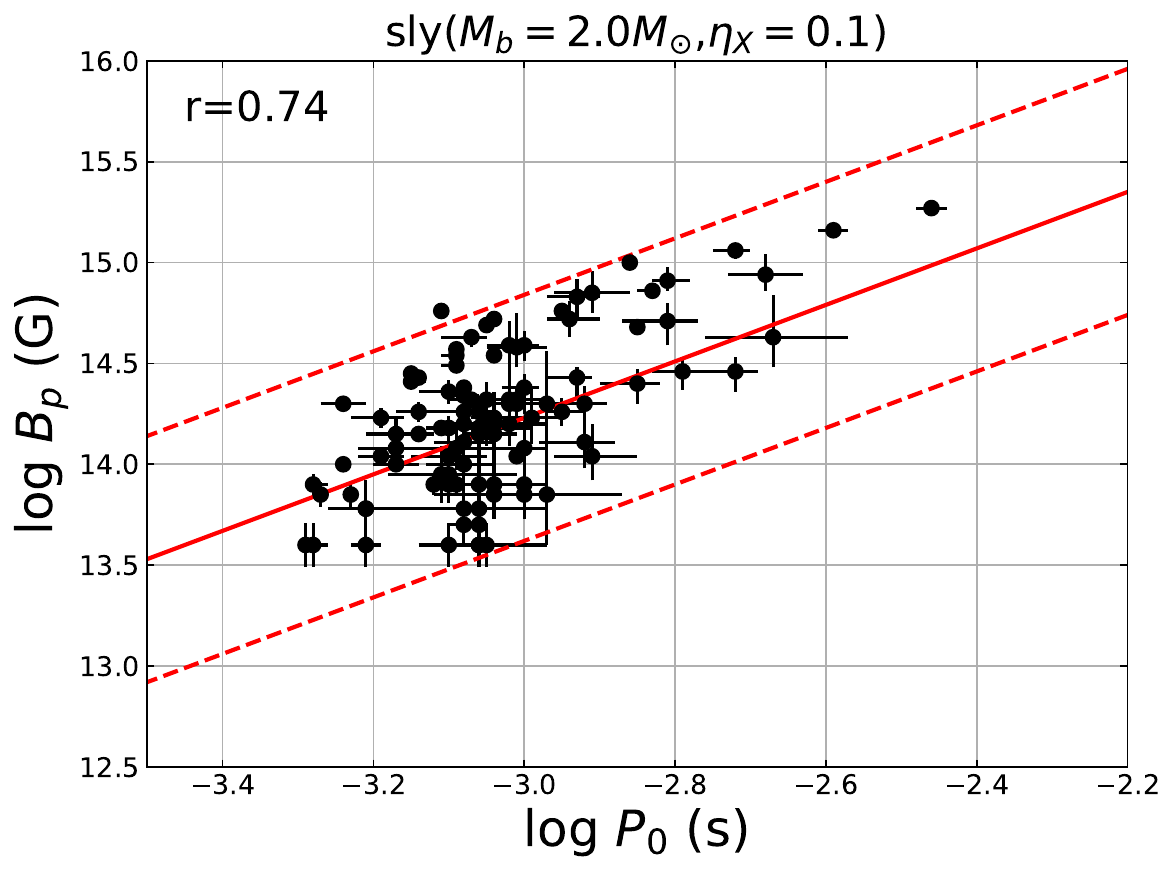}\\
\includegraphics [angle=0,scale=0.29]  {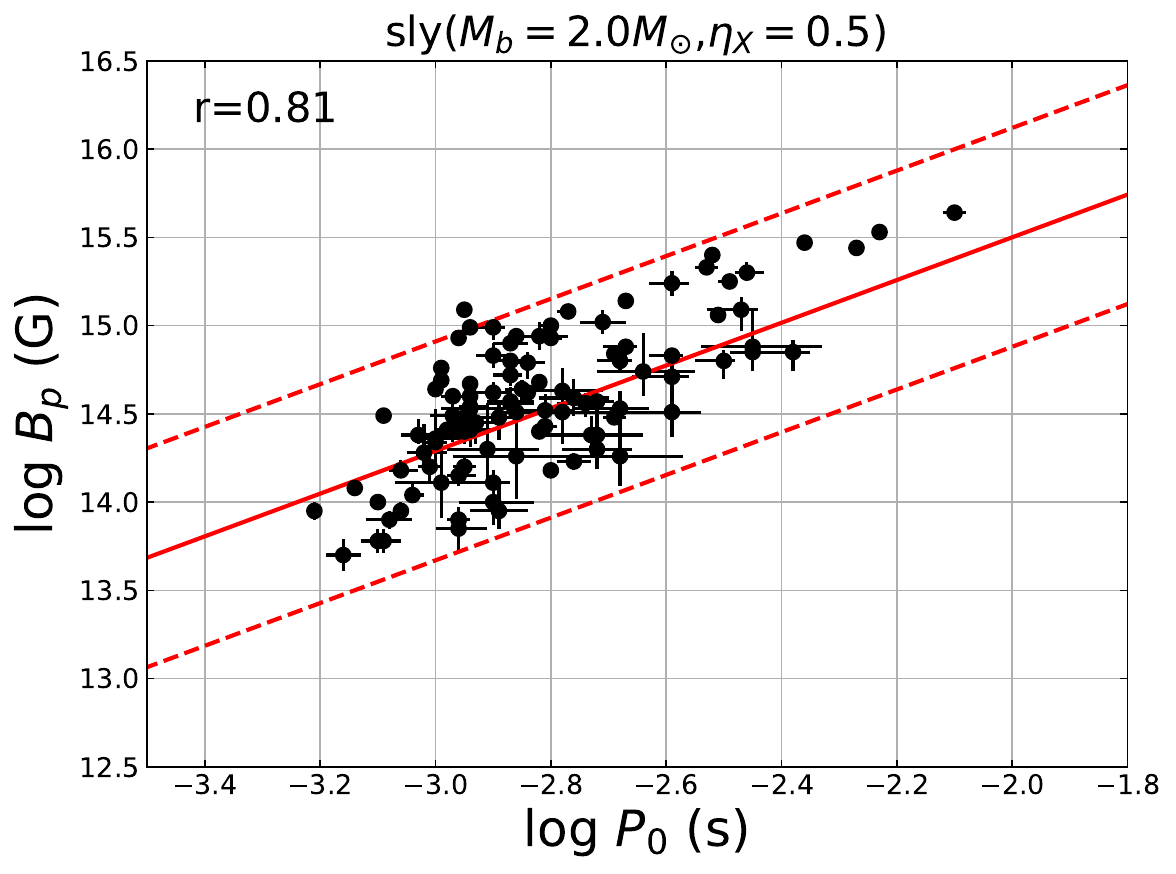}
\includegraphics [angle=0,scale=0.29]  {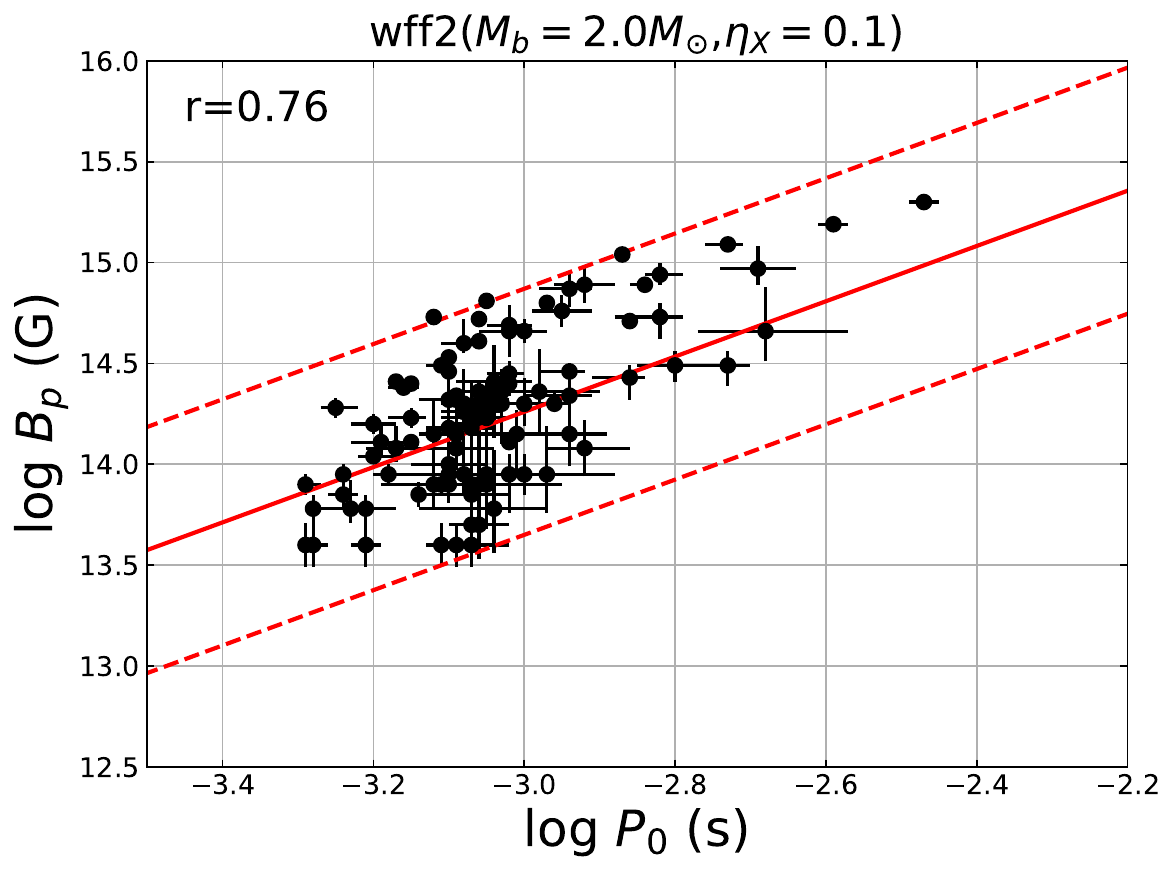}
\includegraphics [angle=0,scale=0.29]  {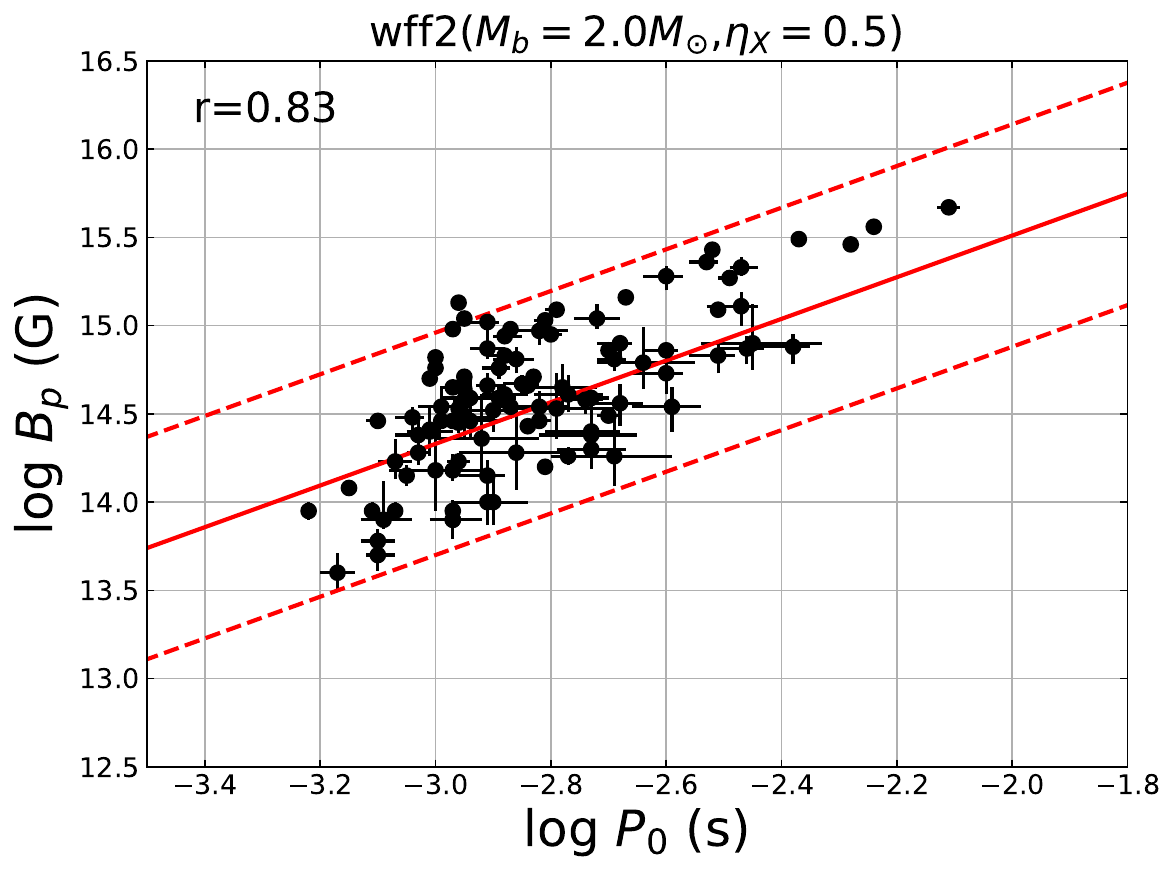}\\
\includegraphics [angle=0,scale=0.29]  {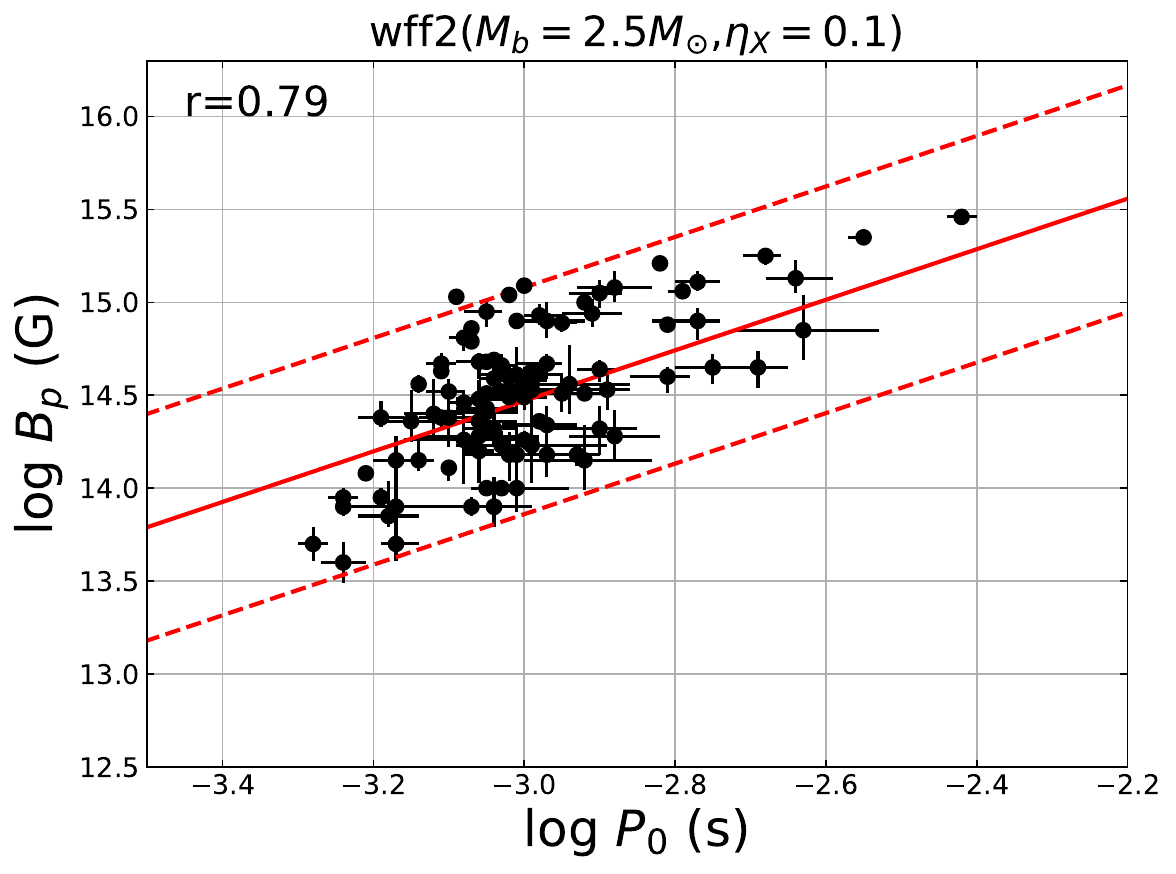}
\includegraphics [angle=0,scale=0.29]  {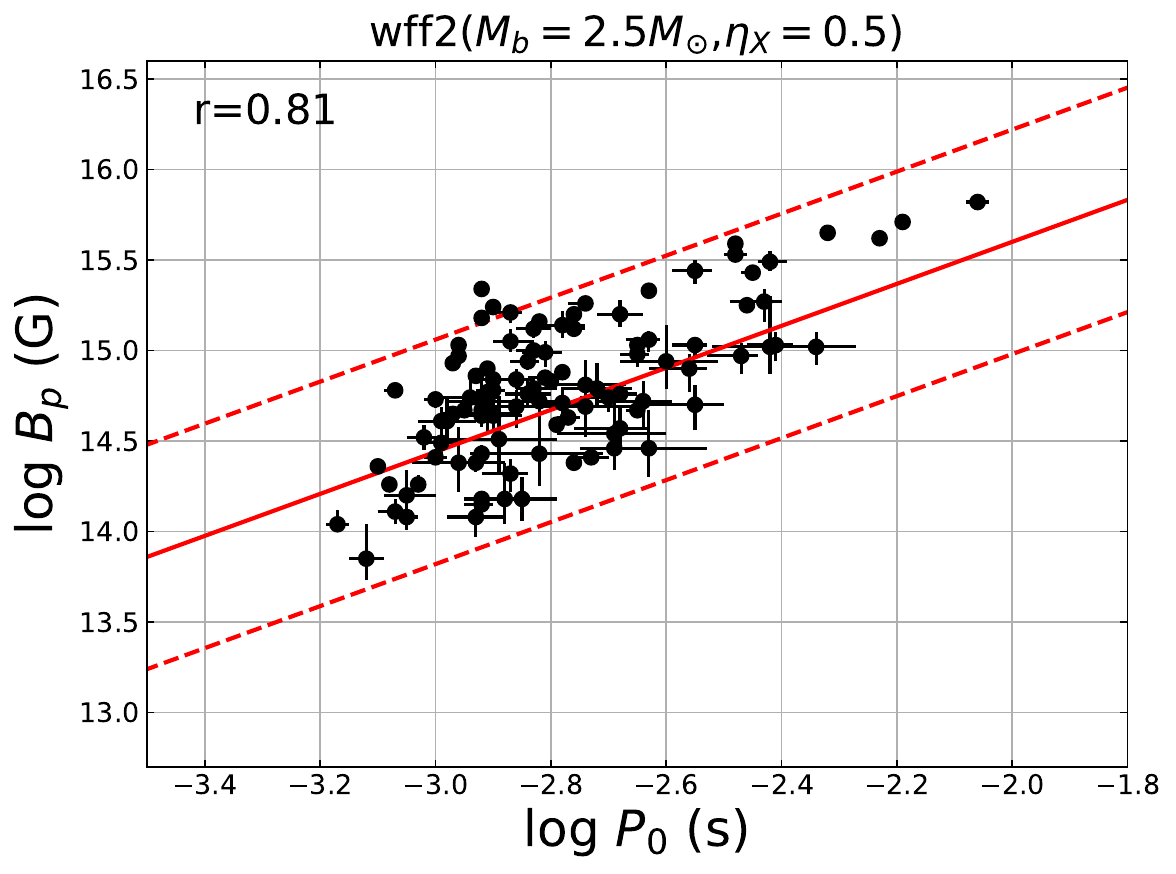}
\caption{The correlations between the $B_p$ and $P_0$ in four samples of EoSs with $M_{b}=2.0~M_{\odot},~2.5~M_{\odot}$ and $\eta_{\rm X}=0.1,~0.5$. The red solid and red dashed lines are the best-fitting results and the 95\% confidence level, respectively.}
\label{fig:Bp-P0}
\end{figure*}

In Figure \ref{fig:Bp-P0}, we also showed a series of the $B_p-P_0$ scatter diagrams for different EoSs and baryonic masses with two $\eta_{\rm x}$ values based on our LGRB sample. Similarly, we found that there are tight correlations between $B_p$ and $P_0$ for our selected EoSs. Most of the GRBs in our sample fall into the $3\sigma$ deviation region of the best-fitting power-law model for all EoSs scenarios. Our best-fitting correlations for various cases are reported in Table \ref{table-13}. The strong correlations of $B_p-P_0$ also appeared to be universal for all of our selected EoSs, which may indicate that the magnetic field of newly born magnetar is coupled to the initial period. The strong correlations of $B_p-P_0$ can be approximately described as $B_p\propto P_0^{1.30\pm0.16}$ for our selected EoSs, with the 1$\sigma$ deviation included. The correlation seems to be highly consistent with the magnetar reaching an equilibrium spin period, i.e., $B_p\propto P_{\rm eq}^{7/6}$ relation for the magnetar propeller mechanism with a given accretion rate, which accelerates local materials via magnetic centrifugal force throwing \citep{Piro2011,Gompertz2014,Gibson2017,Gibson2018,Li2021,Lin2021b,Yu2024}. Within the propeller model, the propeller regime of the magnetar and its surrounding accretion disk is basically defined by the relative position between the Alfvén radius (the radius at which the dynamics of the gas within the disk is strongly influenced by the magnetic field, $r_{\rm m}$) and the corotation radius (the radius at which material in the disk orbits at the same rate as the magnetar surface, $r_{\rm c}$). These two radii can be defined as follows:
\begin{eqnarray}
r_{\rm m}=\mu^{4/7}(GM)^{-1/7}\left(\frac{3M_{\rm d}}{t_{\nu}}\right)^{-2/7},
\end{eqnarray}
\begin{eqnarray}
r_\mathrm{c}=(GM/\Omega^2)^{1/3},
\end{eqnarray}
where $\mu=B_{\rm p}R^3$ is the magnetic dipole moment, $t_\nu=R_{\rm d}/ \alpha c_{\rm s}$ is the viscous time-scale, and $\alpha$ and $c_{\rm s}$ are the viscosity prescription and the speed of sound in the disk, respectively. $\Omega$ is the angular frequency of the magnetar central engine, and $M_{\rm d}$ and $R_{\rm d}$ are the mass and radius of disk, respectively. The interaction process between the magnetar and the material of surrounding accretion disk depends on the relative magnitude between $r_{\rm m}$ and $r_{\rm c}$. In general, when $r_{\rm c}>r_{\rm m}$, the accretion disk is rotating more rapidly than the magnetic field, and the effect of the interaction is to slow the material and allow it to accrete, i.e., the so-called accretion regime. Conversely, if $r_{\rm c}<r_{\rm m}$, the magnetic field is spinning faster than the material, and the materials already within $r_{\rm c}$ accrete to the surface of the magnetar, while the materials within the range of $r_{\rm c}$ and $r_{\rm m}$ will be accelerated to super-Keplerian velocity and propelled away the system. If the propeller power is not strong enough for the materials, it cannot reach the potential well. Then, the materials will return to the disk without any emission signals to be detected, i.e., the so-called propeller regime. It should be noted that $r_{\rm c}$ and $r_{\rm m}$ are always dynamically changing in the accretion regime and the propeller regime. In the accretion regime, the magnetar is spun up by gaining angular momentum from accretion, and the increased spin velocity will cause $r_{\rm c}$ to shrink. Then, as the accretion rate falls off, the Alfvén radius $r_{\rm m}$ will expand until $r_{\rm c}<r_{\rm m}$ thus switching on the propeller regime. In the propeller regime, the materials within the range of $r_{\rm c}$ and $r_{\rm m}$ will be accelerated to super-Keplerian velocity by the magnetar, which results in mass ejection from disk and substantial spin-down of the magnetar. The spin-down of the magnetar, in return, will lead $r_{\rm c}$ to expand. Ultimately, such a magnetar-disk system tends to evolve toward $r_{\rm c} \simeq r_{\rm m}$ if the spin evolution of the magnetar is dominated by the interaction with its accretion disk. When $r_{\rm c}$ equals $r_{\rm m}$, the accreting magnetar would reach an equilibrium spin period, which can be expressed as \citep{Piro2011}
\begin{equation}
P_\mathrm{eq}=2\pi(GM)^{-5/7}R^{18/7}B^{6/7}\dot{M}^{-3/7},
\label{eq: Peq}
\end{equation}
where $\dot{M}$ is the fallback accretion rate of the magnetar. Distinctly, the equilibrium spin period is independent of the initial spin period of the magnetar but is related to the magnetar's surface magnetic field strength and fallback accretion rate. Such a $B_p\propto P_{0}^{1.30\pm0.16}$ correlation in our GRB sample may suggest that the initial spin period derived from observations could deviate from that of the magnetar at birth, but possibly corresponds to the equilibrium spin period as a result of interaction between the magnetar and its accretion disk, and the result is consistent with the statistical results in \cite{Lin2021b} and \cite{Xie2022b}.

\subsection{Relation of Energy Releases between the Jet and Magnetar Wind} 
The energy partition between jet prompt emission and magnetar wind afterglow has been suggested to exhibit a quasi-universal behavior, i.e., a GRB with more energetic prompt emission can be powered by a more energetic newborn magnetar, resulting in stronger afterglow radiation \citep{Zou2019}. Here, it is of interest to investigate the relationships between the energy released fraction in jet prompt emission ($E_{\rm jet,iso}$) and wind afterglow emission ($E_{\rm wind}$), as well as to elucidate the connection between the properties of the magnetar central engine and GRB jet emission.

\begin{figure}
\includegraphics[angle=0,scale=0.45]{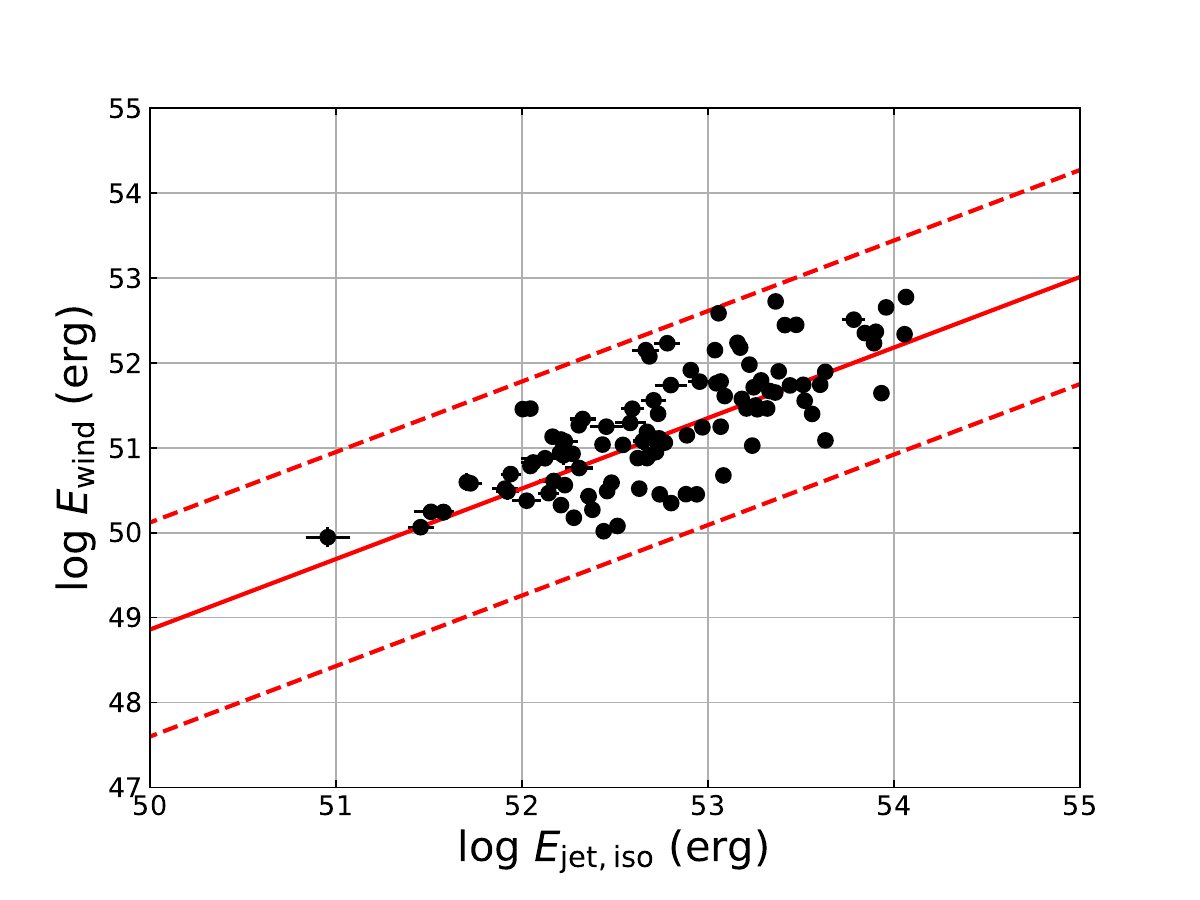}
\includegraphics[angle=0,scale=0.45]{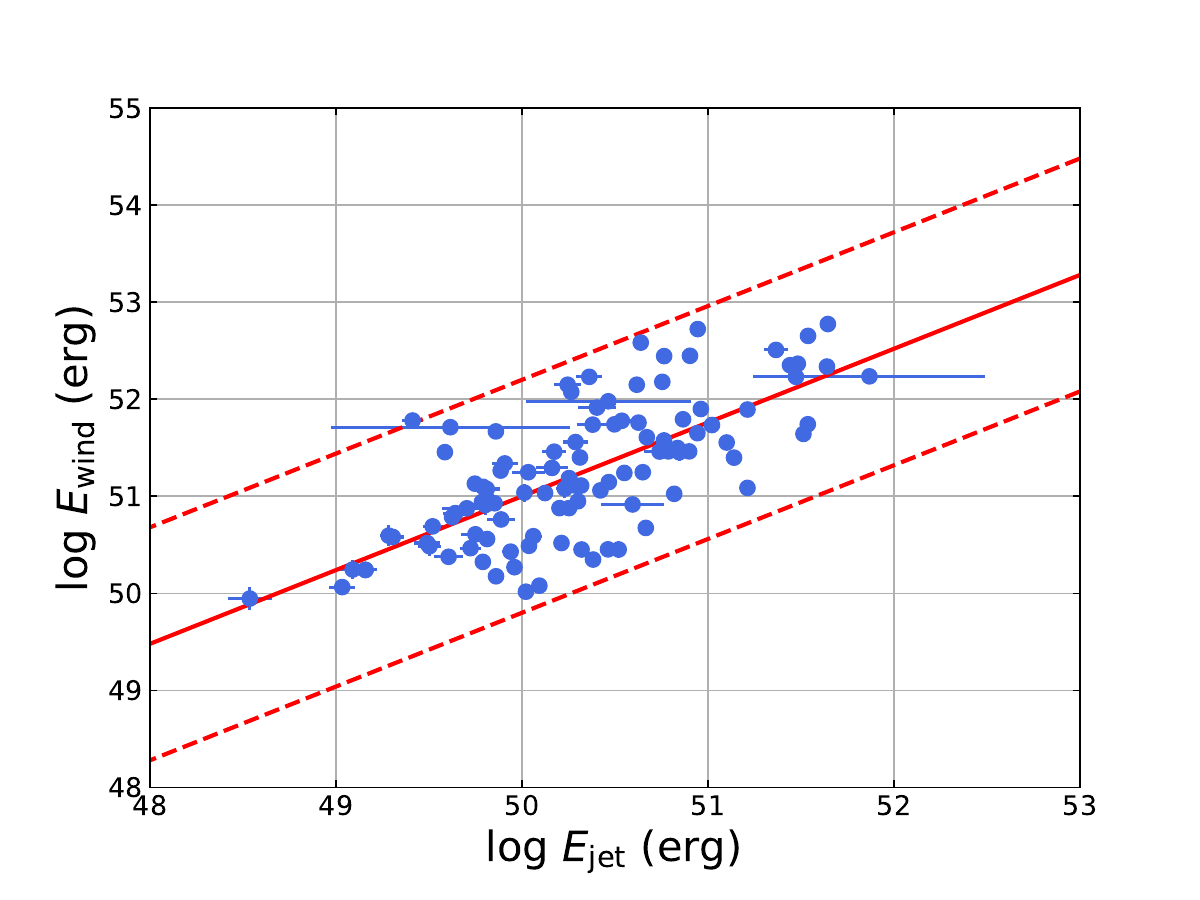}
\caption{The correlations between the isotropic jet energy $E_{\rm jet,iso}$ and the magnetar wind energy $E_{\rm wind}$ (left panel), as well as between the corrected jet energy $E_{\rm jet}$ and the magnetar wind energy $E_{\rm wind}$ (right panel). The red solid line and dashed lines are the least-square fit and the 95\% confidence level of the fits, respectively.}
\label{Ejet-Ewind}
\end{figure}

We first investigated the energy partition between the GRB jet and magnetar wind. The left panel of Figure \ref{Ejet-Ewind} showed $E_{\rm wind}$ as a function of $E_{\rm jet,iso}$. We performed the ordinary least-squares fitting to estimate the scaling relation of between the jet energy and magnetar wind energy, and one can observe a strong correlation between these two quantities, a Pearson correlation analysis gives the dependence
\begin{eqnarray}
\log E_{\rm wind}=(7.36\pm3.71)+(0.83\pm 0.07)\log E_{\rm jet,iso},
\end{eqnarray}
with the Pearson linear correlation coefficient $r = 0.80$, and a chance probability $p<10^{-12}$. The result is consistent with the previous statistical results within permissible systematic error range \citep{Zou2019,Lan2023}. The strong correlation indeed implied that the energy partition between the GRB jet and magnetar wind is quasi-universal among these LGRBs with plateau emission. Note that the GRB jet is highly collimated, and the jet radiation is concentrated within the jet-opening angle rather than isotropic emission, whereas the magnetar wind should be quasi-isotropic. Therefore, we need to correct the isotropic jet energy by considering the collimation effect of jet-opening angle to derive the true jet energy. A jet break was detected in some GRBs in our sample, and we can correct the isotropic jet energy of GRBs. The true jet energy is defined as \citep{Frail2001}
\begin{eqnarray}
E_{\rm jet} &=& f_{\rm b}E_{\rm jet,iso} \nonumber \\
&=& E_{\rm jet,iso}(1-\cos\theta_{j})\simeq\frac{1}{2}E_{\rm jet,iso}\theta_{j}^2,
\end{eqnarray}
where $f_b$ is the correction factor for jet beaming, and $\theta_{j}$ is the jet-opening angle in units of rad. First, we argued that the magnetar wind is isotropic, so that $f_b = 1$. We then corrected the jet prompt emission energy for each GRB. If $\theta_j$ is measured, we simply adopt the value. Otherwise, we choose $\theta_j = 5^{\circ}$, which is a typical jet opening angle for bright long GRBs \citep{Frail2001,Liang2008}. By making a geometrical correction, we reexamined the relation between the energy released fraction between the jet and magnetar wind. In Figure \ref{Ejet-Ewind} right panel, we found that this correlation still exists, which can be expressed as
\begin{eqnarray}
\log E_{\rm wind}=(13.00\pm3.71)+(0.76\pm 0.06)\log E_{\rm jet},
\end{eqnarray}
with the Pearson linear correlation coefficient $r = 0.77$, and a chance probability $p<10^{-9}$. These correlation results suggested that the magnetar wind energy $E_{\rm wind}$ is roughly proportional to isotropic jet energy $E_{\rm jet,iso}$ and true jet energy $E_{\rm jet}$, and we quantified the fraction of energy partition by defining the ratio parameters $\Re\equiv E_{\rm wind}/E_{\rm jet,iso}$ and $\Re'\equiv E_{\rm wind}/E_{\rm jet}$. The distributions of $E_{\rm wind}$, $E_{\rm jet,iso}$, $E_{\rm jet}$, $\Re$ and $\Re'$ together with the Gaussian fits are shown in Figure \ref{Ejet,wind&R}. The Gaussian fit yields $\log E_{\rm wind}/{\rm erg}=(51.25\pm0.70)$, $\log E_{\rm jet,iso}/{\rm erg}=(52.76\pm0.64)$, $\log E_{\rm jet}/{\rm erg}=(50.32\pm0.66)$, $\log \Re=(-1.50\pm 0.47)$, $\log \Re'=(0.93\pm 0.52)$, respectively. Typically, the isotropic jet energy $E_{\rm jet, iso}$ is about two orders of magnitude larger than $E_{\rm wind}$ and the derived $\Re$ typical value is $\Re = 0.03$, yet the true jet energy $E_{\rm jet}$ is about 2 orders of magnitude smaller than $E_{\rm wind}$ and the derived $\Re$ typical value is $\Re' = 8.51$. This result hints that the energies of the jet and magnetar wind are comparable.

\begin{figure}
\includegraphics[angle=0,scale=0.45]{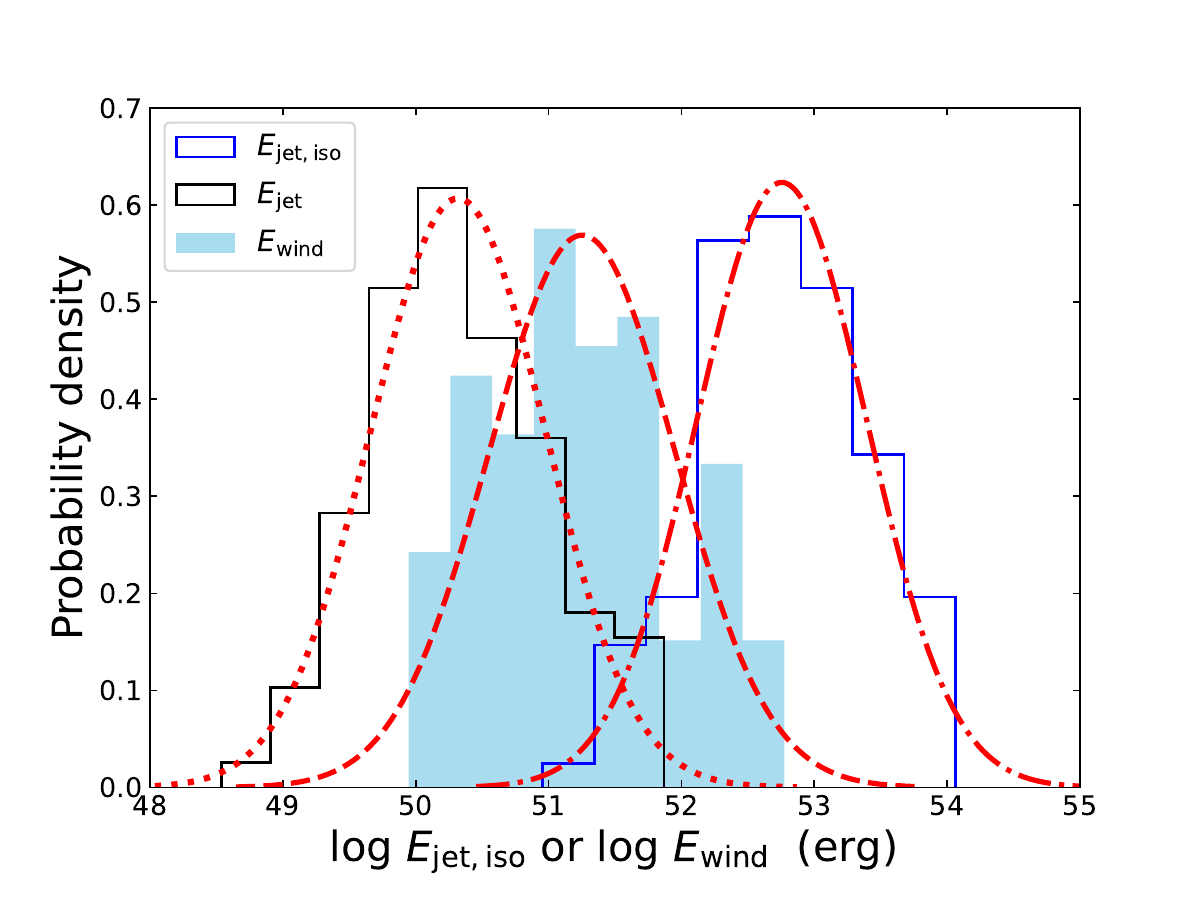}
\includegraphics[angle=0,scale=0.45]{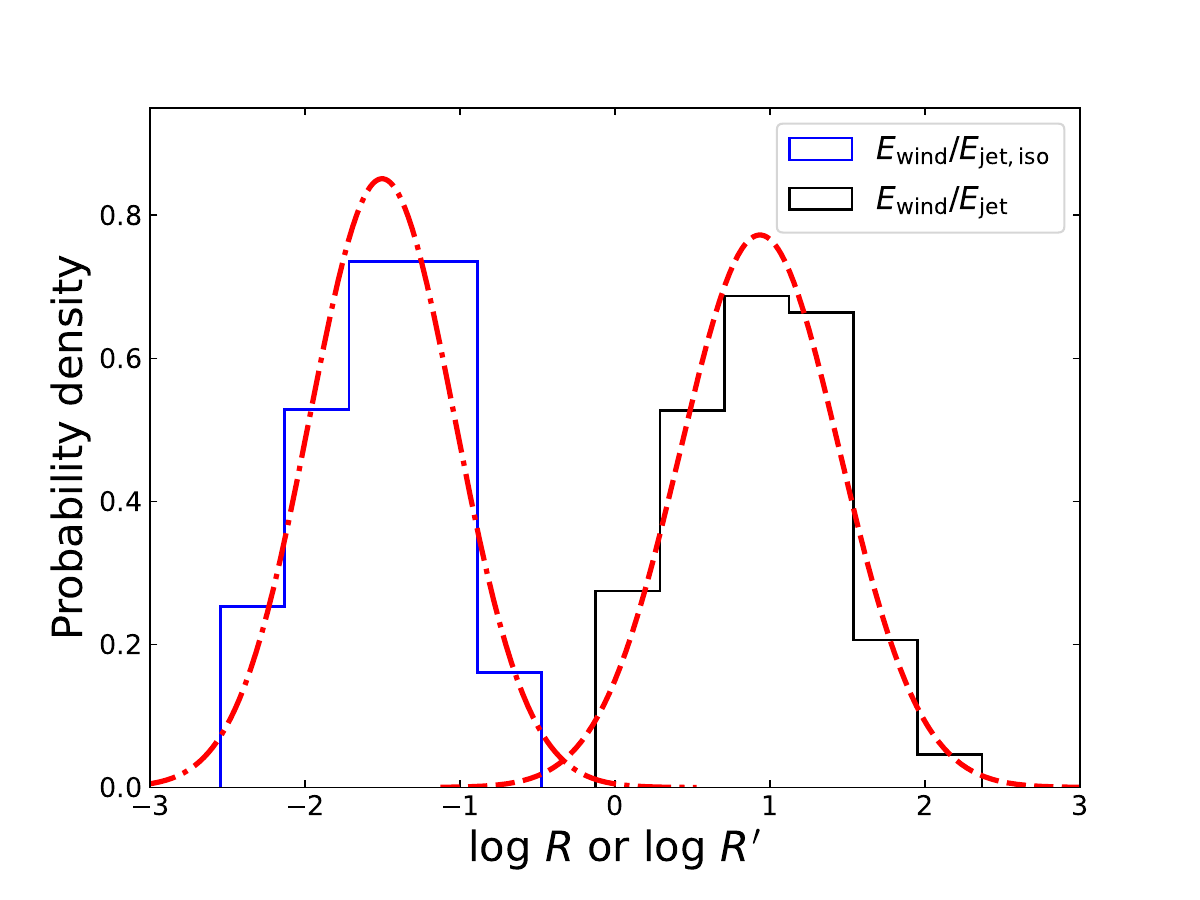}
\caption{Left panel: The distributions of isotropic jet prompt gamma-ray/X-ray energy ($E_{\rm jet, iso}$), corrected jet prompt gamma-ray/X-ray energy ($E_{\rm jet}$), and X-ray energy release of the magnetar wind ($E_{\rm wind}$) for the LGRBs in our sample. Right panel: The distribution of the energy partition ratio $R$ and $R^{\prime}$ for the LGRBs in our sample. The red dashed lines are the best Gaussian fits.}
\label{Ejet,wind&R}
\end{figure}

\begin{figure*}
\centering
\includegraphics [angle=0,scale=0.29]  {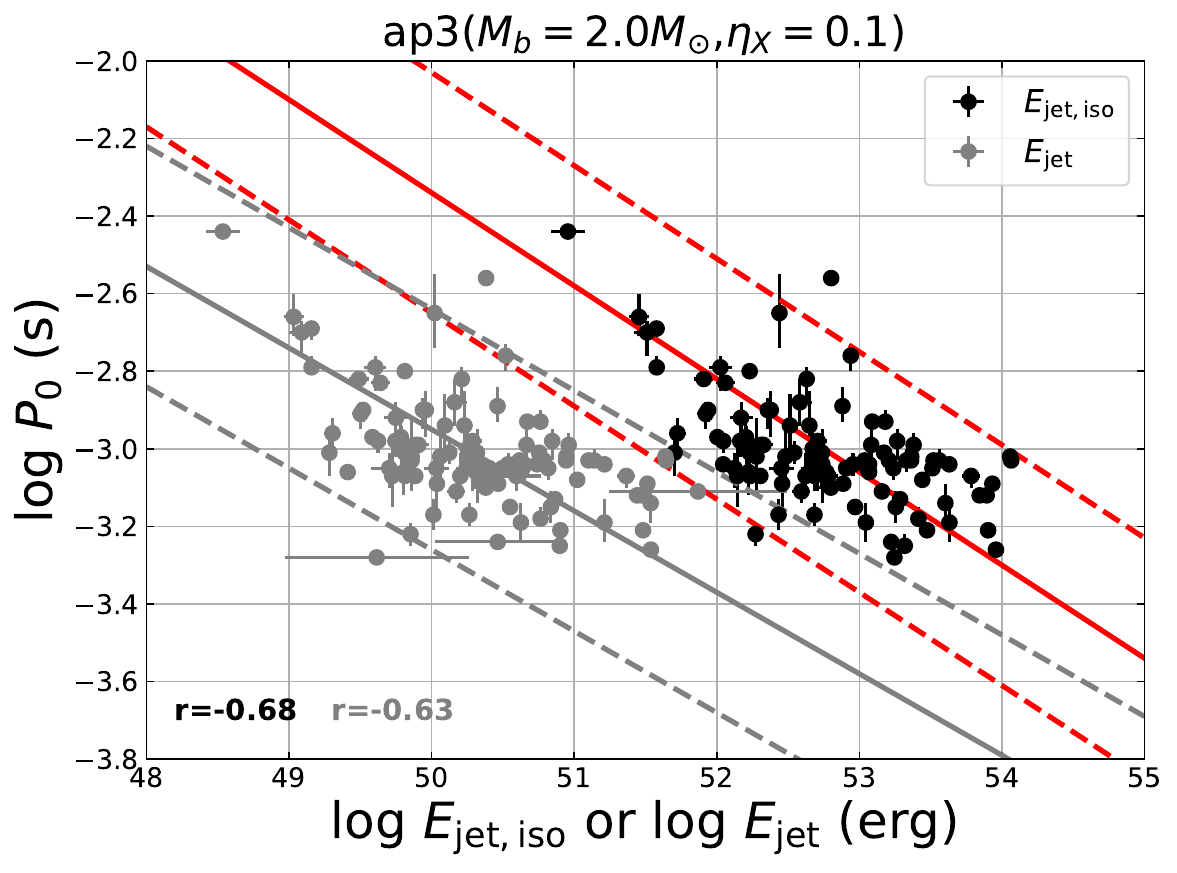}
\includegraphics [angle=0,scale=0.29]  {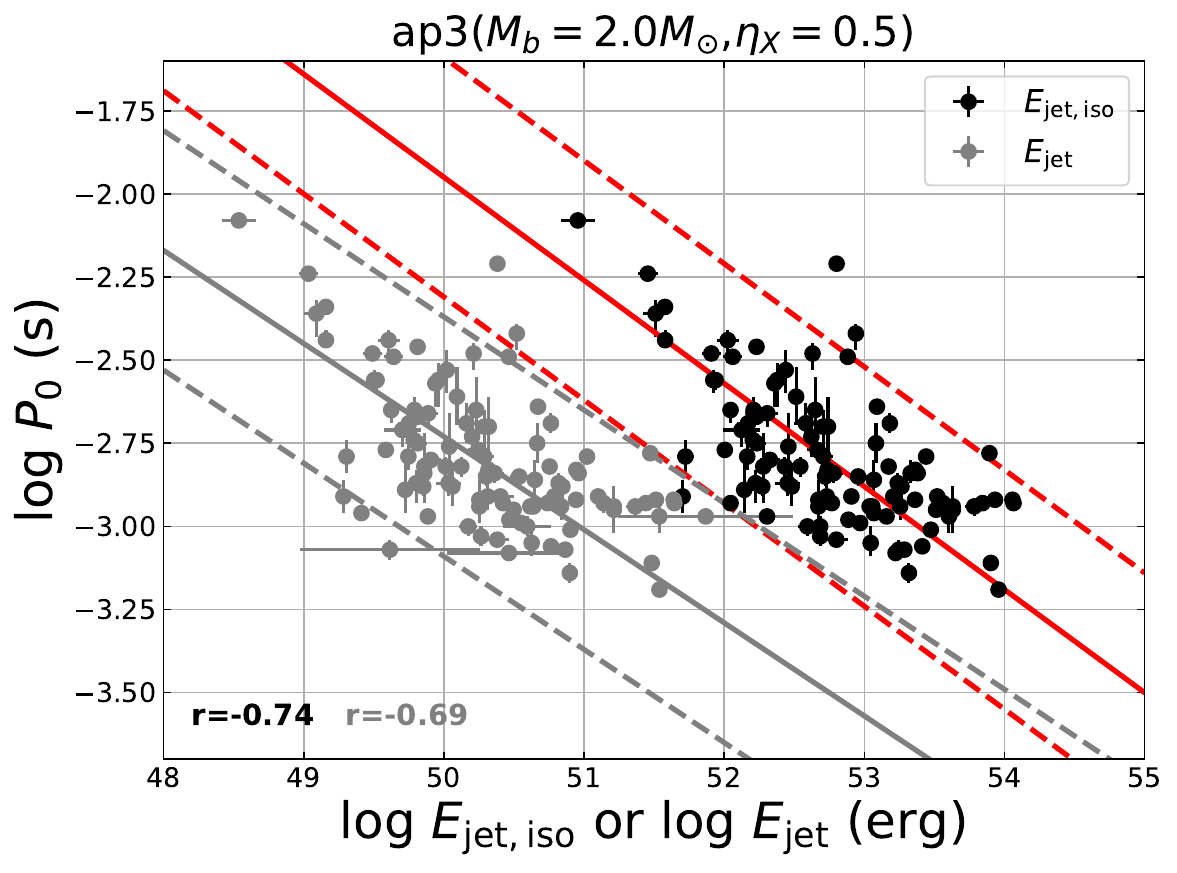}
\includegraphics [angle=0,scale=0.29]  {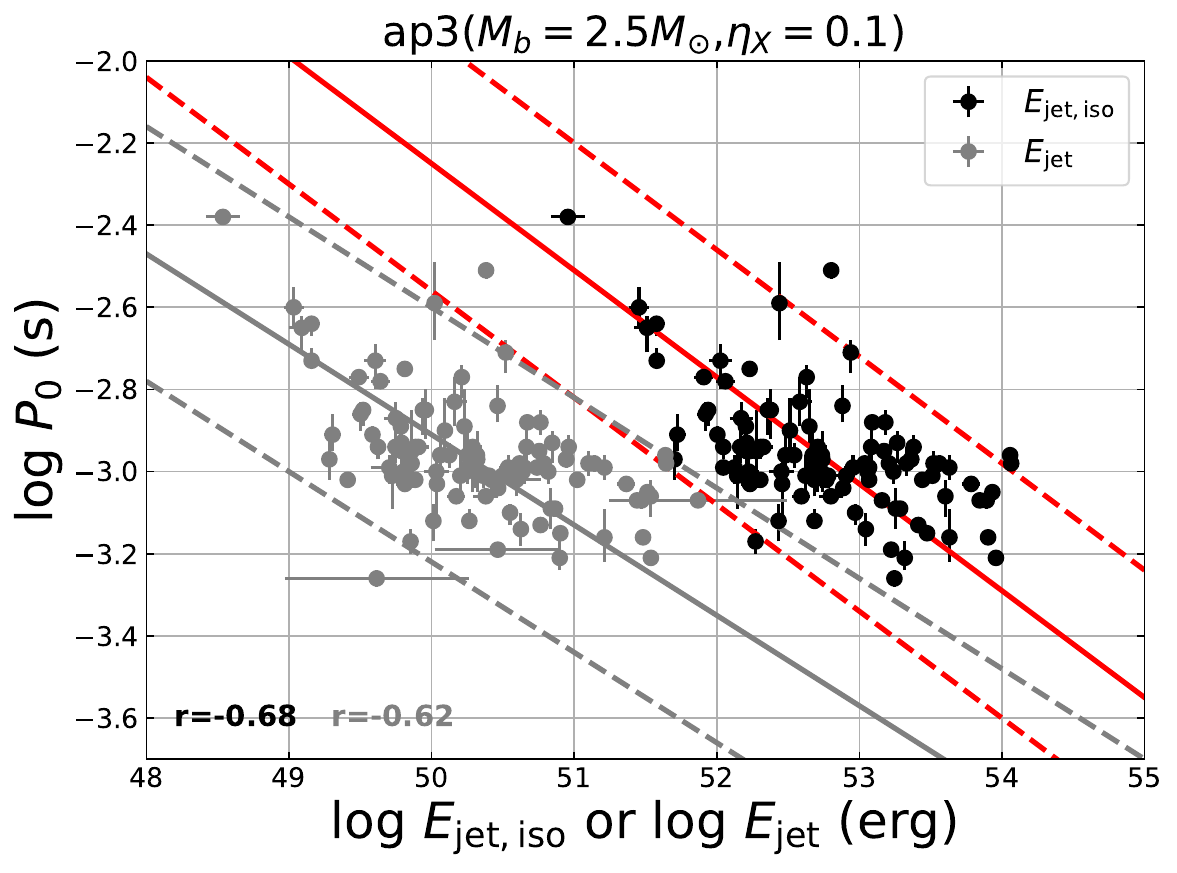}\\
\includegraphics [angle=0,scale=0.29]  {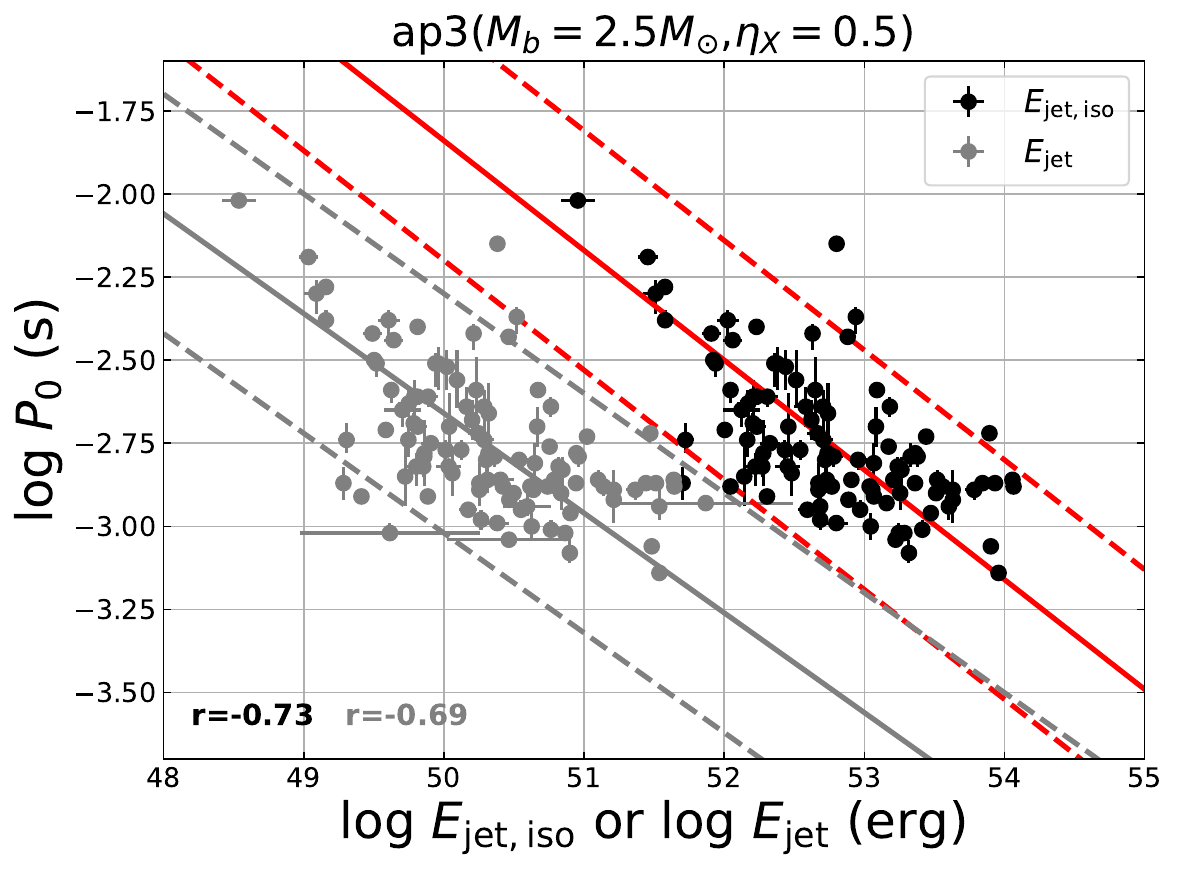}
\includegraphics [angle=0,scale=0.29]  {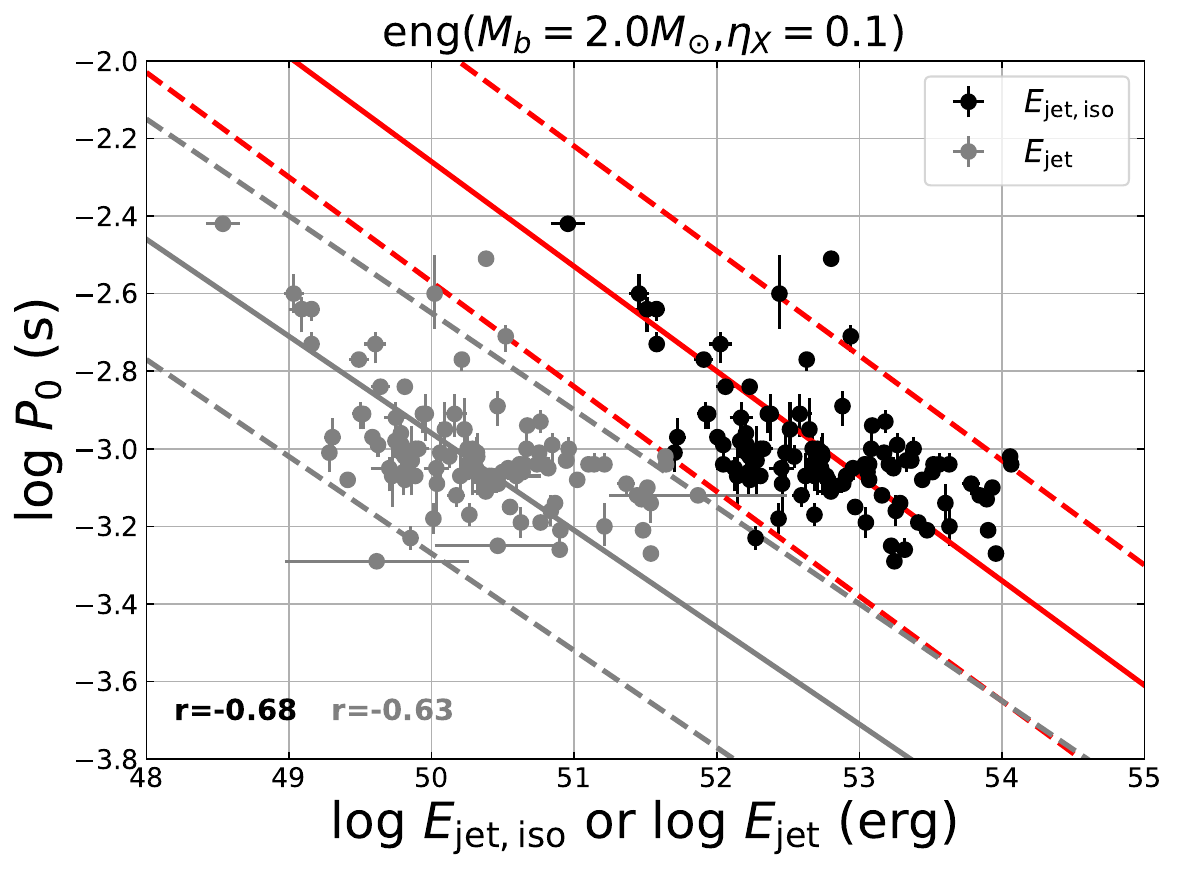}
\includegraphics [angle=0,scale=0.29]  {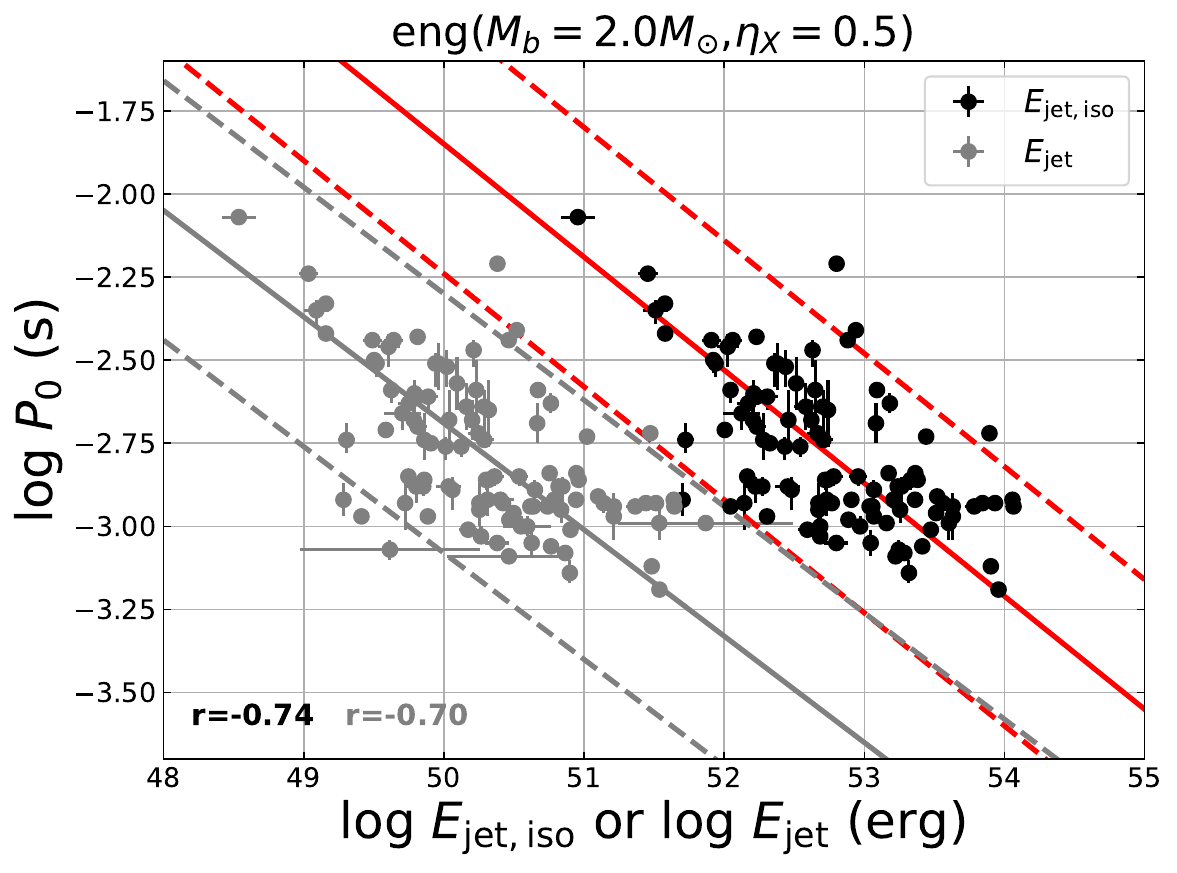}\\
\includegraphics [angle=0,scale=0.29]  {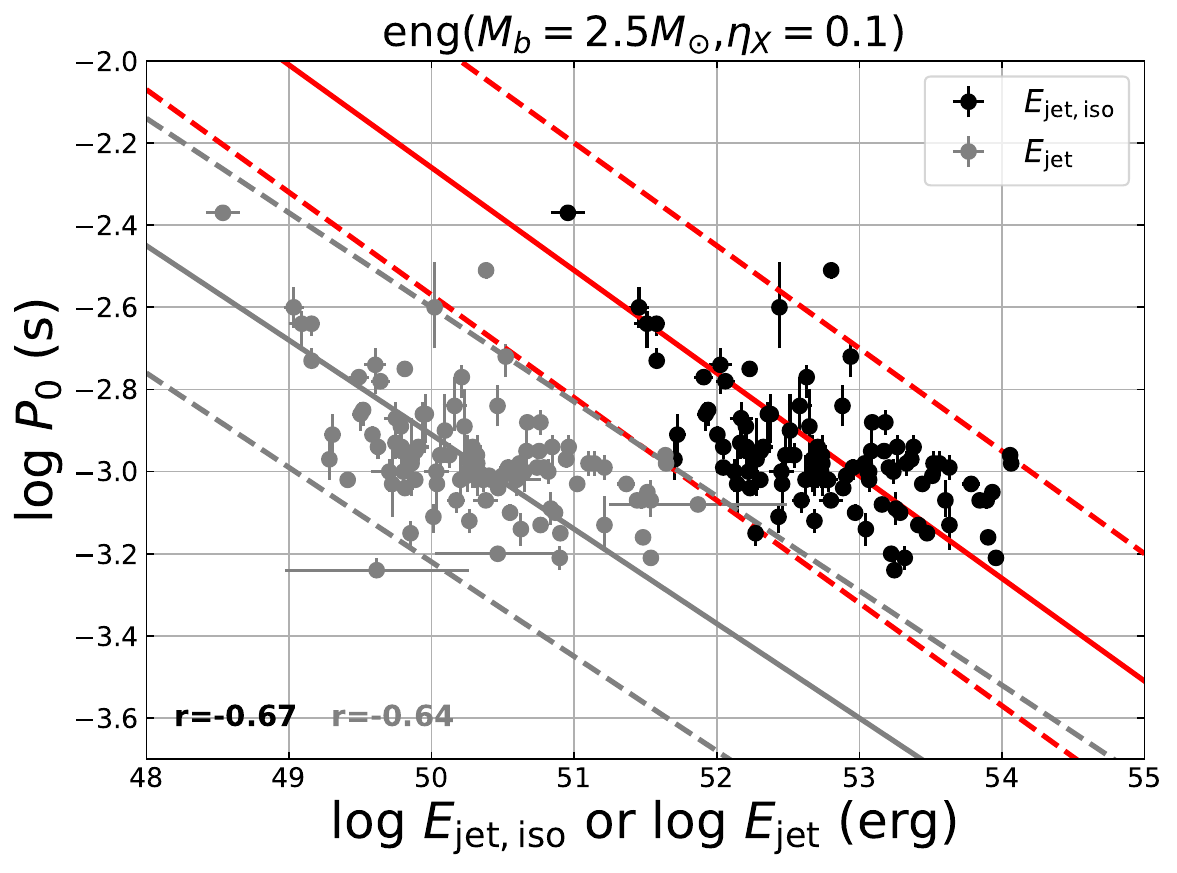}
\includegraphics [angle=0,scale=0.29]  {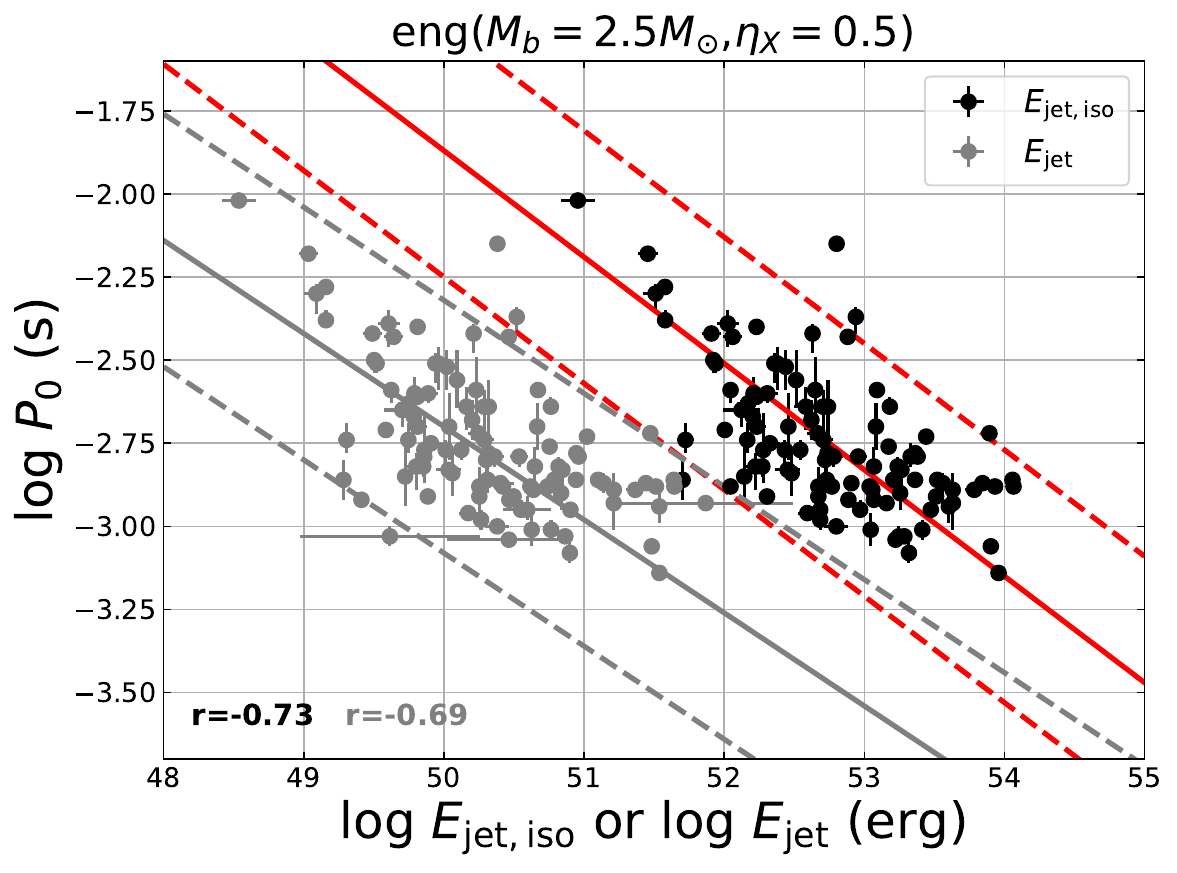}
\includegraphics [angle=0,scale=0.29]  {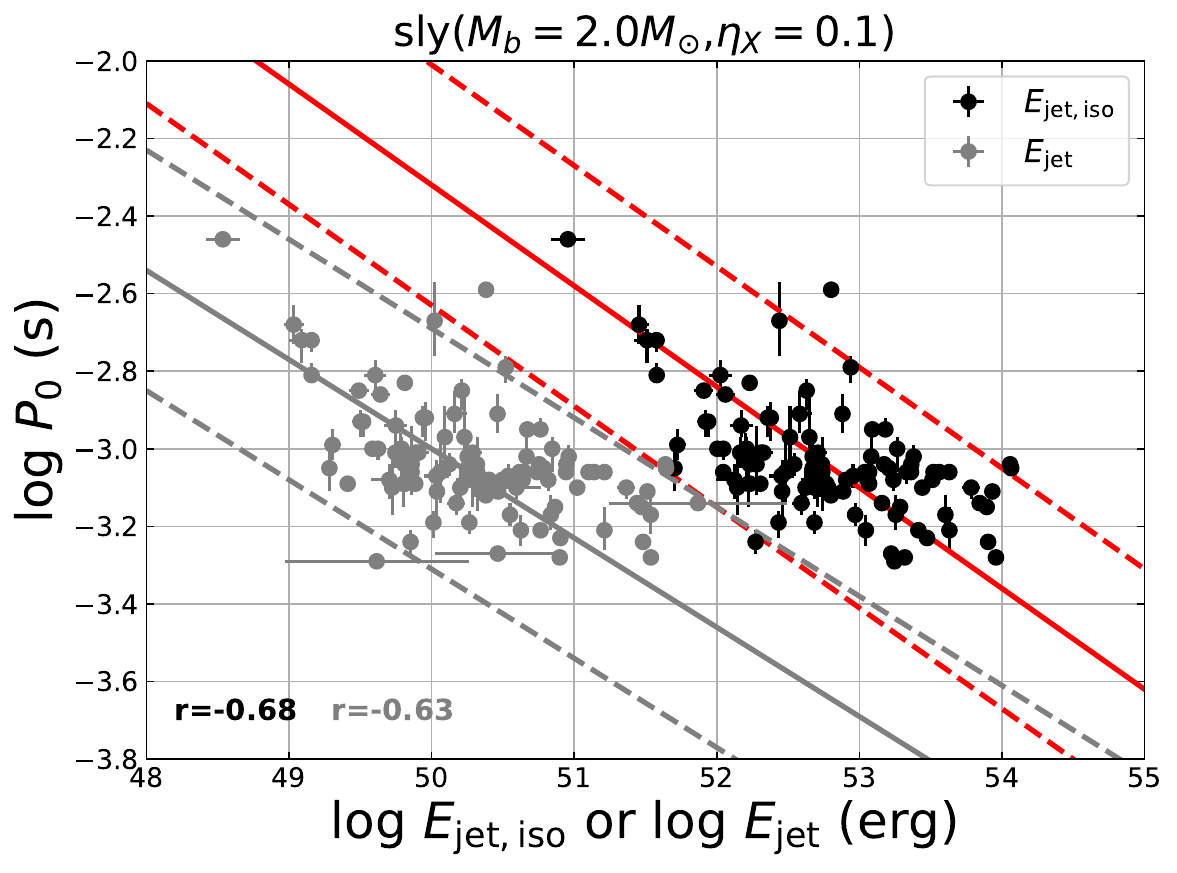}\\
\includegraphics [angle=0,scale=0.29]  {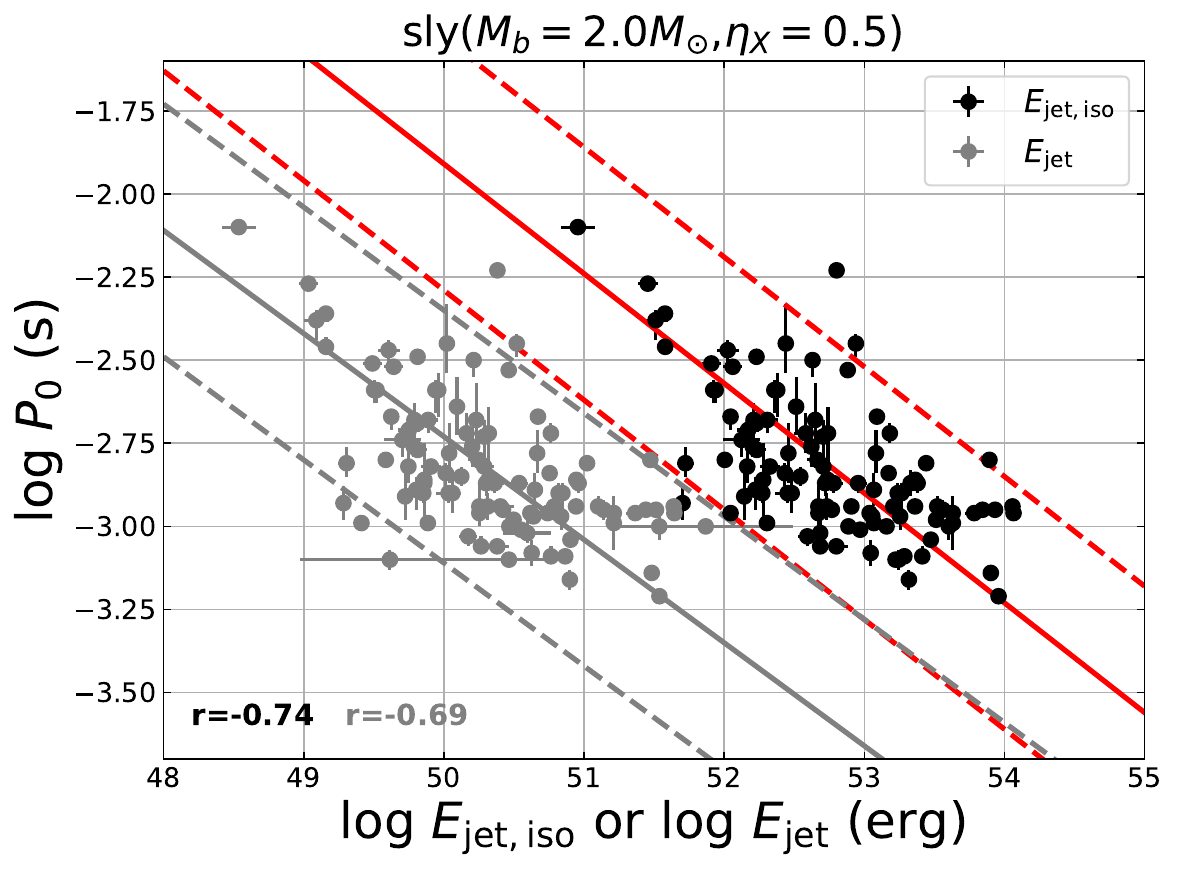}
\includegraphics [angle=0,scale=0.29]  {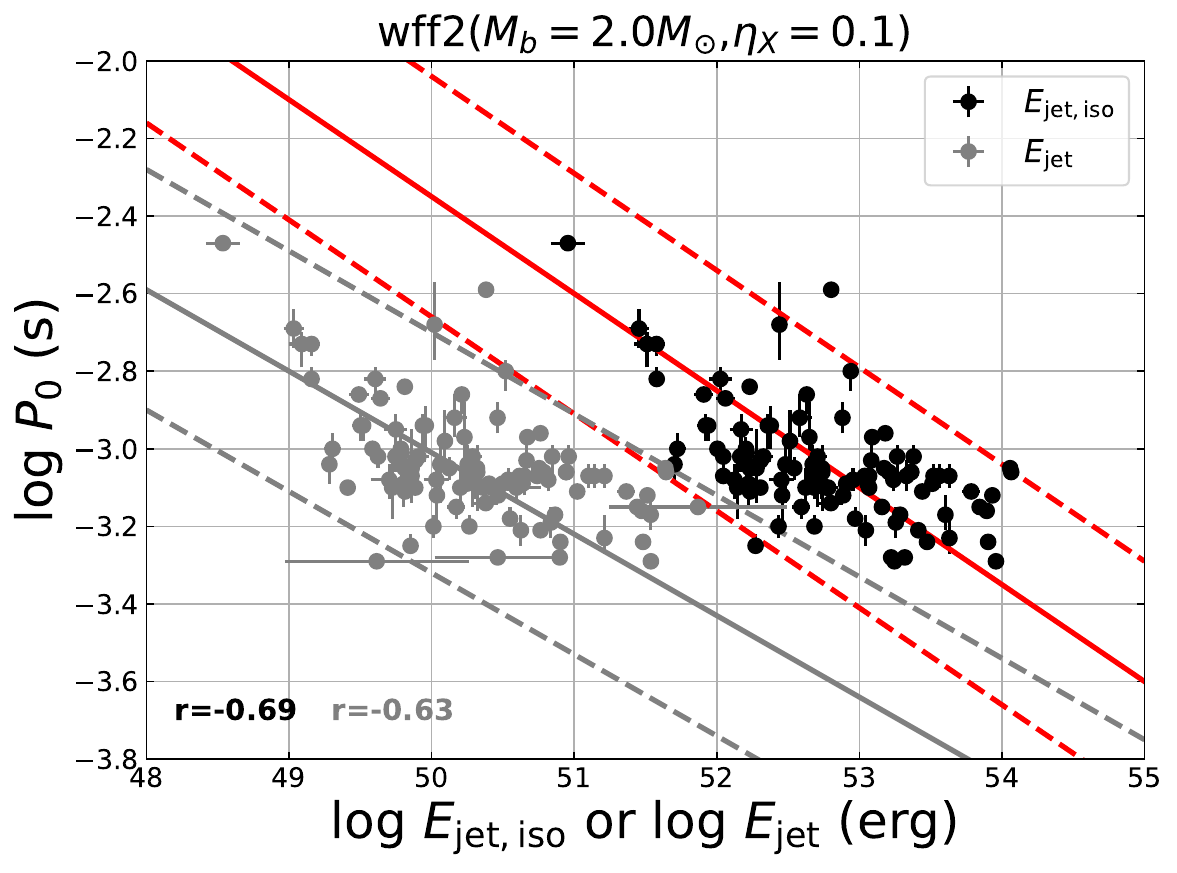}
\includegraphics [angle=0,scale=0.29]  {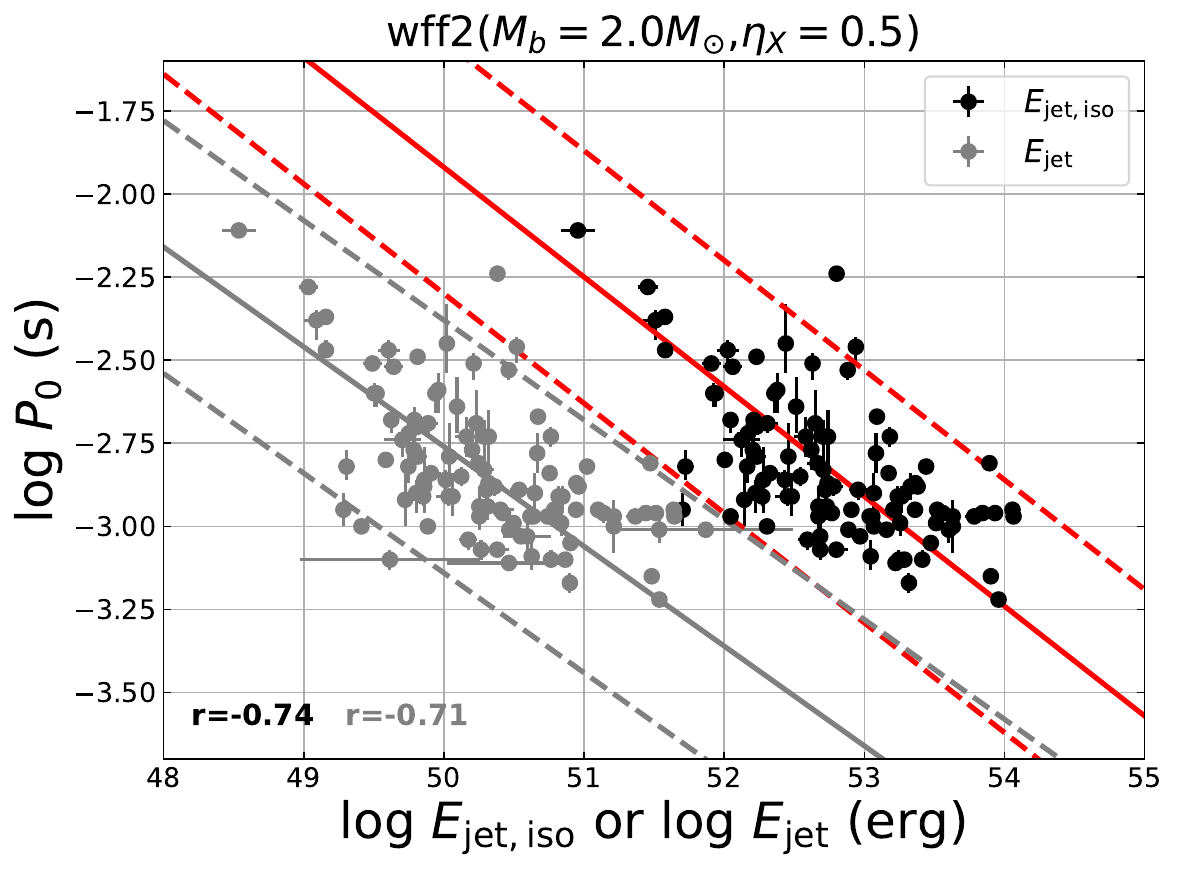}\\
\includegraphics [angle=0,scale=0.29]  {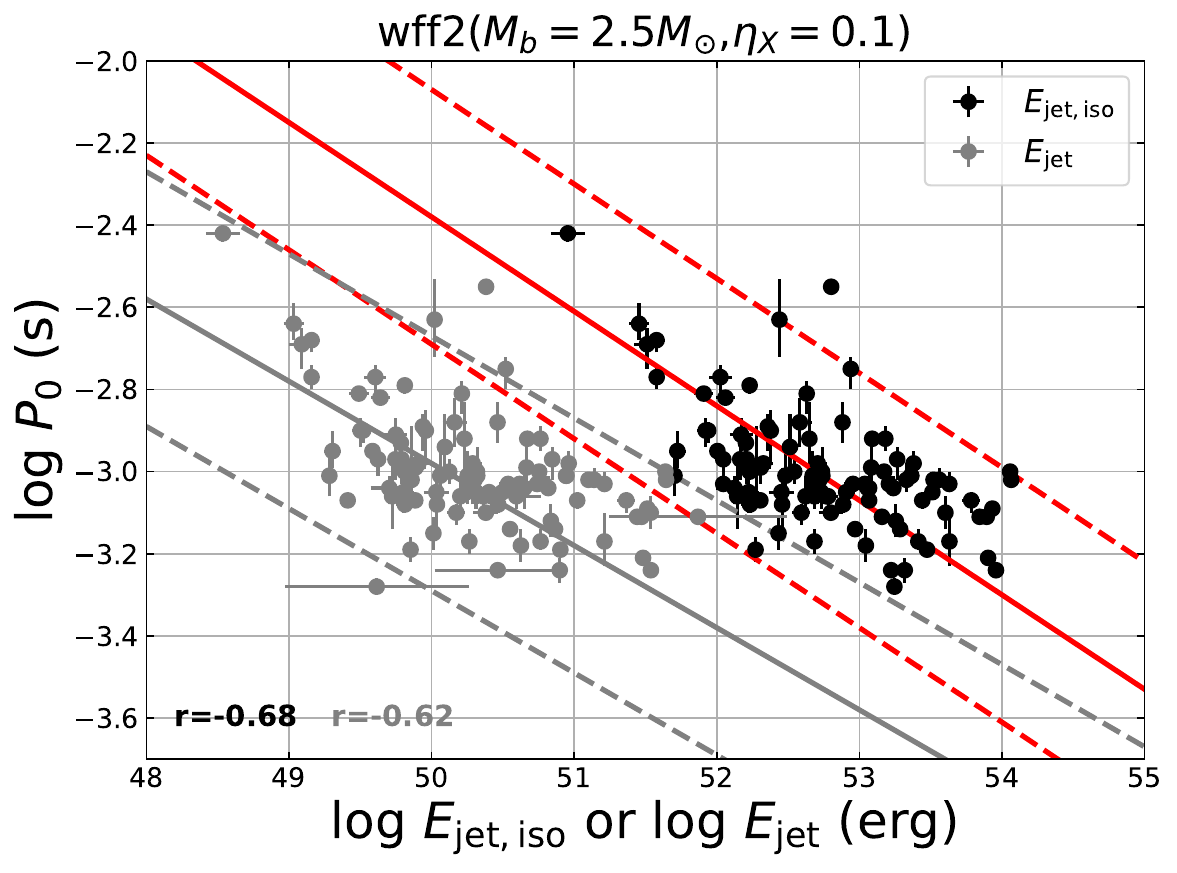}
\includegraphics [angle=0,scale=0.29]  {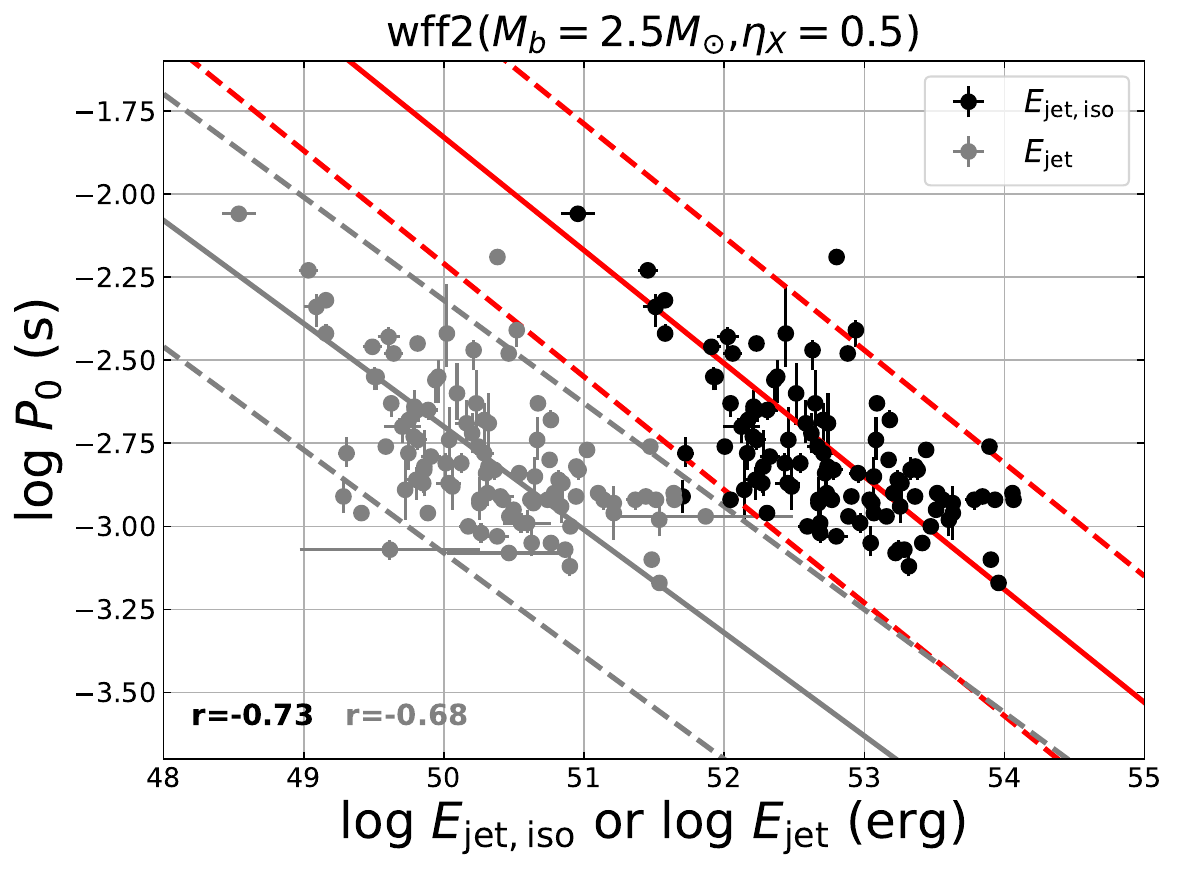}
\caption{The correlations between the $P_0$ and $E_{\rm jet,iso}$ (black circles), and $E_{\rm jet}$ (gray circles) in four samples of EoSs with $M_{b}=2.0~M_{\odot},~2.5~M_{\odot}$ and $\eta_{\rm X}=0.1,~0.5$. The red solid and red dashed lines are the best-fitting results and the 95\% confidence level for $P_0-E_{\rm jet,iso}$, respectively. The gray solid and gray dashed lines are the best-fitting results and the 95\% confidence level for $P_0-E_{\rm jet}$, respectively.}
\label{fig:P0-Ejet}
\end{figure*}

Next, we further investigated how the GRB isotropic jet energy $E_{\rm jet, iso}$ and true jet energy $E_{\rm jet}$ are related to the newborn magnetar physical parameters, and presented a series of scatter plots. In Figure \ref{fig:P0-Ejet}, we presented a series of $P_{0}-E_{\rm jet}$ scatter diagrams for different EoSs and baryonic masses with two $\eta_{\rm X}$ values based on our LGRB sample. Interestingly, we found that there are tight anticorrelations between $E_{\rm jet, iso}$ ($E_{\rm jet}$) and $P_{0}$ with a large systematic error for all of our selected EoSs by using the ordinary least-squares analysis. Most of the GRBs in our sample fall into the $3\sigma$ deviation region of the best-fitting power-law model for all selected EoS cases. Our best-fitting correlations for various cases are reported in Table \ref{table-14}. 

\begin{figure*}
\centering
\includegraphics [angle=0,scale=0.29]  {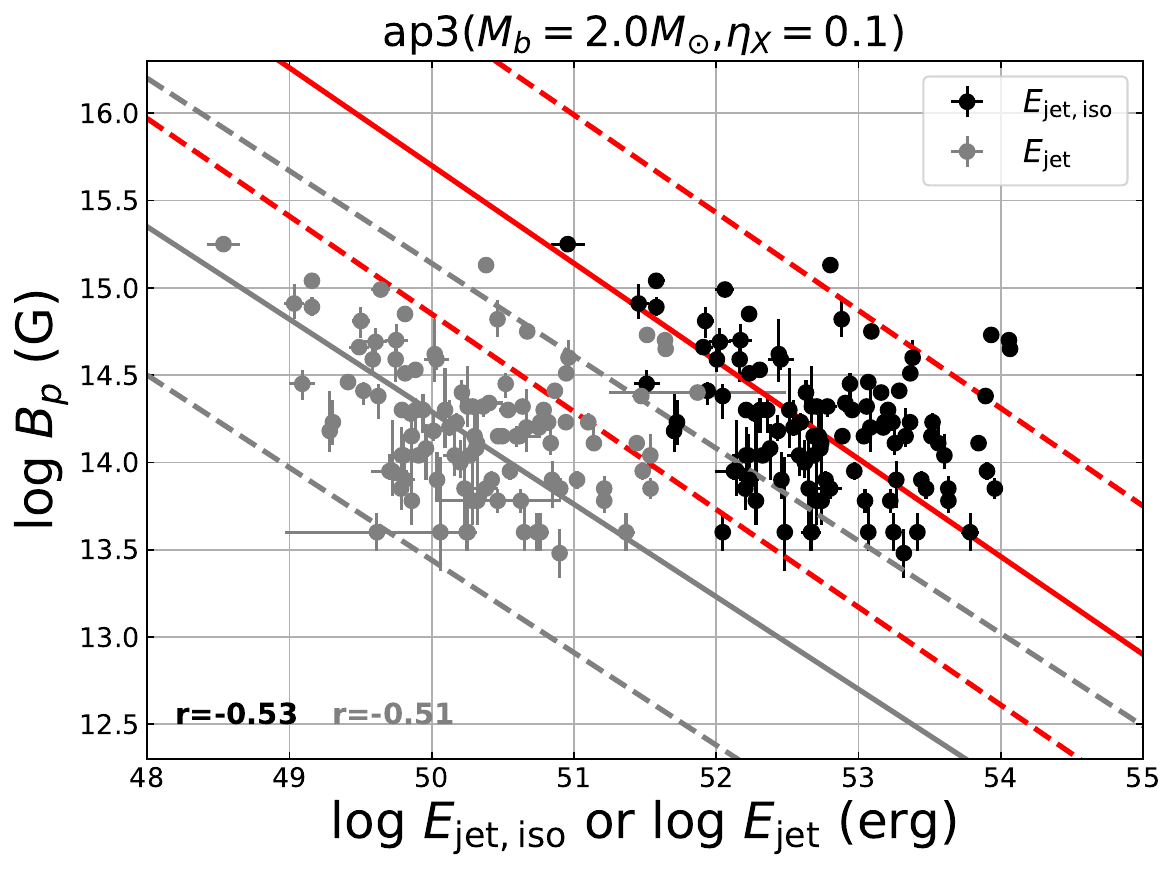}
\includegraphics [angle=0,scale=0.29]  {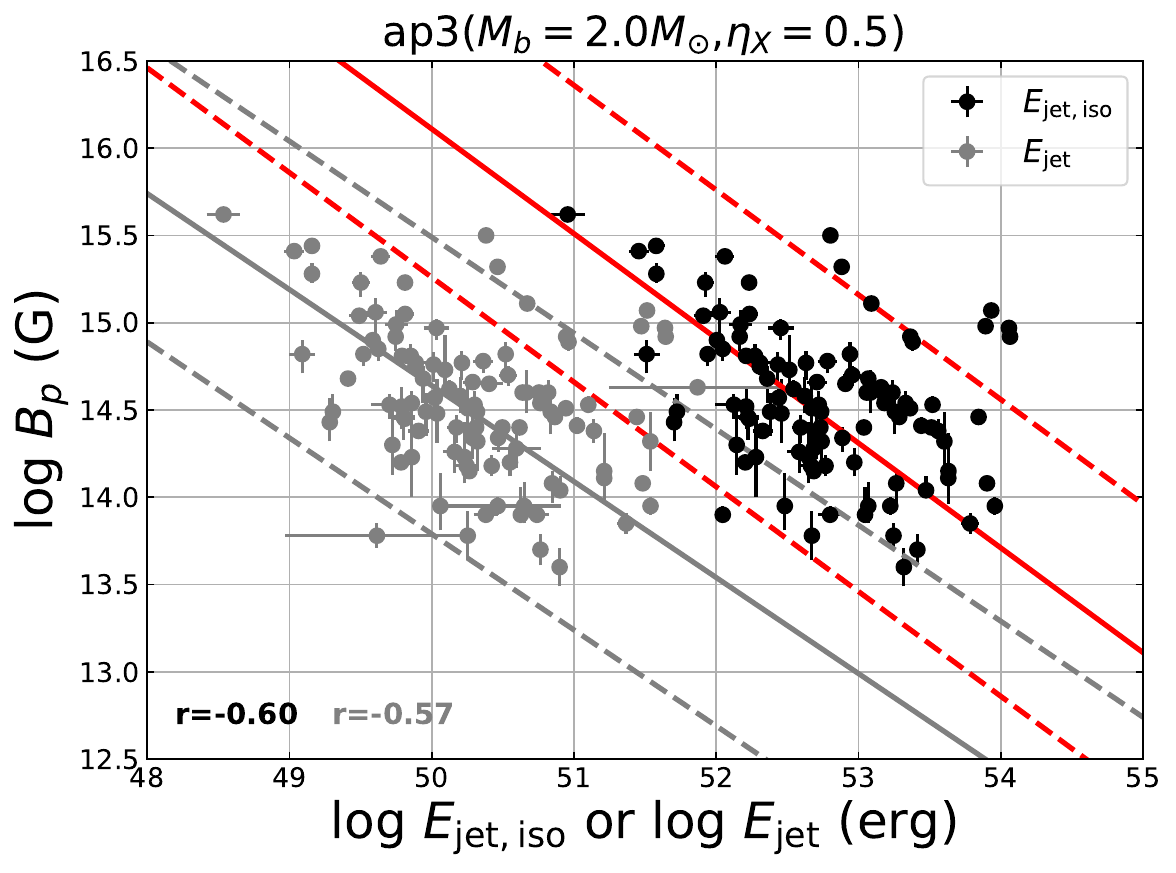}
\includegraphics [angle=0,scale=0.29]  {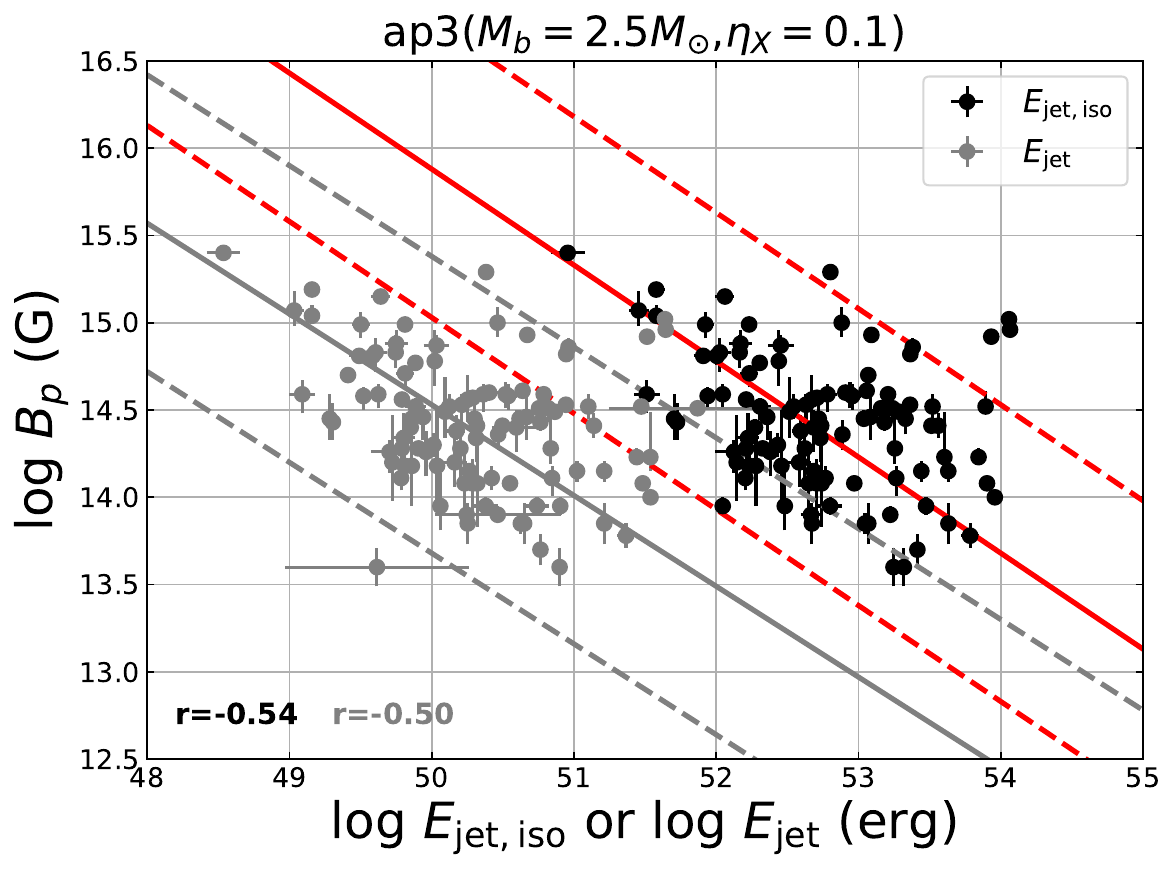}\\
\includegraphics [angle=0,scale=0.29]  {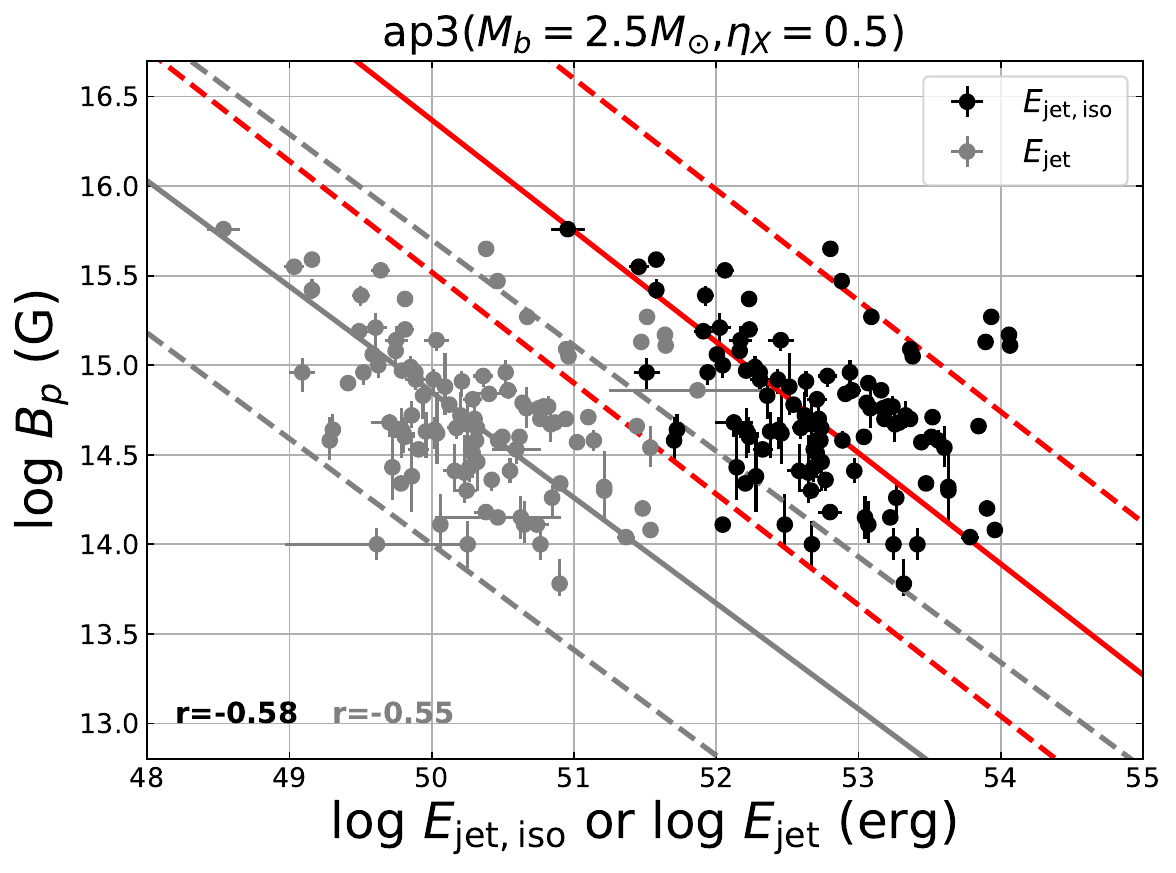}
\includegraphics [angle=0,scale=0.29]  {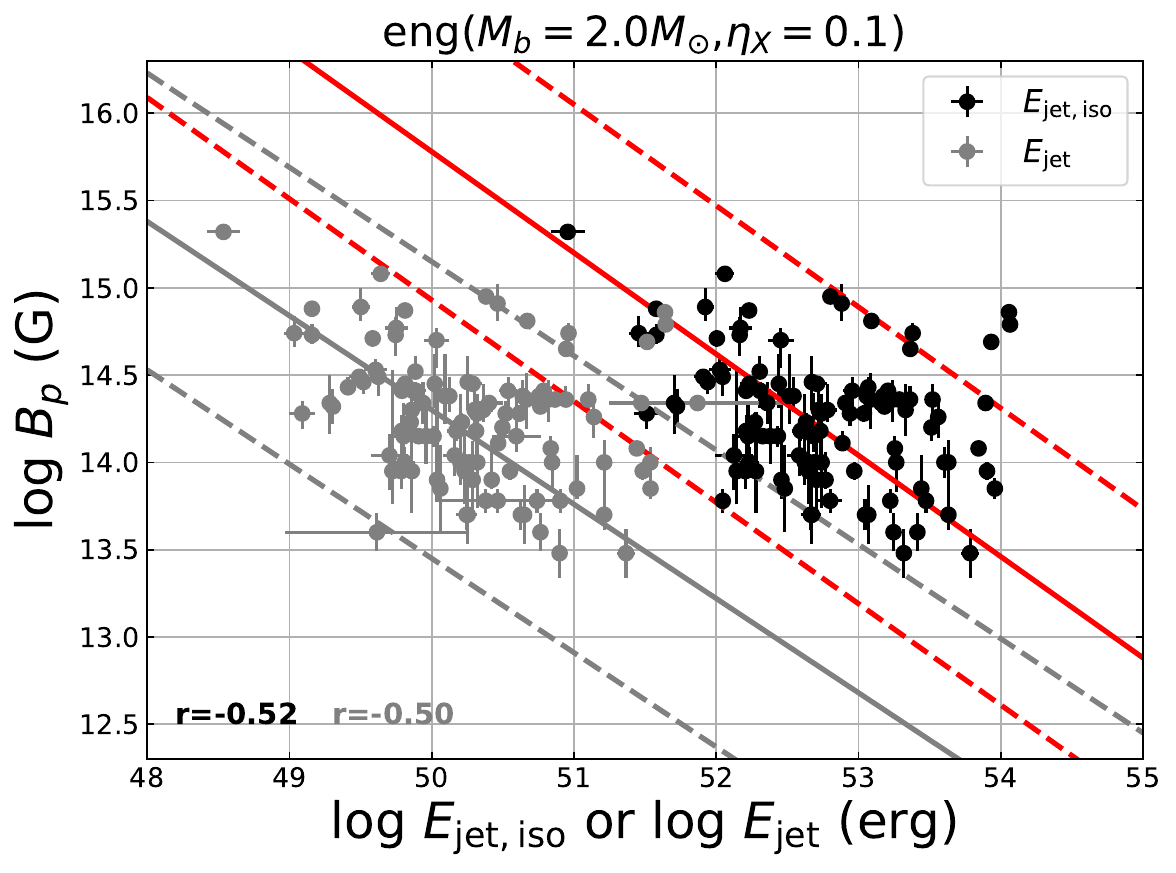}
\includegraphics [angle=0,scale=0.29]  {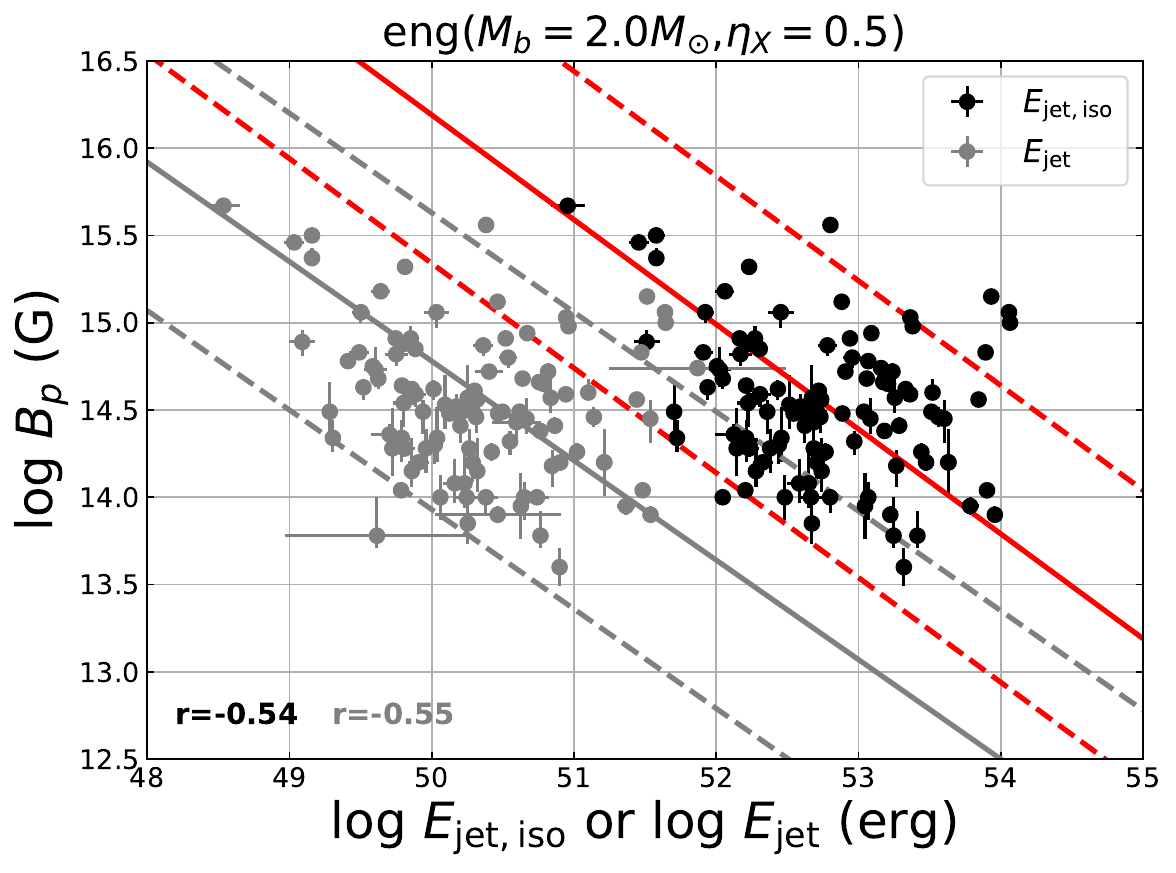}\\
\includegraphics [angle=0,scale=0.29]  {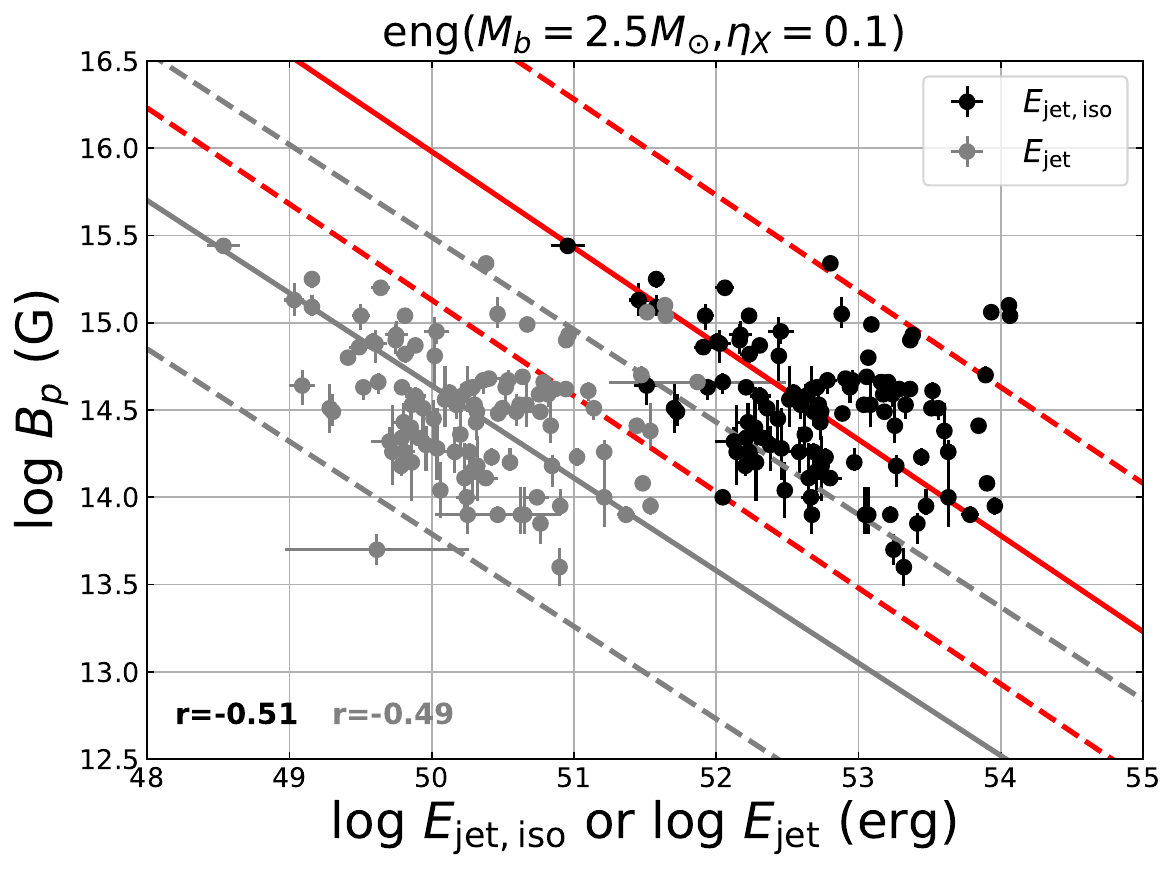}
\includegraphics [angle=0,scale=0.29]  {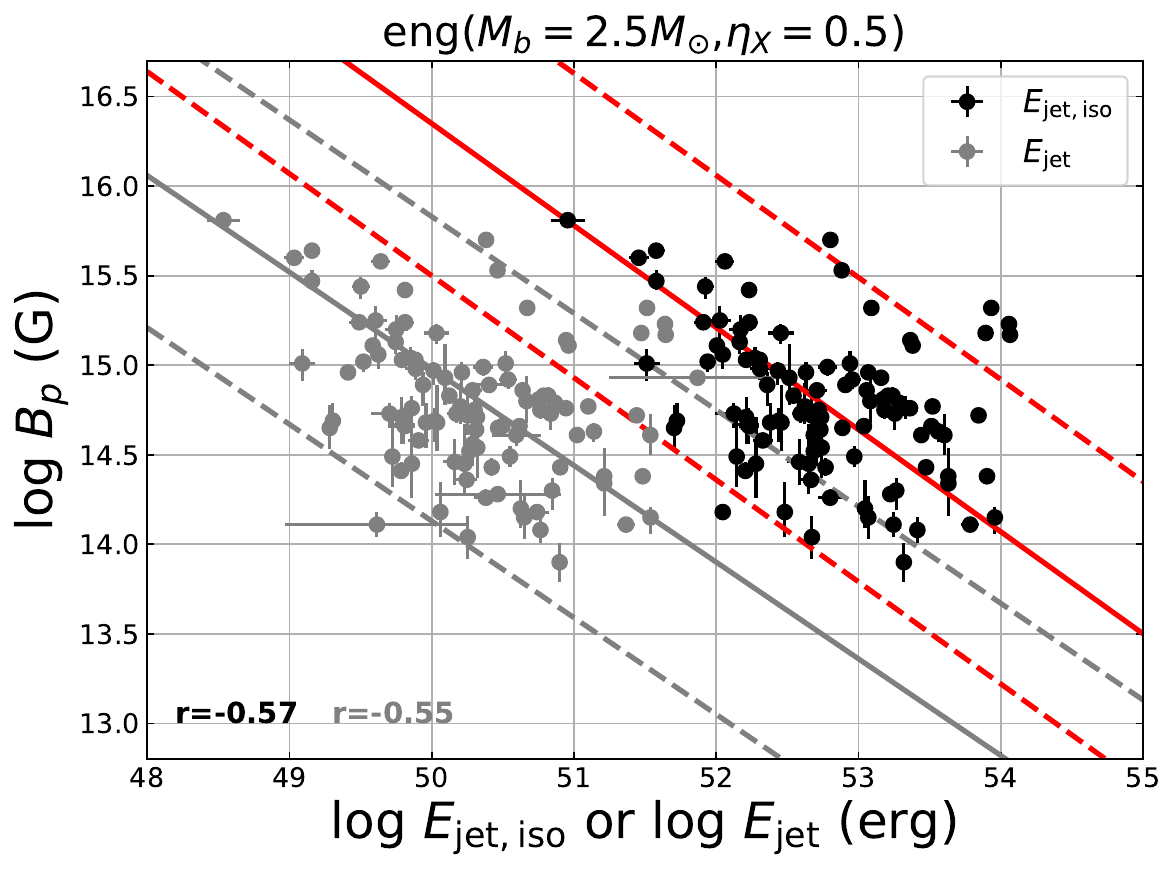}
\includegraphics [angle=0,scale=0.29]  {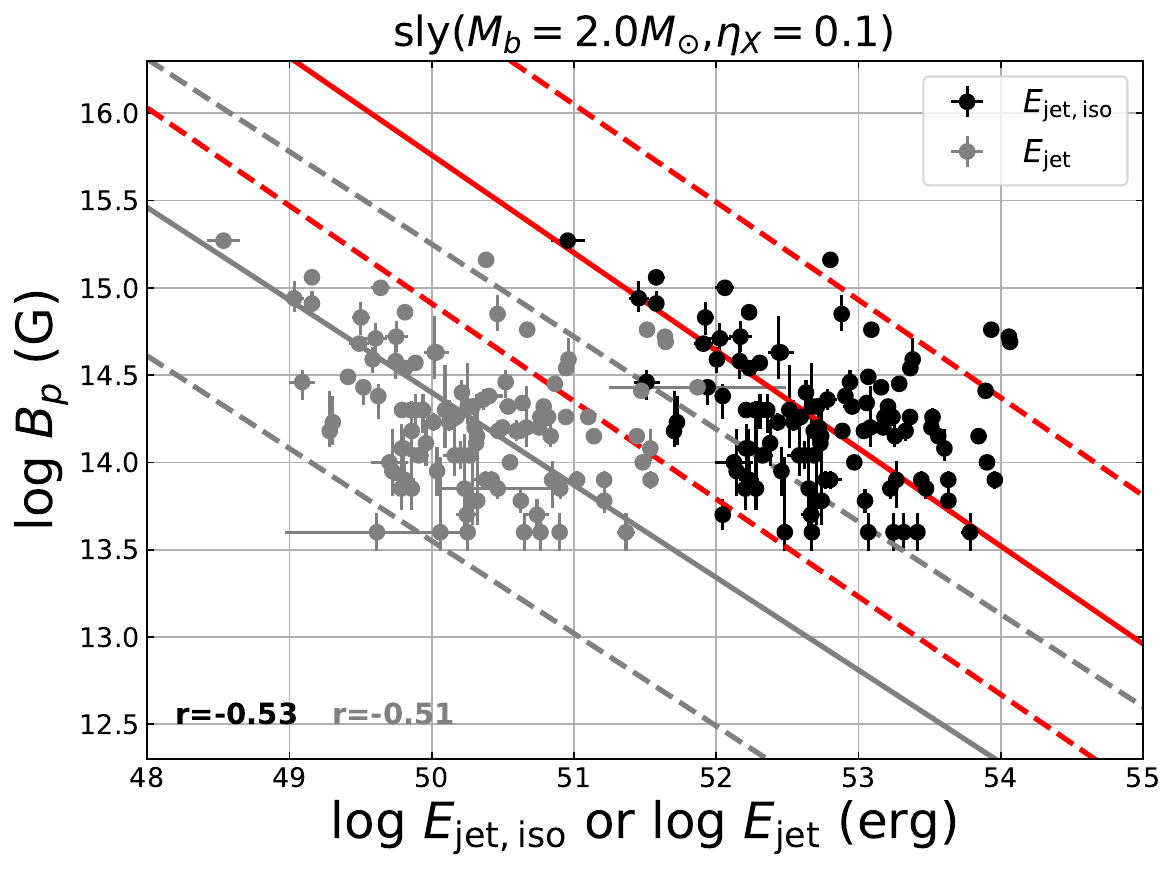}\\
\includegraphics [angle=0,scale=0.29]  {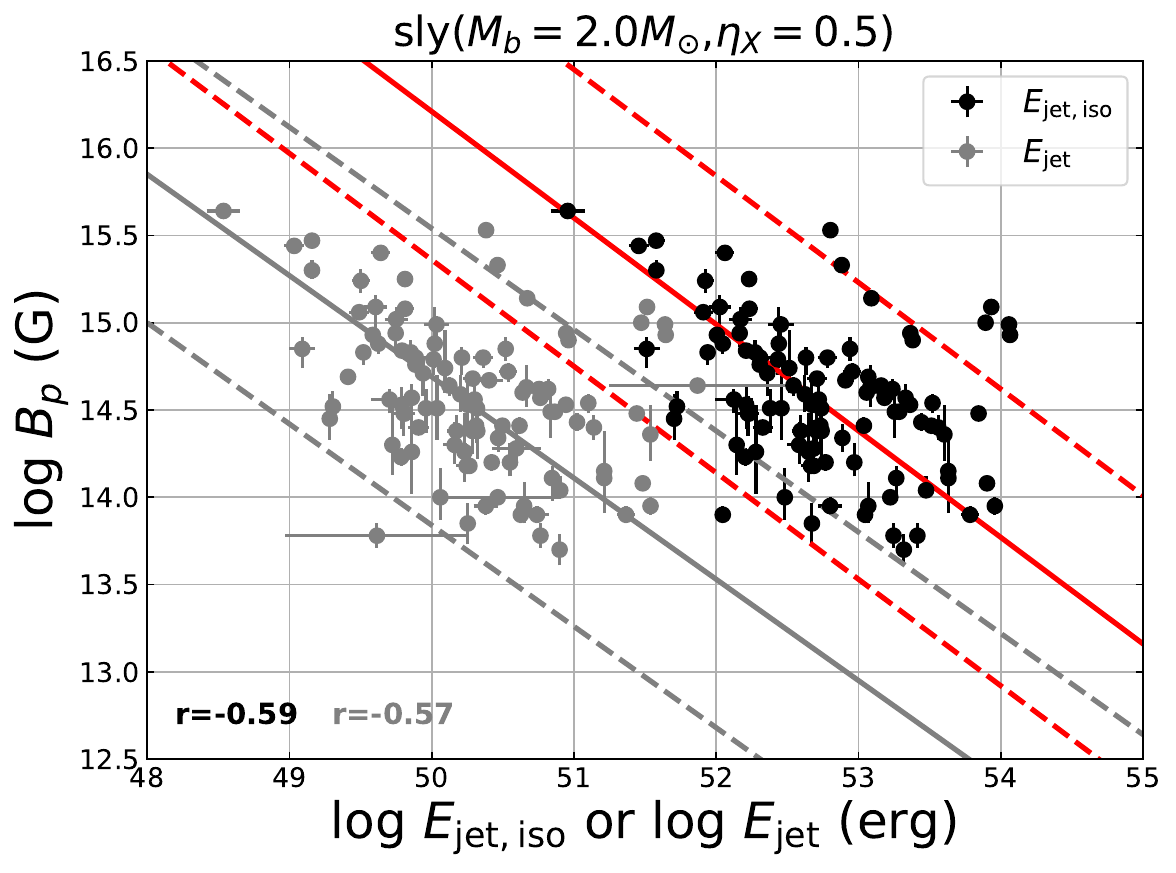}
\includegraphics [angle=0,scale=0.29]  {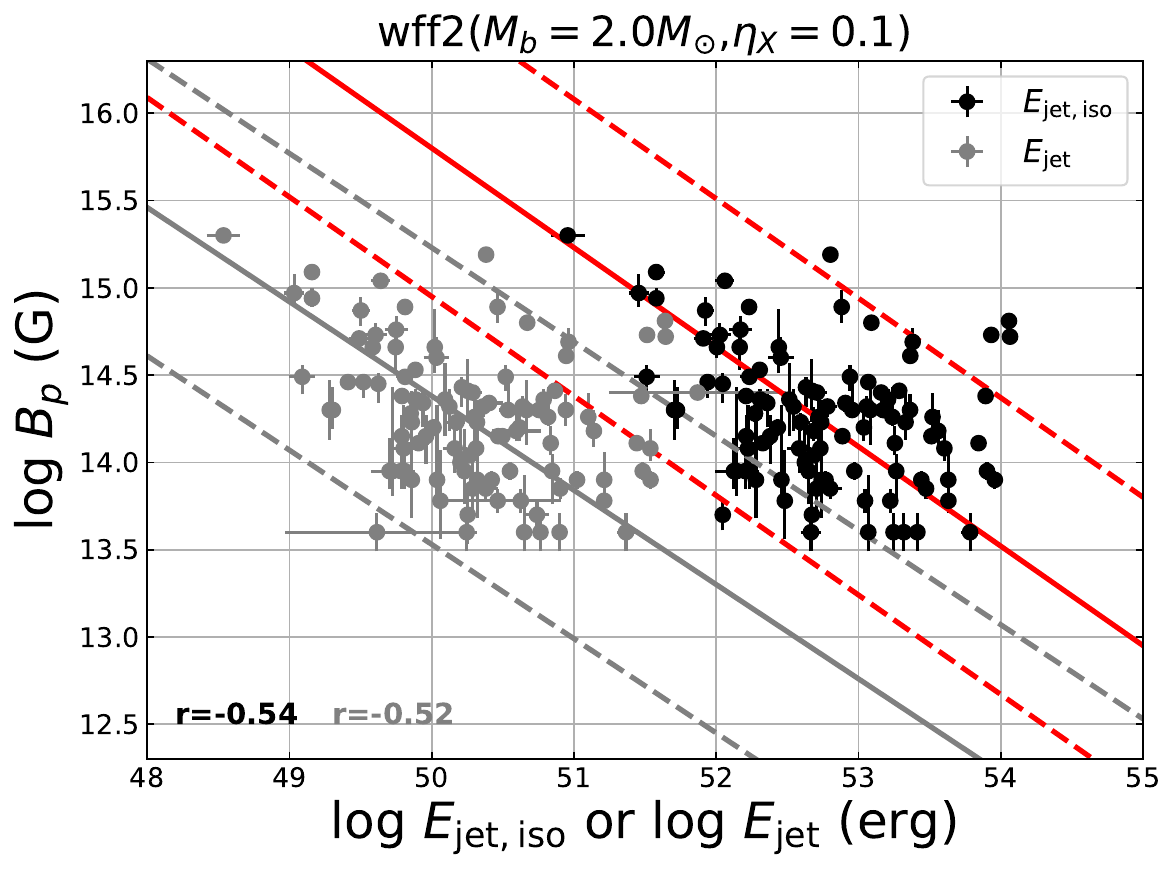}
\includegraphics [angle=0,scale=0.29]  {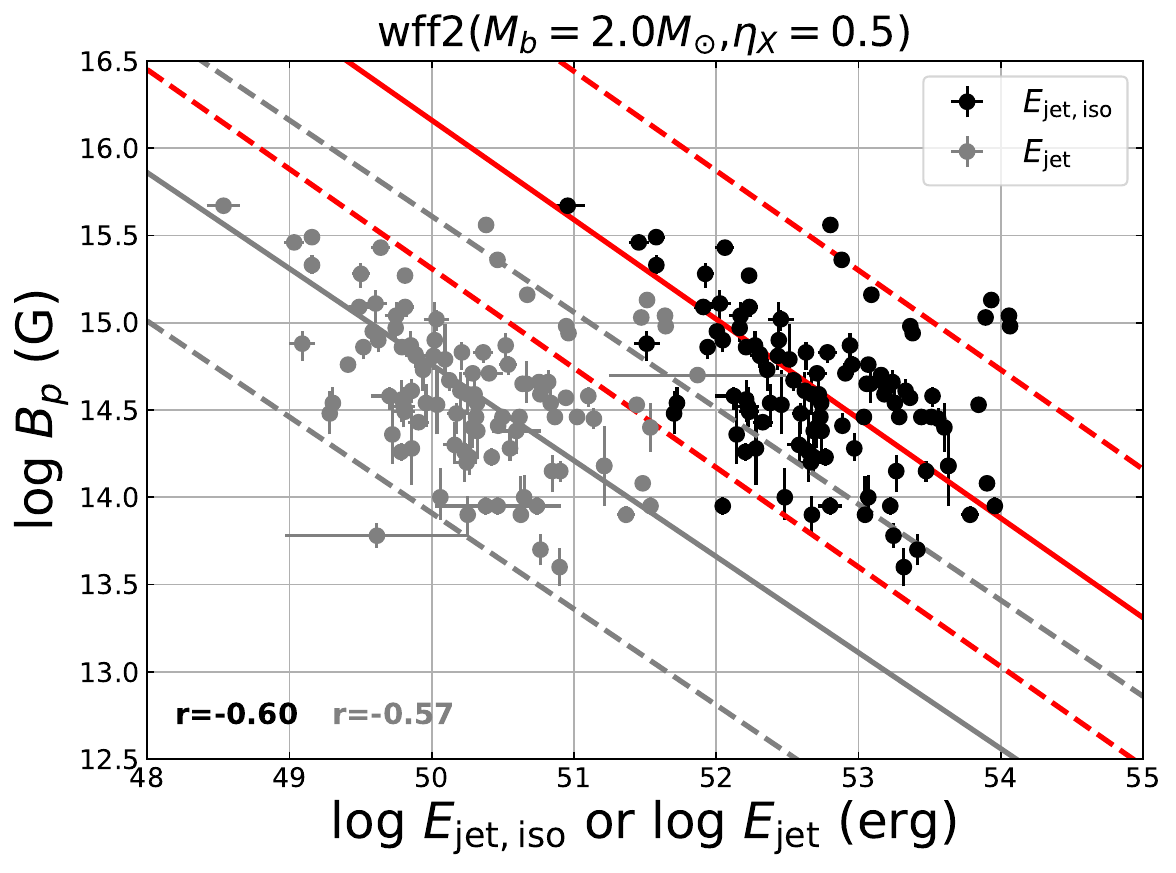}\\
\includegraphics [angle=0,scale=0.29]  {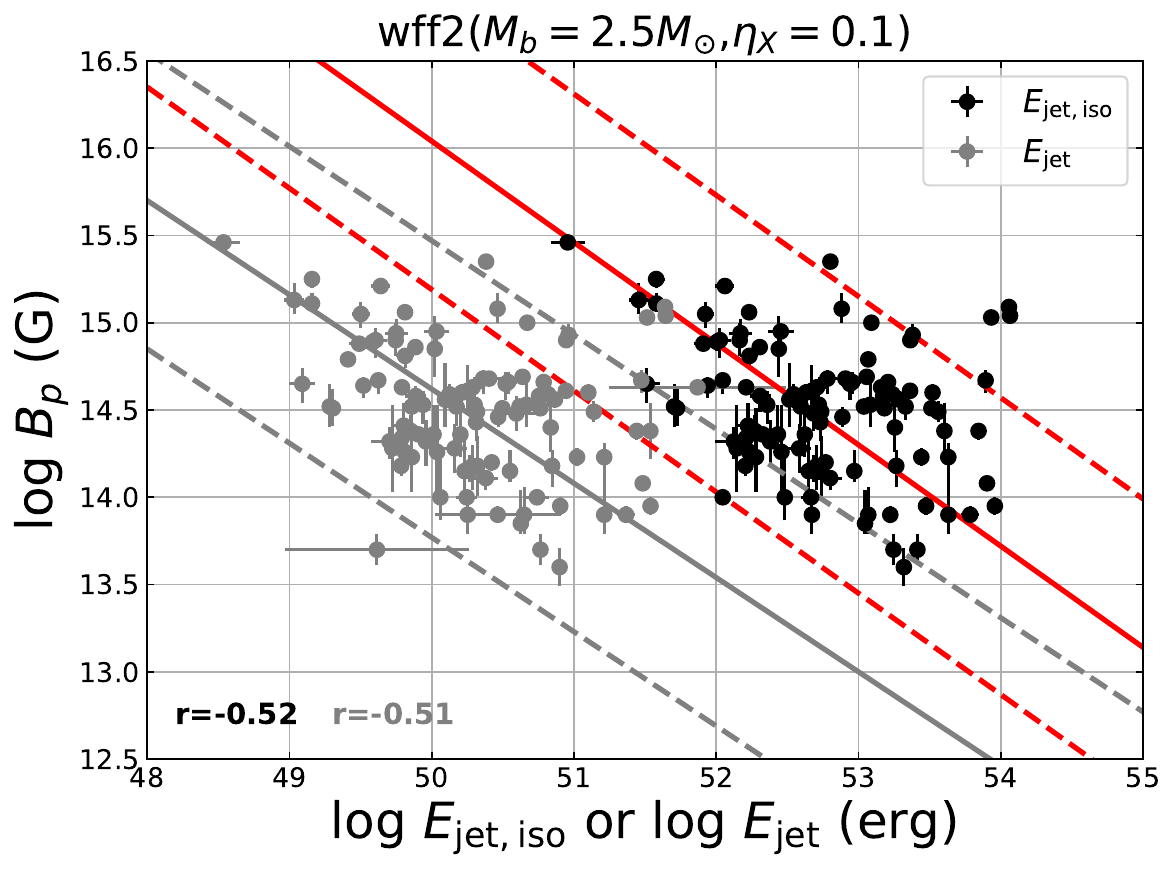}
\includegraphics [angle=0,scale=0.29]  {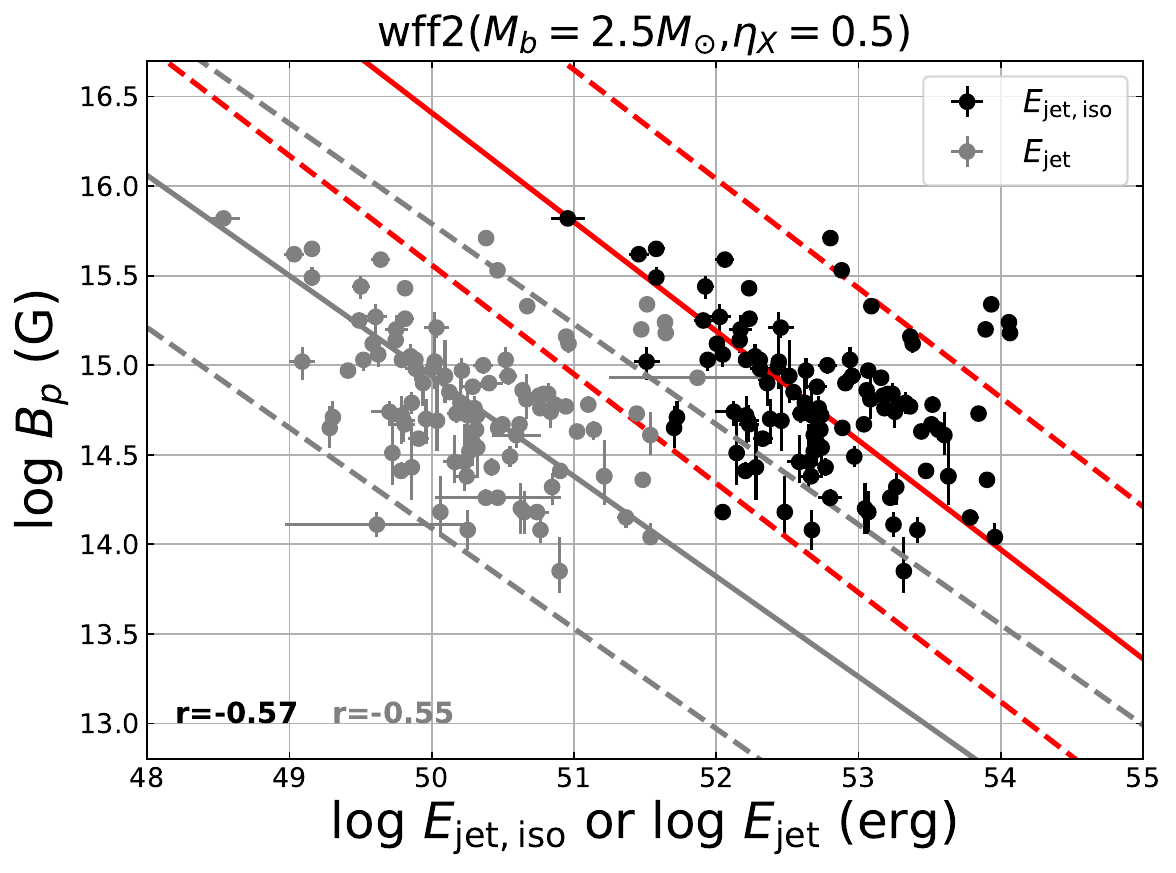}
\caption{The correlations between the $B_p$ and $E_{\rm jet,iso}$ (black circles), and $E_{\rm jet}$ (gray circles) in four samples of EoSs with $M_{b}=2.0~M_{\odot},~2.5~M_{\odot}$ and $\eta_{\rm X}=0.1,~0.5$. The red solid and red dashed lines are the best-fitting results and the 95\% confidence level for $P_0-E_{\rm jet,iso}$, respectively. The gray solid and gray dashed lines are the best-fitting results and the 95\% confidence level for $P_0-E_{\rm jet}$, respectively.}
\label{fig:Bp-Ejet}
\end{figure*}

In Figure \ref{fig:Bp-Ejet}, we also plotted a series of $B_p-E_{\rm jet}$ scatter diagrams for different EoSs and baryonic masses with two $\eta_{\rm X}$ values based on our LGRB sample. Analogously, we found that there are relatively weak anticorrelations between $E_{\rm jet, iso}$ ($E_{\rm jet}$) and $B_p$ with a large systematic error for all of our selected EoSs by using the ordinary least-squares analysis. Most of the GRBs in our sample fall into the $3\sigma$ deviation region of the best-fitting power-law model for all selected EoS cases. Our best-fitting correlations for various cases are reported in Table \ref{table-14}. 

\begin{figure*}
\centering
\includegraphics [angle=0,scale=0.29]  {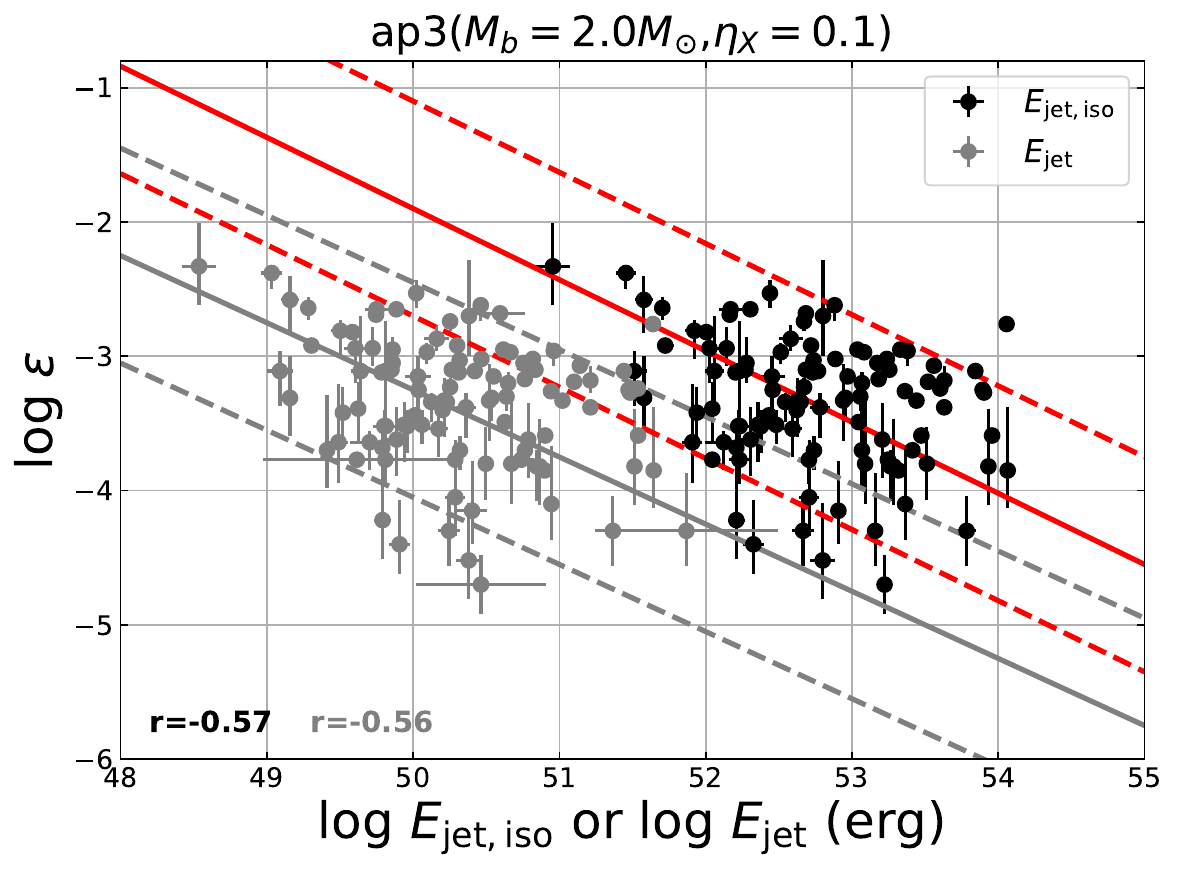}
\includegraphics [angle=0,scale=0.29]  {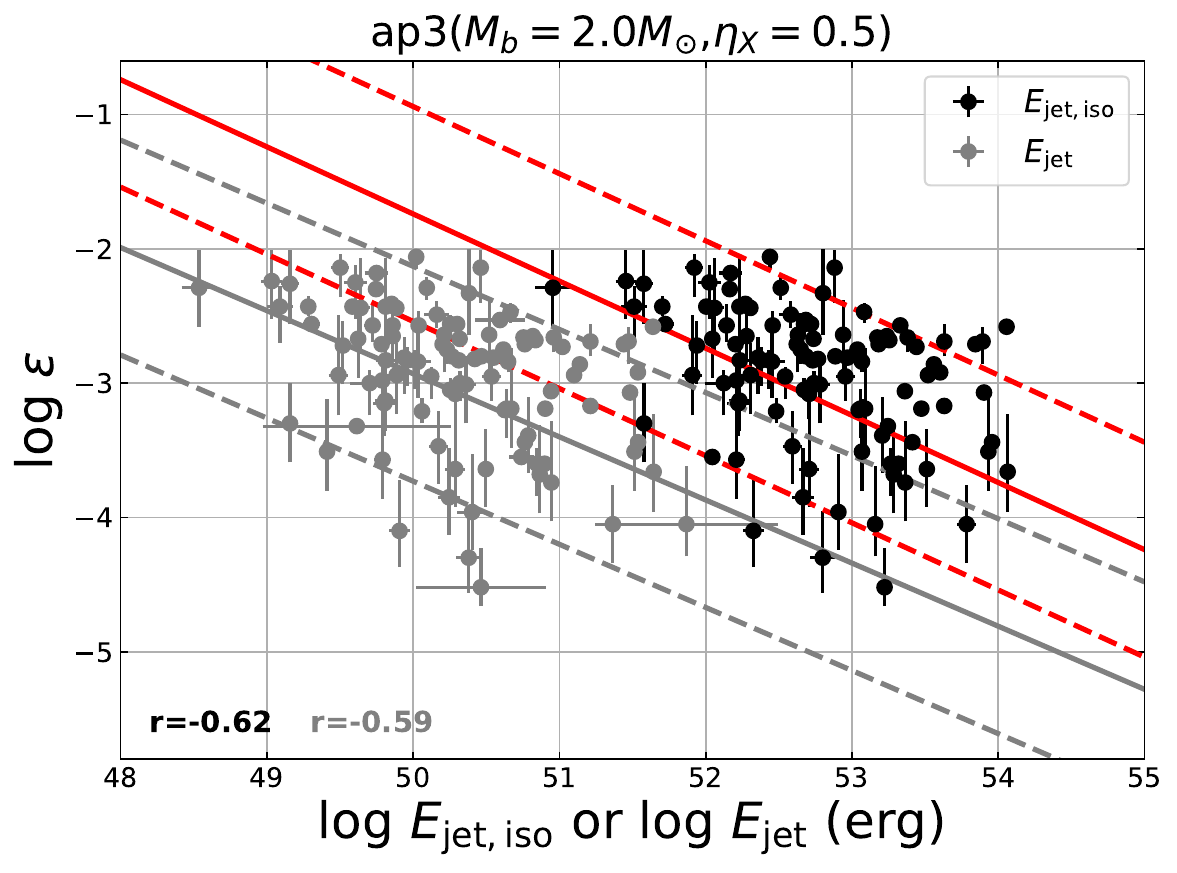}
\includegraphics [angle=0,scale=0.29]  {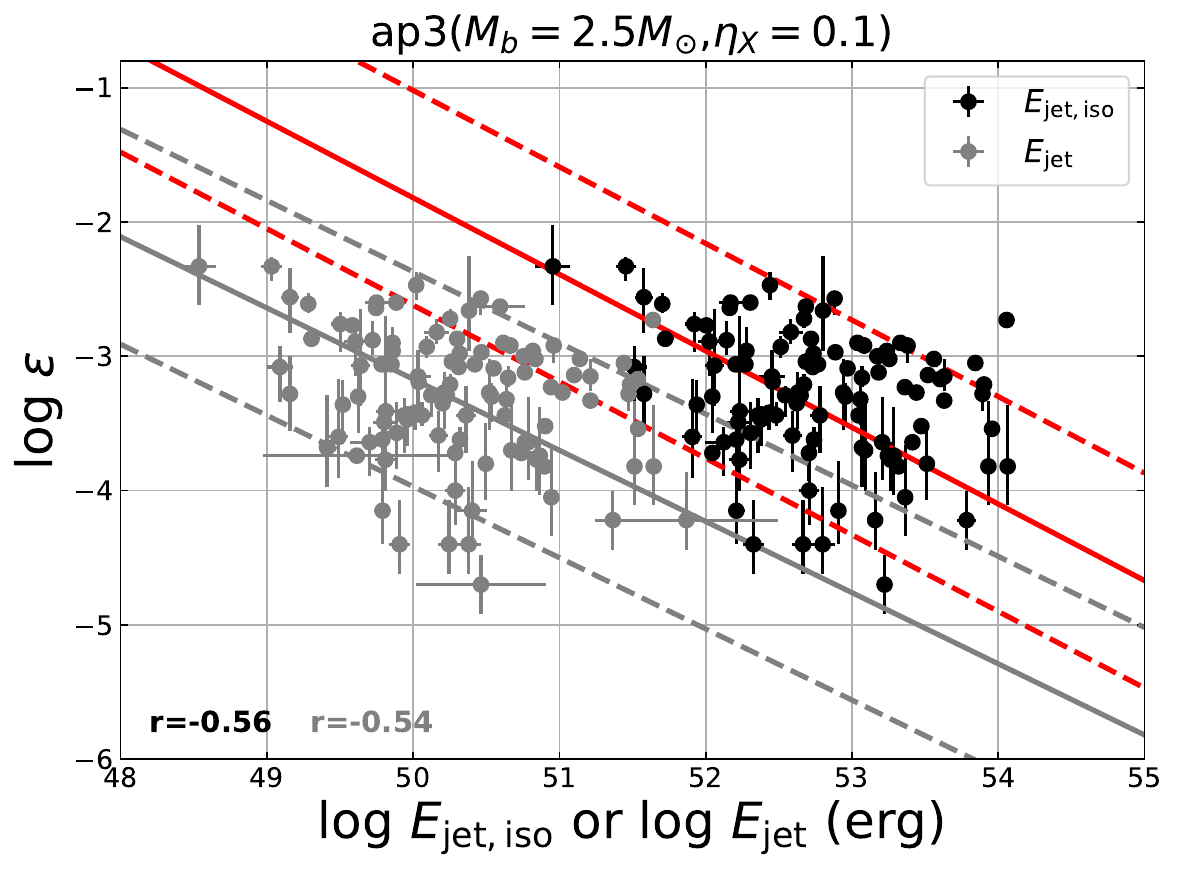}\\
\includegraphics [angle=0,scale=0.29]  {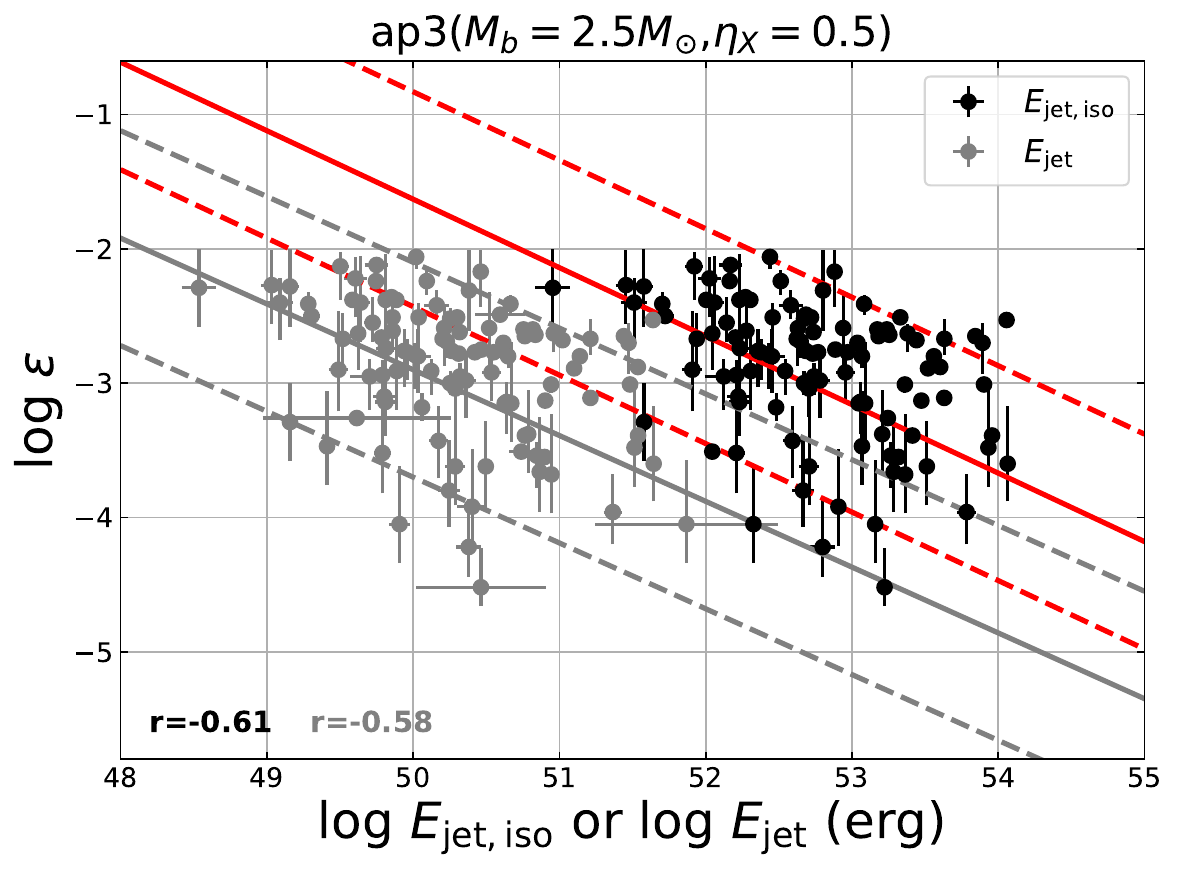}
\includegraphics [angle=0,scale=0.29]  {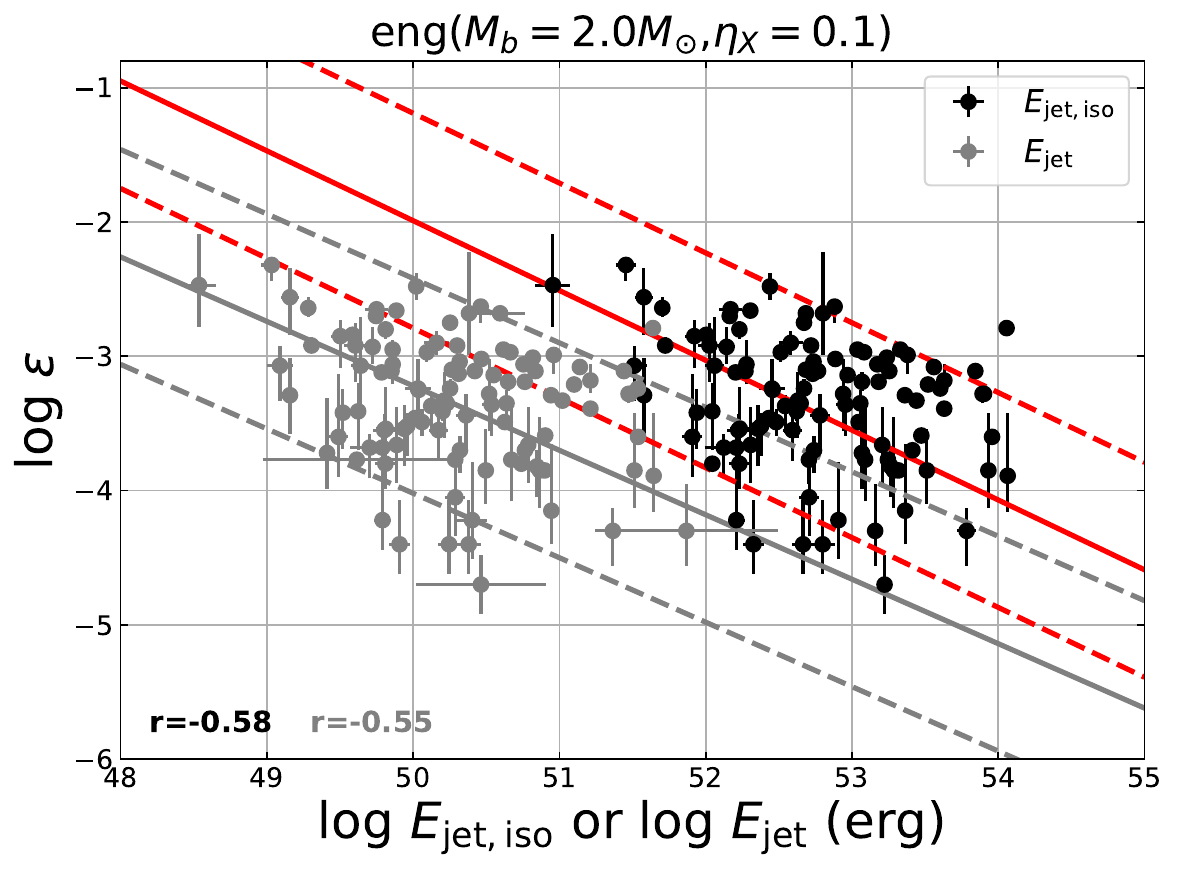}
\includegraphics [angle=0,scale=0.29]  {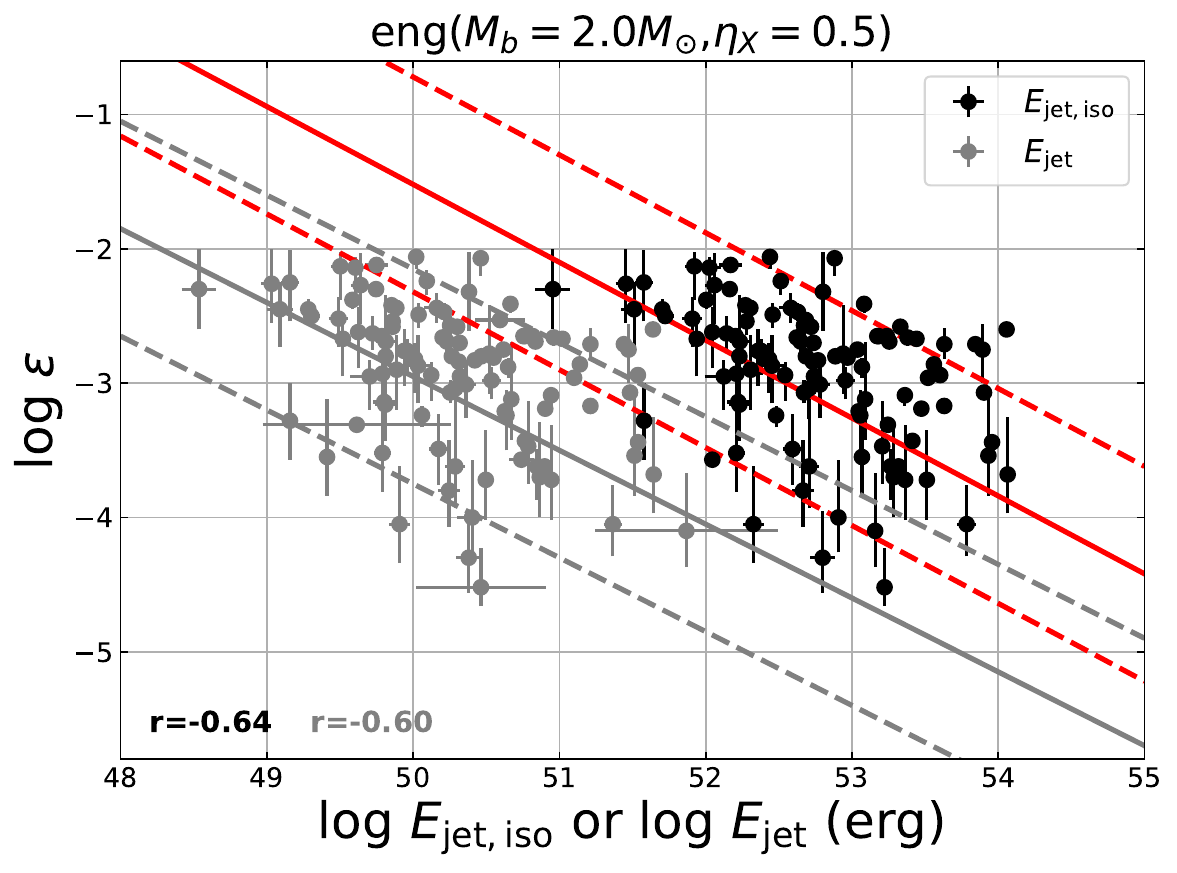}\\
\includegraphics [angle=0,scale=0.29]  {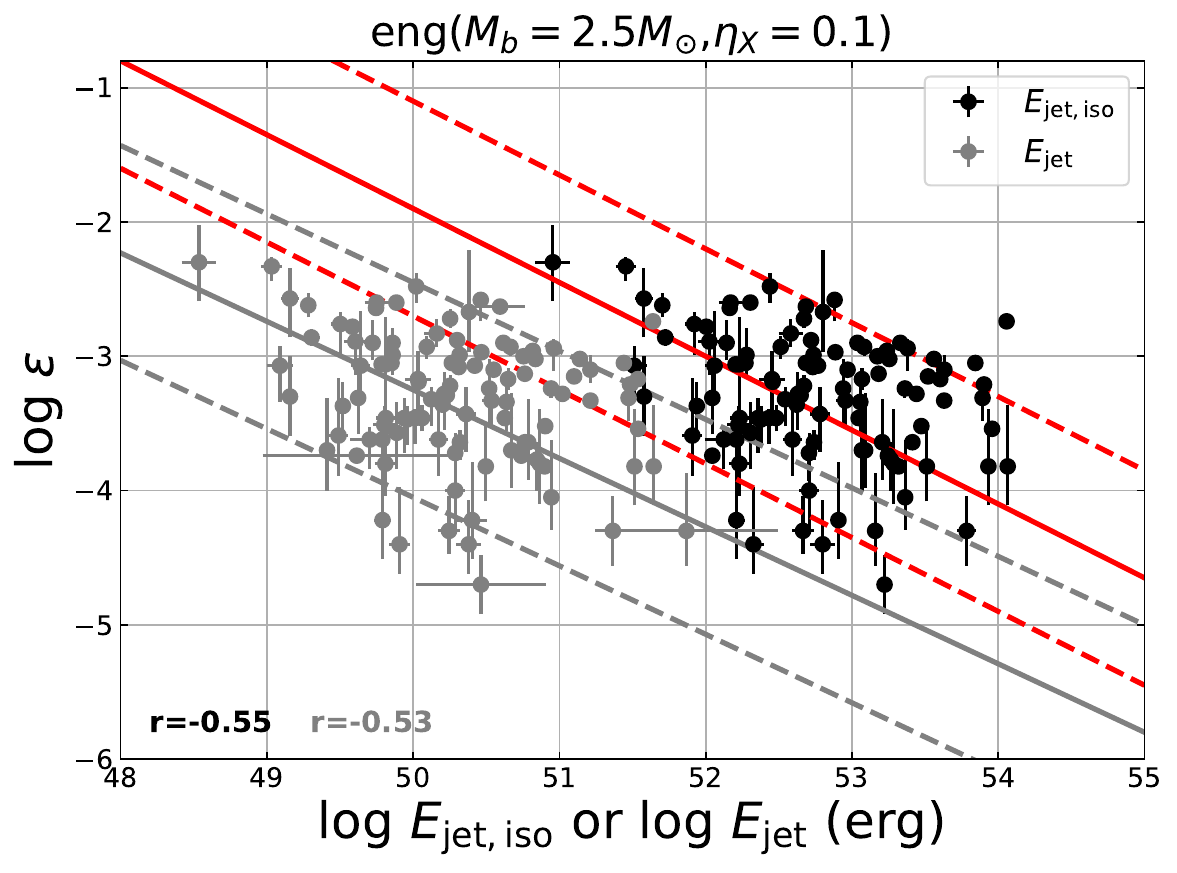}
\includegraphics [angle=0,scale=0.29]  {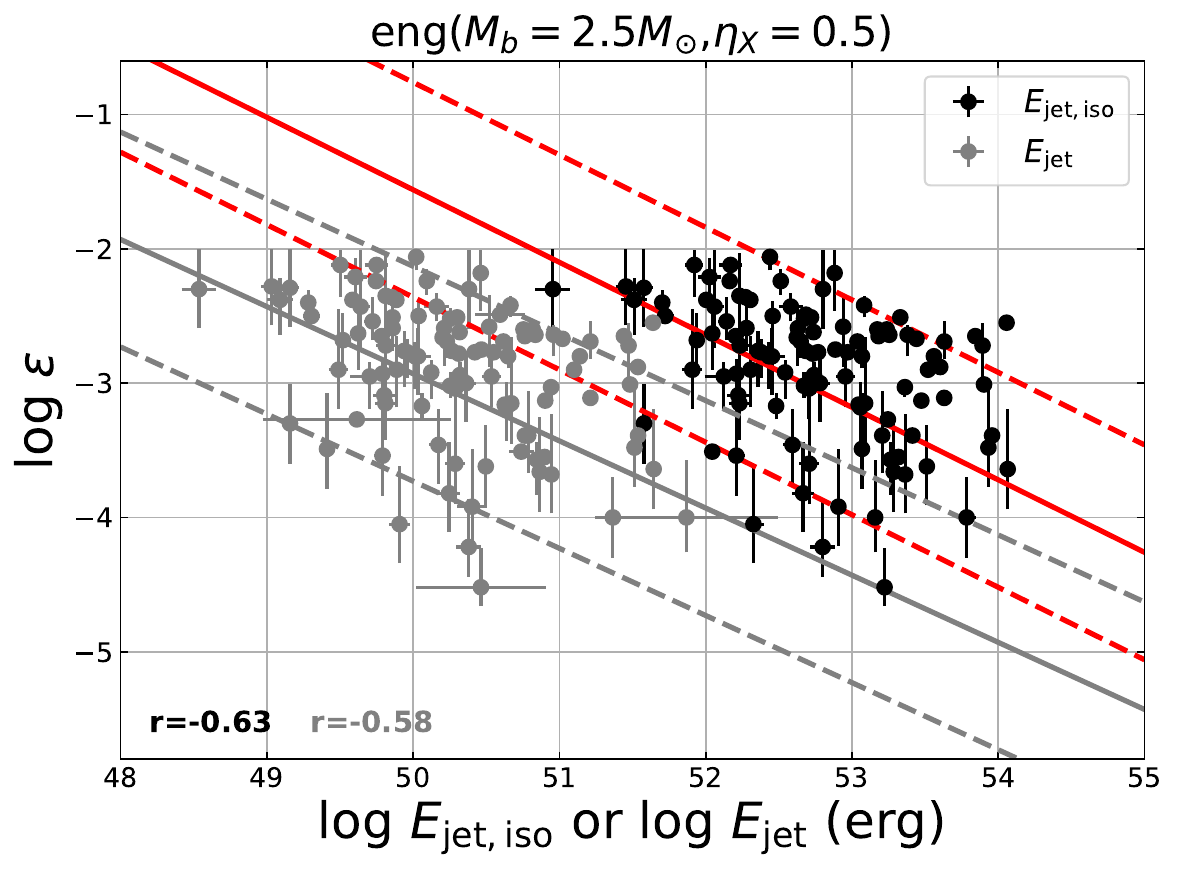}
\includegraphics [angle=0,scale=0.29]  {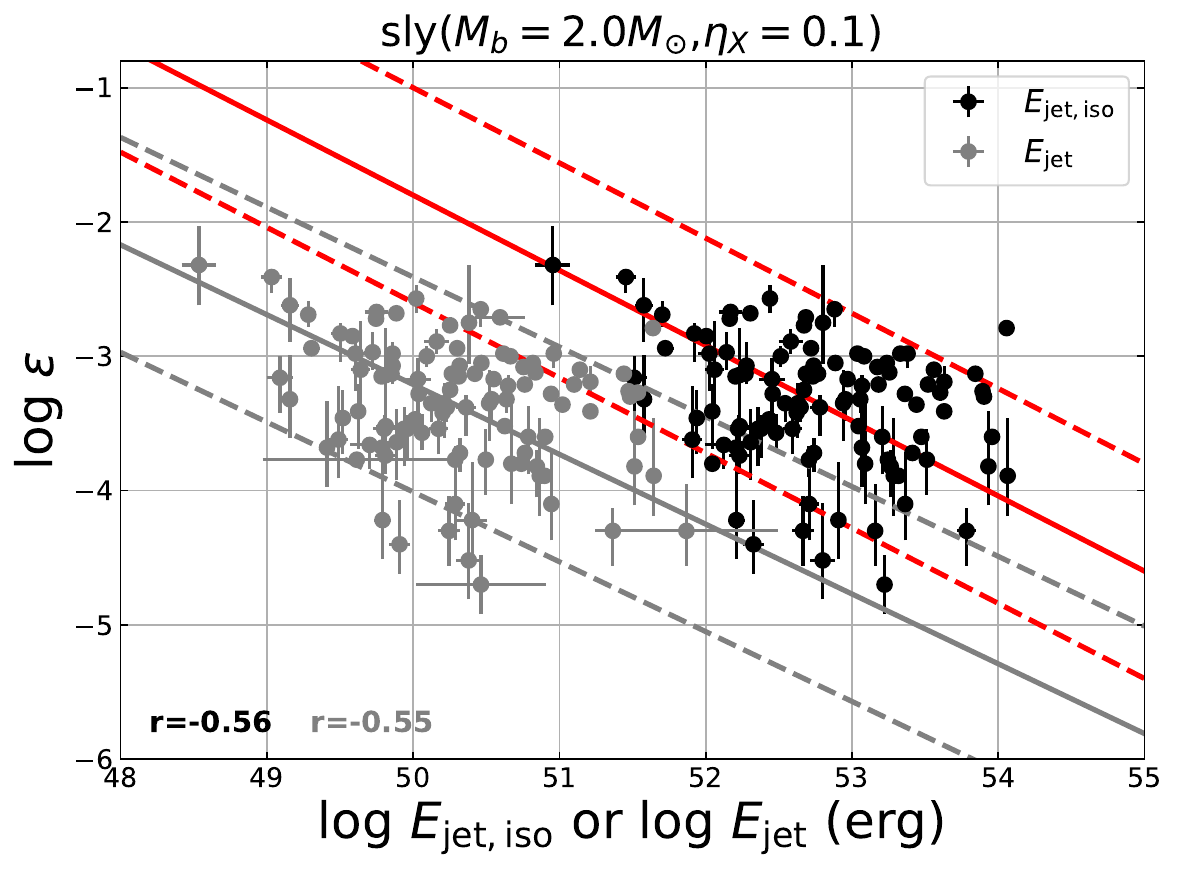}\\
\includegraphics [angle=0,scale=0.29]  {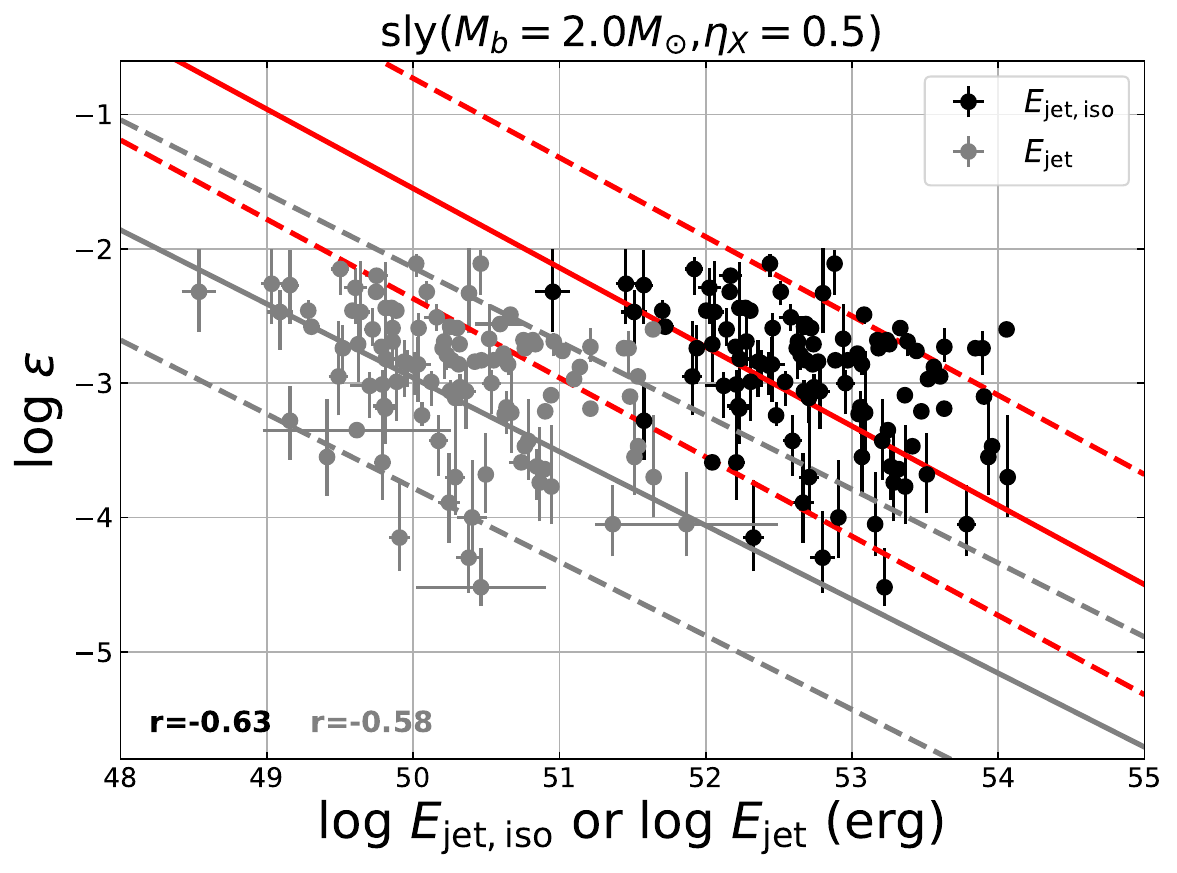}
\includegraphics [angle=0,scale=0.29]  {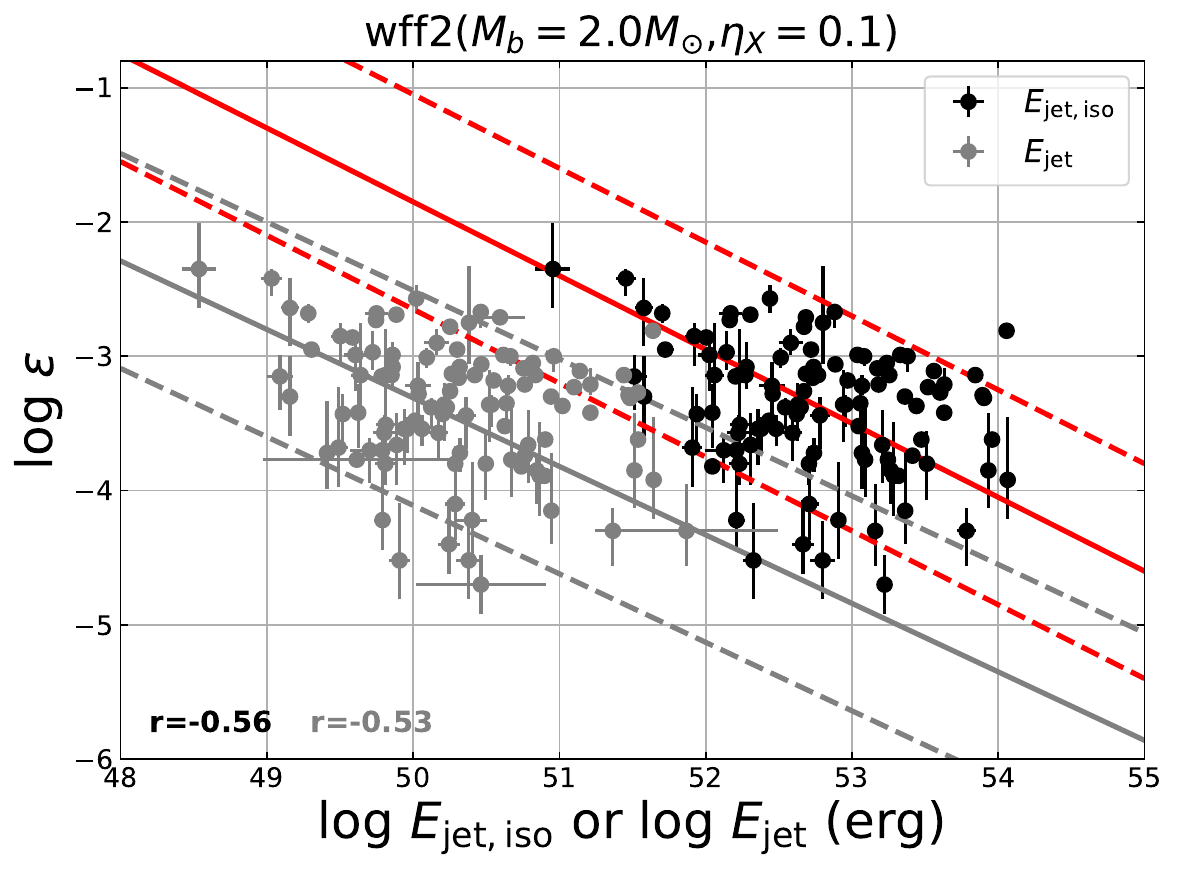}
\includegraphics [angle=0,scale=0.29]  {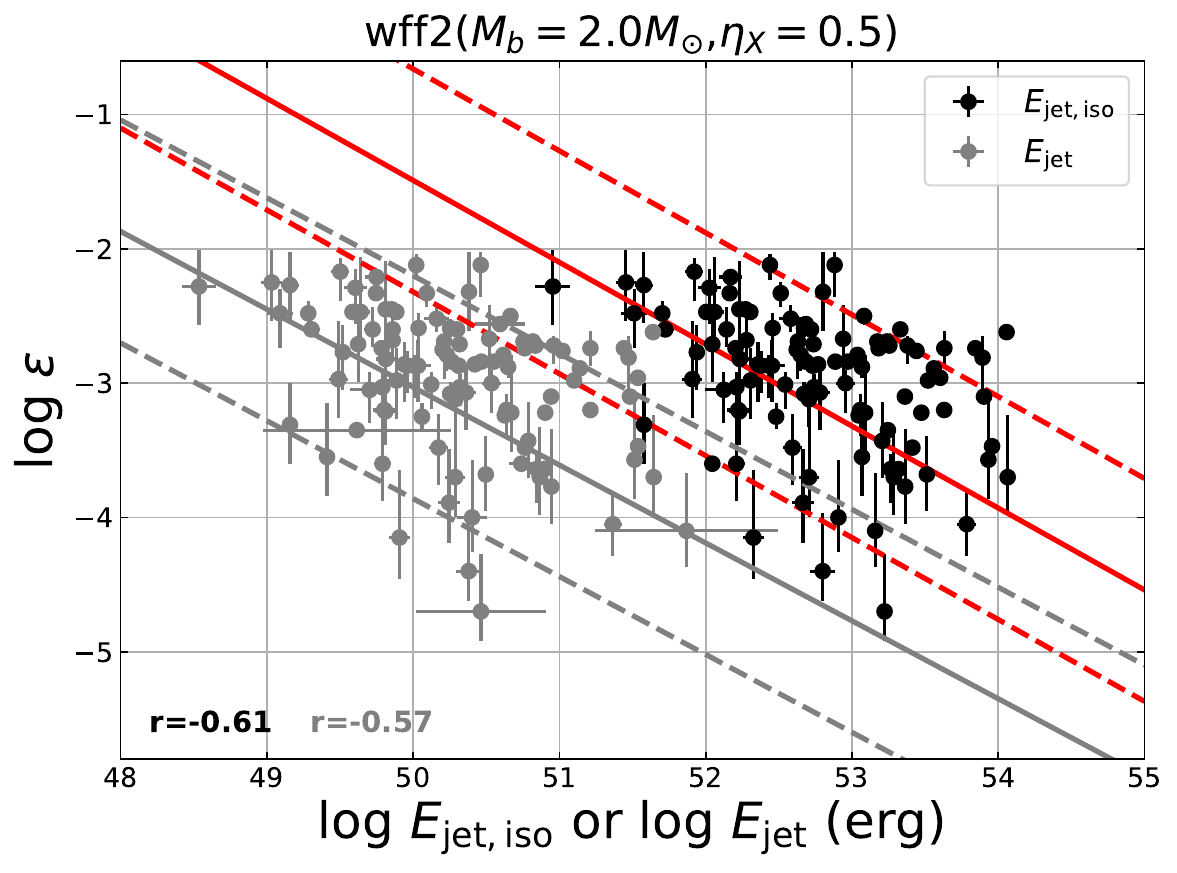}\\
\includegraphics [angle=0,scale=0.29]  {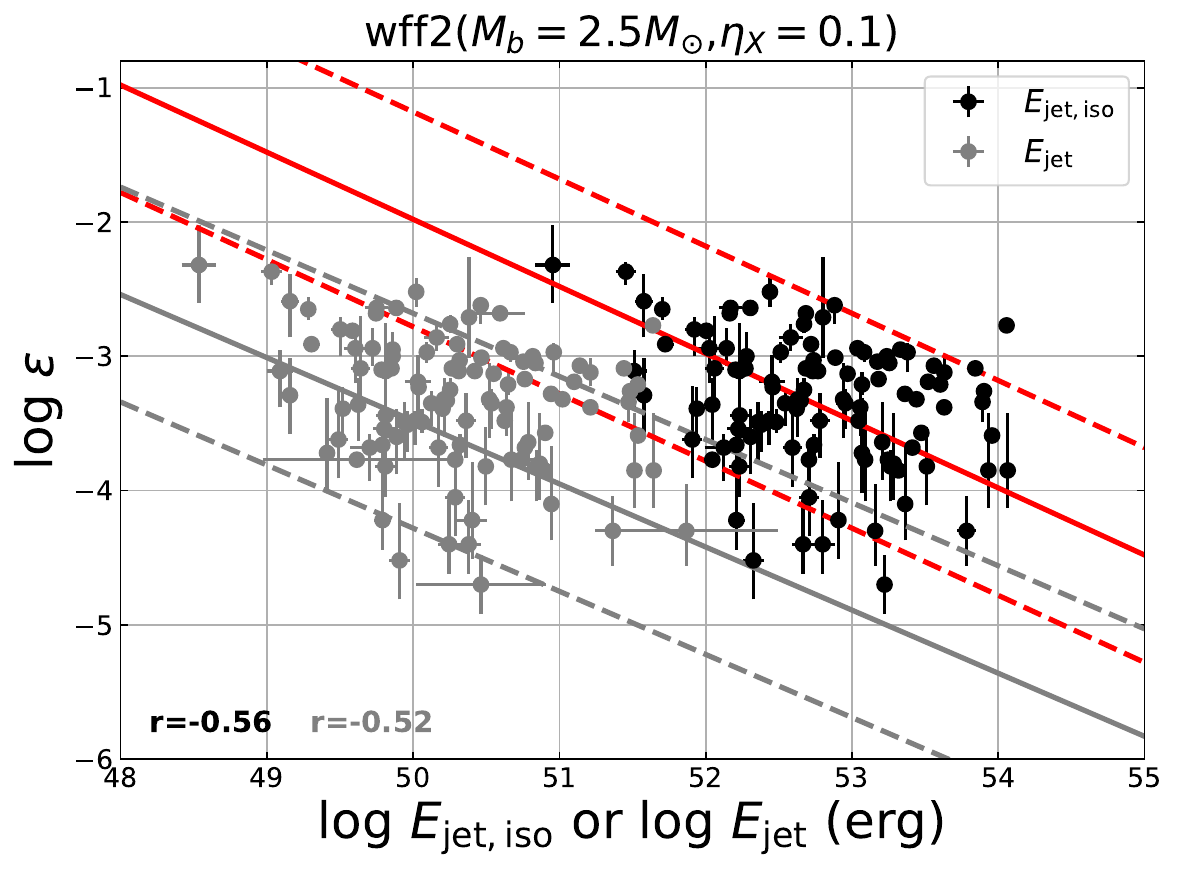}
\includegraphics [angle=0,scale=0.29]  {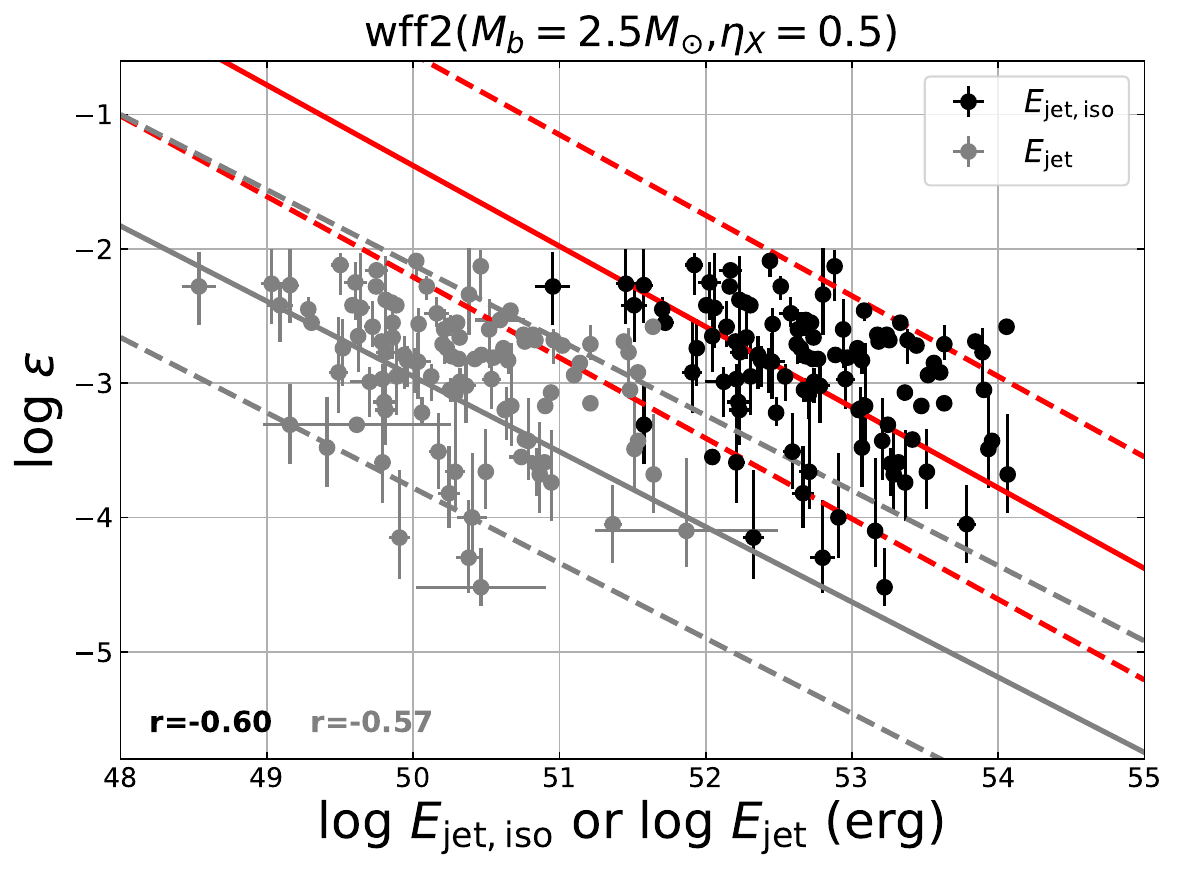}
\caption{The correlations between the $P_0$ and $E_{\rm jet,iso}$ (black circles), and $E_{\rm jet}$ (gray circles) in four samples of EoSs with $M_{b}=2.0~M_{\odot},~2.5~M_{\odot}$ and $\eta_{\rm X}=0.1,~0.5$. The red solid and red dashed lines are the best-fitting results and the 95\% confidence level for $P_0-E_{\rm jet,iso}$, respectively. The gray solid and gray dashed lines are the best-fitting results and the 95\% confidence level for $P_0-E_{\rm jet}$, respectively.}
\label{fig:epsilon-Ejet}
\end{figure*}

In Figure \ref{fig:epsilon-Ejet}, we further plotted a series of $\epsilon-E_{\rm jet}$ scatter diagrams for different EoSs and baryonic masses with two $\eta_{\rm X}$ values based on our LGRB sample. Homogeneously, we found that there are still strong anticorrelations between $E_{\rm jet, iso}$ ($E_{\rm jet}$) and $\epsilon$ with a large systematic error for all of our selected EoSs by using the ordinary least-squares analysis. Most of the GRBs in our sample fall into the $3\sigma$ deviation region of the best-fitting power-law model for all selected EoS cases. Likewise, our best-fitting correlations for various cases are reported in Table \ref{table-14}.

These anticorrelations between the GRB jet energy and newborn magnetar physical parameters appeared to be universal for all of our selected EoSs, which may indicate that the jet energy powered by a newborn magnetar is likely correlated with the magnetar nature properties. Within an admissible systematic error range, the anticorrelations of $P_0-E_{\rm jet,iso}(E_{\rm jet})$, $B_p-E_{\rm jet,iso}(E_{\rm jet})$ and $\epsilon-E_{\rm jet,iso}(E_{\rm jet})$ can be approximately described as $P_0\propto E_{\rm jet,iso}^{-0.29\pm0.03}(E_{\rm jet}^{-0.26\pm0.02})$, $B_p\propto E_{\rm jet,iso}^{-0.58\pm0.06}(E_{\rm jet}^{-0.55\pm0.05})$ and $\epsilon\propto E_{\rm jet,iso}^{-0.55\pm0.07}(E_{\rm jet}^{-0.52\pm0.06})$ for most of our selected EoSs, respectively. The results suggested that a newly formed magnetar with the faster rotational speed, lower surface magnetic field strength, as well as lower ellipticity deformation is more inclined to power a more energetic GRB jet. More importantly, the anticorrelations between GRB jet and magnetar physical parameters may imply that the connections of these parameters are the magnetar intrinsic property, and our derived ellipticity and initial spin period of the newborn magnetar are likely to originate from the magnetically induced distortion mechanism and correspond to the equilibrium spin period as a result of interaction between the magnetar and its accretion disk, respectively. Moreover, we further investigated how the energy ratio parameters ($\Re$ and $\Re'$) are related to the physical parameters of the newborn magnetar. No statistical correlation between $\Re$($\Re'$) and magnetar physical parameters can be claimed for all of our selected EoSs.

\subsection{Detectability of Magnetar Dipole Radiation with \emph{EP} and \emph{SVOM}}
The Einstein Probe (\emph{EP}) mission is designed to discover more interesting high-energy transient and monitor variable objects in the 0.5-4 keV soft X-ray band \citep{Yuan2015,Yuan2025}. The payload of \emph{EP} consists of a Wide-field X-ray Telescope (WXT) and a follow-up X-ray telescope (FXT) with a larger light-collecting capability than WXT, both based on the novel lobster-eye Micro-Pore Optics technology. The WXT has a large instantaneous Field-of-View (FoV, $60^{\circ}\times60^{\circ}$) and high sensitivity for X-ray all-sky monitors, which is about $F_{\rm th}^{\rm WXT}=1\times10^{-10}~\rm erg~s^{-1}~cm^{-2}$ in a 100 s exposure. The FXT has a narrow FoV ($1^{\circ}\times1^{\circ}$) and higher sensitivity for fast X-ray follow-up monitors, which is about $F_{\rm th}^{\rm FXT}=1\times10^{-11}~\rm erg~s^{-1}~cm^{-2}$ in a 100 s exposure. The Space-based multi-band astronomical Variable Objects Monitor (\emph{SVOM}) mission is to continue the exploration of the transient Universe with a set of space-based multi-wavelength instruments, following the way opened by \emph{Swift} mission \citep{Wei2016}. The \emph{SVOM} satellite carries two wide FoV high-energy prompt emission detectors and two narrow FoV low-energy afterglow emission telescopes: a coded-mask gamma-ray imager (ECLAIRs), a gamma-ray spectrometer (GRM), a microchannel X-ray telescope (MXT), and a visible-band telescope (VT). The ground follow-up equipment include: a wide angle optical camera (GWAC) for the observation of the optical prompt emission, and two 1 m class robotic follow-up telescopes (C-GFT and F-GFT) for the follow-up observations in the visible and near-infrared domains. The MXT is an X-ray telescope designed to observe and measure the properties of the GRB X-ray afterglow in the 0.2-10 keV energy range. The MXT has a narrow FoV ($64\times64$ arcmin) and high sensitivity for triggering X-ray afterglow of GRB and providing detailed X-ray spectra, which is about $F_{\rm th}^{\rm MXT}=7\times10^{-11}~\rm erg~s^{-1}~cm^{-2}$ in a 100 s exposure. 

\begin{figure}
\centering
\includegraphics[angle=0,scale=0.22]{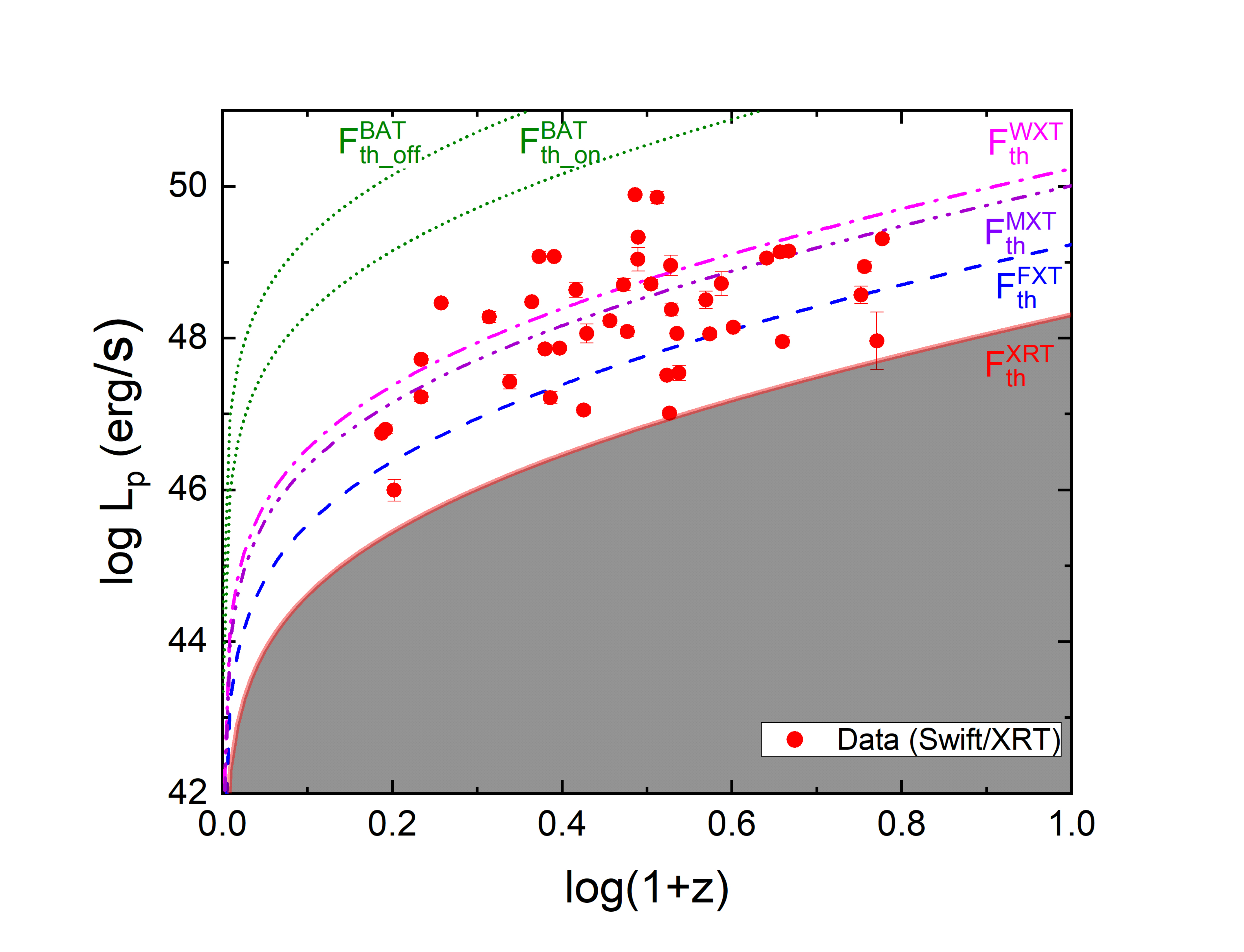}
\caption{The GRB Luminosity thresholds of \emph{Swift}/XRT, \emph{EP}/WXT, \emph{EP}/FXT, \emph{SVOM}/MXT in the flux limit $F_{\rm th}^{\rm XRT}=2\times10^{-12}~\rm erg~s^{-1}~cm^{-2}$, $F_{\rm th}^{\rm WXT}=1\times10^{-10}~\rm erg~s^{-1}~cm^{-2}$, $F_{\rm th}^{\rm FXT}=1\times10^{-11}~\rm erg~s^{-1}~cm^{-2}$, $F_{\rm th}^{\rm MXT}=7\times10^{-11}~\rm erg~s^{-1}~cm^{-2}$, respectively. The data of the X-ray plateau in our sample are shown with red dots, and the thresholds of different instruments are marked by different colored lines to examine the detection probability of magnetar dipole radiation.}
\label{fig:EP_SVOM sensitivity}
\end{figure}

In this section, we investigated the possible detection for the magnetar dipole radiation with the \emph{EP}/WXT, FXT and \emph{SVOM}/MXT based on 19 years of \emph{Swift}/XRT observations (43 GRBs with plateau emission and redshift information in our sample). It should be noted that, due to the many uncertainties in the follow-up observations of \emph{EP} and \emph{SVOM} for the GRB afterglow, here, we ignored the limitations of the various follow-up observations and simply used the sensitivity threshold of different instruments to estimate the ideal detection rate. Figure \ref{fig:EP_SVOM sensitivity} showed our sample in the log(1+z)-log$L_{\rm p}$ plane. Adopting the sensitivity of \emph{EP}/WXT as $F_{\rm th}^{\rm WXT}=1\times10^{-10}~\rm erg~s^{-1}~cm^{-2}$ in a 100 s exposure, we can find that the ideal detection rate of magnetar dipole radiation with WXT is only $\sim$30\% (13 GRBs). Adopting the sensitivity of \emph{EP}/FXT as $F_{\rm th}^{\rm FXT}=1\times10^{-11}~\rm erg~s^{-1}~cm^{-2}$ in a 100s exposure, we can find that the prompt follow-up observations with FXT could increase the percentage to $\sim$81\% (35 GRBs). Adopting the sensitivity of \emph{SVOM}/MXT as $F_{\rm th}^{\rm MXT}=7\times10^{-11}~\rm erg~s^{-1}~cm^{-2}$ in a 100s exposure, we can find that the ideal detection rate of magnetar dipole radiation with MXT is $\sim$40\% (17 GRBs). Notably, the major factor affecting the identification of magnetar wind radiation is the GRB afterglow emission, not the instrument sensitivity. \cite{Zou2019} proposed that about 60\% of magnetar dipole radiation may be obscured by the forward shock afterglows of GRB. Although the \emph{EP}/WXT, FXT, \emph{SVOM}/MXT detectable magnetar wind X-ray events are bright, the obscuring effect by bright X-ray afterglows would also significantly reduce the detection rate of the magnetar wind events. Moreover, the GRB magnetar dipole radiation may be able to generate an orphan X-ray transient when the GRB jet prompt emission is off-axis relative to our line of sight, which is consistent with an NS-NS postmerger product as predicted by \cite{Zhang2013} and \cite{Sun2017}. Some studies have reported the orphan magnetar-powered X-ray transients by analyzing the observational data from Chandra X-ray telescope \citep{Bauer2017,Xue2019,Sun2019,Lin2021a,Ai2021}. Thus, the \emph{EP} and \emph{SVOM} joint survey may be interesting for searching for similar orphan long-lasting X-ray transients, which would help us to provide unprecedented insight into the physics of the newborn magnetar.

\subsection{Detectability of GW Radiation from the Newborn Magnetar with $R$ and $I$ Evolution}
\begin{figure}
\centering
\includegraphics[angle=0,scale=0.45]{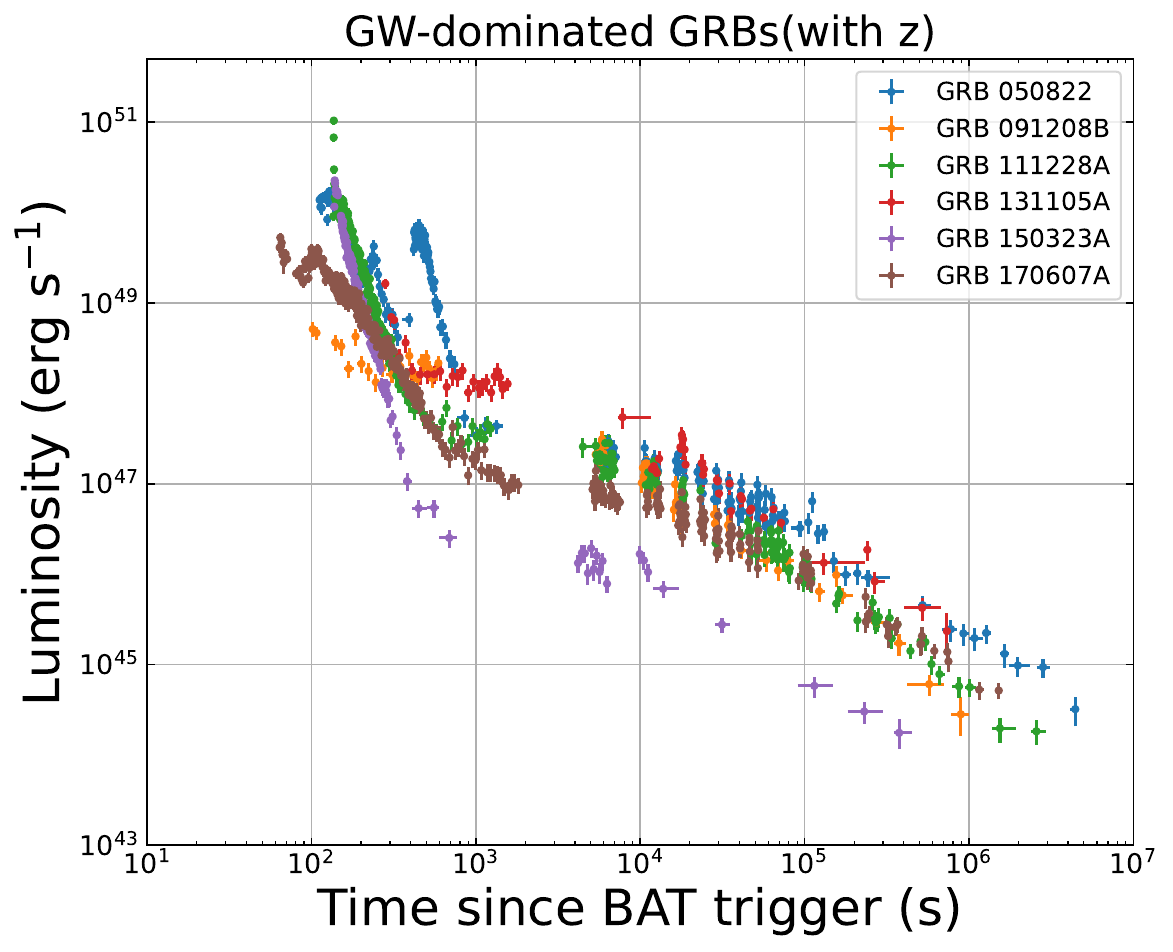}
\includegraphics[angle=0,scale=0.45]{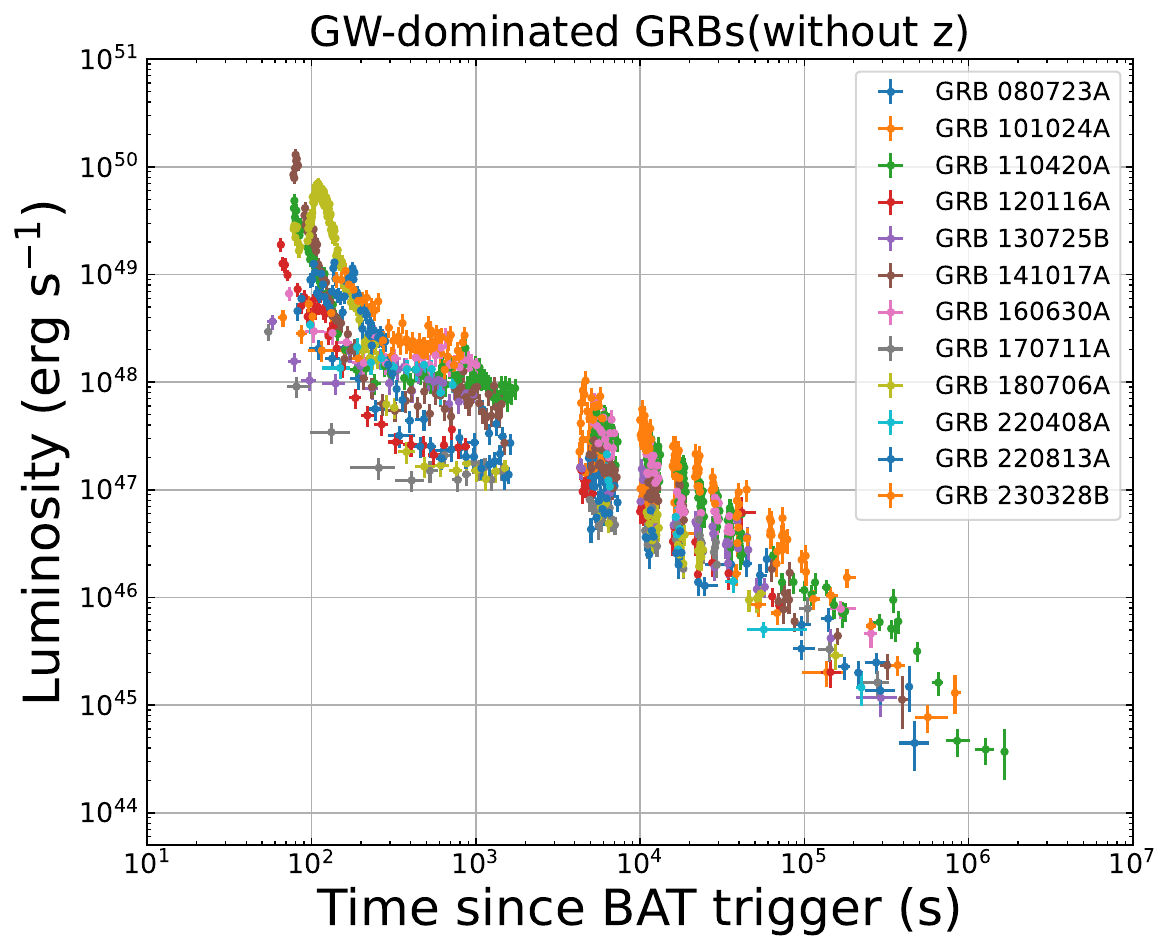}
\caption{Left panel: the X-ray luminosity light curves of GW-dominated GRBs with $z$ measurement in our sample. Right panel: the X-ray luminosity light curves of GW-dominated GRBs without $z$ measurement in our sample, and we adopt $z=1$ to estimate the luminosity of these GRBs.}
\label{fig:GW_GRBs_LC}
\end{figure}

\begin{figure}
\centering
\includegraphics[angle=0,scale=0.44]{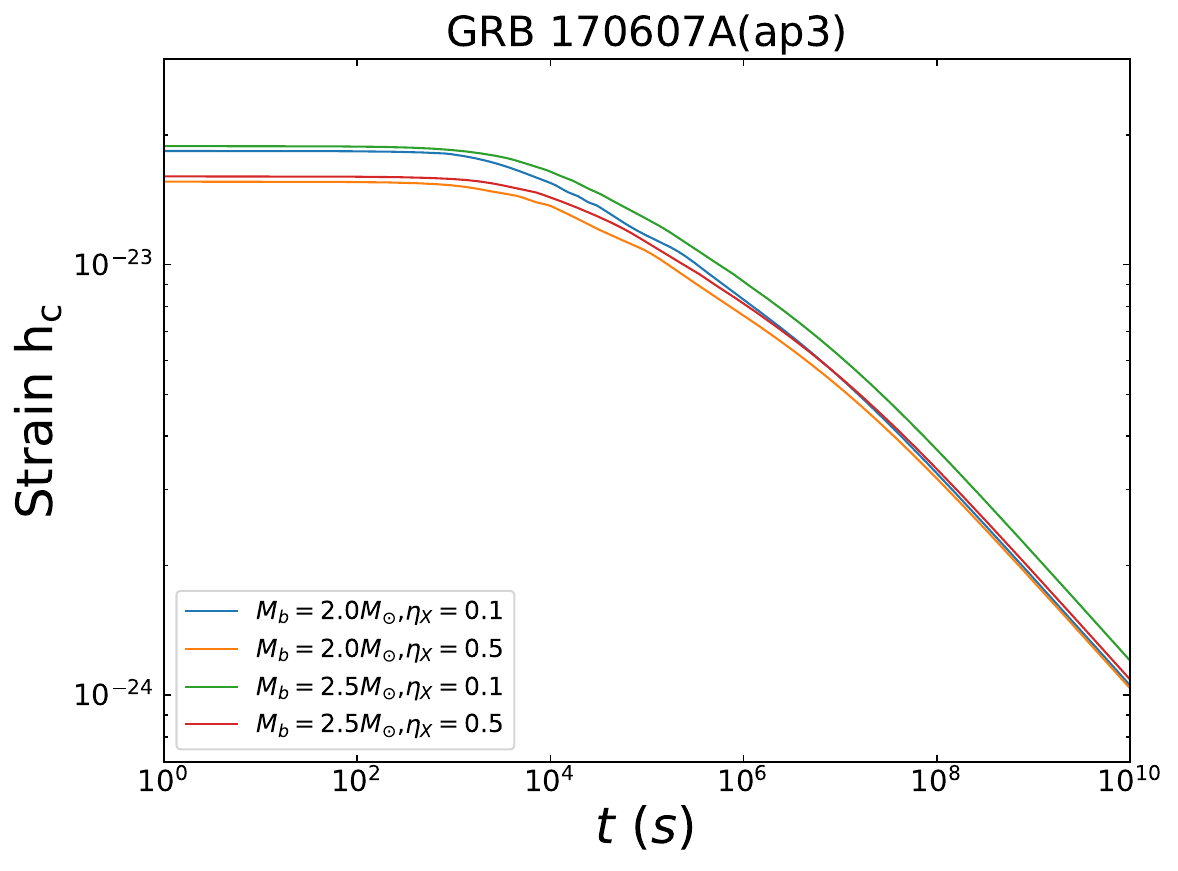}
\includegraphics[angle=0,scale=0.44]{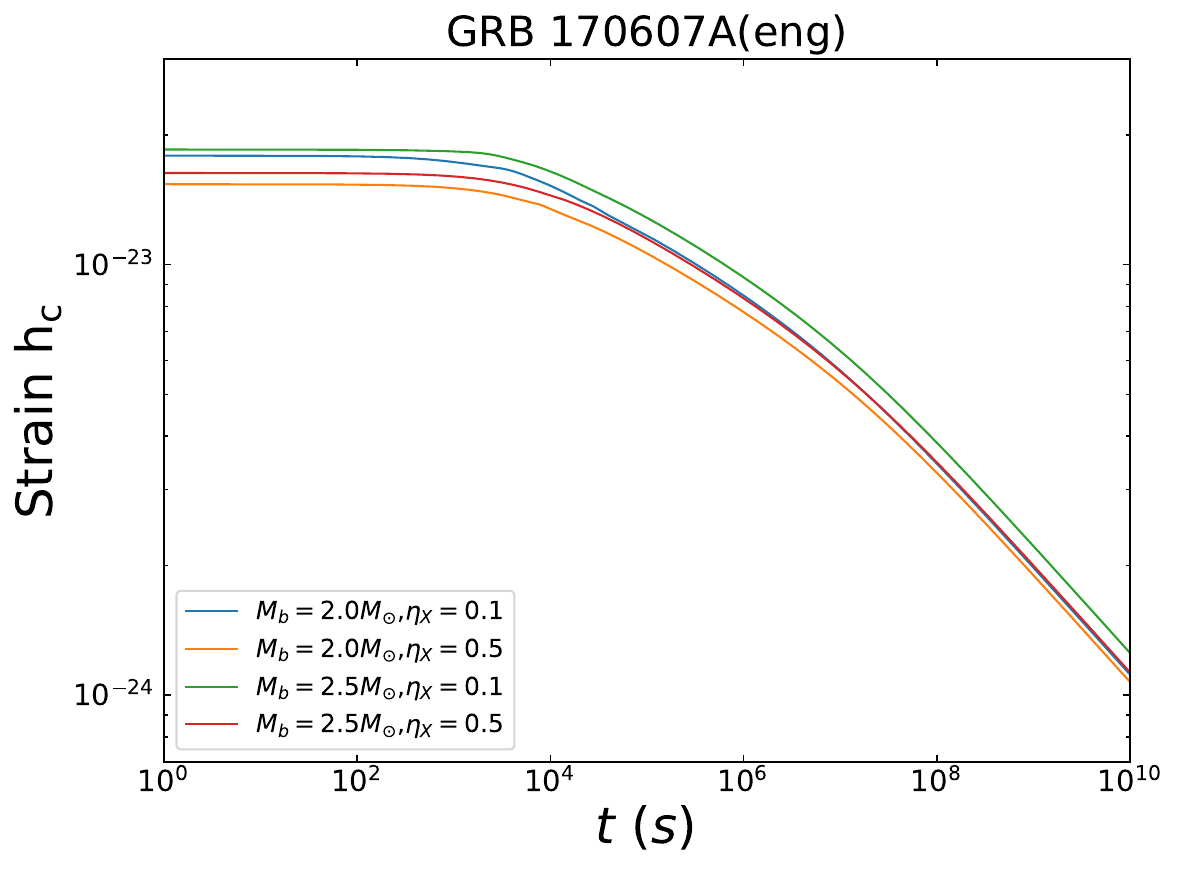}
\includegraphics[angle=0,scale=0.44]{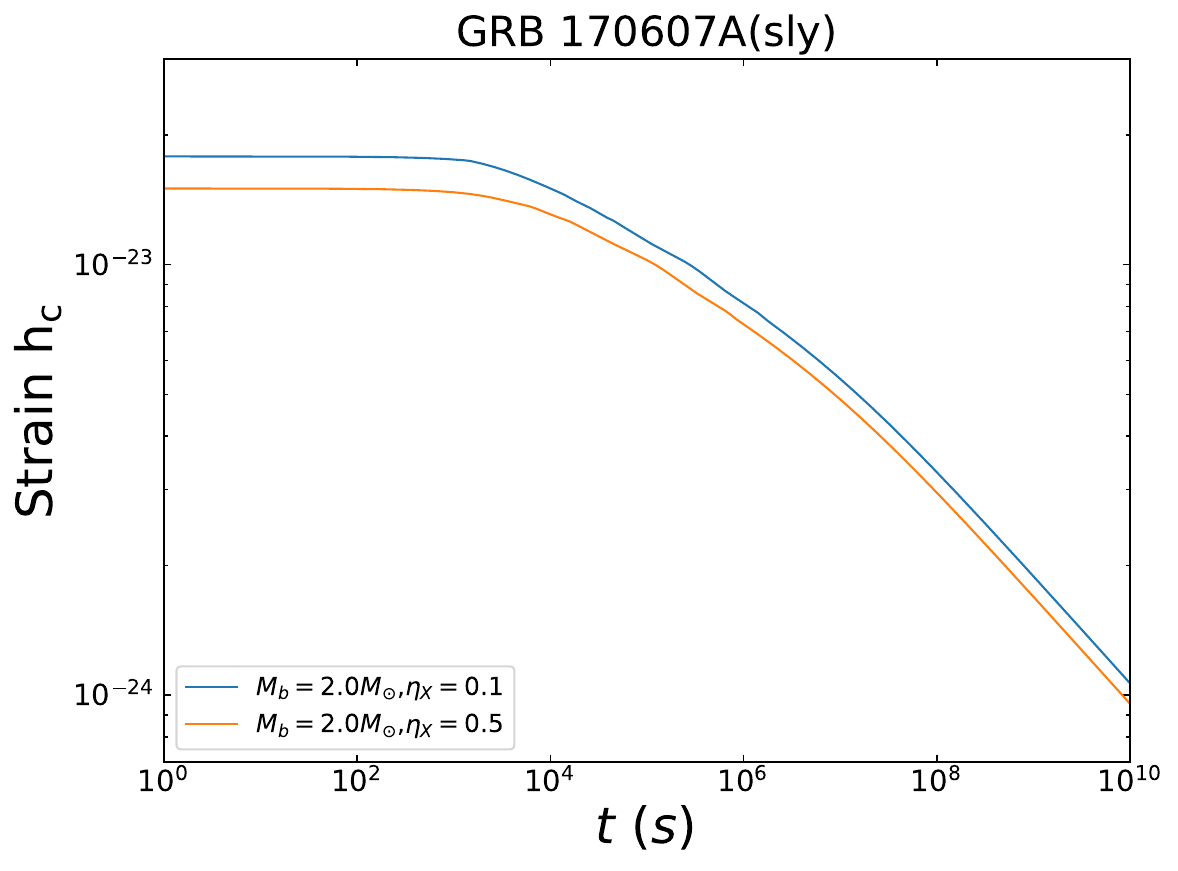}
\includegraphics[angle=0,scale=0.44]{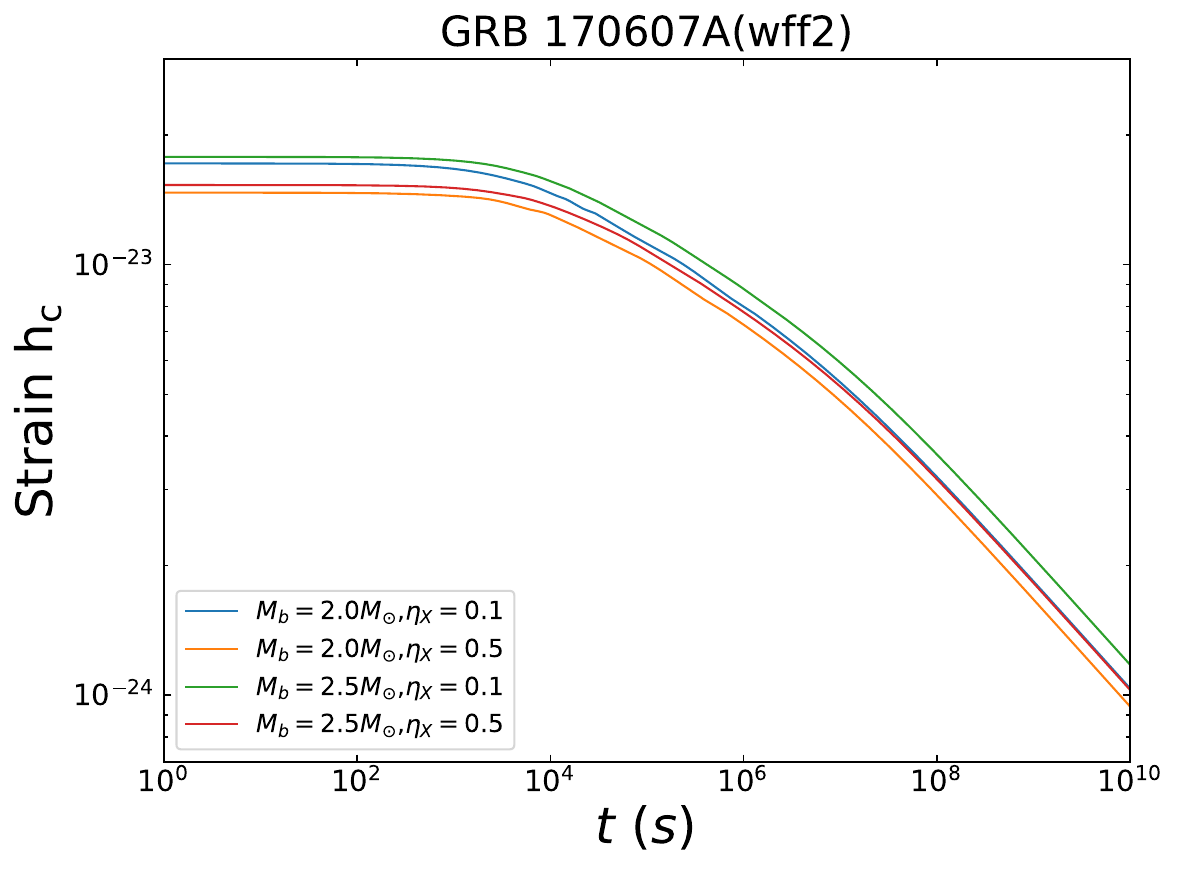}
\caption{The GW characteristic amplitude of GRB 170607A in four samples of EoSs with $M_{b}=2.0~M_{\odot},~2.5~M_{\odot}$ and $\eta_{\rm X}=0.1,~0.5$, respectively.}
\label{fig:170607A_GWhc}
\end{figure}

\begin{figure}
\centering
\includegraphics[angle=0,scale=0.44]{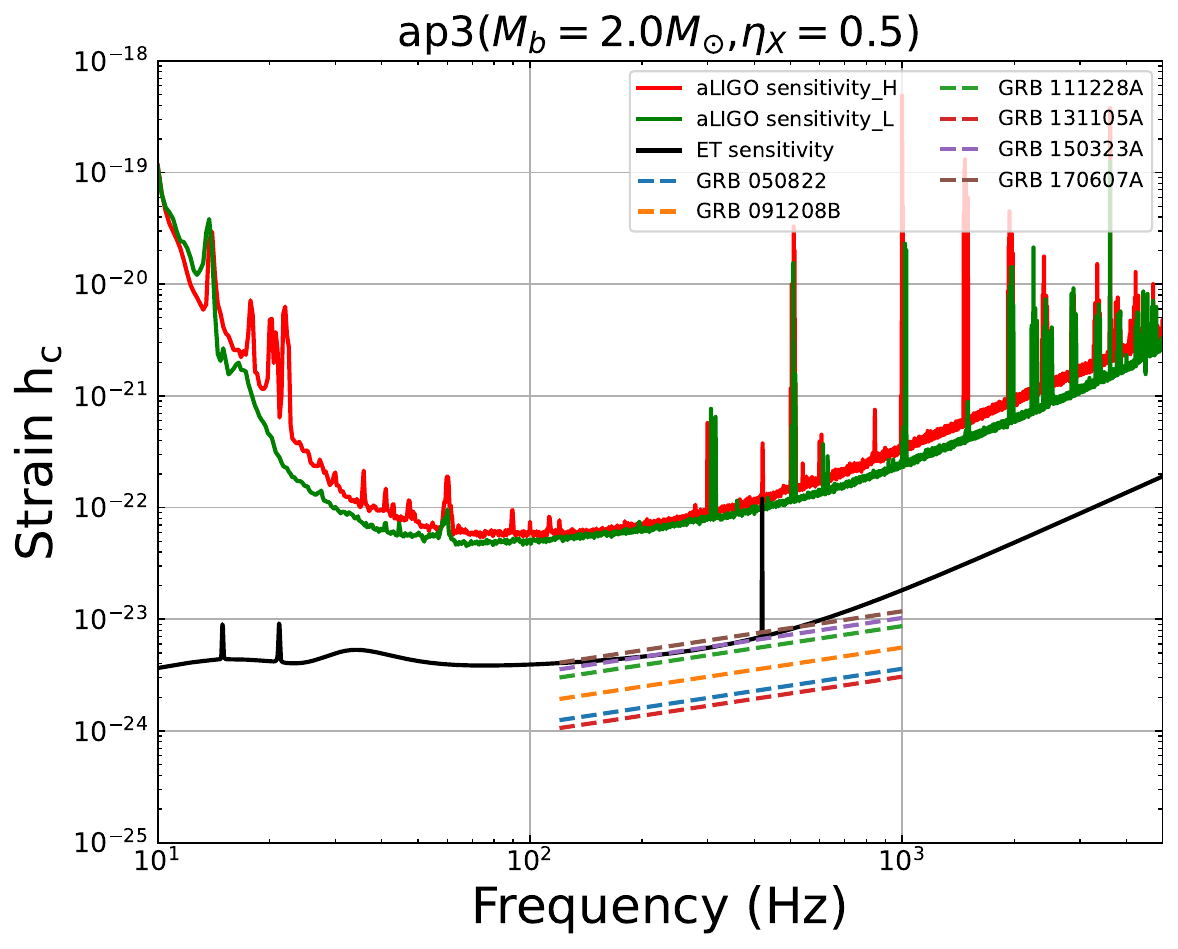}
\includegraphics[angle=0,scale=0.44]{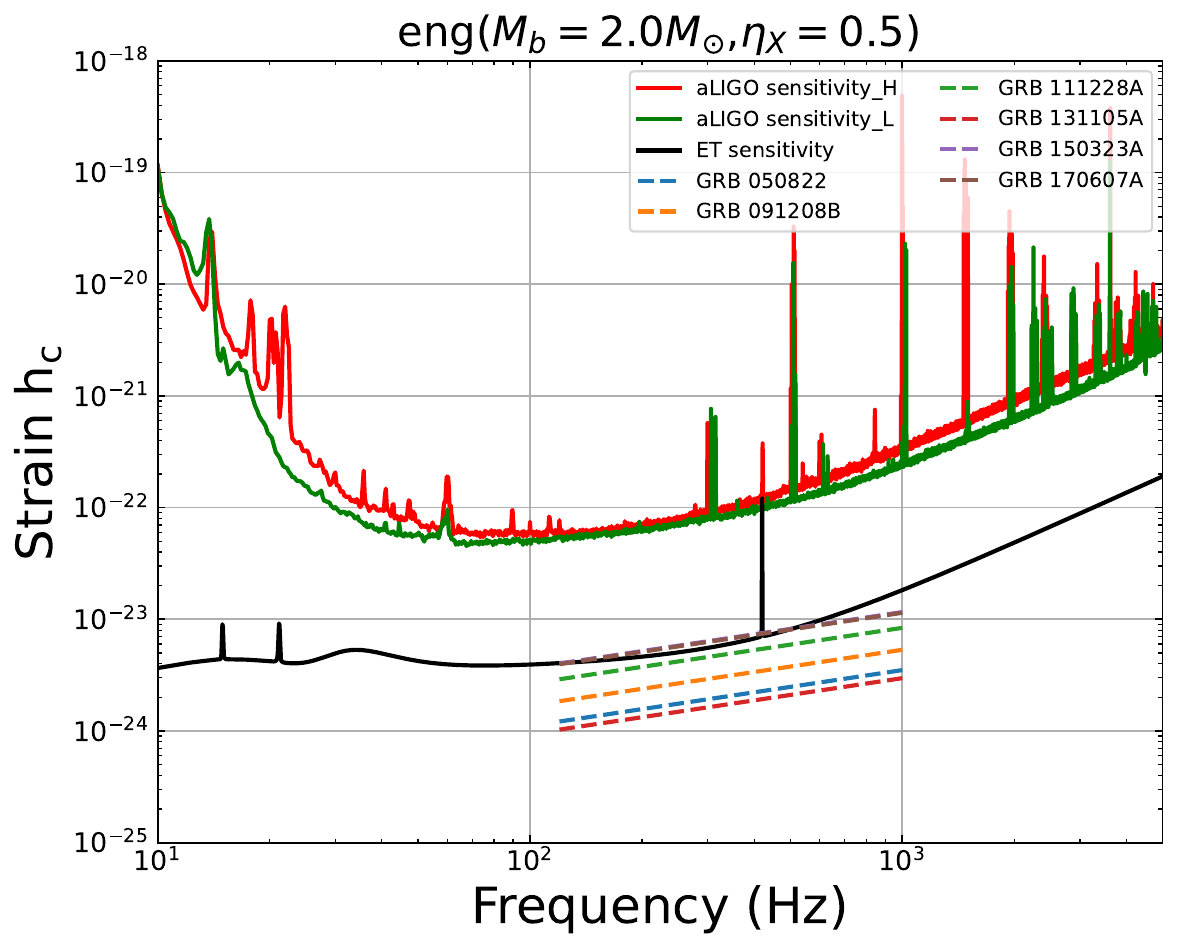}
\includegraphics[angle=0,scale=0.44]{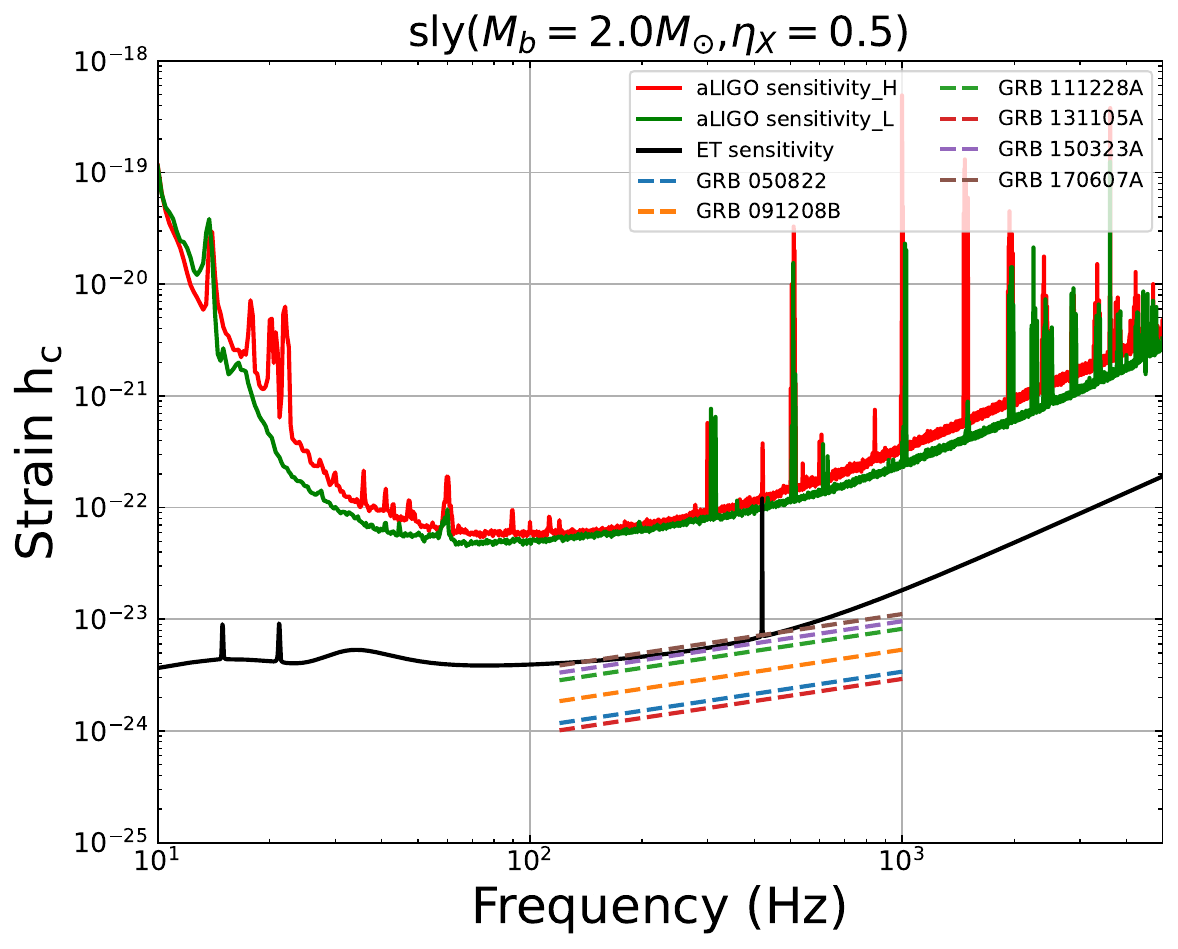}
\includegraphics[angle=0,scale=0.44]{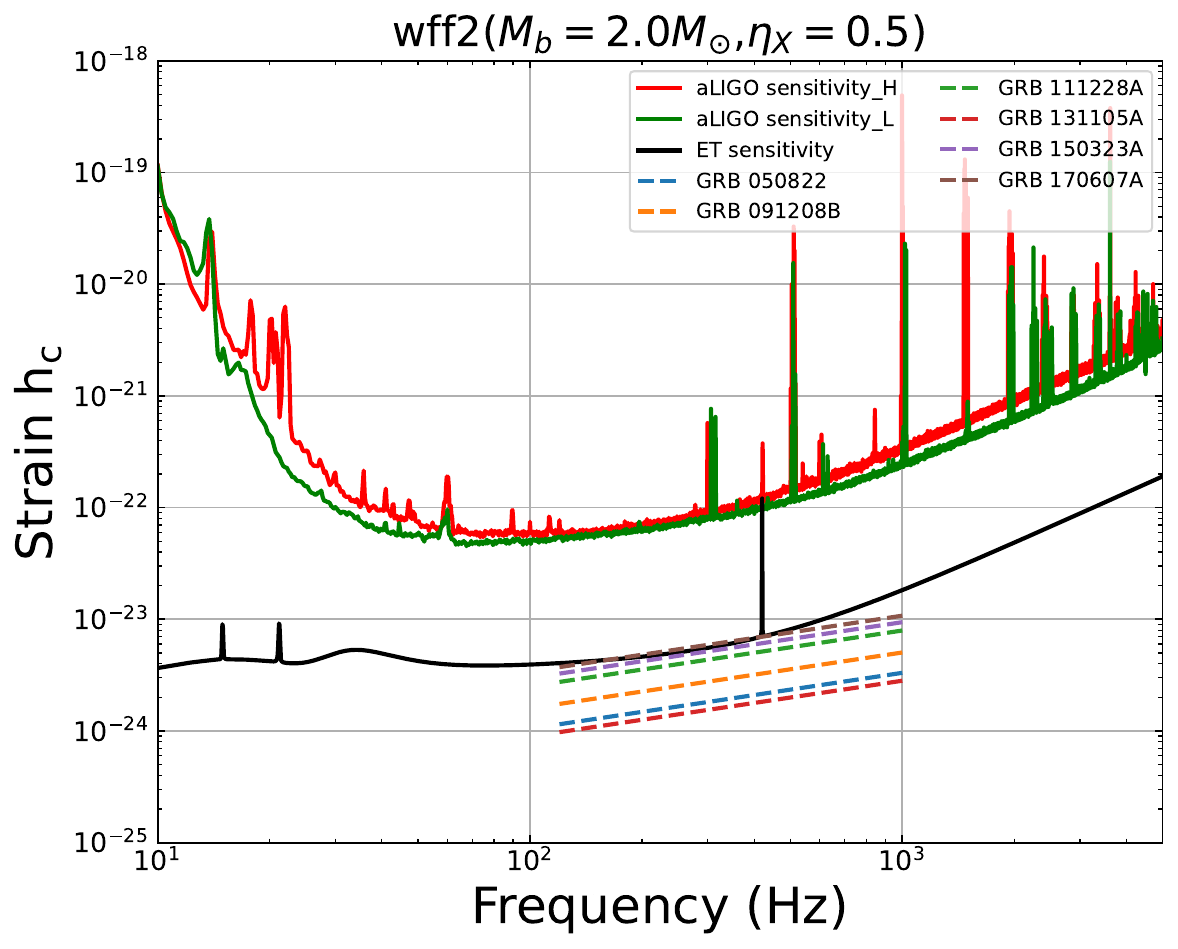}
\caption{Detectability test of GW signals with aLIGO and ET detectors for our GW-dominated GRBs with $z$ measurement in four samples of EoSs with $M_{b}=2.0M_{\odot}$ and $\eta_{X}=0.5$.
The black solid line is the sensitivity limits for ET, and the red and green solid lines are the sensitivity limits for aLIGO-Hanford and aLIGO-Livingston, respectively.}
\label{fig:GW_detectability(with z)}
\end{figure}

\begin{figure}
\centering
\includegraphics[angle=0,scale=0.44]{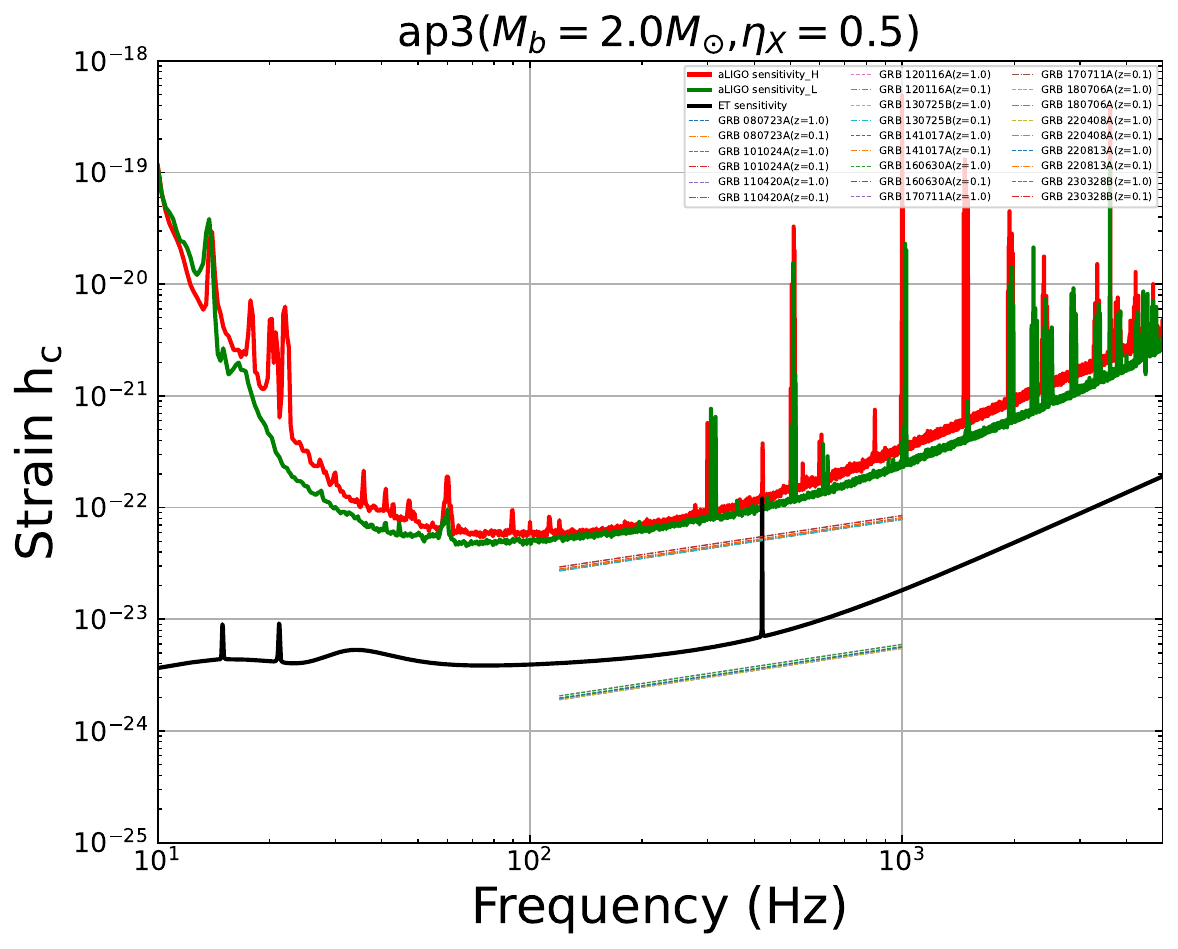}
\includegraphics[angle=0,scale=0.44]{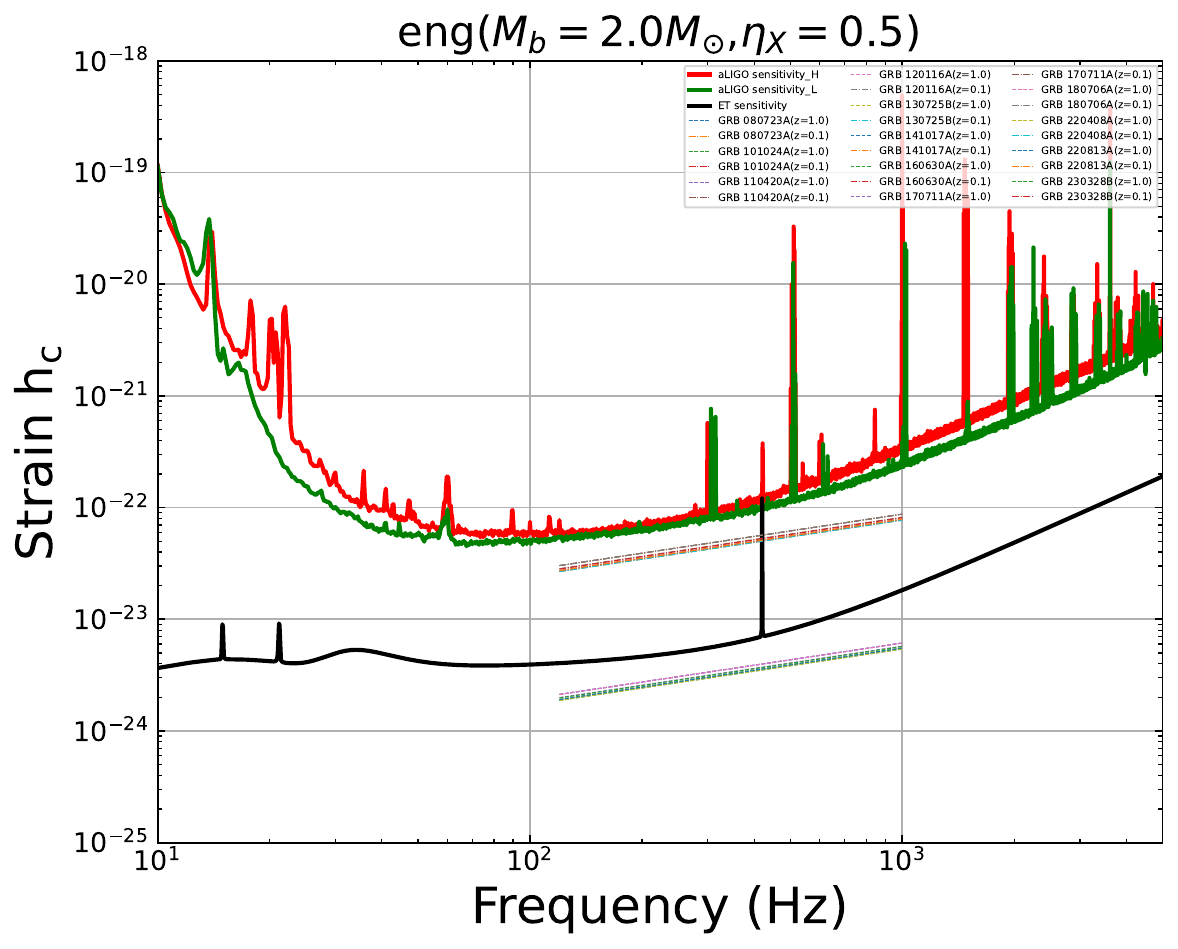}
\includegraphics[angle=0,scale=0.44]{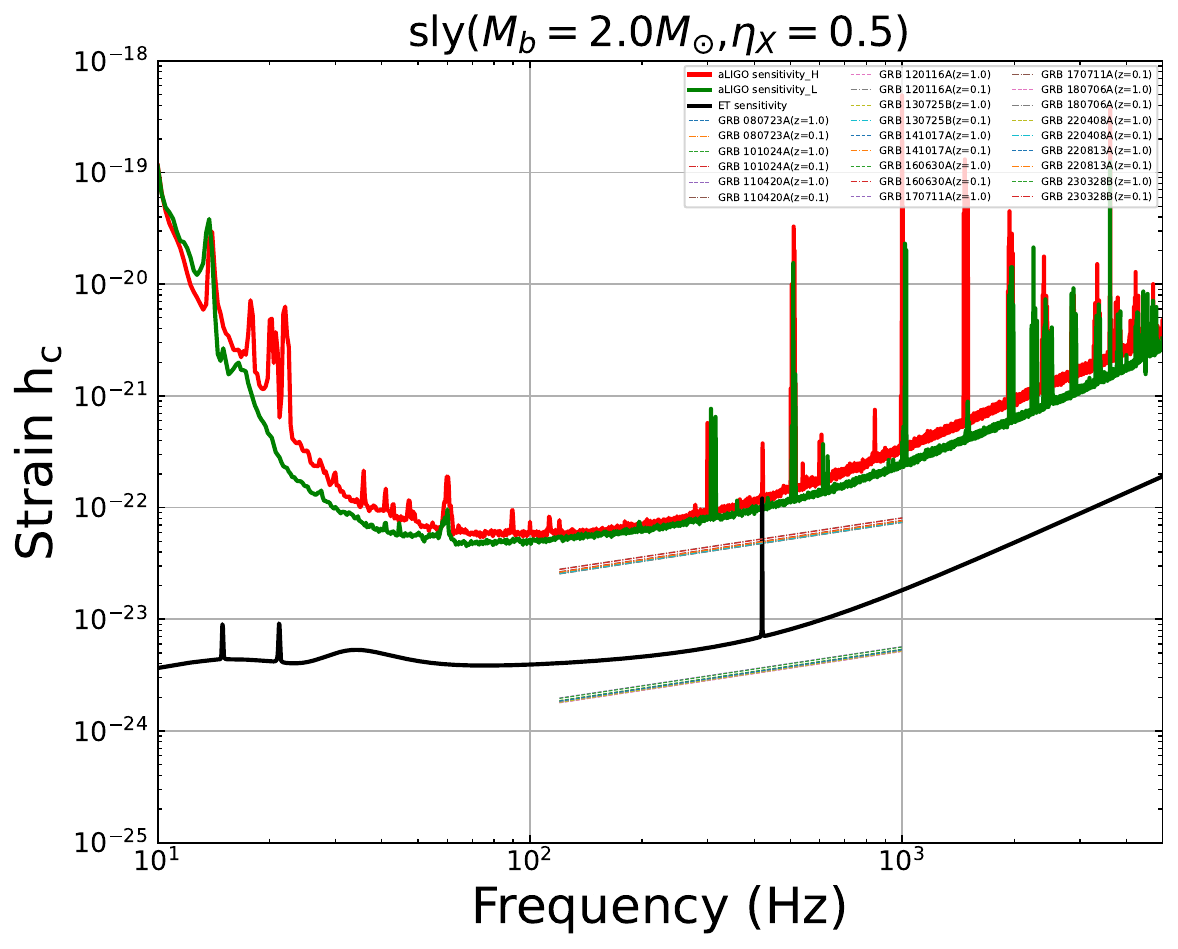}
\includegraphics[angle=0,scale=0.44]{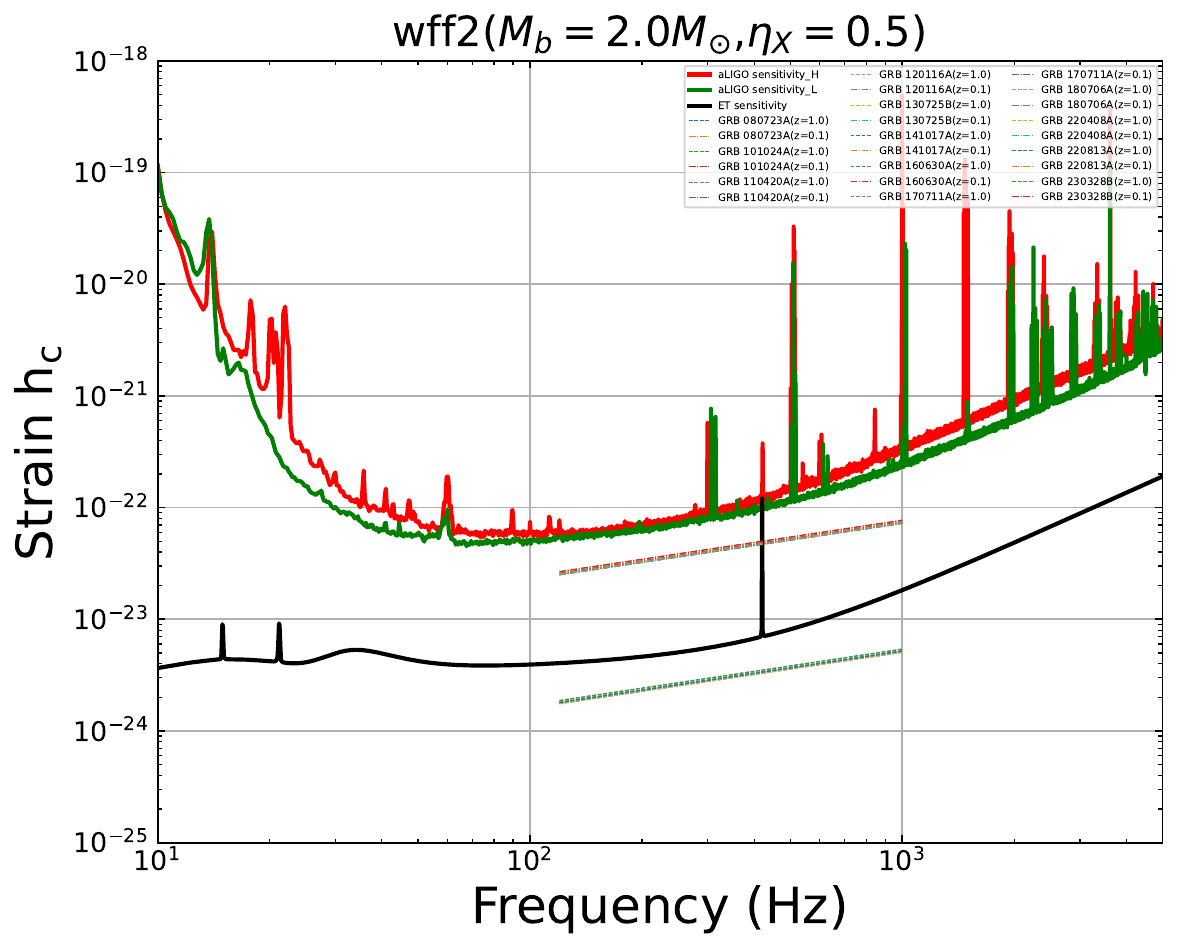}
\caption{Similar to Figure \ref{fig:GW_detectability(with z)}, but for our GW-dominated GRBs without $z$ measurement in four samples of EoSs with $M_{b}=2.0M_{\odot}$ and $\eta_{X}=0.5$. The different color dashed and dashed-dotted lines represent $z = 1$ and $z = 0.1$, respectively.}
\label{fig:GW_detectability(without z)}
\end{figure}

The newborn millisecond magnetars formed from the core collapse of massive stars have been proposed as potential sources of continuous GW emission for Advanced-LIGO (aLIGO, \citep{Aasi2015}) and Einstein Telescope (ET, \citep{Punturo2010}). If the energy loss of newborn magnetar spin-down is dominated by GW radiation, one potential question is that how strong is the GW signal of magnetar spin-down within the context of its $R$ and $I$ evolution. In response to this, we collected all GRBs with GW-dominated energy loss during the spin-down process from our available sample. In the left and right panels of Figure \ref{fig:GW_GRBs_LC}, we showed the XRT light curves for the GW-dominated GRBs with a redshift measurement and without a redshift measurement in our sample, respectively. We can obviously observe that these GW-dominated GRBs are highly consistent in their XRT light-curve morphology, in particular, the postplateau decay behavior corresponds to the GW-dominated spin-down $L_{\rm dip}\propto t^{-1}$. Based on Equations (\ref{E_rot dot}) and (\ref{I_formula}), one can derive $\Omega(t)$ evolution as a function of time during the GW-dominated spin-down in the frame of $R$ and $I$ evolution
\begin{eqnarray}
\Omega(t)=\Omega_0\left[1+\frac{t}{\tau_{\rm sd,gw}}\right]^{-\frac{1}{k+4}},
\label{Omega_GW}
\end{eqnarray}
hence, the GW frequency $f=\Omega/\pi$ can be expressed as
\begin{eqnarray}
f(t)=f_0\left[1+\frac{t}{\tau_{\rm sd,gw}}\right]^{-\frac{1}{k+4}},
\label{f_GW}
\end{eqnarray}
where $f_0$ is initial GW frequency. If most of the rotation energy is released via GW radiation with frequency $f$, the GW strain for a rotating neutron star at distance $D_L$ can be expressed as
\begin{eqnarray}
h(t)&=&\frac{4GI\epsilon}{D_Lc^4}\Omega(t)^2 \nonumber \\
&=&\frac{4GI\epsilon}{D_Lc^4}\Omega_0^2\left[1+\frac{t}{\tau_{\rm sd,gw}}\right]^{-\frac{2}{k+4}},
\label{h_GW}
\end{eqnarray}
The signal-to-noise ratio (SNR) of optimal matched filter can be expressed as \citep{Corsi2009}
\begin{eqnarray}
S/N&=&\int^{+\infty}_{0}\frac{f^2 h^2(t)(dt/df)}{f S_h(f)} d(\ln f) \nonumber \\ 
&=&\int^{+\infty}_{0} \left(\frac{h_c}{h_{\rm rms}}\right)^{2} d(\ln f),
\label{GW_SNR}
\end{eqnarray}
where $h_{\rm rms}=\sqrt{f S_h(f)}$ is the detector noise curve, $S_h(f)$ is the noise power spectral density of the detector, $h_c=f h(t)\sqrt{dt/df}$ is the characteristic strain amplitude of GW signal. Based on Equation (\ref{h_GW}), the characteristic strain amplitude of GW signal from a rotating NS with $R$ and $I$ evolution can be estimated as \citep{Corsi2009,Hild2011,Lasky2016,Lv2017,Abbott2020a}
\begin{eqnarray}
h_{\rm c}& = & f h(t)\sqrt{\frac{dt}{df}}=\frac{1}{D_{\rm L}}\sqrt{\frac{5GIf}{2c^{3}}} \nonumber \\
&=& \frac{1}{D_{\rm L}}\sqrt{\frac{5GIf_{0}}{2c^{3}}}(1+\frac{t}{\tau_{\rm sd,gw}})^{-\frac{1}{2(k+4)}}
\label{GWsignal}
\end{eqnarray}
Following Equation (\ref{GWsignal}), we know that the characteristic amplitude of GW signal depends on the magnetar angular frequency of the evolution with time and the distance to the source. Based on the frequency evolution form of the rotating NS with $R$ and $I$ evolutionary effects in Equation (\ref{f_GW}), one can derive the temporal behavior of GW characteristic strain $h_{\rm c}$ for different EoSs and baryonic masses with two $\eta_{\rm X}$ values. Figure \ref{fig:170607A_GWhc} exhibited a case study for the $h_{\rm c}$ evolution of GRB 170607A with $z=0.557$ by considering the $R$ and $I$ evolution in different EoSs and baryonic masses with two $\eta_{\rm X}$ values scenarios. Apparently, we can find that the GW signal intensity of the newborn magnetar at $t\sim 0$ is the strongest, and the initial $h_{\rm c}$ at different EoSs, baryonic masses and radiation efficiencies are comparable in order of magnitude. We therefore adopted $M_{b}=2.0M_{\odot}$ and $\eta_{X}=0.5$ to test the GW detectability with aLIGO and ET for all GW-dominated GRBs in our sample. For $t\sim 0$, Equation (\ref{GWsignal}) can be approximate to
\begin{eqnarray}
h_{\rm c}&\approx& \frac{1}{D_{\rm L}}\sqrt{\frac{5GIf_{0}}{2c^{3}}} \nonumber \\
&\approx& 2.56\times 10^{-22} \biggl(\frac{I}{10^{45}~\rm g\,cm^{2}}\frac{f_{0}}{1~\rm
kHz}\biggr)^{\!1/2}\biggl(\frac{D_{\rm
L}}{100~\rm Mpc}\biggr)^{\!-1}~\rm.
\label{signal}
\end{eqnarray}
We calculated the GW amplitude ($h_{\rm c}$) for all GW-dominated GRBs in our sample (including 6 GRBs with redshift measurement and 12 GRBs without redshift measurement) from $f_0=120$ Hz to $f_0=1000$ Hz (advanced-LIGO detector sensitive band) by setting $M_{b}=2.0M_{\odot}$ and $\eta_{X}=0.5$. In Figure \ref{fig:GW_detectability(with z)}, we plotted the GW strain sensitivities for aLIGO and ET detectors, as well as the GW strain amplitude for these GRBs with a redshift measurement. Compared with the sensitivities for aLIGO and ET, it is clear that GW signals from these GRBs with a redshift measurement cannot reach the sensitivity threshold of the current aLIGO detector and two GRBs can reach the sensitivity threshold of the prospective ET detector, i.e. GRB 150323A with redshift $z=0.593$ and GRB 170607A with redshift $z=0.557$, corresponding to the luminosity distances $D_{\rm L}\sim3592~\rm Mpc$ and $D_{\rm L}\sim3331~\rm Mpc$, respectively. In Figure \ref{fig:GW_detectability(without z)}, we also plotted the GW strain amplitude for these GRBs without redshift measurement by adopting $z=1.0$ ($D_{\rm L}\sim6801~\rm Mpc$) and $z=0.1$ ($D_{\rm L}\sim477~\rm Mpc$). For the LGRBs without redshift measurements, assuming $z=1.0$, we can find that the GW signals from their remnants cannot reach the sensitivity threshold of the either current aLIGO or the future ET. Assuming $z=0.1$, we can find that the GW signals from their remnants cannot reach the sensitivity threshold of the current aLIGO but can reach the sensitivity threshold of the future ET. Furthermore, from Equation (\ref{GW_SNR}) and Equation (\ref{signal}), we tried to estimate the GW detection horizon of the aLIGO and ET detectors for the GW signal from the remnants of LGRBs. Utilizing the SNR threshold $\sim8$ for aLIGO, the SNR threshold $\sim5$ for ET, as well as the derived statistical values for the physical parameters of magnetars within our LGRB sample ($P_0\sim1$ ms, $I\sim2.5\times10^{45}~\rm g~cm^{-2}$), one can estimate the detection ranges, which are approximately $\sim40$ Mpc for the current aLIGO and $\sim1020$ Mpc for the prospective ET. If future observations allow aLIGO or ET to detect the GW signal from the remnants of LGRBs, it would not only offer the first definitive proof that a magnetar can serve as the central engine of GRBs but also play a crucial role in precisely constraining the NS EoS.

\section{Conclusion and Discussion}
The long-lasting X-ray plateau emission of LGRBs is believed to be from internal dissipation, which can be interpreted as energy injection from the newborn magnetar central engine \citep{Dai1998a,Dai1998b,Zhang2001,Troja2007,Rowlinson2014,Lv2014,Zou2019,Xie2022b}. In this paper, we systematically analyzed the \emph{Swift}/XRT light curves of LGRBs with X-ray plateau emission that were detected before 2023 December. Assuming that the X-ray plateau emission of these LGRBs derive from isotropic magnetic dipole wind energy injection from the newly born magnetar, we reestimated the physical parameters of magnetar and investigated possible relations among these parameters and their relations to the GRB jet and magnetar wind radiation by taking into account the $R$ and $I$ evolutionary effects. In order to precisely diagnose the properties of the nascent NS, we invoked four samples of EoSs with two bayonic masses $M_{b}=2.0M_{\odot}$, $2.5M_{\odot}$ and two X-ray radiation efficiencies $\eta_{X}=0.1$, $0.5$ to simultaneously constrain magnetar physical parameters. Our main results can be summarized as follows:

\begin{itemize}[leftmargin=*]
\item We found a sample of 105 LGRBs with X-ray plateau emission, followed by a decay segment with a decay index between -1 and -2, by systematically searching the current \emph{Swift}/XRT LGRB data, including 43 LGRBs with redshift information. Their joint BAT+XRT light curves can be separated into GRB jet and wind emission epochs by defining $t_{\rm jet}$ as the boundary. We performed a temporal and spectral analysis for the 105 LGRBs and obtained their $E_{\rm jet,iso}$, $E_{\rm jet}$, $E_{\rm wind}$, $t_{\rm jet}$, $t_{\rm wind}$, and $L_{\rm p}$ values.

\item We found that neglecting the evolutionary effects of $R/I$ can lead to systematic overestimation or underestimation of magnetar physical parameters. We averaged the results from the evolved $R/I$ scenarios and compared them with those in the constant $R/I$ scenarios. We found that the derived $B_p$, $P_0$ and $\epsilon$ in the constant $R/I$ cases are overestimated by $\bigtriangleup(\log B_p)\sim0.17~\rm G$, underestimated by $\bigtriangleup P_0\sim0.26~\rm ms$, and essentially equivalent with $\bigtriangleup(\log \epsilon)\sim0.09$. These differences correspond to factors of 1.5, 0.7, and 1.2, respectively, compared to the constant $R/I$ case. Furthermore, for a given EoS, one can find that, with higher X-ray radiation efficiency and baryonic mass, the derived $B_p$, $P_0$, and $\epsilon$ will become larger. For the given baryonic mass and X-ray radiation efficiency, different EoSs could also overall change the derived magnetar physical parameters, but the constraints on magnetar physical parameter magnitude seem to be independent of the EoS stiffness.

\item We found that there are universal correlations among magnetar physical parameters as well as between the GRB jet and magnetar wind radiation based on our curated GRB sample. Within the systematic error range, the correlations of magnetar physical parameters can be approximately expressed as $\epsilon\propto P_0^{1.57\pm0.22}$, $\epsilon\propto B_p^{0.97\pm13}$, $B_p\propto P_0^{1.30\pm0.16}$, with the 1$\sigma$ deviation included. Moreover, the correlations between the GRB jet and magnetar wind can be approximately described as $E_{\rm wind}\propto E_{\rm jet,iso}^{0.83\pm0.07}(E_{\rm jet}^{0.76\pm0.06})$, $P_0\propto E_{\rm jet,iso}^{-0.29\pm0.03}(E_{\rm jet}^{-0.26\pm0.02})$, $B_p\propto E_{\rm jet,iso}^{-0.58\pm0.06}(E_{\rm jet}^{-0.55\pm0.05})$ and $\epsilon\propto E_{\rm jet,iso}^{-0.55\pm0.07}(E_{\rm jet}^{-0.52\pm0.06})$ for most of our selected EoSs. The tight correlations suggested that a newly-formed magnetar with the faster $P_0$, lower $B_p$, as well as lower $\epsilon$ is more inclined to power a more energetic GRB jet, and the ellipticity deformation and initial spin period of the newborn magnetar are likely to originate from the magnetically induced distortion mechanism and correspond to the equilibrium spin period as a result of interaction between the magnetar and its accretion disk, respectively.

\item We found that the ideal detection rate of magnetar dipole radiation with \emph{EP}/WXT is only $\sim$30\% (13 GRBs), the prompt follow-up observations with \emph{EP}/FXT may increase the percentage to $\sim$81\% (35 GRBs), and with \emph{SVOM}/MXT, it is $\sim$40\% (17 GRBs) based on 19 years of \emph{Swift}/XRT observations (43 GRBs with plateau emission and redshift information in our sample) by adopting the sensitivity of $F_{\rm th}^{\rm WXT}=1\times10^{-10}~\rm erg~s^{-1}~cm^{-2}$, $F_{\rm th}^{\rm FXT}=3\times10^{-11}~\rm erg~s^{-1}~cm^{-2}$, $F_{\rm th}^{\rm MXT}=7\times10^{-11}~\rm erg~s^{-1}~cm^{-2}$, respectively. 

\item We found that the GW signals from the remnants of those GW-dominated GRBs with redshift measurements cannot reach the sensitivity threshold of the current aLIGO detector, but only two GRBs can reach the sensitivity threshold of the prospective ET detector, i.e. GRB 150323A and GRB 170607A. For the LGRBs without redshift measurements, assuming $z=1.0$, the GW signals from their remnants cannot reach the sensitivity threshold of the either current aLIGO or the prospective ET; assuming $z = 0.1$, the GW signals from their remnants cannot reach the sensitivity threshold of the current aLIGO but can reach the sensitivity threshold of the prospective ET. Utilizing the SNR threshold $\sim8$ for aLIGO, SNR threshold $\sim5$ for ET, as well as the derived statistical values for the physical parameters of magnetars within our LGRB sample ($P_0\sim1$ ms, $I\sim2.5\times10^{45}~\rm g~cm^{-2}$), one can estimate the detection horizons as $\sim40$ Mpc for the current aLIGO and $\sim1020$ Mpc for the prospective ET.
\end{itemize}

In the end, we would like to point out several caveats of our approach. Our results are based on a sample of LGRBs with X-ray plateau emission, followed by a decay segment with a decay index between -1 and -2, in their XRT light curves. This sample could suffer observational biases from the XRT fluctuation thresholds, and the magnetic dipole emission may be covered by bright jet afterglow emission, leading us to collect an incomplete sample. In other words, the currently available data for LGRBs with plateau emission are still limited. Statistical errors may potentially bias our results. The adopted method for selecting the LGRBs with X-ray plateau emission sample may also involve statistical uncertainties. The specific numbers may differ, but the overall conclusions drawn in this paper would remain valid. In addition, the physical conditions for a nascent NS would be very complicated, and some conditions may significantly alter the dipole radiation light curve, so that the $R/I$ evolution effect we discussed here would be reduced or even completely suppressed. For instance, it has been proposed that the evolution of the inclination angle between the rotation and magnetic axes of the NS could markedly revise the X-ray emission \citep{Cikintoglu2020}. Moreover, the nascent NS is likely to undergo free precession in the early stages of its lifetime when the rotation and magnetic axes of the system are not orthogonal to each other, which would lead to systematic fluctuations in the X-ray light curve \citep{Suvorov2020,Suvorov2021,Zou2021b,Zou2022,Zou2024,Zhang2024}. The impact of these conditions on our constraints of magnetar physical parameters remain to be further investigated in the future.

Furthermore, the differential rotation and monopolar wind emission may also exist at the early stage of the newborn magnetar \citep{Duncan1992,Thompson1993,Thompson2004}. Theoretically, the differential rotation and monopolar wind emission occur only in the early stage of magnetic field amplification and significant mass loss at the magnetar birth. The transition from differential to rigid rotation for a GRB newborn magnetar with a strong magnetic field ($B_{p}\sim10^{14}-10^{15}$ G) is often on the order of millisecond to second \citep{Dai2006}, and the timescale of existence of the monopole wind emission depends mainly on the early evolutionary stage of the nascent magnetar, i.e., the Kelvin-Helmholtz cooling timescale, which is typically tens of seconds (Thompson et al. 2004). However, the plateau emission of all GRBs in our sample starts from hundreds of seconds, and it is clear that such a timescale of plateau emission is in the evolutionary phase following the differential rotation and monopolar wind emission of the newborn magnetar, and not at the very early stage (millisecond to second scale) when the magnetic field amplification is underway. Moreover, the slopes of the steeper decay segment following the plateau segment are in the range of -1 to -2 for our GRB sample, which are typical decay slopes when the newborn magnetar spin-down by losing its rotational energy through EM dipole radiation or GW radiation. It is therefore a reasonable approximation to consider only the rigid rotation and dipole spin-down of the nascent magnetar at this later plateau emission stage. Of course, the $P_0$ that we have constrained using the X-ray plateau data may not be the true initial spin period of the newborn magnetar, which may have a faster rotation rate due to the presence of differential rotation and monopole wind radiation in the early stage of magnetar birth. We hope that the GW radiation associated with the X-ray plateau emission can be detected by Advanced LIGO and ET in the future, which could not only offer the first smoking gun that a protomagnetar can serve as the central engine of GRBs, but could also play a crucial role in precisely constraining the true $P_0$.

\begin{acknowledgements}
We thank the anonymous referee for a very thorough analysis of the original version and extremely helpful comments that have helped us to significantly improve the presentation of the paper. We thank Lang Xie, Yun Wang, and Jun-Xiang Huang for the helpful discussions. This work made use of data supplied by the UK \emph{Swift} Science Data Centre at the University of Leicester. L.L. is supported by the China Postdoctoral Science Foundation (grant No. GZB20230765). S.A. has received support from the Villum Foundation (Project No.~13164, PI: I. Tamborra). This work is supported by the National Natural Science Foundation of China (Projects 12373040, 12021003, 12303050, 12494573), the National SKA Program of China (2022SKA0130100) from the Fundamental Research Funds for the Central Universities, the Strategic Priority Research Program of the Chinese Academy of Sciences, Grant No.XDB0550401.

\end{acknowledgements}

\begin{figure*}
\centering
\includegraphics  [angle=0,scale=0.2]   {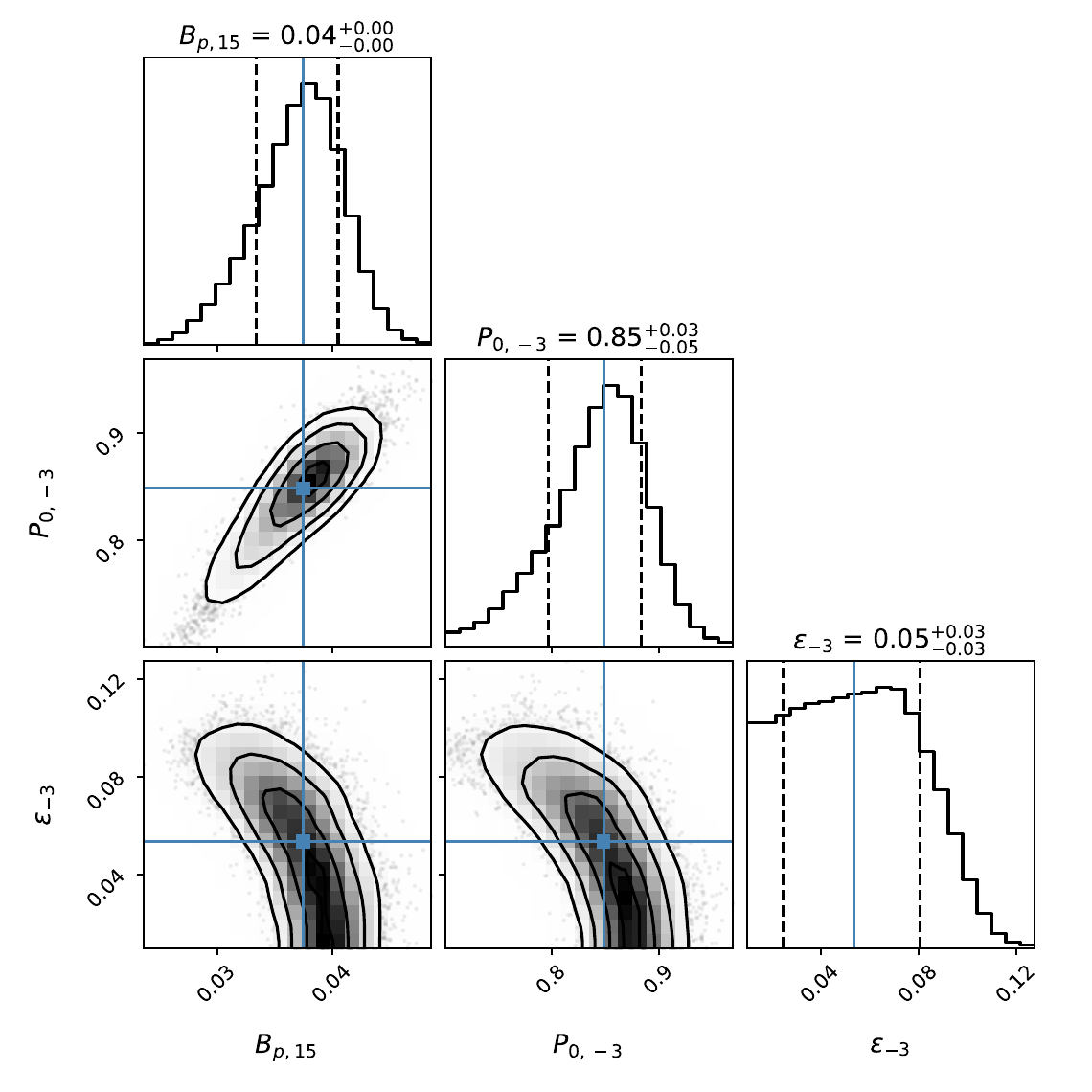}
\includegraphics  [angle=0,scale=0.2]   {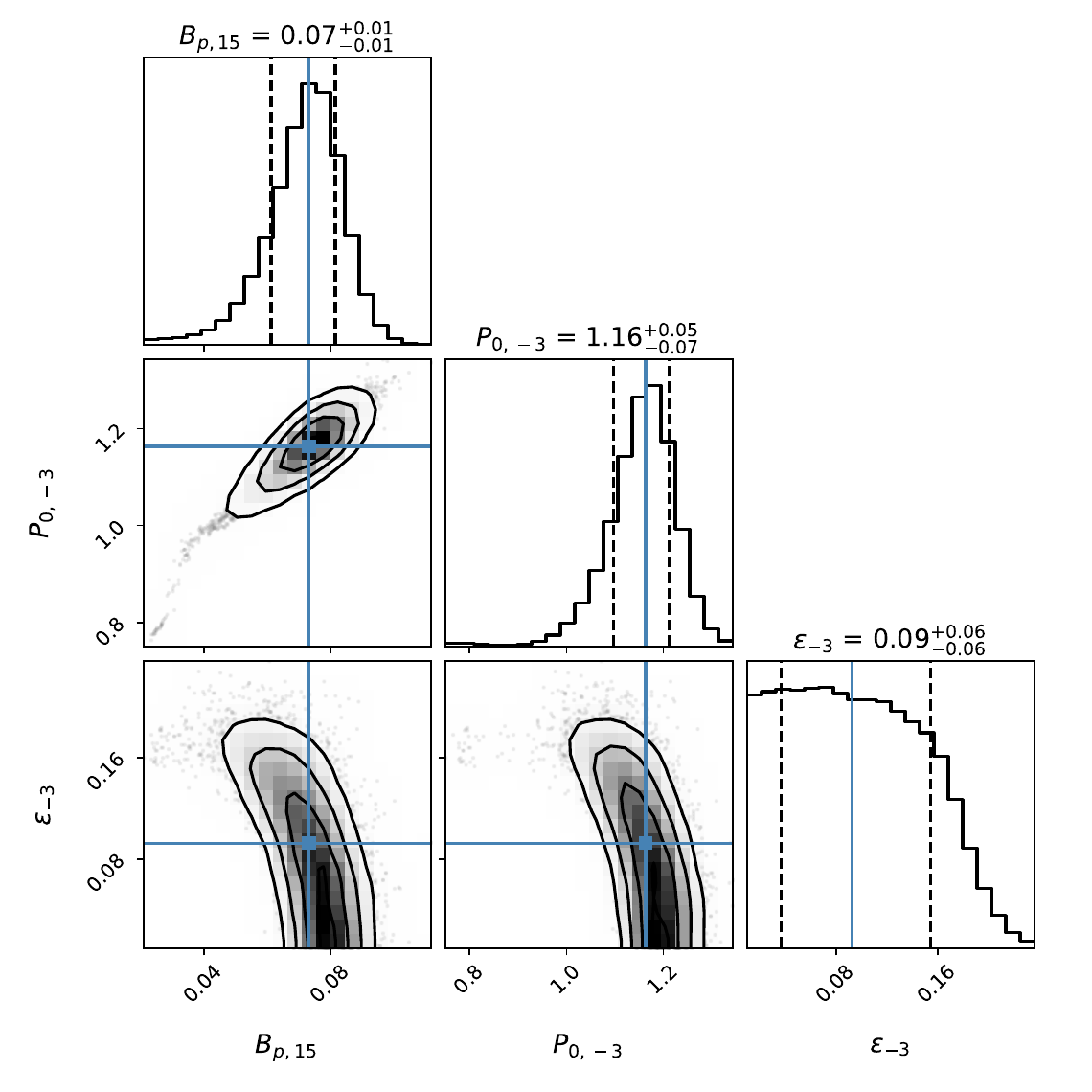}
\includegraphics  [angle=0,scale=0.2]   {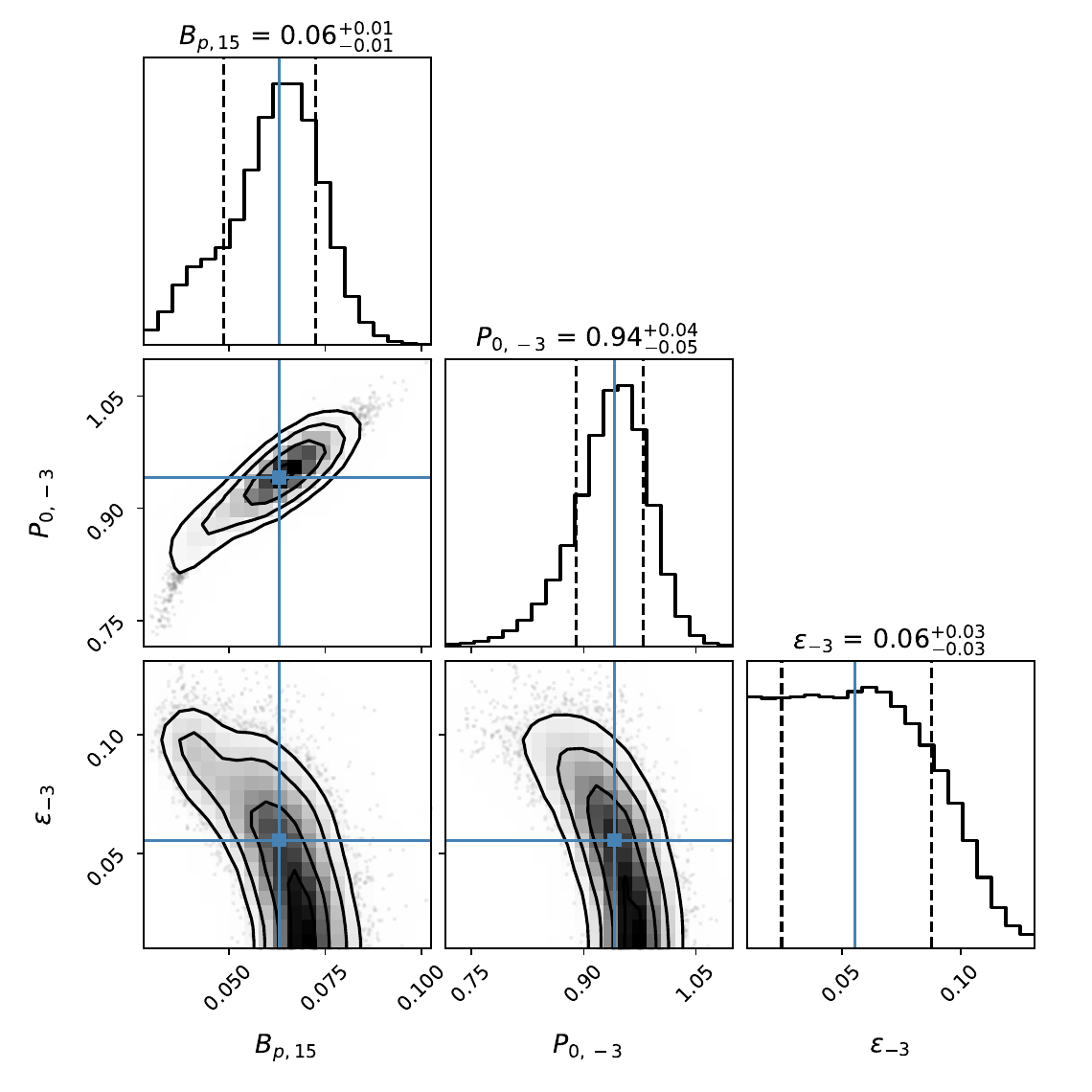} 
\includegraphics  [angle=0,scale=0.2]   {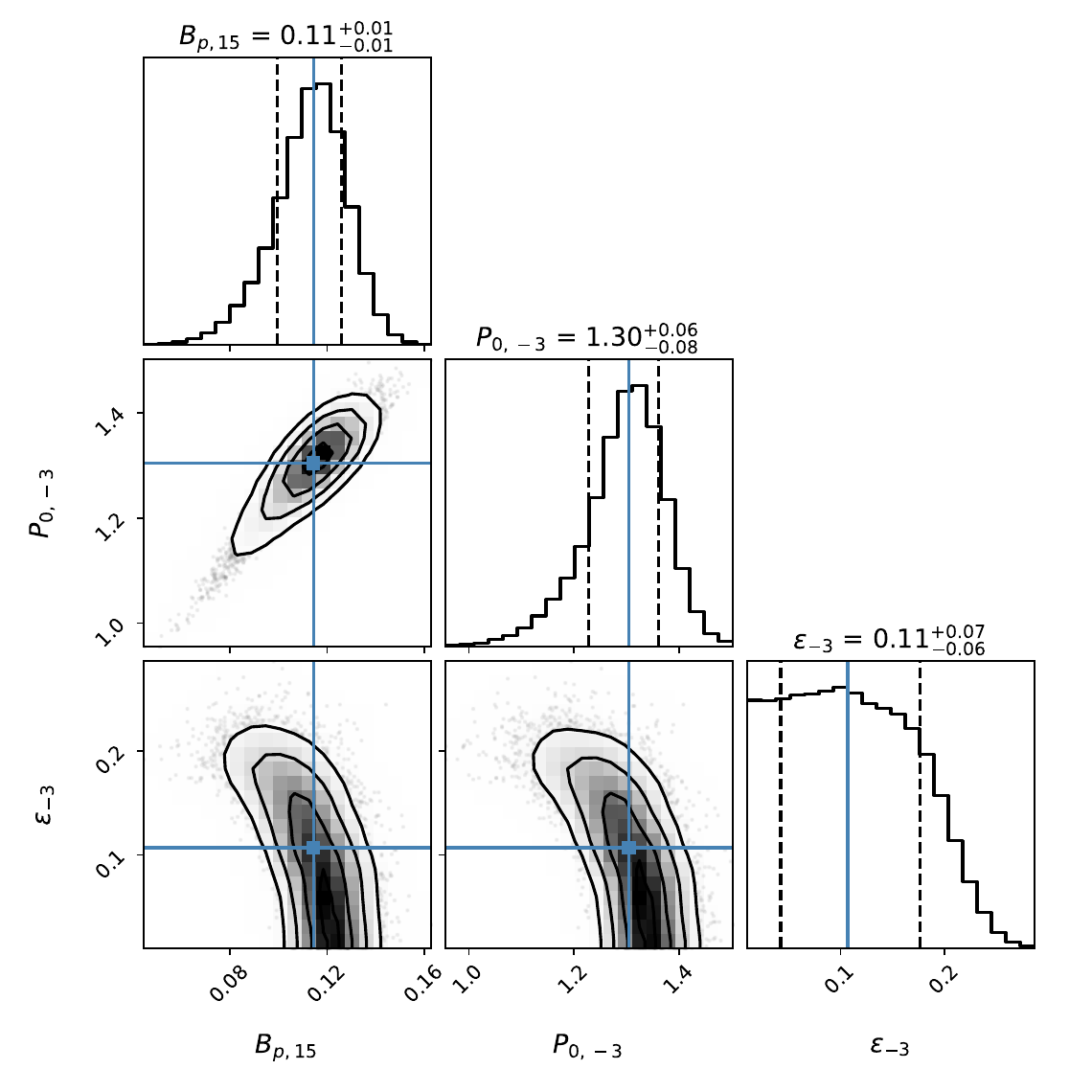} \\
\includegraphics  [angle=0,scale=0.24]   {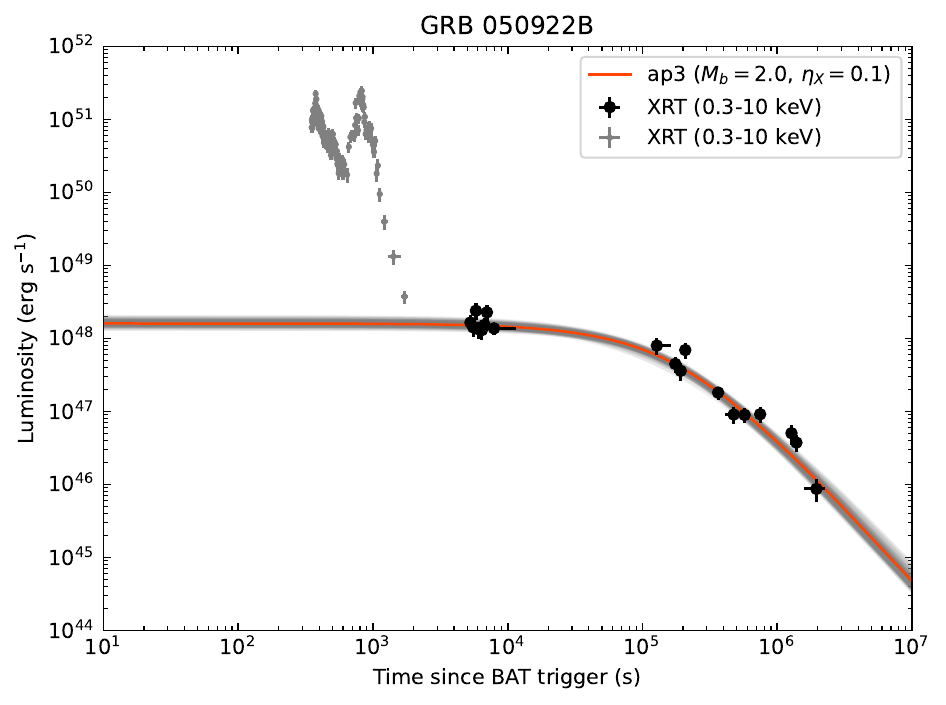}
\includegraphics  [angle=0,scale=0.24]   {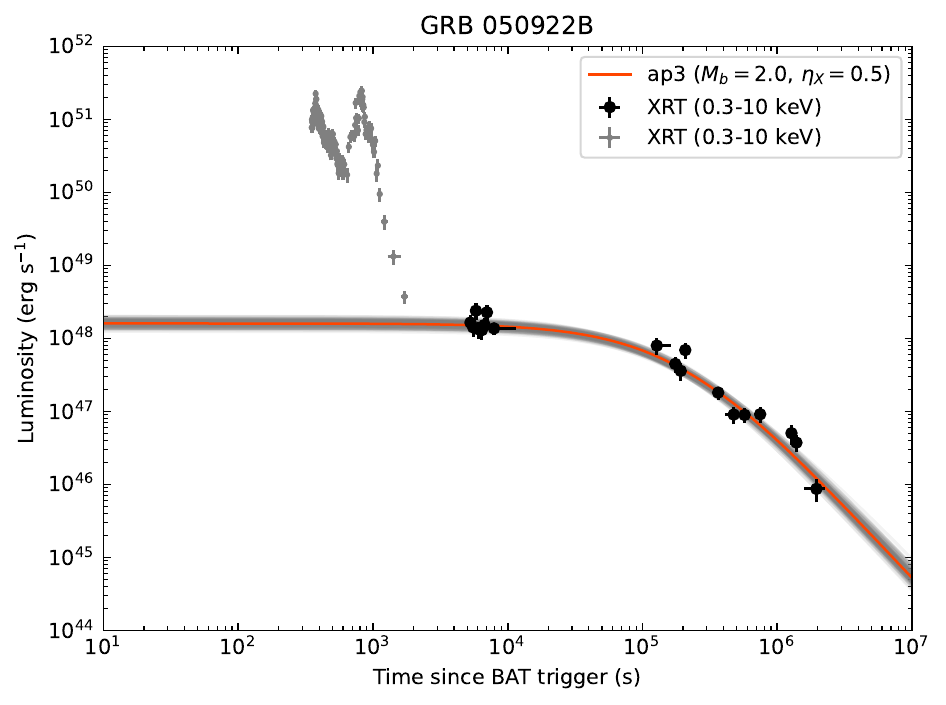}
\includegraphics  [angle=0,scale=0.24]   {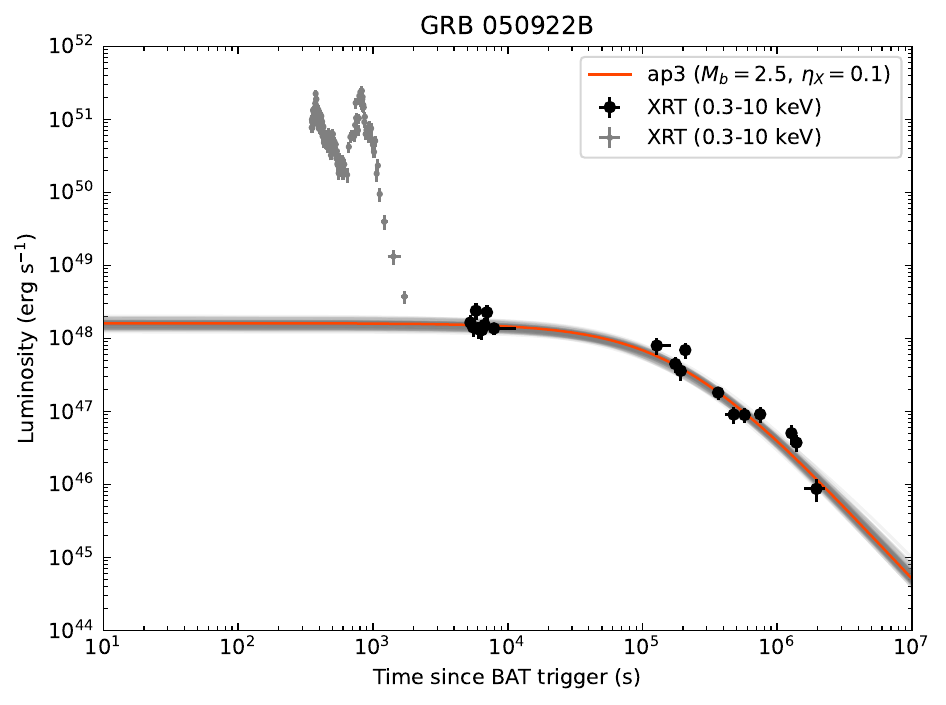}
\includegraphics  [angle=0,scale=0.24]   {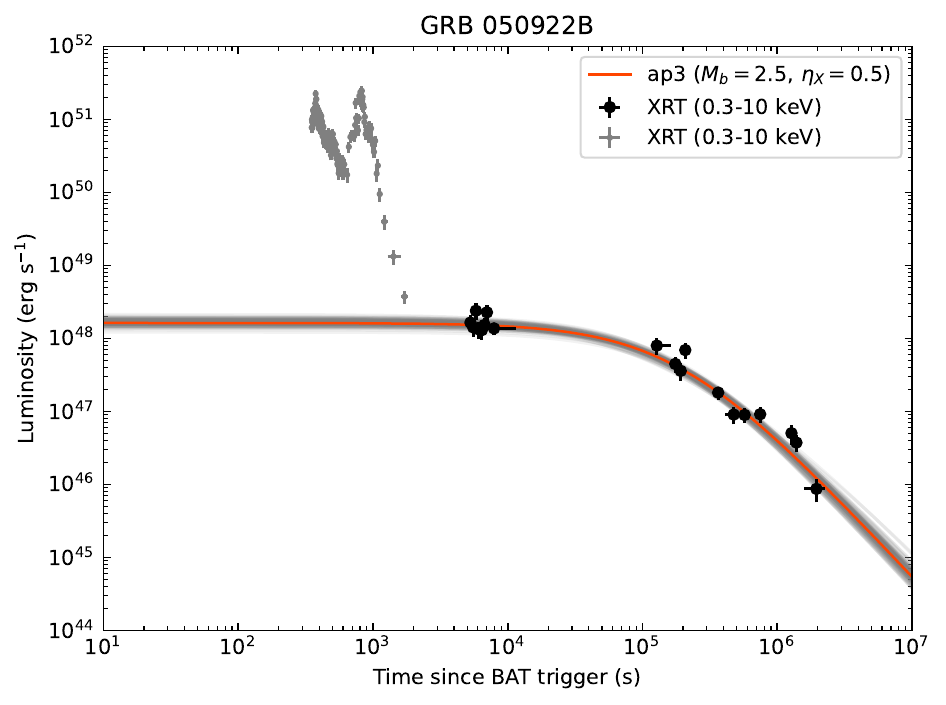} \\
\includegraphics  [angle=0,scale=0.2]   {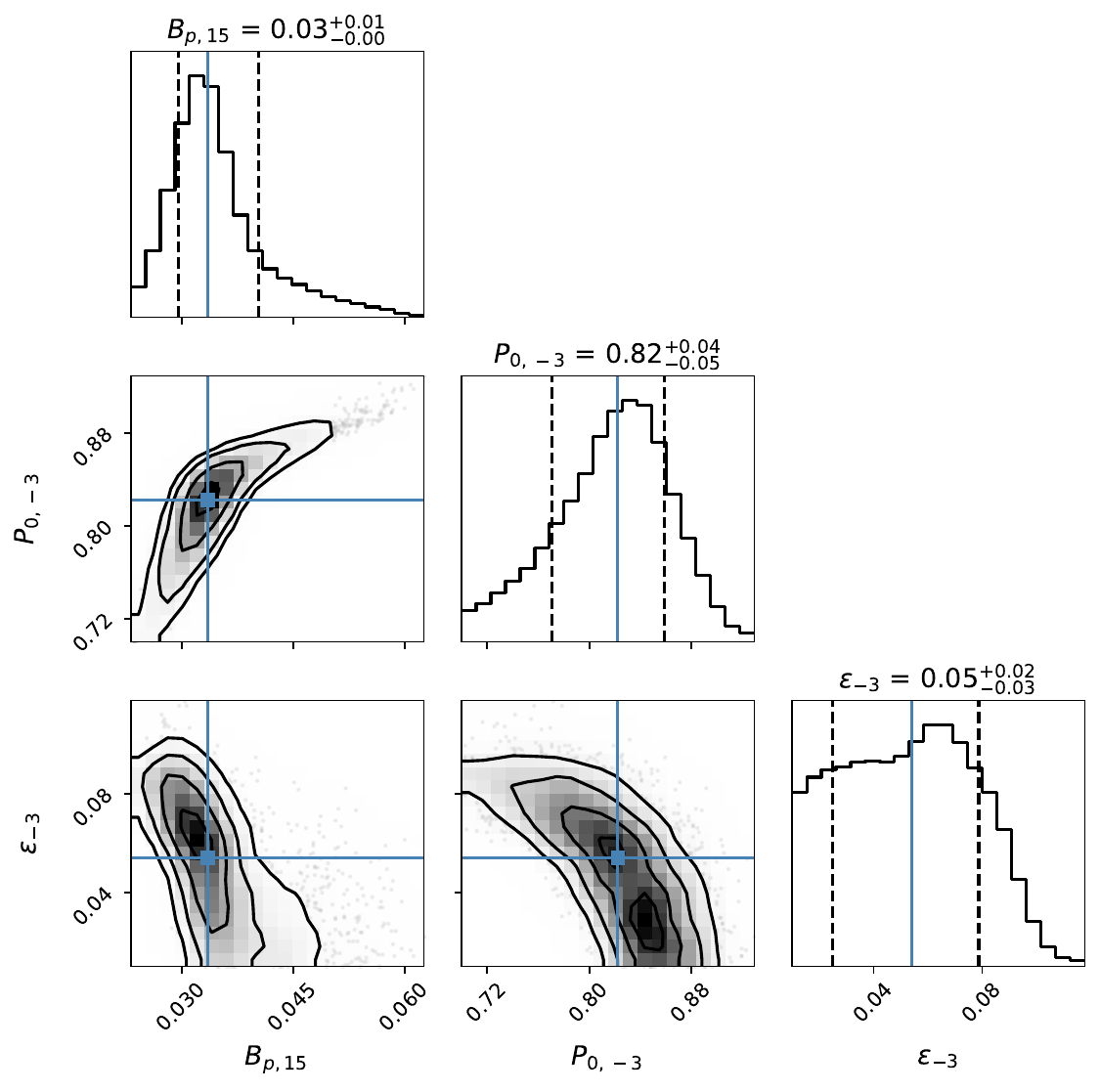}
\includegraphics  [angle=0,scale=0.2]   {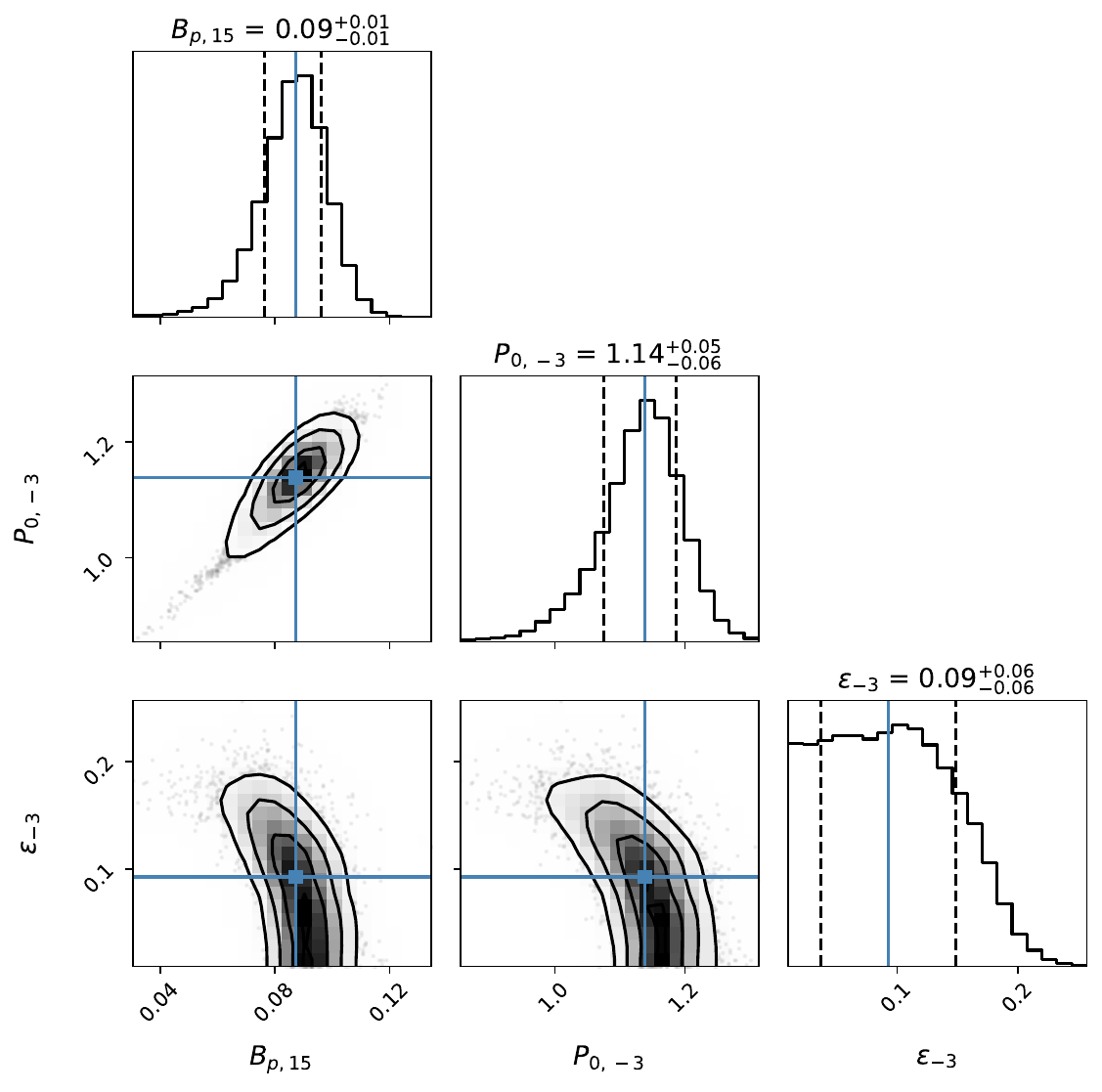}
\includegraphics  [angle=0,scale=0.2]   {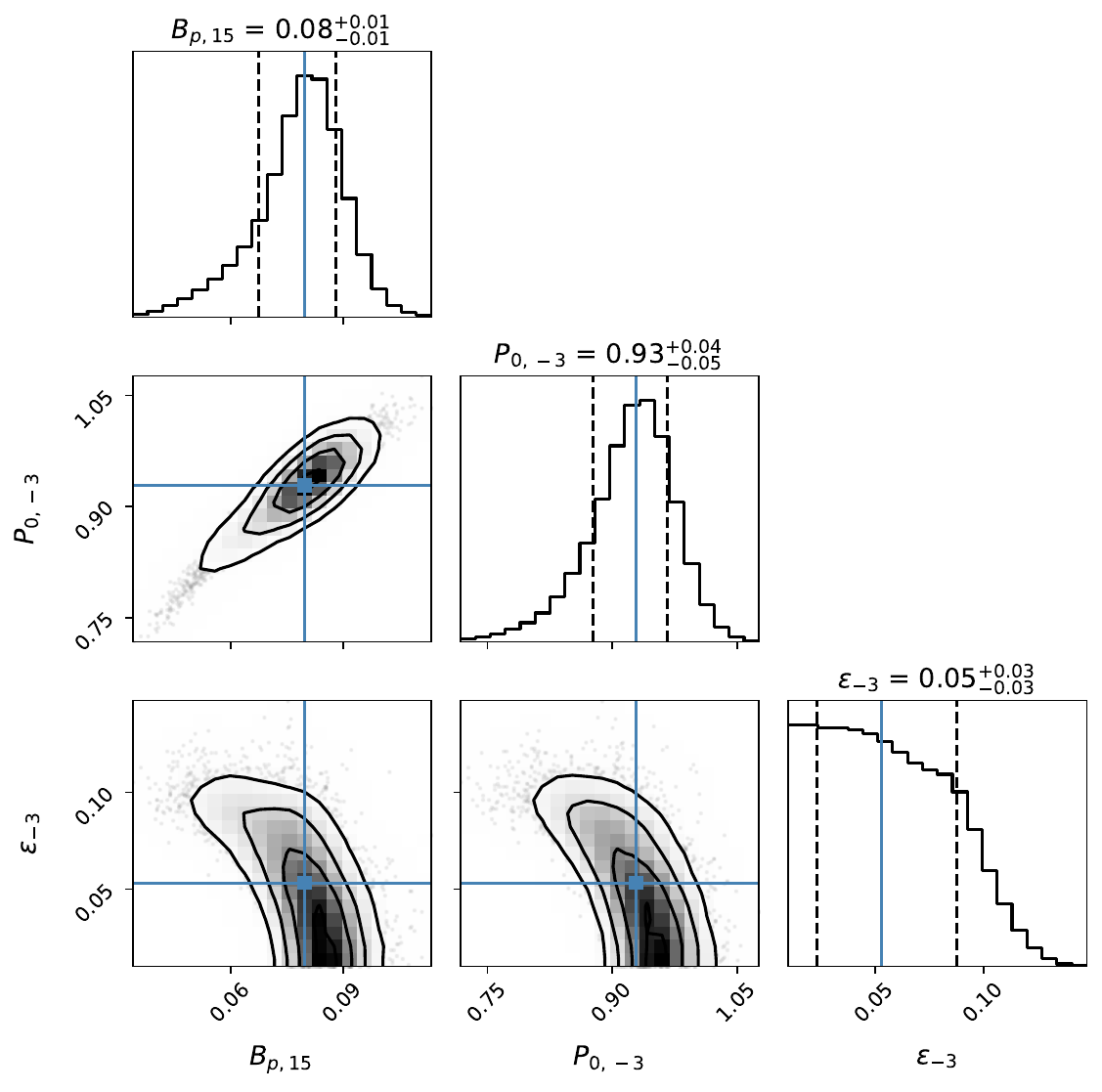}
\includegraphics  [angle=0,scale=0.2]   {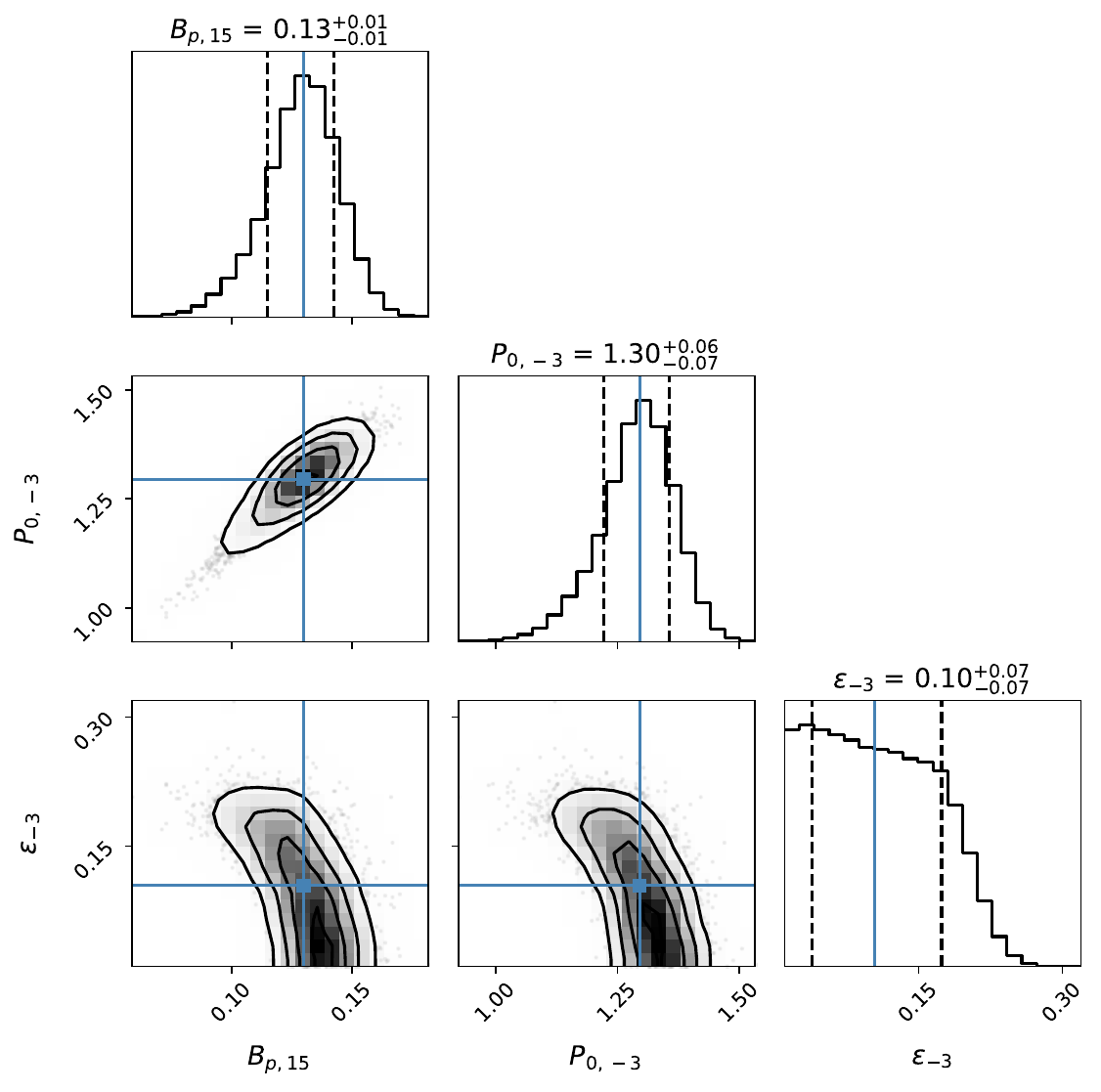} \\
\includegraphics  [angle=0,scale=0.24]   {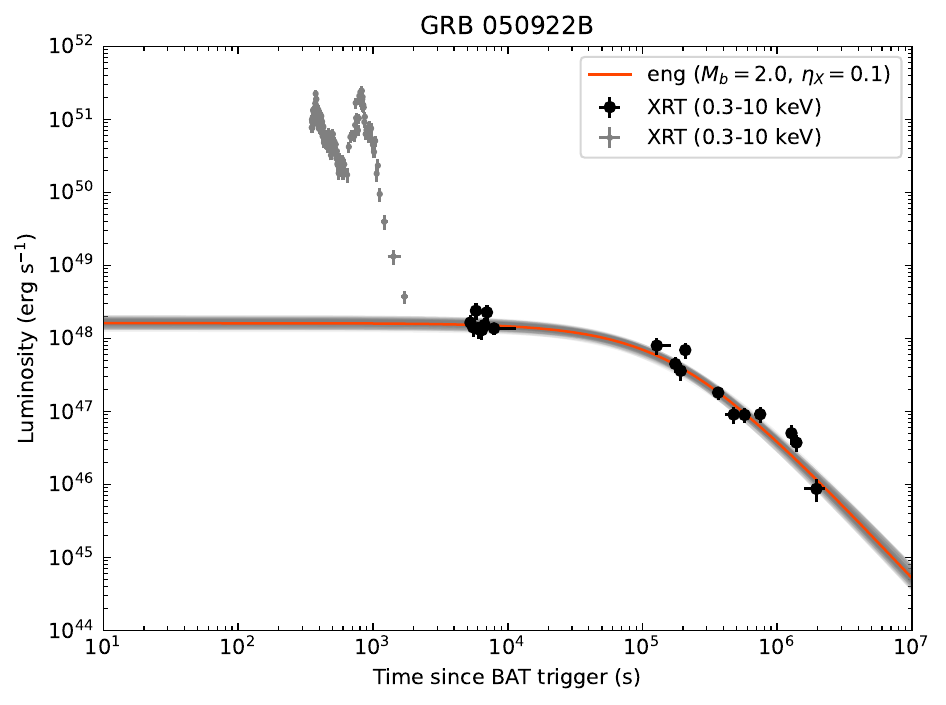}
\includegraphics  [angle=0,scale=0.24]   {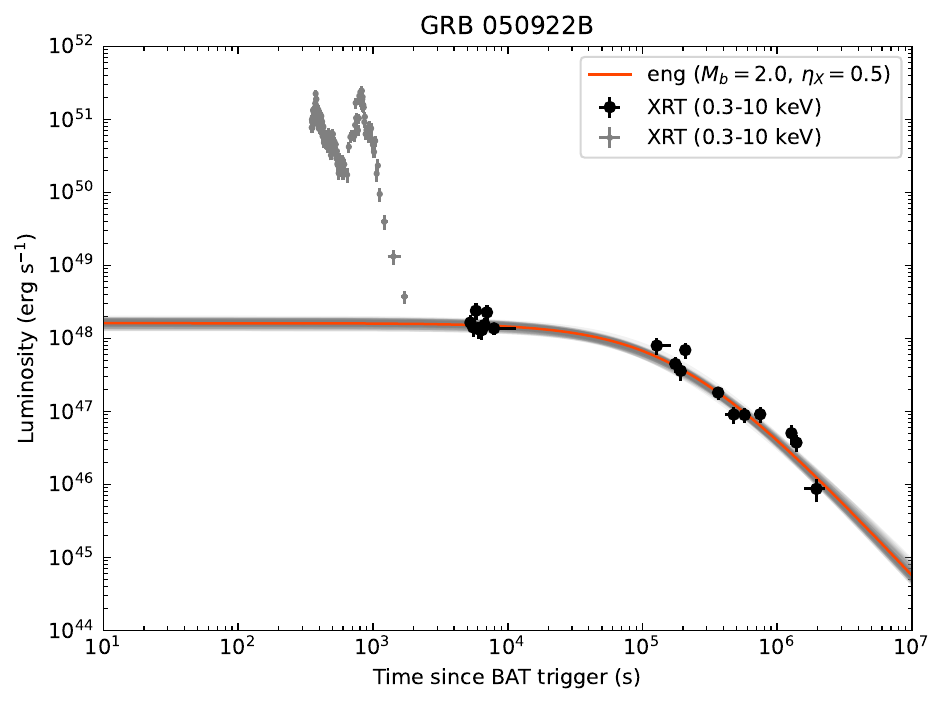}
\includegraphics  [angle=0,scale=0.24]   {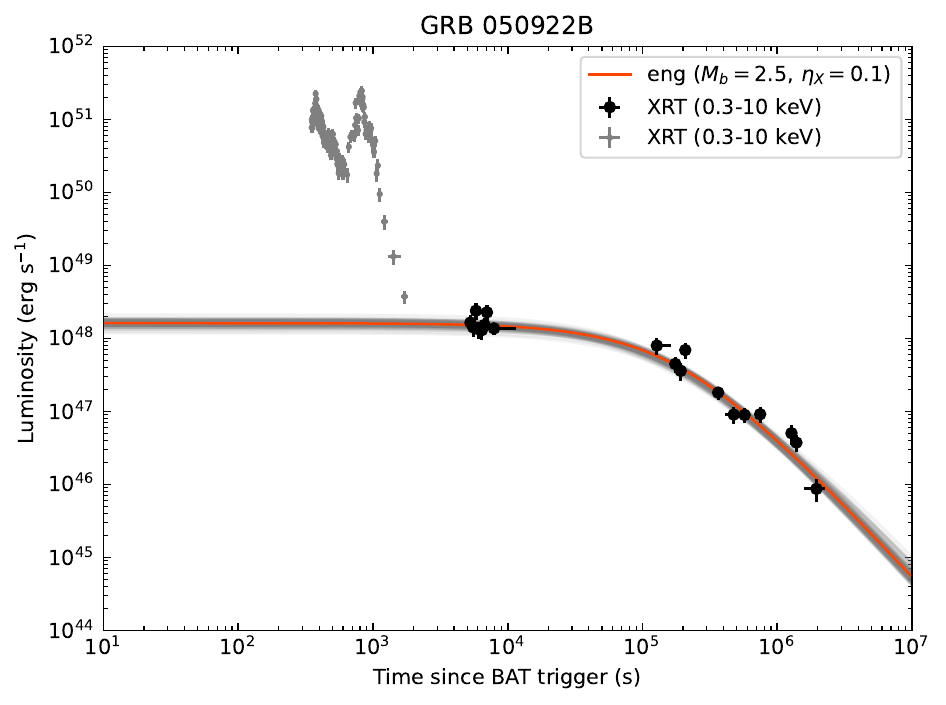}
\includegraphics  [angle=0,scale=0.24]   {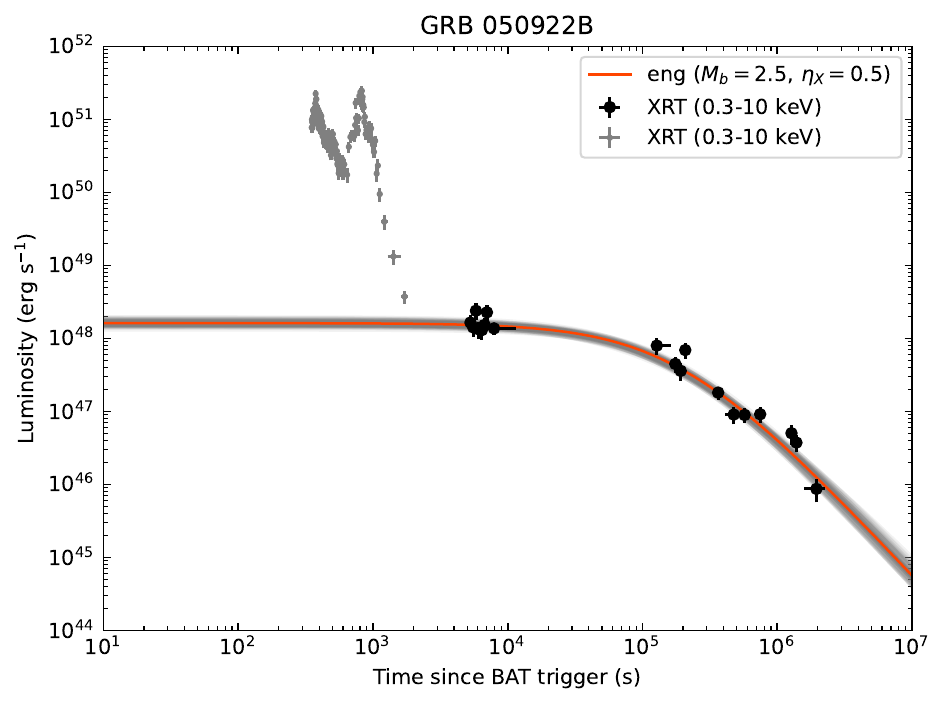} \\
\includegraphics  [angle=0,scale=0.2]   {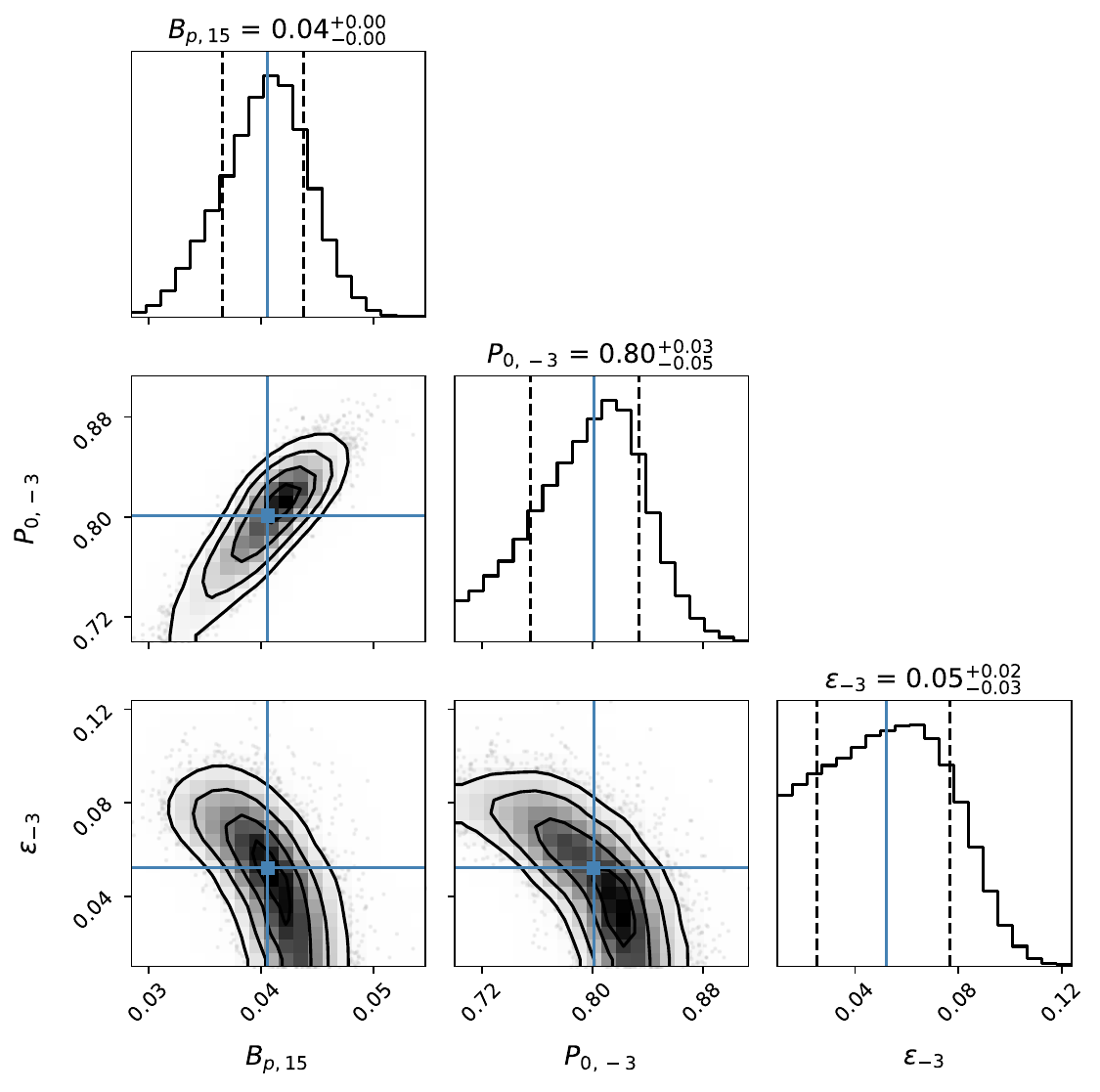}
\includegraphics  [angle=0,scale=0.2]   {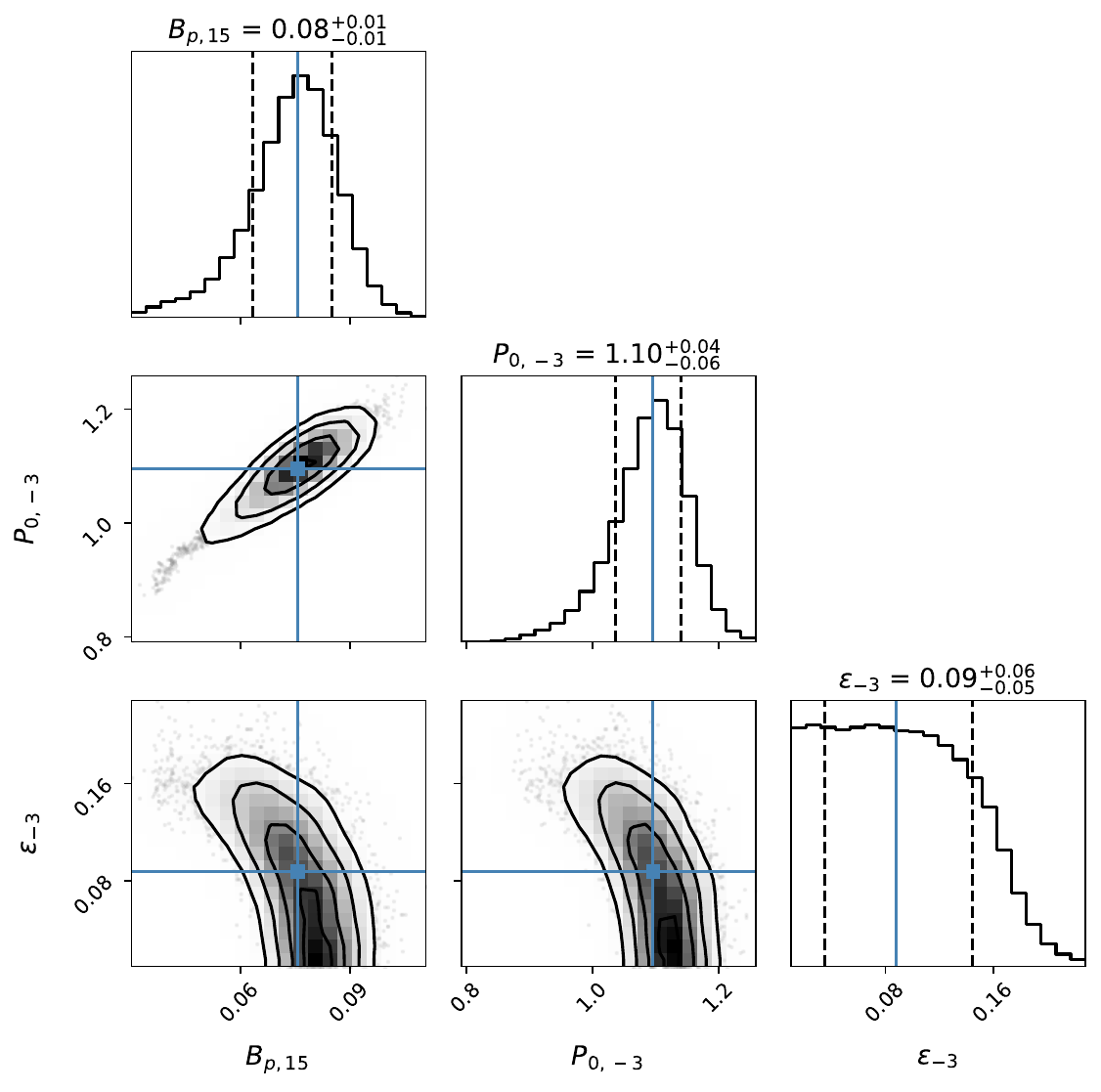}
\includegraphics  [angle=0,scale=0.2]   {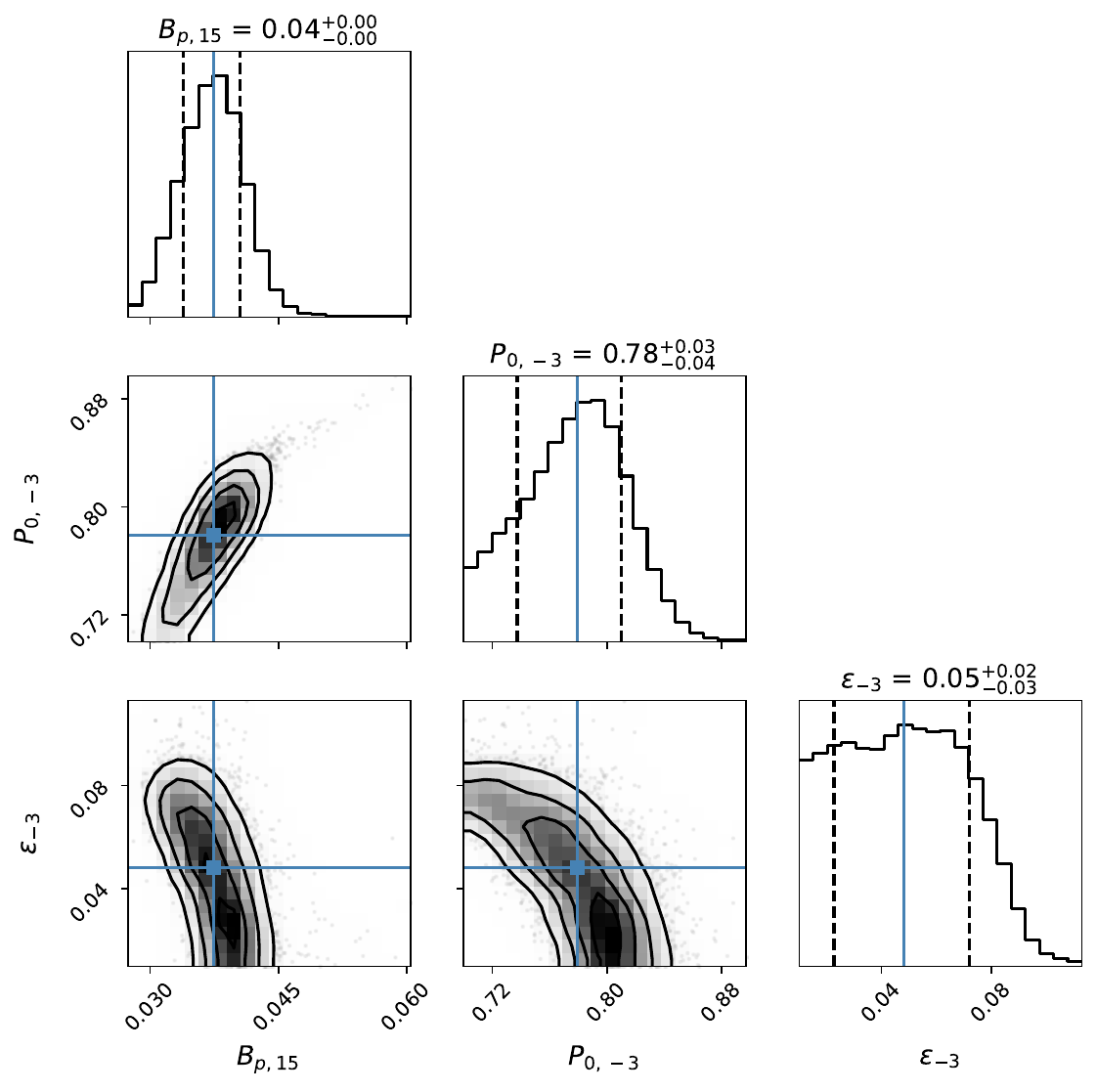}
\includegraphics  [angle=0,scale=0.2]   {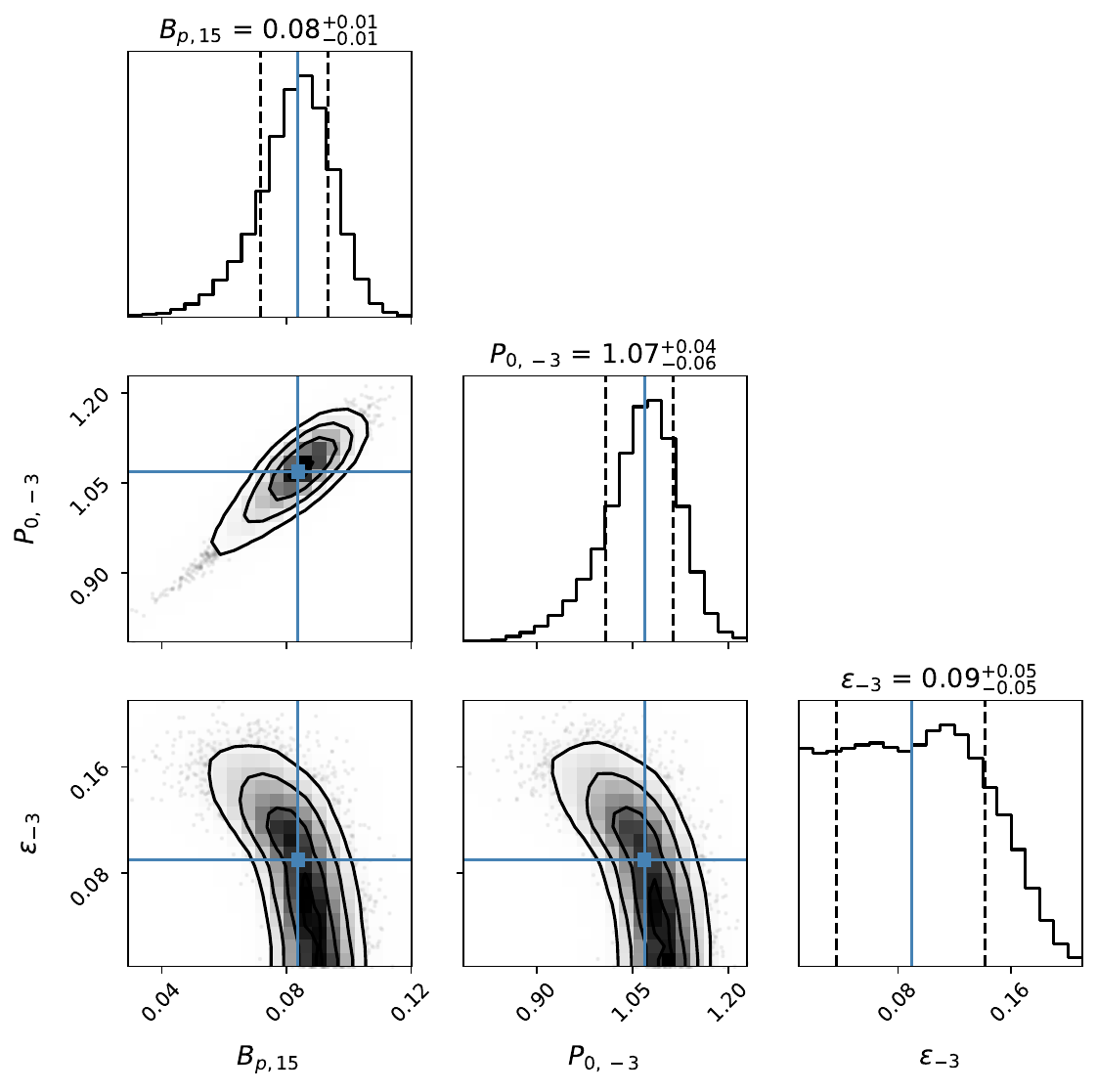} \\
\includegraphics  [angle=0,scale=0.24]   {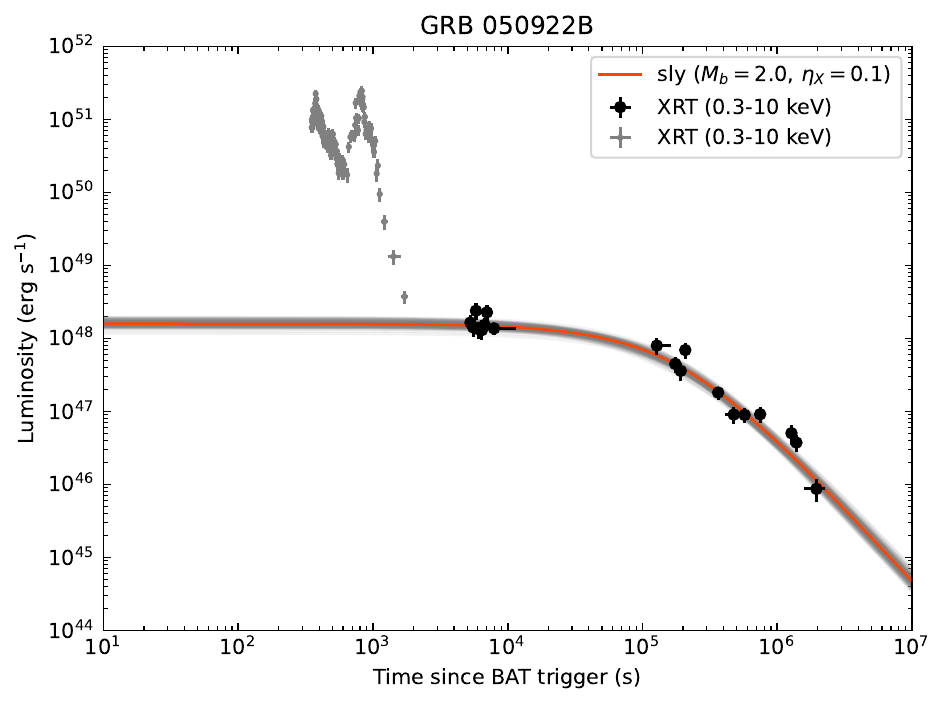}
\includegraphics  [angle=0,scale=0.24]   {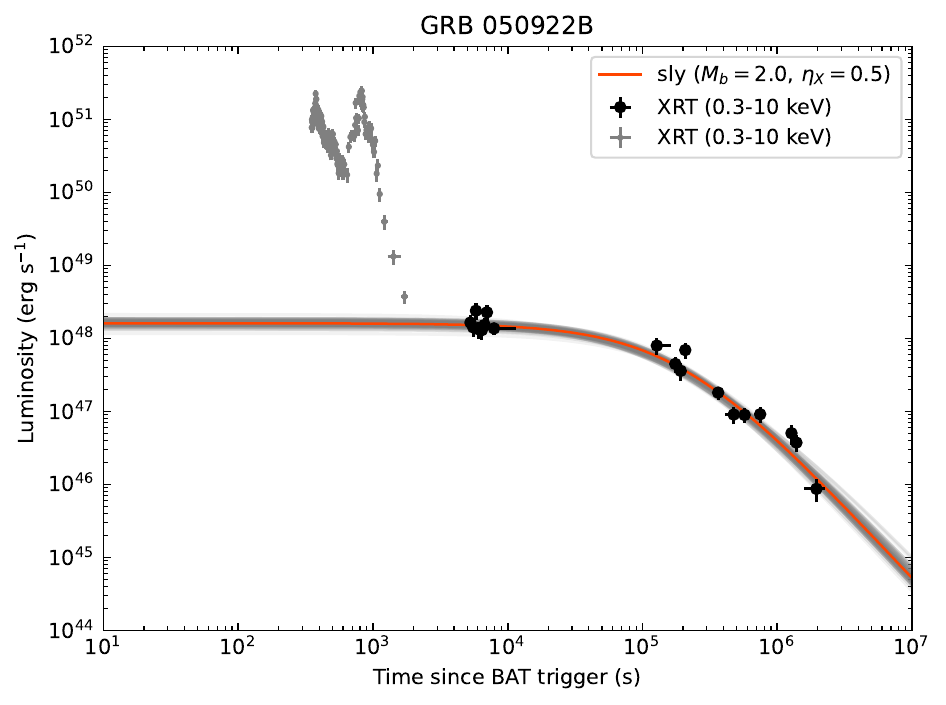}
\includegraphics  [angle=0,scale=0.24]   {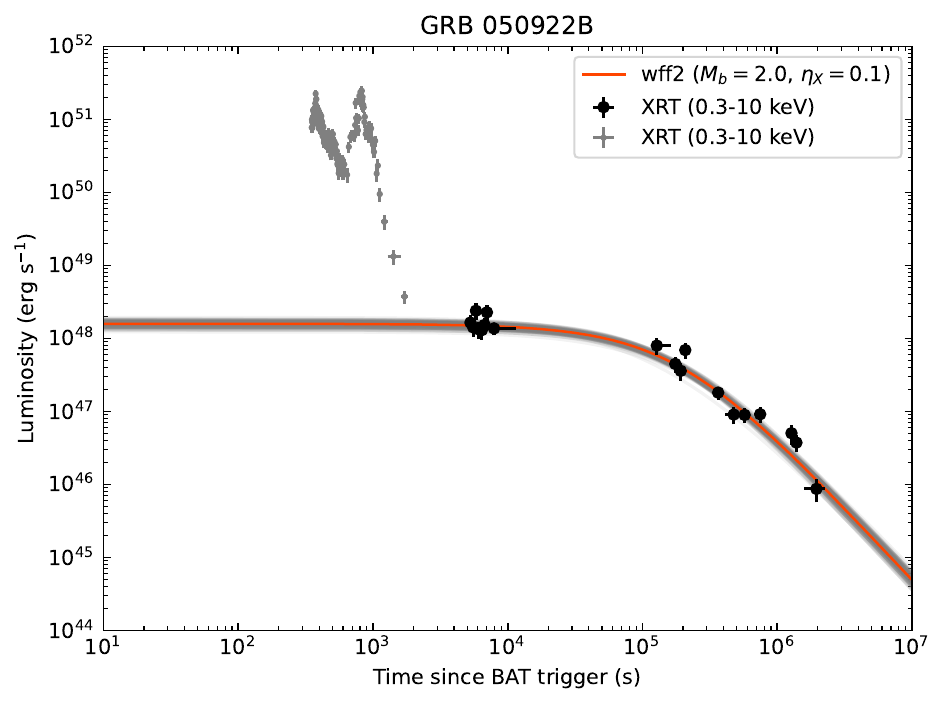}
\includegraphics  [angle=0,scale=0.24]   {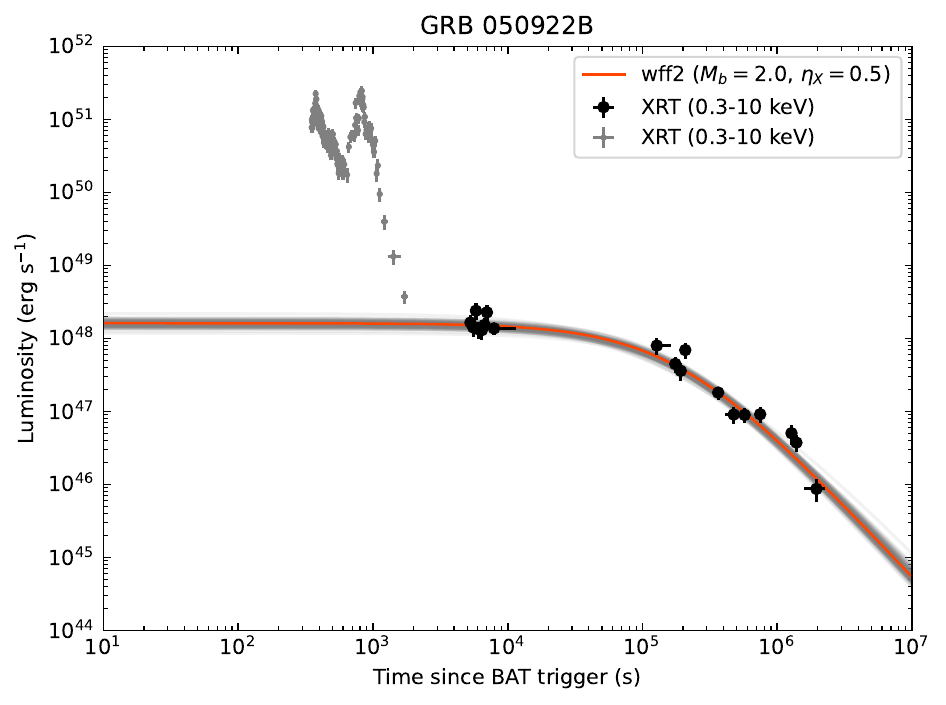} \\
\includegraphics  [angle=0,scale=0.2]   {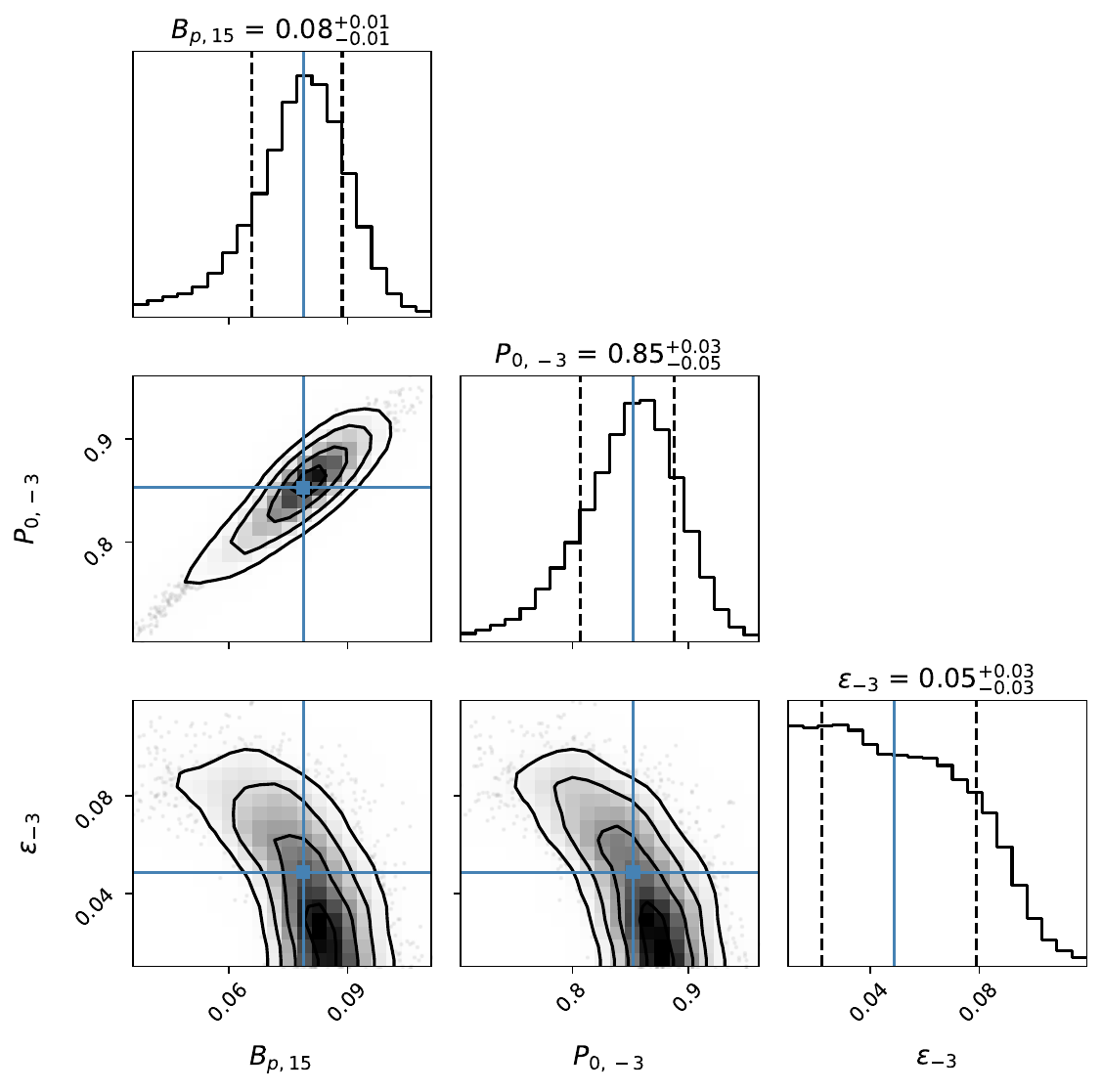}
\includegraphics  [angle=0,scale=0.24]   {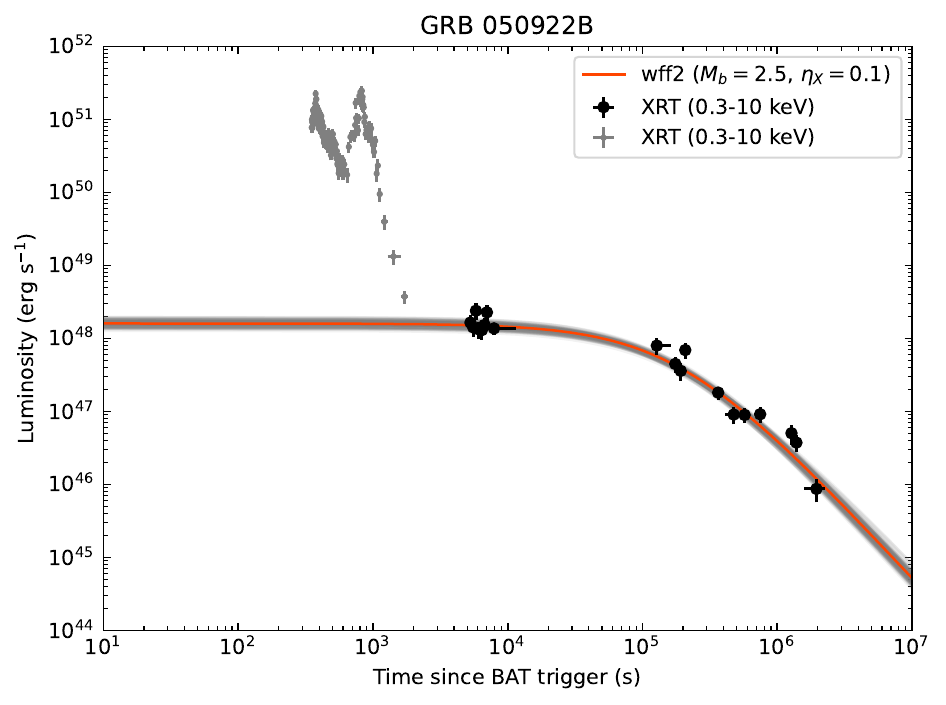}
\includegraphics  [angle=0,scale=0.2]   {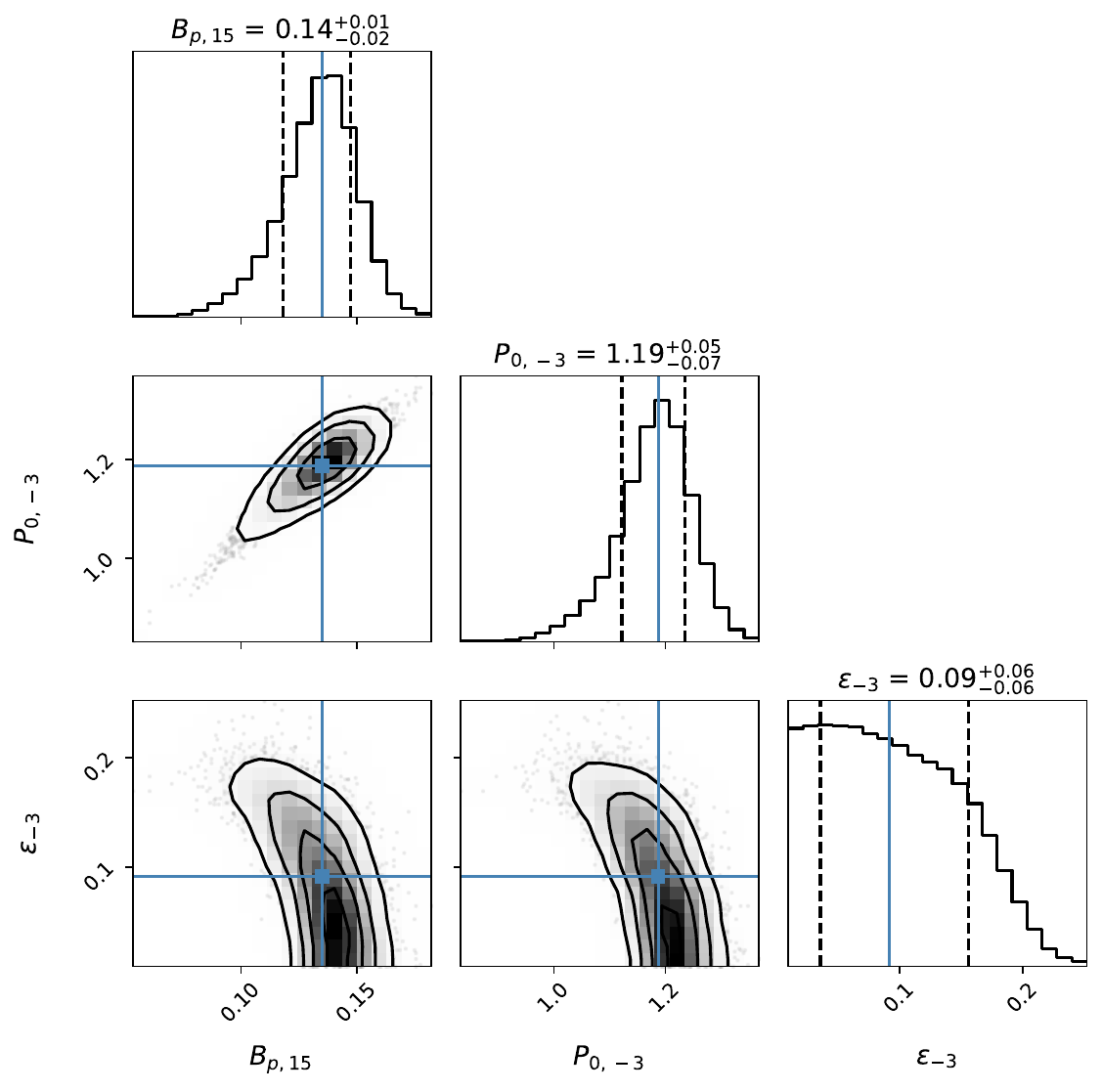}
\includegraphics  [angle=0,scale=0.24]   {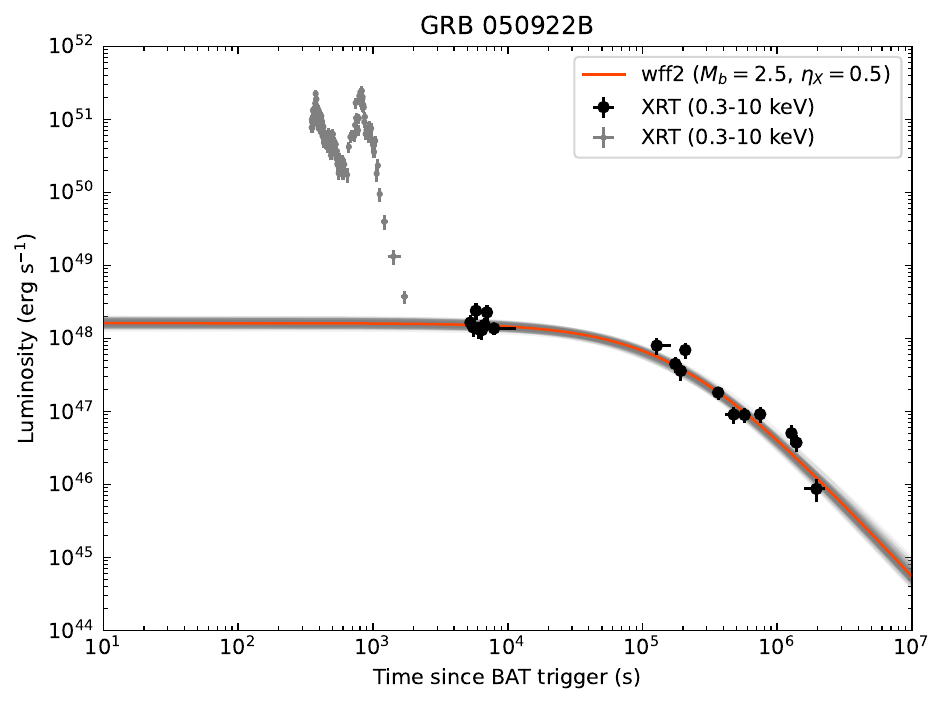} \\
\caption{EM-dominated case: the corner plots and best-fitting results of GRB 050922B in four samples of EoSs with $M_{b}=2.0~M_{\odot},~2.5~M_{\odot}$ and $\eta_{\rm X}=0.1,~0.5$, respectively.} 
\label{fig:EM_mcmc}
\end{figure*}

\begin{figure*}
\centering
\includegraphics  [angle=0,scale=0.2]   {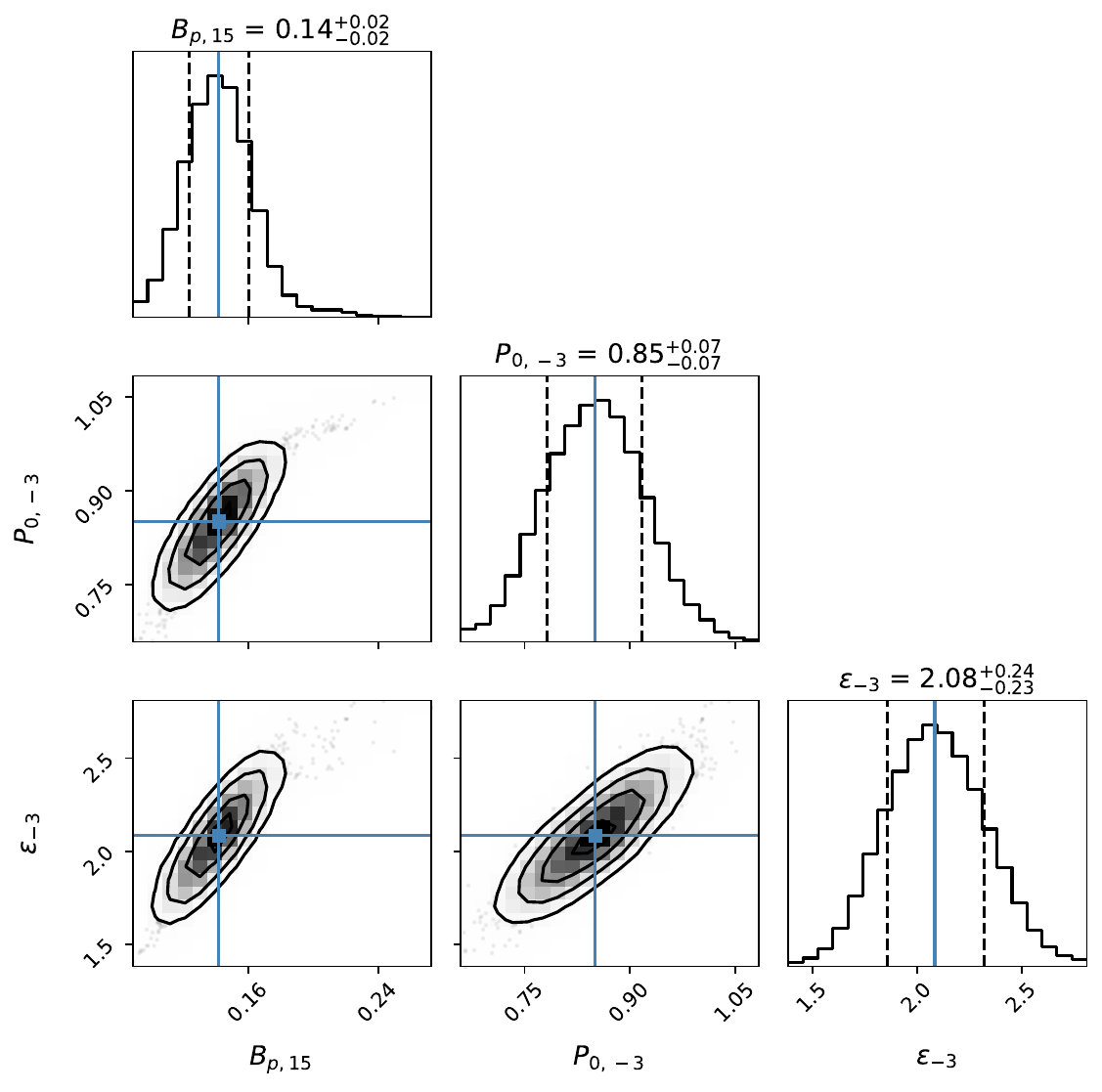}
\includegraphics  [angle=0,scale=0.2]   {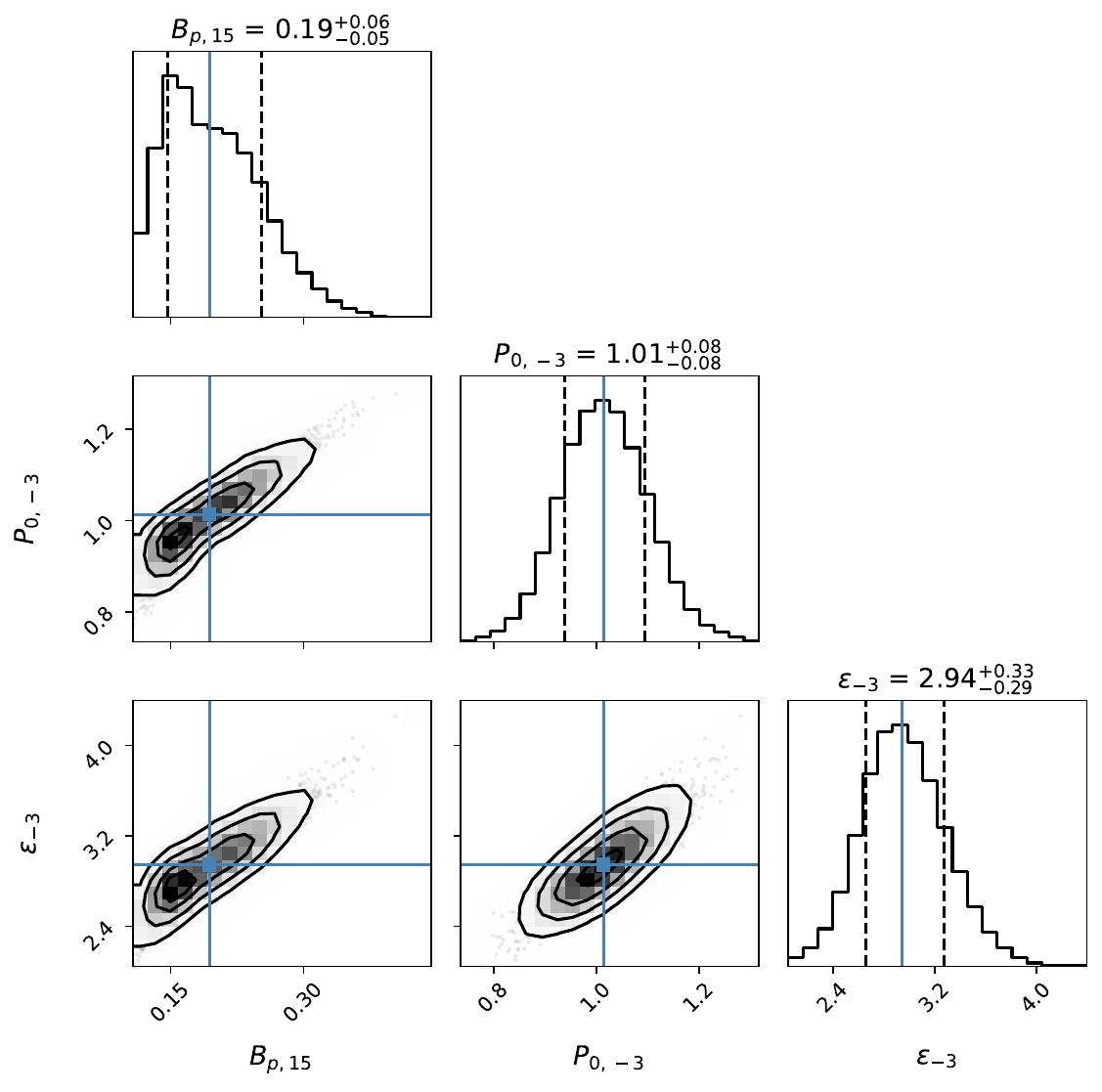}
\includegraphics  [angle=0,scale=0.2]   {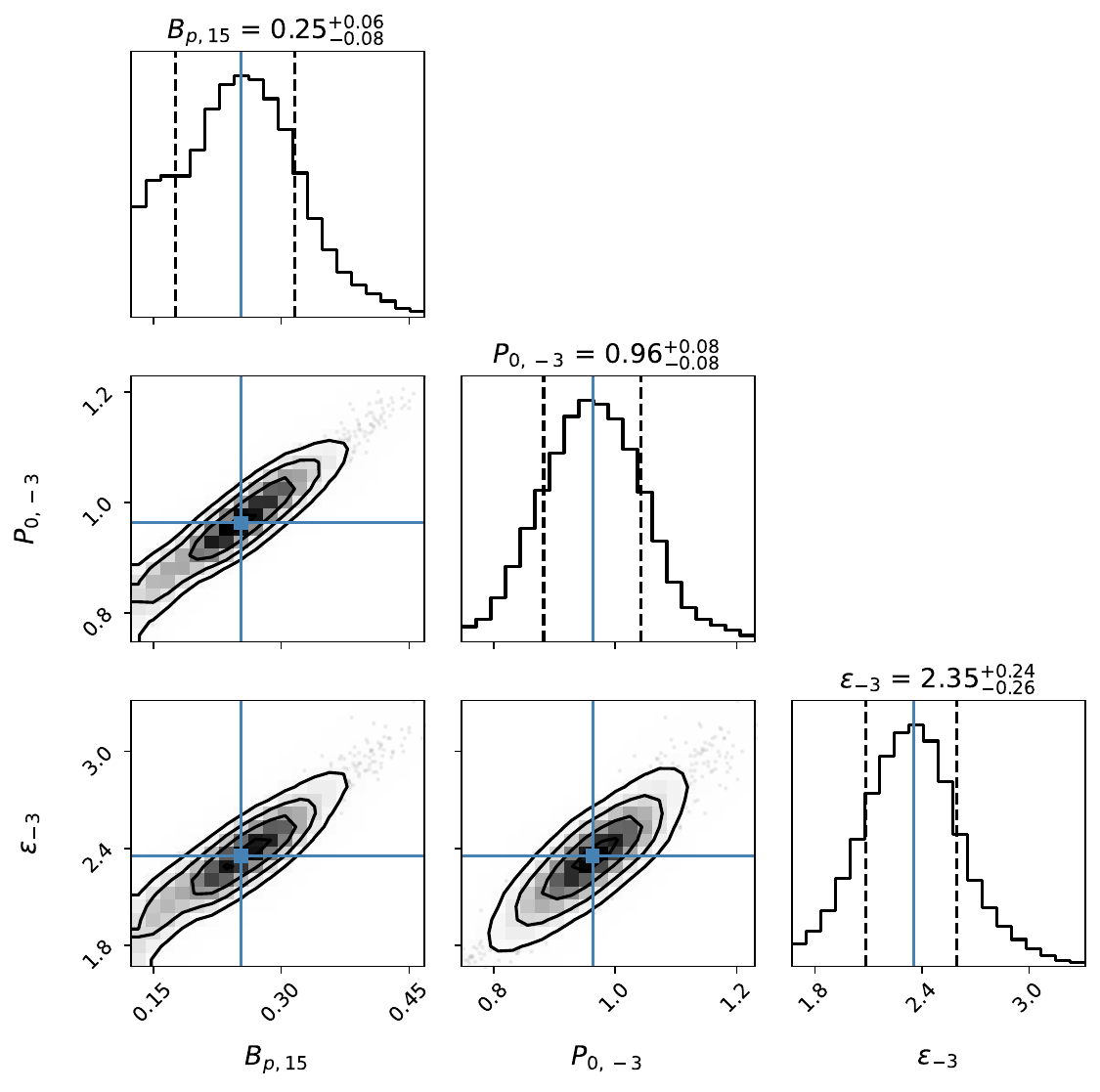}
\includegraphics  [angle=0,scale=0.2]   {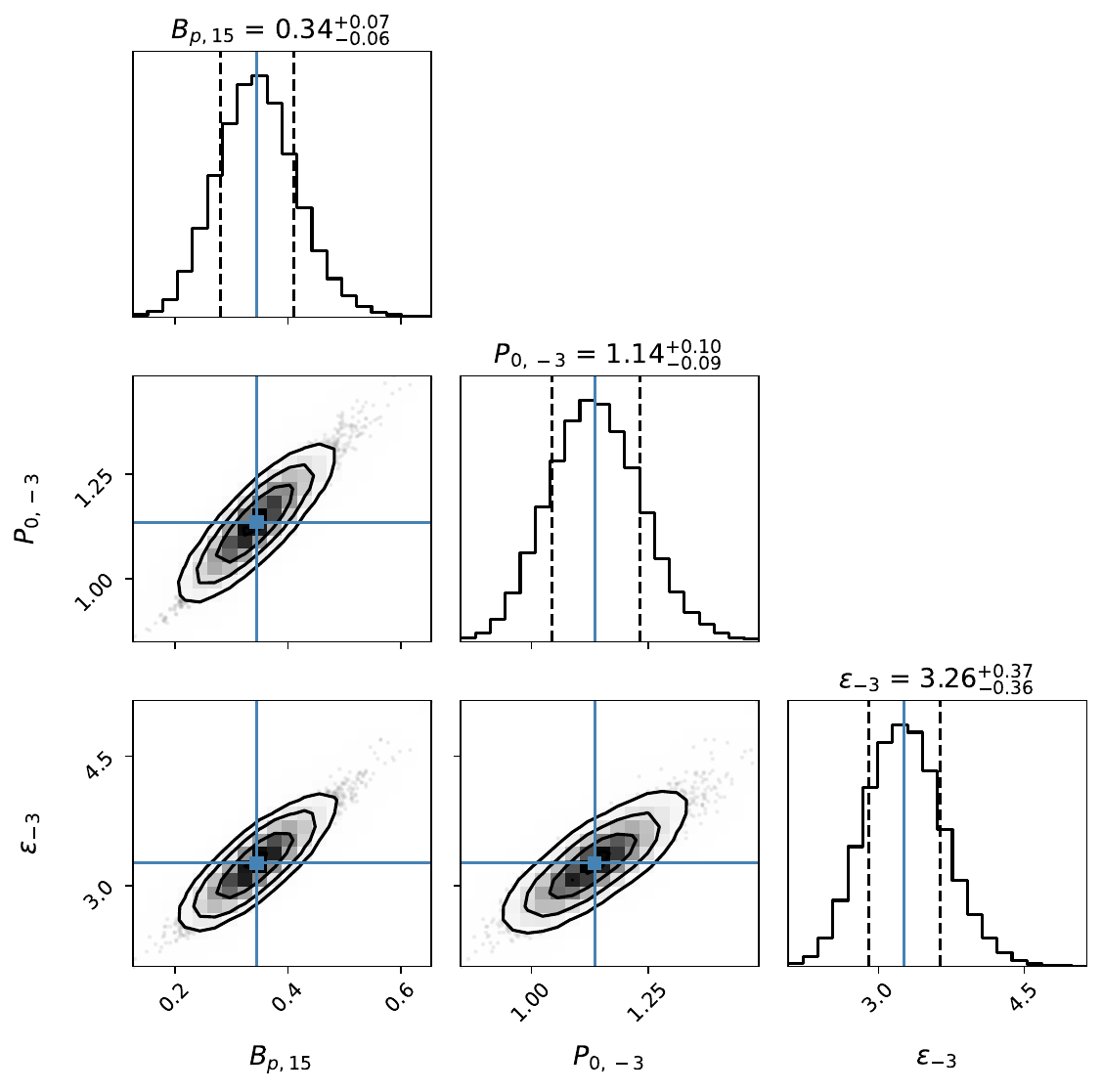} \\
\includegraphics  [angle=0,scale=0.24]   {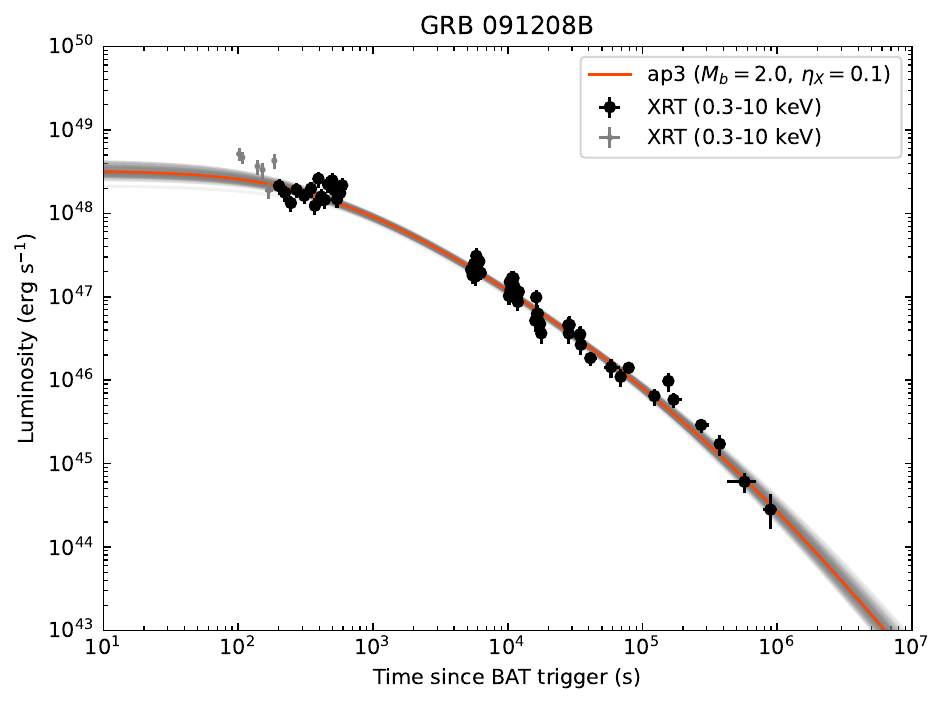}
\includegraphics  [angle=0,scale=0.24]   {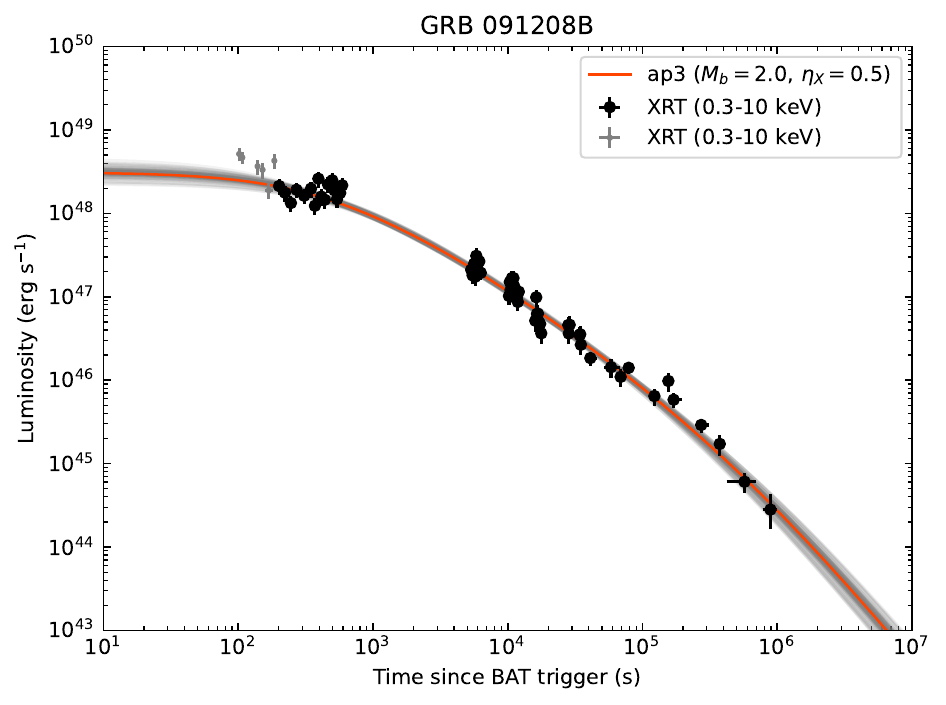}
\includegraphics  [angle=0,scale=0.24]   {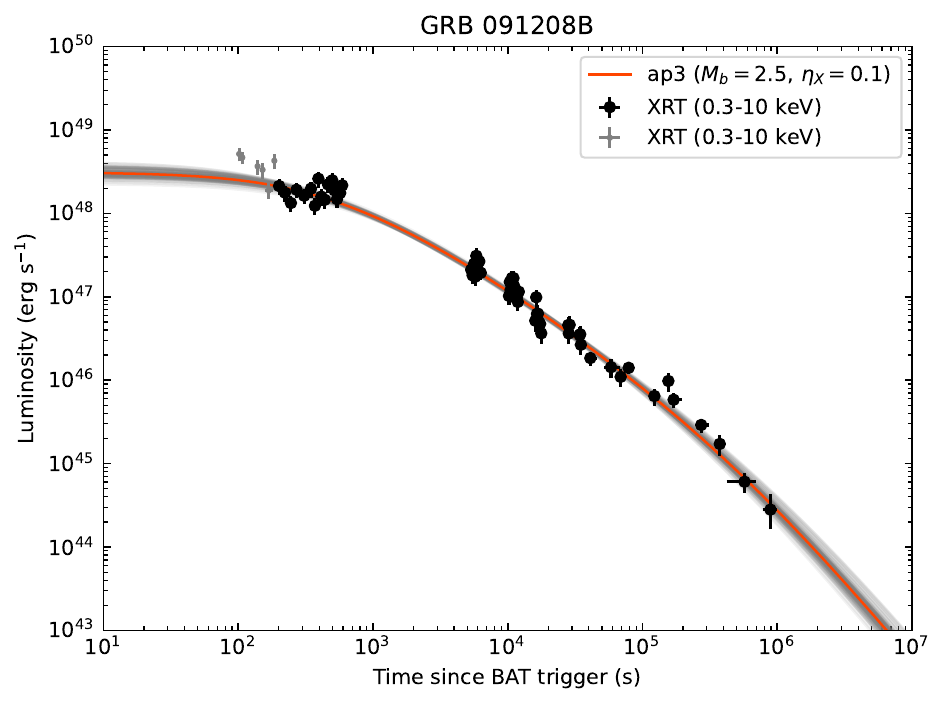}
\includegraphics  [angle=0,scale=0.24]   {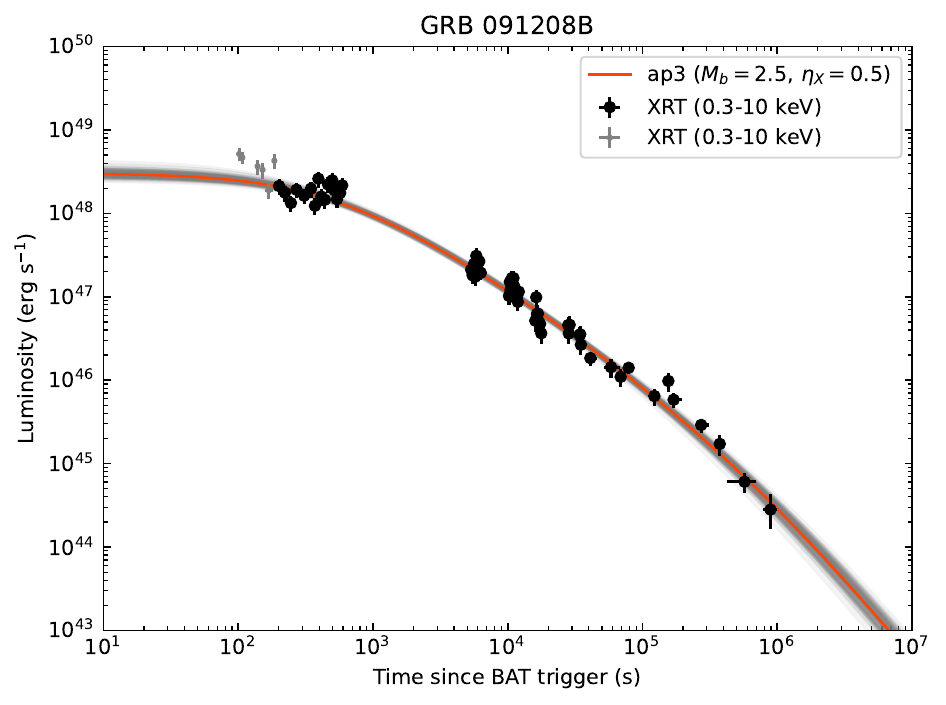} \\
\includegraphics  [angle=0,scale=0.2]   {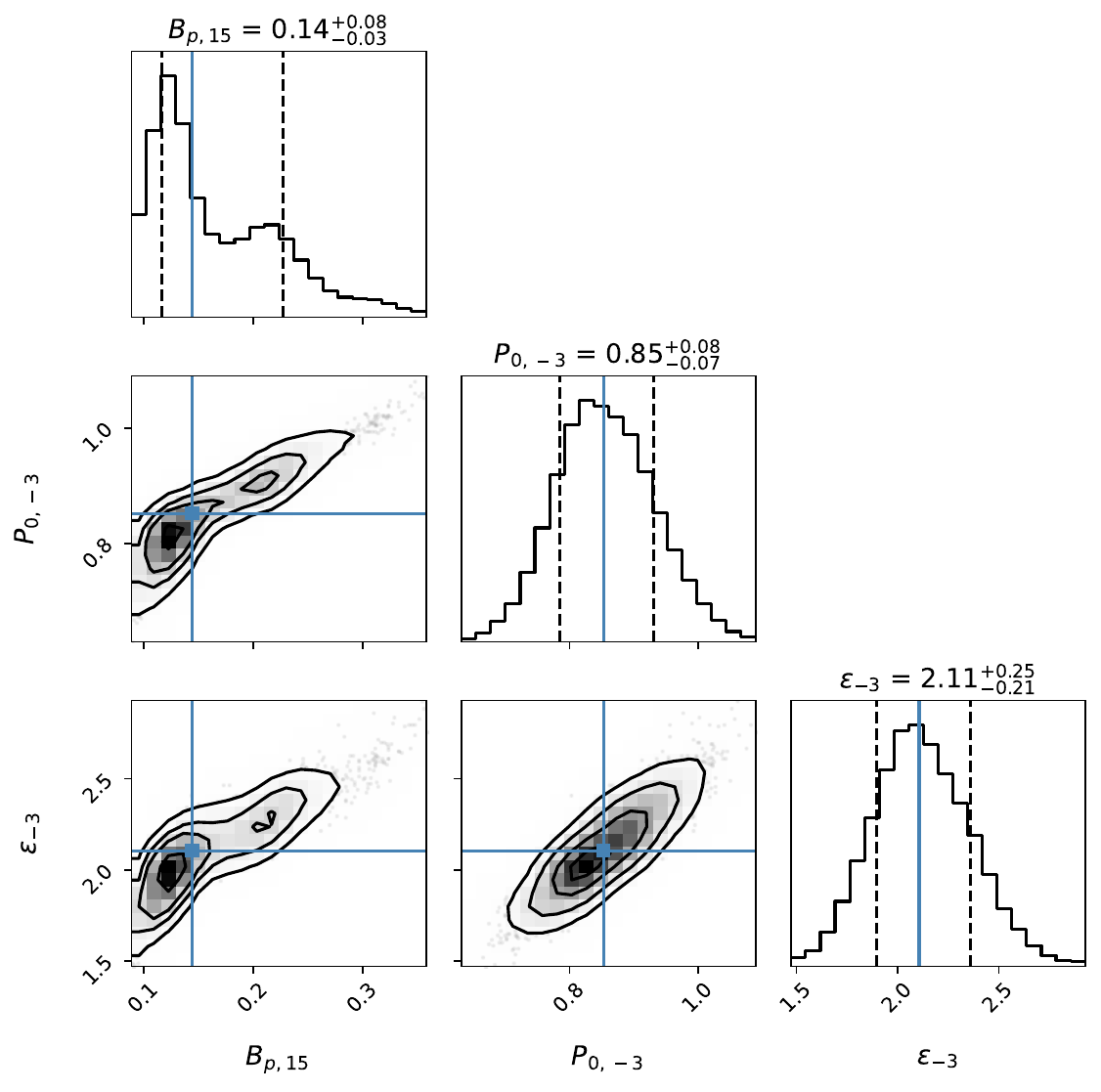}
\includegraphics  [angle=0,scale=0.2]   {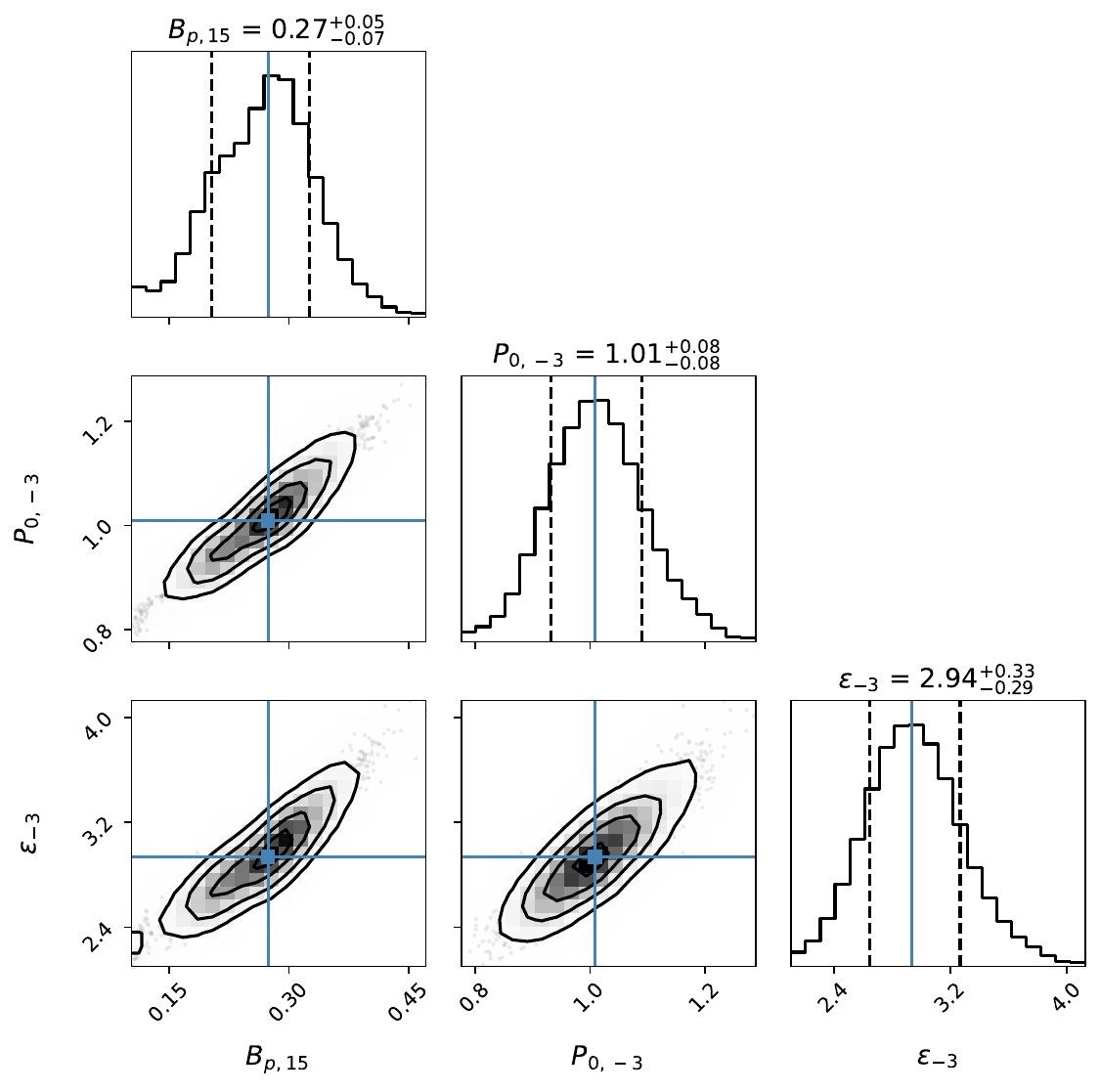}
\includegraphics  [angle=0,scale=0.2]   {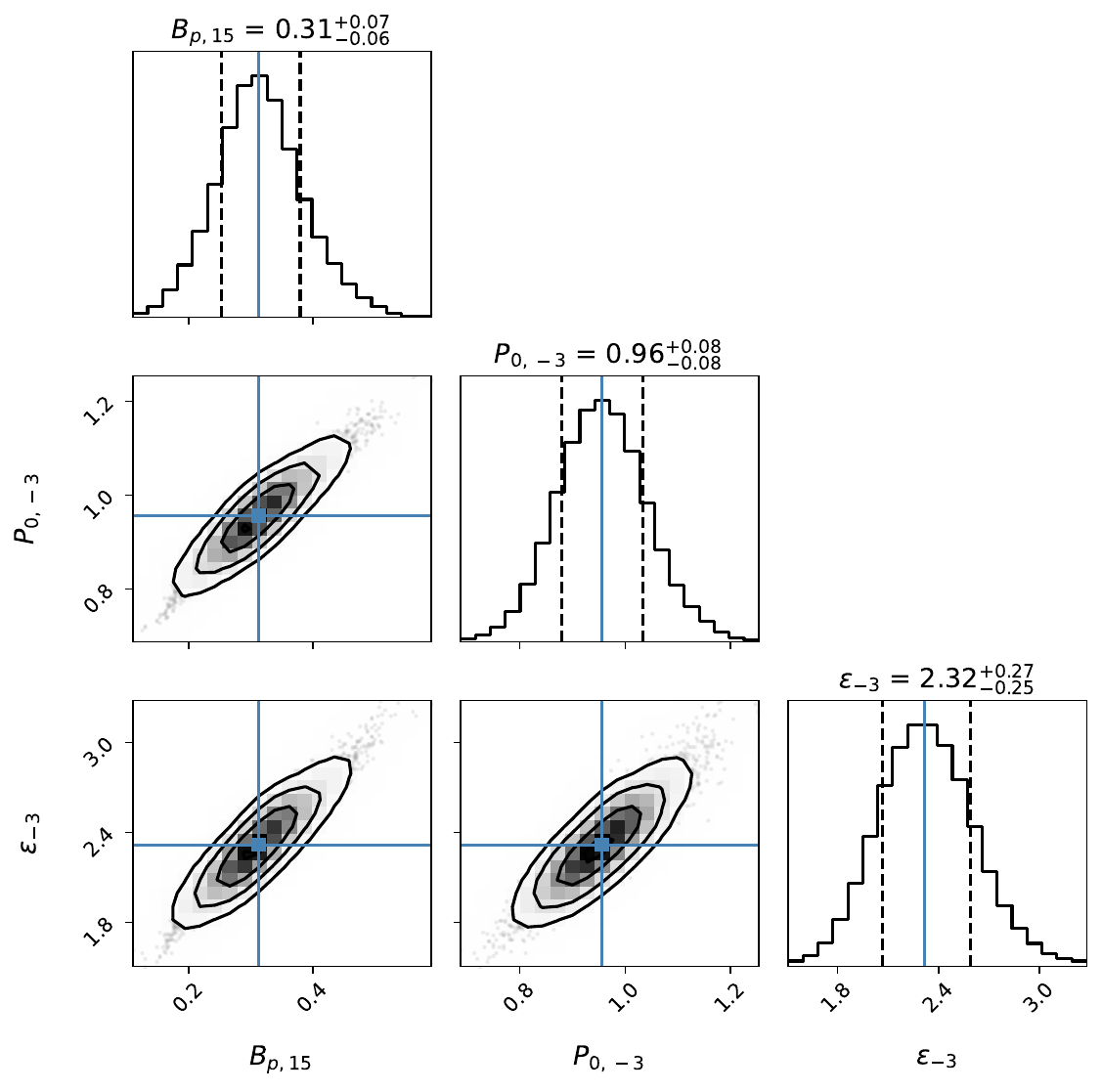}
\includegraphics  [angle=0,scale=0.2]   {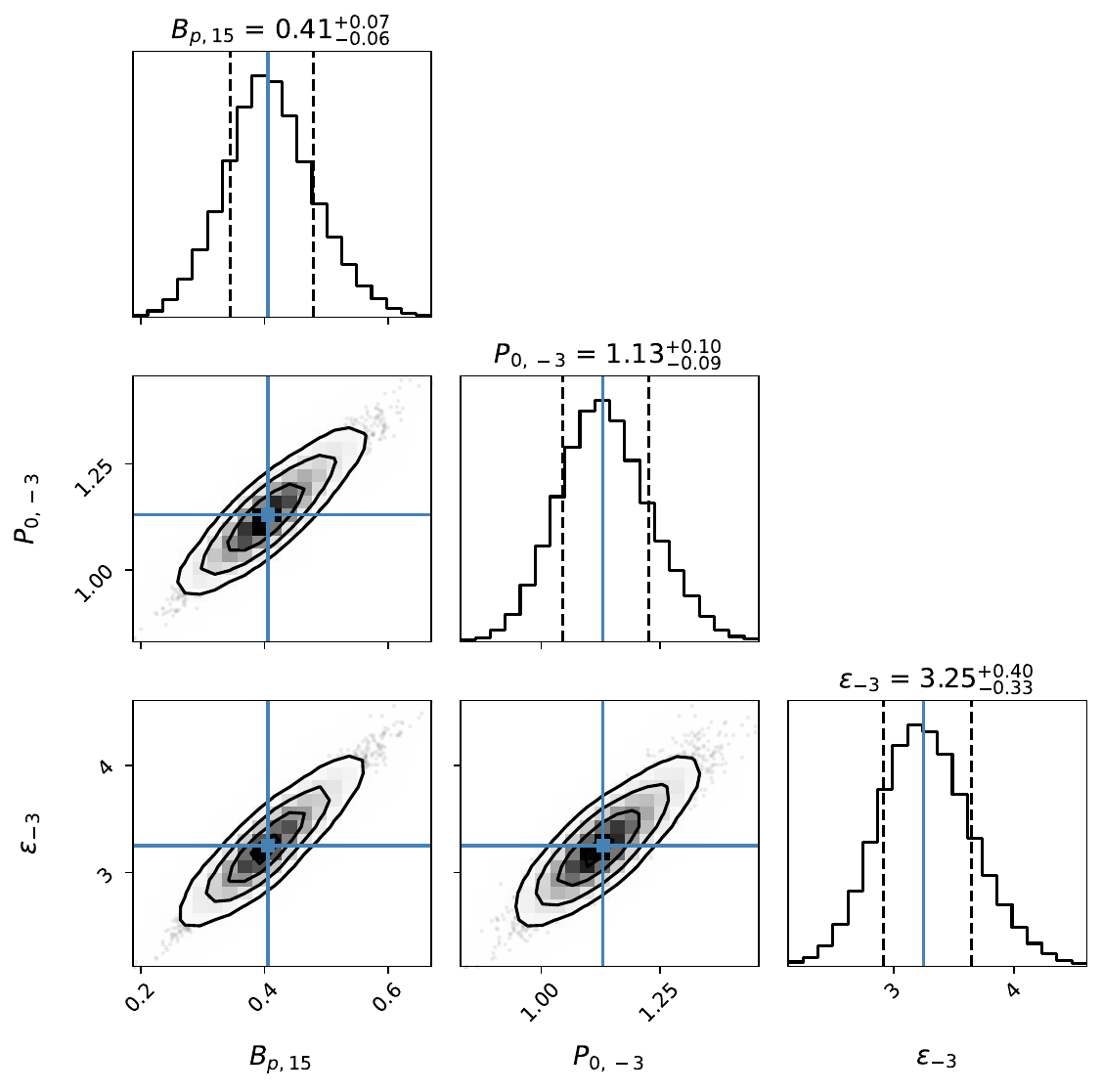} \\
\includegraphics  [angle=0,scale=0.24]   {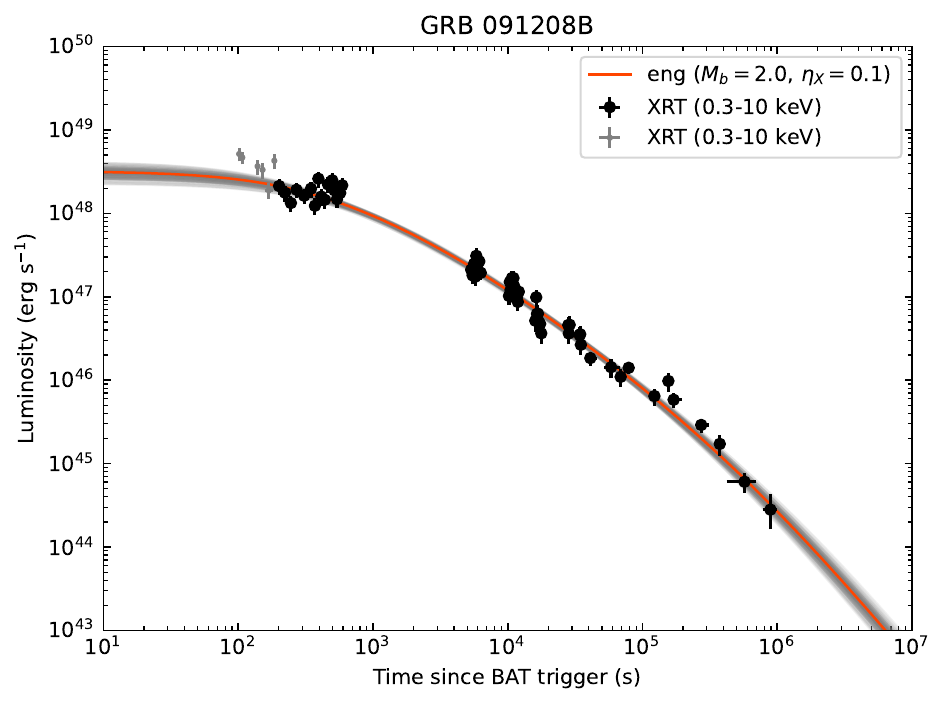}
\includegraphics  [angle=0,scale=0.24]   {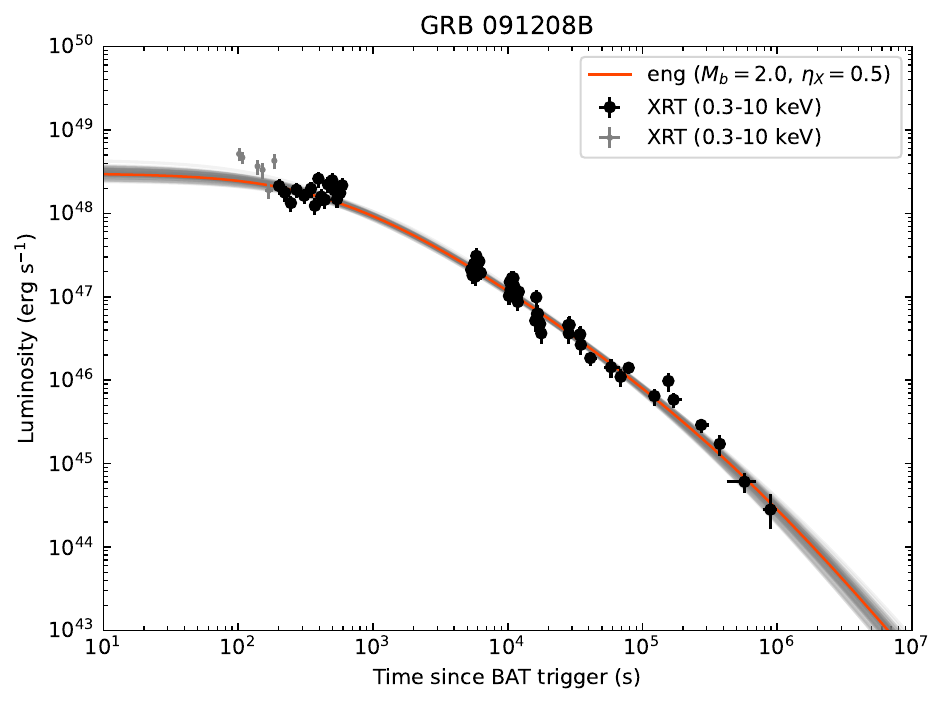}
\includegraphics  [angle=0,scale=0.24]   {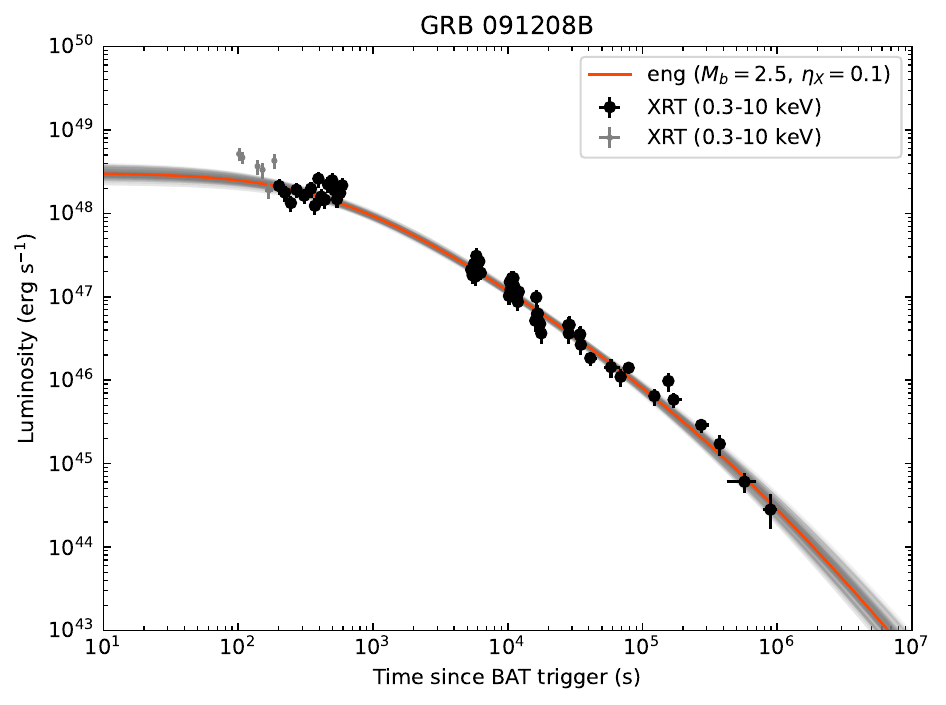}
\includegraphics  [angle=0,scale=0.24]   {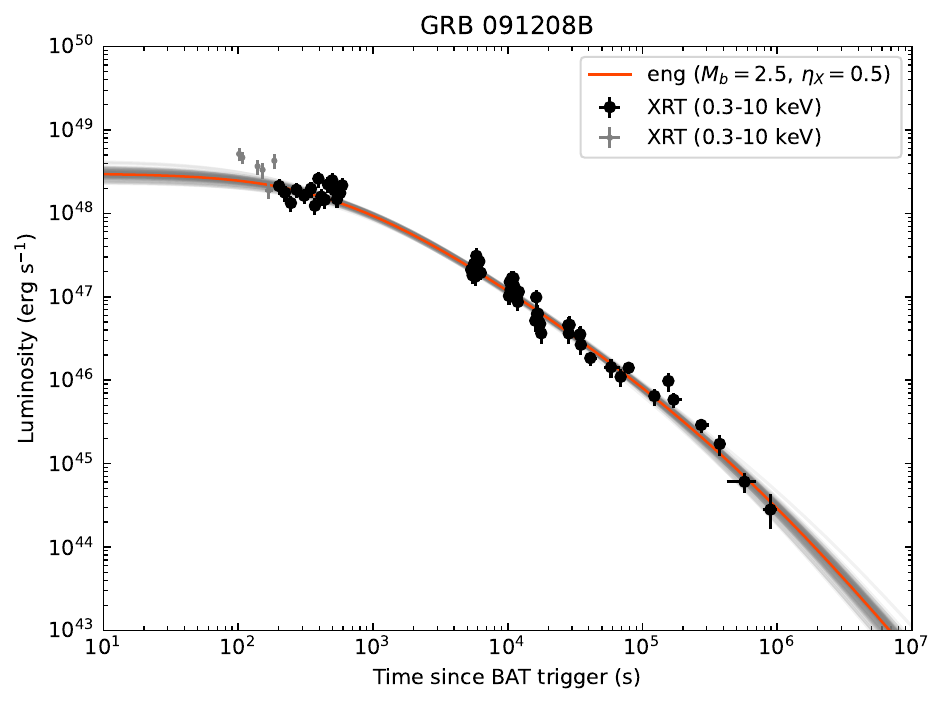} \\
\includegraphics  [angle=0,scale=0.2]   {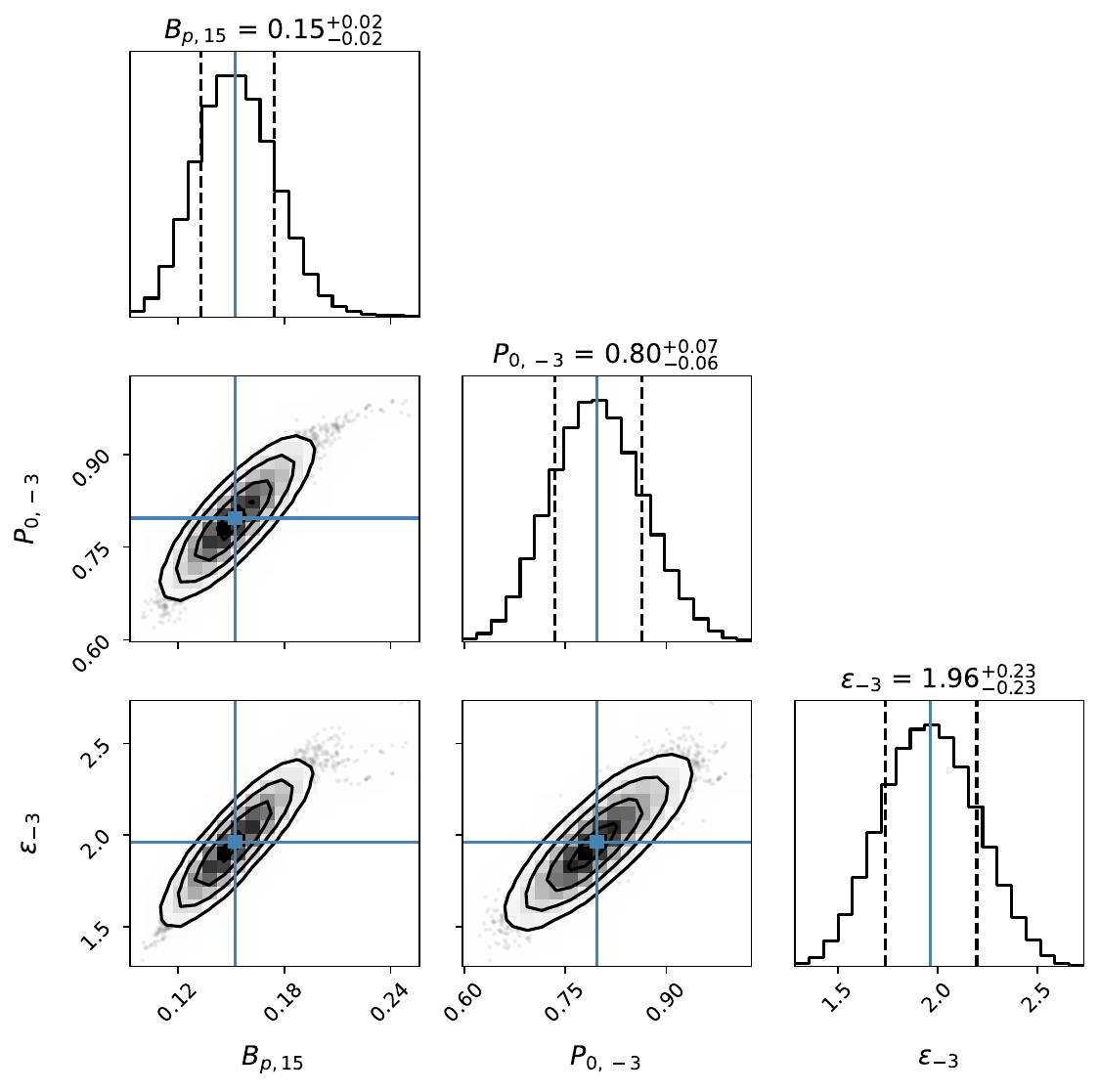}
\includegraphics  [angle=0,scale=0.2]   {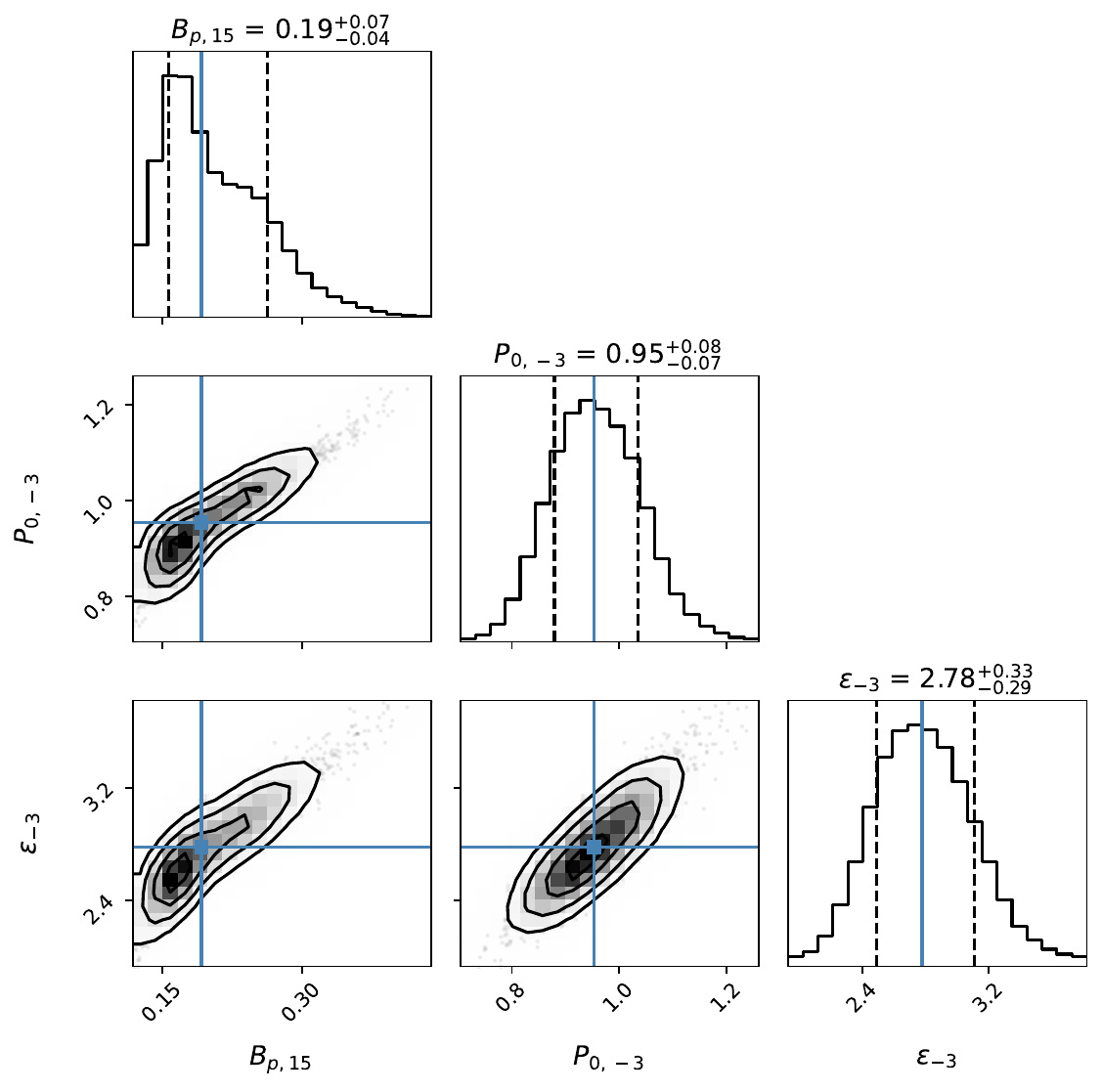}
\includegraphics  [angle=0,scale=0.2]   {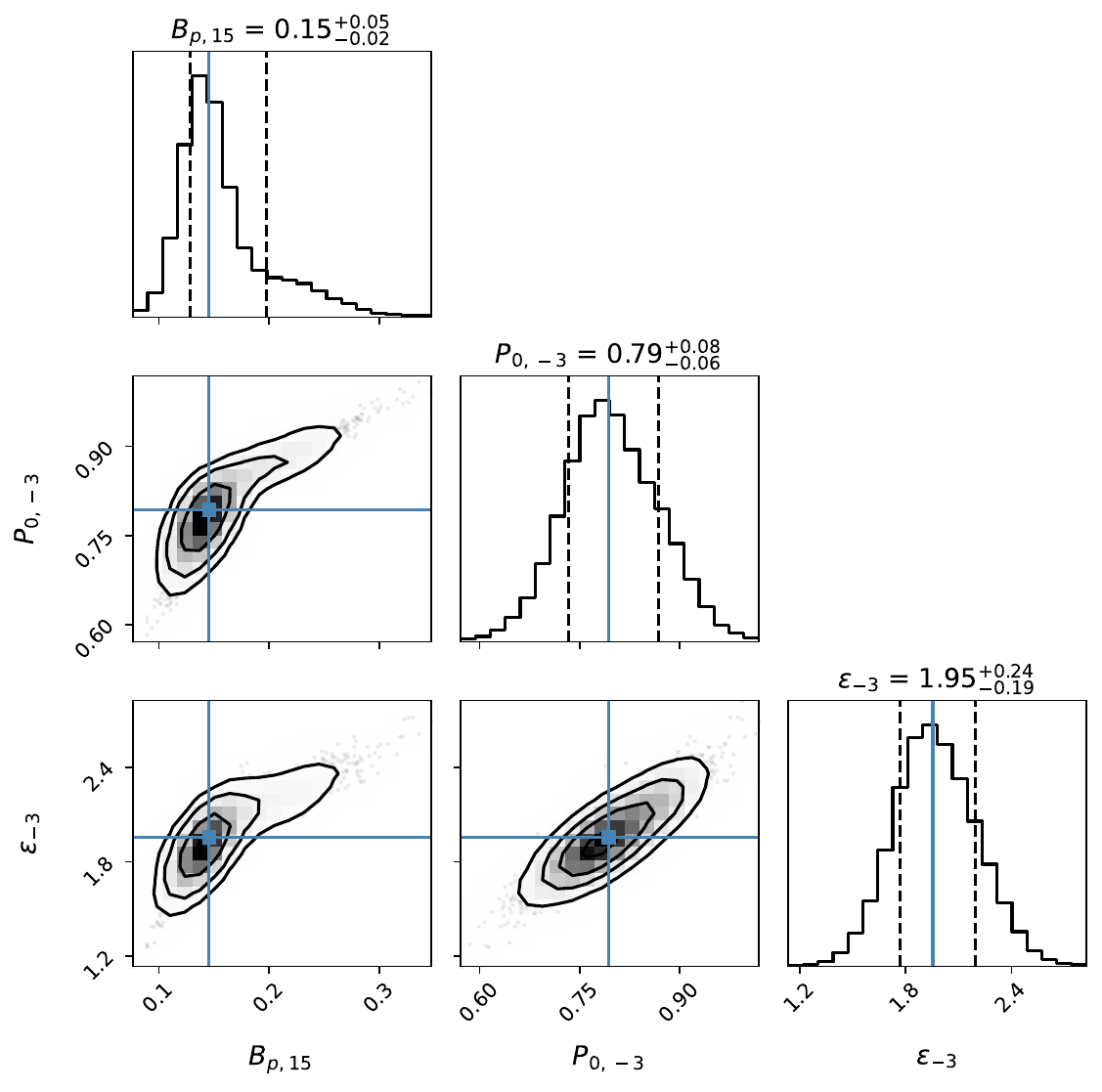}
\includegraphics  [angle=0,scale=0.2]   {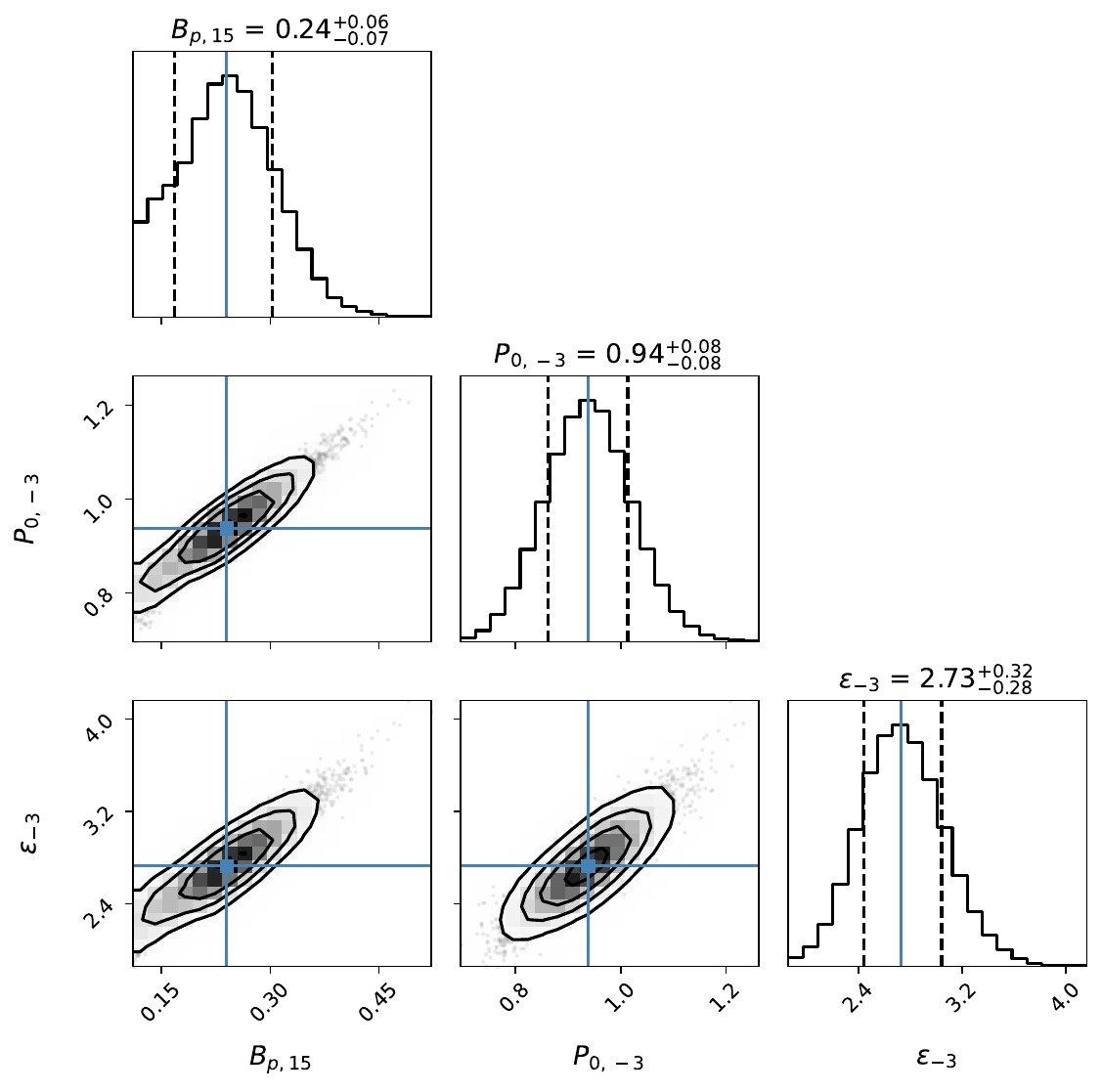} \\
\includegraphics  [angle=0,scale=0.24]   {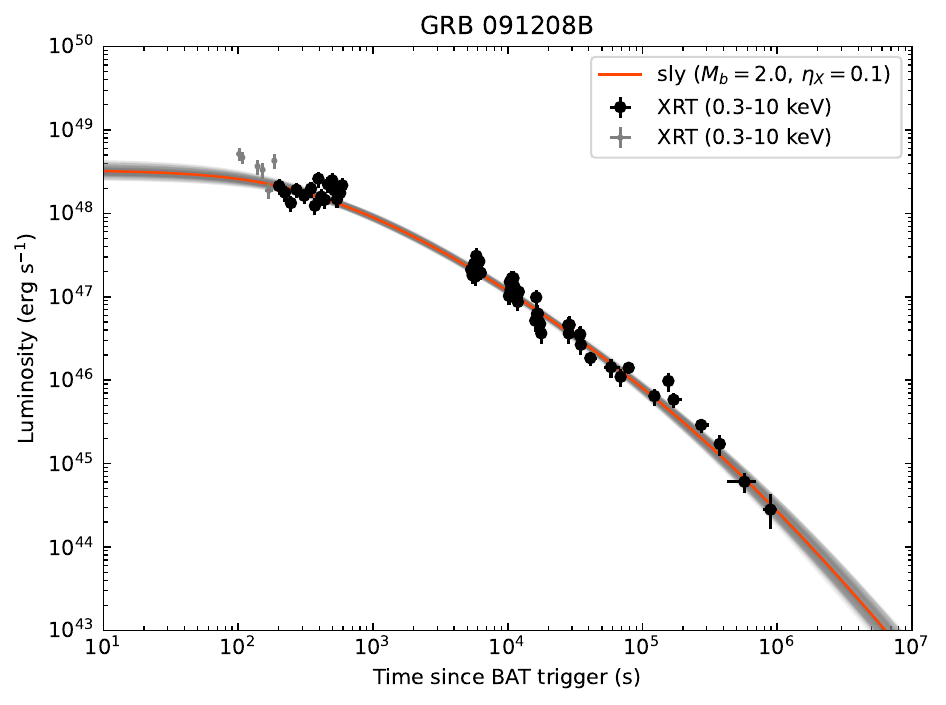}
\includegraphics  [angle=0,scale=0.24]   {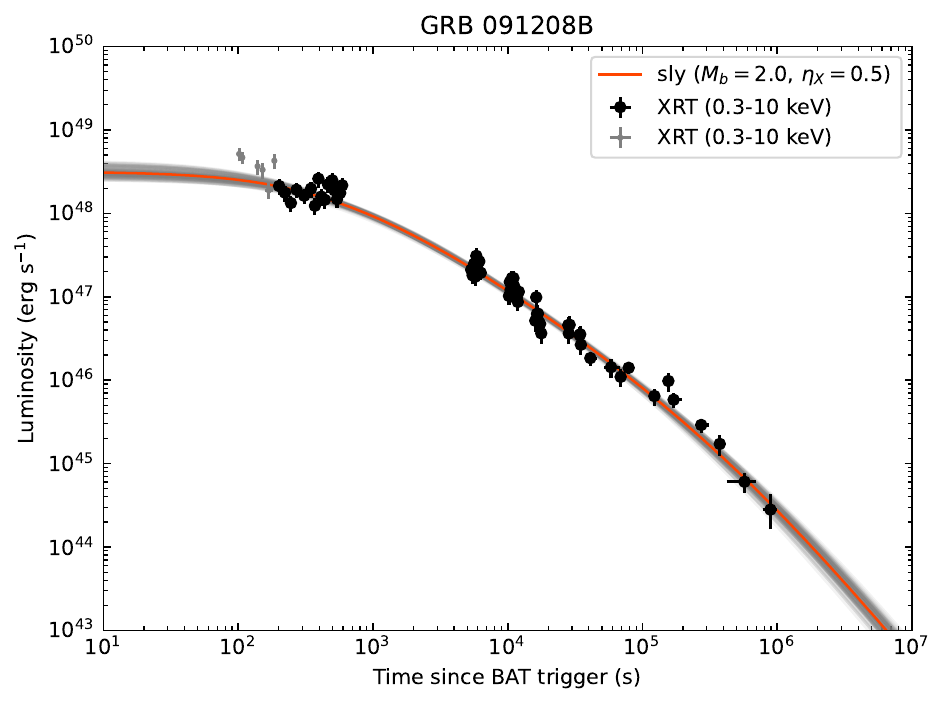}
\includegraphics  [angle=0,scale=0.24]   {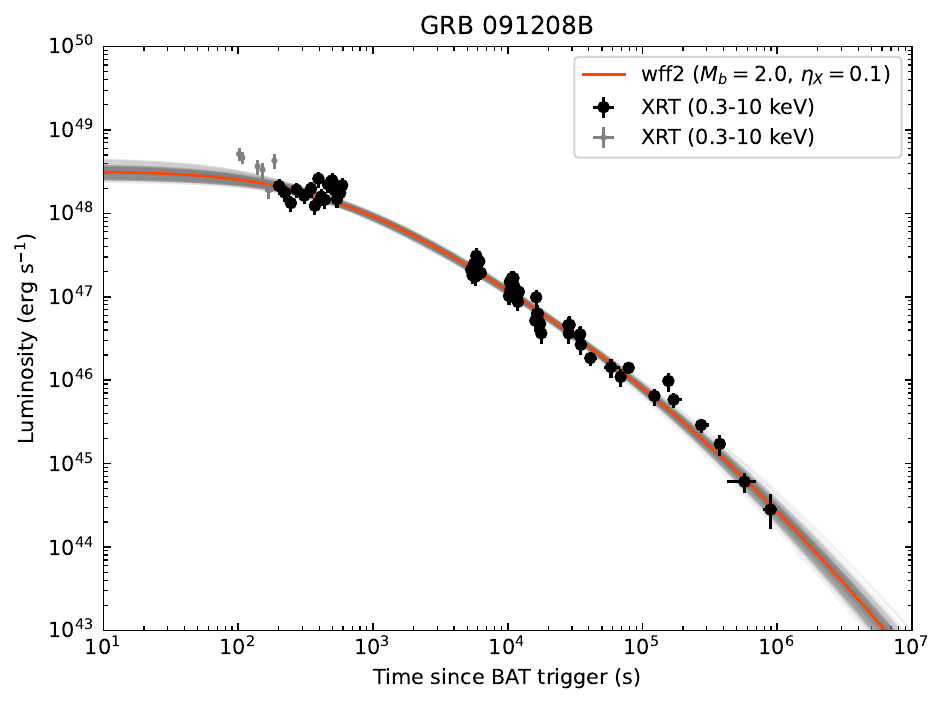}
\includegraphics  [angle=0,scale=0.24]   {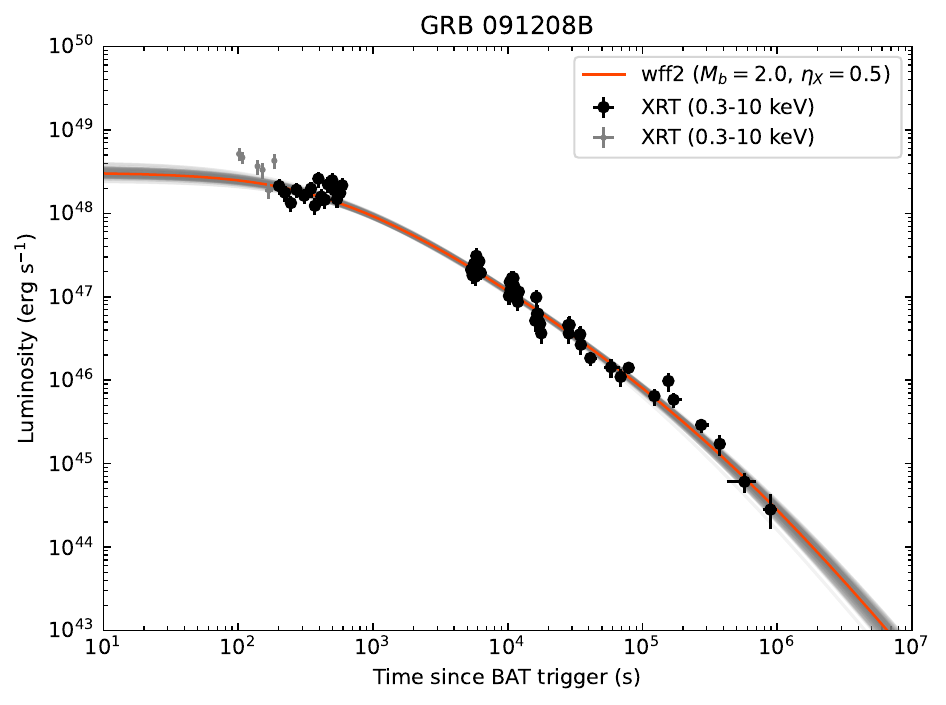} \\
\includegraphics  [angle=0,scale=0.2]   {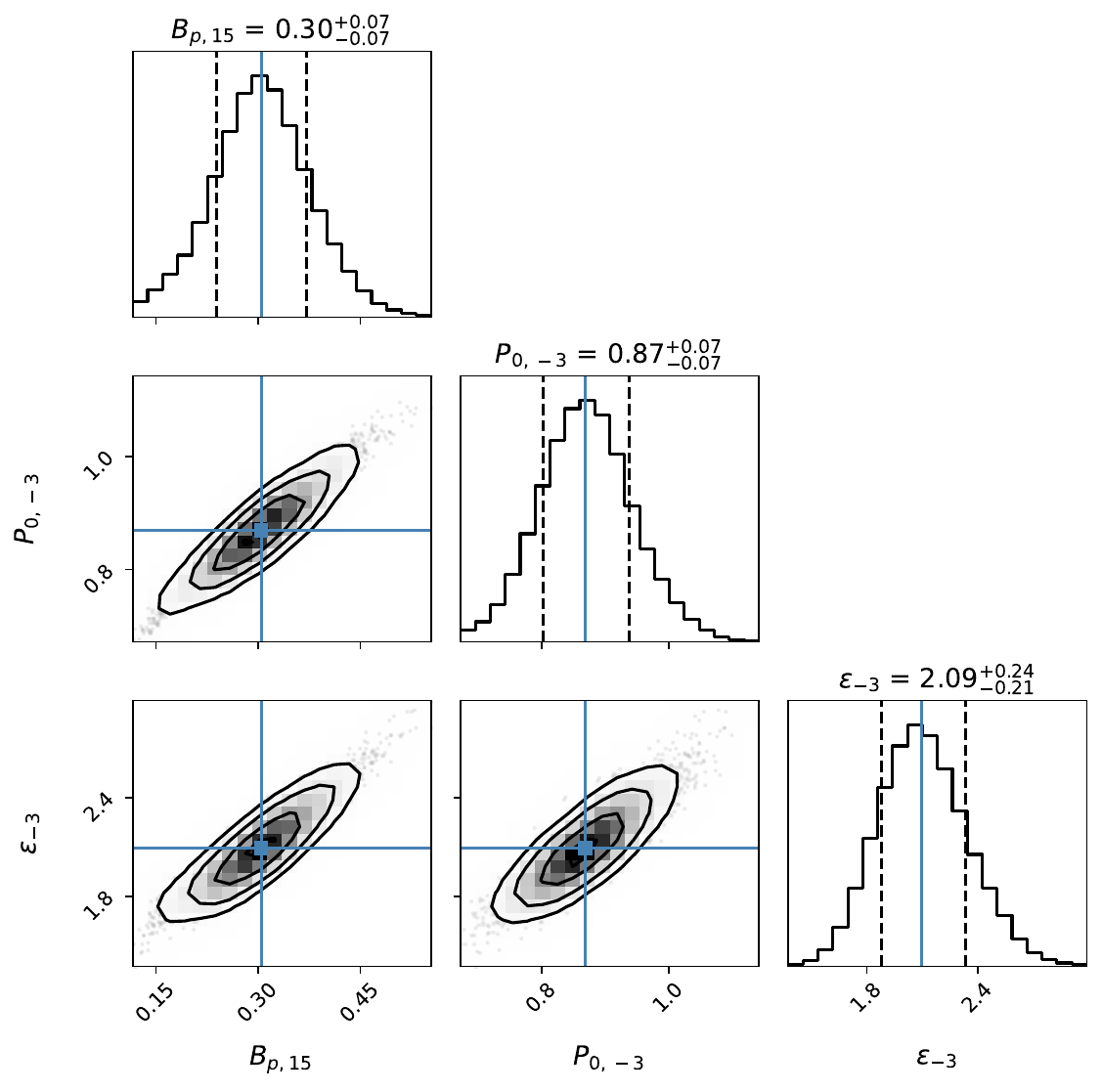}
\includegraphics  [angle=0,scale=0.24]   {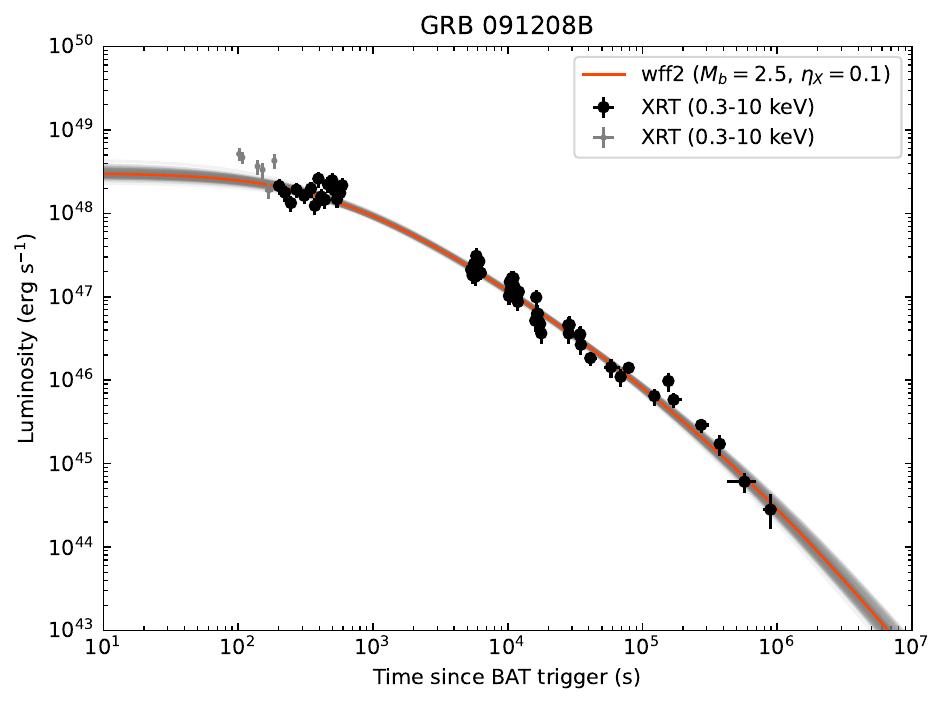}
\includegraphics  [angle=0,scale=0.2]   {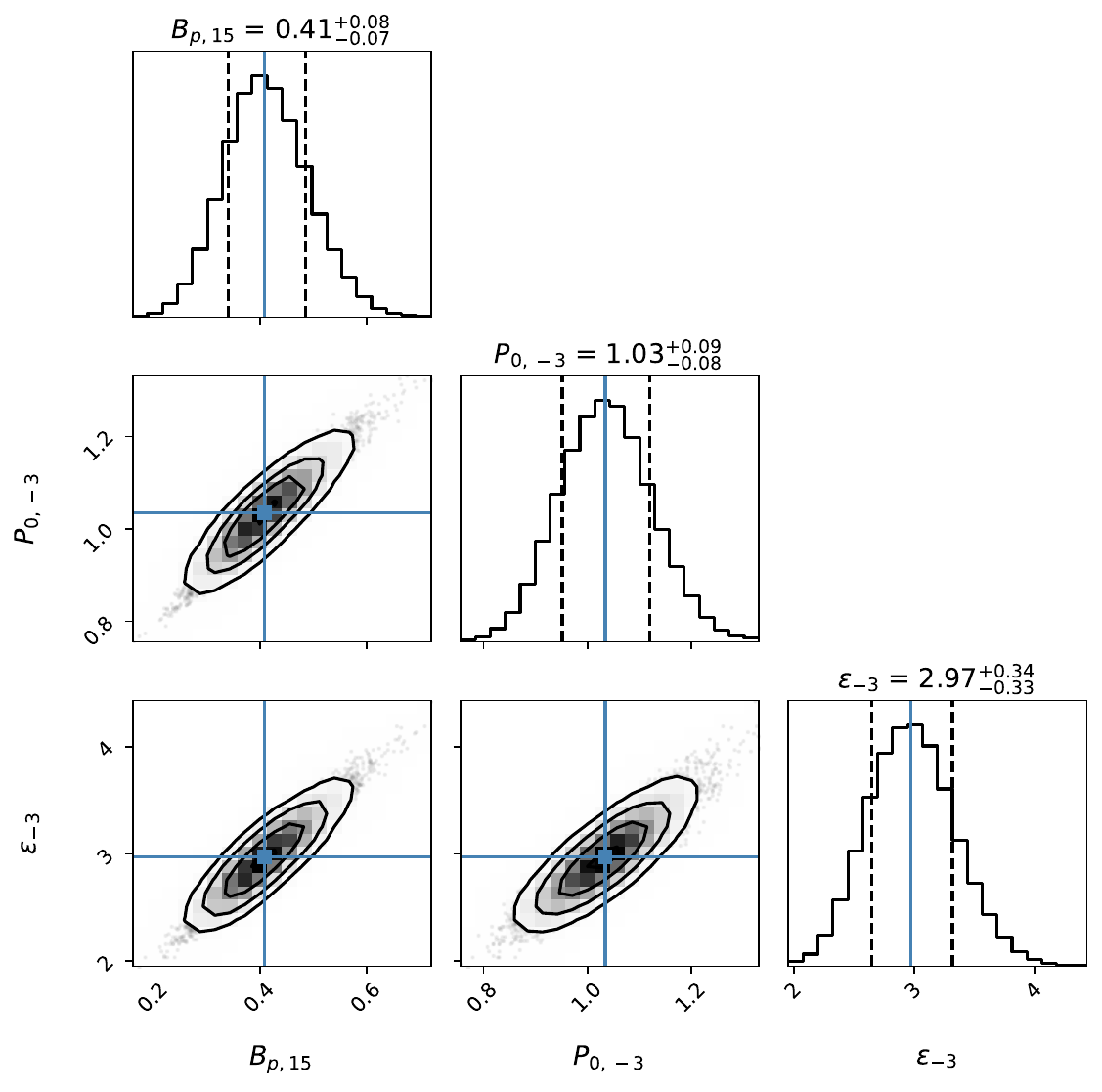}
\includegraphics  [angle=0,scale=0.24]   {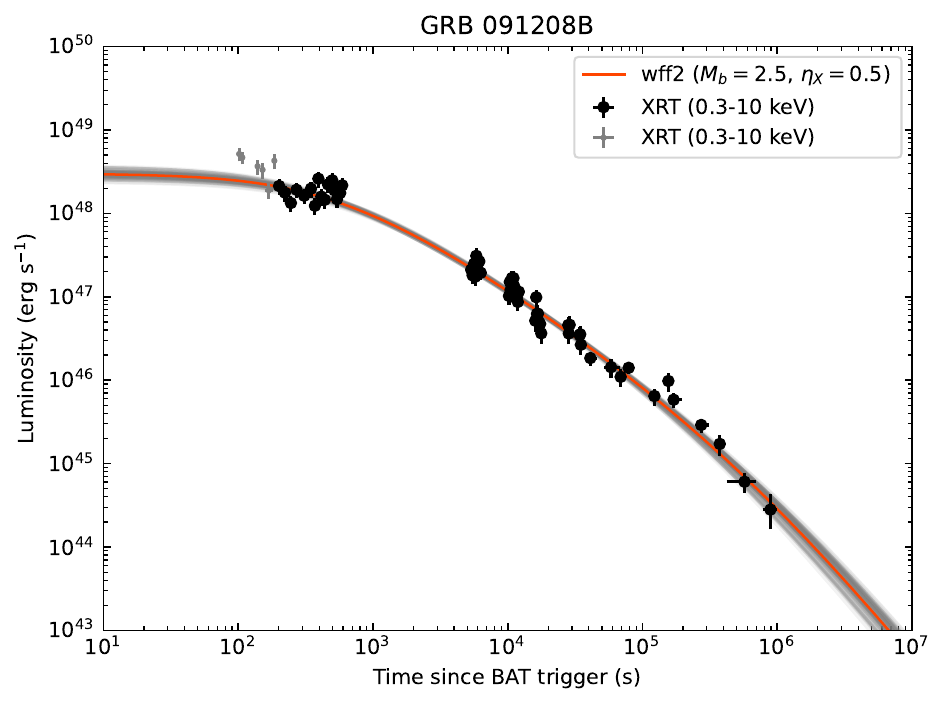} \\
\caption{GW-dominated case: the corner plots and best-fitting results of GRB 091208B in four samples of EoSs with $M_{b}=2.0~M_{\odot},~2.5~M_{\odot}$ and $\eta_{\rm X}=0.1,~0.5$, respectively.}
\label{fig:GW_mcmc}
\end{figure*}

\begin{figure*}
\centering
\includegraphics  [angle=0,scale=0.2]   {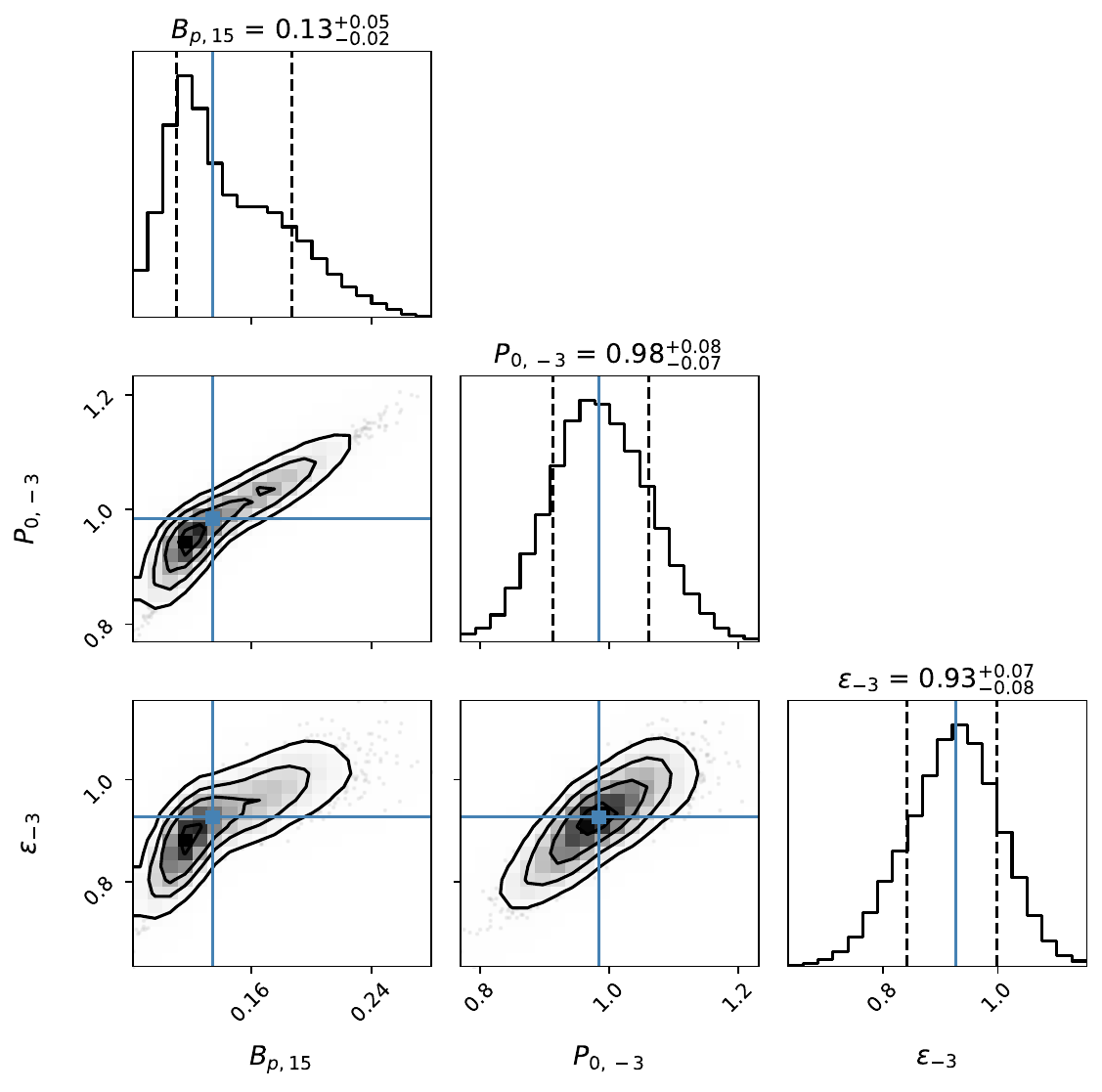}
\includegraphics  [angle=0,scale=0.2]   {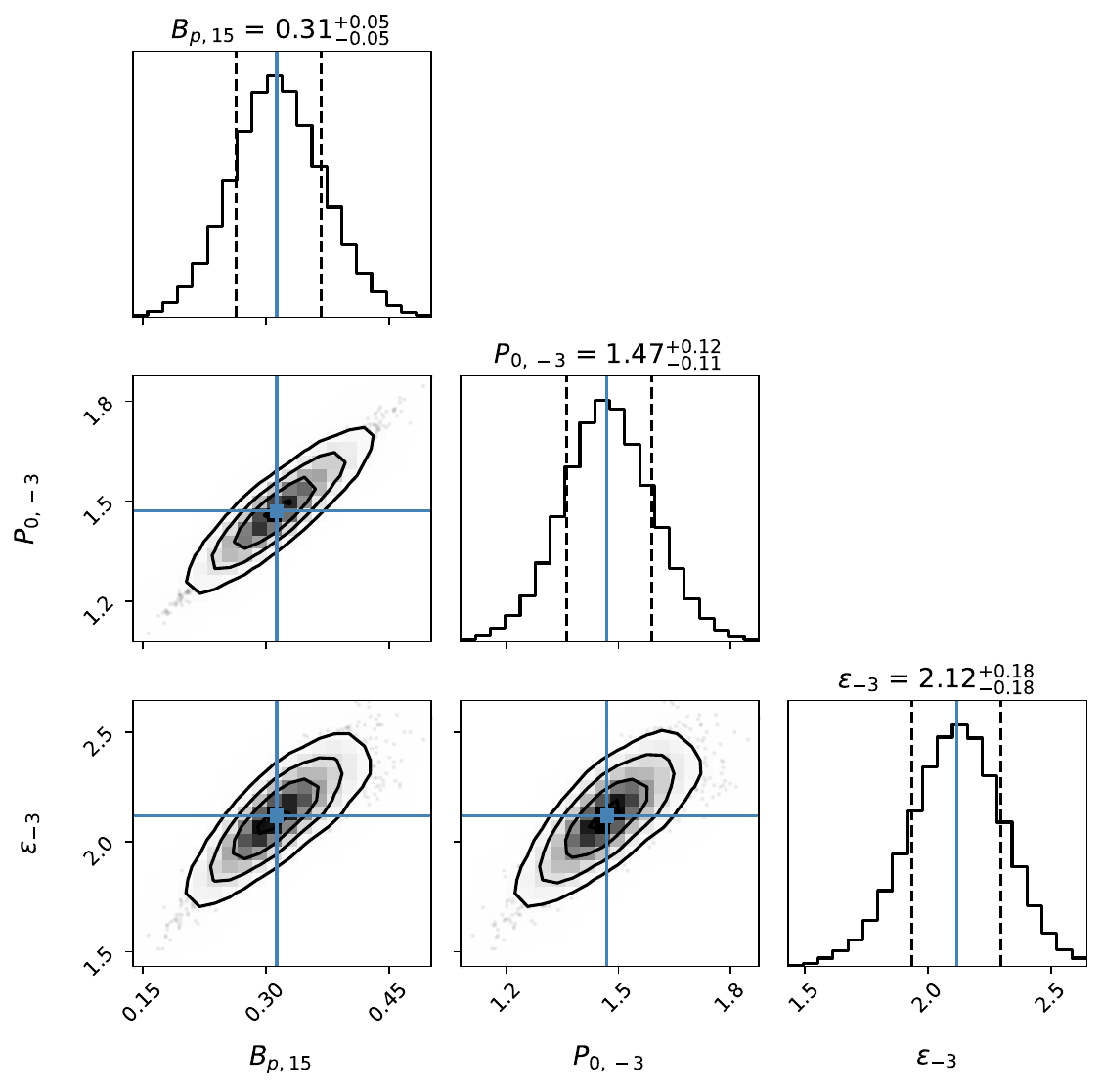}
\includegraphics  [angle=0,scale=0.2]   {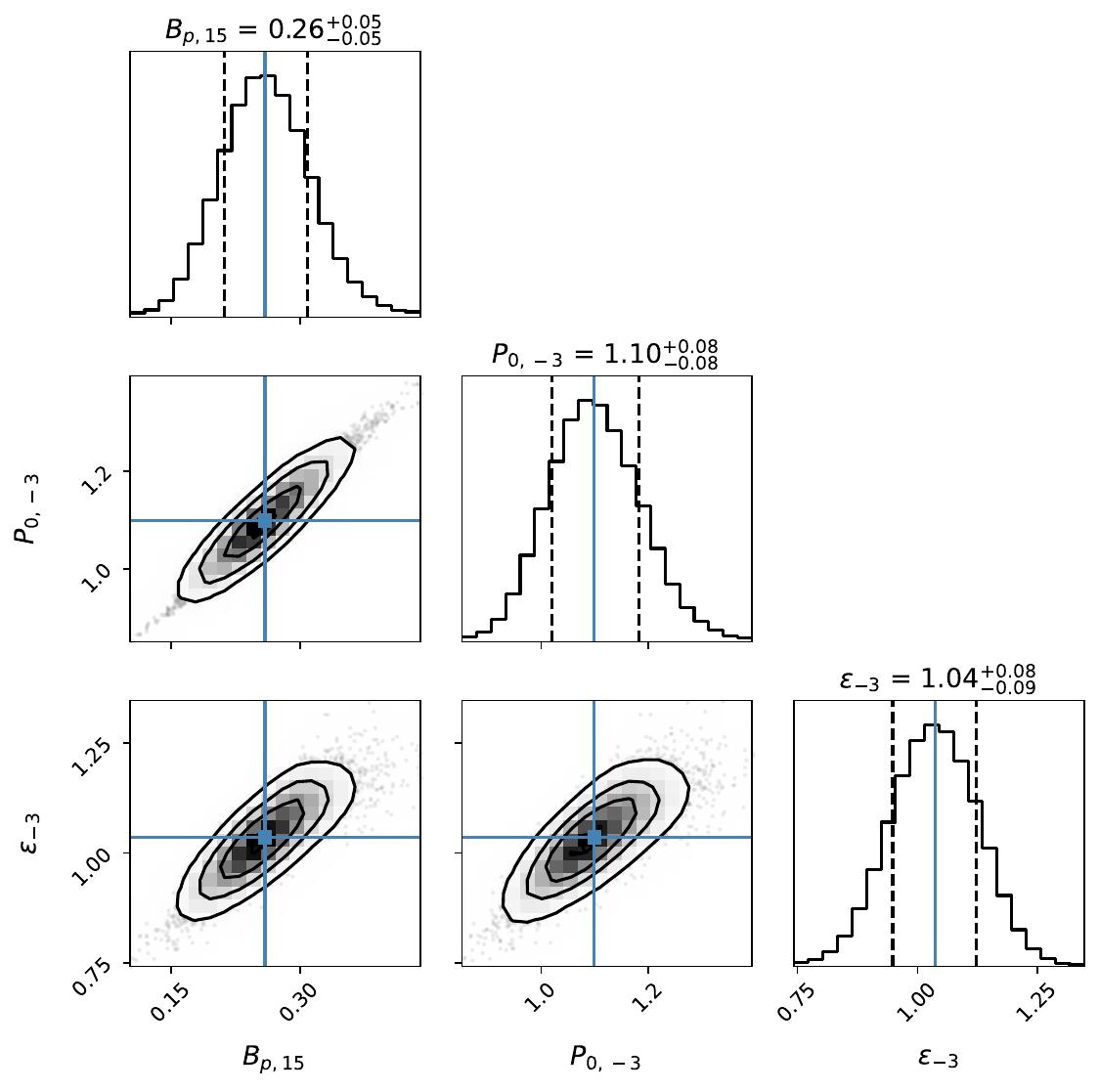}
\includegraphics  [angle=0,scale=0.2]   {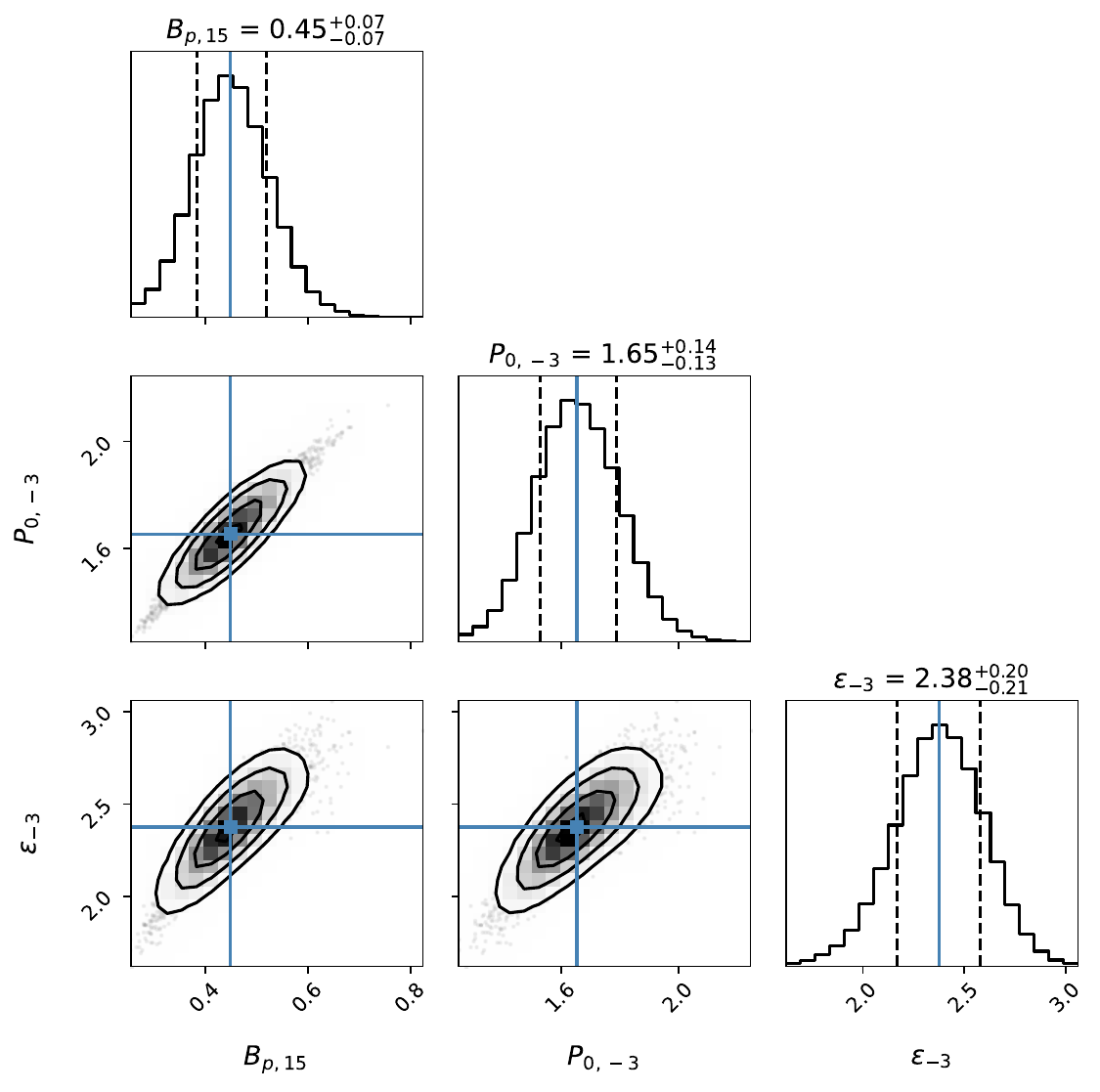} \\
\includegraphics  [angle=0,scale=0.24]   {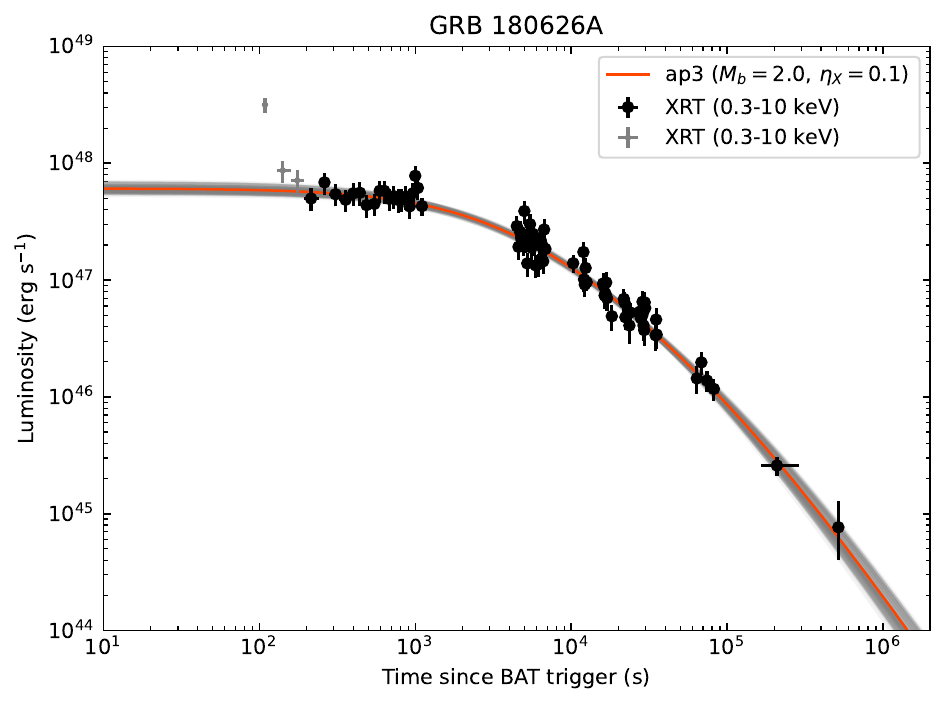}
\includegraphics  [angle=0,scale=0.24]   {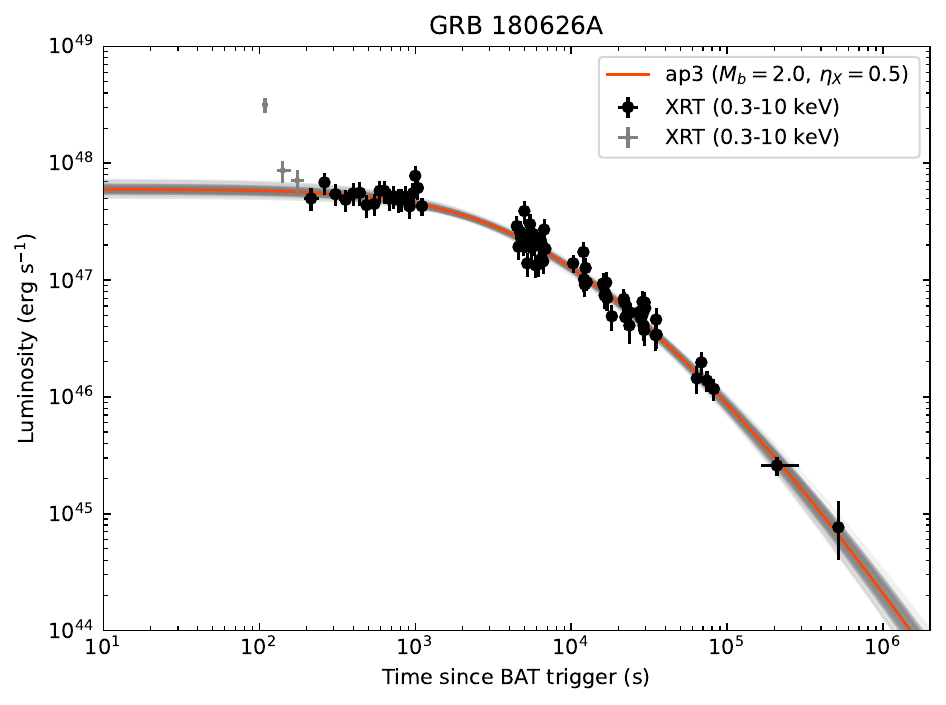}
\includegraphics  [angle=0,scale=0.24]   {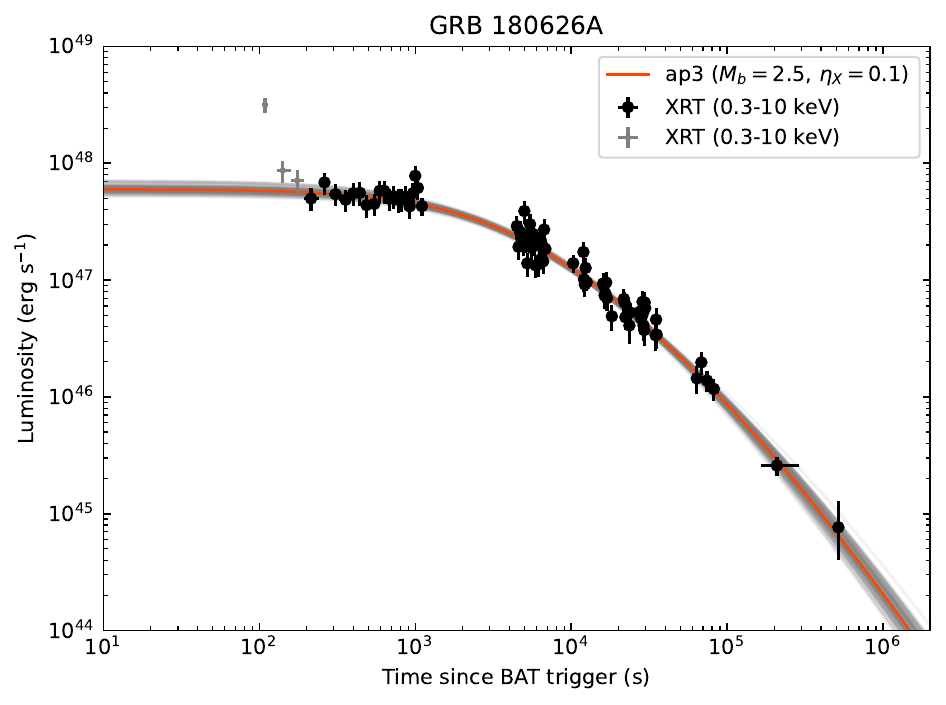}
\includegraphics  [angle=0,scale=0.24]   {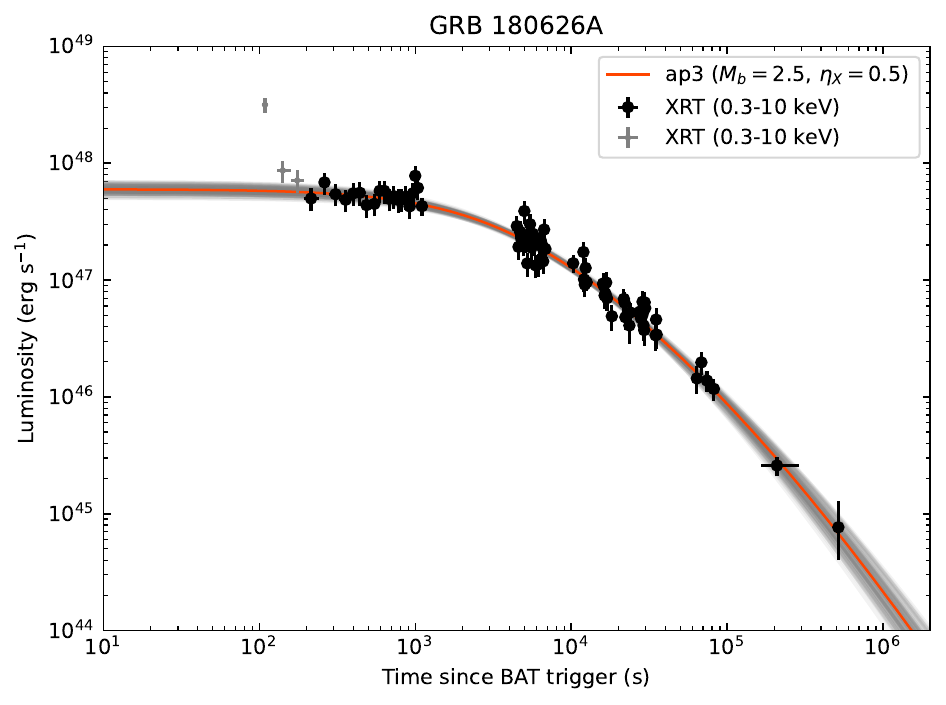} \\
\includegraphics  [angle=0,scale=0.2]   {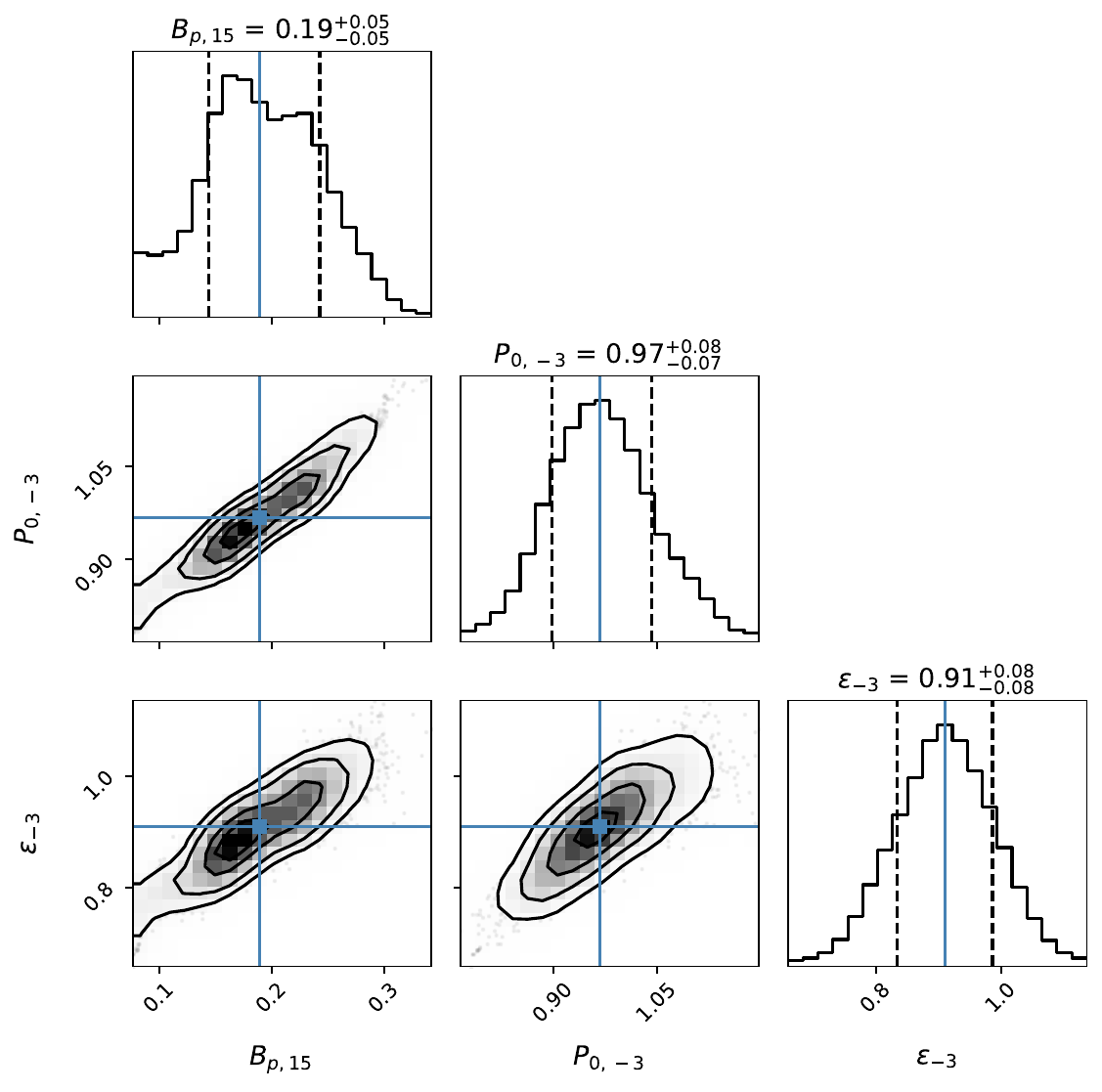}
\includegraphics  [angle=0,scale=0.2]   {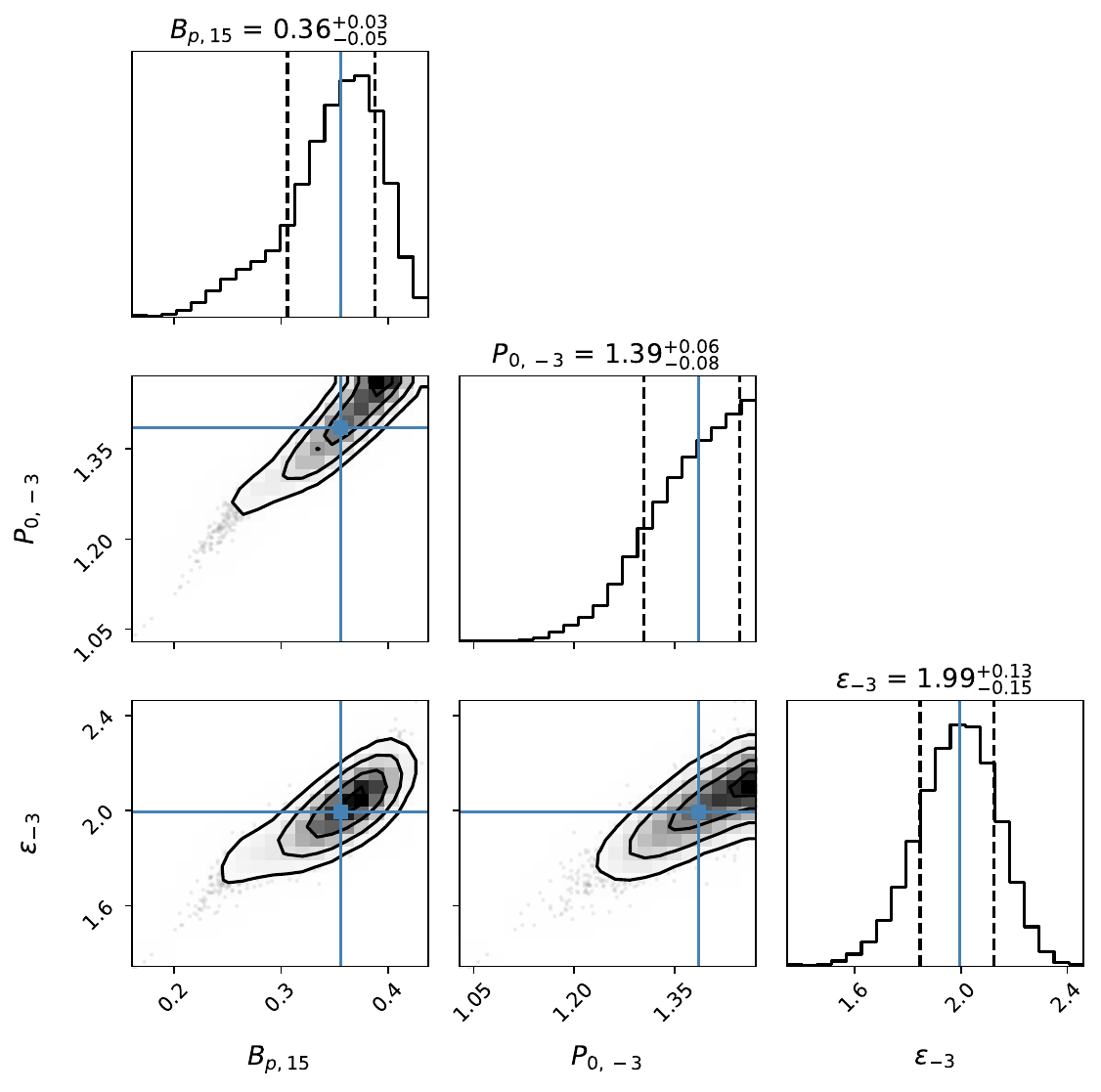}
\includegraphics  [angle=0,scale=0.2]   {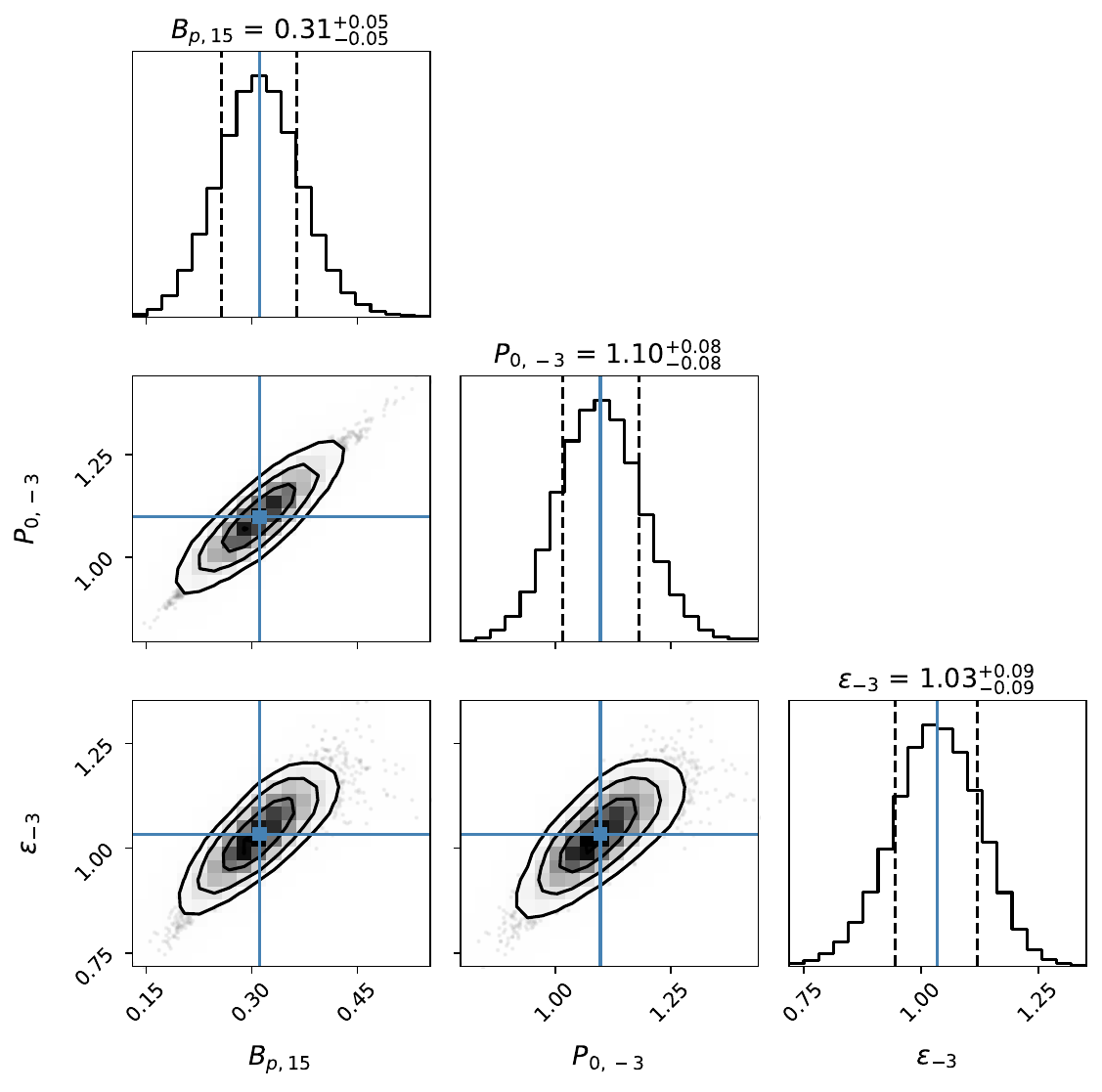}
\includegraphics  [angle=0,scale=0.2]   {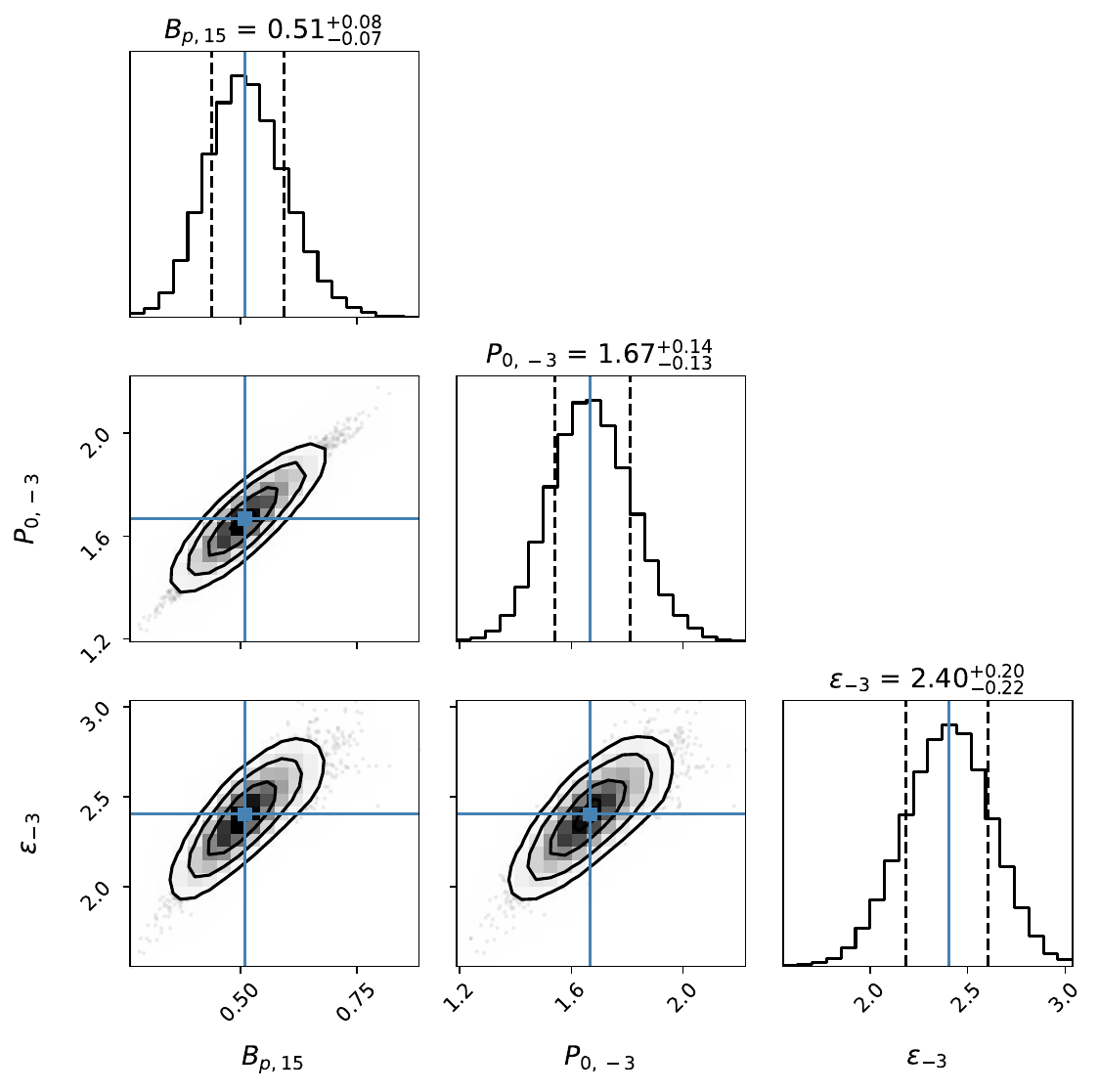} \\
\includegraphics  [angle=0,scale=0.24]   {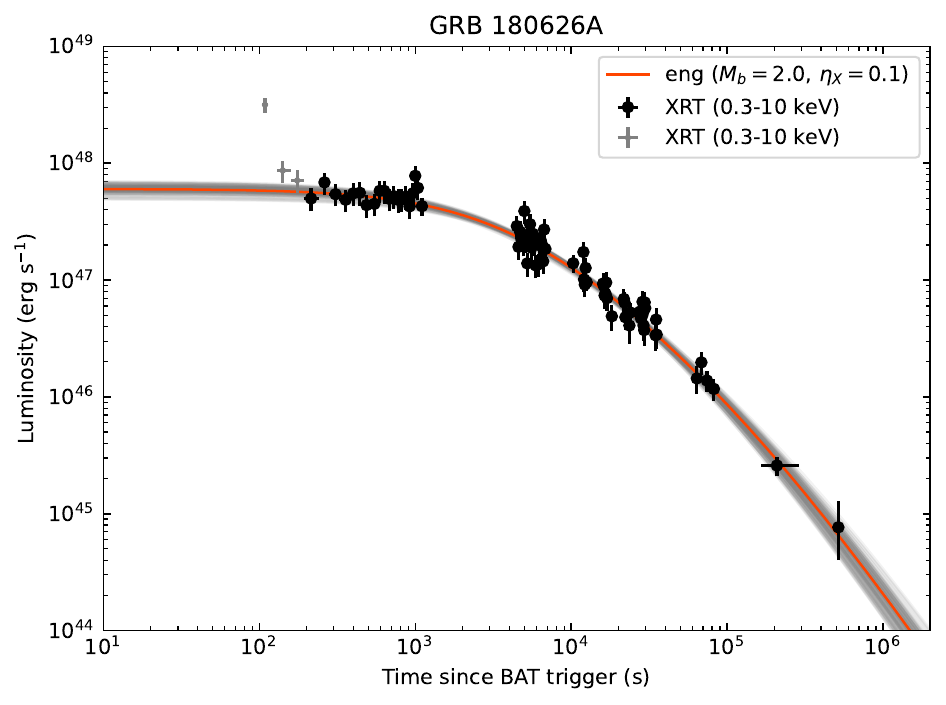}
\includegraphics  [angle=0,scale=0.24]   {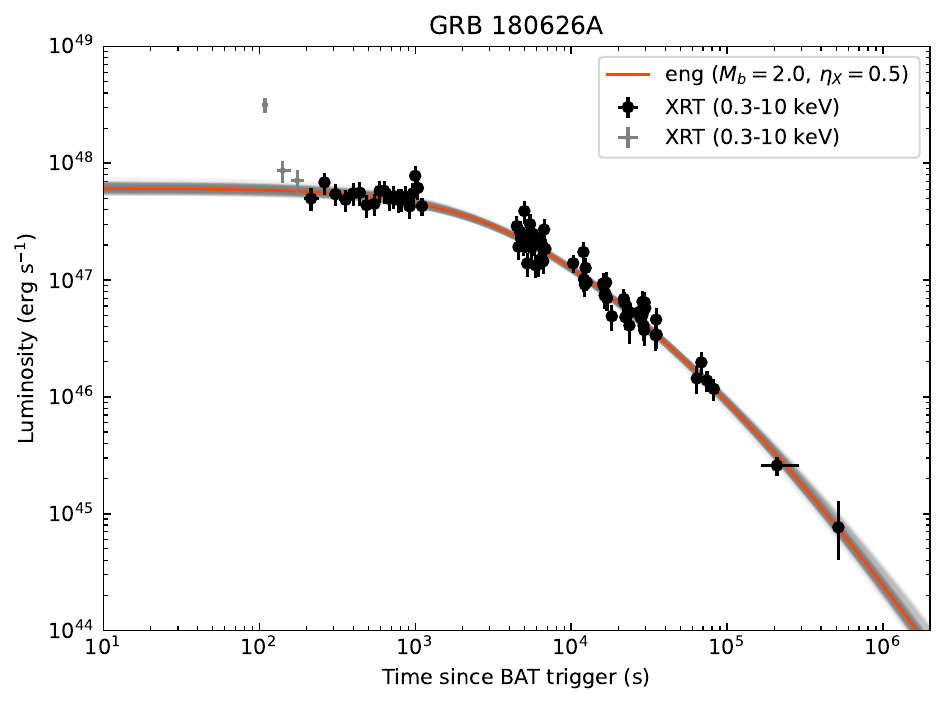}
\includegraphics  [angle=0,scale=0.24]   {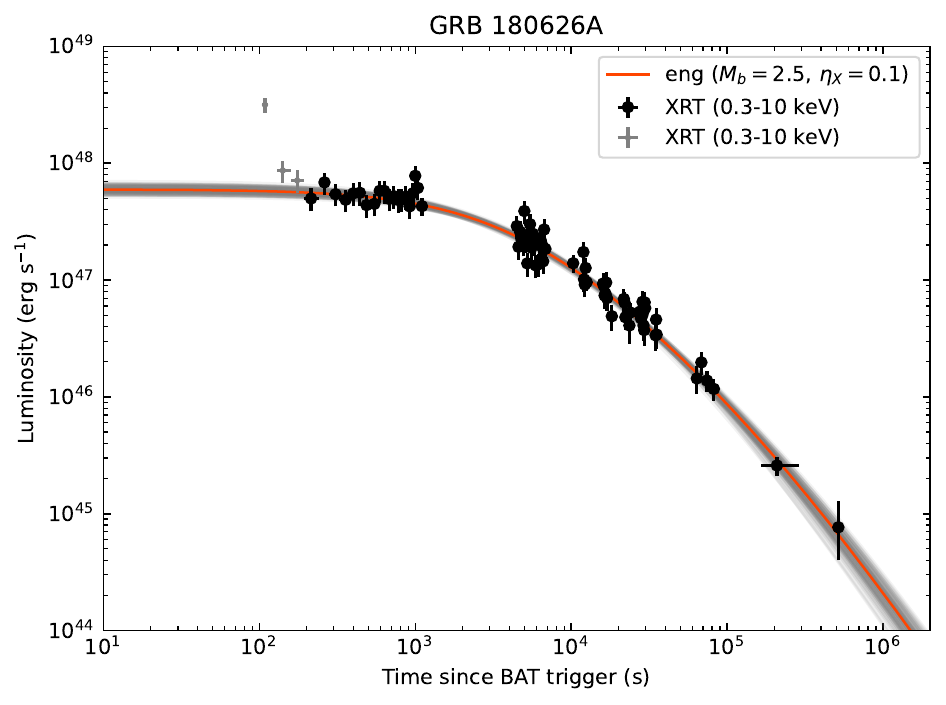}
\includegraphics  [angle=0,scale=0.24]   {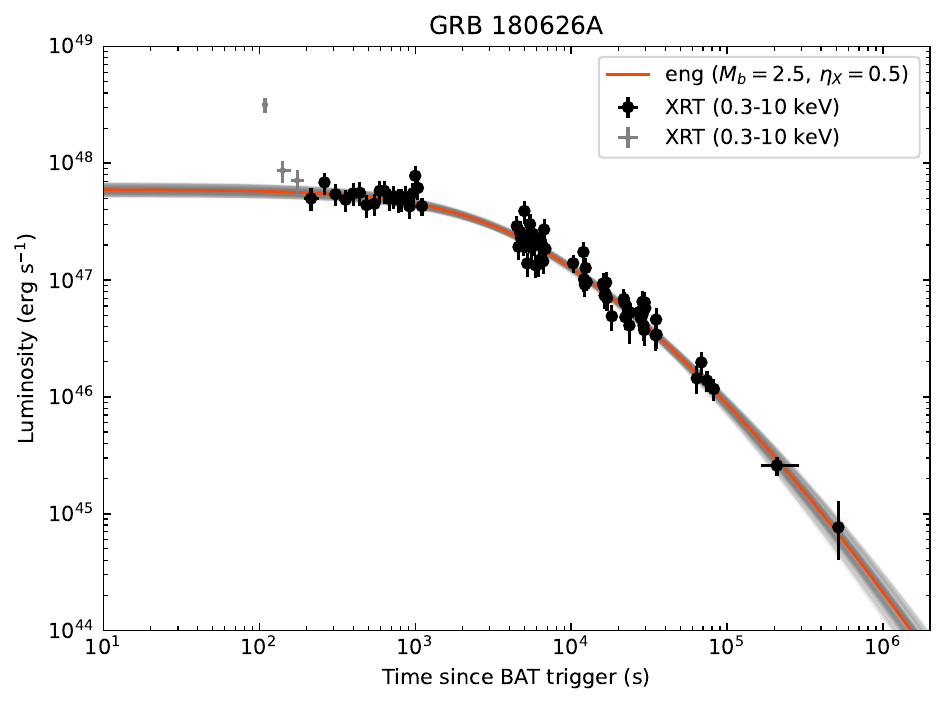} \\
\includegraphics  [angle=0,scale=0.2]   {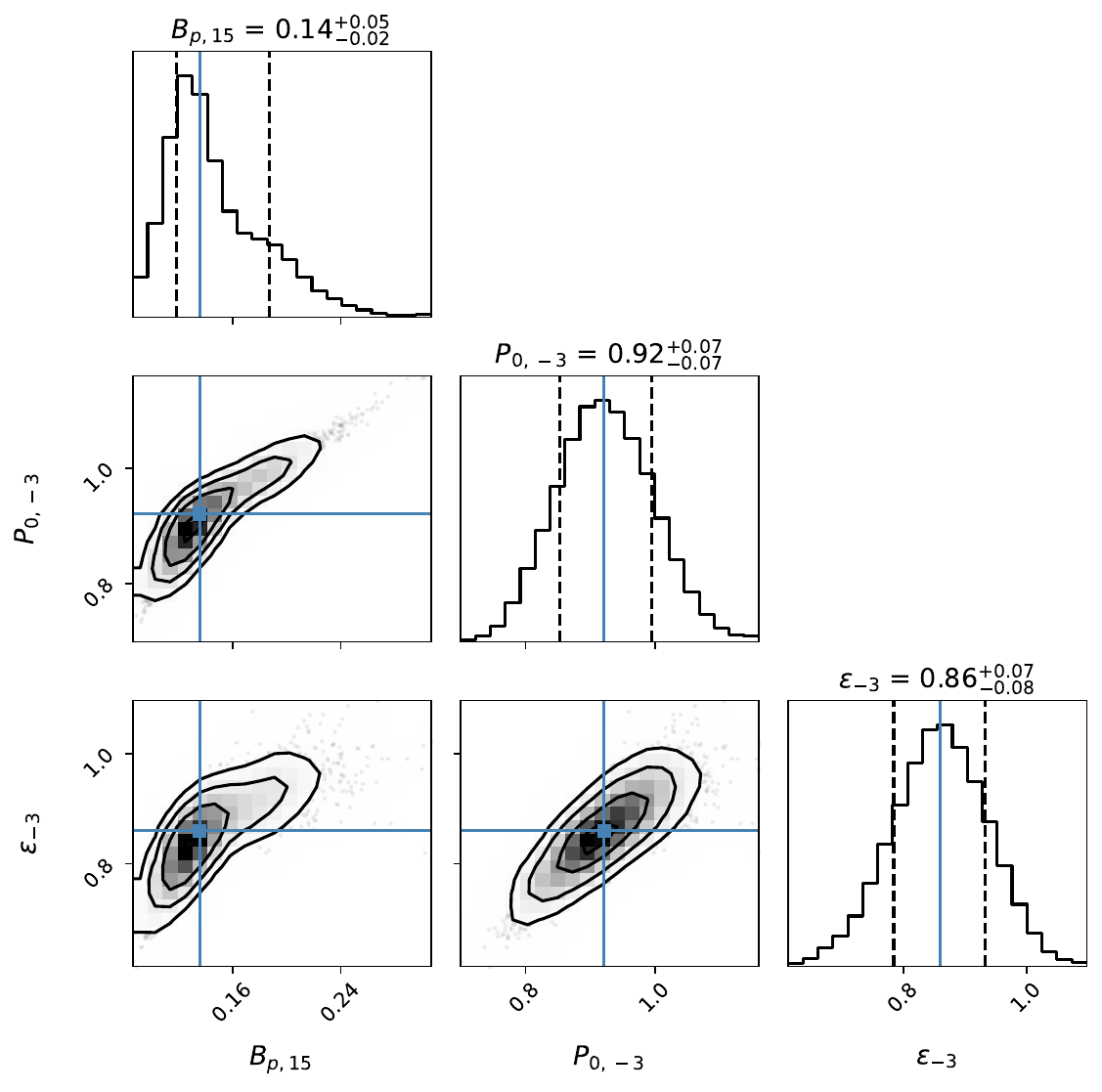}
\includegraphics  [angle=0,scale=0.2]   {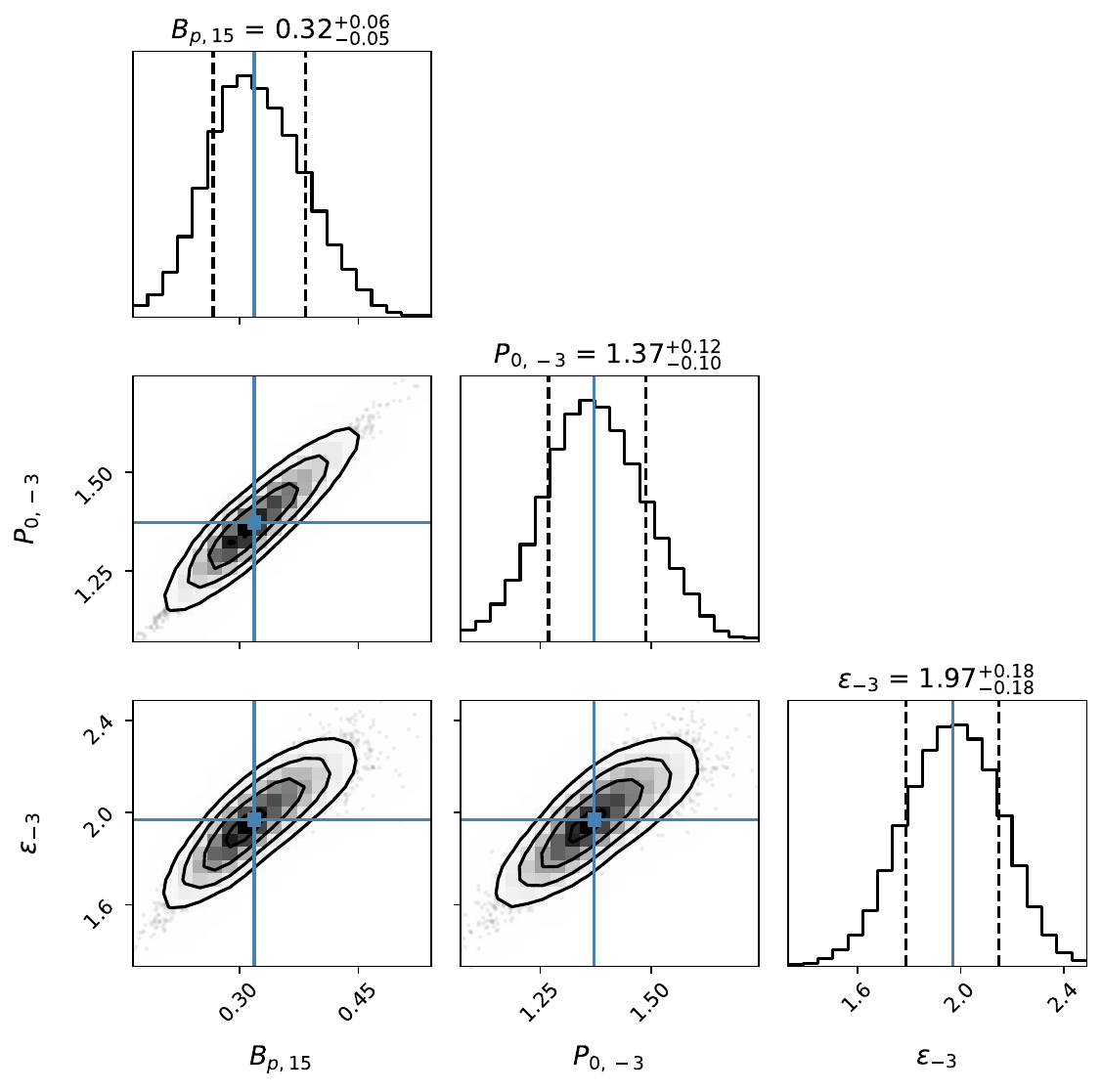}
\includegraphics  [angle=0,scale=0.2]   {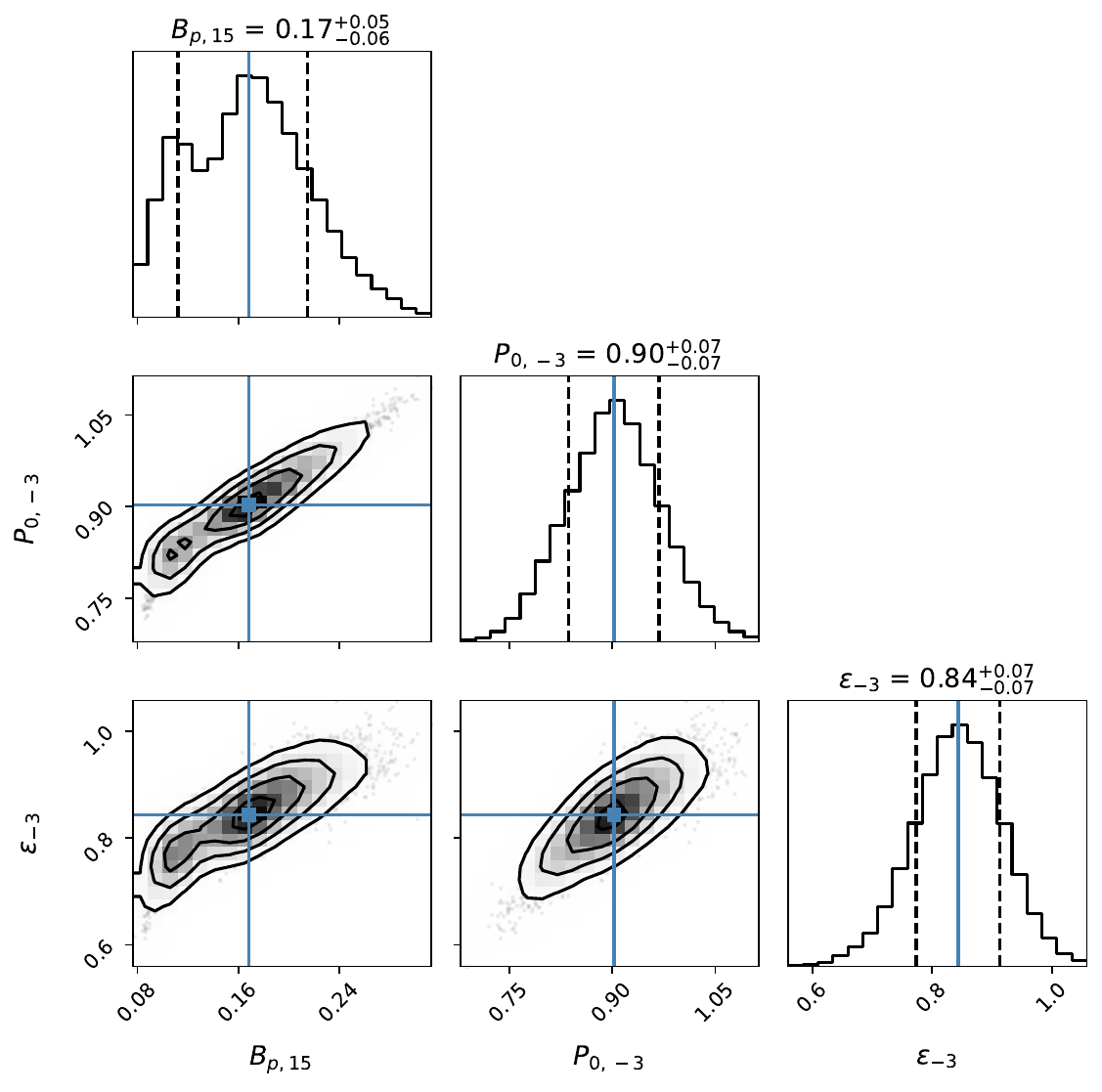}
\includegraphics  [angle=0,scale=0.2]   {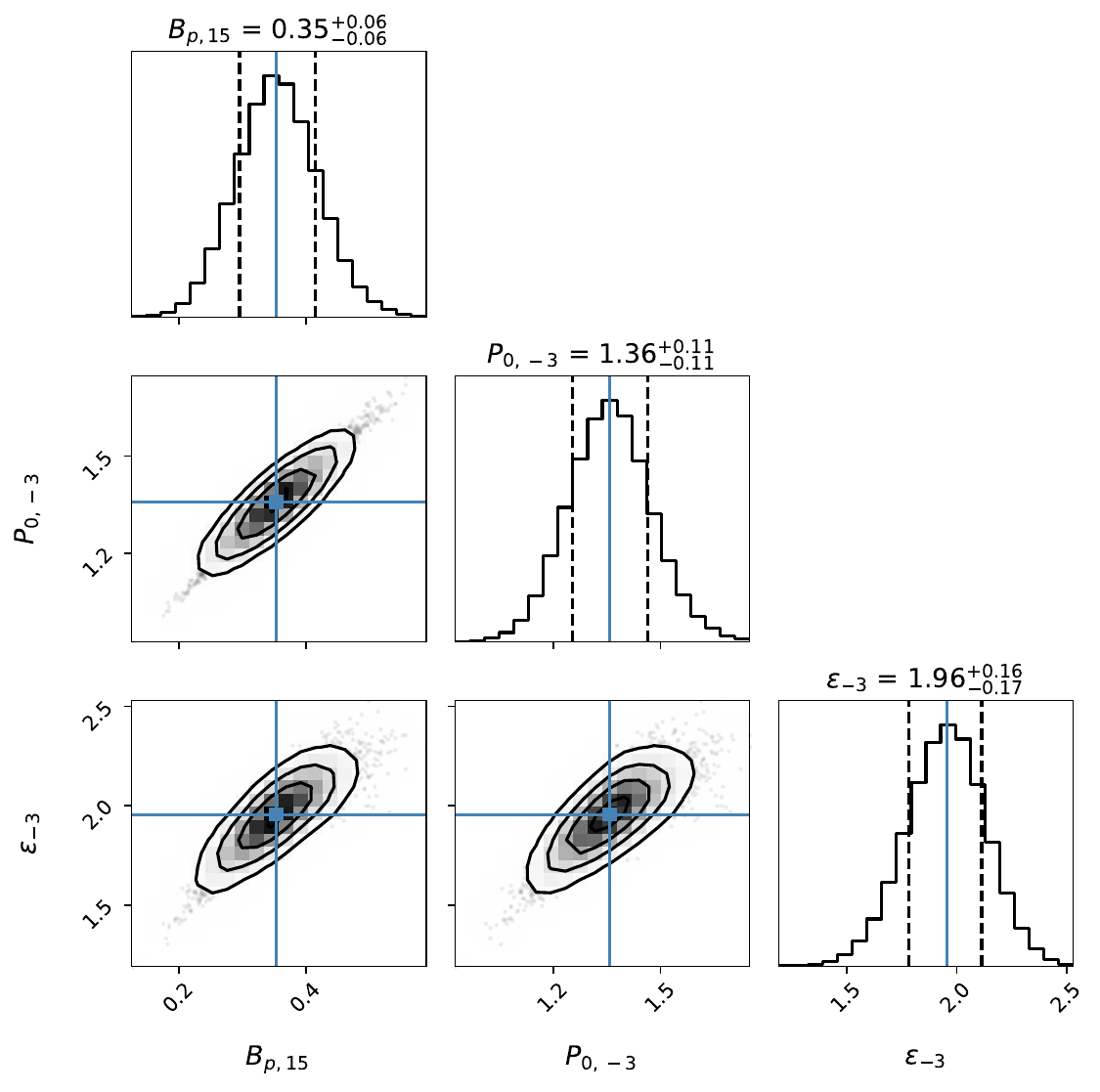} \\
\includegraphics  [angle=0,scale=0.24]   {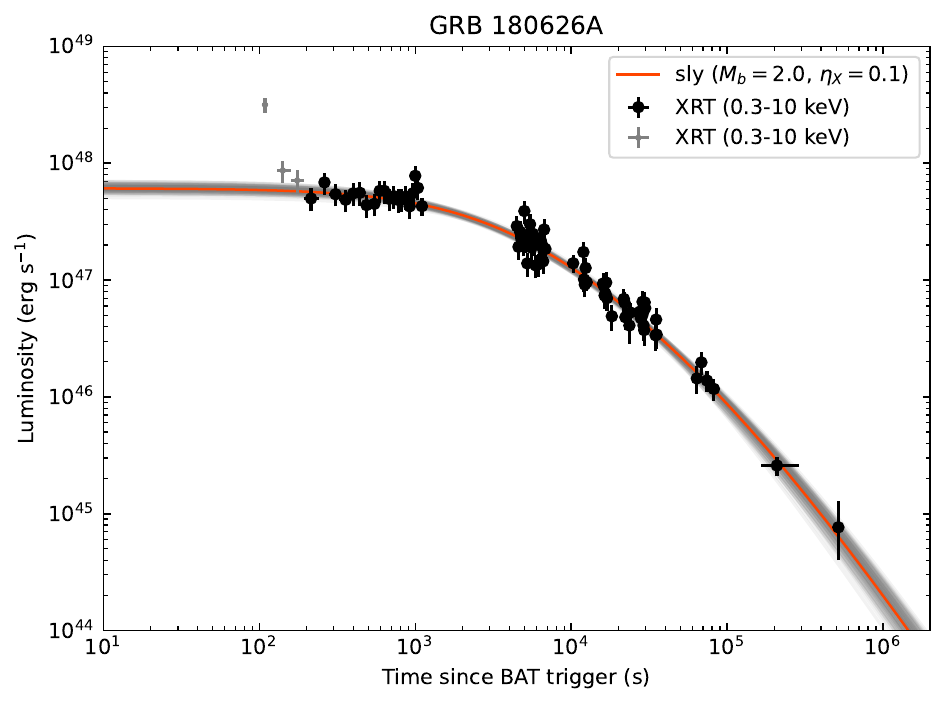}
\includegraphics  [angle=0,scale=0.24]   {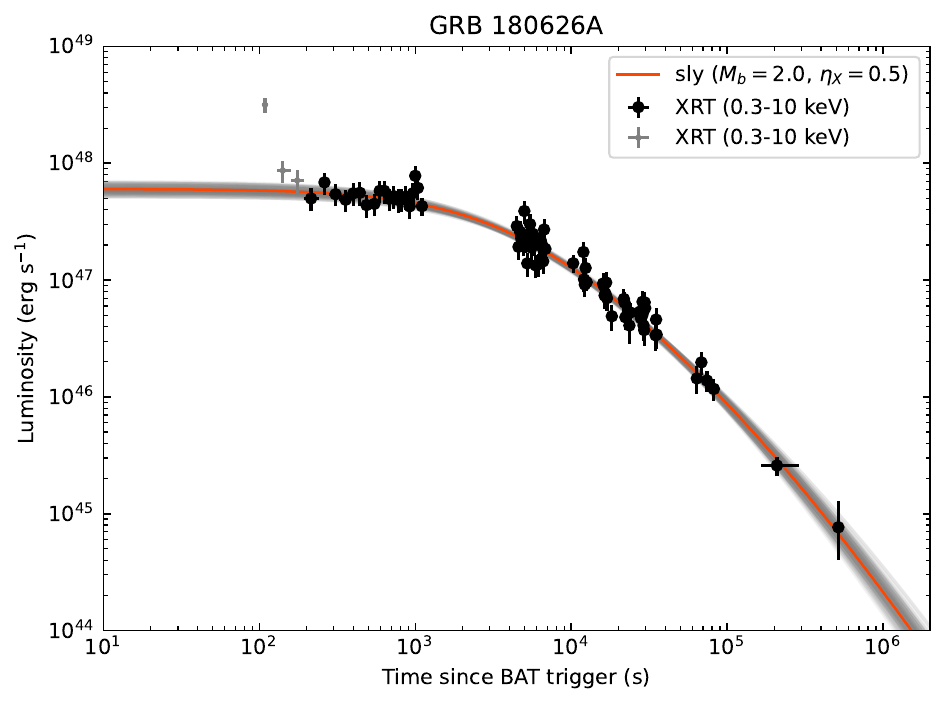}
\includegraphics  [angle=0,scale=0.24]   {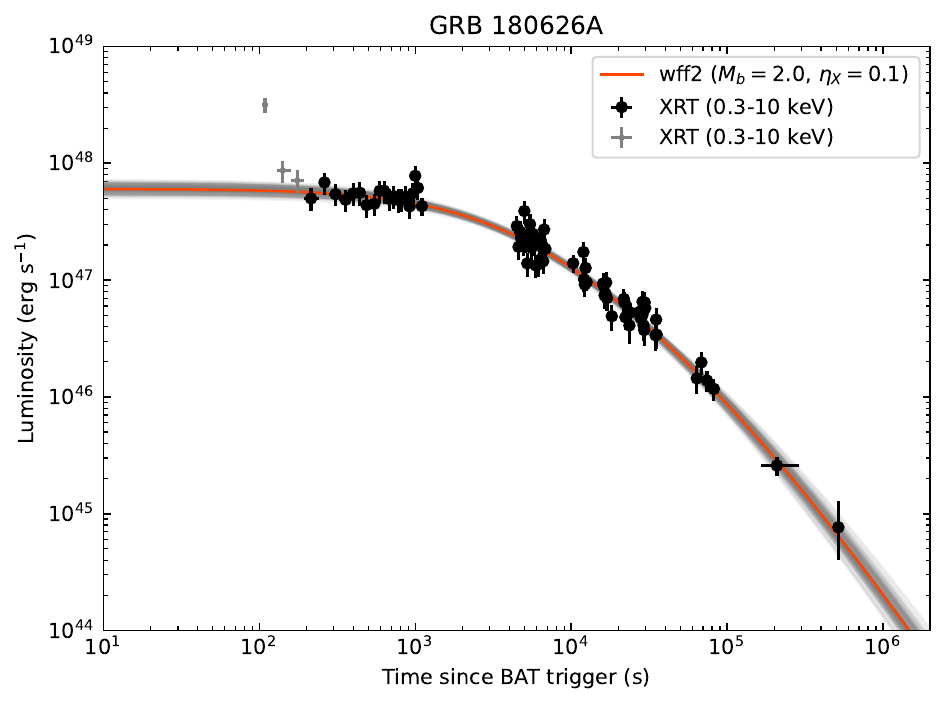}
\includegraphics  [angle=0,scale=0.24]   {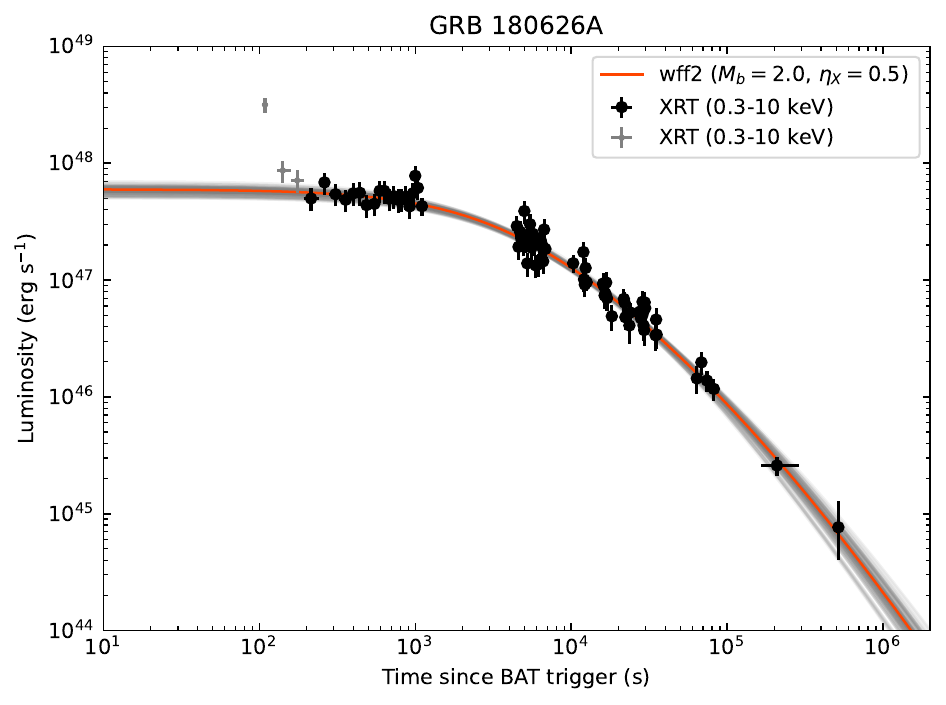} \\
\includegraphics  [angle=0,scale=0.2]   {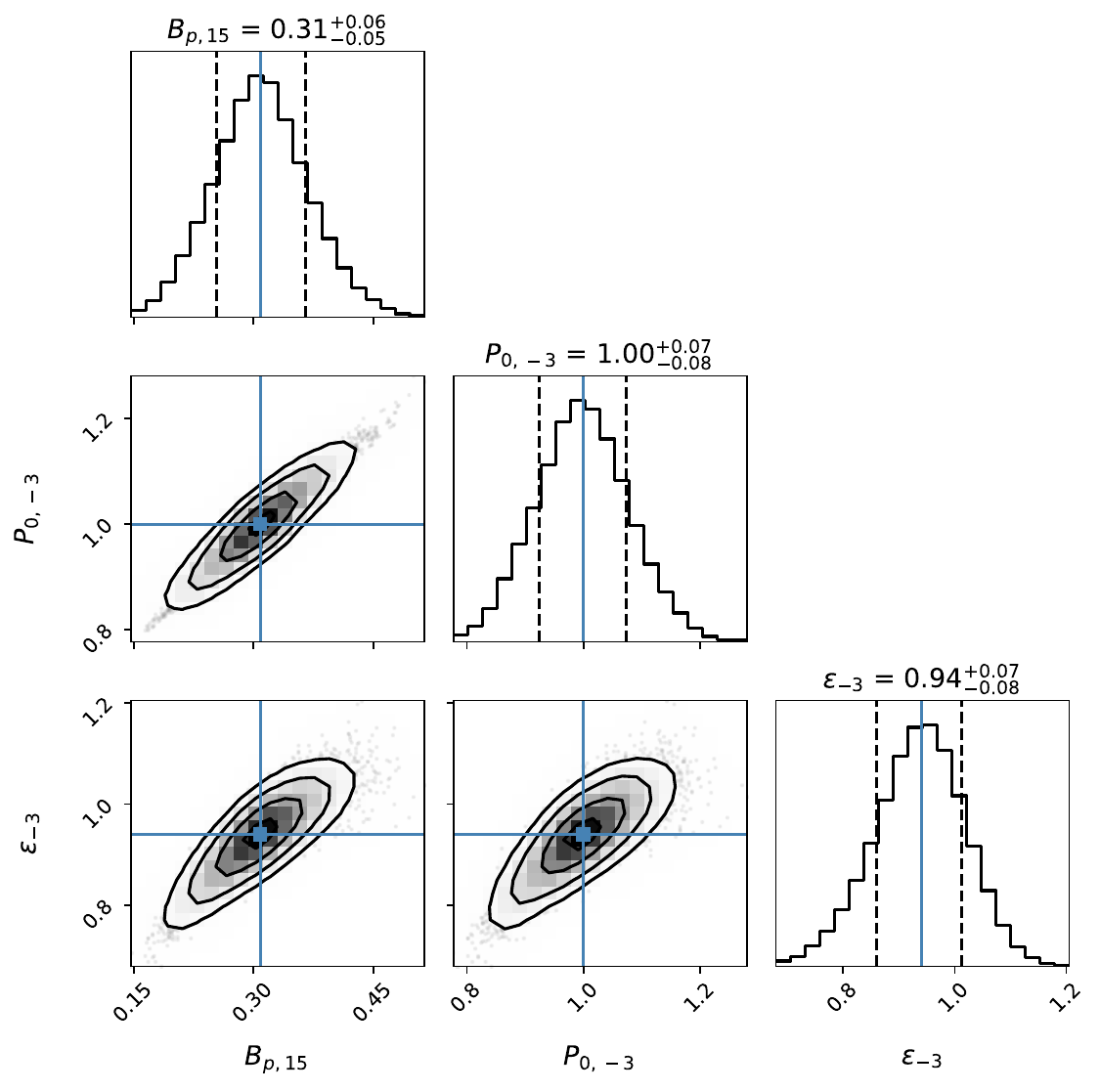}
\includegraphics  [angle=0,scale=0.24]   {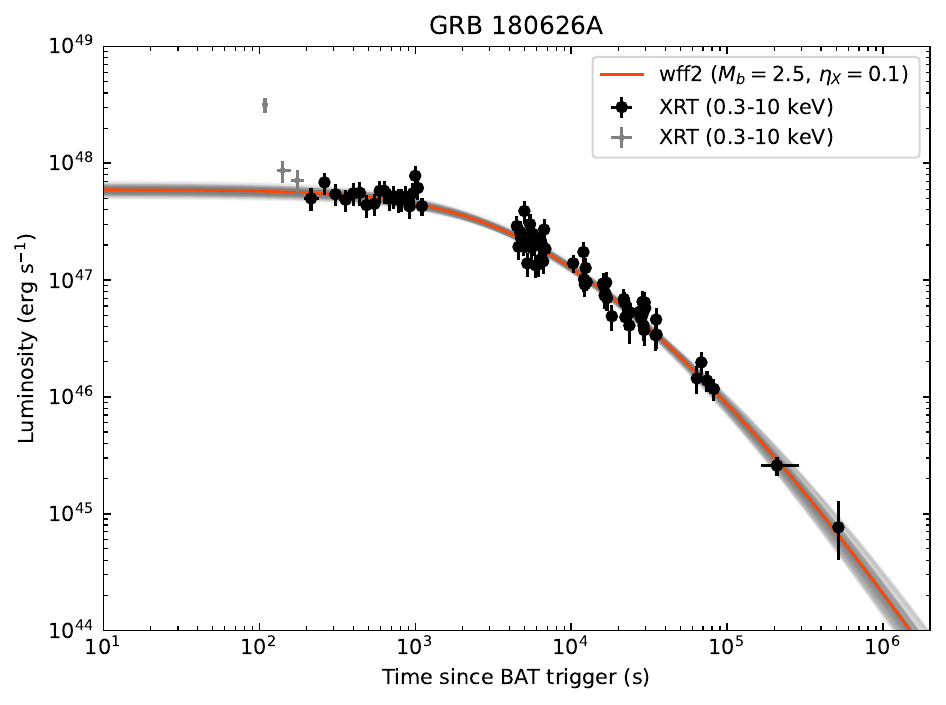}
\includegraphics  [angle=0,scale=0.2]   {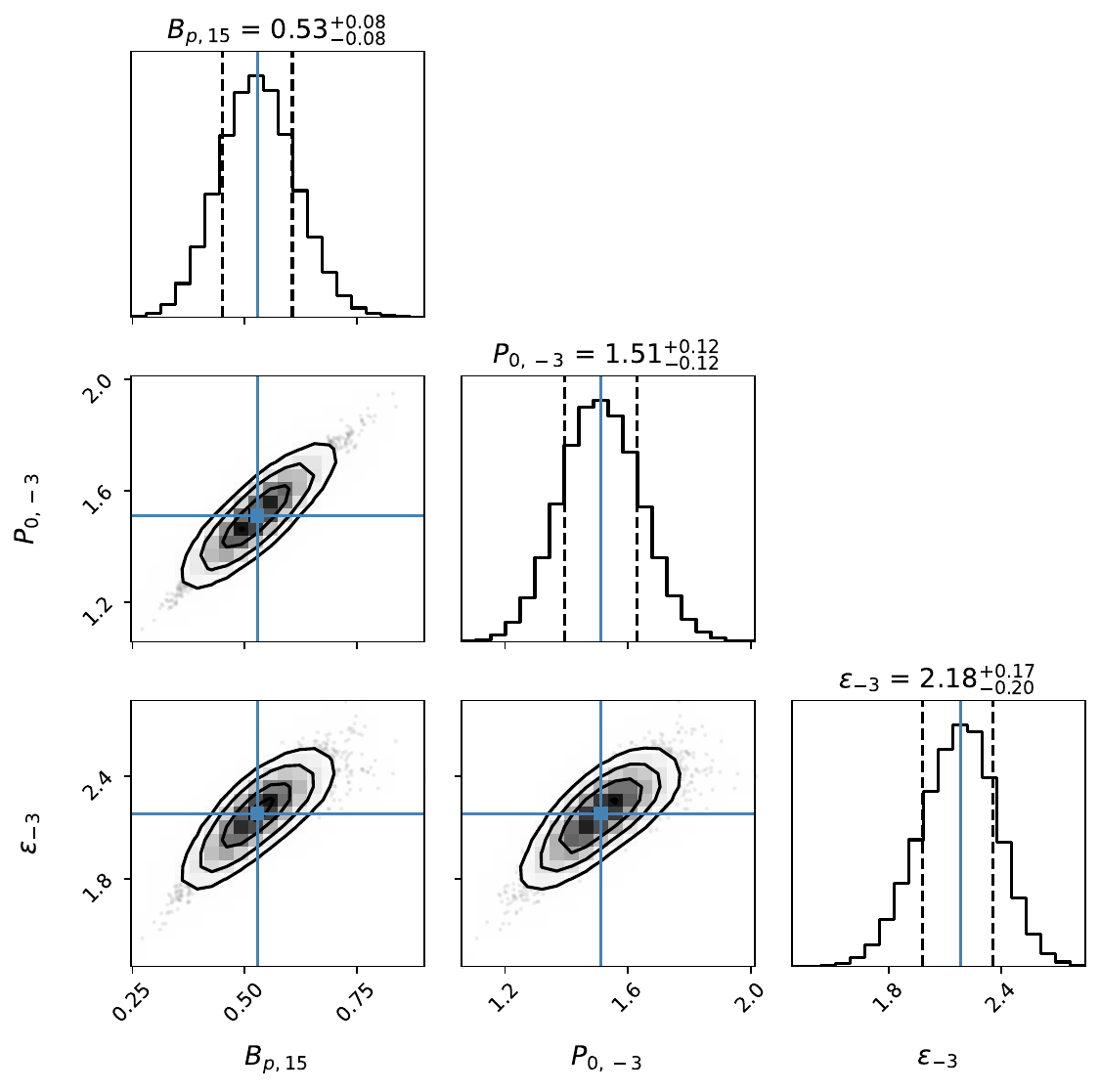}
\includegraphics  [angle=0,scale=0.24]   {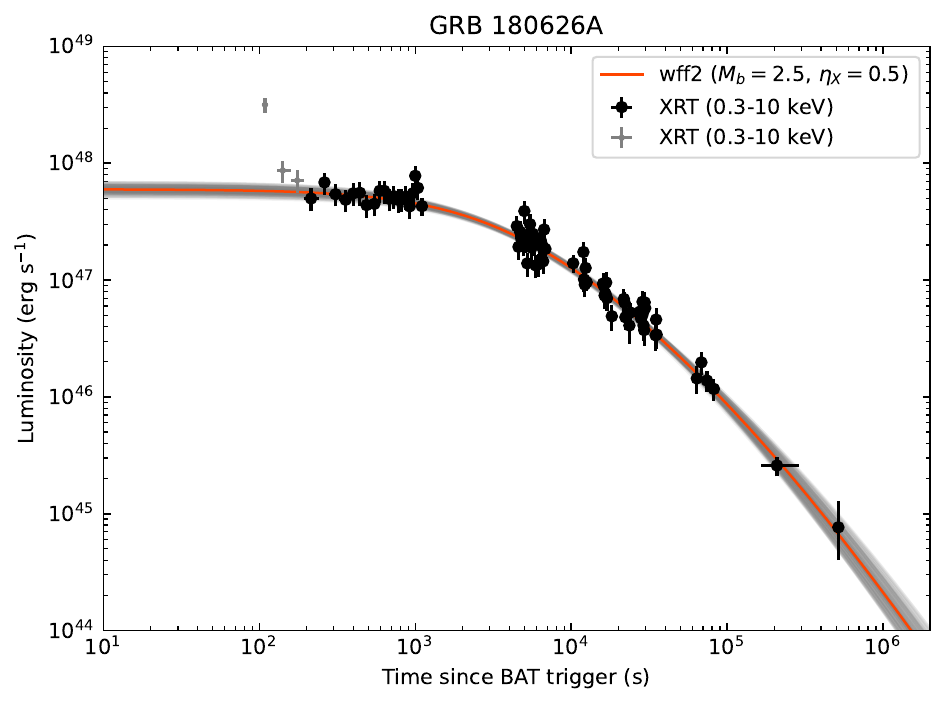} \\
\caption{EM+GW-dominated case: the corner plots and best-fitting results of GRB 180626A in four samples of EoSs with $M_{b}=2.0~M_{\odot},~2.5~M_{\odot}$ and $\eta_{\rm X}=0.1,~0.5$, respectively.}
\label{fig:EM+GW_mcmc}
\end{figure*}

\begin{figure*}
\centering
\includegraphics  [angle=0,scale=0.25] {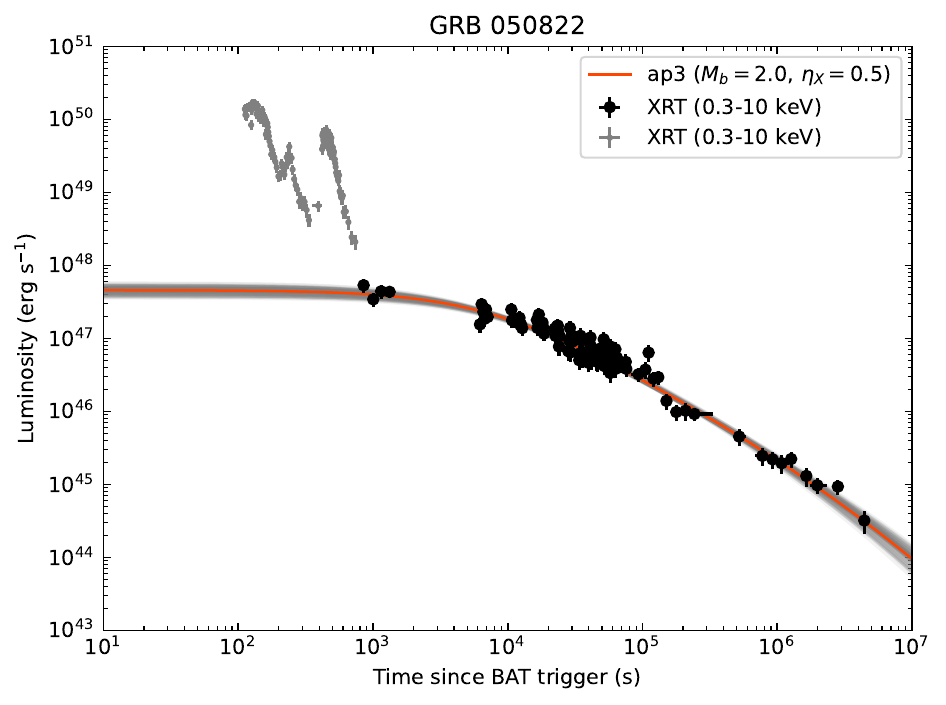}  
\includegraphics  [angle=0,scale=0.25] {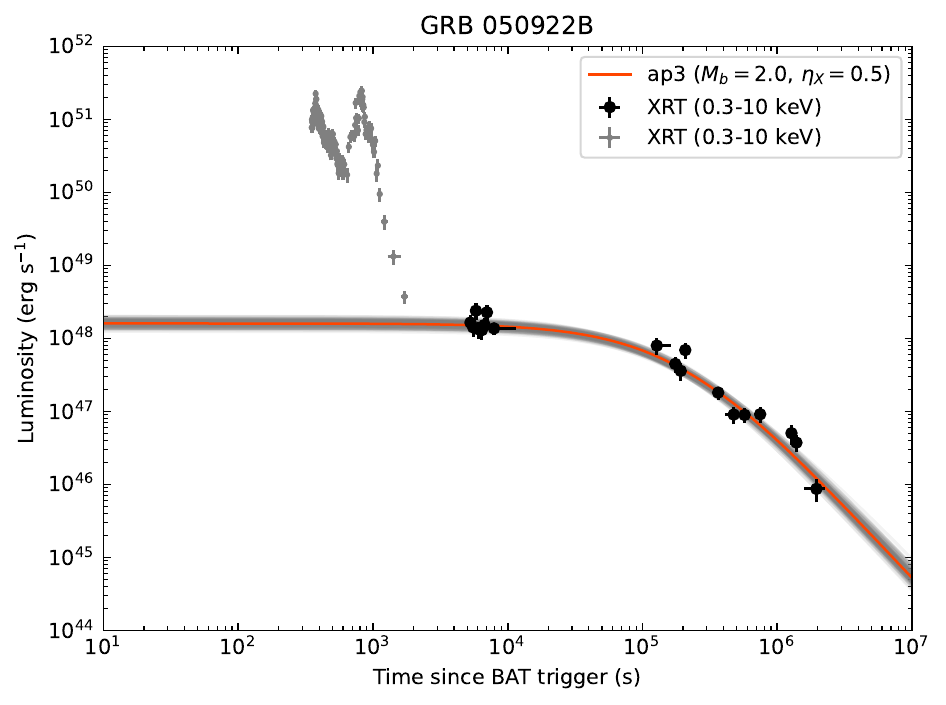}
\includegraphics  [angle=0,scale=0.25] {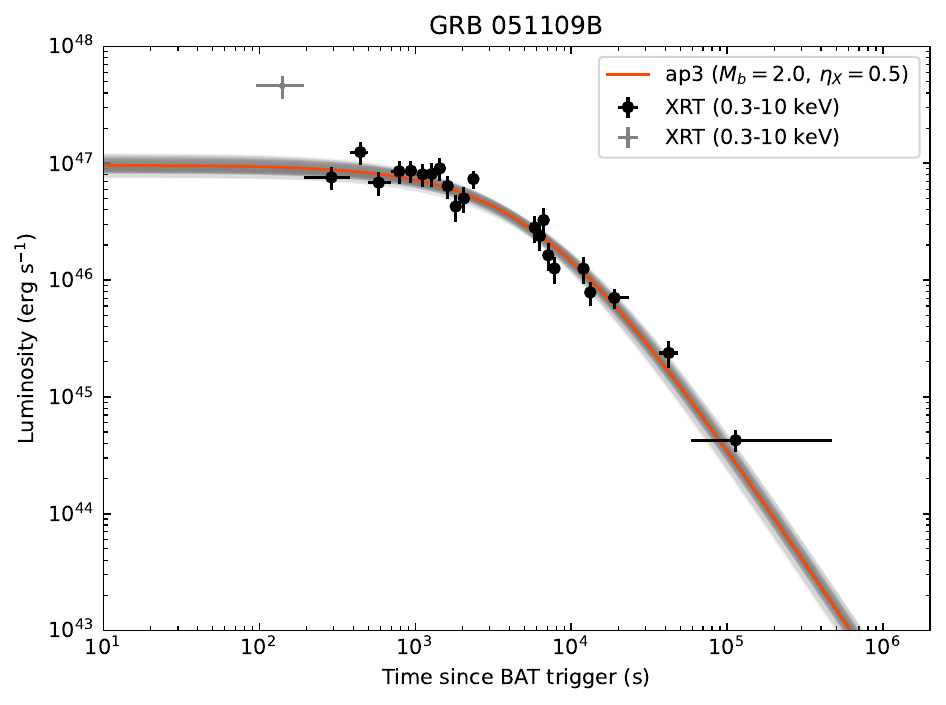}
\includegraphics  [angle=0,scale=0.25] {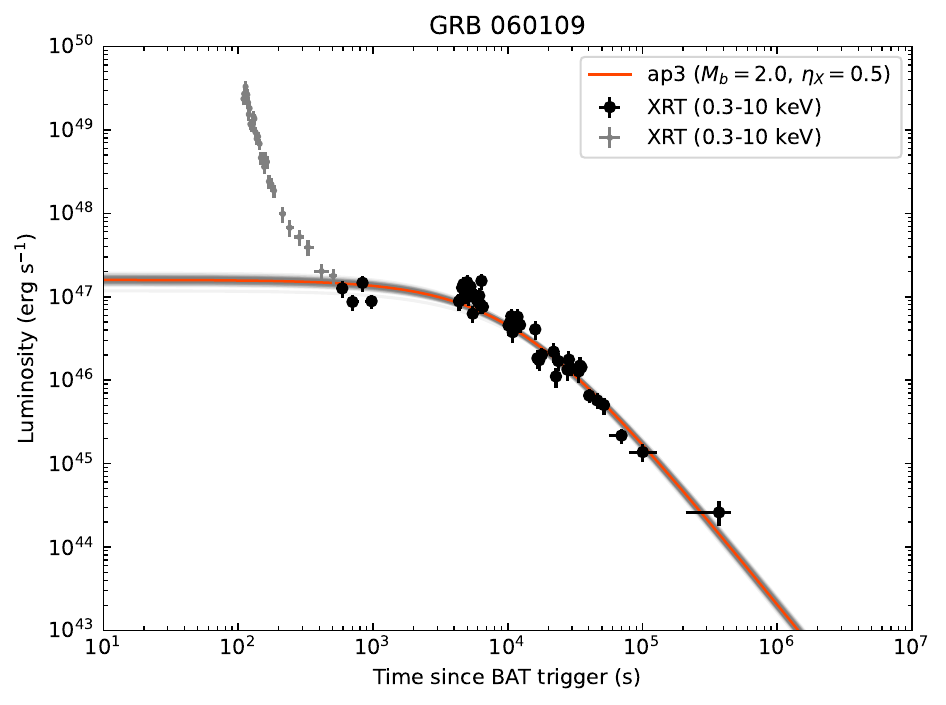}\\
\includegraphics  [angle=0,scale=0.25] {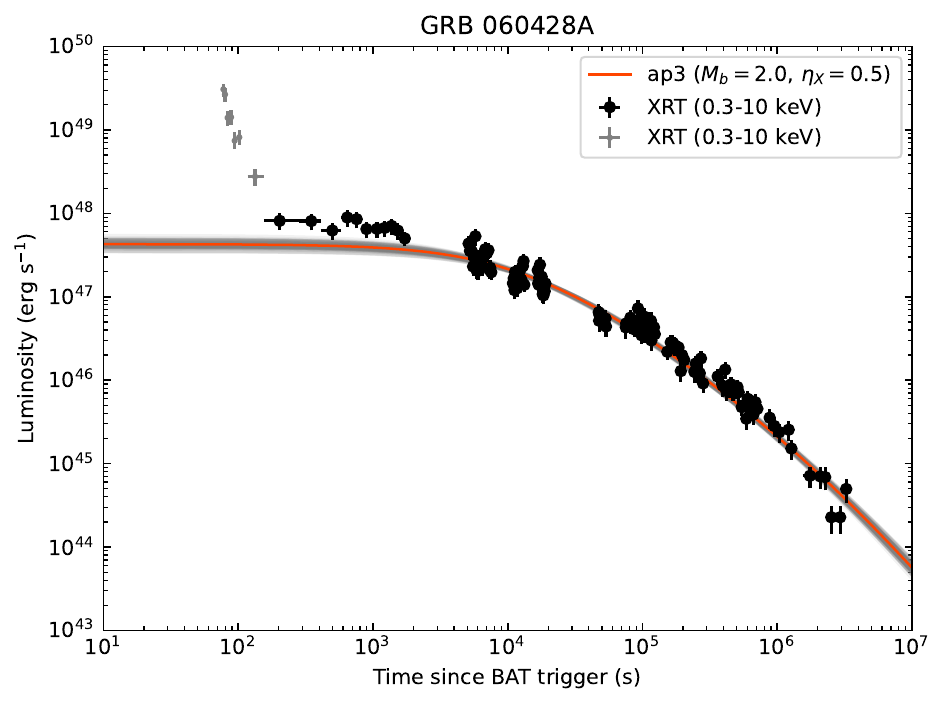}
\includegraphics  [angle=0,scale=0.25] {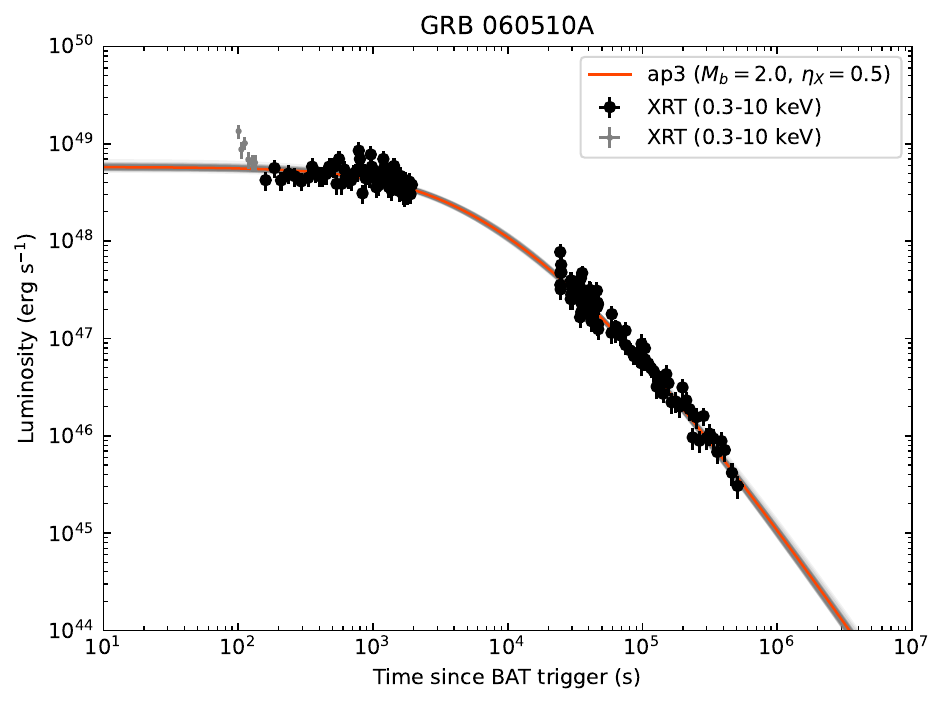}
\includegraphics  [angle=0,scale=0.25] {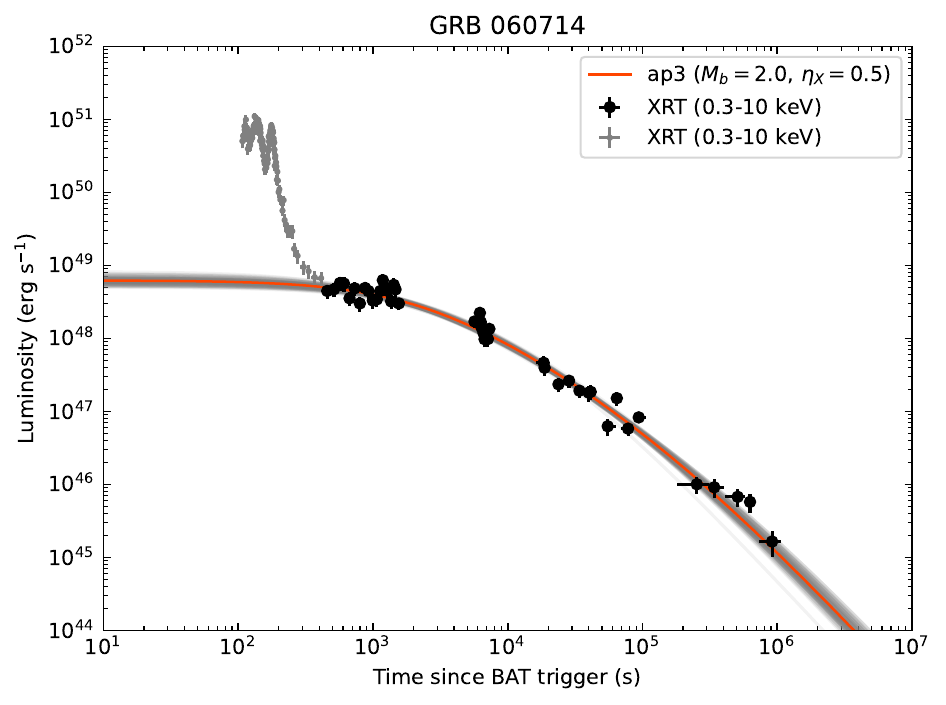}
\includegraphics  [angle=0,scale=0.25] {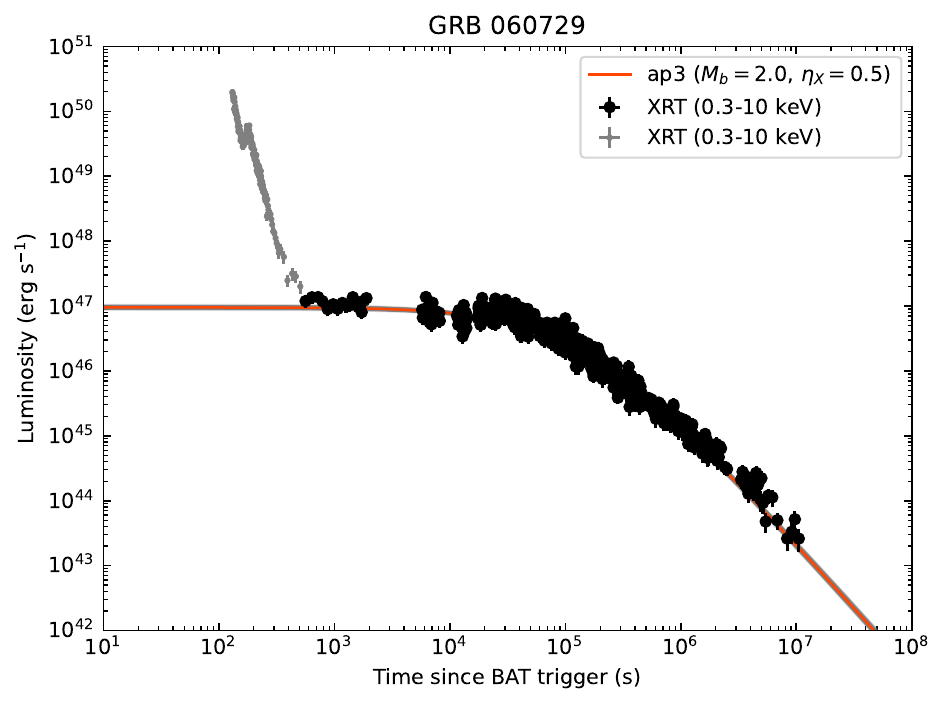}\\
\includegraphics  [angle=0,scale=0.25] {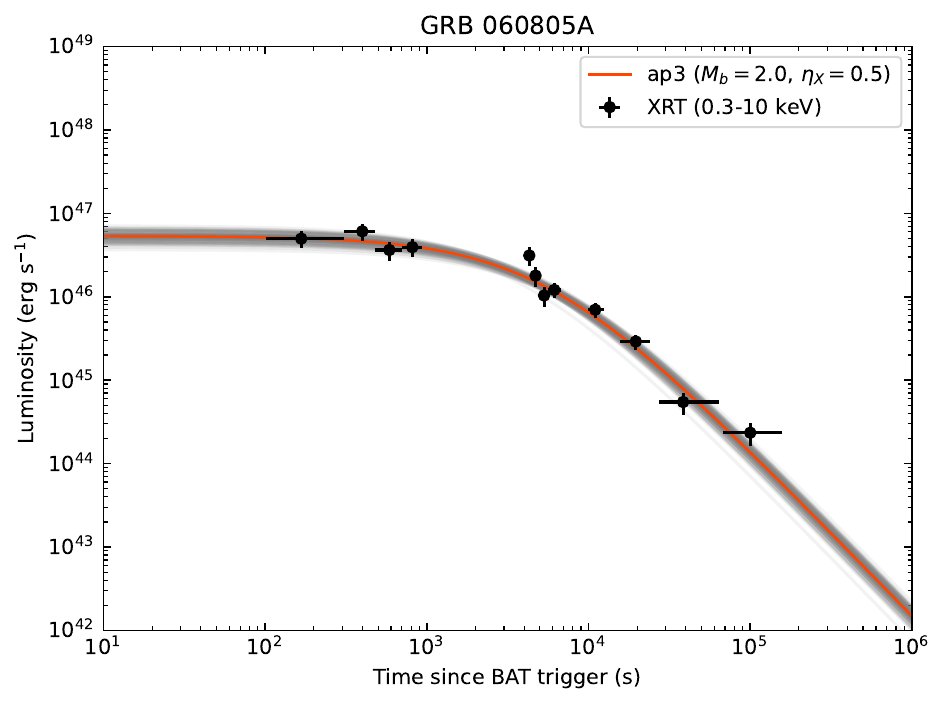}
\includegraphics  [angle=0,scale=0.25] {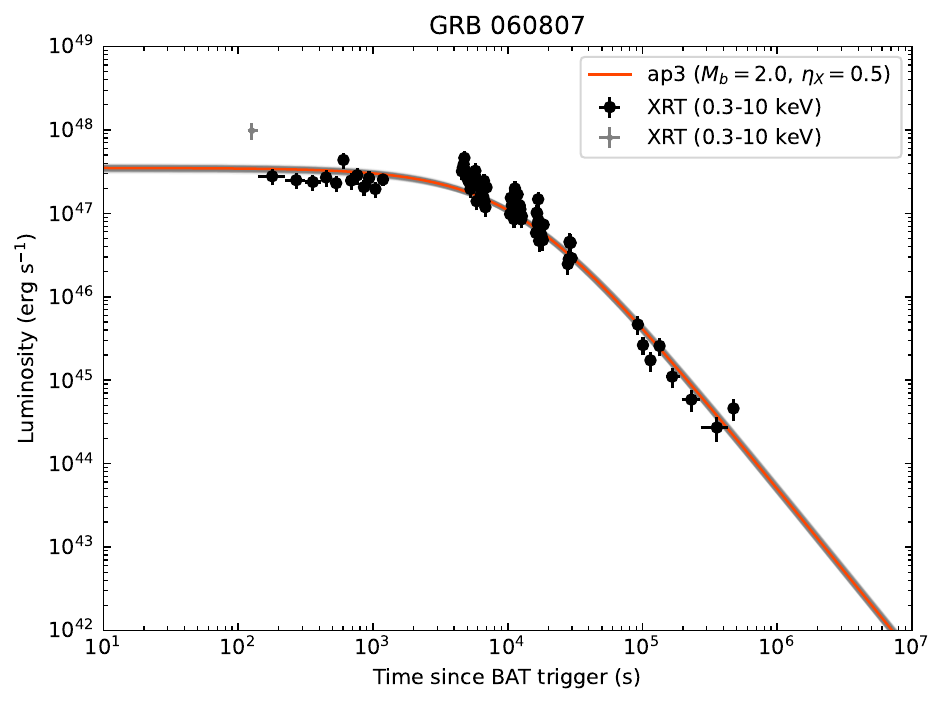}
\includegraphics  [angle=0,scale=0.25] {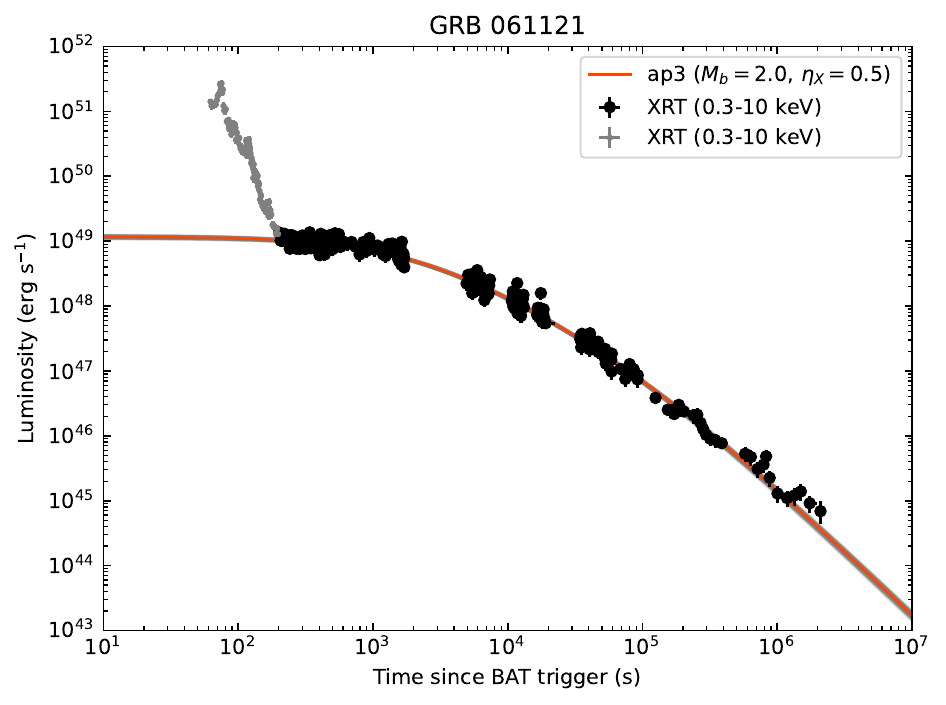}
\includegraphics  [angle=0,scale=0.25] {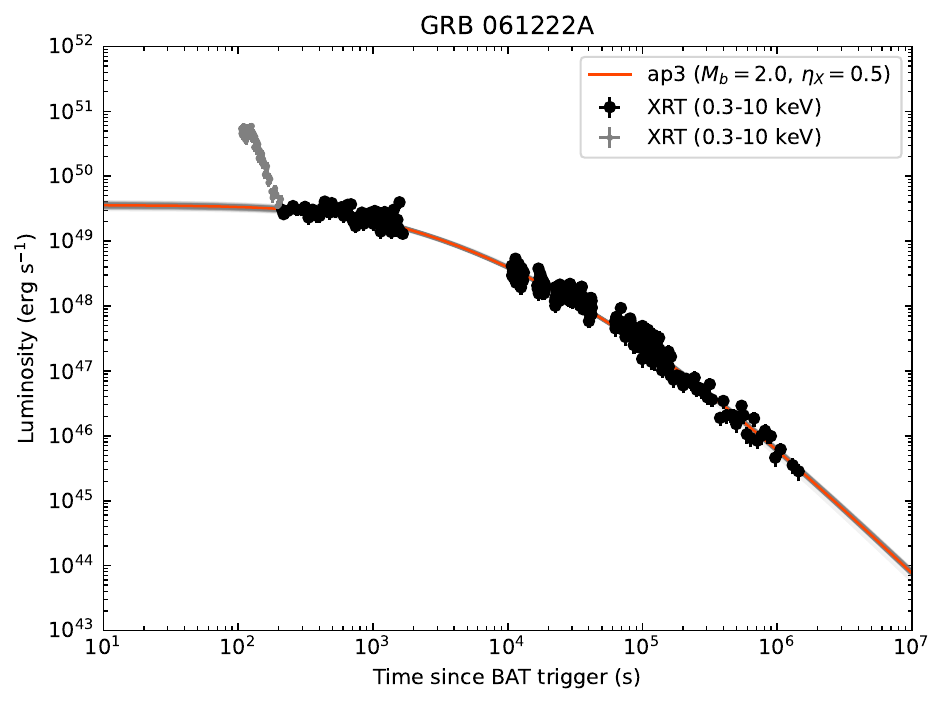}\\
\includegraphics  [angle=0,scale=0.25] {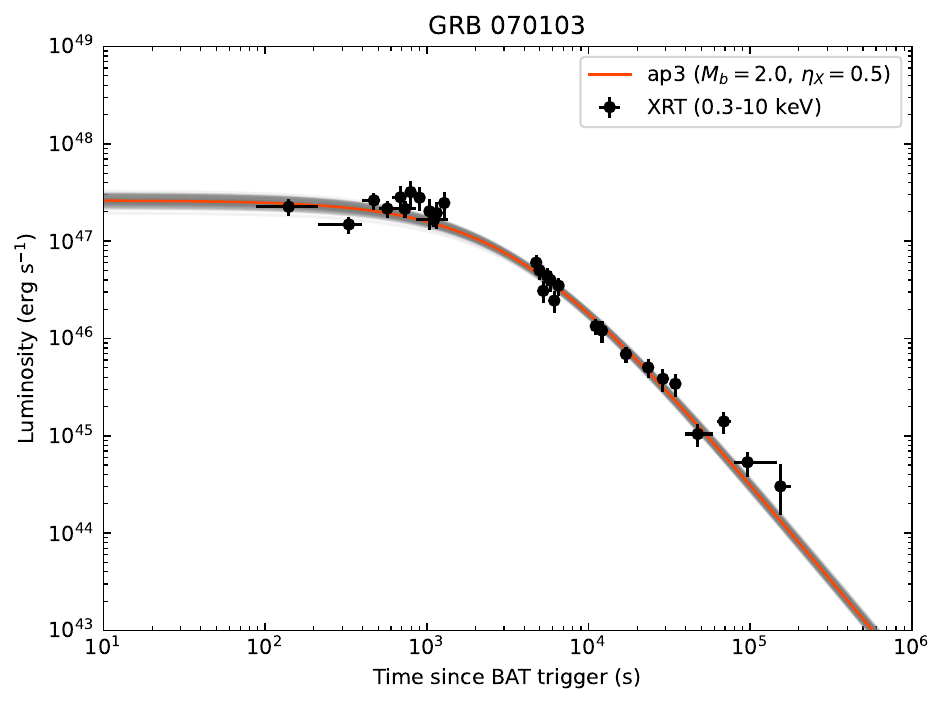}
\includegraphics  [angle=0,scale=0.25] {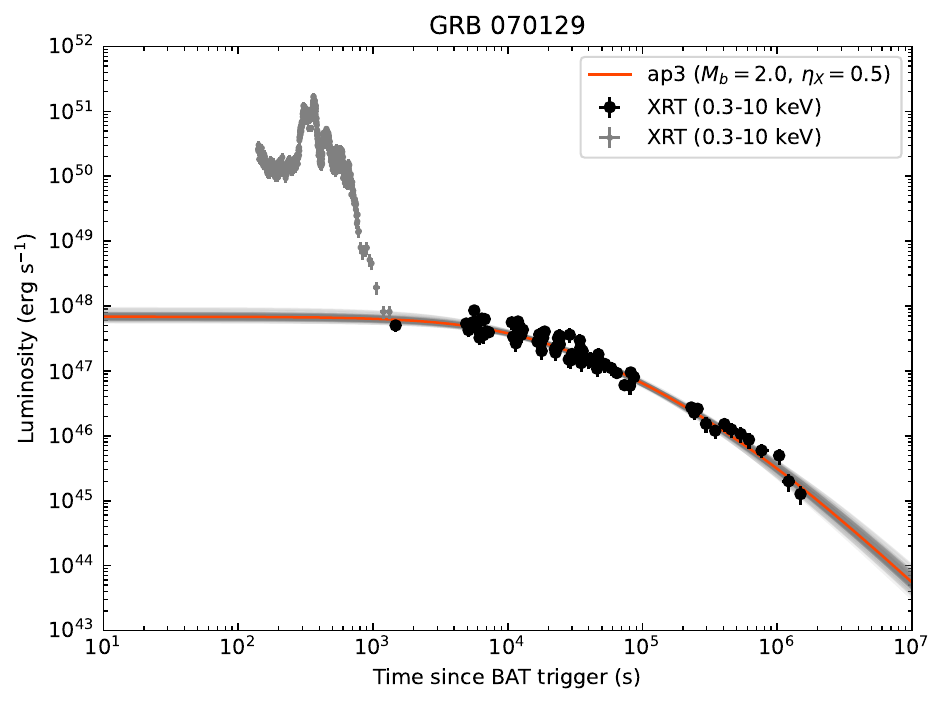}
\includegraphics  [angle=0,scale=0.25] {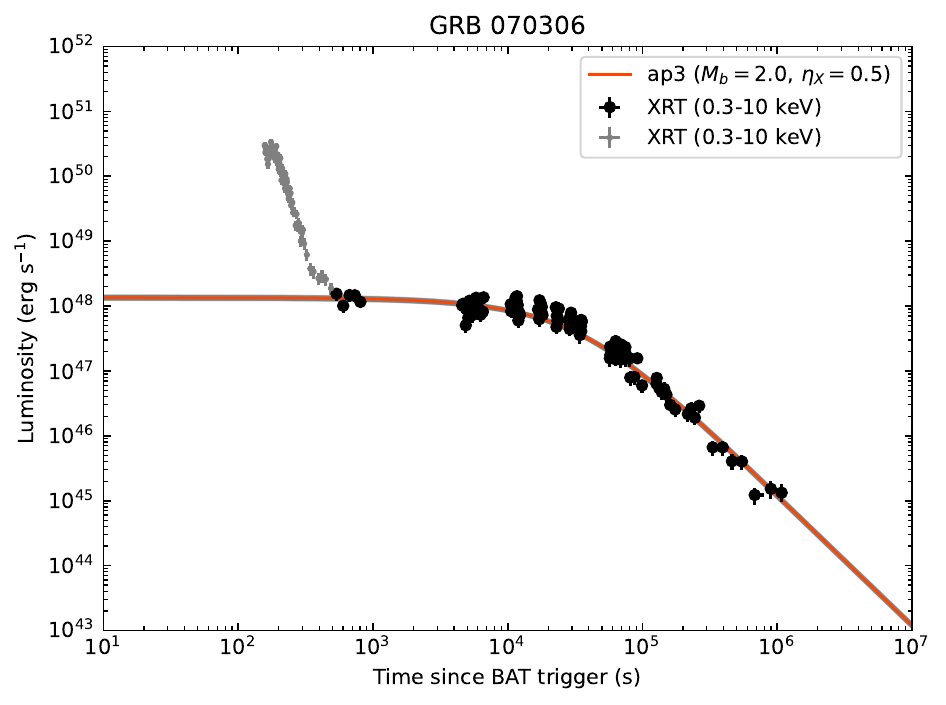}
\includegraphics  [angle=0,scale=0.25] {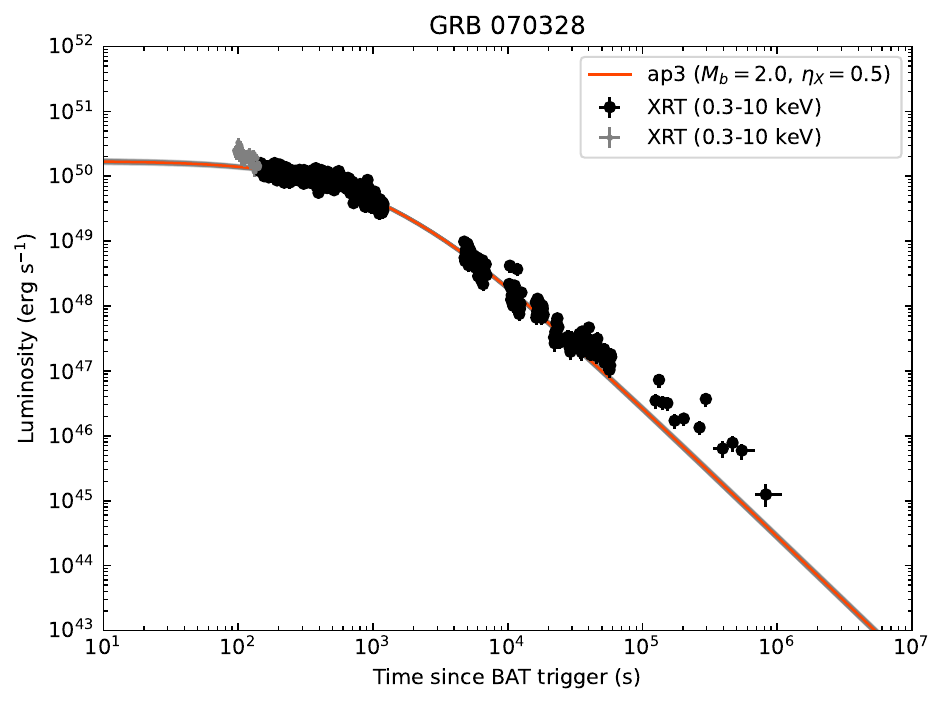}\\
\includegraphics  [angle=0,scale=0.25] {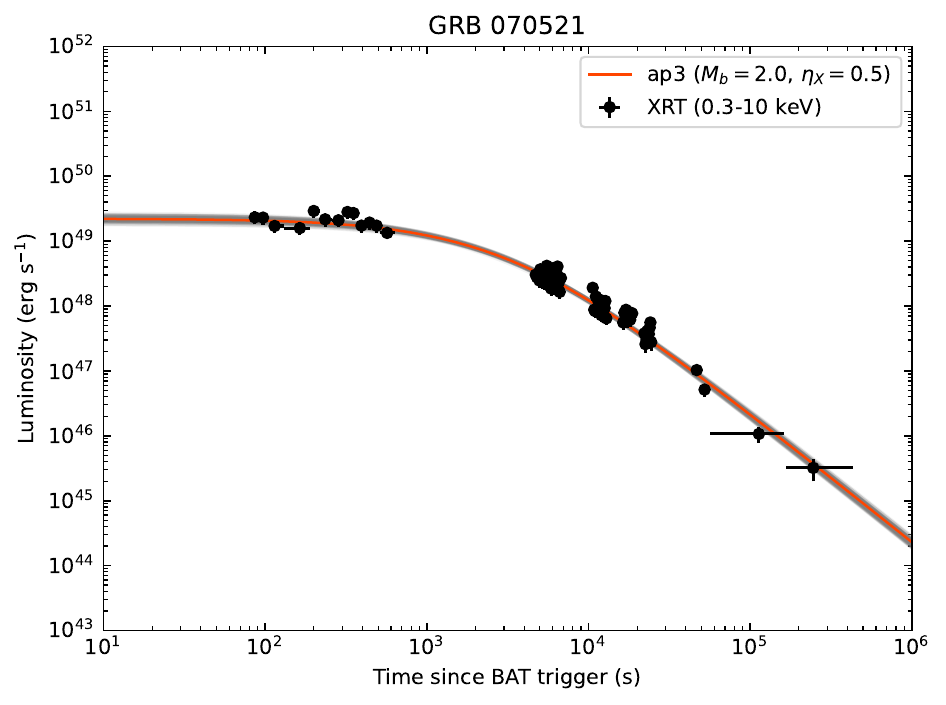}
\includegraphics  [angle=0,scale=0.25] {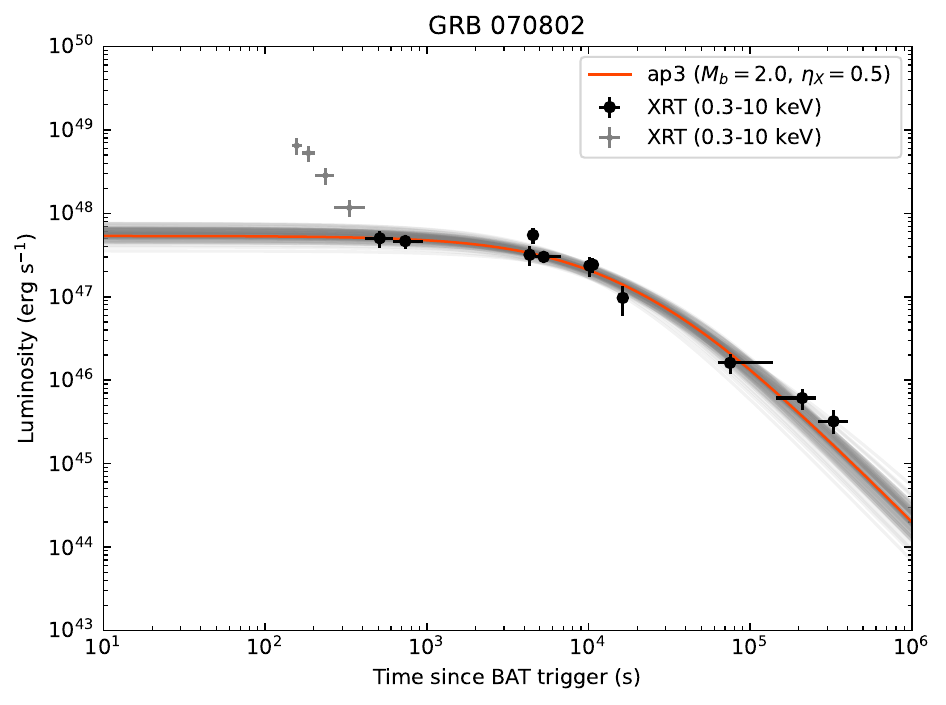}
\includegraphics  [angle=0,scale=0.25] {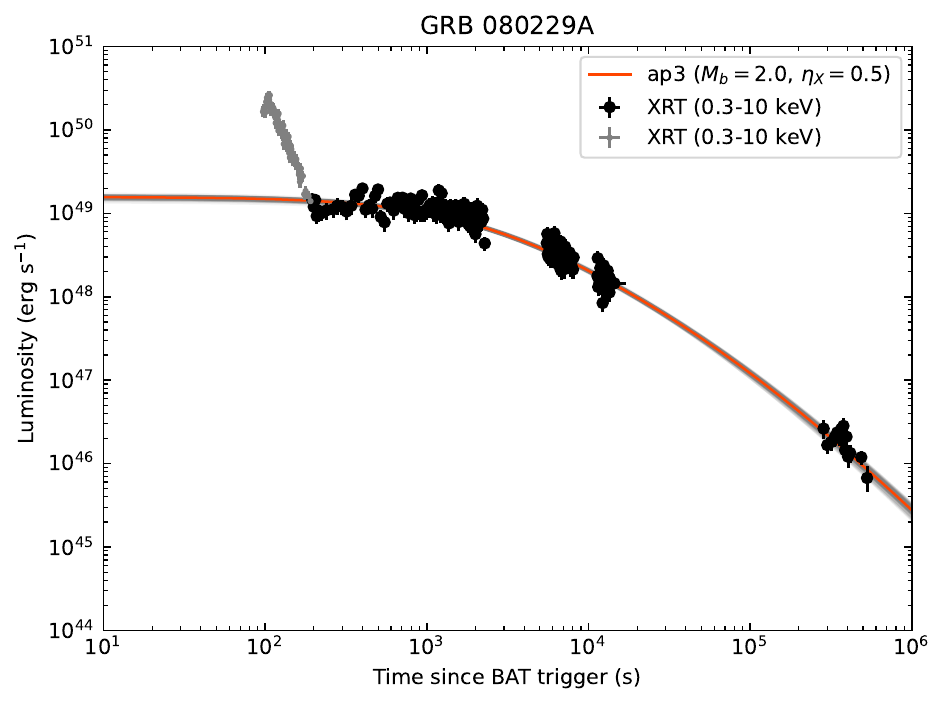}
\includegraphics  [angle=0,scale=0.25] {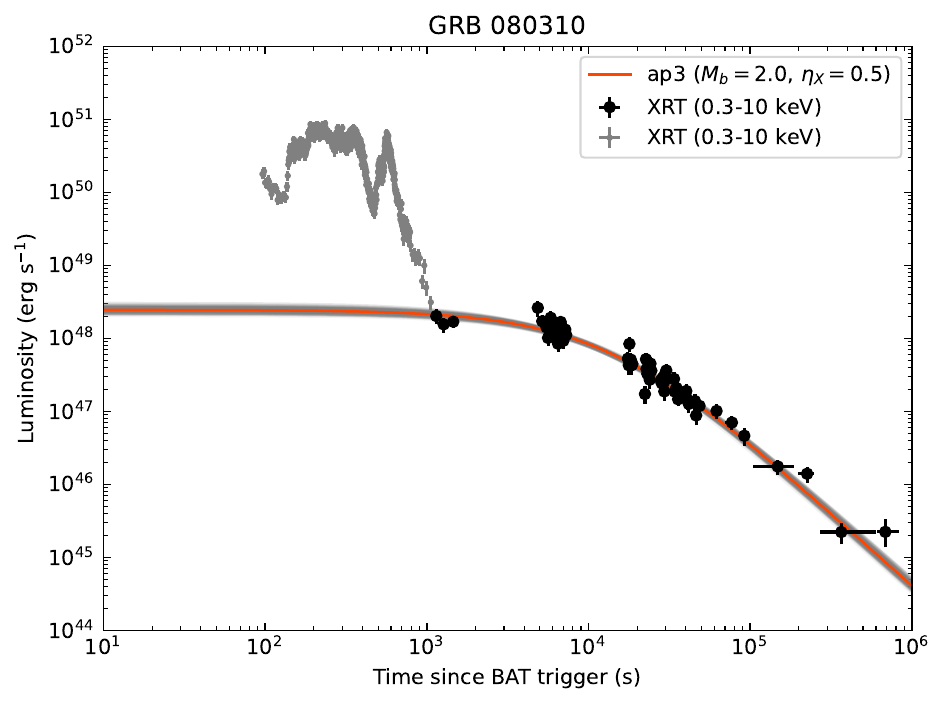}\\
\includegraphics  [angle=0,scale=0.25] {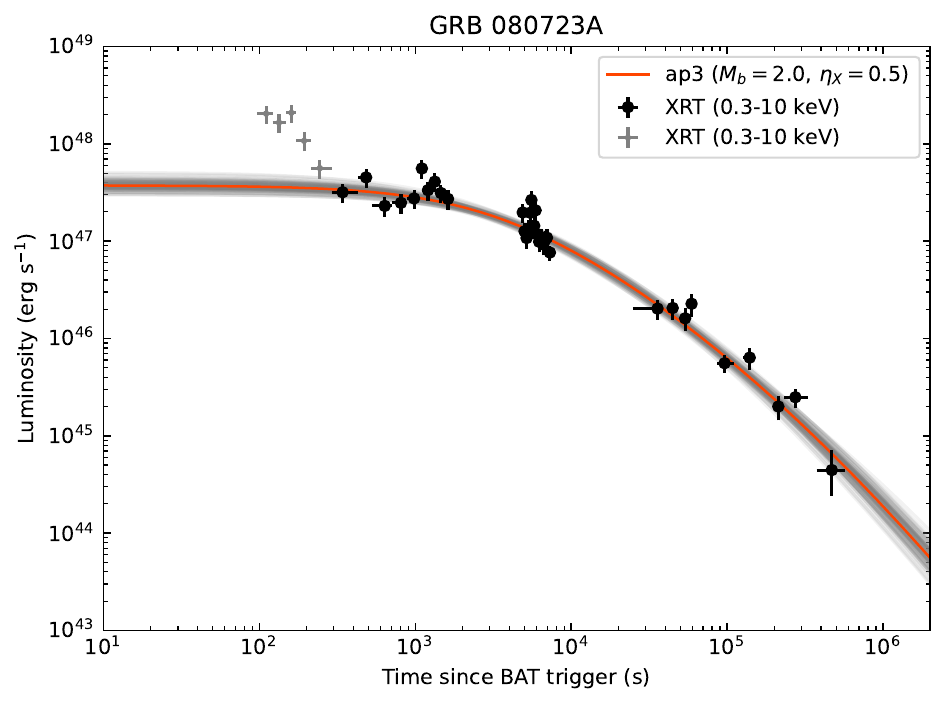}
\includegraphics  [angle=0,scale=0.25] {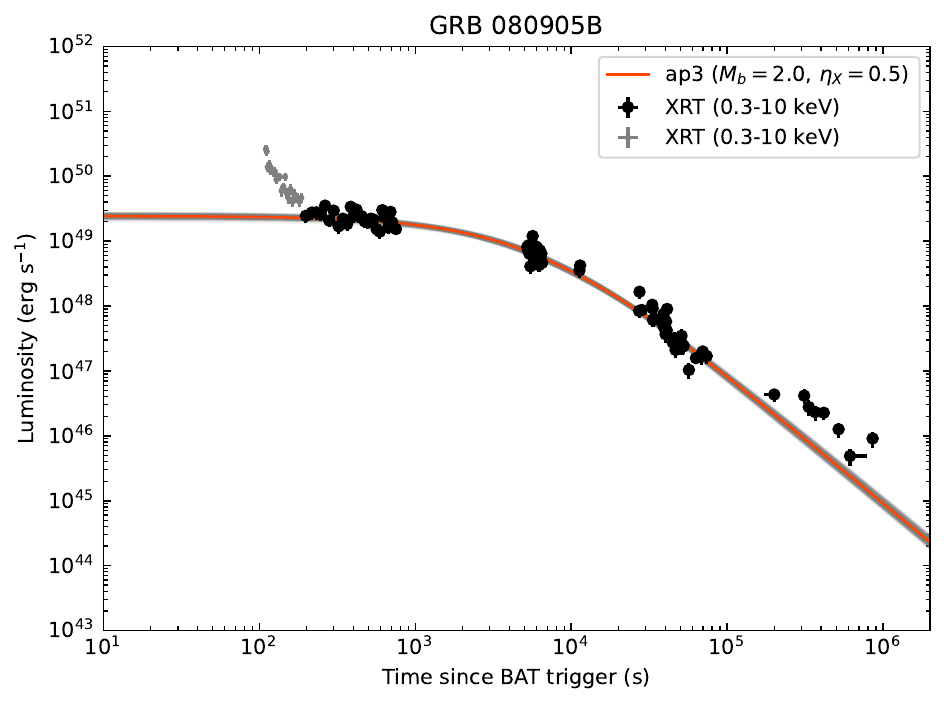}
\includegraphics  [angle=0,scale=0.25] {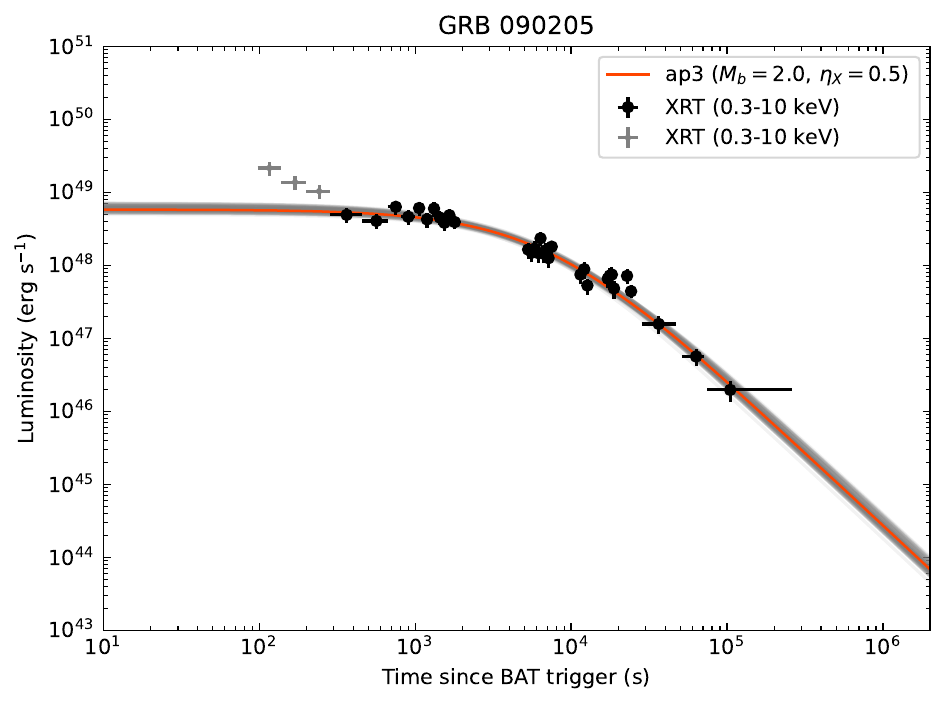}
\includegraphics  [angle=0,scale=0.25] {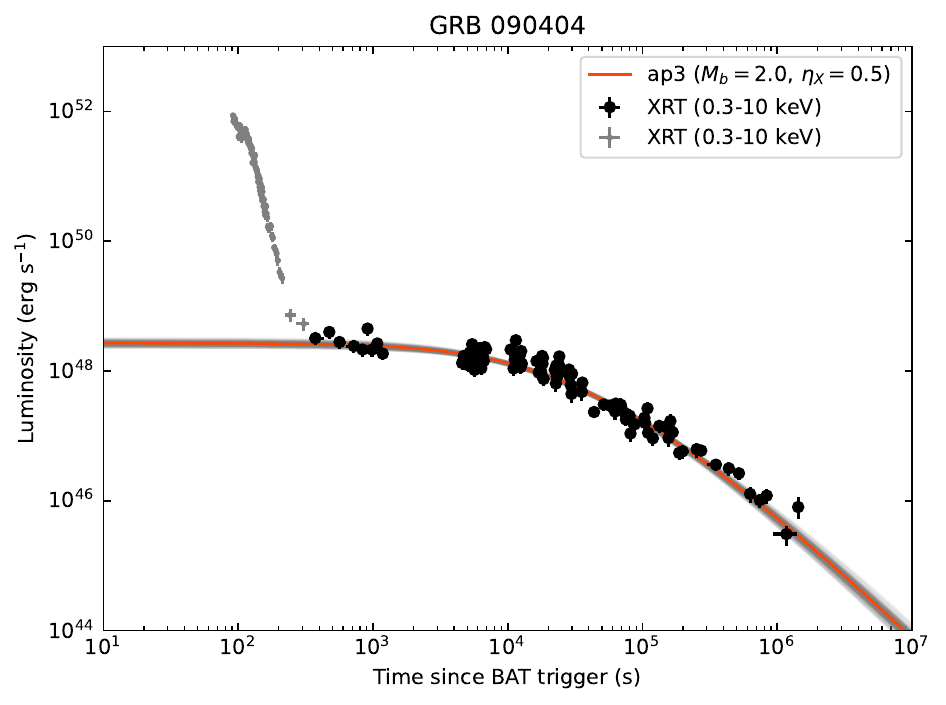}\\
\includegraphics  [angle=0,scale=0.25] {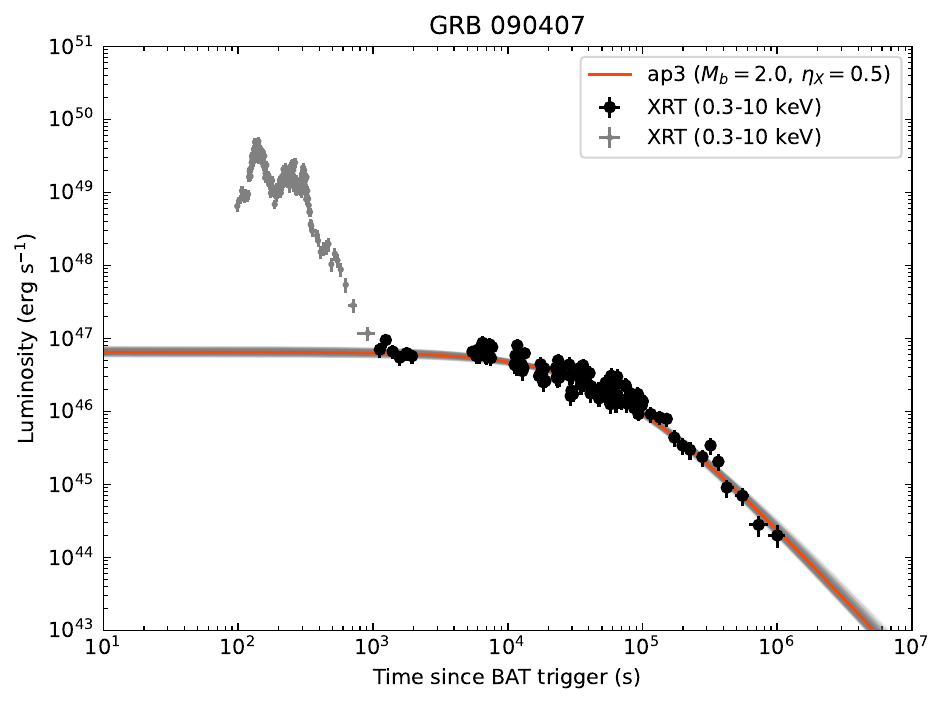}
\includegraphics  [angle=0,scale=0.25] {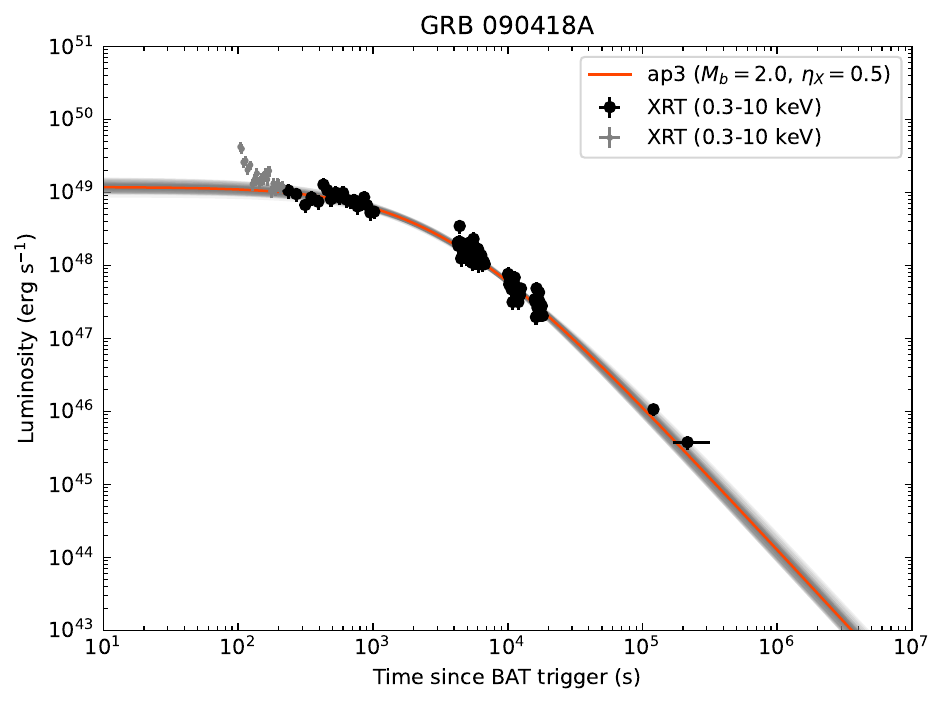}
\includegraphics  [angle=0,scale=0.25] {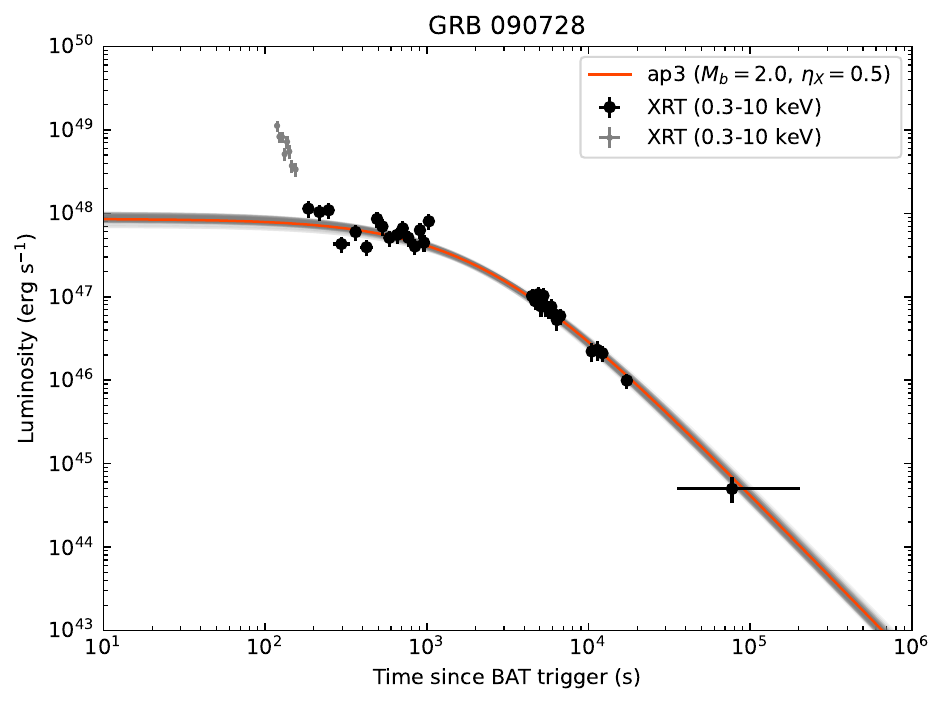}
\includegraphics  [angle=0,scale=0.25] {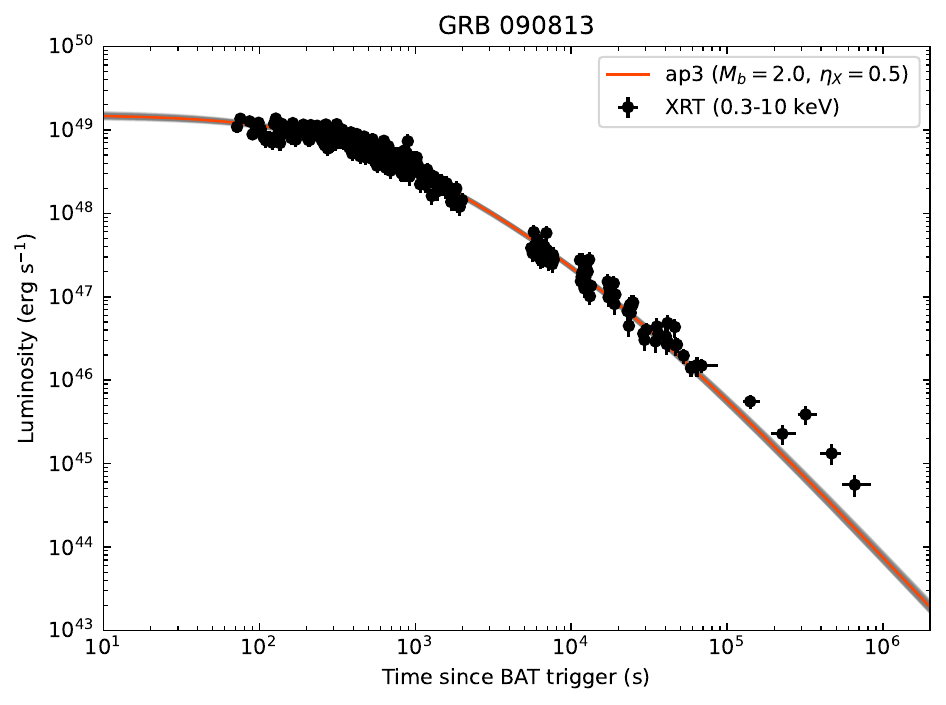}\\
\includegraphics  [angle=0,scale=0.25] {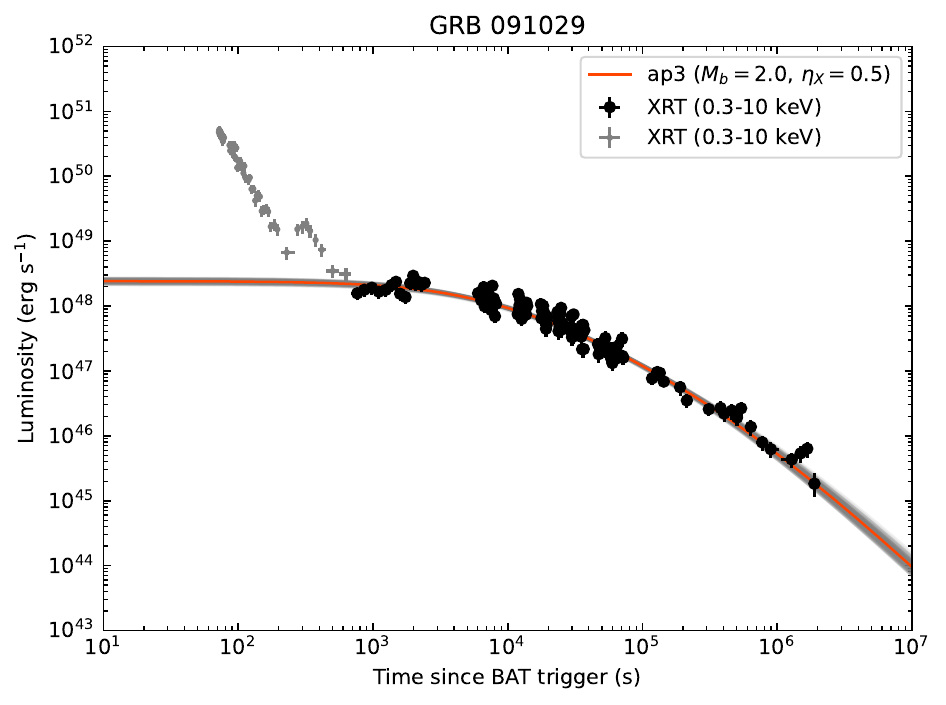}
\includegraphics  [angle=0,scale=0.25] {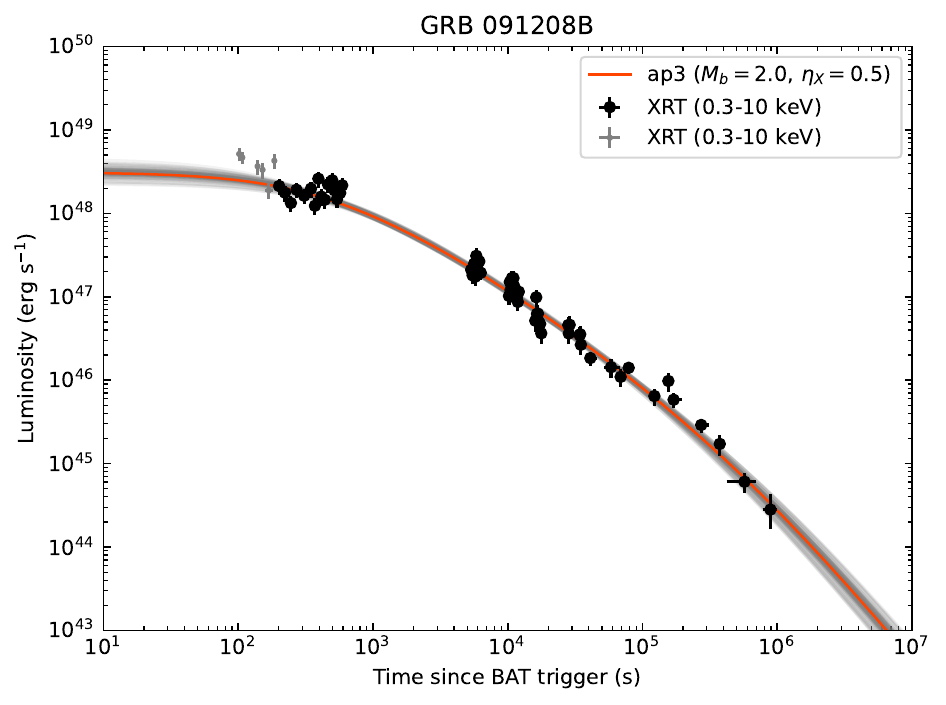}
\includegraphics  [angle=0,scale=0.25] {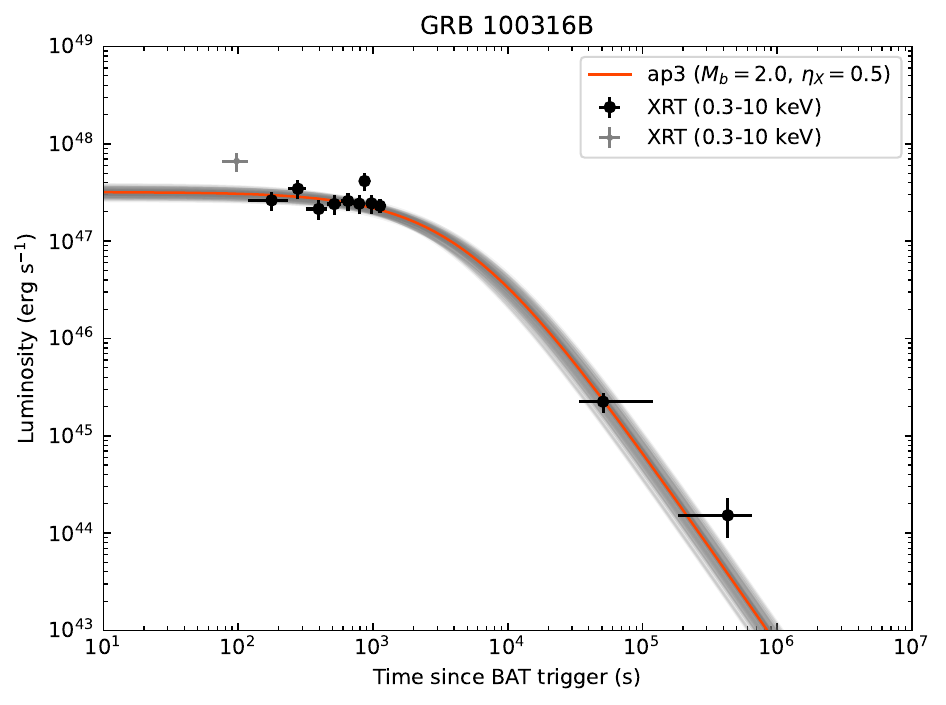}
\includegraphics  [angle=0,scale=0.25] {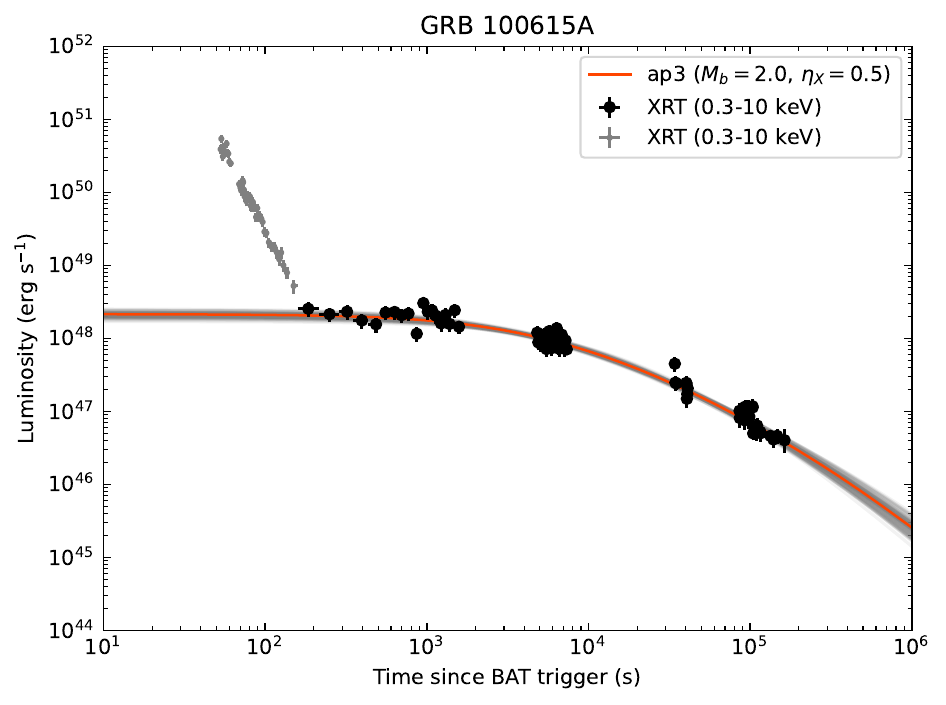}\\
\caption{The best-fitting results of our entire GRB sample in AP3 EoS with $M_{b}=2.0~M_{\odot}$ and $\eta_{\rm X}=0.5$.}
\label{fig:allGRBs_LC}
\end{figure*}

\appendix
In this section, we showed the best-fitting results and corner plots for EM-dominated case, GW-dominated case, and EM+GW co-dominated case in four samples of EoSs with $M_{b}=2.0~M_{\odot},~2.5~M_{\odot}$ and $\eta_{\rm X}=0.1,~0.5$, respectively. Figures \ref{fig:EM_mcmc}-\ref{fig:EM+GW_mcmc} showed the best-fitting results and corner plots for EM-dominated case, GW-dominated case, and EM+GW co-dominated case in four samples of EoSs with $M_{b}=2.0~M_{\odot},~2.5~M_{\odot}$ and $\eta_{\rm X}=0.1,~0.5$, respectively. In Figure \ref{fig:allGRBs_LC}, we selected the AP3 EoS with $M_{b}=2.0~M_{\odot}$ and $\eta_{\rm X}=0.5$ as an example to show the best-fitting results for X-ray light curves in our entire GRB sample.

\begin{figure}
\centering
\includegraphics  [angle=0,scale=0.25] {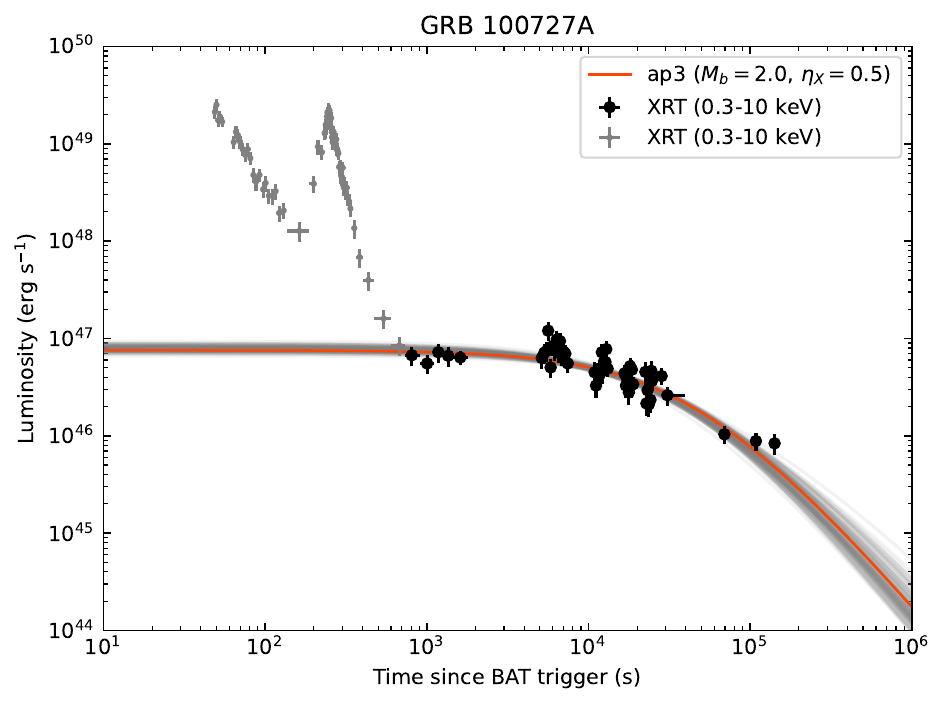}
\includegraphics  [angle=0,scale=0.25] {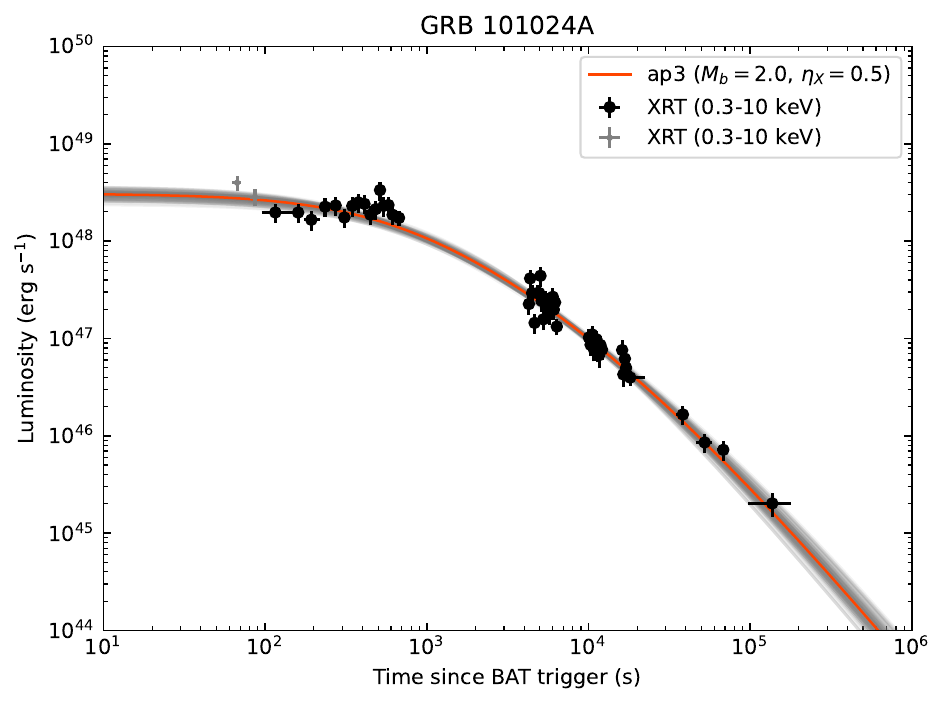}
\includegraphics  [angle=0,scale=0.25] {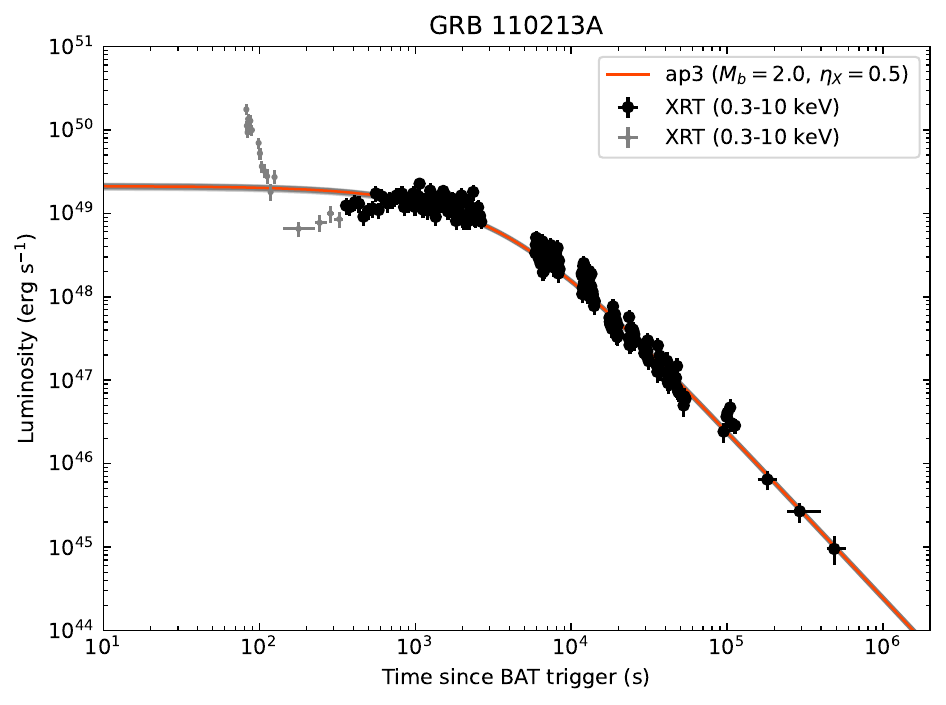}
\includegraphics  [angle=0,scale=0.25] {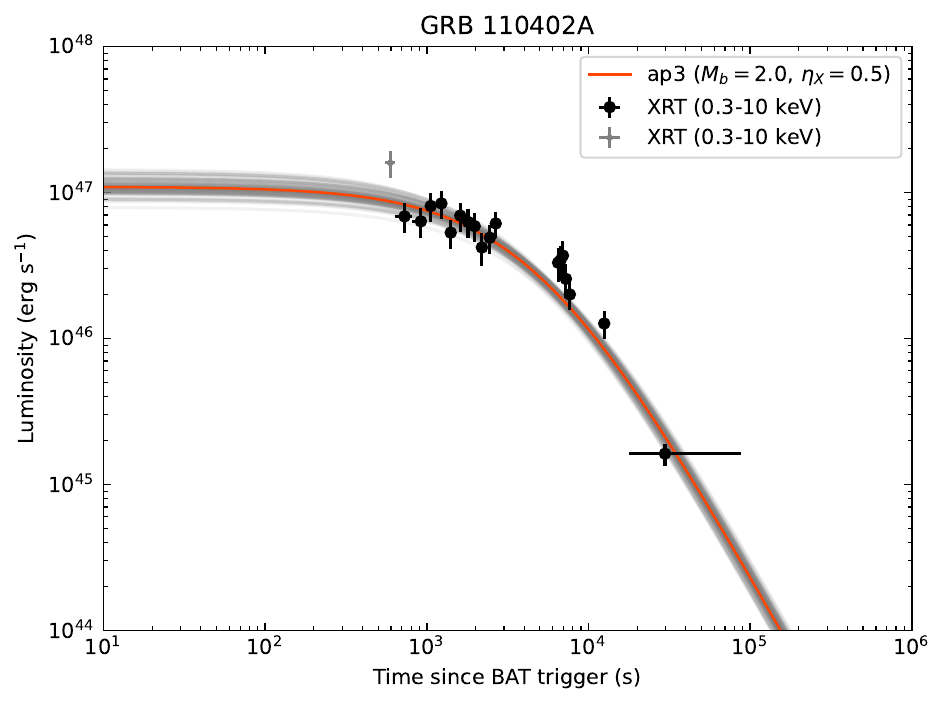}\\
\includegraphics  [angle=0,scale=0.25] {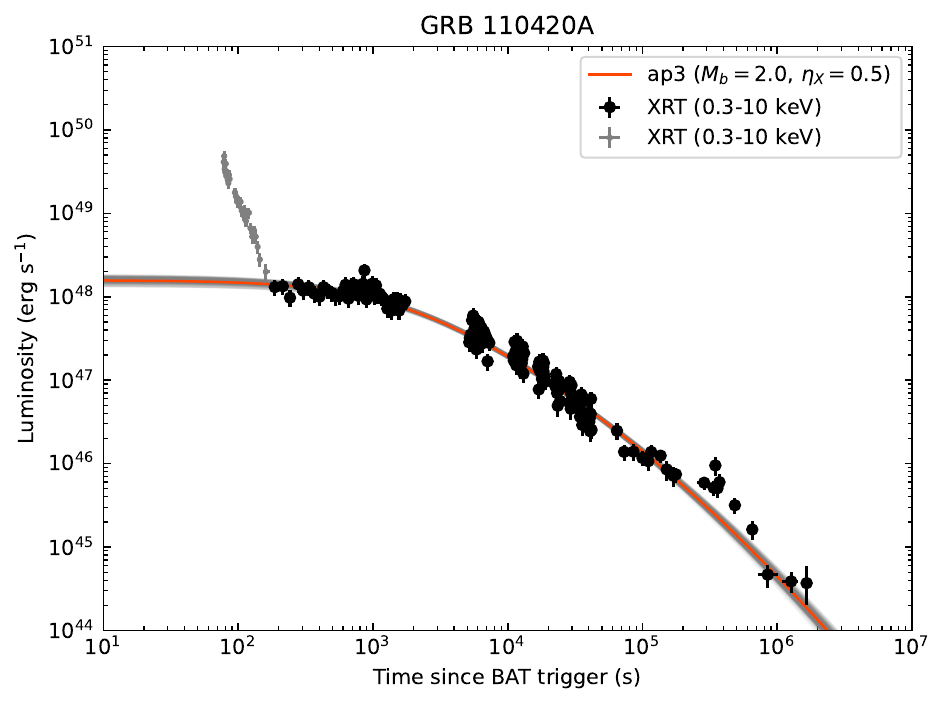}
\includegraphics  [angle=0,scale=0.25] {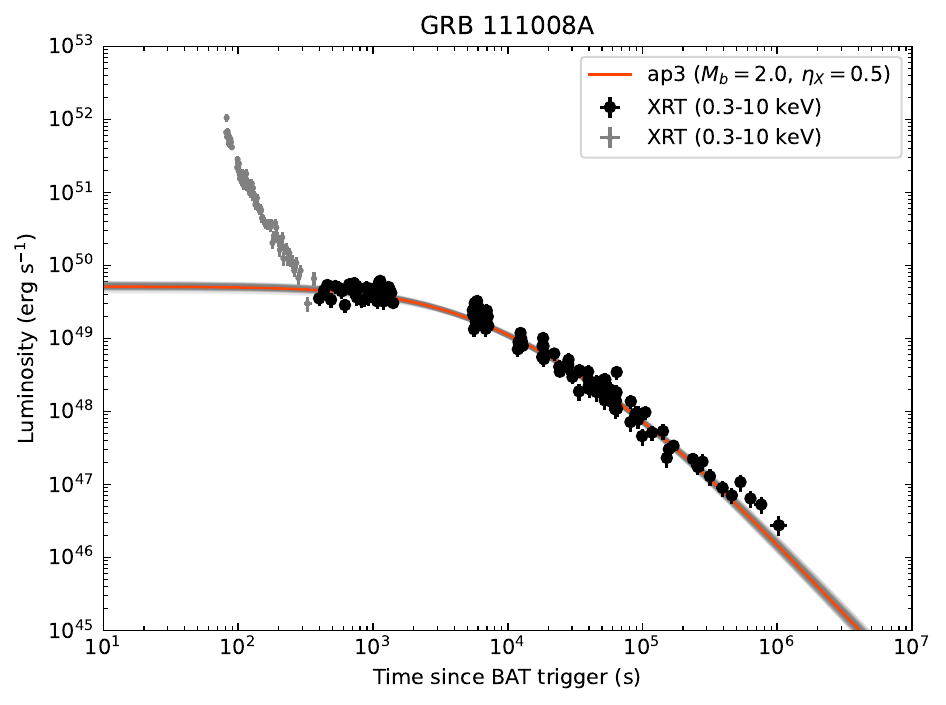}
\includegraphics  [angle=0,scale=0.25] {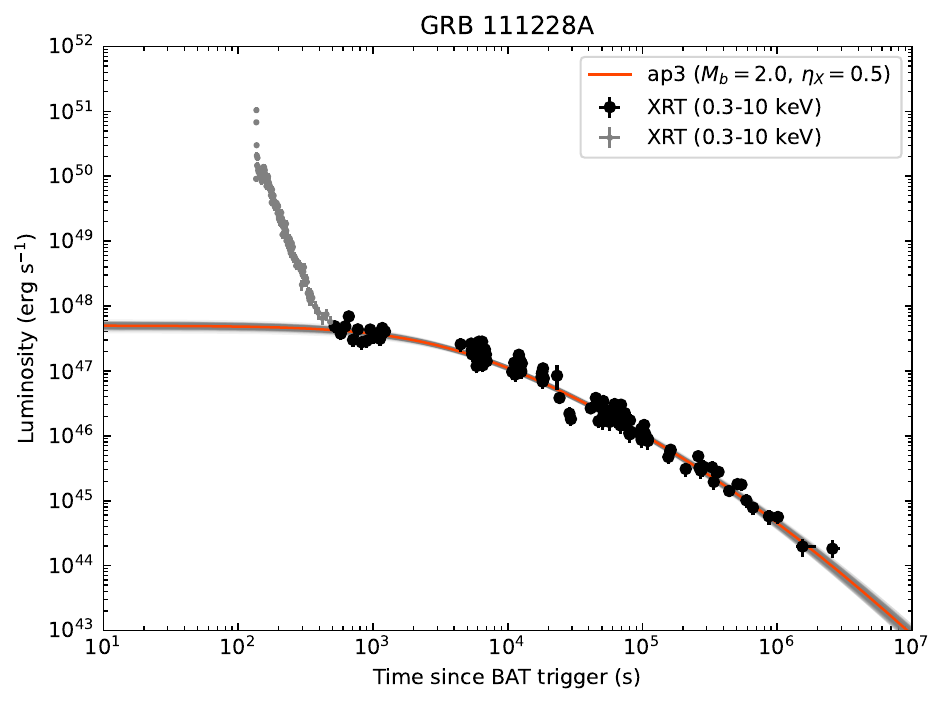}
\includegraphics  [angle=0,scale=0.25] {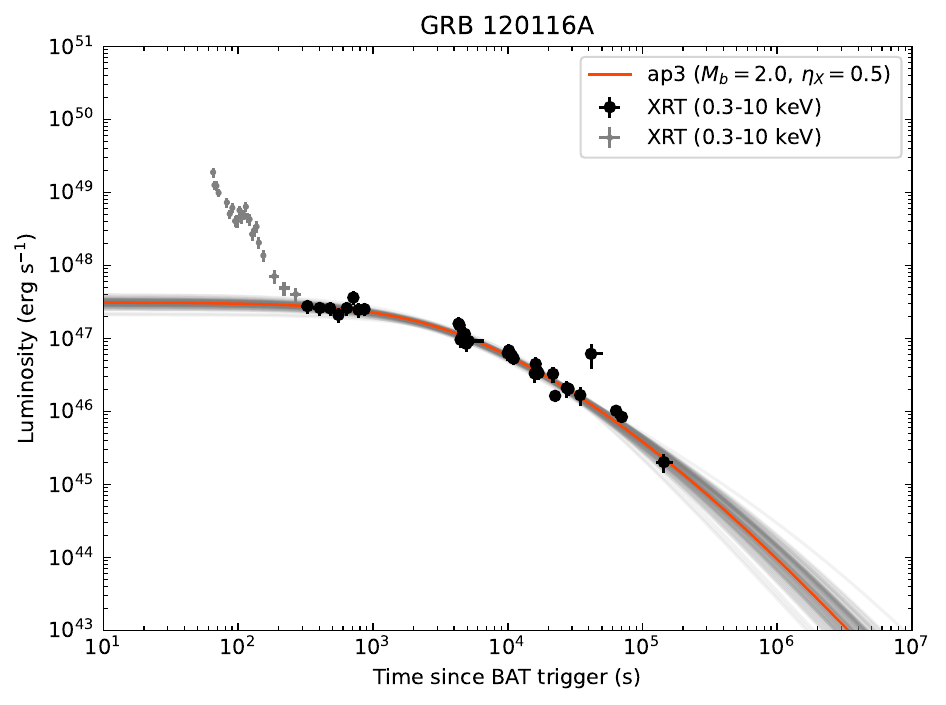}\\
\includegraphics  [angle=0,scale=0.25] {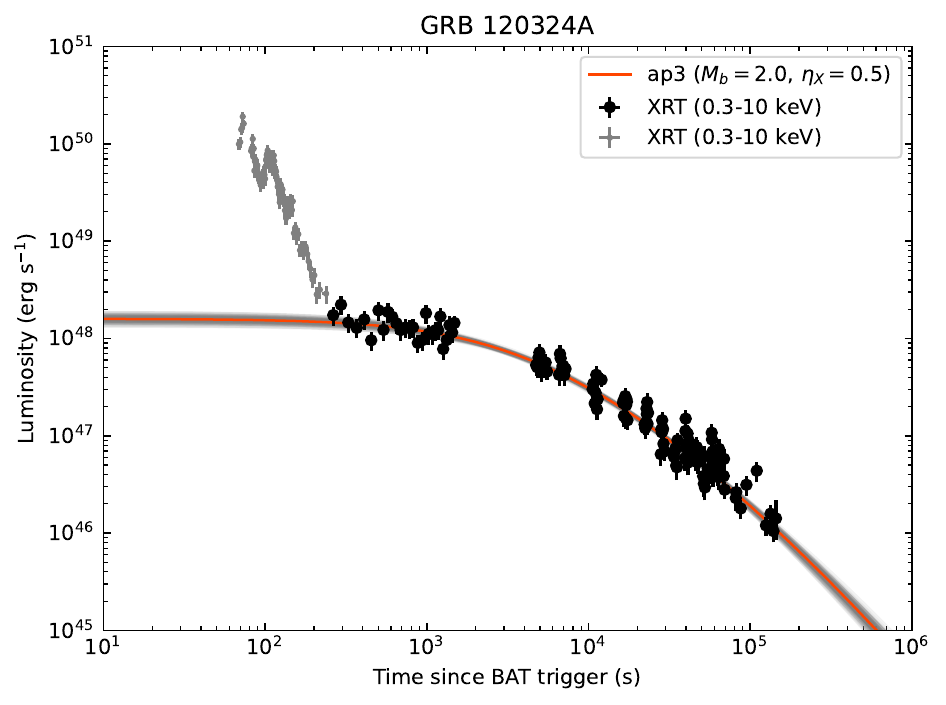}
\includegraphics  [angle=0,scale=0.25] {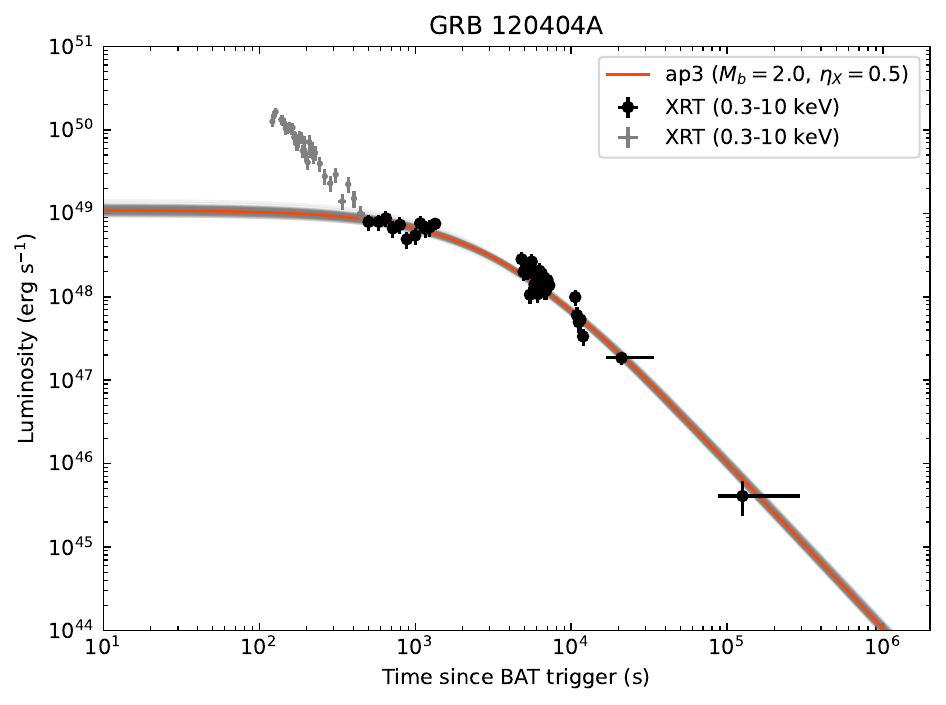}
\includegraphics  [angle=0,scale=0.25] {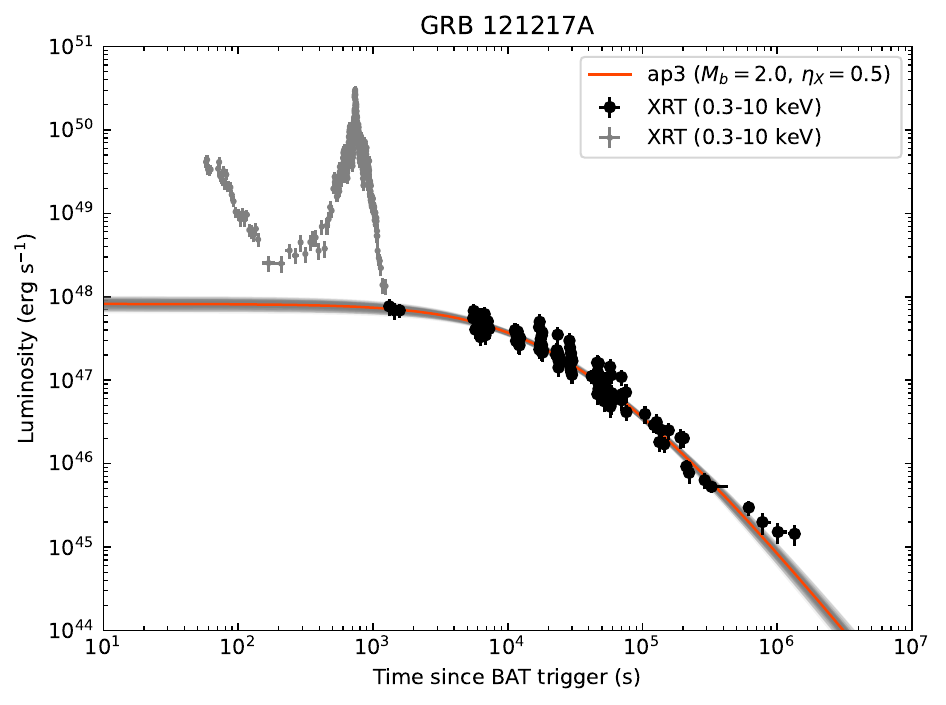}
\includegraphics  [angle=0,scale=0.25] {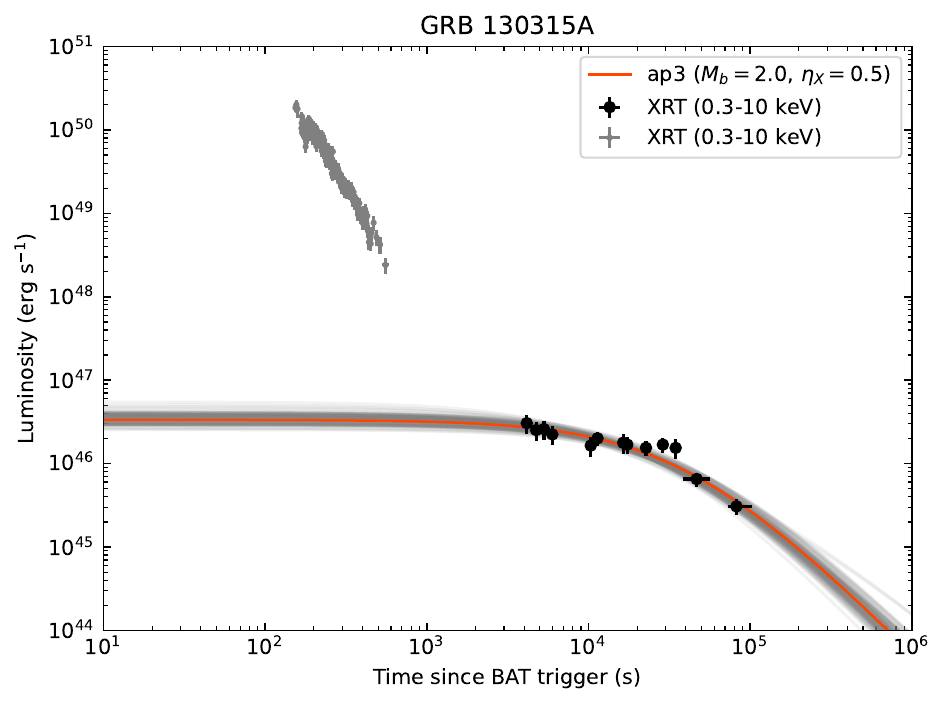}\\
\includegraphics  [angle=0,scale=0.25] {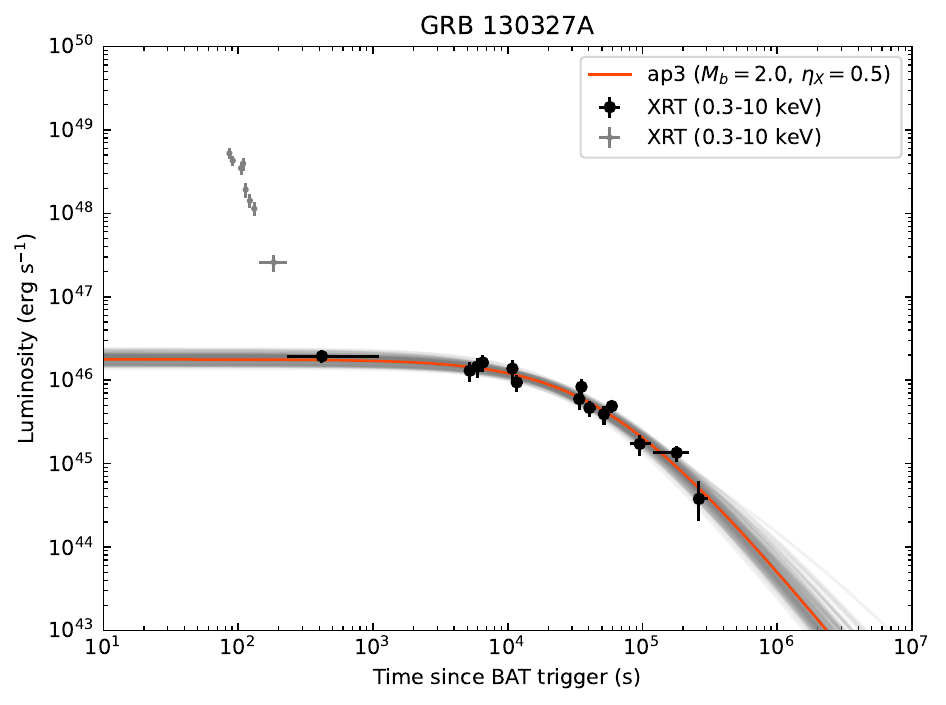}
\includegraphics  [angle=0,scale=0.25] {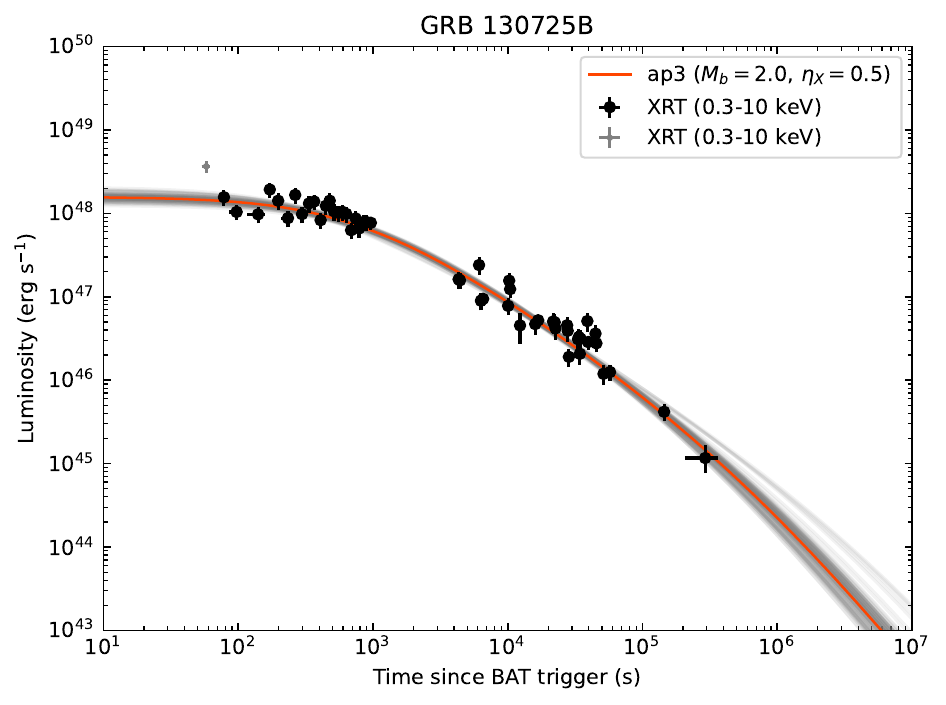}
\includegraphics  [angle=0,scale=0.25] {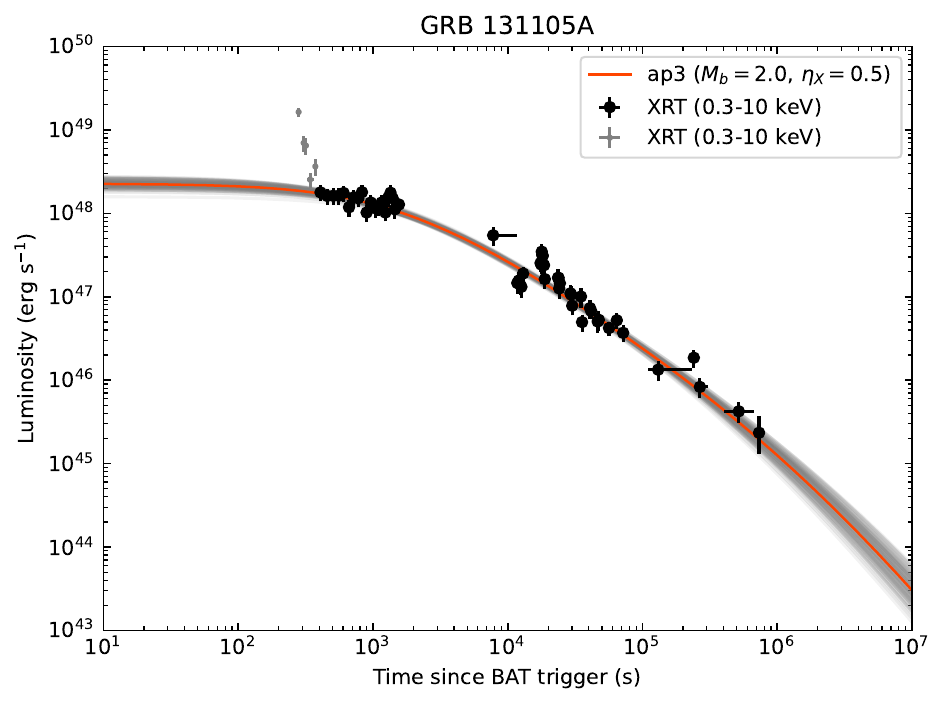}
\includegraphics  [angle=0,scale=0.25] {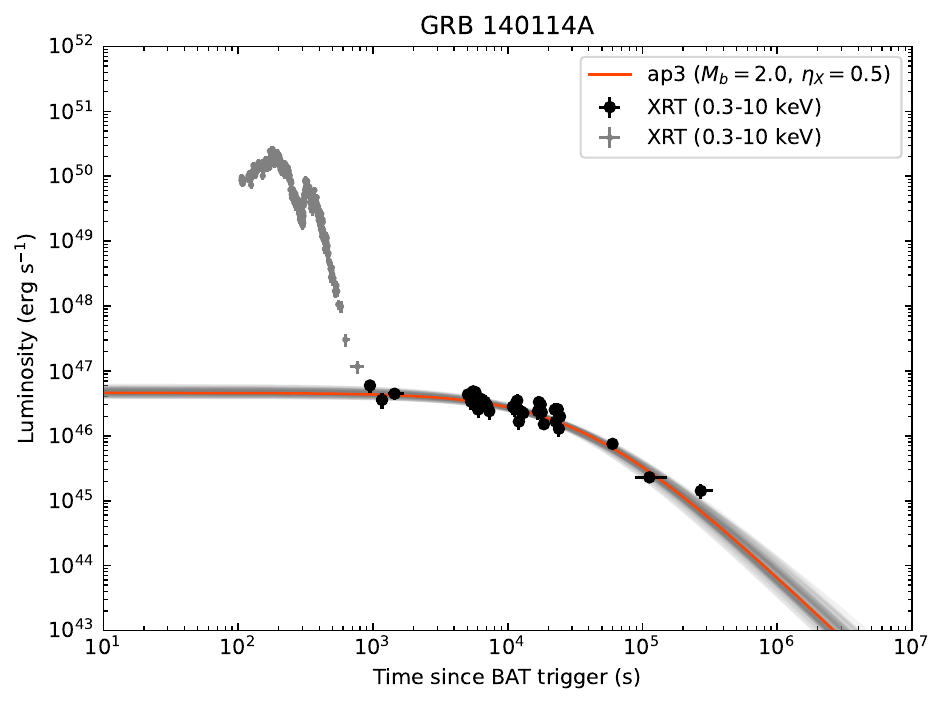}\\
\includegraphics  [angle=0,scale=0.25] {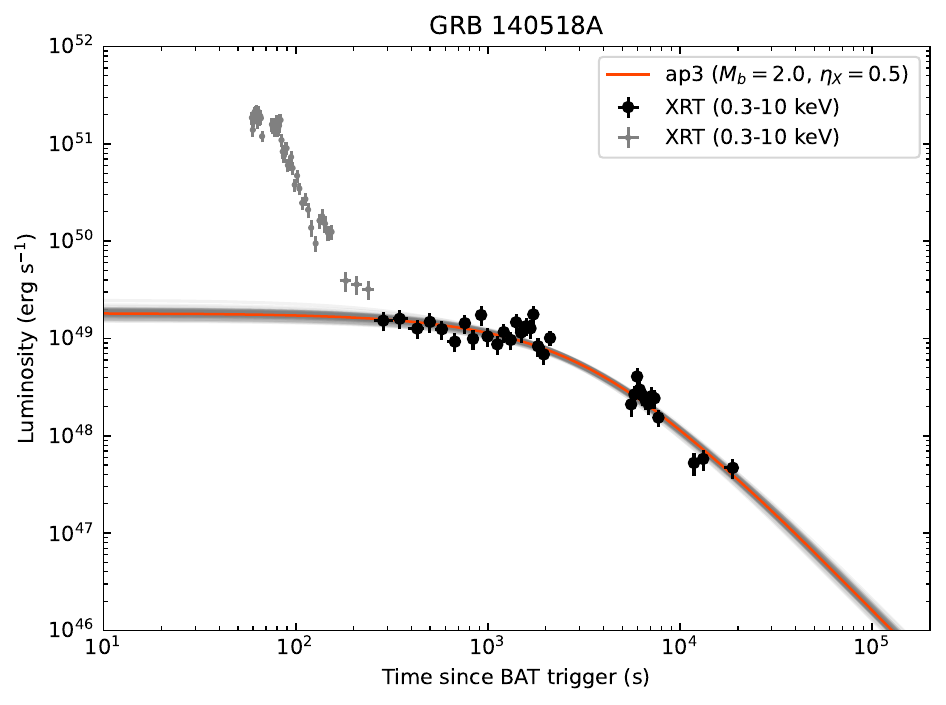}
\includegraphics  [angle=0,scale=0.25] {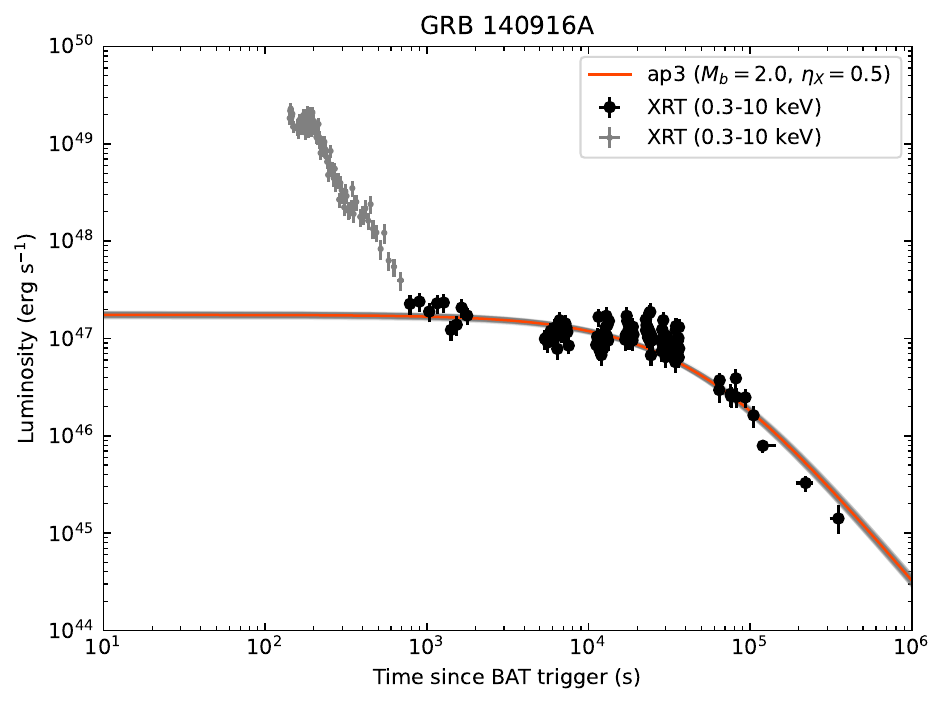}
\includegraphics  [angle=0,scale=0.25] {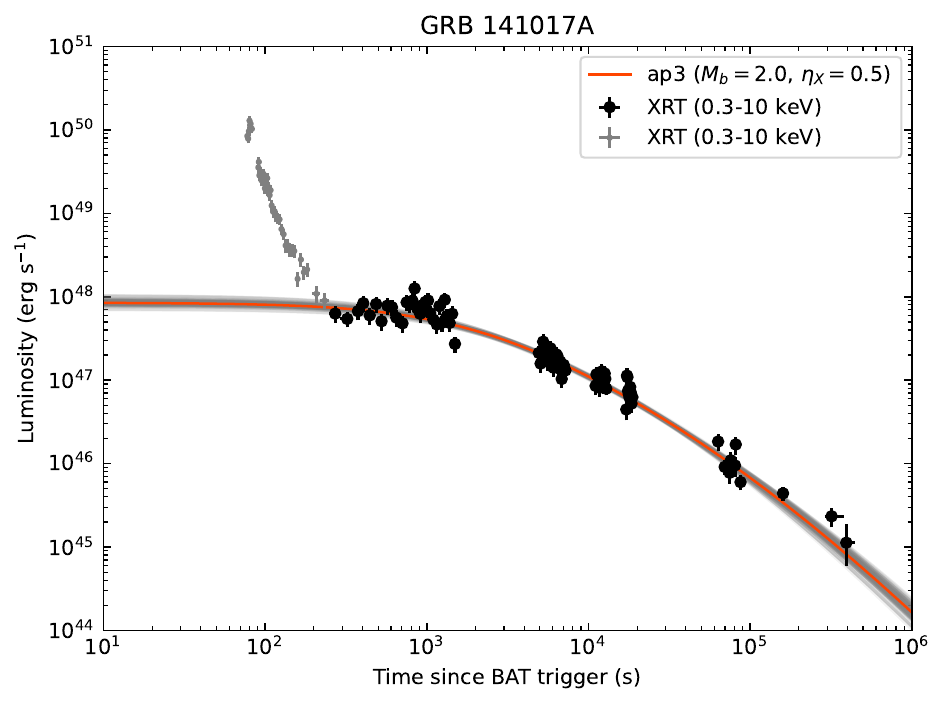}
\includegraphics  [angle=0,scale=0.25] {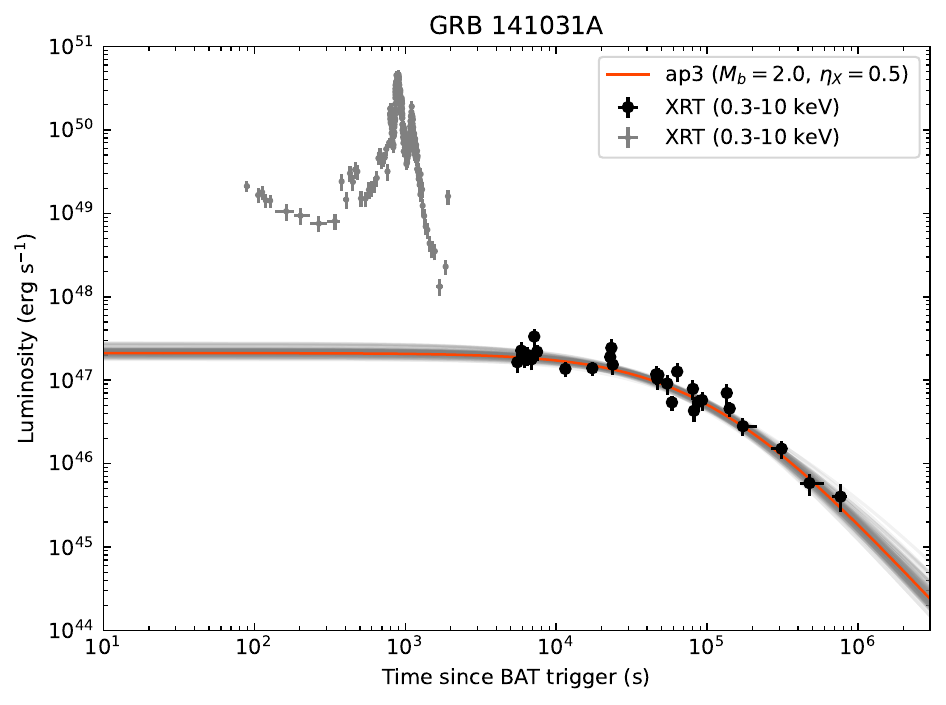}\\
\includegraphics  [angle=0,scale=0.25] {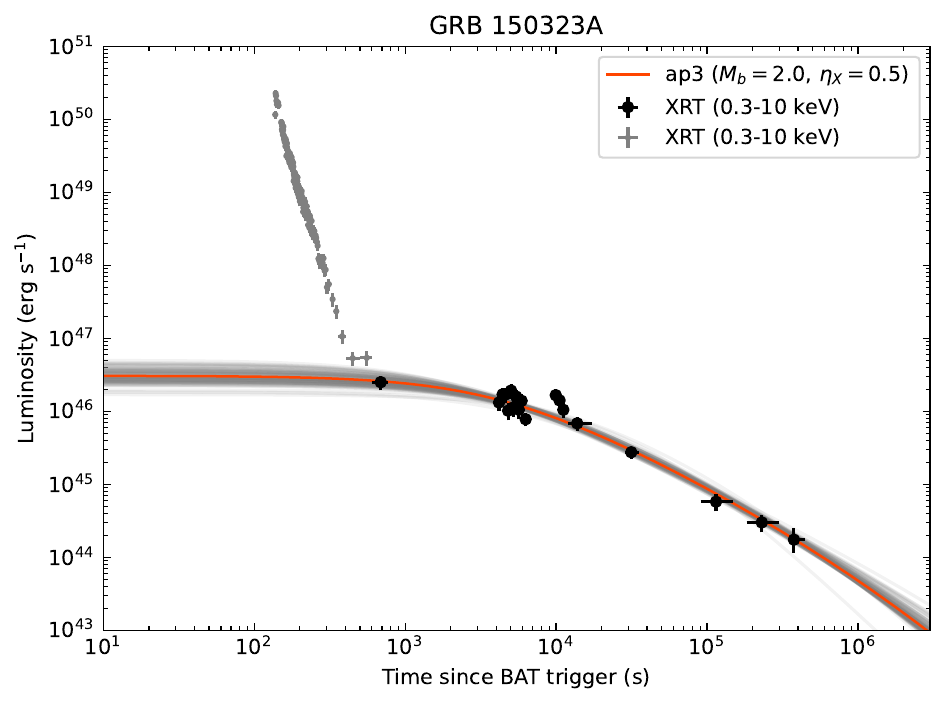}
\includegraphics  [angle=0,scale=0.25] {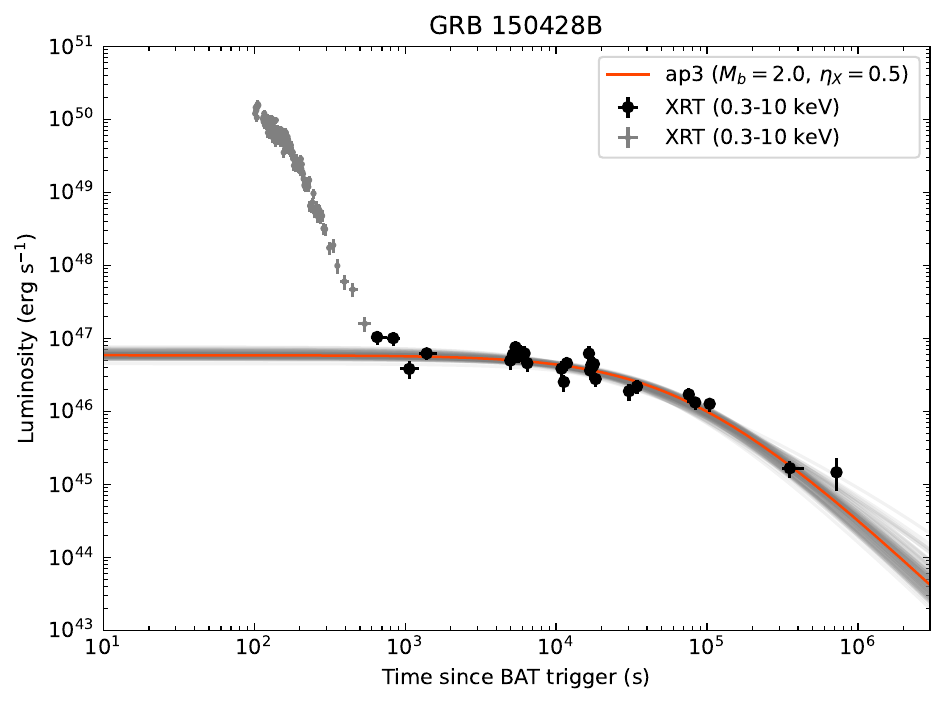}
\includegraphics  [angle=0,scale=0.25] {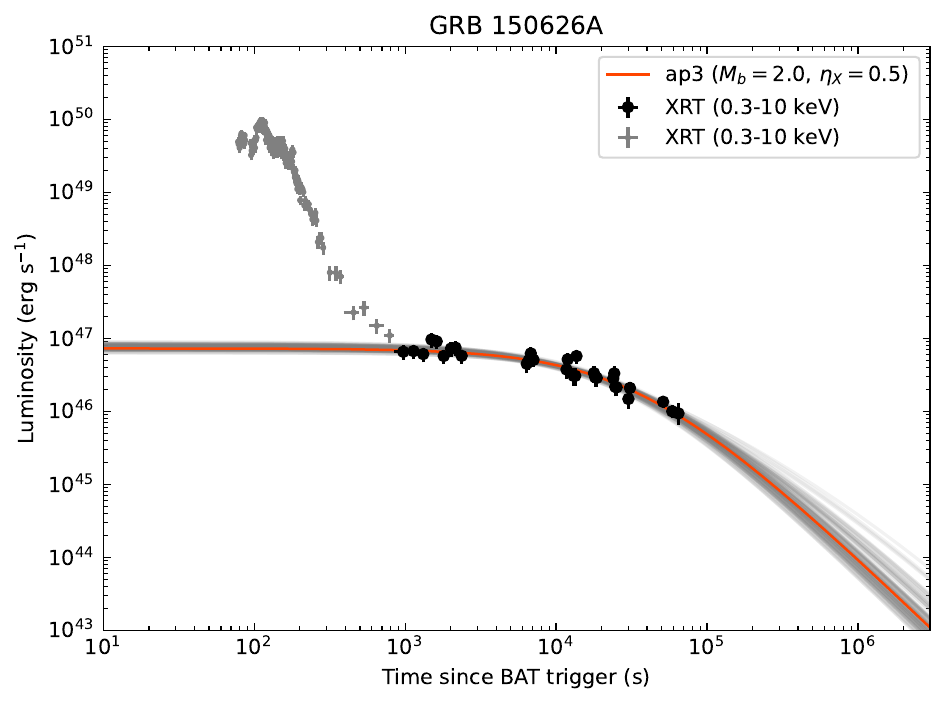}
\includegraphics  [angle=0,scale=0.25] {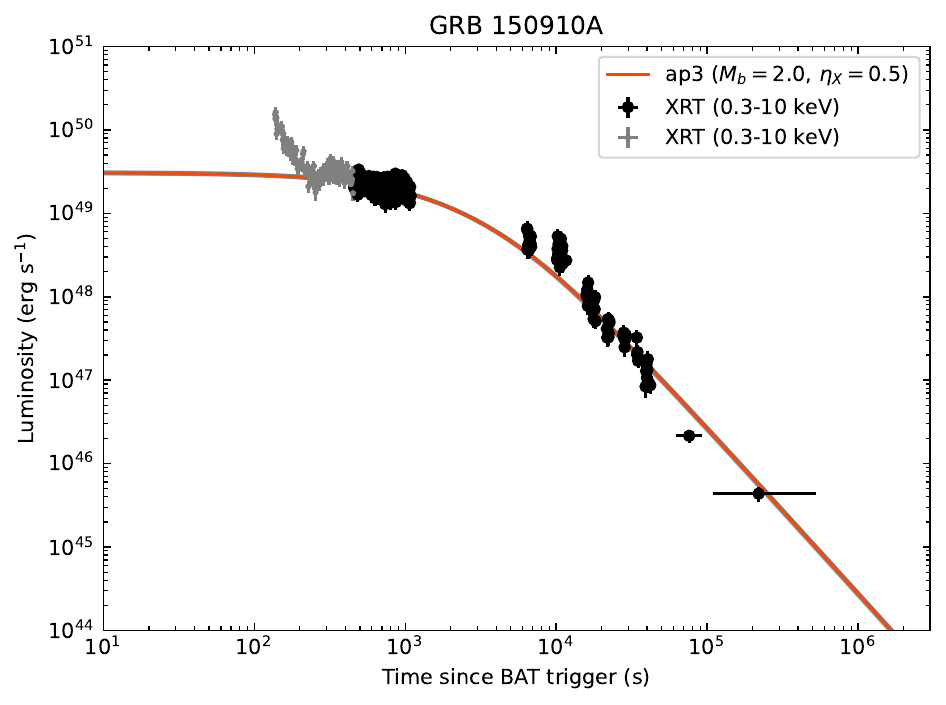}\\
\includegraphics  [angle=0,scale=0.25] {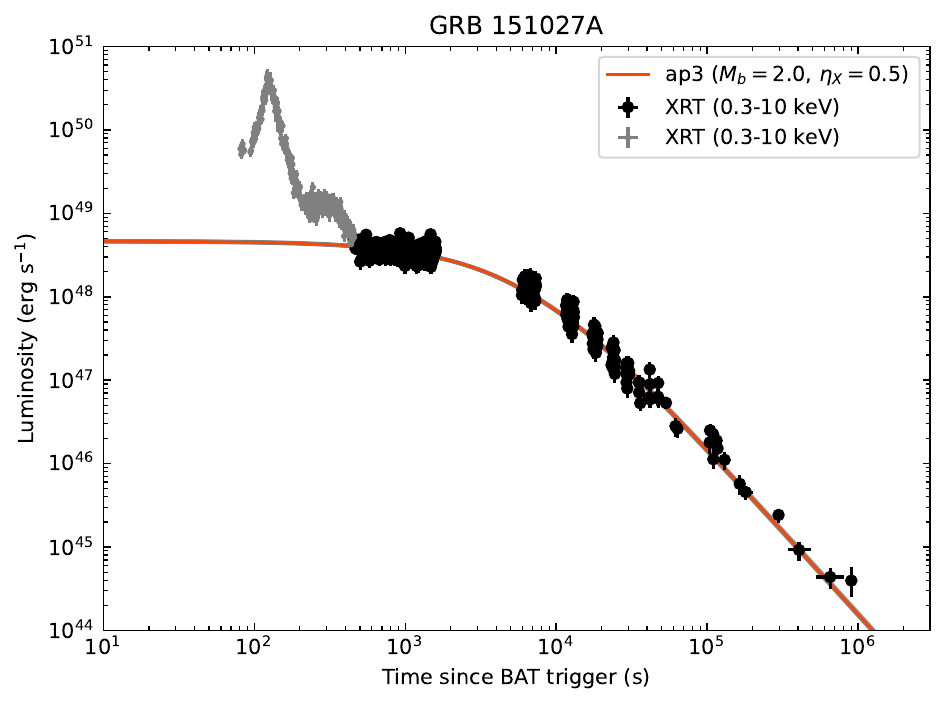}
\includegraphics  [angle=0,scale=0.25] {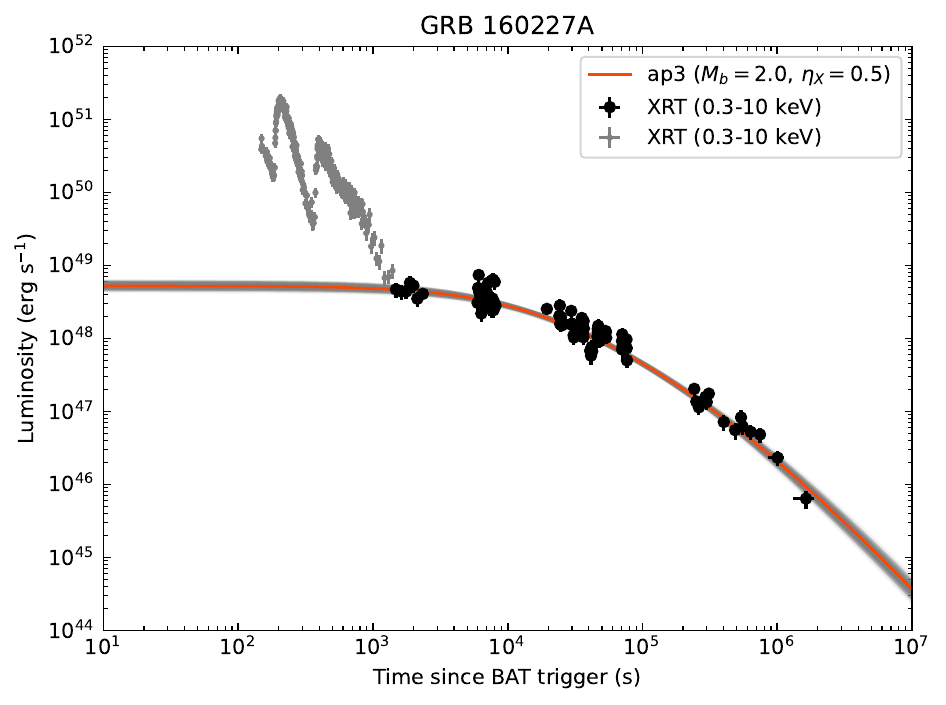}
\includegraphics  [angle=0,scale=0.25] {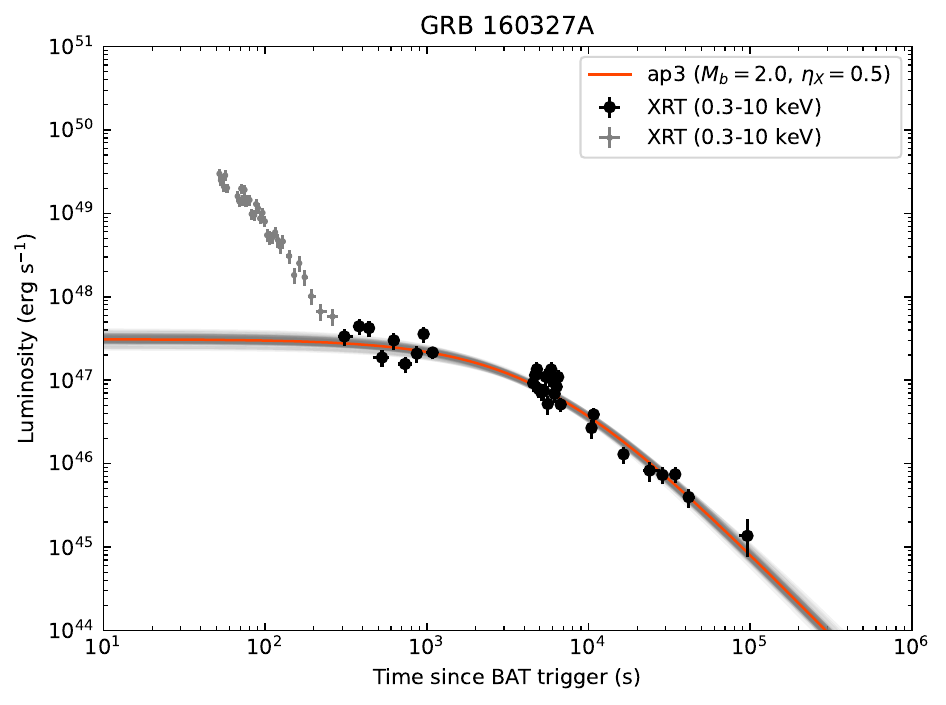}
\includegraphics  [angle=0,scale=0.25] {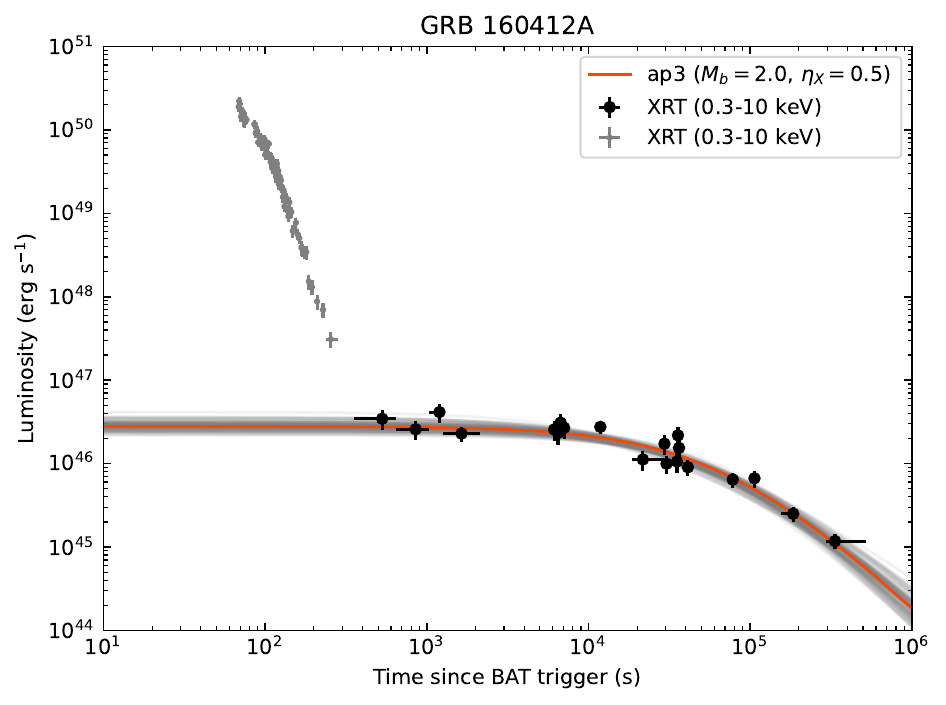}\\
\includegraphics  [angle=0,scale=0.25] {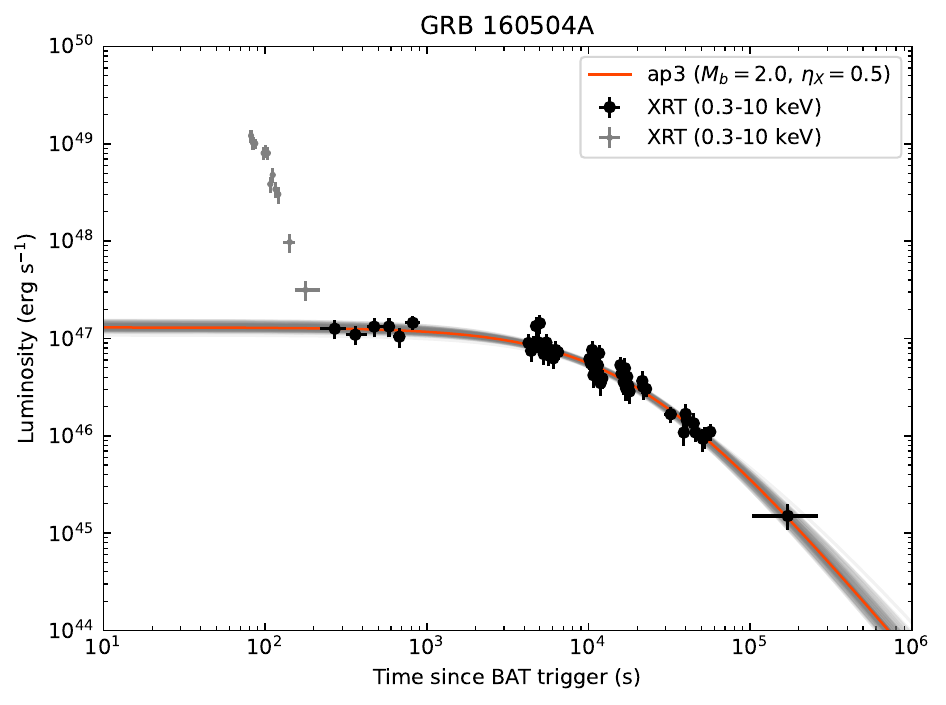}
\includegraphics  [angle=0,scale=0.25] {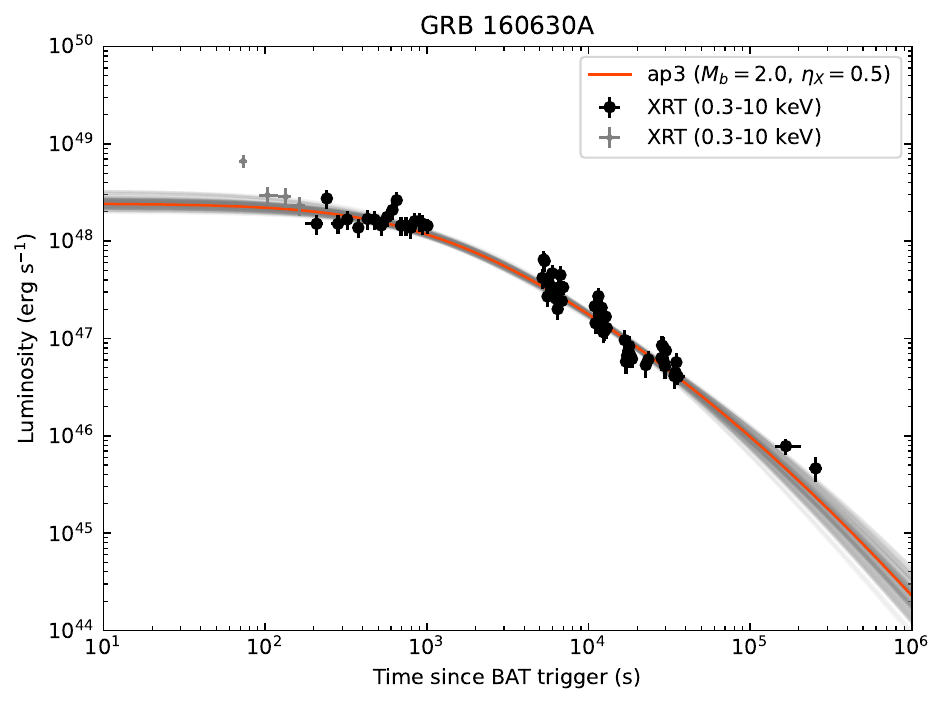}
\includegraphics  [angle=0,scale=0.25] {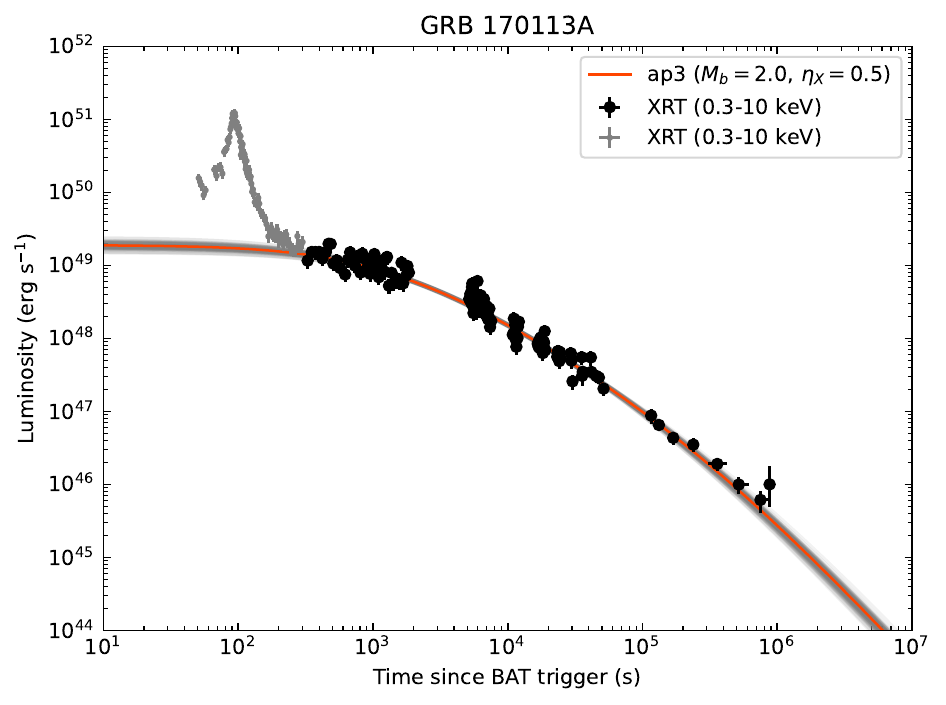}
\includegraphics  [angle=0,scale=0.25] {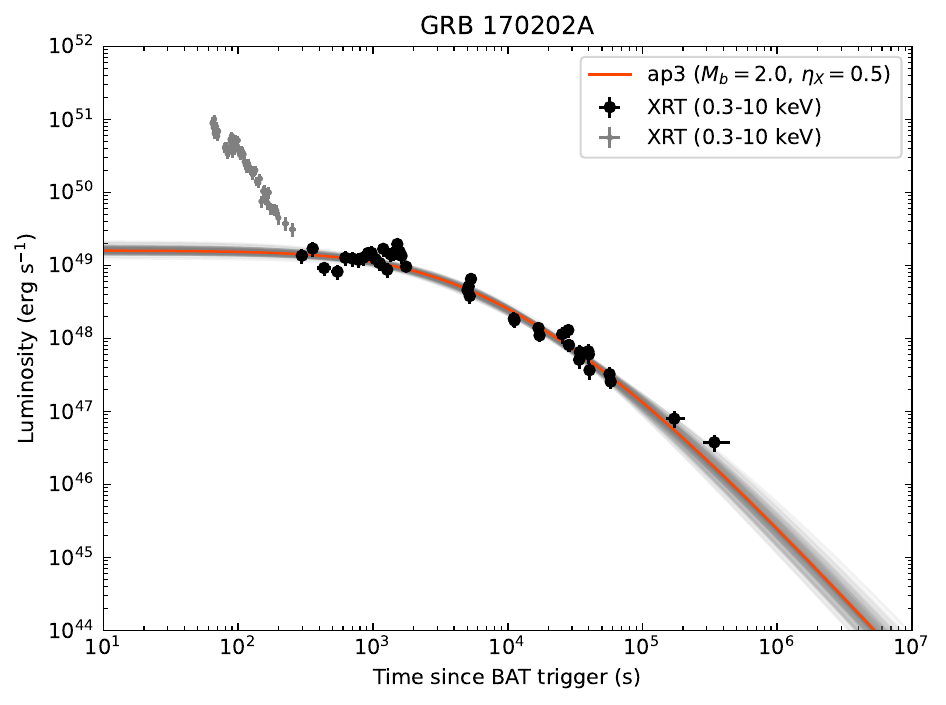}\\
\center{\textbf{Figure 27.} --- continued.}
\end{figure}

\begin{figure}
\centering
\includegraphics  [angle=0,scale=0.25] {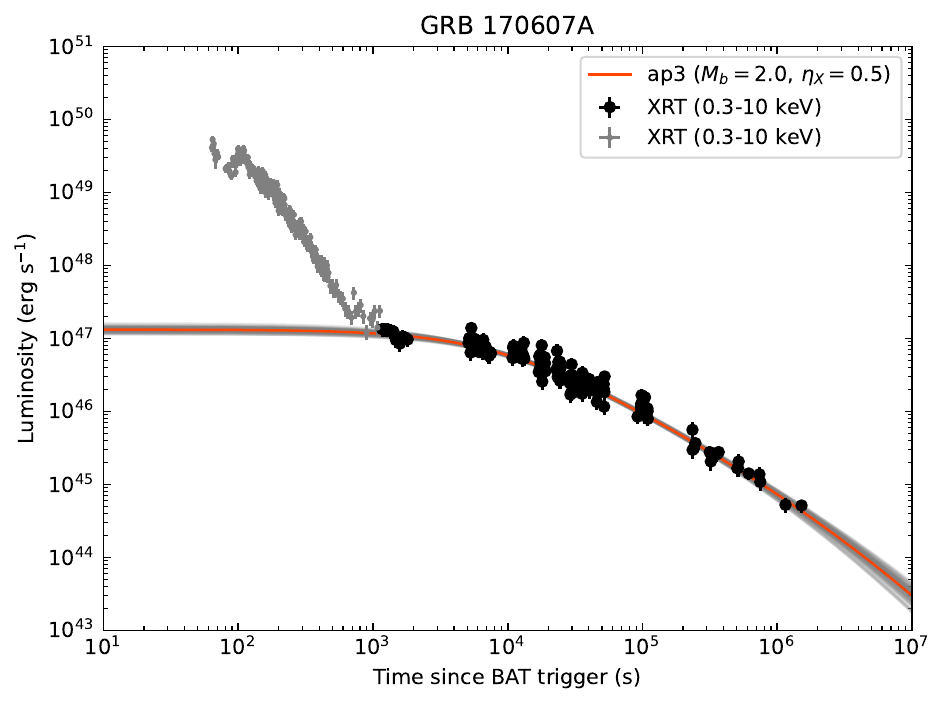}
\includegraphics  [angle=0,scale=0.25] {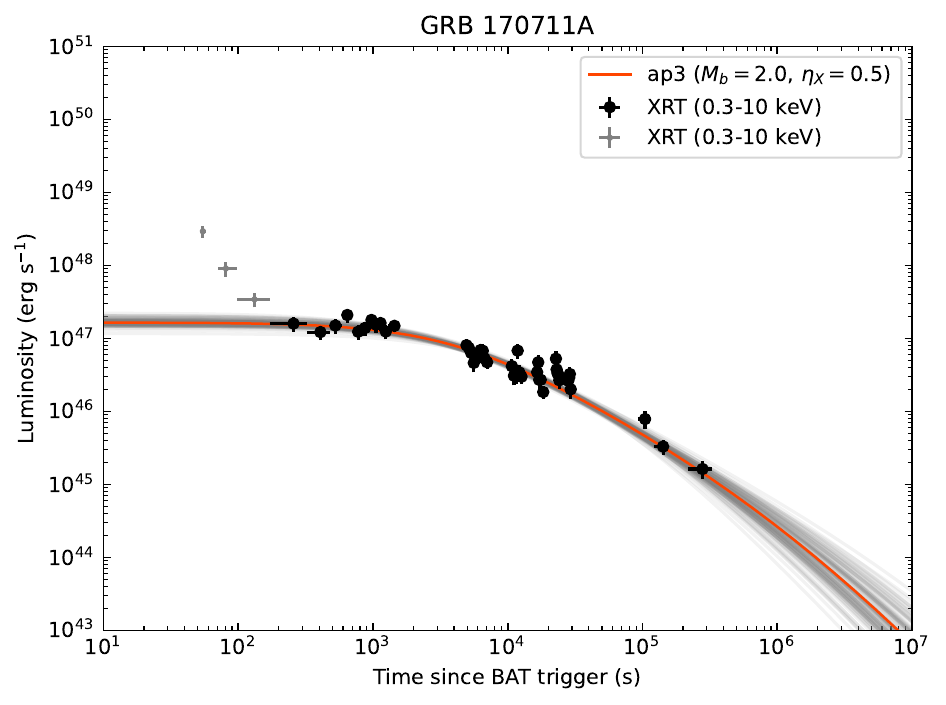}
\includegraphics  [angle=0,scale=0.25] {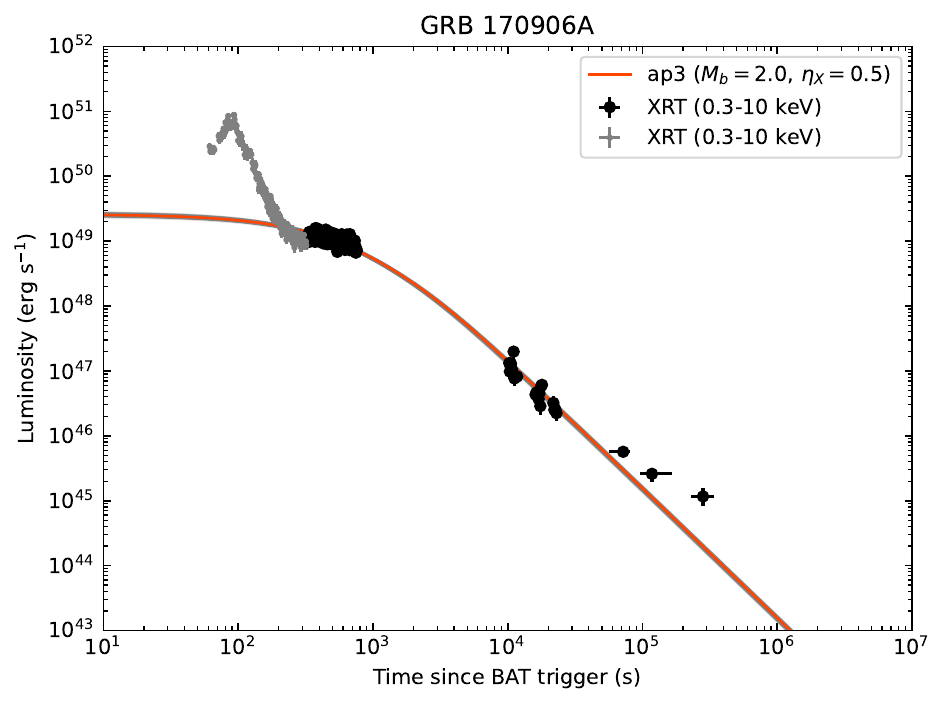}
\includegraphics  [angle=0,scale=0.25] {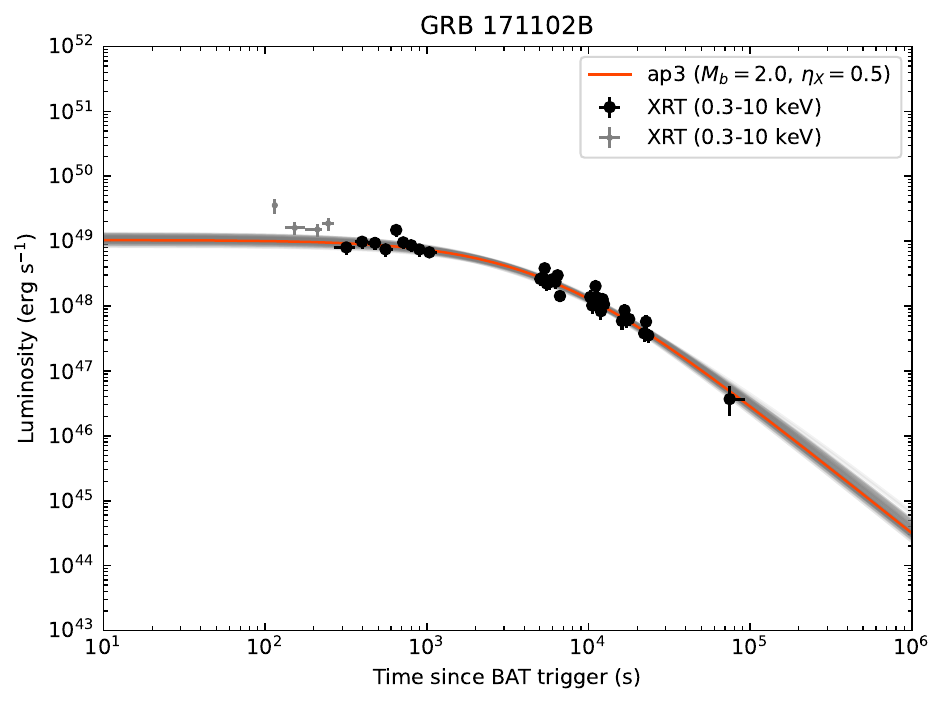}\\
\includegraphics  [angle=0,scale=0.25] {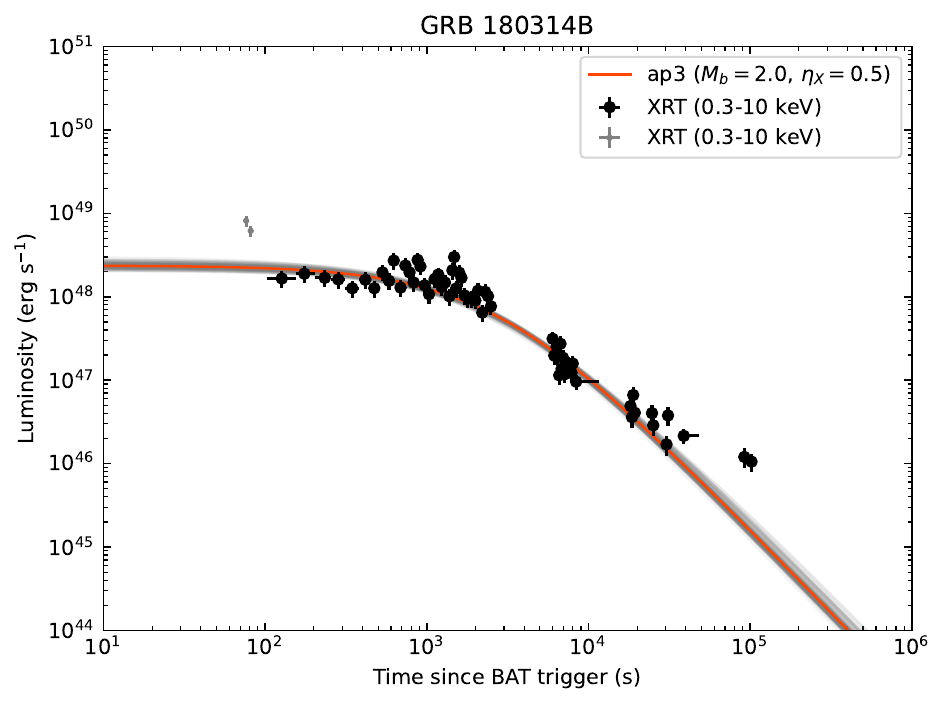}
\includegraphics  [angle=0,scale=0.25] {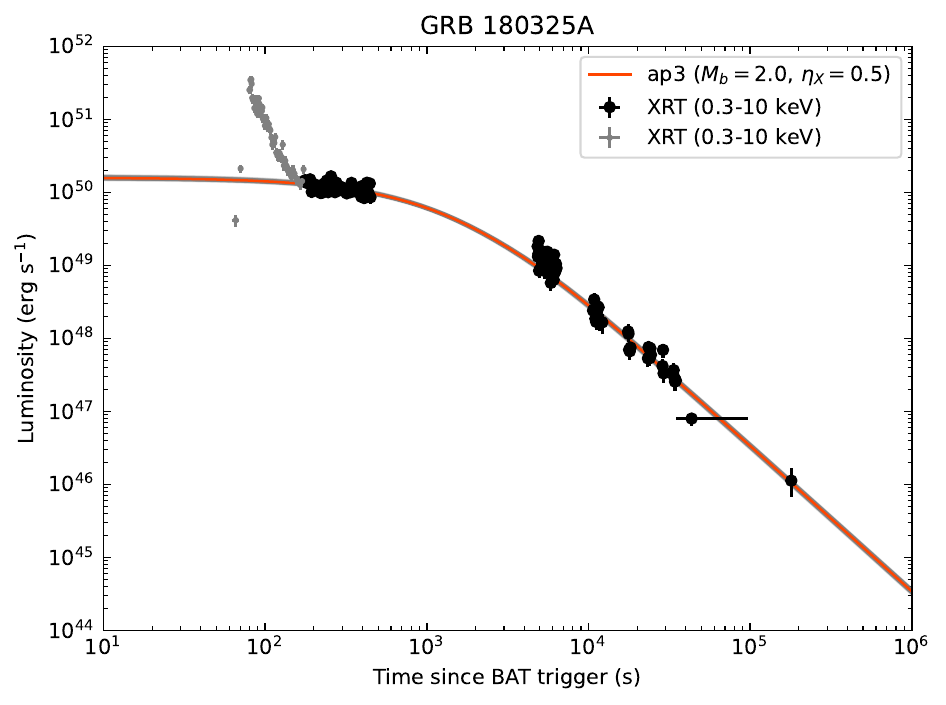}
\includegraphics  [angle=0,scale=0.25] {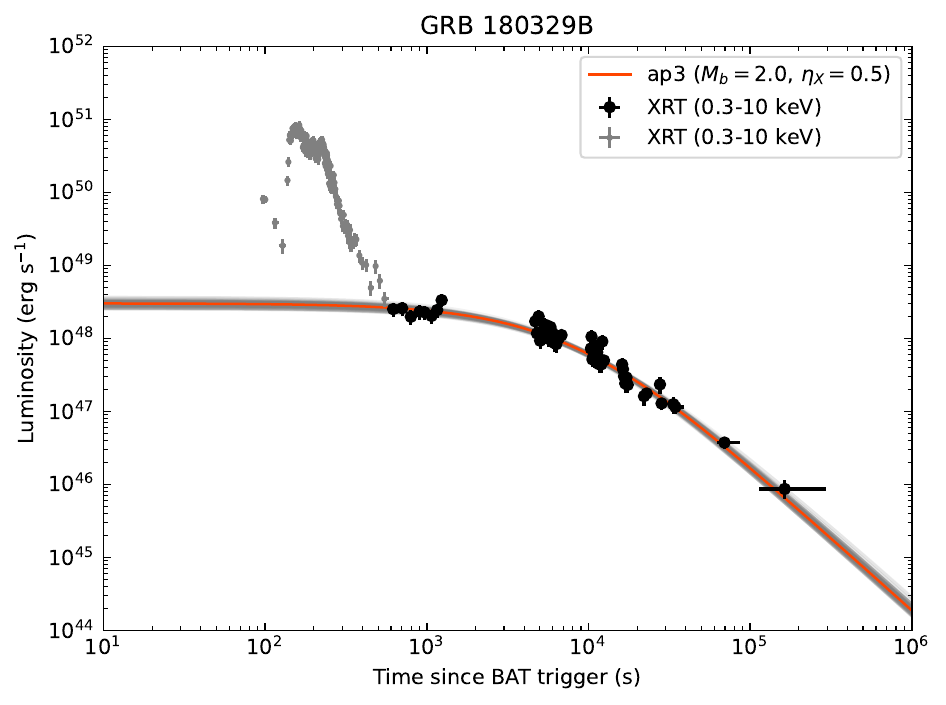}
\includegraphics  [angle=0,scale=0.25] {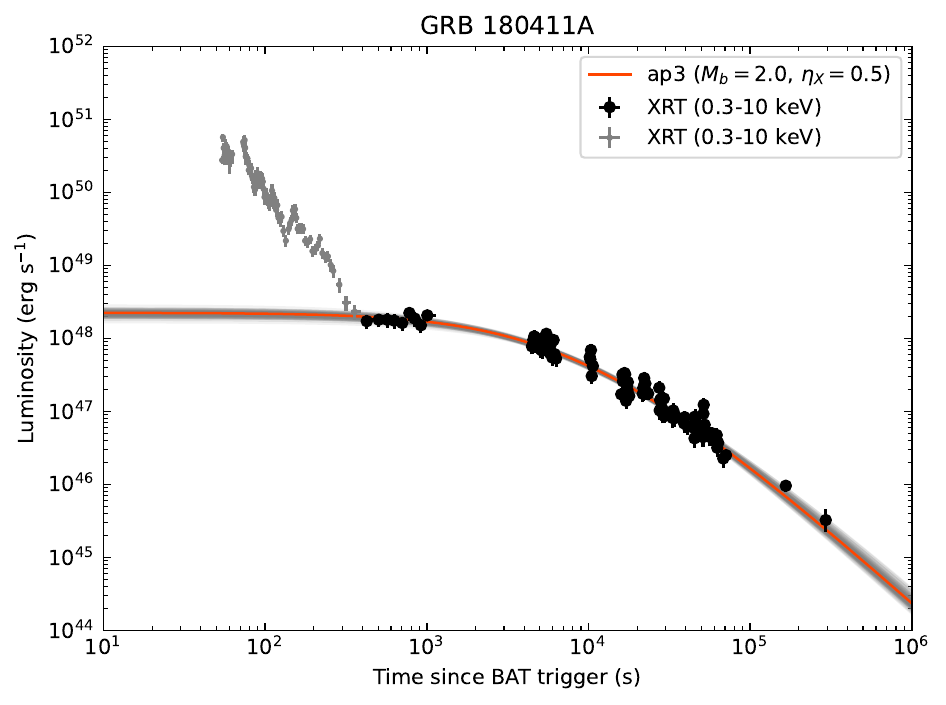}\\
\includegraphics  [angle=0,scale=0.25] {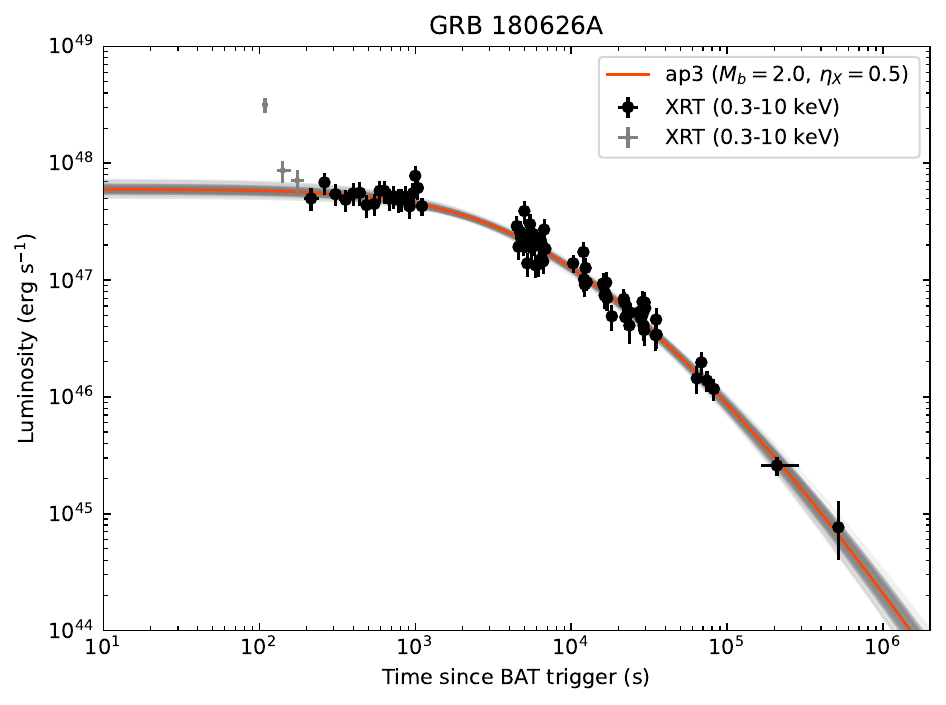}
\includegraphics  [angle=0,scale=0.25] {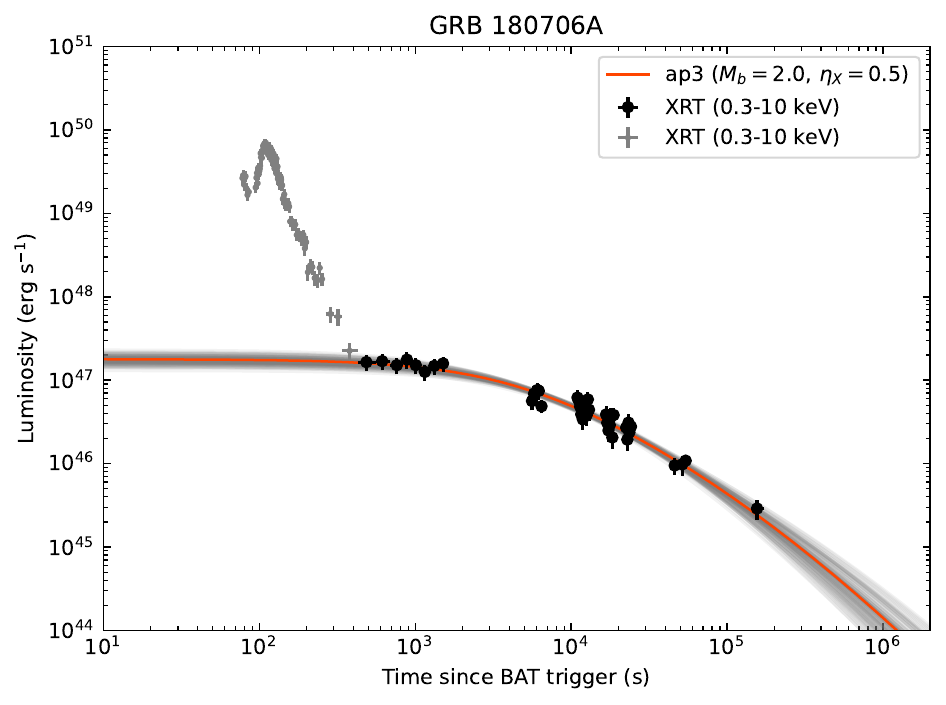}
\includegraphics  [angle=0,scale=0.25] {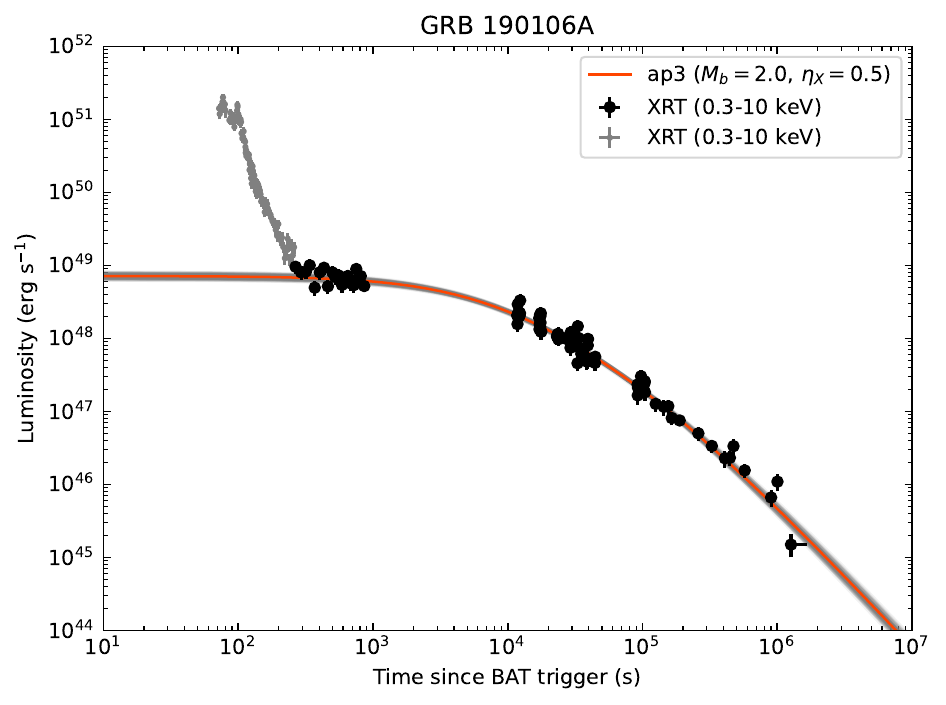}
\includegraphics  [angle=0,scale=0.25] {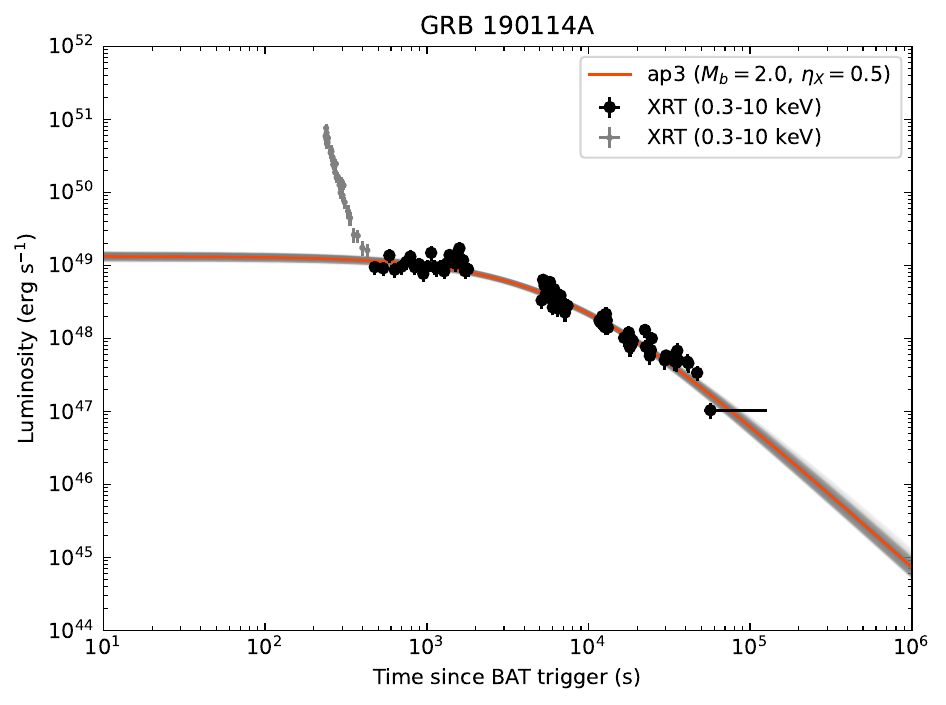}\\
\includegraphics  [angle=0,scale=0.25] {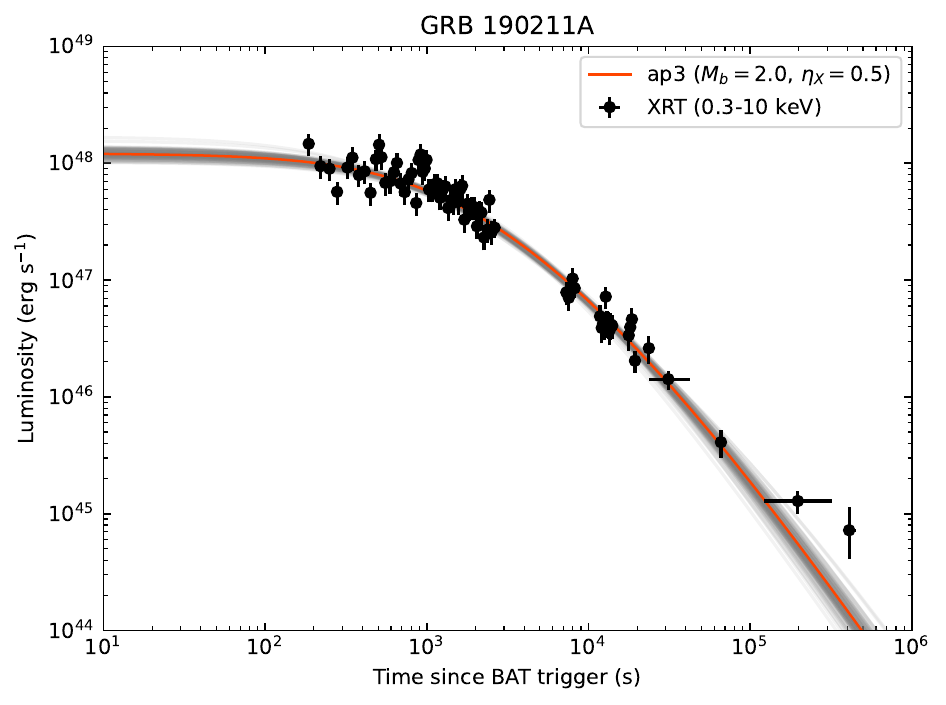}
\includegraphics  [angle=0,scale=0.25] {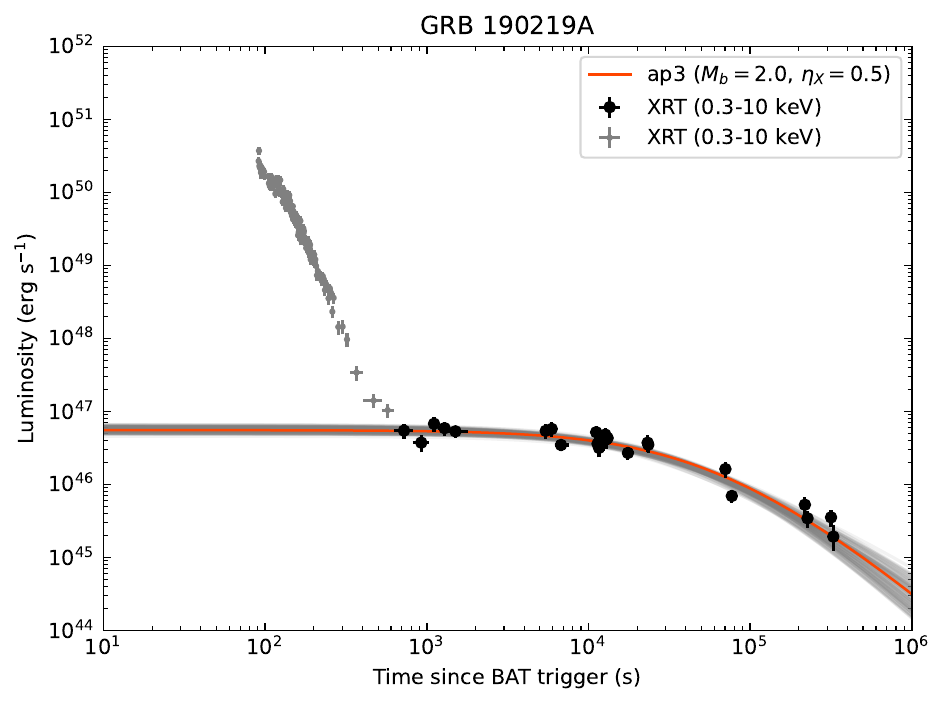}
\includegraphics  [angle=0,scale=0.25] {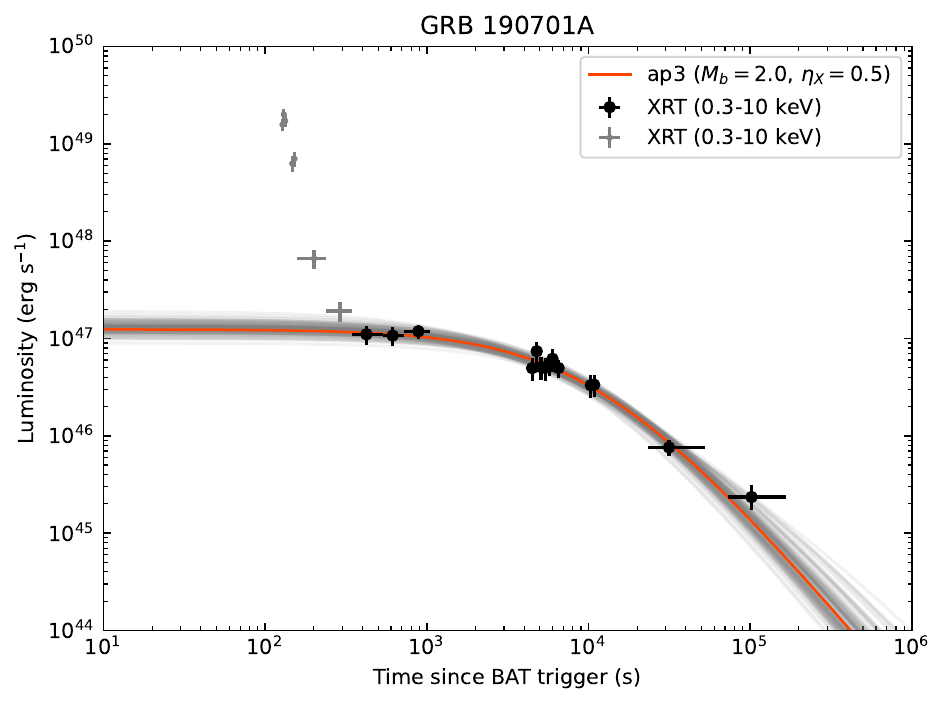}
\includegraphics  [angle=0,scale=0.25] {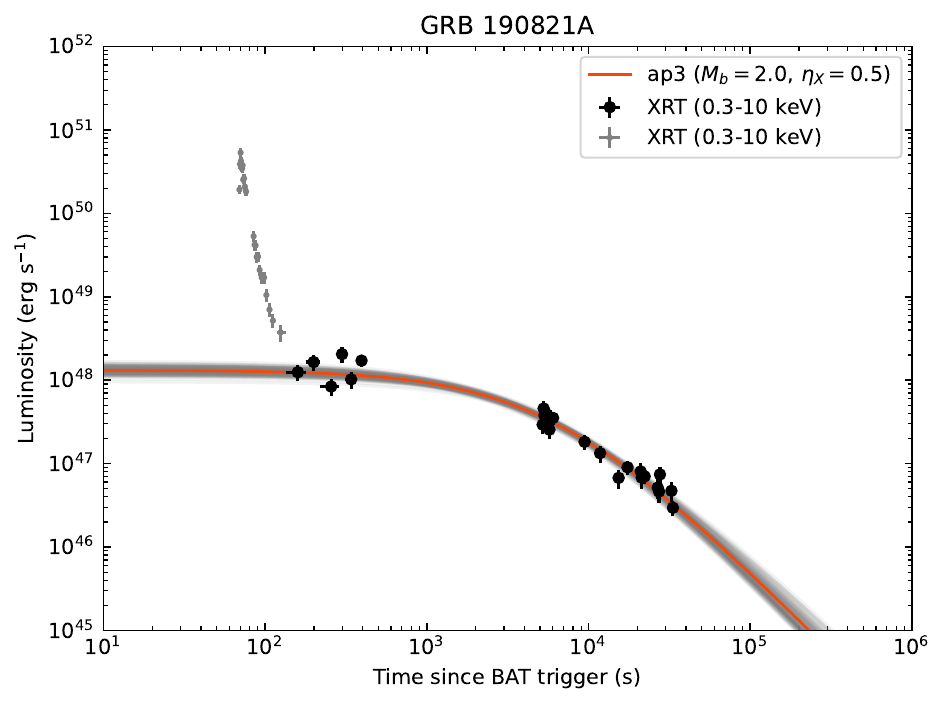}\\
\includegraphics  [angle=0,scale=0.25] {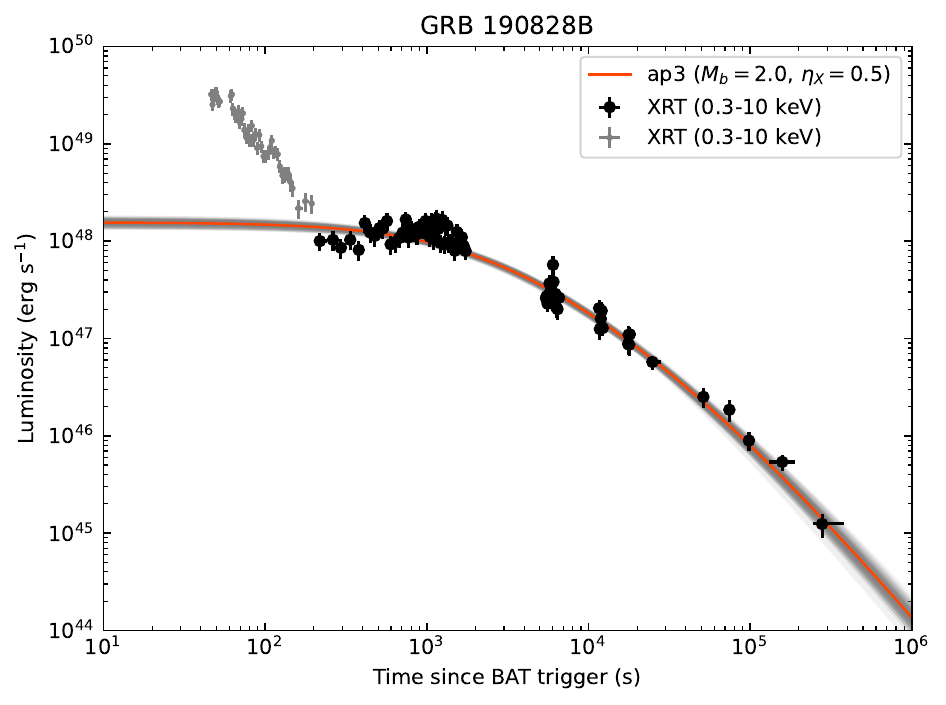}
\includegraphics  [angle=0,scale=0.25] {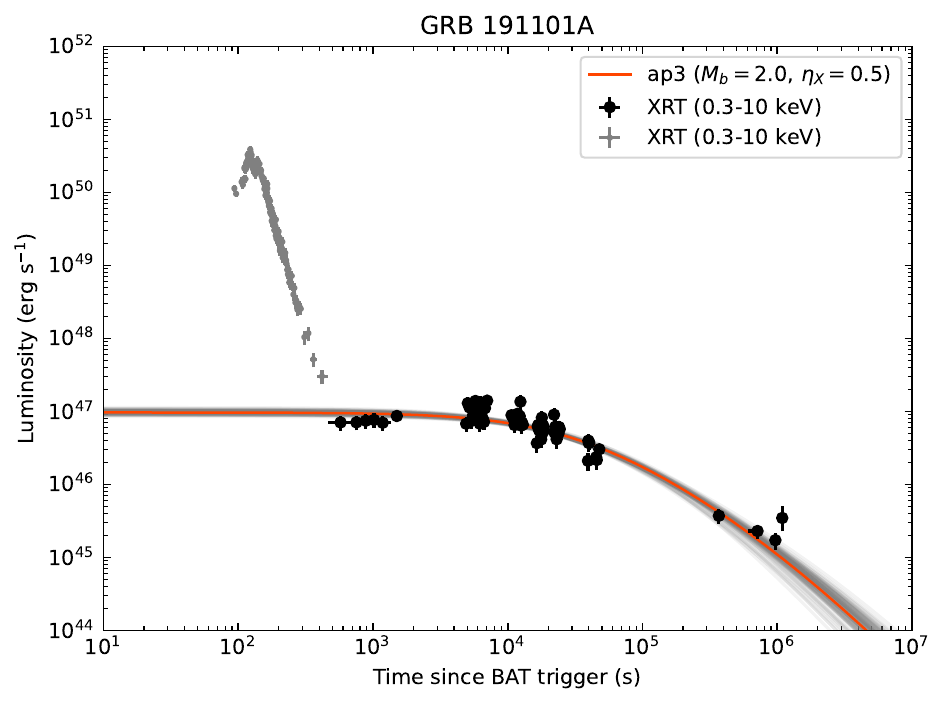}
\includegraphics  [angle=0,scale=0.25] {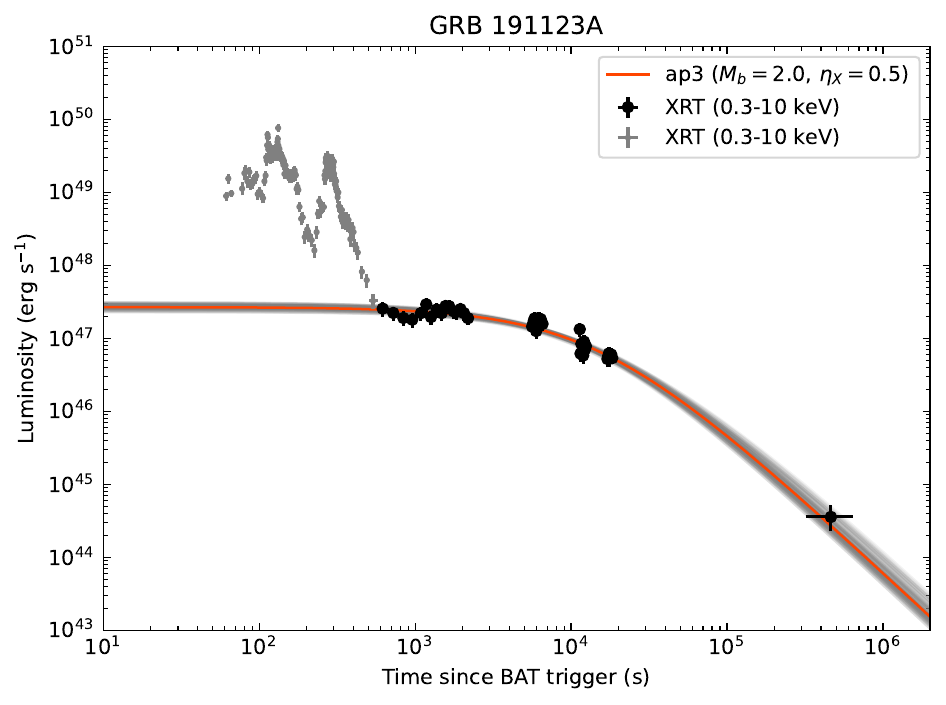}
\includegraphics  [angle=0,scale=0.25] {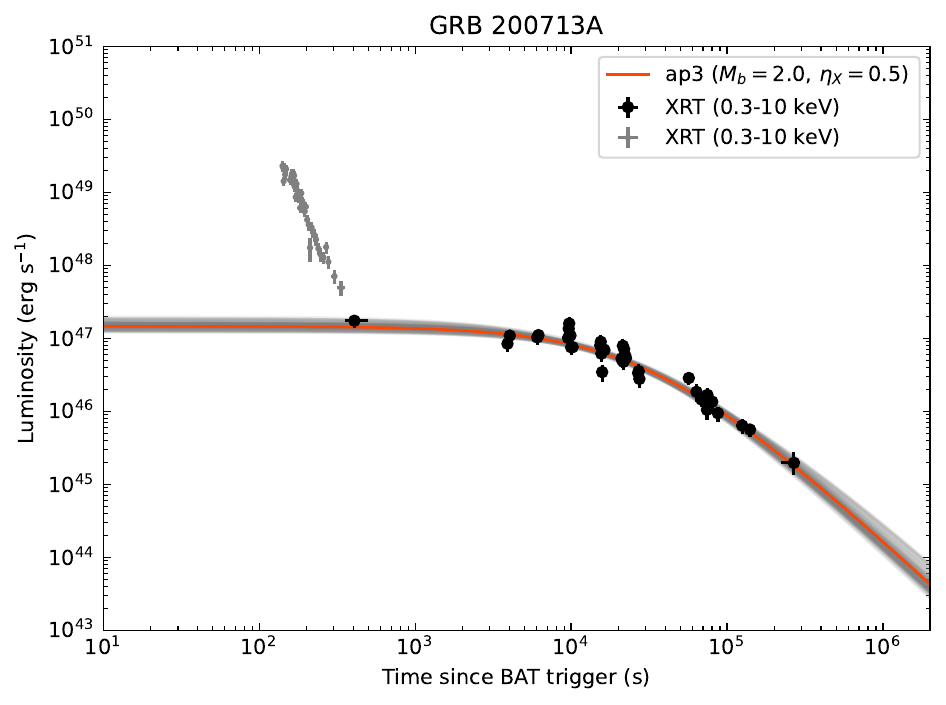}\\
\includegraphics  [angle=0,scale=0.25] {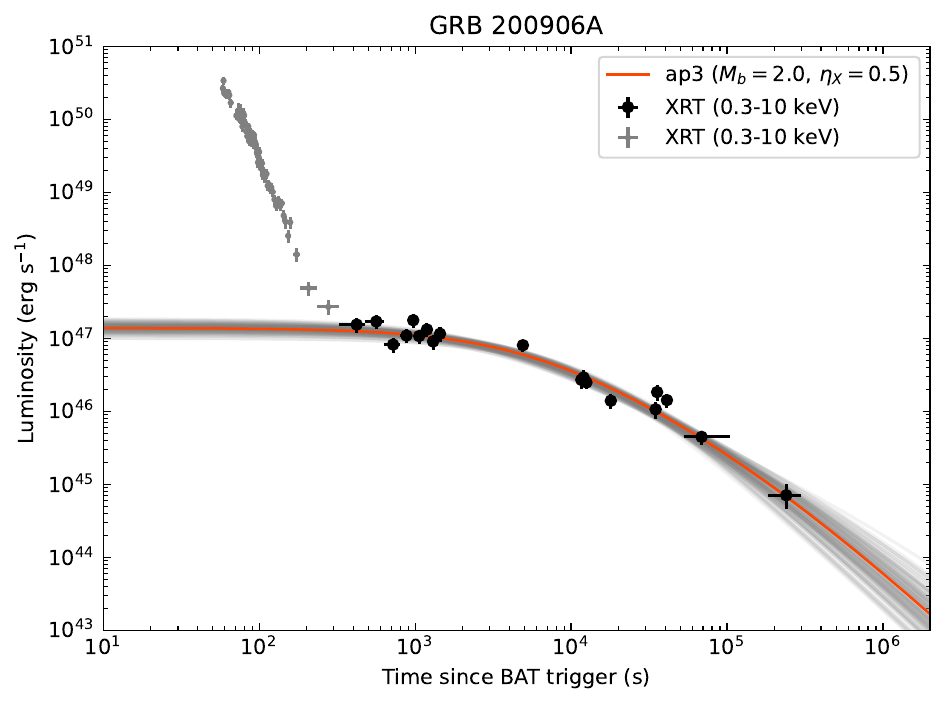}
\includegraphics  [angle=0,scale=0.25] {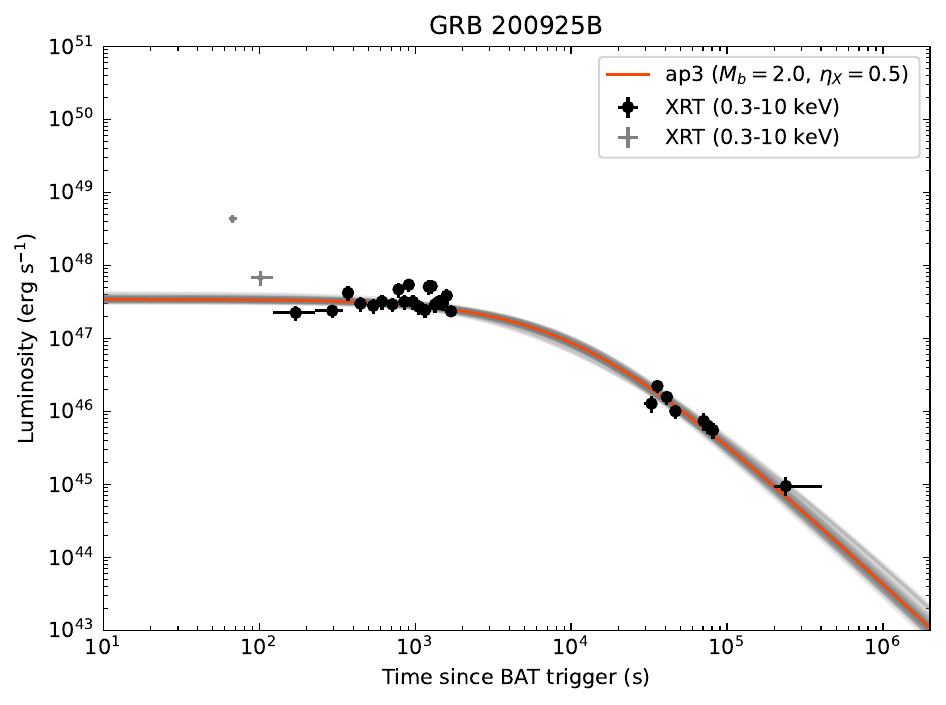}
\includegraphics  [angle=0,scale=0.25] {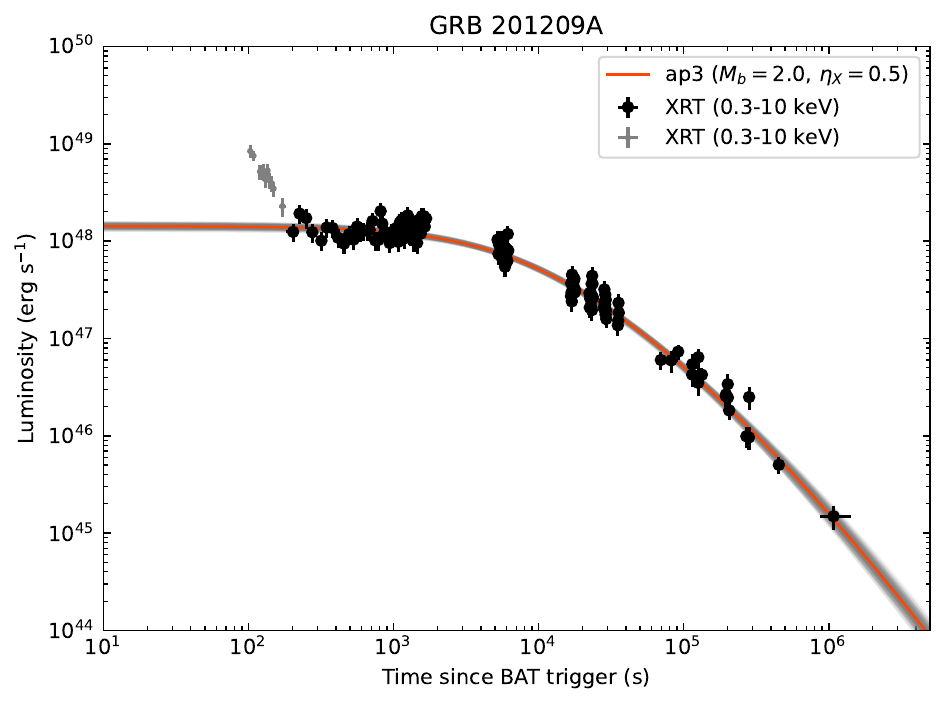}
\includegraphics  [angle=0,scale=0.25] {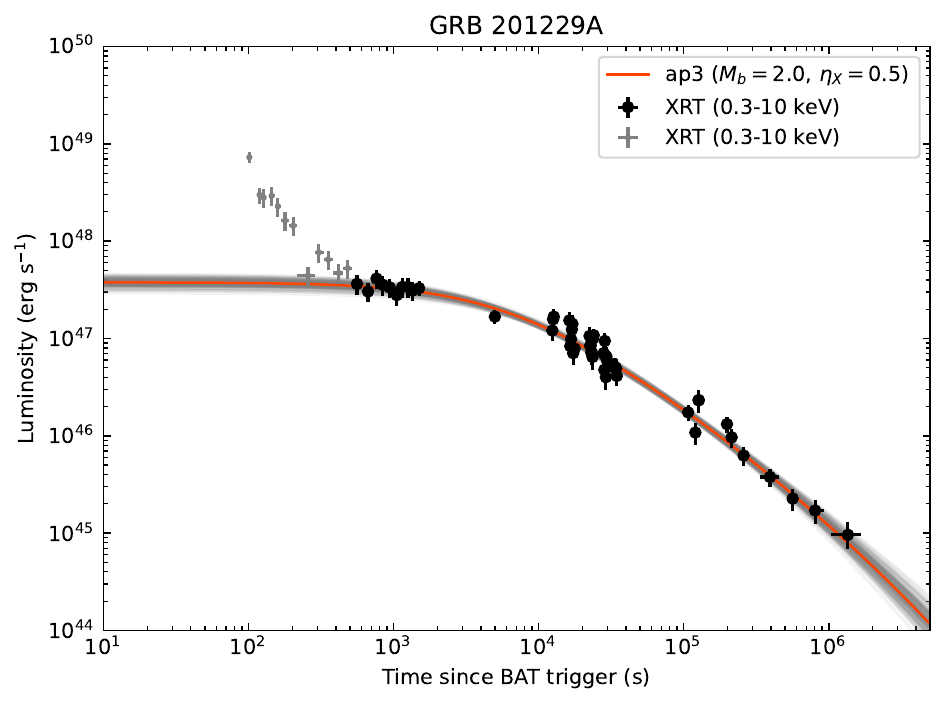}\\
\includegraphics  [angle=0,scale=0.25] {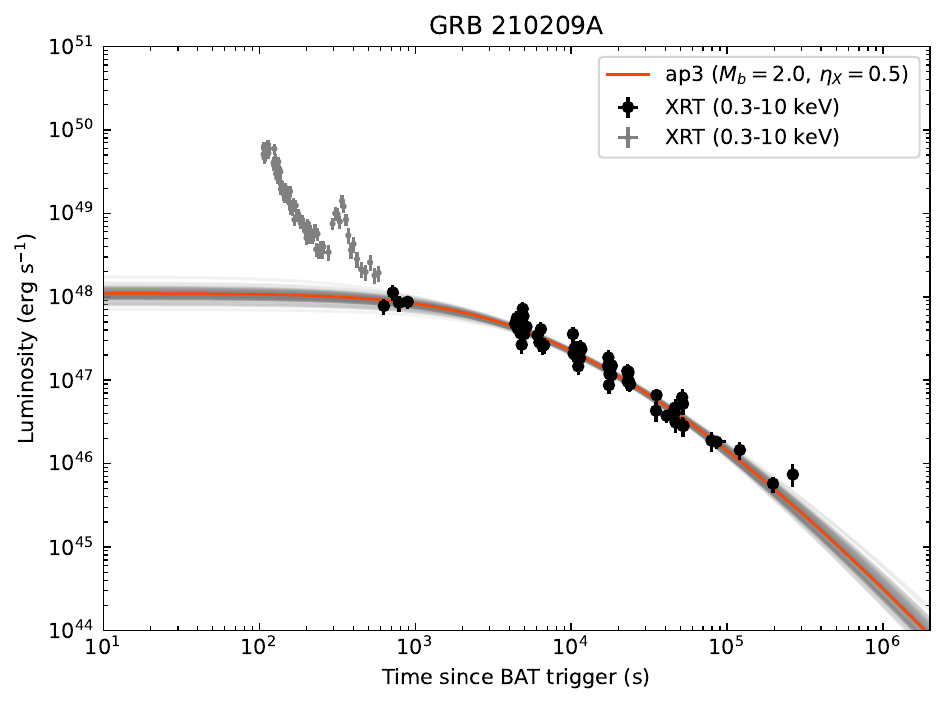}
\includegraphics  [angle=0,scale=0.25] {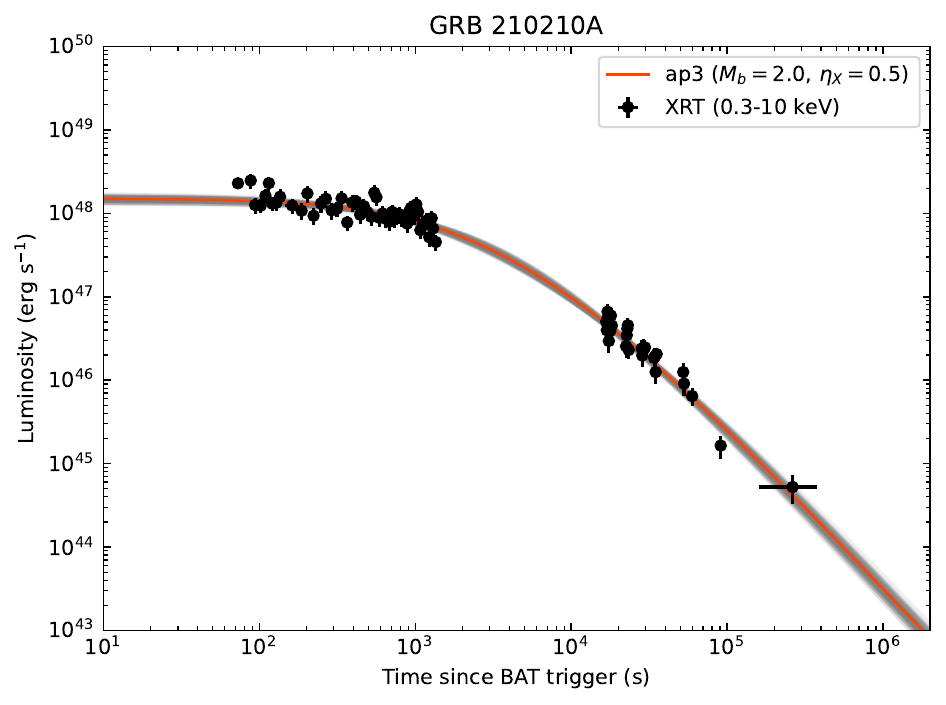}
\includegraphics  [angle=0,scale=0.25] {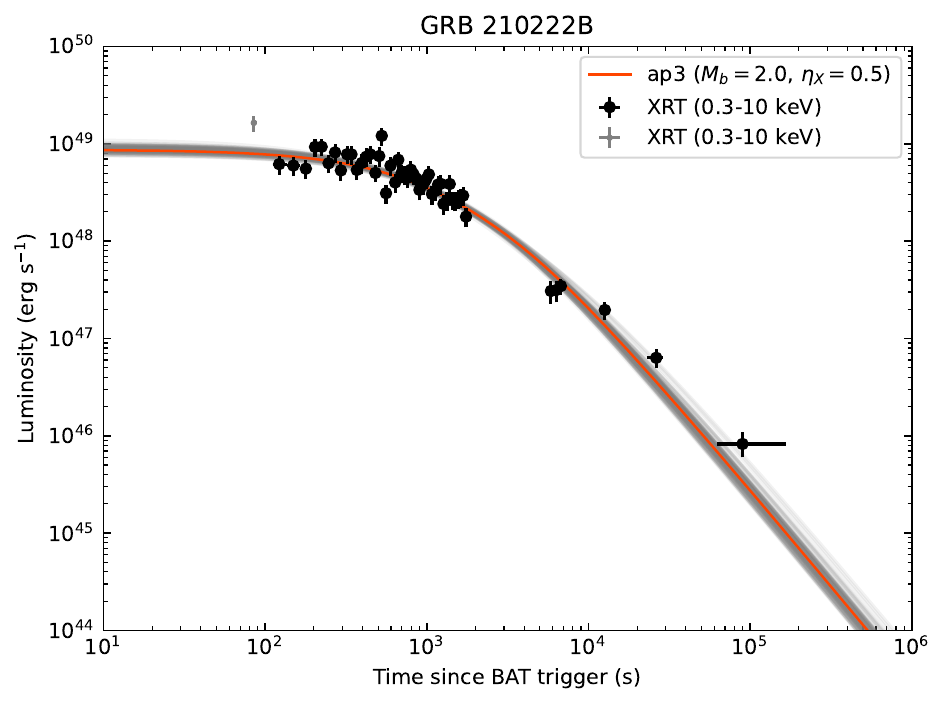}
\includegraphics  [angle=0,scale=0.25] {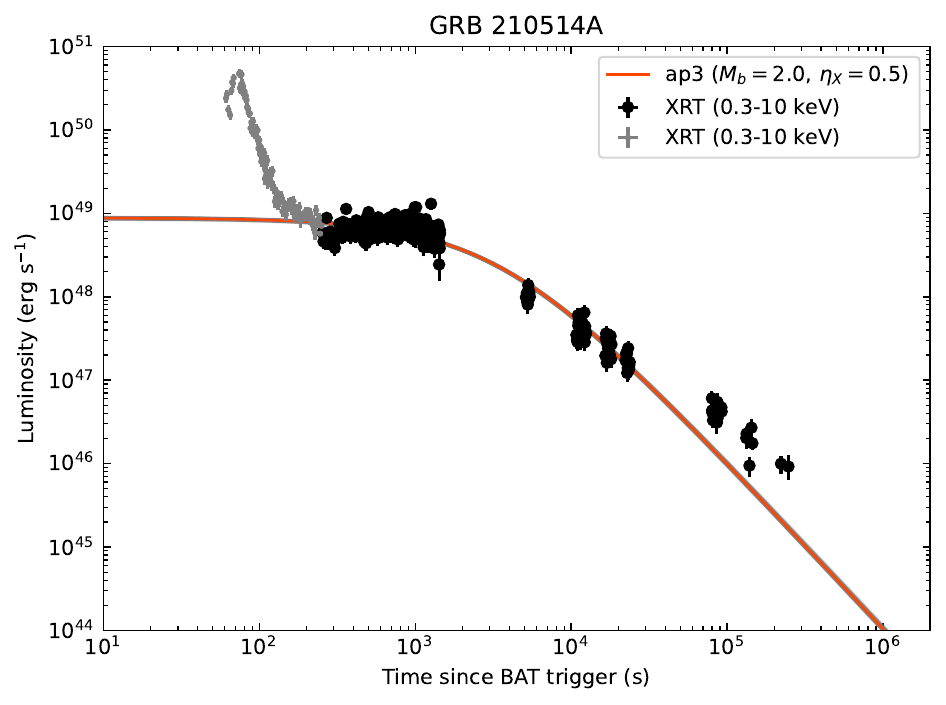}\\
\includegraphics  [angle=0,scale=0.25] {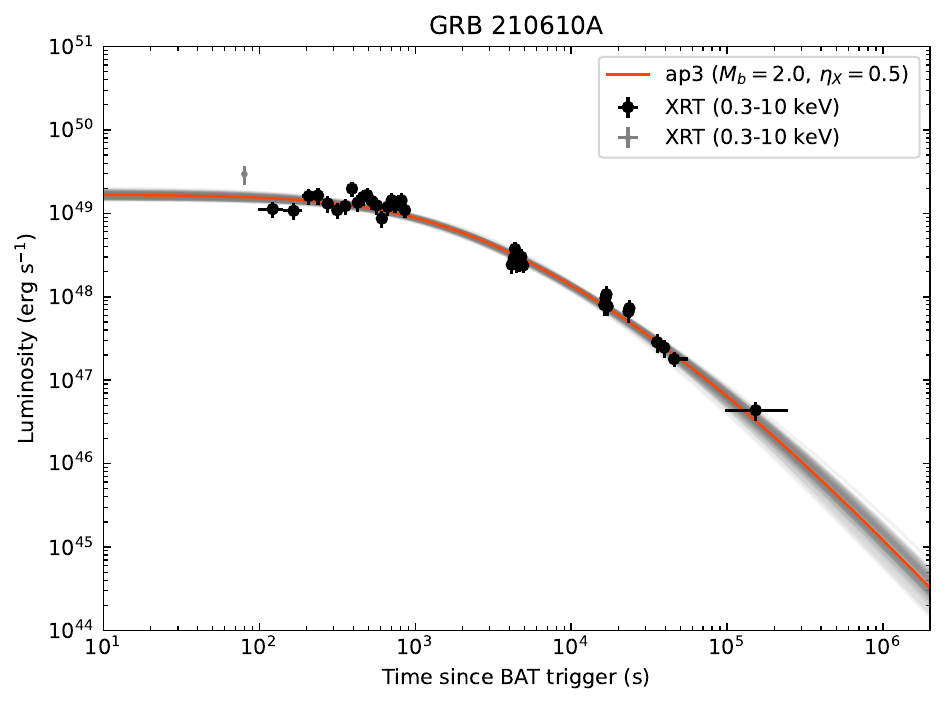}
\includegraphics  [angle=0,scale=0.25] {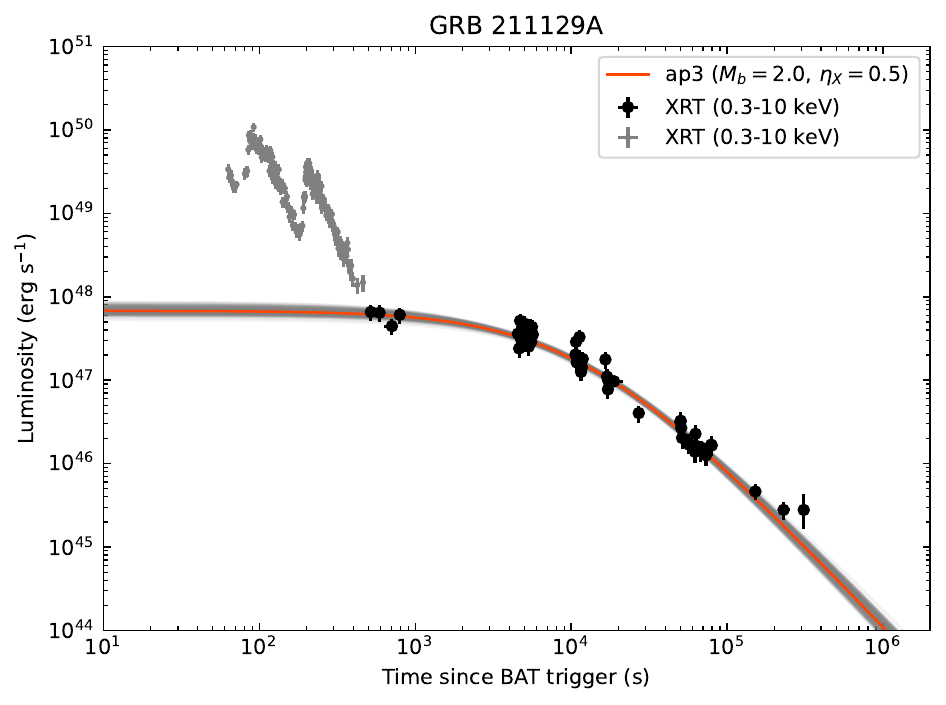}
\includegraphics  [angle=0,scale=0.25] {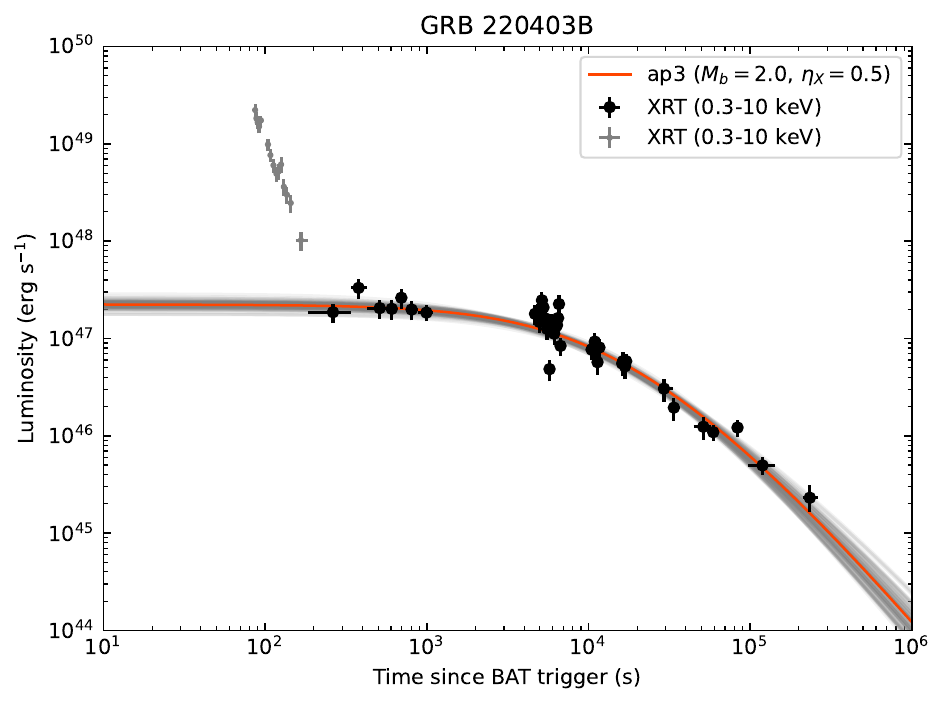}
\includegraphics  [angle=0,scale=0.25] {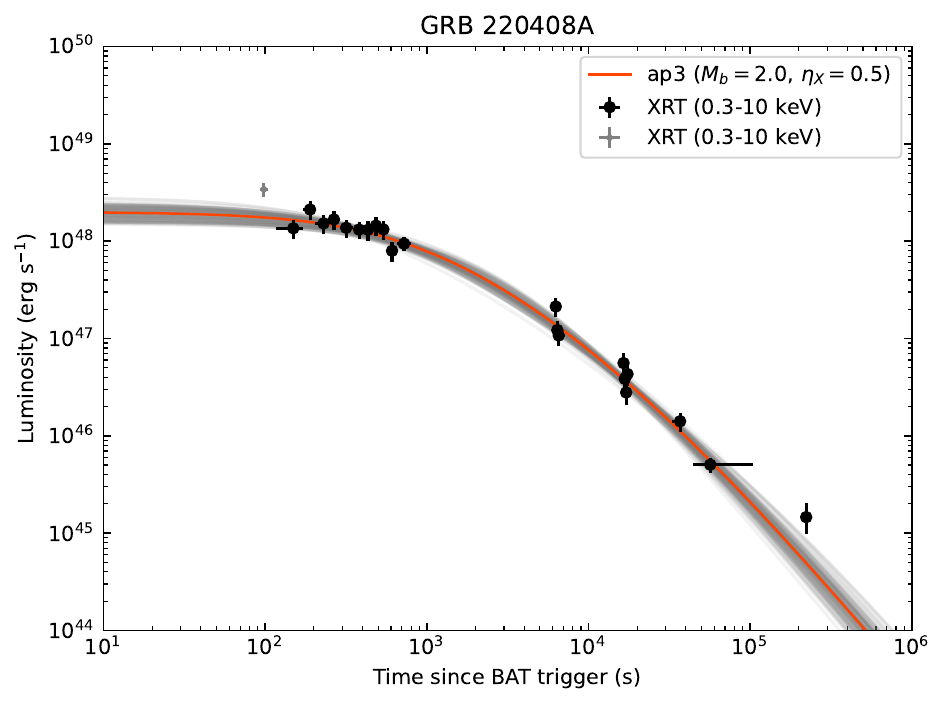}\\
\center{\textbf{Figure 27.} --- continued.}
\end{figure}

\begin{figure}
\centering
\includegraphics  [angle=0,scale=0.25] {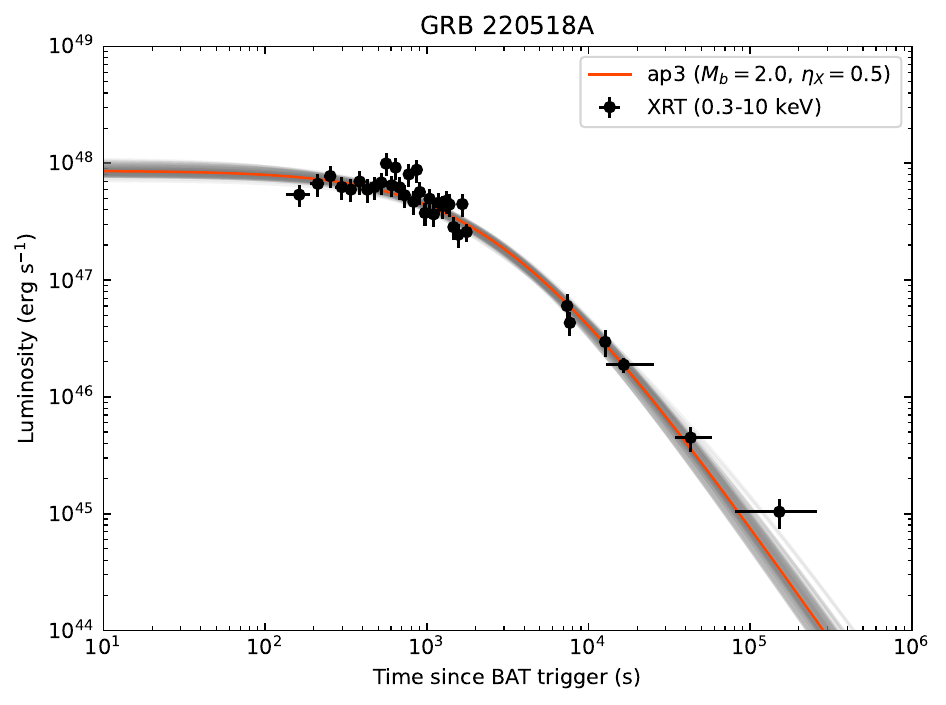}
\includegraphics  [angle=0,scale=0.25] {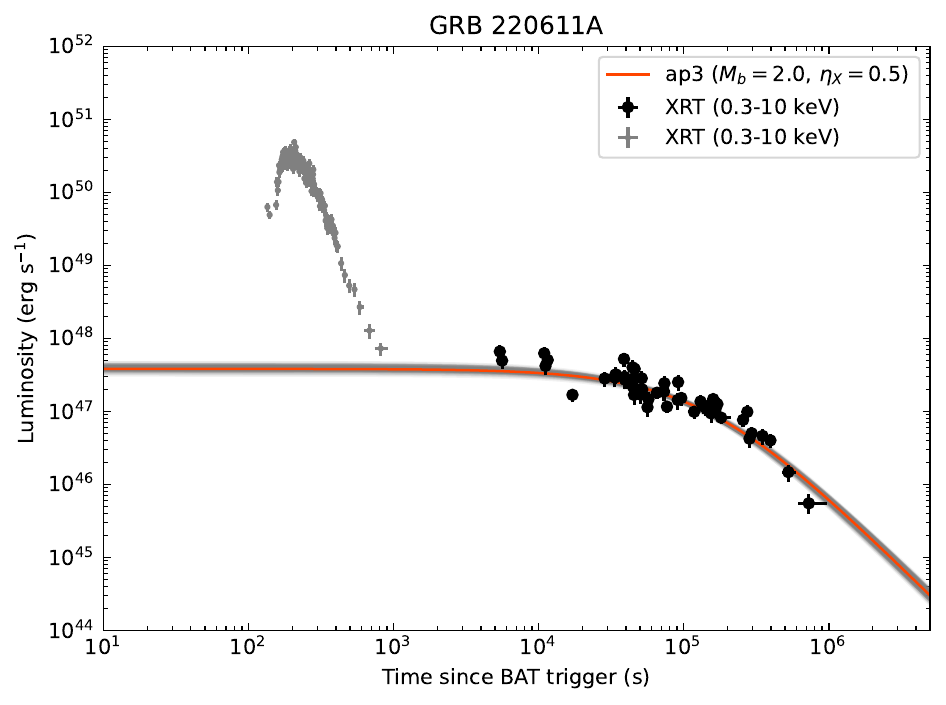}
\includegraphics  [angle=0,scale=0.25] {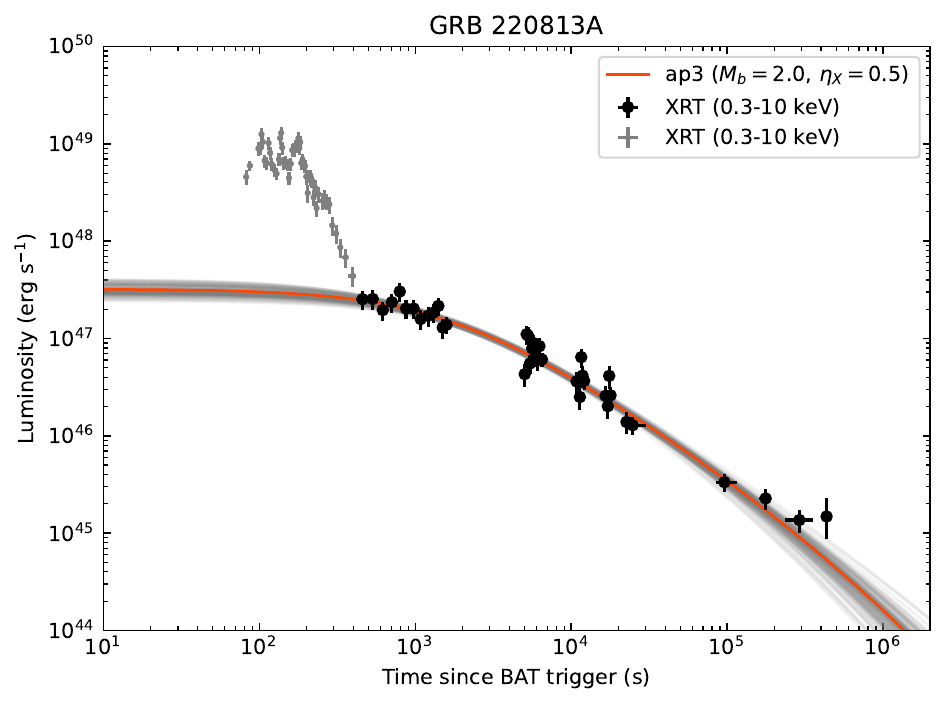}
\includegraphics  [angle=0,scale=0.25] {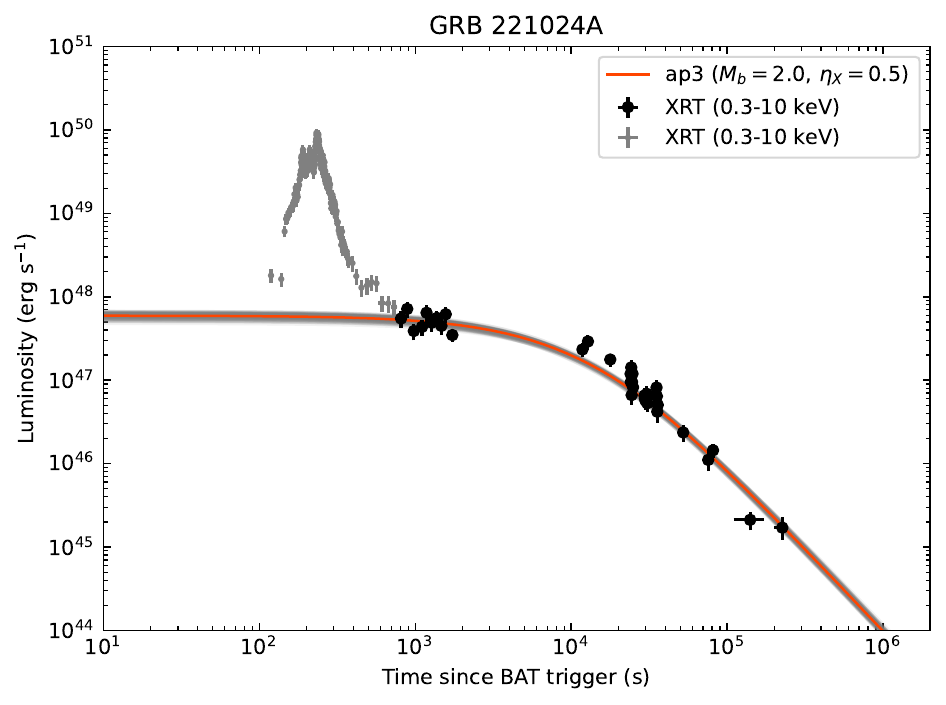}\\
\includegraphics  [angle=0,scale=0.25] {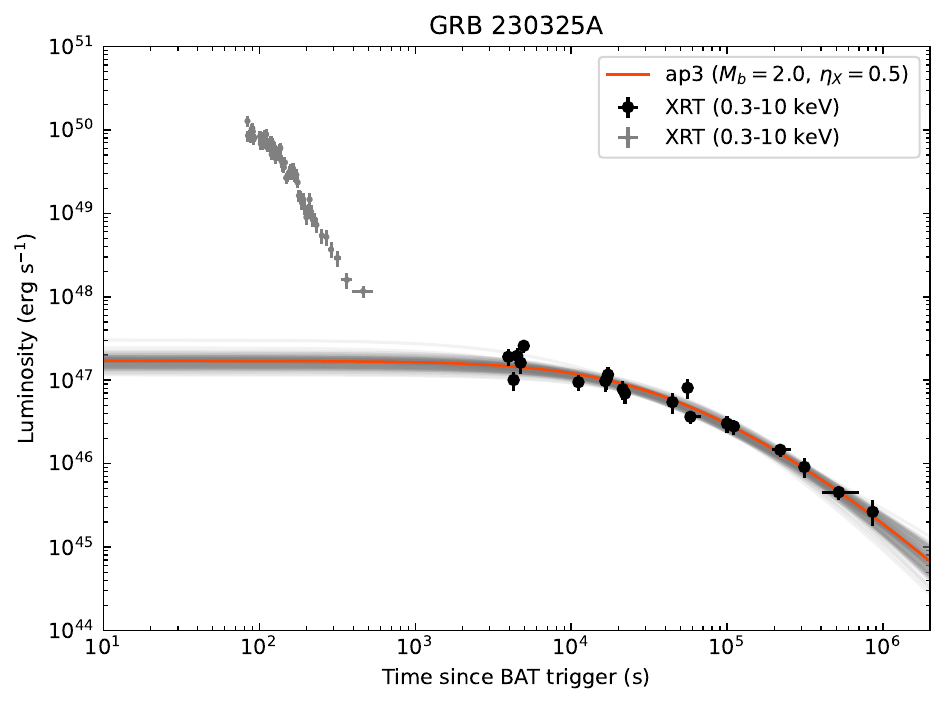}
\includegraphics  [angle=0,scale=0.25] {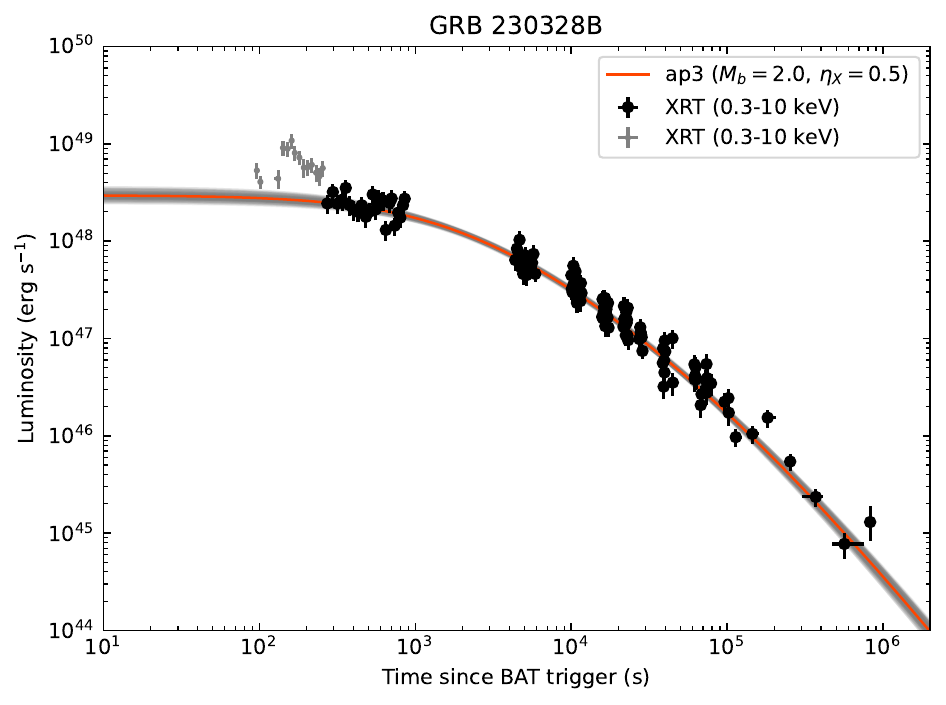}
\includegraphics  [angle=0,scale=0.25] {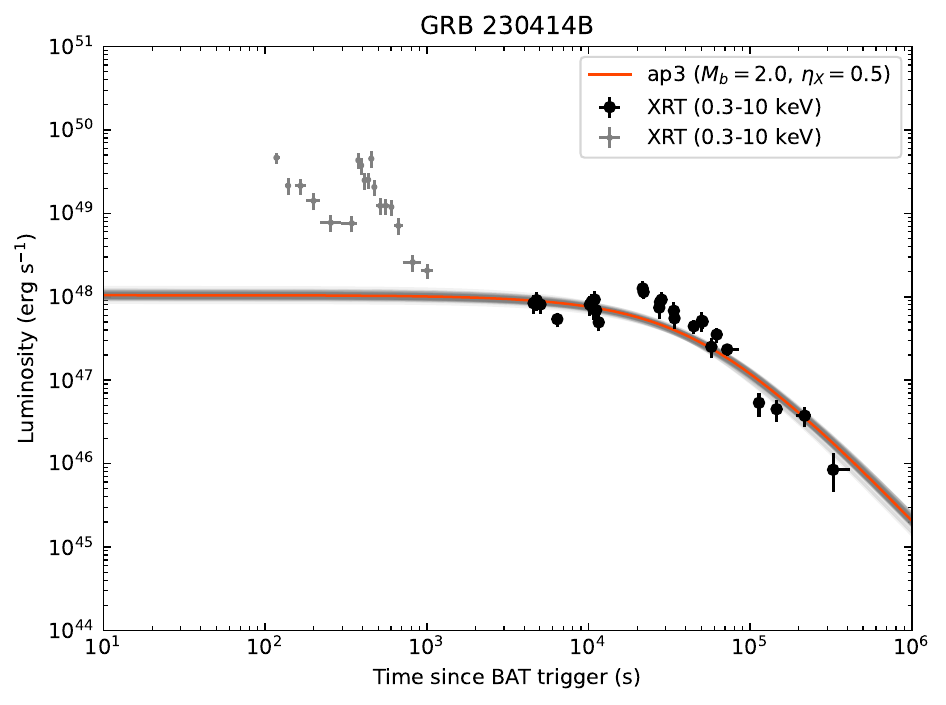}
\includegraphics  [angle=0,scale=0.25] {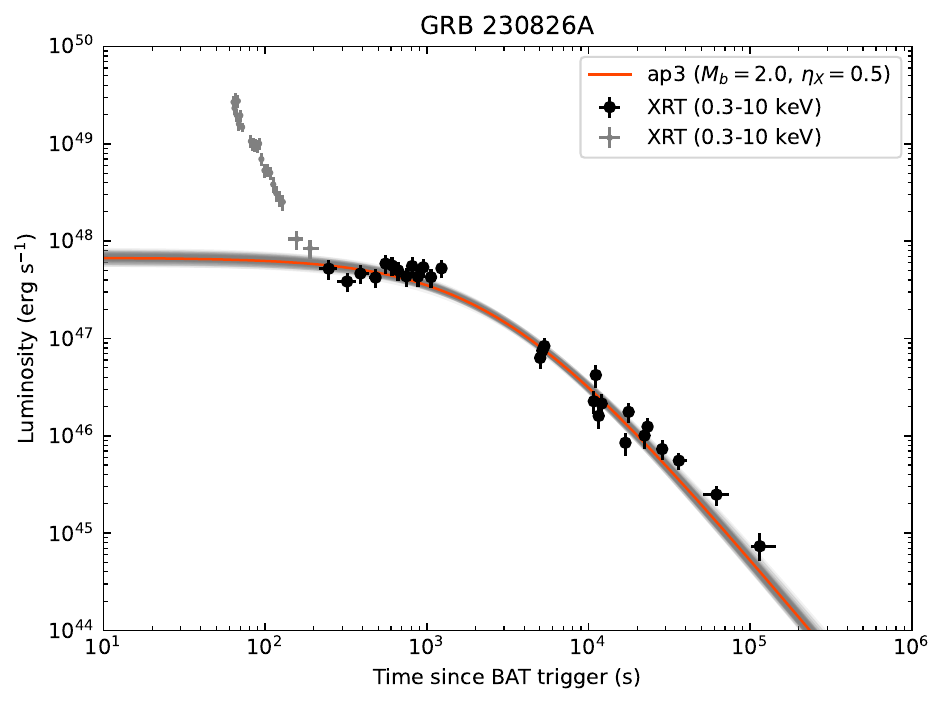}\\
\includegraphics  [angle=0,scale=0.25] {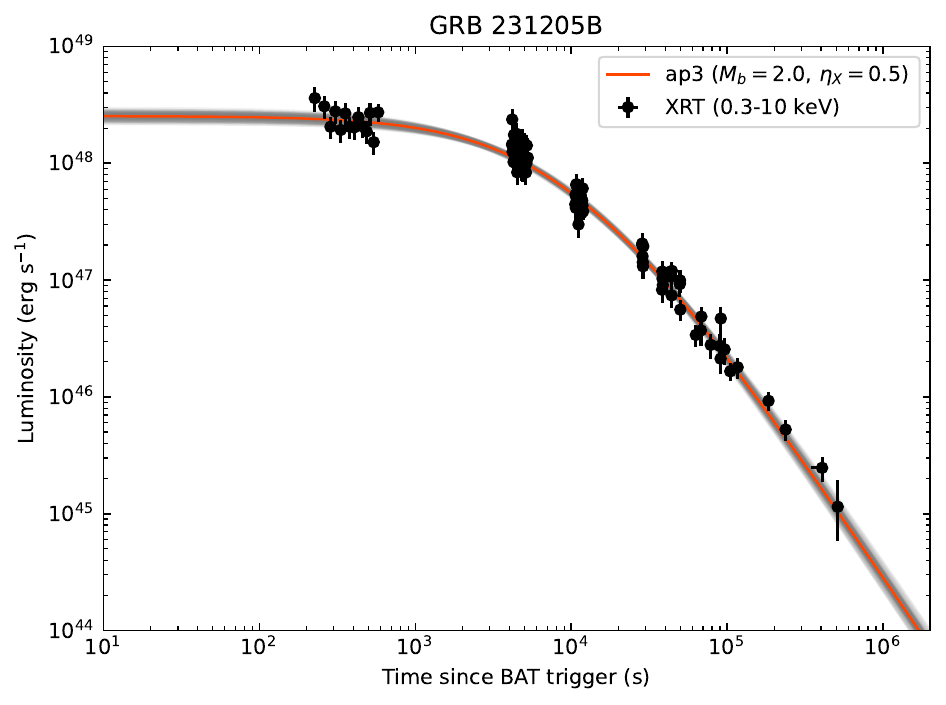}
\center{\textbf{Figure 27.} --- continued.}
\end{figure}



\setlength{\tabcolsep}{4mm}
\begin{deluxetable}{cccccccccc}
\tabletypesize{\scriptsize}
\tablecolumns{9}
\tablewidth{0pc}
\tablecaption{The observational properties for the long GRBs in our sample. \label{table-1}}
\tablenum{1}
\tablehead{
\colhead{GRB$^a$} & \colhead{$z^b$} & \colhead{$T_{\rm 90}^c$} & \colhead{$\theta_{j}^d$} & \colhead{$E_{\rm jet,iso}^e$} & \colhead{$E_{\rm jet}^f$} & \colhead{$E_{\rm wind}^g$} & \colhead{$L_{p}^h$} & \colhead{Refs$^i$}\\
\colhead{}    &  \colhead{}  & \colhead{(s)}   &  \colhead{($^\circ$)}  & \colhead{($10^{52}$ erg)} & \colhead{($10^{50}$ erg)}  & \colhead{($10^{51}$ erg)}   & \colhead{($10^{47} \rm erg~s^{-1}$)}  & \colhead{}}
\startdata
050822  & 1.434  & 103.4  & 5   & $11.70\pm0.76$  & $4.46\pm0.29$ &  $1.77\pm0.18$ & $1.64\pm0.29$  & (1) &\\
050922B & 4.9    & 150.8  & 5   & $60.80\pm8.80$  & $23.20\pm3.35$ &  $32.20\pm6.11$ & $9.14\pm8.00$  & (2) &\\
051109B & (1.0)  & 14.2   & 5   & $0.29\pm0.05$  & $0.11\pm0.02$ &  $0.12\pm0.02$ & $0.62\pm0.12$  & ... &\\
060109  & (1.0)  & 115.4  & 5   & $0.81\pm0.13$  & $0.31\pm0.05$ &  $0.33\pm0.04$ & $1.10\pm0.12$  & ... &\\
060428A & (1.0)  & 39.5   & 5   & $1.60\pm0.10$  & $0.61\pm0.04$ &  $0.87\pm0.14$ & $9.52\pm0.63$  & ... &\\
060510A & (1.0)  & 20.4   & 5   & $14.90\pm0.58$  & $5.68\pm0.22$ &  $15.10\pm0.74$ & $32.59\pm2.90$  & ... &\\
060714  & 2.71   & 115.0  & $1.49\pm0.06$  & $21.40\pm1.25$ &  $0.72\pm0.07$ & $4.66\pm0.54$  & $31.68\pm8.36$ & (3) &\\
060729  & 0.54   & 115.3  & $18\pm1.61$   & $1.11\pm0.08$ & $5.49\pm1.06$  & $2.89\pm0.11$  & $0.56\pm0.02$ & (4) &\\
\enddata
\end{deluxetable}

\setlength{\tabcolsep}{4mm}
\begin{deluxetable}{cccccccccc}
\tabletypesize{\scriptsize}
\tablecolumns{9}
\tablewidth{0pc}
\tablecaption{--- continued \label{table-1}}
\tablenum{1}
\tablehead{
\colhead{GRB$^a$} & \colhead{$z^b$} & \colhead{$T_{\rm 90}^c$} & \colhead{$\theta_{j}^d$} & \colhead{$E_{\rm jet,iso}^e$} & \colhead{$E_{\rm jet}^f$} & \colhead{$E_{\rm wind}^g$} & \colhead{$L_{p}^h$} & \colhead{Refs$^i$}\\
\colhead{}      &  \colhead{}     & \colhead{(s)}   &  \colhead{($^\circ$)} & 
\colhead{($10^{52}$ erg)} & \colhead{($10^{50}$ erg)} & \colhead{($10^{51}$ erg)}   & \colhead{($10^{47} \rm erg~s^{-1}$)}  & \colhead{}}
\startdata
170607A & 0.557  & 320.0  & 5   & $4.70\pm0.20$  & $1.79\pm0.07$ & $0.76\pm0.04$  & $0.63\pm0.08$  & (32) &\\
170711A & (1.0)  & 31.3   & 5   & $1.90\pm0.11$  & $0.73\pm0.04$ & $0.15\pm0.03$  & $1.49\pm0.33$  & ... &\\
170906A & (1.0)  & 88.1   & 5   & $85.60\pm1.10$  & $32.60\pm0.42$ & $4.39\pm0.13$  & $118.38\pm8.06$  & ... &\\
171102B & (1.0)  & 17.8   & 5   & $6.04\pm0.96$  & $2.30\pm0.37$ & $17.00\pm2.75$  & $45.37\pm18.10$  & ... &\\
180314B & (1.0)  & 73.0   & 5   & $1.71\pm0.25$  & $0.65\pm0.10$ & $0.93\pm0.10$  & $17.77\pm2.25$  & ... &\\
180325A & 2.25   & 94.1   & 5   & $116.00\pm3.11$  & $44.10\pm1.18$ & $59.50\pm3.66$  & $708.88\pm132.91$  & (33) &\\
180329B & 1.998  & 210.0  & 5   & $16.10\pm1.21$  & $6.12\pm0.46$ & $2.90\pm0.29$  & $12.10\pm1.88$  & (34) &\\
180411A & (1.0)  & 78.0   & 5   & $33.10\pm0.55$  & $12.60\pm0.21$ & $3.58\pm0.38$  & $7.79\pm1.11$  & ... &\\
\enddata
\tablenotetext{a}{ The GRB name.}
\tablenotetext{b}{ The GRB redshift measurement, and we adopt $z = 1.0$ for these LGRBs without redshift measurement.}
\tablenotetext{c}{ The value of $T_{90}$.}
\tablenotetext{d}{ The jet opening angle measured from afterglow observations \citep{Racusin2009,Lu2012,Nemmen2012,Wang2018,Zhao2020}, or assumed as $\theta_{j}=5^{\circ}$ if no observation is available.}
\tablenotetext{e}{ The isotropic jet energy.}
\tablenotetext{f}{ The corrected jet energy.}
\tablenotetext{g}{ The magnetar wind energy.}
\tablenotetext{h}{ The X-ray plateau luminosity.}
\tablenotetext{i}{ The reference of redshift for our GRB sample.}
\tablecomments{References: (1) \cite{Hjorth2012}; (2) \cite{Perley2016}; (3) \cite{Jakobsson2006}; (4) \cite{Thoene2006}; (5) \cite{Bloom2006}; (6) \cite{Perley2009}; (7) \cite{Kruehler2012}; (8) \cite{Jaunsen2007}; (9) \cite{Kruehler2015}; (10) \cite{Prochaska2007}; (11) \cite{Prochaska2008}; (12) \cite{Vreeswijk2008}; (13) \cite{Fugazza2009}; (14) \cite{Perley2013}; (15) \cite{Chornock2009a}; (16) \cite{Chornock2009b}; (17) \cite{Wiersema2009}; (18) \cite{Vergani2010}; (19) \cite{Kruehler2013}; (20) \cite{Milne2011}; (21) \cite{Wiersema2011}; (22) \cite{Dittmann2011}; (23) \cite{Cucchiara2012}; (24) \cite{Xu2013}; (25) \cite{Chornock2014}; (26) \cite{Perley2015a}; (27) \cite{de-Ugarte-Postigo2015}; (28) \cite{Perley2015b}; (29) \cite{Xu2016}; (30) \cite{Xu2017}; (31) \cite{de-Ugarte-Postigo2017a}; (32) \cite{de-Ugarte-Postigo2017b}; (33) \cite{Heintz2018}; (34) \cite{Izzo2018}; (35) \cite{Xu2019}; (36) \cite{de-Ugarte-Postigo2019}; (37) \cite{de-Ugarte-Postigo2021}; (38) \cite{Agui-Fernandez2021}; (39) \cite{Zhu2021}; (40) \cite{Schneider2022}; (41) \cite{Pieterse2023}; (42) \cite{Agui-Fernandez2023}.}
\end{deluxetable}

\clearpage
\setlength{\tabcolsep}{5.5mm}
\begin{deluxetable}{cccccccccc}
\tablecolumns{9}
\tablewidth{0pc}
\tabletypesize{\scriptsize}
\tablecaption{The derived magnetar physical parameters for our LGRB sample in the AP3 EoS with $M_{b}=2.0~M_{\odot}$. \label{table-2}}
\tablenum{2}
\tablehead{
\colhead{GRB} & \colhead{$B_{p,15}^a$(G)} & \colhead{$P_{0}^a$(ms)}  & \colhead{$\epsilon_{-3}^a$}  &  \colhead{$\chi^2$/dof}  &  \colhead{$B_{p,15}^b$(G)}  & \colhead{$P_{0}^b$(ms)}  &  \colhead{$\epsilon_{-3}^b$}  &  \colhead{$\chi^2$/dof} }
\startdata
050822  & $0.04^{+0.02}_{-0.01}$  & $0.92^{+0.10}_{-0.09}$ & $0.63^{+0.12}_{-0.09}$  & 98/92   & $0.09^{+0.03}_{-0.02}$ & $1.37^{+0.18}_{-0.16}$  & $1.43^{+0.32}_{-0.26}$ & 97/92 \\
050922B & $0.04^{+0.01}_{-0.01}$  & $0.85^{+0.03}_{-0.05}$ & $0.05^{+0.03}_{-0.03}$  & 20/16   & $0.07^{+0.01}_{-0.01}$ & $1.16^{+0.05}_{-0.07}$  & $0.09^{+0.06}_{-0.06}$ & 20/16 \\
051109B & $0.81^{+0.21}_{-0.16}$  & $2.19^{+0.28}_{-0.24}$ & $4.19^{+0.60}_{-1.12}$  & 21/19   & $2.58^{+0.20}_{-0.19}$ & $5.71^{+0.19}_{-0.18}$  & $5.76^{+3.00}_{-3.76}$ & 25/19 \\
060109  & $0.46^{+0.02}_{-0.02}$  & $1.52^{+0.03}_{-0.03}$ & $0.23^{+0.23}_{-0.16}$  & 89/48   & $1.10^{+0.04}_{-0.04}$ & $3.32^{+0.06}_{-0.06}$  & $1.14^{+1.18}_{-0.80}$ & 90/48 \\
060428A & $0.07^{+0.02}_{-0.02}$  & $1.08^{+0.10}_{-0.10}$ & $0.75^{+0.06}_{-0.06}$  & 201/127 & $0.16^{+0.01}_{-0.01}$ & $1.83^{+0.11}_{-0.11}$  & $1.96^{+0.11}_{-0.11}$ & 204/127 \\
060510A & $0.16^{+0.02}_{-0.01}$  & $0.98^{+0.03}_{-0.03}$ & $0.89^{+0.02}_{-0.02}$  & 155/163 & $0.40^{+0.02}_{-0.02}$ & $1.53^{+0.05}_{-0.05}$  & $2.21^{+0.06}_{-0.06}$ & 154/163 \\
060714  & $0.14^{+0.05}_{-0.02}$  & $0.94^{+0.07}_{-0.07}$ & $1.12^{+0.09}_{-0.09}$  & 59/44   & $0.35^{+0.06}_{-0.06}$ & $1.43^{+0.12}_{-0.11}$  & $2.71^{+0.23}_{-0.22}$ & 57/44 \\
060729  & $0.04^{+0.01}_{-0.01}$  & $0.92^{+0.02}_{-0.02}$ & $0.17^{+0.01}_{-0.01}$  & 917/651 & $0.08^{+0.01}_{-0.01}$ & $1.17^{+0.02}_{-0.02}$  & $0.28^{+0.01}_{-0.01}$ & 921/651 \\
\enddata
\end{deluxetable}

\setlength{\tabcolsep}{5.5mm}
\begin{deluxetable}{cccccccccc}
\tablecolumns{9}
\tablewidth{0pc}
\tabletypesize{\scriptsize}
\tablecaption{--- continued \label{table-2}}
\tablenum{2}
\tablehead{
\colhead{GRB} & \colhead{$B_{p,15}^a$(G)} & \colhead{$P_{0}^a$(ms)}  & \colhead{$\epsilon_{-3}^a$}  &  \colhead{$\chi^2$/dof}  &  \colhead{$B_{p,15}^b$(G)}  & \colhead{$P_{0}^b$(ms)}  &  \colhead{$\epsilon_{-3}^b$}  &  \colhead{$\chi^2$/dof} }
\startdata
160630A & $0.21^{+0.11}_{-0.06}$  & $1.00^{+0.12}_{-0.12}$ & $1.80^{+0.22}_{-0.28}$  & 91/66   & $0.32^{+0.13}_{-0.10}$ & $1.21^{+0.17}_{-0.15}$  & $2.75^{+0.46}_{-0.44}$ & 89/66 \\
170113A & $0.11^{+0.01}_{-0.01}$  & $0.68^{+0.04}_{-0.04}$ & $0.80^{+0.07}_{-0.06}$  & 162/139 & $0.14^{+0.03}_{-0.01}$ & $0.93^{+0.06}_{-0.06}$  & $1.57^{+0.12}_{-0.13}$ & 159/139 \\
170202A & $0.11^{+0.03}_{-0.02}$  & $0.72^{+0.09}_{-0.09}$ & $0.57^{+0.06}_{-0.07}$  & 57/39   & $0.21^{+0.08}_{-0.08}$ & $1.06^{+0.11}_{-0.12}$  & $1.19^{+0.10}_{-0.11}$ & 57/39 \\
170607A & $0.04^{+0.02}_{-0.01}$  & $0.94^{+0.09}_{-0.08}$ & $0.59^{+0.09}_{-0.09}$  & 144/148 & $0.06^{+0.02}_{-0.02}$ & $1.15^{+0.13}_{-0.12}$  & $0.90^{+0.19}_{-0.15}$ & 144/148 \\
170711A & $0.06^{+0.07}_{-0.02}$  & $0.95^{+0.21}_{-0.19}$ & $0.89^{+0.39}_{-0.31}$  & 49/40   & $0.17^{+0.09}_{-0.09}$ & $1.50^{+0.34}_{-0.38}$  & $2.24^{+0.93}_{-0.92}$ & 50/40 \\
170906A & $0.54^{+0.01}_{-0.01}$  & $0.82^{+0.01}_{-0.01}$ & $0.15^{+0.15}_{-0.10}$  & 214/199 & $1.18^{+0.03}_{-0.02}$ & $1.19^{+0.01}_{-0.01}$  & $0.31^{+0.32}_{-0.21}$ & 237/199 \\
171102B & $0.21^{+0.02}_{-0.03}$  & $0.85^{+0.05}_{-0.07}$ & $0.42^{+0.11}_{-0.22}$  & 27/33   & $0.60^{+0.06}_{-0.09}$ & $1.44^{+0.06}_{-0.11}$  & $0.97^{+0.46}_{-0.64}$ & 26/33 \\
180314B & $0.32^{+0.01}_{-0.02}$  & $0.85^{+0.02}_{-0.03}$ & $0.30^{+0.23}_{-0.20}$  & 145/62  & $1.13^{+0.05}_{-0.09}$ & $1.79^{+0.04}_{-0.09}$  & $1.49^{+1.18}_{-0.99}$ & 148/62 \\
\enddata
\tablenotetext{a}{ The X-ray radiation efficiency of the magnetar dipole radiation as $\eta = 0.1$.}
\tablenotetext{b}{ The X-ray radiation efficiency of the magnetar dipole radiation as $\eta =0.5$.}
\end{deluxetable}

\clearpage
\setlength{\tabcolsep}{5.5mm}
\begin{deluxetable}{cccccccccc}
\tablecolumns{9}
\tablewidth{0pc}
\tabletypesize{\scriptsize}
\tablecaption{The derived magnetar physical parameters for our LGRB sample in the AP3 EoS with $M_{b}=2.5~M_{\odot}$. \label{table-3}}
\tablenum{3}
\tablehead{
\colhead{GRB} & \colhead{$B_{p,15}^a$(G)} & \colhead{$P_{0}^a$(ms)}  & \colhead{$\epsilon_{-3}^a$}  &  \colhead{$\chi^2$/dof}  &  \colhead{$B_{p,15}^b$(G)}  & \colhead{$P_{0}^b$(ms)}  &  \colhead{$\epsilon_{-3}^b$}  &  \colhead{$\chi^2$/dof} }
\startdata
050822  & $0.07^{+0.02}_{-0.02}$  & $1.02^{+0.12}_{-0.10}$ & $0.69^{+0.14}_{-0.11}$  & 98/92   & $0.13^{+0.04}_{-0.03}$ & $1.54^{+0.21}_{-0.19}$  & $1.60^{+0.39}_{-0.32}$ & 97/92 \\
050922B & $0.06^{+0.01}_{-0.01}$  & $0.94^{+0.04}_{-0.05}$ & $0.06^{+0.03}_{-0.03}$  & 20/16   & $0.11^{+0.01}_{-0.01}$ & $1.30^{+0.06}_{-0.08}$  & $0.11^{+0.07}_{-0.06}$ & 21/16 \\
051109B & $1.18^{+0.29}_{-0.24}$  & $2.50^{+0.30}_{-0.28}$ & $4.71^{+0.71}_{-1.23}$  & 21/19   & $3.56^{+0.29}_{-0.26}$ & $6.53^{+0.21}_{-0.20}$  & $5.39^{+3.23}_{-3.63}$ & 25/19 \\
060109  & $0.65^{+0.02}_{-0.03}$  & $1.71^{+0.03}_{-0.03}$ & $0.25^{+0.25}_{-0.18}$  & 88/48   & $1.54^{+0.06}_{-0.06}$ & $3.77^{+0.07}_{-0.07}$  & $1.27^{+1.30}_{-0.90}$ & 90/48 \\
060428A & $0.13^{+0.02}_{-0.02}$  & $1.29^{+0.10}_{-0.10}$ & $0.88^{+0.06}_{-0.06}$  & 204/127 & $0.22^{+0.02}_{-0.02}$ & $2.05^{+0.14}_{-0.13}$  & $2.21^{+0.14}_{-0.13}$ & 206/127 \\
060510A & $0.32^{+0.02}_{-0.02}$  & $1.11^{+0.03}_{-0.03}$ & $1.01^{+0.03}_{-0.03}$  & 154/163 & $0.58^{+0.03}_{-0.03}$ & $1.75^{+0.06}_{-0.05}$  & $2.50^{+0.07}_{-0.07}$ & 153/163 \\
060714  & $0.28^{+0.06}_{-0.06}$  & $1.04^{+0.09}_{-0.08}$ & $1.25^{+0.10}_{-0.10}$  & 58/44   & $0.52^{+0.09}_{-0.08}$ & $1.63^{+0.14}_{-0.13}$  & $3.08^{+0.26}_{-0.26}$ & 57/44 \\
060729  & $0.09^{+0.01}_{-0.01}$  & $1.03^{+0.02}_{-0.02}$ & $0.19^{+0.01}_{-0.01}$  & 916/651 & $0.13^{+0.01}_{-0.01}$ & $1.32^{+0.03}_{-0.02}$  & $0.31^{+0.01}_{-0.01}$ & 915/651 \\
\enddata
\end{deluxetable}

\setlength{\tabcolsep}{5.5mm}
\begin{deluxetable}{cccccccccc}
\tablecolumns{9}
\tablewidth{0pc}
\tabletypesize{\scriptsize}
\tablecaption{--- continued \label{table-3}}
\tablenum{3}
\tablehead{
\colhead{GRB} & \colhead{$B_{p,15}^a$(G)} & \colhead{$P_{0}^a$(ms)}  & \colhead{$\epsilon_{-3}^a$}  &  \colhead{$\chi^2$/dof}  &  \colhead{$B_{p,15}^b$(G)}  & \colhead{$P_{0}^b$(ms)}  &  \colhead{$\epsilon_{-3}^b$}  &  \colhead{$\chi^2$/dof} }
\startdata
160630A & $0.36^{+0.12}_{-0.11}$  & $1.09^{+0.14}_{-0.13}$ & $1.92^{+0.29}_{-0.29}$  & 89/66   & $0.47^{+0.17}_{-0.14}$ & $1.34^{+0.20}_{-0.18}$  & $3.00^{+0.56}_{-0.57}$ & 88/66 \\
170113A & $0.14^{+0.01}_{-0.01}$  & $0.76^{+0.05}_{-0.04}$ & $0.91^{+0.07}_{-0.07}$  & 160/139 & $0.27^{+0.04}_{-0.05}$ & $1.04^{+0.07}_{-0.07}$  & $1.75^{+0.14}_{-0.14}$ & 157/139 \\
170202A & $0.17^{+0.10}_{-0.03}$  & $0.87^{+0.09}_{-0.10}$ & $0.68^{+0.06}_{-0.06}$  & 58/39   & $0.35^{+0.10}_{-0.09}$ & $1.16^{+0.14}_{-0.12}$  & $1.32^{+0.11}_{-0.13}$ & 56/39 \\
170607A & $0.07^{+0.02}_{-0.02}$  & $1.02^{+0.10}_{-0.10}$ & $0.62^{+0.11}_{-0.09}$  & 144/148 & $0.10^{+0.03}_{-0.03}$ & $1.30^{+0.16}_{-0.15}$  & $1.01^{+0.21}_{-0.20}$ & 143/148 \\
170711A & $0.15^{+0.09}_{-0.08}$  & $1.12^{+0.25}_{-0.23}$ & $1.10^{+0.46}_{-0.37}$  & 50/40   & $0.24^{+0.13}_{-0.11}$ & $1.66^{+0.40}_{-0.39}$  & $2.44^{+1.15}_{-0.92}$ & 49/40 \\
170906A & $0.83^{+0.03}_{-0.03}$  & $0.90^{+0.01}_{-0.01}$ & $0.15^{+0.17}_{-0.10}$  & 229/199 & $1.85^{+0.04}_{-0.04}$ & $1.34^{+0.01}_{-0.01}$  & $0.33^{+0.35}_{-0.22}$ & 242/199 \\
171102B & $0.39^{+0.05}_{-0.09}$  & $0.97^{+0.04}_{-0.07}$ & $0.36^{+0.18}_{-0.23}$  & 26/33   & $0.87^{+0.07}_{-0.12}$ & $1.63^{+0.06}_{-0.12}$  & $1.04^{+0.53}_{-0.67}$ & 26/33 \\
180314B & $0.51^{+0.05}_{-0.09}$  & $0.93^{+0.02}_{-0.04}$ & $0.39^{+0.23}_{-0.25}$  & 146/62  & $1.57^{+0.08}_{-0.14}$ & $2.01^{+0.05}_{-0.10}$  & $1.83^{+1.25}_{-1.20}$ & 147/62 \\
\enddata
\tablenotetext{a}{ The X-ray radiation efficiency of the magnetar dipole radiation as $\eta = 0.1$.}
\tablenotetext{b}{ The X-ray radiation efficiency of the magnetar dipole radiation as $\eta =0.5$.}
\end{deluxetable}

\clearpage
\setlength{\tabcolsep}{5.5mm}
\begin{deluxetable}{cccccccccc}
\tablecolumns{9}
\tablewidth{0pc}
\tabletypesize{\scriptsize}
\tablecaption{The derived magnetar physical parameters for our LGRB sample in the ENG EoS with $M_{b}=2.0~M_{\odot}$. \label{table-4}}
\tablenum{4}
\tablehead{
\colhead{GRB} & \colhead{$B_{p,15}^a$(G)} & \colhead{$P_{0}^a$(ms)}  & \colhead{$\epsilon_{-3}^a$}  &  \colhead{$\chi^2$/dof}  &  \colhead{$B_{p,15}^b$(G)}  & \colhead{$P_{0}^b$(ms)}  &  \colhead{$\epsilon_{-3}^b$}  &  \colhead{$\chi^2$/dof} }
\startdata
050822  & $0.05^{+0.02}_{-0.02}$  & $0.92^{+0.10}_{-0.09}$ & $0.65^{+0.11}_{-0.10}$  & 98/92   & $0.10^{+0.02}_{-0.03}$ & $1.30^{+0.10}_{-0.16}$  & $1.32^{+0.21}_{-0.25}$ & 97/92 \\
050922B & $0.03^{+0.01}_{-0.01}$  & $0.82^{+0.04}_{-0.05}$ & $0.05^{+0.02}_{-0.03}$  & 20/16   & $0.09^{+0.01}_{-0.01}$ & $1.14^{+0.05}_{-0.06}$  & $0.09^{+0.06}_{-0.05}$ & 21/16 \\
051109B & $0.55^{+0.13}_{-0.10}$  & $2.49^{+0.31}_{-0.28}$ & $4.76^{+0.67}_{-1.30}$  & 21/19   & $2.91^{+0.24}_{-0.22}$ & $5.75^{+0.19}_{-0.18}$  & $5.44^{+3.24}_{-3.62}$ & 25/19 \\
060109  & $0.31^{+0.01}_{-0.01}$  & $1.69^{+0.03}_{-0.03}$ & $0.25^{+0.27}_{-0.17}$  & 90/48   & $0.67^{+0.02}_{-0.02}$ & $3.65^{+0.02}_{-0.05}$  & $3.05^{+1.11}_{-1.44}$ & 93/48 \\
060428A & $0.09^{+0.01}_{-0.02}$  & $1.09^{+0.09}_{-0.10}$ & $0.75^{+0.05}_{-0.06}$  & 205/127 & $0.11^{+0.01}_{-0.01}$ & $2.10^{+0.13}_{-0.14}$  & $2.25^{+0.13}_{-0.14}$ & 207/127 \\
060510A & $0.24^{+0.03}_{-0.03}$  & $0.98^{+0.03}_{-0.03}$ & $0.87^{+0.03}_{-0.03}$  & 154/163 & $0.46^{+0.01}_{-0.01}$ & $1.46^{+0.01}_{-0.02}$  & $2.22^{+0.05}_{-0.05}$ & 157/163 \\
060714  & $0.20^{+0.07}_{-0.07}$  & $0.93^{+0.07}_{-0.07}$ & $1.11^{+0.09}_{-0.08}$  & 58/44   & $0.42^{+0.04}_{-0.07}$ & $1.37^{+0.07}_{-0.09}$  & $2.63^{+0.18}_{-0.19}$ & 56/44 \\
060729  & $0.06^{+0.01}_{-0.01}$  & $0.91^{+0.02}_{-0.02}$ & $0.16^{+0.01}_{-0.01}$  & 905/651 & $0.10^{+0.01}_{-0.01}$ & $1.16^{+0.02}_{-0.02}$  & $0.27^{+0.01}_{-0.01}$ & 908/651 \\
\enddata
\end{deluxetable}

\setlength{\tabcolsep}{5.5mm}
\begin{deluxetable}{cccccccccc}
\tablecolumns{9}
\tablewidth{0pc}
\tabletypesize{\scriptsize}
\tablecaption{--- continued \label{table-4}}
\tablenum{4}
\tablehead{
\colhead{GRB} & \colhead{$B_{p,15}^a$(G)} & \colhead{$P_{0}^a$(ms)}  & \colhead{$\epsilon_{-3}^a$}  &  \colhead{$\chi^2$/dof}  &  \colhead{$B_{p,15}^b$(G)}  & \colhead{$P_{0}^b$(ms)}  &  \colhead{$\epsilon_{-3}^b$}  &  \colhead{$\chi^2$/dof} }
\startdata
160630A & $0.29^{+0.10}_{-0.11}$  & $0.99^{+0.12}_{-0.11}$ & $1.77^{+0.22}_{-0.24}$  & 89/66   & $0.37^{+0.16}_{-0.09}$ & $1.18^{+0.15}_{-0.14}$  & $2.69^{+0.41}_{-0.42}$ & 87/66 \\
170113A & $0.10^{+0.01}_{-0.01}$  & $0.67^{+0.04}_{-0.04}$ & $0.80^{+0.06}_{-0.06}$  & 161/139 & $0.19^{+0.05}_{-0.05}$ & $0.93^{+0.06}_{-0.05}$  & $1.57^{+0.12}_{-0.11}$ & 158/139 \\
170202A & $0.10^{+0.02}_{-0.02}$  & $0.72^{+0.08}_{-0.08}$ & $0.57^{+0.05}_{-0.07}$  & 58/39   & $0.28^{+0.07}_{-0.09}$ & $1.03^{+0.11}_{-0.11}$  & $1.16^{+0.09}_{-0.11}$ & 57/39 \\
170607A & $0.05^{+0.02}_{-0.02}$  & $0.91^{+0.09}_{-0.08}$ & $0.57^{+0.09}_{-0.08}$  & 144/148 & $0.07^{+0.02}_{-0.02}$ & $1.11^{+0.14}_{-0.12}$  & $0.86^{+0.18}_{-0.15}$ & 144/148 \\
170711A & $0.09^{+0.07}_{-0.05}$  & $0.94^{+0.20}_{-0.20}$ & $0.88^{+0.37}_{-0.31}$  & 49/40   & $0.14^{+0.06}_{-0.03}$ & $1.83^{+0.35}_{-0.24}$  & $2.91^{+0.97}_{-0.64}$ & 51/40 \\
170906A & $0.49^{+0.01}_{-0.01}$  & $0.80^{+0.01}_{-0.01}$ & $0.14^{+0.14}_{-0.09}$  & 222/199 & $1.42^{+0.03}_{-0.03}$ & $1.17^{+0.01}_{-0.01}$  & $0.29^{+0.31}_{-0.20}$ & 249/199 \\
171102B & $0.20^{+0.07}_{-0.03}$  & $0.84^{+0.04}_{-0.08}$ & $0.36^{+0.13}_{-0.21}$  & 27/33   & $0.74^{+0.06}_{-0.10}$ & $1.41^{+0.05}_{-0.09}$  & $0.98^{+0.39}_{-0.56}$ & 26/33 \\
180314B & $0.28^{+0.02}_{-0.02}$  & $0.83^{+0.02}_{-0.03}$ & $0.29^{+0.21}_{-0.19}$  & 143/62  & $0.76^{+0.04}_{-0.07}$ & $1.99^{+0.06}_{-0.11}$  & $2.03^{+1.16}_{-1.37}$ & 147/62 \\
\enddata
\tablenotetext{a}{ The X-ray radiation efficiency of the magnetar dipole radiation as $\eta = 0.1$.}
\tablenotetext{b}{ The X-ray radiation efficiency of the magnetar dipole radiation as $\eta = 0.5$.}
\end{deluxetable}

\clearpage
\setlength{\tabcolsep}{5.5mm}
\begin{deluxetable}{cccccccccc}
\tablecolumns{9}
\tablewidth{0pc}
\tabletypesize{\scriptsize}
\tablecaption{The derived magnetar physical parameters for our LGRB sample in the ENG EoS with $M_{b}=2.5~M_{\odot}$. \label{table-5}}
\tablenum{5}
\tablehead{
\colhead{GRB} & \colhead{$B_{p,15}^a$(G)} & \colhead{$P_{0}^a$(ms)}  & \colhead{$\epsilon_{-3}^a$}  &  \colhead{$\chi^2$/dof}  &  \colhead{$B_{p,15}^b$(G)}  & \colhead{$P_{0}^b$(ms)}  &  \colhead{$\epsilon_{-3}^b$}  &  \colhead{$\chi^2$/dof} }
\startdata
050822  & $0.08^{+0.03}_{-0.02}$  & $1.00^{+0.12}_{-0.11}$ & $0.68^{+0.13}_{-0.12}$  & 97/92   & $0.14^{+0.04}_{-0.04}$ & $1.51^{+0.22}_{-0.20}$  & $1.57^{+0.39}_{-0.34}$ & 96/92 \\
050922B & $0.08^{+0.01}_{-0.01}$  & $0.93^{+0.04}_{-0.05}$ & $0.05^{+0.03}_{-0.03}$  & 20/16   & $0.13^{+0.01}_{-0.01}$ & $1.30^{+0.06}_{-0.07}$  & $0.10^{+0.07}_{-0.07}$ & 21/16 \\
051109B & $1.34^{+0.31}_{-0.29}$  & $2.51^{+0.29}_{-0.30}$ & $4.71^{+0.72}_{-1.22}$  & 21/19   & $4.01^{+0.32}_{-0.29}$ & $6.55^{+0.21}_{-0.20}$  & $5.28^{+3.35}_{-3.55}$ & 25/19 \\
060109  & $0.72^{+0.03}_{-0.03}$  & $1.70^{+0.03}_{-0.03}$ & $0.26^{+0.26}_{-0.18}$  & 89/48   & $1.73^{+0.06}_{-0.07}$ & $3.79^{+0.07}_{-0.07}$  & $1.25^{+1.29}_{-0.84}$ & 90/48 \\
060428A & $0.15^{+0.02}_{-0.02}$  & $1.29^{+0.10}_{-0.10}$ & $0.88^{+0.06}_{-0.06}$  & 206/127 & $0.26^{+0.03}_{-0.02}$ & $2.12^{+0.14}_{-0.14}$  & $2.26^{+0.14}_{-0.14}$ & 207/127 \\
060510A & $0.39^{+0.02}_{-0.02}$  & $1.11^{+0.04}_{-0.03}$ & $1.00^{+0.03}_{-0.03}$  & 153/163 & $0.65^{+0.03}_{-0.03}$ & $1.75^{+0.05}_{-0.05}$  & $2.51^{+0.07}_{-0.07}$ & 153/163 \\
060714  & $0.34^{+0.06}_{-0.06}$  & $1.04^{+0.08}_{-0.08}$ & $1.25^{+0.10}_{-0.10}$  & 57/44   & $0.58^{+0.09}_{-0.08}$ & $1.63^{+0.14}_{-0.13}$  & $3.08^{+0.28}_{-0.25}$ & 56/44 \\
060729  & $0.10^{+0.01}_{-0.01}$  & $1.03^{+0.02}_{-0.02}$ & $0.18^{+0.01}_{-0.01}$  & 907/651 & $0.15^{+0.01}_{-0.01}$ & $1.32^{+0.03}_{-0.02}$  & $0.31^{+0.01}_{-0.01}$ & 912/651 \\
\enddata
\end{deluxetable}

\setlength{\tabcolsep}{5.5mm}
\begin{deluxetable}{cccccccccc}
\tablecolumns{9}
\tablewidth{0pc}
\tabletypesize{\scriptsize}
\tablecaption{--- continued \label{table-5}}
\tablenum{5}
\tablehead{
\colhead{GRB} & \colhead{$B_{p,15}^a$(G)} & \colhead{$P_{0}^a$(ms)}  & \colhead{$\epsilon_{-3}^a$}  &  \colhead{$\chi^2$/dof}  &  \colhead{$B_{p,15}^b$(G)}  & \colhead{$P_{0}^b$(ms)}  &  \colhead{$\epsilon_{-3}^b$}  &  \colhead{$\chi^2$/dof} }
\startdata
160630A & $0.42^{+0.13}_{-0.13}$  & $1.08^{+0.14}_{-0.14}$ & $1.90^{+0.30}_{-0.32}$  & 88/66   & $0.53^{+0.17}_{-0.14}$ & $1.33^{+0.19}_{-0.18}$  & $3.00^{+0.56}_{-0.54}$ & 87/66 \\
170113A & $0.18^{+0.02}_{-0.05}$  & $0.75^{+0.05}_{-0.05}$ & $0.90^{+0.07}_{-0.07}$  & 159/139 & $0.33^{+0.05}_{-0.05}$ & $1.04^{+0.07}_{-0.06}$  & $1.75^{+0.15}_{-0.14}$ & 157/139 \\
170202A & $0.24^{+0.09}_{-0.06}$  & $0.85^{+0.09}_{-0.08}$ & $0.67^{+0.06}_{-0.06}$  & 57/39   & $0.41^{+0.11}_{-0.10}$ & $1.16^{+0.14}_{-0.14}$  & $1.31^{+0.12}_{-0.14}$ & 56/39 \\
170607A & $0.08^{+0.02}_{-0.02}$  & $1.00^{+0.11}_{-0.11}$ & $0.60^{+0.11}_{-0.10}$  & 144/148 & $0.11^{+0.03}_{-0.03}$ & $1.24^{+0.16}_{-0.15}$  & $0.95^{+0.21}_{-0.19}$ & 143/148 \\
170711A & $0.16^{+0.09}_{-0.08}$  & $1.07^{+0.25}_{-0.22}$ & $1.02^{+0.47}_{-0.35}$  & 49/40   & $0.28^{+0.17}_{-0.12}$ & $1.71^{+0.45}_{-0.41}$  & $2.57^{+1.20}_{-0.99}$ & 51/40 \\
170906A & $1.14^{+0.03}_{-0.03}$  & $0.89^{+0.01}_{-0.01}$ & $0.15^{+0.15}_{-0.10}$  & 233/199 & $2.11^{+0.04}_{-0.04}$ & $1.33^{+0.01}_{-0.01}$  & $0.33^{+0.37}_{-0.22}$ & 248/199 \\
171102B & $0.47^{+0.05}_{-0.09}$  & $0.95^{+0.04}_{-0.07}$ & $0.37^{+0.17}_{-0.24}$  & 26/33   & $0.97^{+0.07}_{-0.11}$ & $1.62^{+0.06}_{-0.10}$  & $0.99^{+0.54}_{-0.65}$ & 26/33 \\
180314B & $0.66^{+0.04}_{-0.08}$  & $0.92^{+0.02}_{-0.04}$ & $0.35^{+0.25}_{-0.24}$  & 145/62  & $1.75^{+0.09}_{-0.16}$ & $2.01^{+0.05}_{-0.11}$  & $1.89^{+1.27}_{-1.25}$ & 147/62 \\
\enddata
\tablenotetext{a}{ The X-ray radiation efficiency of the magnetar dipole radiation as $\eta = 0.1$.}
\tablenotetext{b}{ The X-ray radiation efficiency of the magnetar dipole radiation as $\eta = 0.5$.}
\end{deluxetable}

\clearpage
\setlength{\tabcolsep}{5.5mm}
\begin{deluxetable}{cccccccccc}
\tablecolumns{9}
\tablewidth{0pc}
\tabletypesize{\scriptsize}
\tablecaption{The derived magnetar physical parameters for our LGRB sample in the SLy EoS with $M_{b}=2.0~M_{\odot}$. \label{table-6}}
\tablenum{6}
\tablehead{
\colhead{GRB} & \colhead{$B_{p,15}^a$(G)} & \colhead{$P_{0}^a$(ms)}  & \colhead{$\epsilon_{-3}^a$}  &  \colhead{$\chi^2$/dof}  &  \colhead{$B_{p,15}^b$(G)}  & \colhead{$P_{0}^b$(ms)}  &  \colhead{$\epsilon_{-3}^b$}  &  \colhead{$\chi^2$/dof} }
\startdata
050822  & $0.04^{+0.01}_{-0.01}$  & $0.87^{+0.09}_{-0.07}$ & $0.60^{+0.11}_{-0.09}$  & 99/92   & $0.09^{+0.03}_{-0.02}$ & $1.30^{+0.16}_{-0.15}$  & $1.38^{+0.29}_{-0.26}$ & 97/92 \\
050922B & $0.04^{+0.01}_{-0.01}$  & $0.80^{+0.03}_{-0.05}$ & $0.05^{+0.02}_{-0.03}$  & 20/16   & $0.08^{+0.01}_{-0.01}$ & $1.10^{+0.04}_{-0.06}$  & $0.09^{+0.06}_{-0.05}$ & 20/16 \\
051109B & $0.87^{+0.21}_{-0.17}$  & $2.07^{+0.26}_{-0.23}$ & $3.93^{+0.56}_{-1.08}$  & 20/19   & $2.73^{+0.22}_{-0.20}$ & $5.40^{+0.17}_{-0.17}$  & $5.49^{+3.23}_{-3.77}$ & 25/19 \\
060109  & $0.48^{+0.02}_{-0.02}$  & $1.42^{+0.03}_{-0.03}$ & $0.24^{+0.22}_{-0.16}$  & 88/48   & $1.16^{+0.05}_{-0.05}$ & $3.12^{+0.06}_{-0.06}$  & $1.12^{+1.05}_{-0.76}$ & 89/48 \\
060428A & $0.07^{+0.02}_{-0.02}$  & $1.00^{+0.09}_{-0.09}$ & $0.70^{+0.06}_{-0.05}$  & 199/127 & $0.17^{+0.02}_{-0.02}$ & $1.73^{+0.11}_{-0.11}$  & $1.85^{+0.11}_{-0.11}$ & 205/127 \\
060510A & $0.16^{+0.02}_{-0.01}$  & $0.92^{+0.03}_{-0.03}$ & $0.84^{+0.02}_{-0.02}$  & 155/163 & $0.42^{+0.03}_{-0.03}$ & $1.45^{+0.04}_{-0.04}$  & $2.08^{+0.06}_{-0.06}$ & 154/163 \\
060714  & $0.15^{+0.03}_{-0.02}$  & $0.88^{+0.07}_{-0.07}$ & $1.05^{+0.09}_{-0.08}$  & 59/44   & $0.37^{+0.07}_{-0.06}$ & $1.35^{+0.12}_{-0.11}$  & $2.57^{+0.21}_{-0.21}$ & 57/44 \\
060729  & $0.05^{+0.01}_{-0.01}$  & $0.87^{+0.02}_{-0.02}$ & $0.16^{+0.01}_{-0.01}$  & 918/651 & $0.08^{+0.01}_{-0.01}$ & $1.10^{+0.02}_{-0.02}$  & $0.26^{+0.01}_{-0.01}$ & 917/651 \\
\enddata
\end{deluxetable}

\setlength{\tabcolsep}{5.5mm}
\begin{deluxetable}{cccccccccc}
\tablecolumns{9}
\tablewidth{0pc}
\tabletypesize{\scriptsize}
\tablecaption{--- continued \label{table-6}}
\tablenum{6}
\tablehead{
\colhead{GRB} & \colhead{$B_{p,15}^a$(G)} & \colhead{$P_{0}^a$(ms)}  & \colhead{$\epsilon_{-3}^a$}  &  \colhead{$\chi^2$/dof}  &  \colhead{$B_{p,15}^b$(G)}  & \colhead{$P_{0}^b$(ms)}  &  \colhead{$\epsilon_{-3}^b$}  &  \colhead{$\chi^2$/dof} }
\startdata
160630A & $0.21^{+0.12}_{-0.05}$  & $0.95^{+0.12}_{-0.11}$ & $1.68^{+0.22}_{-0.25}$  & 91/66   & $0.34^{+0.12}_{-0.10}$ & $1.16^{+0.15}_{-0.13}$  & $2.63^{+0.42}_{-0.39}$ & 89/66 \\
170113A & $0.11^{+0.01}_{-0.01}$  & $0.64^{+0.04}_{-0.04}$ & $0.74^{+0.06}_{-0.06}$  & 162/139 & $0.15^{+0.02}_{-0.01}$ & $0.87^{+0.06}_{-0.05}$  & $1.48^{+0.12}_{-0.11}$ & 159/139 \\
170202A & $0.12^{+0.03}_{-0.02}$  & $0.68^{+0.08}_{-0.08}$ & $0.54^{+0.06}_{-0.07}$  & 57/39   & $0.23^{+0.09}_{-0.08}$ & $1.01^{+0.11}_{-0.09}$  & $1.12^{+0.09}_{-0.09}$ & 58/39 \\
170607A & $0.04^{+0.01}_{-0.01}$  & $0.89^{+0.09}_{-0.07}$ & $0.56^{+0.09}_{-0.08}$  & 144/148 & $0.07^{+0.02}_{-0.02}$ & $1.10^{+0.12}_{-0.11}$  & $0.87^{+0.16}_{-0.14}$ & 143/148 \\
170711A & $0.07^{+0.08}_{-0.02}$  & $0.91^{+0.21}_{-0.16}$ & $0.86^{+0.35}_{-0.28}$  & 48/40   & $0.18^{+0.11}_{-0.10}$ & $1.39^{+0.35}_{-0.35}$  & $2.05^{+0.97}_{-0.84}$ & 50/40 \\
170906A & $0.58^{+0.01}_{-0.01}$  & $0.78^{+0.01}_{-0.01}$ & $0.15^{+0.14}_{-0.10}$  & 211/199 & $1.22^{+0.03}_{-0.03}$ & $1.12^{+0.01}_{-0.01}$  & $0.28^{+0.31}_{-0.18}$ & 236/199 \\
171102B & $0.23^{+0.03}_{-0.03}$  & $0.79^{+0.06}_{-0.07}$ & $0.42^{+0.09}_{-0.20}$  & 27/33   & $0.63^{+0.06}_{-0.09}$ & $1.36^{+0.05}_{-0.10}$  & $0.87^{+0.45}_{-0.56}$ & 26/33 \\
180314B & $0.35^{+0.01}_{-0.02}$  & $0.81^{+0.02}_{-0.03}$ & $0.30^{+0.20}_{-0.20}$  & 146/62  & $1.19^{+0.07}_{-0.11}$ & $1.68^{+0.04}_{-0.09}$  & $1.52^{+1.06}_{-1.04}$ & 147/62 \\
\enddata
\tablenotetext{a}{ The X-ray radiation efficiency of the magnetar dipole radiation as $\eta = 0.1$.}
\tablenotetext{b}{ The X-ray radiation efficiency of the magnetar dipole radiation as $\eta = 0.5$.}
\end{deluxetable}

\clearpage
\setlength{\tabcolsep}{5.5mm}
\begin{deluxetable}{cccccccccc}
\tablecolumns{9}
\tablewidth{0pc}
\tabletypesize{\scriptsize}
\tablecaption{The derived magnetar physical parameters for our LGRB sample in the WFF2 EoS with $M_{b}=2.0~M_{\odot}$. \label{table-7}}
\tablenum{7}
\tablehead{
\colhead{GRB} & \colhead{$B_{p,15}^a$(G)} & \colhead{$P_{0}^a$(ms)}  & \colhead{$\epsilon_{-3}^a$}  &  \colhead{$\chi^2$/dof}  &  \colhead{$B_{p,15}^b$(G)}  & \colhead{$P_{0}^b$(ms)}  &  \colhead{$\epsilon_{-3}^b$}  &  \colhead{$\chi^2$/dof} }
\startdata
050822  & $0.04^{+0.03}_{-0.01}$  & $0.86^{+0.10}_{-0.09}$ & $0.60^{+0.11}_{-0.10}$  & 98/92   & $0.10^{+0.03}_{-0.03}$ & $1.27^{+0.17}_{-0.15}$  & $1.33^{+0.31}_{-0.26}$ & 97/92 \\
050922B & $0.04^{+0.01}_{-0.01}$  & $0.78^{+0.03}_{-0.04}$ & $0.05^{+0.02}_{-0.03}$  & 20/16   & $0.08^{+0.01}_{-0.01}$ & $1.07^{+0.04}_{-0.06}$  & $0.09^{+0.05}_{-0.05}$ & 21/16 \\
051109B & $0.93^{+0.23}_{-0.18}$  & $2.05^{+0.24}_{-0.24}$ & $3.79^{+0.59}_{-1.14}$  & 21/19   & $2.88^{+0.23}_{-0.22}$ & $5.30^{+0.17}_{-0.16}$  & $5.66^{+3.13}_{-3.74}$ & 25/19 \\
060109  & $0.51^{+0.02}_{-0.02}$  & $1.39^{+0.02}_{-0.03}$ & $0.21^{+0.22}_{-0.14}$  & 88/48   & $1.24^{+0.05}_{-0.05}$ & $3.08^{+0.06}_{-0.06}$  & $1.07^{+1.06}_{-0.72}$ & 90/48 \\
060428A & $0.09^{+0.02}_{-0.02}$  & $1.01^{+0.08}_{-0.09}$ & $0.70^{+0.05}_{-0.05}$  & 202/127 & $0.18^{+0.02}_{-0.02}$ & $1.69^{+0.11}_{-0.11}$  & $1.81^{+0.11}_{-0.11}$ & 205/127 \\
060510A & $0.20^{+0.02}_{-0.02}$  & $0.90^{+0.03}_{-0.03}$ & $0.82^{+0.02}_{-0.02}$  & 155/163 & $0.46^{+0.03}_{-0.03}$ & $1.43^{+0.05}_{-0.04}$  & $2.03^{+0.06}_{-0.06}$ & 154/163 \\
060714  & $0.17^{+0.07}_{-0.04}$  & $0.86^{+0.07}_{-0.07}$ & $1.03^{+0.08}_{-0.08}$  & 58/44   & $0.41^{+0.07}_{-0.07}$ & $1.33^{+0.11}_{-0.11}$  & $2.52^{+0.20}_{-0.20}$ & 57/44 \\
060729  & $0.05^{+0.01}_{-0.01}$  & $0.85^{+0.02}_{-0.01}$ & $0.15^{+0.01}_{-0.01}$  & 916/651 & $0.09^{+0.01}_{-0.01}$ & $1.08^{+0.02}_{-0.02}$  & $0.25^{+0.01}_{-0.01}$ & 915/651 \\
\enddata
\end{deluxetable}

\setlength{\tabcolsep}{5.5mm}
\begin{deluxetable}{cccccccccc}
\tablecolumns{9}
\tablewidth{0pc}
\tabletypesize{\scriptsize}
\tablecaption{--- continued \label{table-7}}
\tablenum{7}
\tablehead{
\colhead{GRB} & \colhead{$B_{p,15}^a$(G)} & \colhead{$P_{0}^a$(ms)}  & \colhead{$\epsilon_{-3}^a$}  &  \colhead{$\chi^2$/dof}  &  \colhead{$B_{p,15}^b$(G)}  & \colhead{$P_{0}^b$(ms)}  &  \colhead{$\epsilon_{-3}^b$}  &  \colhead{$\chi^2$/dof} }
\startdata
160630A & $0.26^{+0.11}_{-0.11}$  & $0.92^{+0.11}_{-0.11}$ & $1.65^{+0.21}_{-0.23}$  & 90/66   & $0.39^{+0.13}_{-0.11}$ & $1.14^{+0.16}_{-0.14}$  & $2.59^{+0.43}_{-0.42}$ & 89/66 \\
170113A & $0.11^{+0.01}_{-0.01}$  & $0.63^{+0.04}_{-0.03}$ & $0.74^{+0.06}_{-0.06}$  & 162/139 & $0.17^{+0.05}_{-0.04}$ & $0.86^{+0.05}_{-0.06}$  & $1.46^{+0.12}_{-0.12}$ & 158/139 \\
170202A & $0.12^{+0.03}_{-0.02}$  & $0.68^{+0.08}_{-0.07}$ & $0.54^{+0.05}_{-0.06}$  & 59/39   & $0.25^{+0.08}_{-0.08}$ & $0.97^{+0.10}_{-0.10}$  & $1.09^{+0.09}_{-0.10}$ & 57/39 \\
170607A & $0.05^{+0.02}_{-0.02}$  & $0.88^{+0.08}_{-0.08}$ & $0.55^{+0.08}_{-0.08}$  & 144/148 & $0.08^{+0.02}_{-0.02}$ & $1.07^{+0.12}_{-0.11}$  & $0.83^{+0.17}_{-0.14}$ & 143/148 \\
170711A & $0.08^{+0.08}_{-0.04}$  & $0.89^{+0.21}_{-0.19}$ & $0.84^{+0.37}_{-0.31}$  & 49/40   & $0.19^{+0.11}_{-0.09}$ & $1.39^{+0.32}_{-0.32}$  & $2.08^{+0.89}_{-0.76}$ & 50/40 \\
170906A & $0.54^{+0.01}_{-0.01}$  & $0.76^{+0.01}_{-0.01}$ & $0.14^{+0.14}_{-0.09}$  & 217/199 & $1.35^{+0.03}_{-0.03}$ & $1.10^{+0.01}_{-0.01}$  & $0.27^{+0.31}_{-0.18}$ & 239/199 \\
171102B & $0.21^{+0.02}_{-0.03}$  & $0.79^{+0.04}_{-0.07}$ & $0.36^{+0.12}_{-0.21}$  & 27/33   & $0.68^{+0.05}_{-0.10}$ & $1.33^{+0.05}_{-0.09}$  & $0.86^{+0.41}_{-0.56}$ & 26/33 \\
180314B & $0.31^{+0.01}_{-0.02}$  & $0.78^{+0.02}_{-0.03}$ & $0.31^{+0.20}_{-0.20}$  & 145/62  & $1.24^{+0.06}_{-0.10}$ & $1.64^{+0.04}_{-0.08}$  & $1.51^{+0.99}_{-0.99}$ & 146/62 \\
\enddata
\tablenotetext{a}{ The X-ray radiation efficiency of the magnetar dipole radiation as $\eta = 0.1$.}
\tablenotetext{b}{ The X-ray radiation efficiency of the magnetar dipole radiation as $\eta = 0.5$.}
\end{deluxetable}

\clearpage
\setlength{\tabcolsep}{5.5mm}
\begin{deluxetable}{cccccccccc}
\tablecolumns{9}
\tablewidth{0pc}
\tabletypesize{\scriptsize}
\tablecaption{The derived magnetar physical parameters for our LGRB sample in the WFF2 EoS with $M_{b}=2.5~M_{\odot}$. \label{table-8}}
\tablenum{8}
\tablehead{
\colhead{GRB} & \colhead{$B_{p,15}^a$(G)} & \colhead{$P_{0}^a$(ms)}  & \colhead{$\epsilon_{-3}^a$}  &  \colhead{$\chi^2$/dof}  &  \colhead{$B_{p,15}^b$(G)}  & \colhead{$P_{0}^b$(ms)}  &  \colhead{$\epsilon_{-3}^b$}  &  \colhead{$\chi^2$/dof} }
\startdata
050822  & $0.08^{+0.03}_{-0.02}$  & $0.92^{+0.11}_{-0.10}$ & $0.62^{+0.13}_{-0.11}$  & 98/92   & $0.15^{+0.04}_{-0.04}$ & $1.40^{+0.19}_{-0.19}$  & $1.47^{+0.35}_{-0.33}$ & 96/92 \\
050922B & $0.08^{+0.01}_{-0.01}$  & $0.85^{+0.03}_{-0.05}$ & $0.05^{+0.03}_{-0.03}$  & 20/16   & $0.14^{+0.01}_{-0.02}$ & $1.19^{+0.05}_{-0.07}$  & $0.09^{+0.06}_{-0.06}$ & 21/16 \\
051109B & $1.35^{+0.32}_{-0.25}$  & $2.27^{+0.26}_{-0.23}$ & $4.31^{+0.65}_{-1.04}$  & 21/19   & $4.13^{+0.34}_{-0.30}$ & $5.93^{+0.20}_{-0.19}$  & $5.46^{+3.26}_{-3.74}$ & 25/19 \\
060109  & $0.76^{+0.03}_{-0.03}$  & $1.55^{+0.03}_{-0.03}$ & $0.24^{+0.22}_{-0.16}$  & 87/48   & $1.79^{+0.07}_{-0.07}$ & $3.43^{+0.06}_{-0.07}$  & $1.20^{+1.12}_{-0.82}$ & 90/48 \\
060428A & $0.15^{+0.02}_{-0.02}$  & $1.18^{+0.08}_{-0.08}$ & $0.80^{+0.05}_{-0.05}$  & 205/127 & $0.26^{+0.03}_{-0.02}$ & $1.88^{+0.13}_{-0.13}$  & $2.02^{+0.13}_{-0.13}$ & 206/127 \\
060510A & $0.38^{+0.02}_{-0.02}$  & $1.01^{+0.03}_{-0.03}$ & $0.92^{+0.02}_{-0.03}$  & 154/163 & $0.68^{+0.04}_{-0.04}$ & $1.59^{+0.05}_{-0.05}$  & $2.27^{+0.07}_{-0.07}$ & 153/163 \\
060714  & $0.33^{+0.07}_{-0.06}$  & $0.95^{+0.08}_{-0.07}$ & $1.13^{+0.09}_{-0.09}$  & 57/44   & $0.61^{+0.09}_{-0.09}$ & $1.47^{+0.13}_{-0.11}$  & $2.80^{+0.24}_{-0.24}$ & 56/44 \\
060729  & $0.10^{+0.01}_{-0.01}$  & $0.94^{+0.02}_{-0.02}$ & $0.17^{+0.01}_{-0.01}$  & 914/651 & $0.15^{+0.01}_{-0.01}$ & $1.20^{+0.02}_{-0.02}$  & $0.28^{+0.01}_{-0.01}$ & 914/651 \\
\enddata
\end{deluxetable}

\setlength{\tabcolsep}{5.5mm}
\begin{deluxetable}{cccccccccc}
\tablecolumns{9}
\tablewidth{0pc}
\tabletypesize{\scriptsize}
\tablecaption{--- continued \label{table-8}}
\tablenum{8}
\tablehead{
\colhead{GRB} & \colhead{$B_{p,15}^a$(G)} & \colhead{$P_{0}^a$(ms)}  & \colhead{$\epsilon_{-3}^a$}  &  \colhead{$\chi^2$/dof}  &  \colhead{$B_{p,15}^b$(G)}  & \colhead{$P_{0}^b$(ms)}  &  \colhead{$\epsilon_{-3}^b$}  &  \colhead{$\chi^2$/dof} }
\startdata
160630A & $0.41^{+0.14}_{-0.12}$  & $0.98^{+0.13}_{-0.12}$ & $1.72^{+0.27}_{-0.28}$  & 89/66   & $0.53^{+0.19}_{-0.15}$ & $1.20^{+0.19}_{-0.16}$  & $2.70^{+0.51}_{-0.52}$ & 87/66 \\
170113A & $0.14^{+0.03}_{-0.02}$  & $0.68^{+0.04}_{-0.04}$ & $0.82^{+0.06}_{-0.06}$  & 159/139 & $0.33^{+0.05}_{-0.05}$ & $0.95^{+0.06}_{-0.06}$  & $1.58^{+0.14}_{-0.13}$ & 157/139 \\
170202A & $0.24^{+0.09}_{-0.09}$  & $0.79^{+0.08}_{-0.08}$ & $0.61^{+0.05}_{-0.05}$  & 58/39   & $0.41^{+0.12}_{-0.10}$ & $1.05^{+0.12}_{-0.12}$  & $1.19^{+0.11}_{-0.12}$ & 56/39 \\
170607A & $0.08^{+0.02}_{-0.02}$  & $0.92^{+0.10}_{-0.09}$ & $0.56^{+0.11}_{-0.09}$  & 144/148 & $0.12^{+0.03}_{-0.03}$ & $1.17^{+0.14}_{-0.14}$  & $0.90^{+0.19}_{-0.18}$ & 143/148 \\
170711A & $0.17^{+0.11}_{-0.08}$  & $1.02^{+0.23}_{-0.20}$ & $1.01^{+0.42}_{-0.34}$  & 50/40   & $0.27^{+0.14}_{-0.11}$ & $1.50^{+0.37}_{-0.32}$  & $2.18^{+0.96}_{-0.78}$ & 49/40 \\
170906A & $1.07^{+0.03}_{-0.03}$  & $0.81^{+0.01}_{-0.01}$ & $0.14^{+0.14}_{-0.09}$  & 233/199 & $2.17^{+0.05}_{-0.04}$ & $1.21^{+0.01}_{-0.01}$  & $0.32^{+0.33}_{-0.21}$ & 243/199 \\
171102B & $0.48^{+0.05}_{-0.10}$  & $0.87^{+0.03}_{-0.07}$ & $0.33^{+0.16}_{-0.21}$  & 26/33   & $1.01^{+0.08}_{-0.12}$ & $1.48^{+0.06}_{-0.10}$  & $0.96^{+0.46}_{-0.61}$ & 26/33 \\
180314B & $0.64^{+0.05}_{-0.10}$  & $0.84^{+0.02}_{-0.04}$ & $0.36^{+0.22}_{-0.22}$  & 146/62  & $1.82^{+0.09}_{-0.17}$ & $1.83^{+0.05}_{-0.11}$  & $1.68^{+1.20}_{-1.13}$ & 147/62 \\
\enddata
\tablenotetext{a}{ The X-ray radiation efficiency of the magnetar dipole radiation as $\eta = 0.1$.}
\tablenotetext{b}{ The X-ray radiation efficiency of the magnetar dipole radiation as $\eta = 0.5$.}
\end{deluxetable}

\clearpage
\setlength{\tabcolsep}{5.5mm}
\begin{deluxetable}{cccccccccc}
\tablecolumns{9}
\tablewidth{0pc}
\tabletypesize{\scriptsize}
\tablecaption{The derived magnetar physical parameters for our LGRB sample in the constant $R=10~{\rm km},~I=1.5\times10^{45}~{\rm g~cm^2}$. \label{table-9}}
\tablenum{9}
\tablehead{
\colhead{GRB} & \colhead{$B_{p,15}^a$(G)} & \colhead{$P_{0}^a$(ms)}  & \colhead{$\epsilon_{-3}^a$}  &  \colhead{$\chi^2$/dof}  &  \colhead{$B_{p,15}^b$(G)}  & \colhead{$P_{0}^b$(ms)}  &  \colhead{$\epsilon_{-3}^b$}  &  \colhead{$\chi^2$/dof} }
\startdata
050822  & $0.08^{+0.02}_{-0.02}$  & $0.72^{+0.10}_{-0.10}$ & $0.47^{+0.12}_{-0.10}$  & 96/92   & $0.12^{+0.04}_{-0.03}$ & 
$1.13^{+0.17}_{-0.16}$  & $1.17^{+0.29}_{-0.27}$ & 95/92 \\
050922B & $0.09^{+0.01}_{-0.01}$  & $0.68^{+0.03}_{-0.04}$ & $0.04^{+0.02}_{-0.02}$  & 21/16   & $0.12^{+0.01}_{-0.01}$ & 
$0.97^{+0.04}_{-0.05}$  & $0.08^{+0.05}_{-0.05}$ & 20/16 \\
051109B & $1.17^{+0.26}_{-0.22}$  & $1.88^{+0.23}_{-0.21}$ & $3.52^{+0.52}_{-1.04}$  & 21/19   & $3.34^{+0.29}_{-0.26}$ & 
$4.87^{+0.17}_{-0.16}$  & $5.73^{+3.13}_{-3.86}$ & 25/19 \\
060109  & $0.65^{+0.02}_{-0.03}$  & $1.27^{+0.02}_{-0.02}$ & $0.19^{+0.19}_{-0.13}$  & 90/48   & $1.46^{+0.06}_{-0.05}$ & $2.83^{+0.05}_{-0.05}$  & $0.94^{+0.96}_{-0.65}$ & 91/48 \\
060428A & $0.14^{+0.01}_{-0.01}$  & $1.00^{+0.06}_{-0.06}$ & $0.67^{+0.04}_{-0.04}$  & 207/127 & $0.23^{+0.02}_{-0.02}$ & 
$1.57^{+0.11}_{-0.10}$  & $1.68^{+0.10}_{-0.10}$ & 208/127 \\
060510A & $0.38^{+0.02}_{-0.02}$  & $0.83^{+0.03}_{-0.02}$ & $0.75^{+0.02}_{-0.02}$  & 153/163 & $0.59^{+0.03}_{-0.03}$ & $1.31^{+0.04}_{-0.04}$  & $1.87^{+0.06}_{-0.06}$ & 154/163 \\
060714  & $0.33^{+0.05}_{-0.04}$  & $0.77^{+0.07}_{-0.06}$ & $0.92^{+0.08}_{-0.08}$  & 57/44   & $0.52^{+0.09}_{-0.07}$ & 
$1.21^{+0.11}_{-0.10}$  & $2.30^{+0.20}_{-0.19}$ & 56/44 \\
060729  & $0.11^{+0.01}_{-0.01}$  & $0.77^{+0.02}_{-0.02}$ & $0.14^{+0.01}_{-0.01}$  & 910/651 & $0.14^{+0.01}_{-0.01}$ & 
$0.99^{+0.02}_{-0.02}$  & $0.23^{+0.01}_{-0.01}$ & 909/651 \\
\enddata
\end{deluxetable}

\setlength{\tabcolsep}{5.5mm}
\begin{deluxetable}{cccccccccc}
\tablecolumns{9}
\tablewidth{0pc}
\tabletypesize{\scriptsize}
\tablecaption{--- continued \label{table-9}}
\tablenum{9}
\tablehead{
\colhead{GRB} & \colhead{$B_{p,15}^a$(G)} & \colhead{$P_{0}^a$(ms)}  & \colhead{$\epsilon_{-3}^a$}  &  \colhead{$\chi^2$/dof}  &  \colhead{$B_{p,15}^b$(G)}  & \colhead{$P_{0}^b$(ms)}  &  \colhead{$\epsilon_{-3}^b$}  &  \colhead{$\chi^2$/dof} }
\startdata
160630A & $0.37^{+0.11}_{-0.10}$  & $0.75^{+0.11}_{-0.11}$ & $1.30^{+0.26}_{-0.27}$  & 87/66   & $0.47^{+0.16}_{-0.13}$ & 
$0.96^{+0.17}_{-0.15}$  & $2.14^{+0.49}_{-0.49}$ & 86/66 \\
170113A & $0.23^{+0.03}_{-0.02}$  & $0.55^{+0.04}_{-0.03}$ & $0.65^{+0.06}_{-0.05}$  & 157/139 & $0.33^{+0.04}_{-0.04}$ & 
$0.77^{+0.05}_{-0.05}$  & $1.29^{+0.11}_{-0.11}$ & 156/139 \\
170202A & $0.27^{+0.06}_{-0.05}$  & $0.60^{+0.07}_{-0.06}$ & $0.49^{+0.04}_{-0.05}$  & 56/39   & $0.37^{+0.09}_{-0.08}$ & 
$0.83^{+0.11}_{-0.11}$  & $0.95^{+0.10}_{-0.13}$ & 56/39 \\
170607A & $0.08^{+0.02}_{-0.02}$  & $0.71^{+0.10}_{-0.10}$ & $0.42^{+0.10}_{-0.10}$  & 143/148 & $0.10^{+0.03}_{-0.02}$ & 
$0.92^{+0.12}_{-0.13}$  & $0.69^{+0.17}_{-0.16}$ & 144/148 \\
170711A & $0.17^{+0.09}_{-0.06}$  & $0.83^{+0.21}_{-0.18}$ & $0.82^{+0.38}_{-0.29}$  & 50/40   & $0.25^{+0.13}_{-0.11}$ & 
$1.26^{+0.31}_{-0.33}$  & $1.86^{+0.83}_{-0.78}$ & 50/40 \\
170906A & $1.29^{+0.03}_{-0.03}$  & $0.65^{+0.01}_{-0.01}$ & $0.11^{+0.11}_{-0.07}$  & 249/199 & $1.97^{+0.04}_{-0.04}$ & 
$0.99^{+0.01}_{-0.01}$  & $0.25^{+0.27}_{-0.17}$ & 247/199 \\
171102B & $0.51^{+0.04}_{-0.06}$  & $0.70^{+0.03}_{-0.05}$ & $0.24^{+0.14}_{-0.15}$  & 26/33   & $0.87^{+0.07}_{-0.10}$ & 
$1.21^{+0.05}_{-0.08}$  & $0.76^{+0.39}_{-0.48}$ & 25/33 \\
180314B & $0.70^{+0.04}_{-0.07}$  & $0.67^{+0.02}_{-0.04}$ & $0.30^{+0.17}_{-0.19}$  & 147/62  & $1.57^{+0.08}_{-0.15}$ & 
$1.49^{+0.04}_{-0.08}$  & $1.49^{+0.91}_{-1.05}$ & 147/62 \\
\enddata
\tablenotetext{a}{ The X-ray radiation efficiency of the magnetar dipole radiation as $\eta = 0.1$.}
\tablenotetext{b}{ The X-ray radiation efficiency of the magnetar dipole radiation as $\eta = 0.5$.}
\end{deluxetable}

\setlength{\tabcolsep}{11mm}
\begin{deluxetable}{cccccccccc}
\tablewidth{0pt} 
\tabletypesize{\footnotesize}
\tablecaption{The K-S test $p$-value between magnetar physical parameters derived under constant $R/I$ and evolving $R/I$ scenarios. \label{table-10}}
\tablenum{10}
\tablehead{ \colhead{EoS}& \colhead{$p_{\rm KS}~(B_p$)} &
\colhead{$p_{\rm KS}~(P_0$)}& \colhead{$p_{\rm KS}~(\epsilon$)}}
\startdata
$\eta_{\rm X}=0.1$\\
\hline
AP3($M_{b}=2.0M_{\odot}$)  & $6.8\times10^{-8}$ & $2.9\times10^{-10}$ & $4.0\times10^{-2}$\\
ENG($M_{b}=2.0M_{\odot}$)  & $1.6\times10^{-7}$ & $4.9\times10^{-9}$ & $5.0\times10^{-2}$\\
SLy($M_{b}=2.0M_{\odot}$)  & $7.7\times10^{-7}$ & $7.2\times10^{-6}$ & $7.3\times10^{-2}$\\
WFF2($M_{b}=2.0M_{\odot}$) & $7.8\times10^{-7}$ & $5.5\times10^{-5}$ & $8.4\times10^{-2}$\\
AP3($M_{b}=2.5M_{\odot}$)  & $1.6\times10^{-2}$ & $4.5\times10^{-15}$ & $1.7\times10^{-2}$\\
ENG($M_{b}=2.5M_{\odot}$)  & $1.2\times10^{-1}$ & $1.5\times10^{-14}$ & $1.8\times10^{-2}$\\
WFF2($M_{b}=2.5M_{\odot}$) & $1.1\times10^{-1}$ & $1.1\times10^{-10}$ & $4.0\times10^{-2}$\\
\hline
$\eta_{\rm X}=0.5$\\
\hline
AP3($M_{b}=2.0M_{\odot}$)  & $1.8\times10^{-3}$ & $1.1\times10^{-3}$ & $3.1\times10^{-2}$\\
ENG($M_{b}=2.0M_{\odot}$)  & $3.5\times10^{-4}$ & $3.1\times10^{-3}$ & $1.7\times10^{-2}$\\
SLy($M_{b}=2.0M_{\odot}$)  & $5.1\times10^{-3}$ & $2.0\times10^{-3}$ & $6.1\times10^{-2}$\\
WFF2($M_{b}=2.0M_{\odot}$) & $1.3\times10^{-3}$ & $3.0\times10^{-3}$ & $7.3\times10^{-2}$\\
AP3($M_{b}=2.5M_{\odot}$)  & $1.1\times10^{-1}$ & $3.5\times10^{-6}$ & $9.1\times10^{-3}$\\
ENG($M_{b}=2.5M_{\odot}$)  & $9.2\times10^{-2}$ & $7.2\times10^{-6}$ & $1.3\times10^{-2}$\\
WFF2($M_{b}=2.5M_{\odot}$) & $8.4\times10^{-2}$ & $1.1\times10^{-4}$ & $3.1\times10^{-2}$\\
\enddata
\end{deluxetable}

\setlength{\tabcolsep}{11mm}
\begin{deluxetable}{cccccccccc}
\tablewidth{0pt} 
\tabletypesize{\footnotesize}
\tablecaption{The range of magnetar physical parameters for various scenarios. \label{table-11}}
\tablenum{11}
\tablehead{ \colhead{EoS}& \colhead{$B_p$ ($\times10^{15}$ G)}&
\colhead{$P_0$ (ms)}& \colhead{$\epsilon$ ($\times10^{-3}$)}}
\startdata
$\eta_{\rm X}=0.1$\\
\hline
AP3($M_{b}=2.0M_{\odot}$)  &($0.03\sim1.78$) &($0.52\sim3.67$) &($0.02\sim4.69$)\\
ENG($M_{b}=2.0M_{\odot}$)  &($0.03\sim2.11$) &($0.51\sim3.80$) &($0.02\sim4.76$)\\
SLy($M_{b}=2.0M_{\odot}$)  &($0.04\sim1.87$) &($0.51\sim3.46$) &($0.02\sim4.83$)\\
WFF2($M_{b}=2.0M_{\odot}$) &($0.04\sim2.00$) &($0.51\sim3.41$) &($0.02\sim4.43$)\\
AP3($M_{b}=2.5M_{\odot}$)  &($0.04\sim2.49$) &($0.55\sim4.21$) &($0.02\sim4.73$)\\
ENG($M_{b}=2.5M_{\odot}$)  &($0.04\sim2.78$) &($0.57\sim4.22$) &($0.02\sim5.02$)\\
WFF2($M_{b}=2.5M_{\odot}$) &($0.04\sim2.87$) &($0.53\sim3.82$) &($0.02\sim4.75$)\\
Constant $R$ and $I$       &($0.08\sim2.34$) &($0.39\sim3.14$) &($0.01\sim4.11$)\\
\hline
$\eta_{\rm X}=0.5$\\
\hline
AP3($M_{b}=2.0M_{\odot}$)  &($0.04\sim4.13$) &($0.65\sim8.37$) &($0.03\sim8.78$)\\
ENG($M_{b}=2.0M_{\odot}$)  &($0.04\sim4.66$) &($0.64\sim8.44$) &($0.03\sim8.64$)\\
SLy($M_{b}=2.0M_{\odot}$)  &($0.05\sim4.40$) &($0.61\sim7.93$) &($0.03\sim7.85$)\\
WFF2($M_{b}=2.0M_{\odot}$) &($0.04\sim4.66$) &($0.60\sim7.80$) &($0.02\sim7.66$)\\
AP3($M_{b}=2.5M_{\odot}$)  &($0.06\sim5.71$) &($0.73\sim9.56$) &($0.03\sim8.71$)\\
ENG($M_{b}=2.5M_{\odot}$)  &($0.08\sim6.43$) &($0.73\sim9.60$) &($0.03\sim8.67$)\\
WFF2($M_{b}=2.5M_{\odot}$) &($0.07\sim6.65$) &($0.67\sim8.72$) &($0.03\sim8.16$)\\
Constant $R$ and $I$       &($0.10\sim5.43$) &($0.53\sim7.18$) &($0.02\sim8.31$)\\
\enddata
\end{deluxetable}

\setlength{\tabcolsep}{11mm}
\begin{deluxetable}{cccccccccc}
\tablewidth{0pt} 
\tabletypesize{\footnotesize}
\tablecaption{The center values and standard deviations of the
Gaussian fits of various distributions. \label{table-12}}
\tablenum{12}
\tablehead{ \colhead{EoS}& \colhead{log $B_p$ (G)}&
\colhead{log $P_0$ (s)}& \colhead{log $\epsilon$}}
\startdata
$\eta_{\rm X}=0.1$\\
\hline
AP3($M_{b}=2.0M_{\odot}$)  &($14.22\pm0.38$) &($-3.01\pm0.14$) &($-3.33\pm0.48$)\\
ENG($M_{b}=2.0M_{\odot}$)  &($14.26\pm0.37$) &($-3.01\pm0.15$) &($-3.34\pm0.47$)\\
SLy($M_{b}=2.0M_{\odot}$)  &($14.24\pm0.37$) &($-3.03\pm0.14$) &($-3.35\pm0.47$)\\
WFF2($M_{b}=2.0M_{\odot}$) &($14.26\pm0.40$) &($-3.04\pm0.14$) &($-3.37\pm0.47$)\\
AP3($M_{b}=2.5M_{\odot}$)  &($14.43\pm0.38$) &($-2.95\pm0.14$) &($-3.29\pm0.48$)\\
ENG($M_{b}=2.5M_{\odot}$)  &($14.51\pm0.37$) &($-2.96\pm0.14$) &($-3.30\pm0.47$)\\
WFF2($M_{b}=2.5M_{\odot}$) &($14.50\pm0.38$) &($-3.00\pm0.14$) &($-3.33\pm0.48$)\\
Constant $R$ and $I$       &($14.53\pm0.34$) &($-3.09\pm0.15$) &($-3.43\pm0.49$)\\
\hline
$\eta_{\rm X}=0.5$\\
\hline
AP3($M_{b}=2.0M_{\odot}$)  &($14.56\pm0.42$) &($-2.79\pm0.21$) &($-2.93\pm0.51$)\\
ENG($M_{b}=2.0M_{\odot}$)  &($14.54\pm0.41$) &($-2.78\pm0.22$) &($-2.92\pm0.52$)\\
SLy($M_{b}=2.0M_{\odot}$)  &($14.58\pm0.42$) &($-2.82\pm0.21$) &($-2.96\pm0.51$)\\
WFF2($M_{b}=2.0M_{\odot}$) &($14.61\pm0.42$) &($-2.83\pm0.21$) &($-2.97\pm0.52$)\\
AP3($M_{b}=2.5M_{\odot}$)  &($14.73\pm0.40$) &($-2.74\pm0.21$) &($-2.89\pm0.51$)\\
ENG($M_{b}=2.5M_{\odot}$)  &($14.79\pm0.39$) &($-2.73\pm0.21$) &($-2.90\pm0.51$)\\
WFF2($M_{b}=2.5M_{\odot}$) &($14.80\pm0.40$) &($-2.78\pm0.20$) &($-2.93\pm0.51$)\\
Constant $R$ and $I$       &($14.75\pm0.38$) &($-2.87\pm0.22$) &($-3.02\pm0.52$)\\
\enddata
\end{deluxetable}

\setlength{\tabcolsep}{5mm}
\begin{deluxetable}{cccccccccc}
\tablewidth{0pt} 
\tabletypesize{\tiny}
\tablecaption{The best-fitting correlations among the magnetar physical parameters for various scenarios. \label{table-13}}
\tablenum{13}
\tablehead{ \colhead{EoS}& \colhead{log $\epsilon-P_0$}&
\colhead{log $\epsilon-B_p$}& \colhead{log $B_p-P_0$}}
\startdata
$\eta_{\rm X}=0.1$\\
\hline
AP3($M_{b}=2.0M_{\odot}$)  & $\log\epsilon=(1.55\pm0.57) + (1.69\pm0.26)\log P_0$     & $\log\epsilon=(-18.42\pm1.67) + (1.05\pm0.12)\log B_p$     & $\log B_p=(18.39\pm0.65) + (1.41\pm0.21)\log P_0$ \\
ENG($M_{b}=2.0M_{\odot}$)  & $\log\epsilon=(1.53\pm0.48) + (1.68\pm0.22)\log P_0$     & $\log\epsilon=(-17.46\pm1.82) + (0.98\pm0.13)\log B_p$     & $\log B_p=(18.20\pm0.61) + (1.34\pm0.20)\log P_0$ \\
SLy($M_{b}=2.0M_{\odot}$)  & $\log\epsilon=(1.50\pm0.59) + (1.66\pm0.26)\log P_0$     & $\log\epsilon=(-18.28\pm1.71) + (1.04\pm0.11)\log B_p$     & $\log B_p=(18.43\pm0.66) + (1.40\pm0.20)\log P_0$ \\
WFF2($M_{b}=2.0M_{\odot}$) & $\log\epsilon=(1.37\pm0.58) + (1.63\pm0.25)\log P_0$     & $\log\epsilon=(-17.42\pm1.66) + (0.98\pm0.13)\log B_p$     & $\log B_p=(18.37\pm0.64) + (1.37\pm0.21)\log P_0$ \\
AP3($M_{b}=2.5M_{\odot}$)  & $\log\epsilon=(1.35\pm0.55) + (1.64\pm0.25)\log P_0$     & $\log\epsilon=(-17.69\pm1.74) + (0.99\pm0.13)\log B_p$     & $\log B_p=(18.49\pm0.55) + (1.39\pm0.18)\log P_0$ \\
ENG($M_{b}=2.5M_{\odot}$)  & $\log\epsilon=(1.31\pm0.56) + (1.61\pm0.26)\log P_0$     & $\log\epsilon=(-17.98\pm1.89) + (1.01\pm0.15)\log B_p$     & $\log B_p=(18.51\pm0.53) + (1.37\pm0.18)\log P_0$ \\
WFF2($M_{b}=2.5M_{\odot}$) & $\log\epsilon=(1.26\pm0.58) + (1.60\pm0.26)\log P_0$     & $\log\epsilon=(-17.90\pm1.83) + (1.00\pm0.16)\log B_p$     & $\log B_p=(18.55\pm0.55) + (1.36\pm0.18)\log P_0$ \\
\enddata
\end{deluxetable}

\setlength{\tabcolsep}{4mm}
\begin{deluxetable}{cccccccc}
\tablewidth{0pt} 
\tabletypesize{\tiny}
\tablecaption{The best-fitting correlations between $E_{\rm jet, iso}$, $E_{\rm jet}$ and the magnetar physical parameters for various scenarios. \label{table-14}}
\tablenum{14}
\tablehead{ \colhead{EoS}& \colhead{log $P_0-E_{\rm jet,iso}$($P_0-E_{\rm jet}$)}& \colhead{log $B_p-E_{\rm jet,iso}$($B_p-E_{\rm jet}$)}& \colhead{log $\epsilon-E_{\rm jet,iso}$($\epsilon-E_{\rm jet}$)}}
\startdata
\multirow{1}{*}{$\eta_{\rm X}=0.1$} \\
\cline{1-4}
\multirow{2}{*}{AP3($M_{b}=2.0M_{\odot}$)}   
& $\log P_0=(9.66\pm0.94)-(0.24\pm0.02)\log E_{\rm jet,iso}$     & $\log B_p=(43.70\pm2.90)-(0.56\pm0.05)\log E_{\rm jet,iso}$    
& $\log \epsilon=(24.60\pm3.69)-(0.53\pm0.07)\log E_{\rm jet,iso}$  \\
& $\log P_0=(7.55\pm0.91)-(0.21\pm0.02)\log E_{\rm jet}$    
& $\log B_p=(40.79\pm2.71)-(0.53\pm0.05)\log E_{\rm jet}$      
& $\log \epsilon=(21.75\pm3.44)-(0.50\pm0.07)\log E_{\rm jet}$  \\
\cline{1-4} 
\multirow{2}{*}{ENG($M_{b}=2.0M_{\odot}$)}   
& $\log P_0=(11.24\pm1.02)-(0.27\pm0.02)\log E_{\rm jet,iso}$     & $\log B_p=(44.78\pm2.83)-(0.58\pm0.05)\log E_{\rm jet,iso}$    
& $\log \epsilon=(24.01\pm3.72)-(0.52\pm0.07)\log E_{\rm jet,iso}$ \\
& $\log P_0=(9.54\pm0.99)-(0.25\pm0.02)\log E_{\rm jet}$   
& $\log B_p=(41.30\pm2.65)-(0.54\pm0.05)\log E_{\rm jet}$     
& $\log \epsilon=(20.78\pm3.47)-(0.48\pm0.07)\log E_{\rm jet}$  \\
\cline{1-4}
\multirow{2}{*}{SLy($M_{b}=2.0M_{\odot}$)}   
& $\log P_0=(10.68\pm0.94)-(0.26\pm0.03)\log E_{\rm jet,iso}$     & $\log B_p=(43.76\pm2.88)-(0.56\pm0.05)\log E_{\rm jet,iso}$    
& $\log \epsilon=(26.20\pm3.64)-(0.56\pm0.07)\log E_{\rm jet,iso}$  \\
& $\log P_0=(8.50\pm0.91)-(0.23\pm0.02)\log E_{\rm jet}$   
& $\log B_p=(40.90\pm2.70)-(0.53\pm0.05)\log E_{\rm jet}$      
& $\log \epsilon=(22.79\pm3.39)-(0.52\pm0.06)\log E_{\rm jet}$ \\
\cline{1-4}
\multirow{2}{*}{WFF2($M_{b}=2.0M_{\odot}$)}   
& $\log P_0=(10.15\pm0.93)-(0.25\pm0.02)\log E_{\rm jet,iso}$     & $\log B_p=(44.30\pm2.90)-(0.57\pm0.06)\log E_{\rm jet,iso}$    
& $\log \epsilon=(25.65\pm3.68)-(0.55\pm0.06)\log E_{\rm jet, iso}$ \\
\enddata
\end{deluxetable}


\end{document}